\documentclass[twoside, openright, 11pt]{book} 
\usepackage[utf8]{inputenc} 
\usepackage[spanish,english]{babel}
\usepackage{graphicx}
\usepackage[intoc,english]{nomencl} 

\usepackage{geometry} 
\usepackage[all]{nowidow}
\makenomenclature 
\nomrefpage 
\usepackage[sort&compress, numbers]{natbib}

\usepackage{titlesec}
\newcommand*{\justifyheading}{\raggedleft}
\titleformat{\chapter}[display]
  {\normalfont\huge\bfseries\justifyheading}{\chaptertitlename\ \thechapter}
  {20pt}{\Huge}

\usepackage{fancyhdr} 
\usepackage{etoolbox}
\patchcmd{\chapter}{plain}{empty}{}{}
\patchcmd{\part}{plain}{empty}{}{}

\usepackage[margin=10pt,font=small,labelfont=bf]{caption} 
\usepackage{subcaption}

\usepackage{lipsum} 

\usepackage{array}
\newcolumntype{P}[1]{>{\centering\arraybackslash}p{#1}}
\usepackage{comment}
\usepackage{amsmath}
\usepackage{amsfonts}
\usepackage{bm}
\usepackage{multirow}
\usepackage[usenames,dvipsnames]{color}
\usepackage{xcolor}              
\usepackage[toc,page]{appendix}
\usepackage{ulem}
\usepackage{bm}
\usepackage{cases}

\usepackage[section]{placeins}

\usepackage{braket}
\usepackage{diagbox}
\usepackage{lscape}

\usepackage{stackrel}
\usepackage{wasysym} 
\newcommand{\leftrarrows}{\mathrel{\raise.75ex\hbox{\oalign{%
  $\scriptstyle\leftarrow$\cr
  \vrule width0pt height.5ex$\hfil\scriptstyle\relbar$\cr}}}}
\newcommand{\lrightarrows}{\mathrel{\raise.75ex\hbox{\oalign{%
  $\scriptstyle\relbar$\hfil\cr
  $\scriptstyle\vrule width0pt height.5ex\smash\rightarrow$\cr}}}}
\newcommand{\Rrelbar}{\mathrel{\raise.75ex\hbox{\oalign{%
  $\scriptstyle\relbar$\cr
  \vrule width0pt height.5ex$\scriptstyle\relbar$}}}}

\usepackage{amssymb}
\usepackage{tablefootnote}
\usepackage{booktabs}
\usepackage{schemata} 
\usepackage{tabularx}
\usepackage{dcolumn}
\newcolumntype{d}[1]{D{.}{.}{#1}}
\usepackage{url}
\usepackage{doi}
\usepackage[shortcuts]{extdash}
\usepackage{emptypage}
\usepackage{enumitem}

\usepackage{tikz}

\begin{document}
\frontmatter

\graphicspath{{0_portada/img/}}
  \thispagestyle{empty}
  \begin{center}
    {\large \vspace*{5mm} \includegraphics[width=0.9\textwidth]{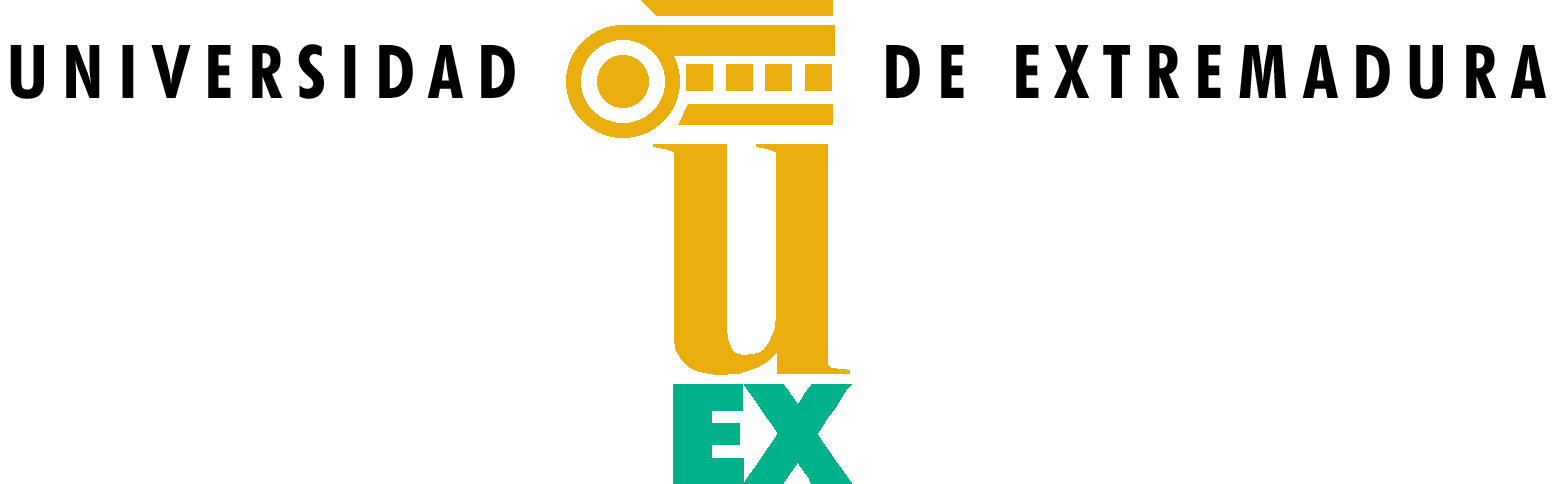} \par} \vspace*{10mm}
    {\large 
        { \bfseries{TESIS DOCTORAL} \par}
    	\vspace*{2ex}
    }
    \end{center}
    {\bfseries {EXPANSIÓN DE FRENTES: SIMULACIONES NUMÉRICAS EN MODELOS DISCRETOS Y ECUACIONES CONTINUAS} \par}
    \vspace*{2ex}
    {\bfseries{JESÚS MARÍA MARCOS MERINO} \par}
    \vspace*{2ex}
    {\bfseries{PROGRAMA DE DOCTORADO EN MODELIZACIÓN Y EXPERIMENTACIÓN EN CIENCIA Y TECNOLOGÍA} \par}
  \begin{center}\vspace*{3ex}
    {Con la conformidad del director y los codirectores \par}
    \vspace*{10ex}
    \begin{tabular}{m{3cm} m{0.1cm} m{3.4cm} m{0.1cm} m{3cm}}
    \centering
    Juan Jesús Ruiz Lorenzo & & \centering Juan José Meléndez Martínez & & \centering Rodolfo Cuerno Rejado   
    \end{tabular}
    \vspace*{1ex}
  \end{center}
  Esta tesis cuenta con la autorización del director y codirectores de la misma y de la Comisión Académica del programa. Dichas autorizaciones constan en el Servicio de la Escuela Internacional de Doctorado de la Universidad de Extremadura. 
  \begin{center}
      \bfseries{2025}
  \end{center}
\newpage
\mbox{}
\thispagestyle{empty}
\newpage
\thispagestyle{empty}

\begin{center}
\noindent\rule{\textwidth}{1pt}\\[0.6cm]
{\huge \bfseries Resumen }\\[0.2cm] 
\noindent\rule{\textwidth}{1pt} \\[1.5cm]
\end{center}

Los sistemas fuera del equilibrio, inherentemente complejos y difíciles de comprender, son comunes en diversas disciplinas, incluida la física, donde surgen en contextos como la dinámica de fluidos. En particular, los sistemas críticos fuera del equilibrio combinan esta complejidad con las leyes de escala y las clases de universalidad observadas en los fenómenos críticos, siendo la rugosidad cinética de superficies, el estudio de cómo una superficie plana se vuelve progresivamente más rugosa con el tiempo, un ejemplo destacado. Este comportamiento se manifiesta en una amplia variedad de contextos, incluyendo la corrosión de metales, la proliferación celular y, notablemente, el crecimiento de películas delgadas, que puede originarse como resultado de procesos de \textit{wetting}. En esta tesis, realizamos extensas simulaciones numéricas para estudiar las fluctuaciones críticas e identificar características universales de varias interfases rugosas, generadas mediante la simulación de modelos discretos de crecimiento de películas delgadas y la integración numérica directa de ecuaciones continuas. Para explorar el comportamiento universal de estas interfases, identificamos los exponentes críticos que caracterizan las fluctuaciones espacio-temporales del frente. Además, analizamos la dinámica de las películas delgadas en diferentes escenarios físicos para profundizar en la comprensión de su comportamiento en condiciones fuera del equilibrio, especialmente en el caso en que estas películas se forman por la acción de una fuerza externa, como las ondas acústicas de superficie.

\newpage
\mbox{}
\thispagestyle{empty}
\newpage
\thispagestyle{empty}
\begin{center}
\textsc{\LARGE University of Extremadura}\\[1.5cm] 

\textsc{\Large Doctoral Thesis}\\[0.5cm] 

\noindent\rule{\textwidth}{1pt}\\[0.6cm]
{\LARGE \bfseries Spreading fronts:   }\\[0.2cm]
{\LARGE \bfseries numerical simulations on discrete}\\[0.2cm]
{\LARGE \bfseries models and continuous equations}\\[0.4cm]
\noindent\rule{\textwidth}{1pt} \\[1.5cm]

\begin{minipage}{0.4\textwidth}
\begin{flushleft} \large
\emph{Author:}\\
{Jesús María MARCOS MERINO} \\
{ \ \\ \ \ } \\
{ \ \\ \  } \\
\end{flushleft}
\end{minipage} 
\begin{minipage}{0.4\textwidth}
\begin{flushright} \large
\emph{Supervisors:} \\
{Juan Jesús RUIZ LORENZO} \\ 
{Juan José MELÉNDEZ MARTÍNEZ} \\ 
{Rodolfo CUERNO REJADO} \\ 
\end{flushright}
\end{minipage}\\[2cm]
\large \textit{A thesis submitted in fulfilment of the requirements\\ for the degree of Doctor of Philosophy}\\[0.3cm] 
\textit{in the}\\[0.4cm]
{Department of Physics}\\[1.0cm] 
{\large 2025} 
\vfill
  \newpage
  \mbox{}
  \thispagestyle{empty}
  \newpage
\end{center} 
\thispagestyle{empty}

\begin{center}
\noindent\rule{\textwidth}{1pt}\\[0.6cm]
{\huge \bfseries Abstract }\\[0.2cm] 
\noindent\rule{\textwidth}{1pt} \\[1.5cm]
\end{center}



Out-of-equilibrium systems, inherently complex and challenging to understand, are prevalent across various disciplines, including physics where they arise in contexts such as fluid dynamics. In particular, critical out-of-equilibrium systems combine this complexity with the scaling laws and universality classes observed in critical phenomena, with kinetic surface roughening, the study of how a flat surface becomes progressively rougher over time, serving as a prime example. This behavior manifests in a wide variety of contexts, including metal corrosion, cell proliferation, and, notably, the growth of thin films, which can emerge as a result of wetting processes. In this thesis, we conduct extensive numerical simulations to study critical fluctuations and identify universal features of several rough interfaces, generated by simulating discrete models of thin film growth and by performing direct numerical integration of continuum equations. To explore the universal behavior of these interfaces, we identify the critical exponents that characterize the spatio-temporal fluctuations of the front. Additionally, we analyze the dynamics of thin films across different physical scenarios to deepen our understanding of their behavior in out-of-equilibrium conditions, especially in the case where these films are formed by the action of an external force such as Surface Acoustic Waves.

\newpage
\mbox{}
\thispagestyle{empty}
\newpage 
\thispagestyle{empty}
\vspace*{1.5cm}
\begin{flushright}
  \it{A mi familia}
\end{flushright}
\newpage
\mbox{}
\thispagestyle{empty}
\newpage
\thispagestyle{empty}
\chapter*{Agradecimientos}

\begin{otherlanguage}{spanish}
Esta tesis no es solo el fruto de un esfuerzo individual, sino también de la contribución de todas las personas que, con su apoyo, la han hecho posible.

Me gustaría empezar estos agradecimientos, como no podría ser de otra manera, con mis directores de tesis: Juan Jesús Ruiz Lorenzo y Juan José Meléndez Martínez, de la Universidad de Extremadura, y Rodolfo Cuerno Rejado, de la Universidad Carlos III de Madrid. Sin su valiosa orientación, apoyo y dedicación a lo largo de este proceso esta tesis no habría sido posible. Cada uno, desde su perspectiva y conocimiento, ha contribuido de manera significativa no sólo al contenido de esta tesis, sino también a mi formación personal y profesional.

Empecé a trabajar con Juanjo y Juan mientras cursaba el Máster en Simulación en Ciencias e Ingeniería, donde ambos fueron tutores de mi Trabajo de Fin de Máster. Desde aquellos primeros momentos, en los que aprendí con ellos a programar en C y a lanzar trabajos en los clústeres del ICCAEx, se han volcado en que mi futuro académico fuera lo más prometedor posible. También de su mano he iniciado mis primeras responsabilidades docentes, una faceta que debo decir me entusiasma tanto, e incluso algunos días más, que la propia investigación.

También he tenido el placer de trabajar con Rodolfo, quien siempre ha estado ahí para, con paciencia y claridad, responder a las preguntas más complicadas. Me acogió durante varios días en Leganés, en una estancia que fue de gran ayuda en un momento clave del desarrollo de esta tesis. Gracias a él, tuve la oportunidad de conocer a Lou Kondic y realizar una estancia muy enriquecedora en Nueva Jersey. Ha sido una pena tenerlo tan lejos durante estos años, pero aun así siempre ha estado disponible y dispuesto a ayudarme en todo lo que necesitara.

Muchas gracias a los tres. Ha sido un auténtico placer trabajar con vosotros.

Más allá de mis directores de tesis, son muchas las personas que han contribuido para que esta tesis sea una realidad. Primero, me gustaría acordarme del resto de miembros “senior” del grupo de investigación SPhinX: Enrique Abad, Antonio Astillero, Santos Bravo, Vicente Garzó, Antonio Gordillo, y Andrés Santos. Muchas gracias a todos por vuestra amabilidad y por estar siempre dispuestos a compartir un café con los jóvenes. En especial me gustaría agradecer a Enrique y a Vicente su labor como coordinadores del grupo durante estos años. Gracias a ellos viajar a congresos ha sido siempre más fácil.

Asimismo, quisiera reconocer y agradecer la labor de Francisco Naranjo y Nuria María García del Moral como gestores del grupo durante el periodo en que he desarrollado esta tesis.

Por supuesto, no puedo olvidarme de dar las gracias a todos los jóvenes, y no tan jóvenes, que han ido pasando por grupo SPhinX durante estos años: Alberto, Ana, Beatriz, Felipe, Javier, Juan, Miguel, Miguel Angel y Rubén. Muchas gracias a todos por compartir cafés y bromas conmigo durante las mañanas de los últimos años.

También quiero agradecer a Lou Kondic su hospitalidad durante mi estancia en el Instituto Tecnológico de Nueva Jersey. Gracias a él y a su grupo de investigación he podido aprender mucho. Además, creo que el trabajar junto a experimentales me ha enriquecido como investigador. Tampoco puedo olvidarme de Mark, Ofer, Yifan, Linda y Javier, con los que he compartido numerosas videollamadas en los últimos dos años.

Del mismo modo, quiero expresar mi sincero agradecimiento a los revisores externos de mi tesis, Mariano López de Haro, Haim Taitelbaum, Isidoro González-Adalid Pemartin, Silvia Noemí Santalla Arribas y Ricardo Gutiérrez Díez por su disposición para asumir esta tarea, por la lectura atenta del manuscrito y por las valiosas aportaciones que han contribuido significativamente a mejorar este trabajo.

También quiero dar las gracias a todas las personas que hacen que el Instituto de Investigación de Computación Científica Avanzada de la Universidad de Extremadura (ICCAEx) funcione correctamente. En especial a Carlos García Orellana por estar siempre pendiente cuando había algún problema técnico. Sin los más de 300 años de cálculo que he utilizado durante estos años en los 3 clúster del instituto, \textit{iccaex}, \textit{grinfis} y \textit{ada}, esta tesis no habría sido posible.

Asimismo, quiero agradecer a Francisco Javier Acero su labor como director del Departamento de Física, sin el cuál hubiera sido mucho más difícil poder cumplir todas las obligaciones de mi contrato FPU.

También me gustaría agradecer a Manuel Antón, y anteriormente a Andrés Santos, como coordinadores del programa de doctorado, por su amabilidad y disposición para ayudarnos con todos los trámites relacionados con la defensa de esta tesis.

Creo que también es digno de mencionar y profundamente agradecer a todos los profesores de Física y Matemáticas que, con su dedicación, han contribuido a mi formación a lo largo de toda mi vida. En especial, me gustaría acordarme de Genoveva, de Vicente y de Juan.

Finalmente no puedo dejar de acordarme de mi familia, por apoyarme durante toda mi vida y por brindarme la mejor educación posible. A mis padres, por enseñarme el valor del esfuerzo y la perseverancia. A mis hermanos Javi y Pepe y a mi cuñada Elena, por su compañía y apoyo durante todo este proceso. A mis sobrinos Julia y Jorge que siempre consiguen alegrarme el día. A mis abuelos, que aunque ninguno haya llegado a verme acabar esta tesis espero que estén orgullosos de mí. A mis tíos: Nines, Felipe, Jaime, Mara, José y Belén. Gracias por vuestro apoyo y cariño a lo largo de estos años. A mis amigos Alberto, Álvaro J, Álvaro R, David y Alejandro. Gracias por saber sacarme una sonrisa siempre. Muchas gracias a todos.


Por último, agradezco la financiación para la realización de esta tesis proveniente del Gobierno de España mediante la ayuda predoctoral FPU2021-01334 y a la Junta de Extremadura por la financiación de algunos contratos previos a esta ayuda.
\end{otherlanguage}

\begin{flushright}
  Badajoz, 2025 \par 
  \it{Jesús María Marcos Merino}
\end{flushright}


\setcounter{secnumdepth}{3} 
\setcounter{tocdepth}{2}    
\tableofcontents            
\newpage


%

\nomenclature{BC}{Boundary Conditions, page 51\footnote[1]{The page number refers to the first occurrence of the term in the text.}\nomnorefpage}
\printnomenclature[2cm] 

\label{nom}



\mainmatter



\markboth{Introduction}{}
\chapter*{Introduction}
\addcontentsline{toc}{chapter}{Introduction}

Systems that operate far from equilibrium are intrinsically intricate and difficult to study, yet they manifest across a vast array of disciplines, from biology and engineering to economics. In physics, nonequilibrium behavior is central to many phenomena, including fluid turbulence, chemical kinetics, and the dynamics of semiconductors. A foundational idea in understanding such systems is criticality: the emergence of collective behavior governed by universal scaling laws. These laws organize seemingly different systems into universality classes, where macroscopic patterns remain consistent despite differences in microscopic details.

A classic illustration of a nonequilibrium critical system is kinetic surface roughening. This process involves the gradual transformation of a smooth surface into a rough one over time, typically driven by random or stochastic events such as the deposition of particles. Remarkably, this seemingly simple evolution underpins a wide variety of complex, real-world phenomena, ranging from the etching and corrosion of metals, to the expansion of biological tissues, and most prominently, to the growth of thin films in materials science.
Thin film growth, in particular, can result from wetting phenomena, where intermolecular forces such as van der Waals attractions and electrostatic interactions dictate how a fluid spreads across a solid substrate. A detailed analysis of these interactions reveals that non-volatile droplets can develop microscopic precursor films, which are of great interest for further study.

In this thesis, we conduct extensive numerical simulations to study critical fluctuations and identify universal features of rough interfaces. This is achieved by simulating discrete models of thin film growth and by performing direct numerical integration of continuum equations. To explore the  universal behavior of these interfaces, we identify the critical exponents that characterize the spatio-temporal fluctuations of the front. Furthermore, we investigate the statistical properties of these fluctuations, such as their correlation functions, with the aim of gaining a deeper understanding of the underlying universal behavior. Additionally, we analyze the dynamics of thin films across different physical scenarios to deepen our understanding of their behavior in out-of-equilibrium conditions, especially in the case where these films are formed by the action of an external force such as Surface Acoustic Waves.

The thesis is organized into eight chapters. Chapters 1 and 2 provide the theoretical framework, while Chapter 3 outlines the methodology applied in the subsequent chapters. The main novel contributions of the thesis are presented in Chapters 4, 5, 6 and 7, which examine the various systems where a growing front or film arise. 
Each of these chapters presents its individual results together with the corresponding conclusions. The final chapter summarizes the thesis results, highlights its contributions to the scientific field, and proposes directions for future research.

A brief explanation of the content of each of the chapters of the thesis follows:
\begin{itemize}
    \item Chapter 1 introduces the fundamental concepts of kinetic surface roughening. We begin by focusing on particle deposition models and formulating stochastic growth equations with time-dependent noise from a continuum viewpoint, with particular emphasis on the Kardar-Parisi-Zhang equation. These frameworks help characterize the various universality classes that will be discussed throughout the thesis. We then briefly examine the tensionless case of the Kardar-Parisi-Zhang equation and the roughening transition this equation exhibits. The chapter concludes with a presentation of experimental studies in which interfaces have been measured and analyzed.
    \item Chapter 2 focuses on describing the physical principles of the spreading model that will be simulated in Chapters 4 and 5 of this thesis. Starting from equilibrium properties, we examine how the various interactions involved naturally lead to the spreading of non-volatile droplets. Finally, we introduce the case of microscopic precursor films, first reviewing the experimental evidence supporting their existence, and then discussing the various theoretical models that have been proposed to understand them.
    \item Chapter 3 outlines the methodology used to study the systems presented in the following chapters. It includes a review of the Monte Carlo method, as well as a definition of the observables to be measured and the procedures for estimating their statistical errors.
    \item Chapters 4 and 5 are devoted to the study of the fronts generated by the precursor films of spreading droplets. Specifically, these chapters examine the front dynamics of the spreading model introduced in Chapter 2, focusing on band and radial geometries, respectively.
    \item Chapter 6 presents the numerical integration of the Kardar-Parisi-Zhang equation and its variants on the Bethe lattice. To this end, various integration schemes developed for application on non-regular lattices are first presented.
    \item Chapter 7 presents a simplified model for the extraction of oil from an oil-in-water emulsion driven by a Surface Acoustic Wave. The focus is placed on the modeling of the wave itself and how variations in this modeling affect the resulting dynamics.
    \item Chapter 8 provides a concise overview of the work carried out throughout the thesis, summarizing the main results, methodologies, and physical insights presented in each chapter. In addition, it discusses potential extensions of the current work and outlines several directions for future research.
\end{itemize}

\textbf{Scientific manuscripts related to this thesis}
\begin{enumerate}
    \item J. M. Marcos,  P. Rodríguez-López, J. J. Meléndez, R. Cuerno, and J. J. Ruiz-Lorenzo, \textit{Spreading fronts of wetting liquid droplets: Microscopic simulations and universal fluctuations}, Phys. Rev. E \textbf{105}, 05480 (2022).
    \item J. M. Marcos, J. J. Meléndez, R. Cuerno, and J. J. Ruiz-Lorenzo, \textit{Microscopic fluctuations in the spreading fronts of circular wetting liquid droplets}, Phys. Rev. E \textbf{111}, 045504 (2025).
    \item J. M. Marcos, J. J. Meléndez, R. Cuerno, and J. J. Ruiz-Lorenzo, \textit{Numerical Integration of the KPZ and Related Equations on Networks: The Case of the Cayley Tree}, J. Stat. Mech, 083203 (2025).
    \item Y. Li, J. M. Marcos, M. Fasano, J. Diez, L. J. Cummings, L. Kondic, O. Manor, \textit{Using wetting and ultrasonic waves to extract oil from oil/water mixtures}, J. Colloid Interface Sci \textbf{700}, 138442 (2025).
    \item J. M. Marcos, Y. Li, M. Fasano, J. Diez, L. J. Cummings, L. Kondic, O. Manor, \textit{Monte Carlo–based model for the extraction of oil from oil-water mixtures using wetting and surface acoustic waves}, Phys. Rev. E \textbf{112}, 025502 (2025).
\end{enumerate}

\textbf{Financial support}

The thesis was supported by Universidad de Extremadura through Programa Propio de Iniciación a la Investigación de la Universidad de Extremadura, Junta de Extremadura (Spain) through Grant No.~IB20079 (partially funded by FEDER) and by  Ministerio de Universidades (Spain) through pre-doctoral Grant No.~FPU2021-01334. 
\graphicspath{{1_capitulo/fig1/}}

\chapter{Surface growth and kinetic roughening}
\label{chap1:intro}


The study of surface growth stands as a cornerstone in the physics of non-equilibrium systems and materials science. It focuses on understanding how surfaces evolve over time under the influence of various physical and chemical processes such as deposition, etching, and epitaxial growth \cite{Barabasi1995}. These dynamic processes not only determine the visual morphology of a surface but also its functional properties, making them crucial for technological applications such as semiconductor manufacturing, protective coatings, thin film fabrication \cite{Evans2006}, and the optimization of properties like electrical conductivity, wear resistance, and optical performance \cite{Evans2006,Krug1997}.

A key feature of surface growth phenomena is the spontaneous emergence of complex patterns and fractal-like structures from simple, local interactions between atoms or molecules \cite{Family1986,Family1991}. For example, during material deposition, the accumulation of particles leads to the development of surface roughness, mounds, and self-similar formations that evolve according to universal statistical laws \cite{Barabasi1995}. These processes are typically modeled through mathematical formalisms, including stochastic partial differential equations, that describe how roughness changes over time.

Notably, the interfaces formed during growth exhibit self-affine properties, meaning that upon anisotropic rescaling, a portion of the interface appears statistically indistinguishable from the whole \cite{Barabasi1995}. These scaling behaviors connect directly with the concept of universality, which is central to the theoretical analysis of such systems. Indeed, simple scaling relationships often link seemingly independent quantities and critical exponents, allowing diverse systems to be classified into universality classes, analogous to those found in equilibrium critical phenomena. Theoretical models and simulation studies are instrumental in this context, providing a bridge between microscopic dynamics and macroscopic observables, and helping to identify which features are essential for a given morphological behavior \cite{Barabasi1995}.

While surface growth is of clear technological relevance, 
its conceptual richness extends far beyond practical applications. It has become a meeting point between disciplines, linking statistical physics with biology, chemistry, and nanotechnology \cite{Meakin1998}. The observation that very different systems can exhibit remarkably similar behavior underlines the power of universality: a central theme in this thesis.

In particular, the phenomenon of kinetic roughening exemplifies non-equilibrium criticality in surface growth. It refers to the progressive increase in surface roughness over time due to the random deposition and movement of particles. This type of dynamic roughening occurs in a wide variety of contexts, including thin film growth, snowflake formation, and metal corrosion, all of which have direct implications for materials science, biomedical applications (such as cell proliferation), and nanofabrication \cite{Barabasi1995}.

From a theoretical standpoint, our focus lies on surface kinetic roughening analyzed through the lens of critical fluctuations at the interface of driven systems subjected to noise. Recent research has shown that the associated universality classes and their properties extend and generalize those of equilibrium critical dynamics to non-equilibrium conditions \cite{Kriecherbauer2010, Takeuchi2018, tauber2014}. These concepts have proven so robust that their applicability now extends to systems without explicit interfaces, broadening their significance across physics and beyond.

To fully appreciate this extension of criticality to non-equilibrium systems, it is essential to understand the foundations of critical phenomena in equilibrium. In physics, criticality refers to the behavior of systems near a critical point, where they undergo profound changes in their macroscopic properties, commonly characterized by scale invariance. These transitions, known as phase transitions, occur when the internal structure or order of a system changes dramatically, and they can be classified as either discontinuous or continuous.

The study of continuous phase transitions has been particularly fruitful in the context of magnetism. Ferromagnetic materials, for instance, exhibit spontaneous magnetization even in the absence of an external magnetic field. However, as the system reaches the Curie temperature, it undergoes a continuous transition from a ferromagnetic to a paramagnetic state, accompanied by the disappearance of spontaneous magnetization and the emergence of critical fluctuations. The behavior of physical quantities near this point, such as magnetization and specific heat, is governed by critical exponents, which encapsulate how these observables diverge or vanish near the transition.

These critical exponents are not unique to ferromagnetic systems. Indeed, they are found in all continuous phase transitions and offer deep insights into the universal behavior exhibited by diverse systems near criticality. For example, the Ising model, a paradigmatic theoretical model for ferromagnetism, predicts precise values for the exponents governing the ferromagnetic–paramagnetic transition. Remarkably, experiments have confirmed that very different systems, such as simple fluids and uniaxial ferromagnets, share exactly the same critical exponents. This observation highlights the profound idea of universality: that microscopic details become irrelevant near the critical point, and only a few key variables determine the large-scale behavior of the system. 

The formalism that made this understanding possible is the Renormalization Group (RG)\nomenclature{RG}{Renormalization Group}, introduced by Wilson in the early 1970s \cite{Wilson1974}. The RG provides a powerful framework for systematically calculating scaling exponents and identifying universality classes. It explains why seemingly unrelated systems can behave identically near criticality: they belong to the same universality class, governed by shared symmetries, dimensions, and conservation laws \cite{Sethna2006}.

In this chapter, we will delve into the interplay between non-equilibrium surface growth and critical phenomena. By analyzing kinetic roughening through the concepts of universality and scaling, we aim to highlight how the language of criticality, originally developed for equilibrium systems, can be successfully extended to understand far-from-equilibrium dynamics. This framework forms the theoretical backbone for the results presented in the subsequent sections of this work.

\section{Fundamental scaling properties}
\label{sec1:scaling_properties}

In this section, we outline a few key scaling behaviors that help characterize kinetically roughened surfaces. A more detailed discussion of this topic will be presented in Chapter \ref{chap3:methods}.

\begin{figure}[ht]
    \centering
    \includegraphics[width=0.9\textwidth]{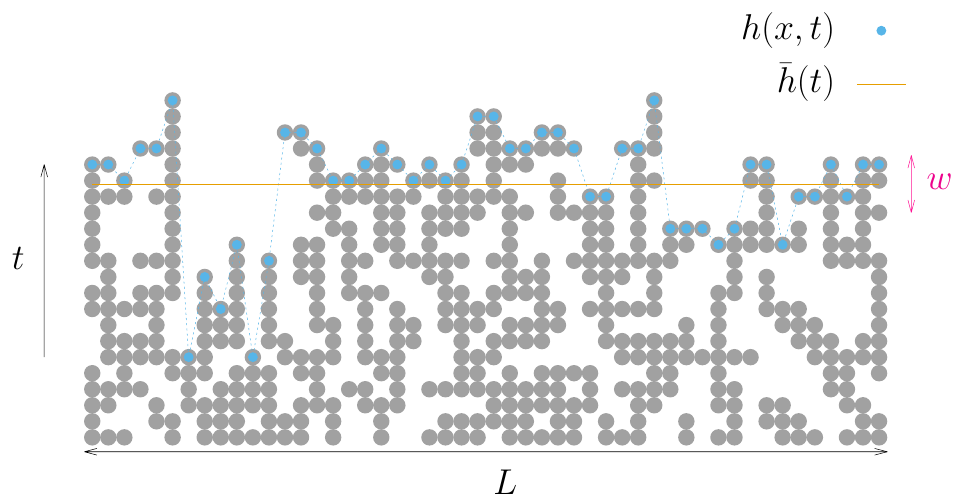}
    \caption[Front example]{Example of a particle aggregate of lateral size $L$. The local heights $h(x,t)$ (blue dots) define the front. The system evolves over time, as more particles are aggregated, and the heights grow in the vertical direction. The mean front $\bar{h}(t)$ and the width $w(L,t)$ are represented by a solid orange line and a magenta arrow, respectively. Reproduced from Ref.~\cite{GarciaBarreales2024}.}
    \label{fig1:ejemplo}
\end{figure}

The fundamental concept is that of a front, which will be referred to interchangeably as a surface or interface throughout this text. In all cases, the front is described by a set of local space-time variables $h(\boldsymbol{x},t)$. In the simplest scenario, the front is defined as the collection of particles in an aggregate that are highest at each position of the substrate, forming a set of height variables $h(\boldsymbol{x},t)$, where $\boldsymbol{x}$ represents the substrate positions, although more complex definitions may exist. Figure~\ref{fig1:ejemplo} illustrates a particle aggregate on a one-dimensional substrate of size $L$. The heights $h(x,t)$, where $x=1,...,L$, are represented by blue points. The mean front, $\bar{h}(t)$, is defined as the average of the local heights $h(\boldsymbol{x},t)$, providing a measure of the mean position of the front.

Besides its position, another key variable used to describe a front is its width, $w(L,t)$, which quantifies the roughness of the interface. This width is defined as the standard deviation of the height $h(\boldsymbol{x},t)$. In Fig.~\ref{fig1:ejemplo}, the mean height $\bar{h}(t)$ is represented by a solid orange line while the front width is indicated by a magenta arrow.

In the context of kinetic roughening, the width $w(L,t)$ of a growing interface is expected to follow the so-called Family-Vicsek (FV)\nomenclature{FV}{Family-Vicsek} scaling law \cite{Family1986,Barabasi1995}:
\begin{equation}
    w(L,t) \sim 
    \left\{
        \begin{array}{lr}
            t^{\beta}, & \text{if } t \ll t_{\mathrm{x}},\\
             L^{\alpha} \equiv w_{\mathrm{sat}}(L), & \text{if } t \gg t_{\mathrm{x}}.
        \end{array}
    \right.
    \label{eq1:wcomportamiento}
\end{equation}
For short times the width grows as a power law, with an exponent $\beta$, called the \textit{growth exponent}, that characterizes the time-dependent dynamics of the roughening process. For longer times this regime transitions into a saturation regime where the front width stabilizes at a saturation value, $w_{\mathrm{sat}}$. This value increases with the system size $L$ following also a power law. The exponent $\alpha$, known as the \textit{roughness exponent}, is a second critical exponent that characterizes the roughness of the interface in its saturated state. The crossover time $t_{\mathrm{x}}$ that separates both regimes depends of the system size as
\begin{equation}
    t_{\mathrm{x}}\sim L^z,
    \label{eq1:z}
\end{equation}
where $z$ is known as the \textit{dynamic exponent}. 

While $z$ provides information about how fast the dynamics of the system is, $\alpha$ provides information about the self-affine structure of the front. Specifically, a front that follows the scaling relation \eqref{eq1:wcomportamiento} remains statistically indistinguishable under the transformations~\cite{Barabasi1995}:
\begin{equation}
    x\rightarrow bx \hspace{1cm} h\rightarrow b^\alpha h.
    \label{eq1:self_affine}
\end{equation}
The scaling exponents $\alpha$, $\beta$, and $z$ are not independent. In fact, by approaching the crossover point from both sides, one obtains $w(t_{\mathrm{x}}) \sim t_{\mathrm{x}}^{\beta}$ on one side and $w(t_{\mathrm{x}}) \sim L^{\alpha}$ on the other. These two relations, together with $t_{\mathrm{x}}\sim L^z$, lead to the expression
\begin{equation}
    z=\frac{\alpha}{\beta},
    \label{eq1:zalphabeta}
\end{equation}
which holds for any growth process that follows the scaling relation \eqref{eq1:wcomportamiento}.
The width of the front evolves over time and eventually saturates due to the points along the front not being independent from each other. In other words, spatial correlations exist because local heights are influenced by the heights of neighboring sites. Although the growth process is inherently local, information about the height of each site propagates laterally along the front. The characteristic distance over which height correlations extend is known as the parallel, or lateral, correlation length, $\xi $.

At the start of the growth process all sites are typically uncorrelated. As time progresses, the parallel correlation length increases as the system evolves. However, in a finite system, this correlation length cannot grow indefinitely, as it is ultimately limited by the system size $L$. When $\xi$ expands to the system size, the entire interface becomes correlated, leading to the saturation of the interface width. Thus, $\xi \sim L $ at saturation, which occurs at a time $t_{\mathrm{x}}$ given by Eq.~\eqref{eq1:z}. By replacing $L$ with $\xi$ in Eq.~\eqref{eq1:z}, we get that $\xi \sim t_{\mathrm{x}}^{1/z}$, which is also valid for $ t < t_{\mathrm{x}} $. In this case,
\begin{equation}
\xi \sim
\left\{
    \begin{array}{lr}
         t^{1/z}, & \text{if } t \ll t_{\mathrm{x}},\\
         L, & \text{if } t \gg t_{\mathrm{x}}.
    \end{array}
\right.
\label{eq1:xi}
\end{equation}
From this perspective, the dynamic exponent $z$ characterizes the power law growth of the parallel correlation length along the surface.

Up to this point, we have assumed that the system size $L$ remains constant. However, this is not always the case, as we will see below. For certain interfaces, particularly those that grow radially, the front length $L$ increases as the interface expands. In such cases, there is a competition between the growth of the lateral correlation length $ \xi$ and the growth of the front length. 
If the correlations grow faster than the length of the front the system will end up fully correlated and will saturate. However, if the length of the front grows faster than the correlations, which is usually the case, the system will never saturate. In that case the roughness exponent must be measured by others means. 

It is important to note that there are other scaling behaviors that deviate from the standard FV scaling behavior, namely the anomalous scaling \cite{Lopez1997}. Anomalous scaling occurs when the roughness exponent $\alpha$, which describes how the global interface width scales with system size differs from the \textit{local roughness exponent} $ \alpha_{\text{loc}}$, which characterizes height fluctuations over small length scales. This difference arises because some interfaces develop correlations at different scales in a non-trivial way, leading to multi-scaling effects. In other words, height fluctuations exhibit different scaling behaviors at different length scales, rather than being described by a single roughness exponent across the entire system. In standard (non-anomalous) scaling, $\alpha = \alpha_{\text{loc}}$, meaning that roughness behaves uniformly across all scales. However, in anomalous scaling, these exponents differ ($\alpha \neq \alpha_{\text{loc}}$), indicating that local fluctuations evolve differently from global ones, often due to complex growth mechanisms or long-range interactions \cite{Lopez1997}. A more detailed explanation of the competition between the lateral correlation length $ \xi$ and the front length, the various methods for measuring the roughness exponent, and the different scaling schemes will be presented in Chapter \ref{chap3:methods}.


Before moving forward, it is important to clarify a potential ambiguity in the notation for dimensions. The dimension of an interface will be denoted by $d$. Therefore, $d=1$ refers to a one-dimensional interface embedded in a two-dimensional plane, while $d=2$ represents a two-dimensional interface embedded in a three-dimensional space.

\section{Deposition models and growth equations}
\label{sec1:dep_models}

In this section, we will explore simple discrete models where particles are added either to a substrate or a cluster of particles. This accumulation of particles forms a front, whose critical exponents we will analyze. For certain models, we will examine how they can be associated with continuous growth equations. These equations will then serve as a basis for introducing the main universality classes of surface kinetic roughening.

\subsection{Discrete deposition models}



\begin{figure}[t]
    \centering
    \includegraphics[width=1\textwidth]{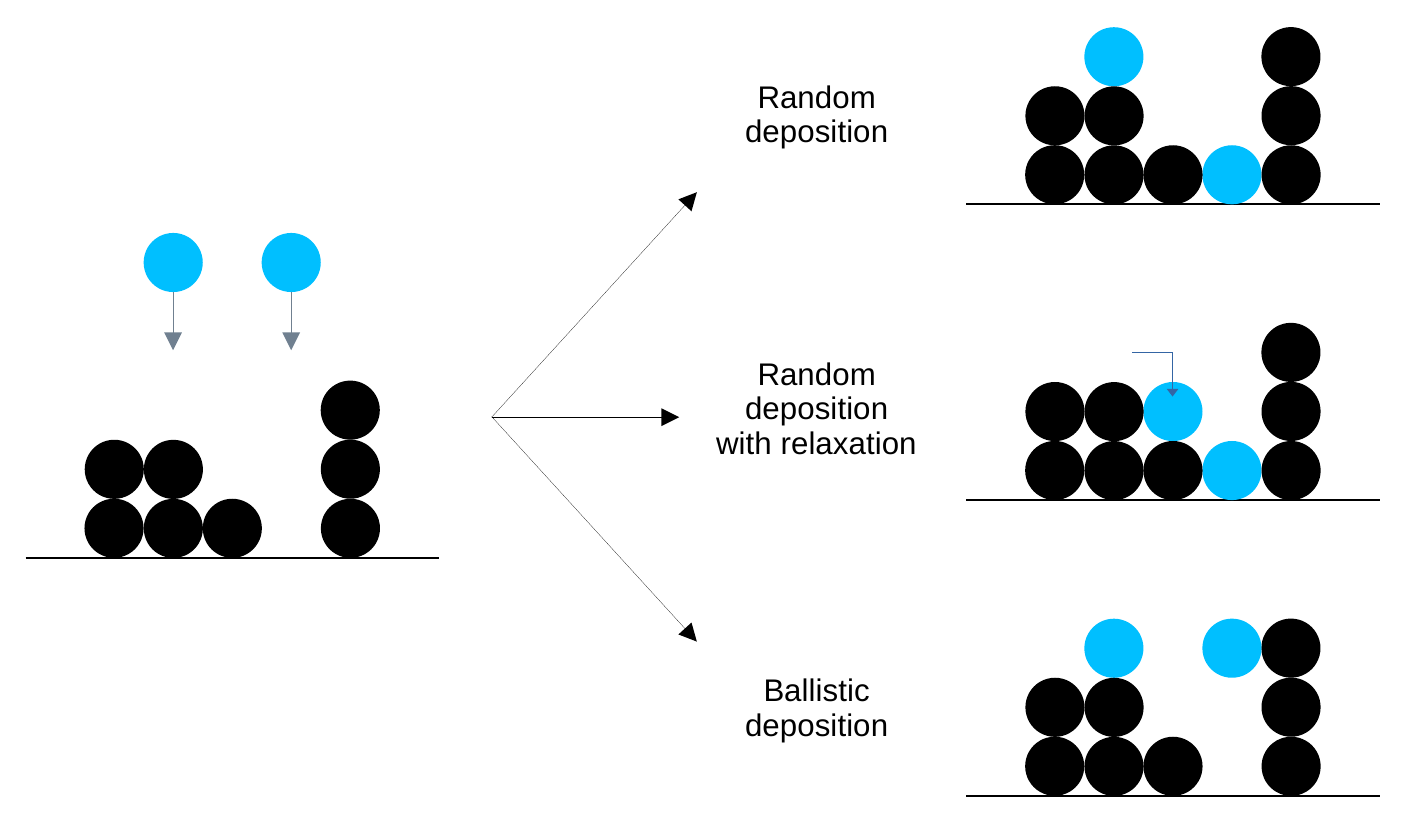}
    \caption[Diagrams of particle sticking rules in deposition models.]{Diagrams of particle sticking rules in deposition models: random deposition, random deposition with relaxation, and ballistic deposition. Reproduced from Ref.~\cite{GarciaBarreales2024}.}
    \label{fig1:sch}
\end{figure}

\begin{figure}[h!t]
     \centering
     \begin{subfigure}[b]{0.8\textwidth}
         \centering
         \includegraphics[width=\textwidth]{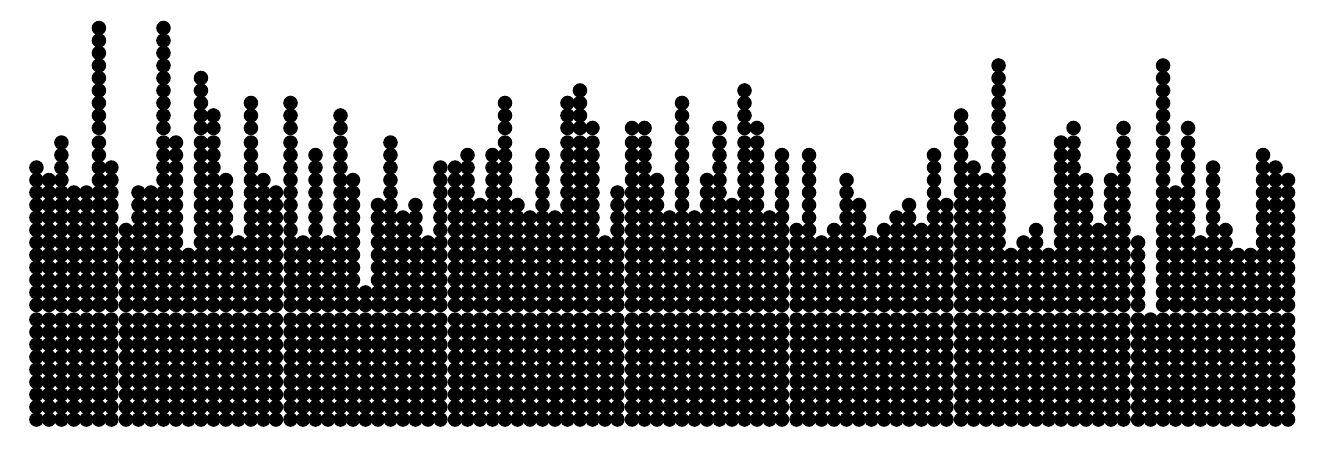}
         \caption{Random deposition morphology.}
         \label{fig1:RD}
     \end{subfigure}
     \hfill
     \begin{subfigure}[b]{0.8\textwidth}
         \centering
         \includegraphics[width=\textwidth]{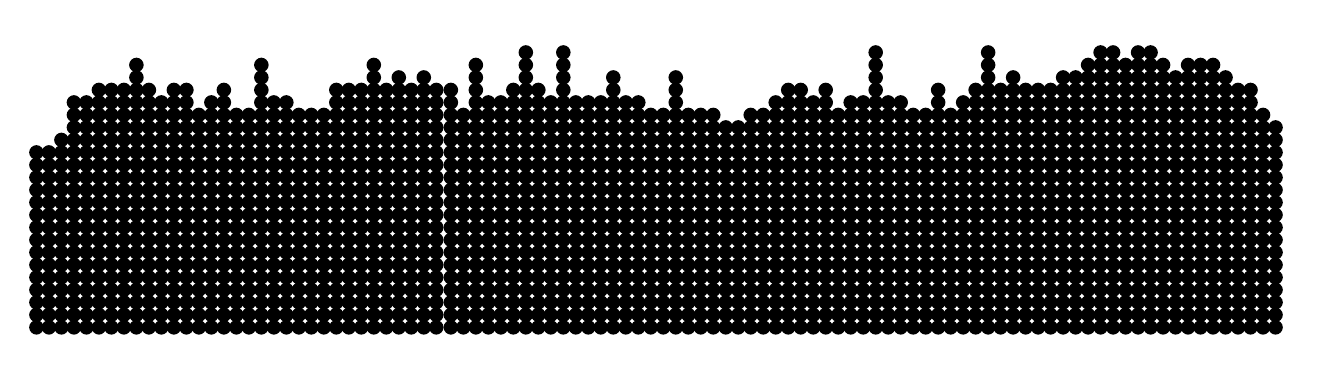}
         \caption{Random deposition with surface relaxation morphology.}
         \label{fig1:RD_relax}
     \end{subfigure}
     \hfill
     \begin{subfigure}[b]{0.8\textwidth}
         \centering
         \includegraphics[width=\textwidth]{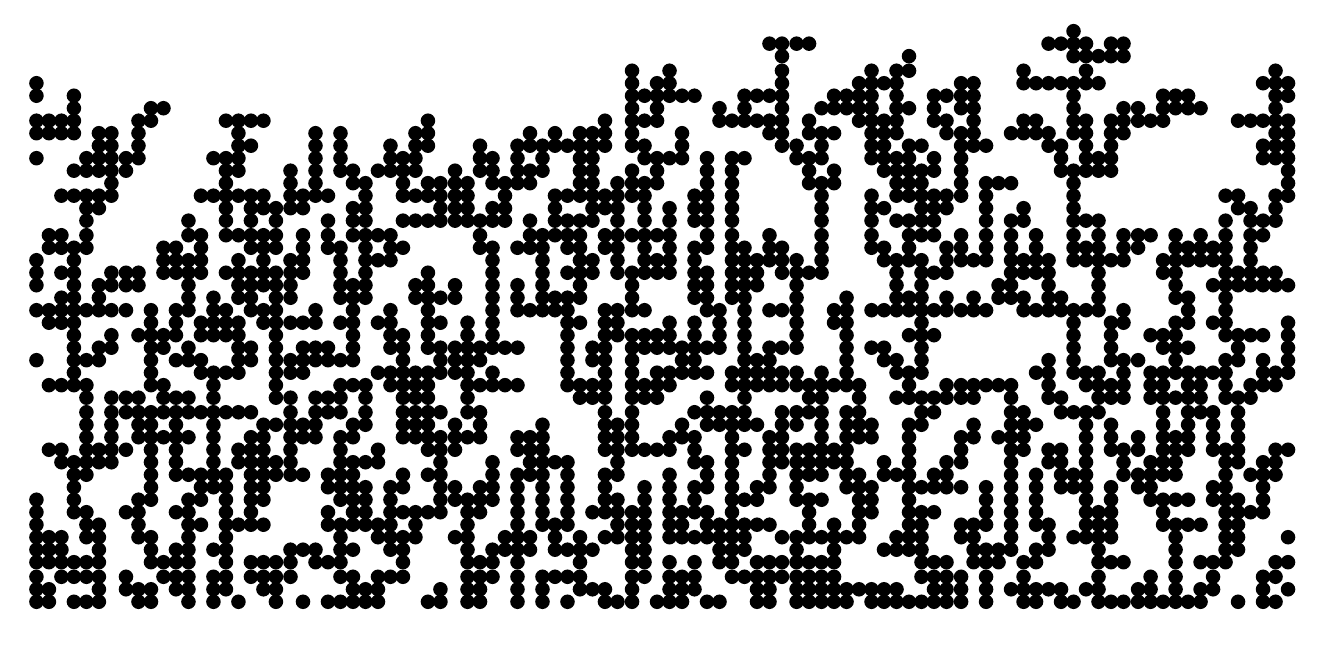}
         \caption{Ballistic deposition morphology.}
         \label{fig1:BD}
     \end{subfigure}
        \caption{Particle aggregates for RD, RDSR, and BD models. As a reference, the substrate size is always $L=100$ in all cases, and the total number of particles in each morphology is 2000. Reproduced from Ref.~\cite{GarciaBarreales2024}.}
        \label{fig1:dep_models}
\end{figure}

The \textit{random deposition} (RD)\nomenclature{RD}{Random Deposition} is the simplest possible surface growth model. In this model, particles descend vertically to a randomly chosen position on the substrate. Upon reaching the surface, they adhere either to the substrate itself or to previously deposited particles at that location. Figure~\ref{fig1:sch} shows the sticking rule for the RD model. The surface height grows as particles stack vertically in individual columns, with no correlation between them. Figure~\ref{fig1:RD} shows the surface morphology of the RD model.

The simplicity of the model enables the exact calculation of the critical exponents. Since each column has an equal probability of growing, given by $p = 1/L$, being $L $ the system size, the probability that a column reaches a height $h$ after the deposition of $N$ particles follows a binomial distribution:
\begin{equation}
    P(h,N)= \binom{N}{h} p^h(1-p)^{N-h}.
\end{equation}
As the moments of the binomial distribution can be calculated exactly,
\begin{equation}
    E[X]=N p, \hspace{1cm} E[X^2]=Np(1-p)+(Np)^2
\end{equation}
then the width is straightforwardly calculated as:
\begin{equation}
    w^2(t)=\langle {h^2} \rangle - \langle {h} \rangle ^2=Np(1-p)=\frac{N}{L}\left(1-\frac{1}{L} \right).
    \label{eq1:wRD}
\end{equation}
If the evolution is defined as $t=N/L$, i.e. the time is updated with an amount $1/L$ each time a new particle is added into the system, then from Eq.~\eqref{eq1:wcomportamiento} and Eq.~\eqref{eq1:wRD} we get $w(t)\sim t^{1/2}$, and thus $\beta=1/2$ for the RD model.

On the other hand, due to the absence of spatial or lateral correlations, the correlation length $\xi$ remains zero at all times. As a result, the interface does not saturate, and the roughness exponent $\alpha$ is not well-defined. Thus, in the RD model, the interface width increases indefinitely without reaching saturation. 

Another important discrete model is \textit{random deposition with surface relaxation} (RDSR)\nomenclature{RDSR}{Random Deposition with Surface Relaxation}. To incorporate surface relaxation into the RD model, each deposited particle is allowed to diffuse along the surface within a limited range (typically only to the nearest neighboring sites), stopping once it reaches a position with a lower height. Figure~\ref{fig1:sch} shows this sticking rule, while Fig.~\ref{fig1:RD_relax} shows an example of the surface morphology of the RDSR model. It is clear that this interface is significantly smoother than that of the RD model.

Note that in the RDSR model, the heights of neighboring columns affect particle placement, leading to the emergence of correlations. As we will see below, this model is associated with a continuum equation that can be solved  analytically. Its solutions in one dimension yield $\beta = 1/4$ and $\alpha = 1/2$, in good agreement with simulation results \cite{Family1986}. 

\textit{Ballistic deposition} (BD)\nomenclature{BD}{Ballistic Deposition} is another modification of the RD model that produces a non-equilibrium interface with interesting growth characteristics. Once again, particles fall vertically onto a random position on the substrate. However, in this case, they attach upon making lateral contact with previously deposited nearest-neighbor particles or upon reaching the substrate. Figure~\ref{fig1:sch} shows once again the sticking rule for the BD model, while Fig.~\ref{fig1:BD} shows the morphology of this model.
Clearly, the aggregate generated is quite different from the previous ones. The critical exponents in one dimension have been determined through numerical simulations, yielding values of $\alpha=0.47(2)$ and $\beta=0.330(6)$ \cite{Baiod1988,Meakin1986}. 

As evident from the previous discussion, these three models, despite small differences in their definitions, produce remarkably distinct interfaces (see Fig.~\ref{fig1:dep_models}). This highlights how small variations in microscopic rules can lead to noticeable differences in kinetic surface roughening behavior. In the next section, we will see that these three models fall into three different universality classes. Before proceeding with that, we will introduce additional models that will be relevant for the analyses presented throughout this thesis.

\begin{figure}[t]
    \centering
    \includegraphics[width=0.8\textwidth]{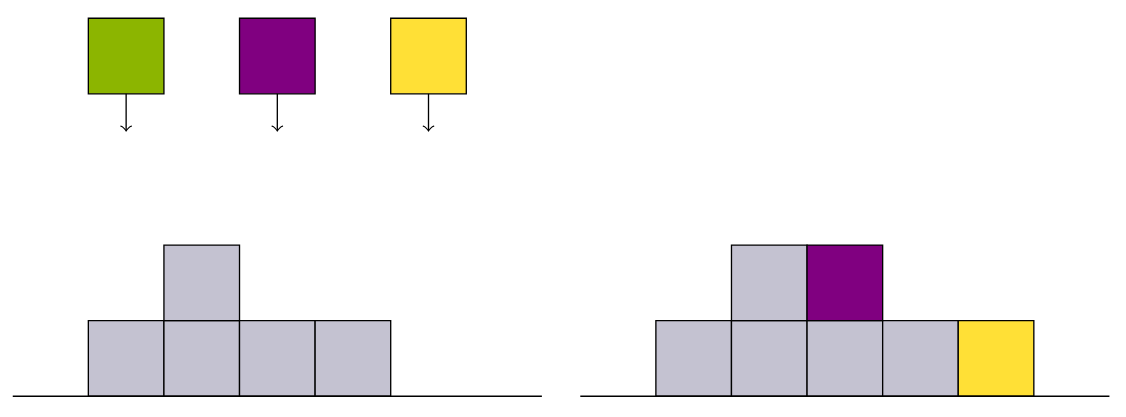}
    \caption{Example of the sticking rule in the RSOS model: Purple and yellow particles adhere to the surface, whereas the green particle does not, as the height difference on the left exceeds one. Reproduced from Ref.~\cite{Sudijono2023}.}
    \label{fig1:RSOS}
\end{figure}

The \textit{Restricted solid-on-solid} (RSOS)\nomenclature{RSOS}{Restricted Solid-on-Solid} model is a straightforward modification of the RD model, where particles adhere to the position where they land only if the resulting height differences between adjacent sites remain bounded by one. Figure~\ref{fig1:RSOS} illustrates an example of this sticking rule. Several simulations of this model have been performed, yielding results comparable to those of the BD model. Namely, the growth exponent is $\beta=0.332(5)$ in $d=1$ and $\beta=0.241(1)$ in $d=2$ \cite{Kim1989,Kelling2016}.

In addition to deposition models on substrates, there exist other aggregation models that also produce fronts of interest. Here, we introduce two very simple models that have been widely used: the \textit{Eden} model and the \textit{Diffusion-limited aggregation} (DLA)\nomenclature{DLA}{Diffusion-Limited Aggregation} model.

\begin{figure}[ht]
    \centering
    \includegraphics[width=0.8\textwidth]{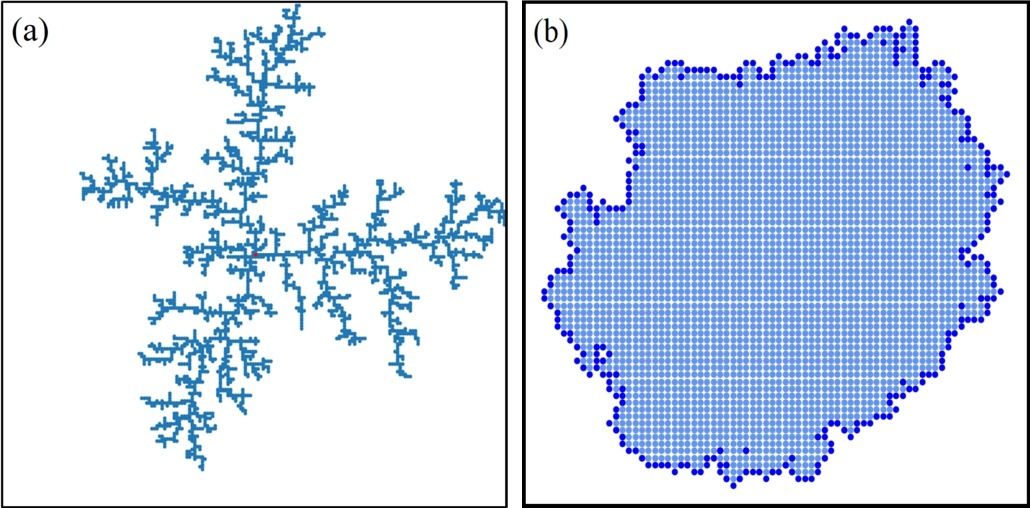}
    \caption{(a) Typical morphology of the DLA model. (b) Typical morphology of the Eden model. Reproduced from Ref.~\cite{Tian2024}.}
    \label{fig1:DLA_Eden}
\end{figure}

The Eden model \cite{Eden1961} was originally introduced to study cell proliferation in biological systems. However, this model and its modifications have also been extensively used to investigate growing interfaces and out-of-equilibrium aggregation processes, such as crystal growth \cite{Mansfield1994}. In its simplest form, the model works as follows: it begins with a single cell or particle at an initial site within the network. At each step, a new particle is randomly added to the interface at a neighboring site adjacent to the existing structure. Only sites directly connected to the aggregate boundary are eligible for occupation in the next growth step. Since growth occurs locally, the aggregate tends to generate relatively compact fronts. The Eden model constitutes a simple system in which the front length $L$  increases over time. Figure~\ref{fig1:DLA_Eden}b shows an example of the morphology of the Eden model.

The DLA model \cite{Witten1981} is a variation of the Eden model designed to describe the growth of branching structures in non-equilibrium systems. Rather than being directly added to the cluster, particles undergo random motion before adhering to the growing aggregate. The model operates as follows: a particle is fixed at the center of the system. A new particle is then introduced at a random location away from the aggregate. This particle undergoes a random walk on the network until it comes into contact with an already attached particle. Upon making contact, the particle irreversibly adheres to the aggregate, becoming part of the expanding structure. Figure~\ref{fig1:DLA_Eden}a shows an example of the morphology of the DLA model. This model has been widely used to describe the formation of fractal patterns in natural systems and to simulate physical phenomena such as electrodeposition \cite{Castro2000}.

\subsection{Continuum growth equations}


In this section, we derive and analyze the properties of several partial differential equations (PDEs)\nomenclature{PDE}{Partial Differential Equation} related to the growth model discussed in the previous ones. Through this section $h(\boldsymbol{x},t)$ will denote the interface height and $\boldsymbol{x}$ will denote a position within a $d$-dimensional substrate.

Let us start with the RD model. As each column grows independently from each other, i.e. there is no spatial correlations in the model, the growth process can then be described by:
\begin{equation}
    \frac{\partial h(\boldsymbol{x},t)}{\partial t}=F+\eta(\boldsymbol{x},t),
    \label{eq1:eq_simple}
\end{equation}
where $F$ is the average number of particles per unit time arriving at site $\boldsymbol{x}$, and $\eta(\boldsymbol{x},t)$ is an uncorrelated space-time noise whose mean and variance verify: 
\begin{equation}
\begin{aligned}
    \langle \eta(\boldsymbol{x},t) \rangle  & = 0, \\ 
    \langle \eta(\boldsymbol{x},t) \eta(\boldsymbol{x'},t') \rangle & =2D \delta ^d (\boldsymbol{x}-\boldsymbol{x'})\delta (t-t'),
    \end{aligned}
    \label{eq1:var_noise}
\end{equation}
with $D$ being a parameter that regulates the noise amplitude. This noise represents the random fluctuations in the deposition process or, more broadly, the time-dependent stochastic variations at the interface. The growth exponent that arises from this equation can be obtained analytically (see Appendix~\ref{app:RD}) and is that of an interface growing through RD, i.e. $\beta=1/2$. This defines the so-called RD universality class, whose exponents are listed in Table~\ref{tab1:dep_models_exp}.

To derive a continuum equation for the RDSR model, we must incorporate a surface relaxation term into Eq.~\eqref{eq1:eq_simple}. Such a term must satisfy specific symmetry constraints. Firstly, the surface evolution should be independent of the origin in the coordinate system and the origin of time. In other words, it must be invariant under the following transformations:
\begin{equation}
\begin{aligned}
    h\rightarrow h+\Delta h,\hspace{1cm}\boldsymbol{x}\rightarrow \boldsymbol{x}+\Delta \boldsymbol{x},\hspace{1cm}t\rightarrow t+\Delta t.
    \end{aligned}
    \label{eq1:sym1}
\end{equation}
Furthermore, the surface should be symmetric with respect to the origin of the coordinate system and the mean height, meaning it must also remain invariant under the following transformations:
\begin{equation}
\begin{aligned}
    \boldsymbol{x}\rightarrow -\boldsymbol{x},\hspace{1cm}h\rightarrow -h.
    \end{aligned}
    \label{eq1:sym2}
\end{equation}
The transformations described in Eq.~\eqref{eq1:sym1} eliminate any explicit dependence on $h$, $\boldsymbol{x}$ and $t$, allowing only derivatives of $h$ to remain. However, the first transformation in Eq.~\eqref{eq1:sym2} (inversion symmetry in the substrate direction) rules out any dependence on odd spatial derivatives of $h$ like $\nabla h$. Finally, the second transformation in Eq.~\eqref{eq1:sym2} (inversion symmetry of the mean height) eliminates derivatives like $(\nabla h)^2$. Keeping only the lowest-order terms, we arrive at the so-called \textit{Edwards-Wilkinson} (EW)\nomenclature{EW}{Edwards-Wilkinson} \textit{equation}.
\begin{equation}
    \frac{\partial h(\boldsymbol{x},t)}{\partial t}=\nu \nabla^2 h+\eta(\boldsymbol{x},t),
    \label{eq1:EW}
\end{equation}
where the parameter $\nu$ is referred to as surface tension, as the Laplacian term $\nabla^2 h$ tends to smooth the interface. Note that the second symmetry condition in Eq.~\eqref{eq1:sym2} assumes that the interface is in equilibrium, where by equilibrium we mean that it is not driven by an external field, i.e. $F=0$ and the interface relaxes around its mean height. That condition, which excludes the term $(\nabla h)^2$, will no longer hold for non-equilibrium interfaces. For this reason, a constant term $v$, representing the average growth velocity of the interface, is sometimes added to this equation. However, it is often omitted, as it can be absorbed through the Galilean transformation to a reference frame that moves with the interface, $h \rightarrow h + v t$. 


The EW equation can be solved exactly (see Appendix~\ref{app:Ew_exp}), resulting in the following critical exponents for dimension $d$:
\begin{equation}
    \alpha=\frac{2-d}{2},  \ \ \ \beta=\frac{2-d}{4}, \ \ \   z=2.
    \label{eq1:EWexponents}
\end{equation}
Therefore, the discrete RDSR model and the continuous EW equation define a distinct universality class from RD, commonly known as the Edwards-Wilkinson universality class. The critical exponents associated with this class are also summarized in Table~\ref{tab1:dep_models_exp}.

For $d=2$, Eq.~\eqref{eq1:EWexponents} yields $\alpha = \beta = 0 $. In this case, the correlations exhibit logarithmic behavior, meaning that the width grows logarithmically with time at early stages, while the saturation width scales with the logarithm of the system size \cite{Barabasi1995}. For $d > 2$, the roughness exponent $\alpha$ becomes negative, indicating that the interface remains flat, i.e. the width remains constant and does not scale with either time or system size. Any noise-induced irregularity resulting in a non-zero width is suppressed by surface tension. Thus, the upper critical dimension, i.e. the largest dimension where the fluctuations are still relevant, of the EW universality class is $d_u^{\mathrm{EW}}=2$
\begin{figure}[ht]
     \centering
     \begin{subfigure}[b]{0.4\textwidth}
         \centering
         \includegraphics[height=4cm]{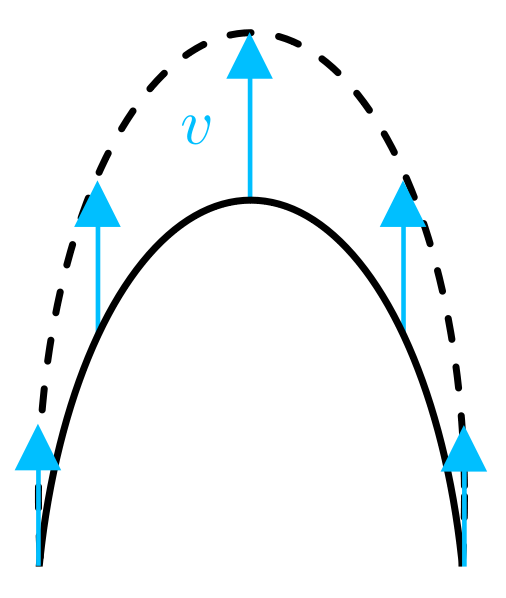}
         \caption{Sketch of an interface that grows according to the EW equation. We assume that the interface grows along the vertical axis with a velocity $v$.}
         \label{fig1:crecimiento_ew}
     \end{subfigure}
     \hspace{0.5cm}
     \begin{subfigure}[b]{0.5\textwidth}
         \centering
         \includegraphics[height=4cm]{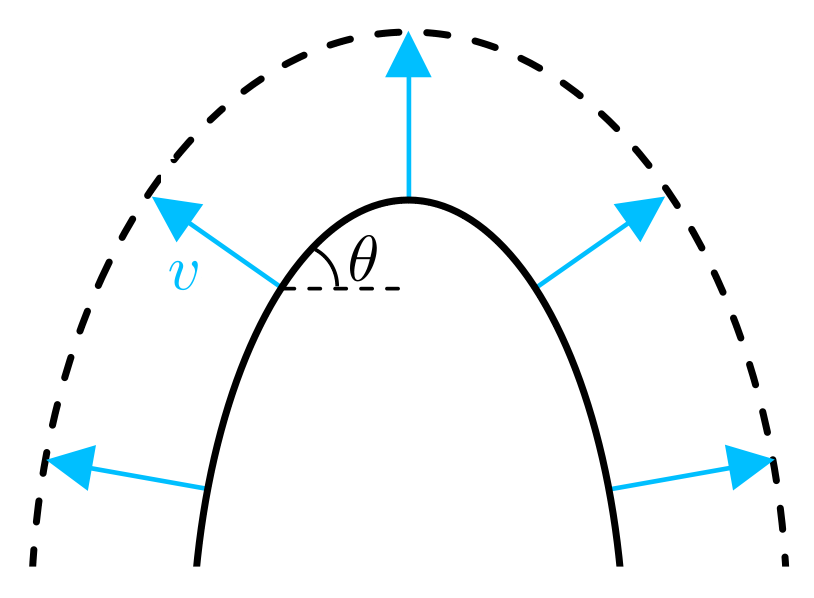}
         \caption{Sketch of an interface that grows along the local normal direction. The interface grows isotropically, so that each local piece of the interface advances in the direction normal to the interface.}
         \label{fig1:crecimiento_kpz}
     \end{subfigure}
     \begin{subfigure}[b]{0.6\textwidth}
         \centering
         \includegraphics[width=0.5\textwidth]{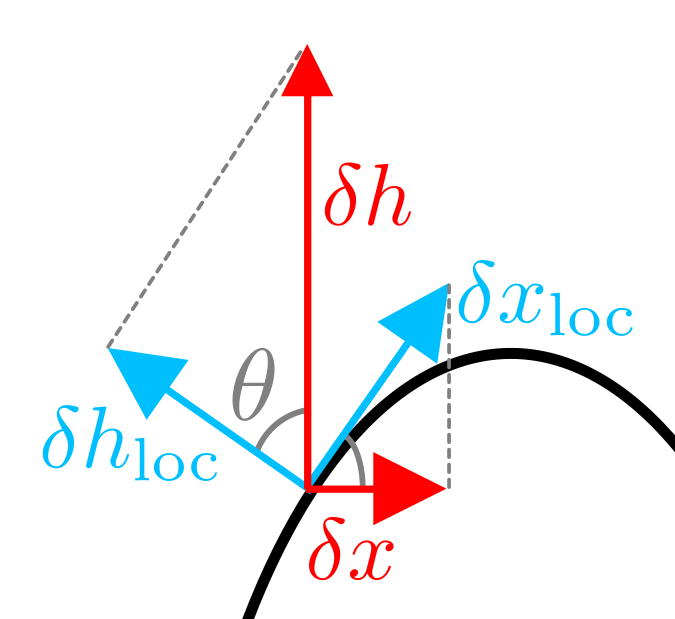}
         \caption{The local growth direction ($\delta h_{loc}$), defined as normal to the interface, is related to the growth along the vertical direction ($\delta h$).}
         \label{fig1:deltahloc}
     \end{subfigure}
        \caption{Interfaces growing along a preferred axis (a) or growing locally in the normal direction (b). Panel (c) shows the definition of the local coordinates for the interface (b). Adapted from Ref.~\cite{GarciaBarreales2024}.}
        \label{fig1:crecimiento_ew_kpz}
\end{figure}

The EW equation is the simplest linear model for describing interface growth driven by particle deposition. A nonlinear extension of this model was first proposed by Kardar, Parisi, and Zhang \cite{Kardar1986}. An example of an interface governed by the EW equation is shown in Fig.~\ref{fig1:crecimiento_ew}. Crucially, the EW equation assumes that growth occurs strictly along the vertical direction, i.e., perpendicular to the substrate (the $h$-direction). However, it is natural to consider that the interface might also advance along the local normal direction, as illustrated in Fig.~\ref{fig1:crecimiento_kpz}. As a first approximation, one can assume that the interface evolves locally according to the EW equation, in the coordinate system defined by the local coordinates $\delta x_{\mathrm{loc}}$ and $\delta h_{\mathrm{loc}}$, while producing a net increase $\delta h$ along the vertical axis (see Fig.~\ref{fig1:deltahloc}) \cite{Takeuchi2018}. Applying basic trigonometric relations, one obtains
\begin{align}
\delta h &= \frac{\delta h_{\mathrm{loc}}}{\cos \theta} = \delta h_{\mathrm{loc}} \sqrt{1 + \tan^2(\theta)} \notag \\
         &= v \delta t \sqrt{1 + \left( \frac{\delta h_{\mathrm{loc}}}{\delta x_{\mathrm{loc}}} \right)^2 } = v \delta t \sqrt{1 + (\nabla h)^2}.
\end{align}
Assuming $|\nabla h| \ll 1$, one may expand the time derivative of the front as
\begin{equation}
    \frac{\partial h(\boldsymbol{x},t)}{\partial t}=v+ \frac{v}{2} (\nabla h)^2 + ...    
\end{equation}
Substituting the right-hand side contribution into the EW equation and retaining only the lowest-order nonlinear term we obtain the so-called \textit{Kardar-Parisi-Zhang} (KPZ)\nomenclature{KPZ}{Kardar-Parisi-Zhang} \textit{equation}:
\begin{equation}
    \frac{\partial h(\boldsymbol{x},t)}{\partial t}=v+\nu \nabla^2 h+ \frac{\lambda}{2} (\nabla h)^2 +\eta(\boldsymbol{x},t),
    \label{eq1:KPZ}
\end{equation}
where $\nu$ and $\lambda$ are constants, while $\eta$ represents a noise term satisfying Eq.~\eqref{eq1:var_noise}, similar to the previous equations. 
The velocity term $v$ is typically omitted, as in the case of the EW equation.

Since $(\nabla h)^2$ is always positive, the inclusion of the new term causes the interface to rise by accumulating material when $\lambda>0$; conversely, if $\lambda <0$, material is locally removed from the interface. This behavior contrasts with the effect of the linear term, which aims to preserve the total mass by redistributing the interface height. As previously mentioned while deriving the EW equation, the inclusion of the term $(\nabla h)^2$ breaks the inversion symmetry of the mean height, which in this case does not hold due to the accumulation/removal of material. In this sense, the KPZ equation represents a genuine growth process, independent of the constant term $v$\cite{Takeuchi2018}. Moreover, it can be shown that higher-order derivatives, such as $(\nabla h)^4$, do not affect the scaling behavior in the hydrodynamic limit (for long times $t \to \infty$ and long distances $ \boldsymbol{x} \to \infty $) \cite{Barabasi1995}.

The KPZ equation has been exactly solved in one dimension, yielding the critical exponents $\beta=1/3$, $\alpha=1/2$ and $z=3/2$ \cite{Kriecherbauer2010,HalpinHealy2015,Takeuchi2018}. Notably, the scaling behavior remains unchanged regardless of the sign of $\lambda$. On the other hand, the exponents obtained numerically for the BD model align well with those of the KPZ equation. In fact, based on physical and symmetry principles, it can be demonstrated that the stochastic growth equation governing BD is precisely the KPZ equation 
\cite{Barabasi1995}. As a result, the BD model and the KPZ equation belong to the same universality class. Furthermore, the RSOS model and the Eden model are also part of this universality class, known as the KPZ universality class, which will be examined in the next section.

As a final remark in this section, it is important to note that universality classes are not solely characterized by the values of their critical exponents but also by other universal properties that help classify systems accordingly. For instance, the one-point statistics of field fluctuations or the height covariance $C_1(r,t)$, quantities that will be defined in Sec.~\ref{sec3:observables}, follow specific characteristic functions. In the next section, we will present those associated with the KPZ universality class.

\subsection{KPZ universality class}
\label{sec1:KPZ}


The KPZ universality class plays a crucial role in statistical physics, particularly in surface growth processes, as its universal behavior frequently manifests across a diverse range of systems \cite{HalpinHealy2015,Takeuchi2018}.

The KPZ exponents, whether derived analytically (for $d=1$) or estimated numerically (for $d>1$), satisfy the scaling relation
\begin{equation}
    \alpha + z=2.
    \label{eq1:KPZaz}
\end{equation}
This relation holds in any dimension and has been derived using RG methods, though it can also be obtained through scaling arguments (see Appendix~\ref{app:KPZalphaz}). Since the relation in Eq.~\eqref{eq1:zalphabeta} also applies, only one exponent is independent.

\begin{figure}[hbt]
        \vspace{0.4cm}
        \centering
        \includegraphics[width=0.75\textwidth]{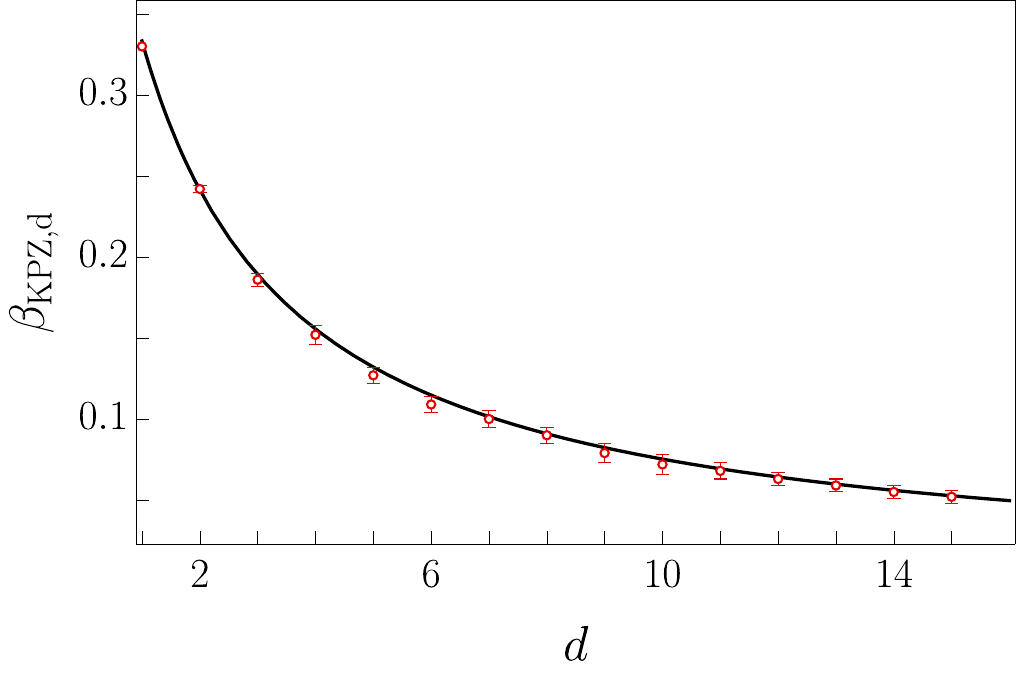}
        \caption[]{$\beta_{\textrm{KPZ},d}$ as predicted by Oliveira in Ref.~\cite{Oliveira2022}. The solid black line is the prediction of Eq.~\eqref{eq1:prediccionOliveira} and the red points are the average of exponents obtained from simulations. Note that for $d=1$, the value of $\beta$ is known exactly and therefore has no associated error. Adapted from Ref.~\cite{Oliveira2022}.}
        \label{fig1:prediccion_beta}
        \vspace{0.3cm}
\end{figure}

Determining the exact exponents of the KPZ class for any substrate dimension $d$ remains a major open challenge in statistical physics. Given the absence of analytical solutions, numerical computations have provided critical exponent values for $d > 1$. Recently, Oliveira computed the exponent $\beta$ up to $d = 15 $ through numerical simulations and using real-space RG calculations, detailed in Ref.~\cite{Oliveira2022}, proposed the following equation
\begin{equation}
    \beta_{\textrm{KPZ},d}=\dfrac{7}{8d+13},
    \label{eq1:prediccionOliveira}
\end{equation}
which holds exceptionally well. Figure~\ref{fig1:prediccion_beta} shows Oliveira's prediction for $ \beta $ alongside numerical estimates for this exponent. 

As mentioned earlier, since there is only one independent exponent, knowing $\beta$ allows the derivation of the remaining exponents for this universality class. The exact critical exponents of the KPZ class, along with their estimates from Ref.~\cite{Oliveira2022}, are also summarized in Table~\ref{tab1:dep_models_exp}.

The lack of exact exponents for the KPZ universality class leaves unresolved the fundamental issue of determining its higher critical dimension $d_u$, at which the width of the front should scale logarithmically and above which the surface should be flat. 

The analysis of the equation yields varying and contradictory predictions for $d_u$. Some studies, based on mode-coupling theory and field-theoretical approaches, suggest that $d_u<4$~\cite{Colaiori2001,Bouchaud1993,Doherty1994,HalpinHealy1990}. In contrast, renormalization group calculations indicate $d_u > 4$ \cite{Kloss2014,Kloss2014-2}, while other studies propose that $d_u$ tends to infinity \cite{Castellano1998,Castellano1998-2}. Simulations of various models within the KPZ class provide strong evidence that, if $d_u$ is finite, it is not small, as no indications of it have been observed in simulations up to $d_u>15$~\cite{Alves2016,Kim2014,Oliveira2022,Alves2014}.

Some years ago, Saberi \cite{Saberi2013} conducted simulations of various KPZ-class models on the Bethe lattice to explore the upper critical dimension of this universality class. The Bethe lattice is often used as an approximation for an infinite-dimensional system in certain cases. Due to its unique topological structure, several statistical models with interactions defined on the Bethe lattice are exactly solvable \cite{Baxter1985}. For instance, the Ising model on the Bethe lattice is exactly solvable and exhibits the same critical exponents as in the mean-field approximation \cite{Kurata1953}. In Ref. \cite{Saberi2013}, Saberi demonstrated that the width of the front followed a logarithmic scaling, leading him to conclude that the KPZ nonlinearity remains relevant even in infinite dimensions, thereby rejecting the existence of a finite upper critical dimension for the KPZ class.

Later, Oliveira \cite{Oliveira2021} re-examined the work of Saberi and concluded that he had mistakenly interpreted the standard deviation of a non-flat surface as the surface width. Furthermore, he demonstrated that certain models within the EW class, for which $ d_u^{\mathrm{EW}} = 2 $, exhibit the same type of scaling, challenging the notion that the Bethe lattice represents an infinite-dimensional system. Oliveira argued that, in the case of non-flat surfaces, height fluctuations should be measured at a single or a few surface points \cite{Alves2013,Saberi2019,Oliveira2013}, as spatial translation symmetry is lost. Consequently, the question of the upper critical dimension of the KPZ class remains unsolved.


\begin{table}[t]
\centering
\renewcommand{\arraystretch}{1.5}
\begin{tabular}{@{}cccc@{}}
\toprule
\textbf{Class} & $\boldsymbol{\beta}$ & $\boldsymbol{\alpha}$ & $\boldsymbol{z}$ \\
\midrule\midrule

RD & $\beta = 1/2$ & not defined & \\
\midrule

EW & $\beta = \dfrac{2 - d}{4}$ & $\alpha = \dfrac{2 - d}{2}$ & $z = 2$ \\
\midrule

KPZ ($d = 1$) & $\beta = 1/3$ & $\alpha = 1/2$ & $z =3/2$ \\
\midrule

KPZ ($d > 1$) & 
\multirow{2}{*}{$\beta = \dfrac{7}{8d + 13}$} & 
\multirow{2}{*}{$\alpha = \dfrac{7}{4d + 10}$} & 
\multirow{2}{*}{$z = \dfrac{8d + 13}{4d + 10}$} \\
Conjecture~\cite{Oliveira2022} & & & \\

\bottomrule
\end{tabular}
\caption{Critical exponents for the RD, EW, and KPZ universality classes.}
\label{tab1:dep_models_exp}
\end{table}

Beyond the values of the KPZ critical exponents, the one-point statistics of field fluctuations are also recognized as another universal characteristic of the KPZ universality class. The probability density function (PDF)\nomenclature{PDF}{Probability Density Function} of the front fluctuations, rescaled by the roughness [see Eq.~\eqref{eq3:flu}], follows the Tracy–Widom (TW)\nomenclature{TW}{Tracy-Widow} distribution for the one-dimensional KPZ class~\cite{Kriecherbauer2010,HalpinHealy2015}. This contrasts with the EW and RD universality classes, for which the PDFs of the rescaled fluctuations are Gaussian~\cite{Krug1997,Barabasi1995}. The TW distribution emerges within the framework of random matrix theory \cite{Anderson2009}, which explores the fluctuation characteristics of eigenvalues in matrices with randomly generated entries. Notably, matrices composed of Gaussian-distributed random numbers represent the most fundamental classes of random matrix ensembles. Gaussian ensembles are groups of randomly generated matrices with normally distributed entries, whose distributions remain invariant under various unitary transformations. These ensembles have been widely studied, not only for their analytical properties but also because their spectral properties closely resemble those of numerous systems with a large number of degrees of freedom. There are three distinct Gaussian ensembles: the Gaussian Unitary Ensemble (GUE)\nomenclature{GUE}{Gaussian Unitary Ensemble}, consisting of real symmetric matrices; the Gaussian Orthogonal Ensemble (GOE)\nomenclature{GOE}{Gaussian Orthogonal Ensemble}, composed of complex Hermitian matrices; and the Gaussian Symplectic Ensemble (GSE)\nomenclature{GSE}{Gaussian Symplectic Ensemble}, which includes quaternionic, self-dual Hermitian matrices. Tracy and Widom \cite{Tracy1994,Tracy1996} explicitly derived the distributions of the largest eigenvalue for these ensembles, denoted as $\chi_{\textrm{TW},\tilde{\beta}}$. These distributions correspond to the three Gaussian ensembles: GOE ($\tilde{\beta}=1$), GUE ($\tilde{\beta}=2$), and GSE ($\tilde{\beta}=4$)\footnote{Here, $\tilde{\beta}$ is associated with the probability density of those random matrices, which is given by: $P(M)=\frac{1}{Z}e^{-\frac{\tilde{\beta}}{2} TrM^2}$ \cite{Takeuchi2018}}. From this point forward, we will simply refer to these distributions as TW-GOE\nomenclature{TW-GOE}{Tracy–Widom distribution for the largest eigenvalue of a random matrix in the Gaussian Orthogonal Ensemble}, TW-GUE\nomenclature{TW-GUE}{Tracy–Widom distribution for the largest eigenvalue of a random matrix in the Gaussian Unitary Ensemble}, and TW-GSE\nomenclature{TW-GSE}{Tracy–Widom distribution for the largest eigenvalue of a random matrix in the Gaussian Symplectic Ensemble}. This distinction is significant because the one-point distribution of height fluctuations in the one-dimensional KPZ universality class varies depending on the global geometry of the interfaces. Specifically, the distribution differs depending on whether the front length $L$ grows or not \cite{Takeuchi2018}. In the case of a flat interface, where the front length $L$ is fixed, the PDF of rescaled front fluctuations follows the TW-GOE distribution, whereas for circular interfaces, where the front length $L$ grows, it follows the TW-GUE distribution.

\begin{figure}[h!t]
    \centering
    \includegraphics[width=1\textwidth]{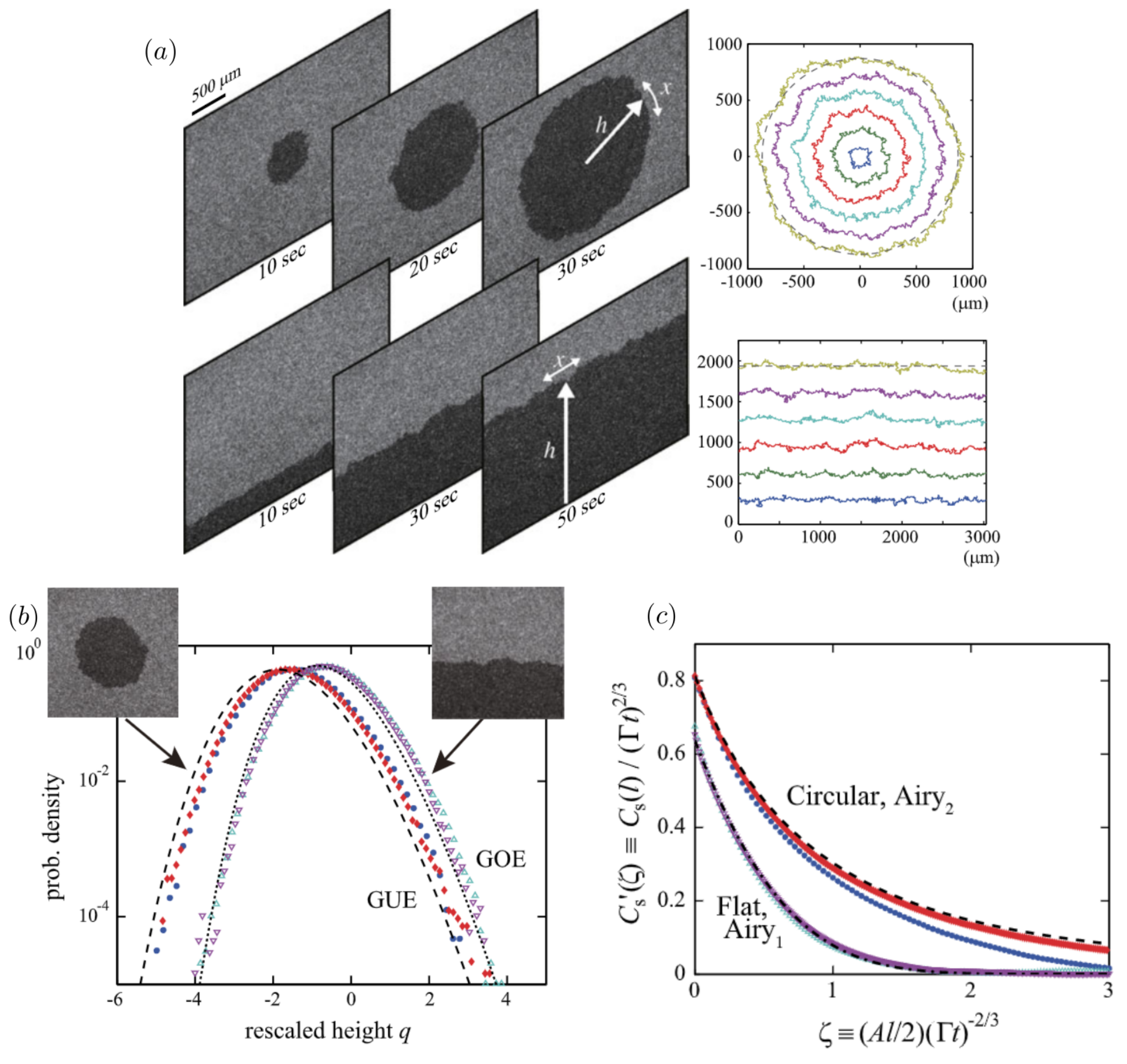}
    \caption[]{Kardar-Parisi-Zhang interfaces in liquid-crystal turbulence. (a)~Growing cluster with a circular (top) and flat (bottom) interface. (b)~The blue and red solid symbols show the histograms for the circular interfaces at \mbox{t = 10~s and 30~s}; the light blue and purple open symbols are for the flat interfaces at t = 20~s and 60~s, respectively. The dashed and dotted curves show the TW-GUE and TW-GOE distributions, respectively. (c) Rescaled correlation function. The symbols indicate the experimental data for the circular and flat interfaces, as explained in (b). The dashed and dashed-dotted lines indicate the correlation function for the Airy$_2$ and Airy$_1$ processes, respectively. See, for more details, Ref.~\cite{Takeuchi2012}. Figure reproduced from Ref.~\cite{GarciaBarreales2024}}
    \label{fig1:takeuchi2012}
\end{figure}

For example, Takeuchi \textit{et al.} \cite{Takeuchi2012} conducted experiments on turbulent liquid crystals, uncovering the influence of geometry on interface behavior. These studies examine the convection of nematic liquid crystals subjected to an electric field applied between two parallel plates. Figure~\ref{fig1:takeuchi2012}a illustrates two interfaces—one circular and one flat—where the two regimes are distinguishable, with the darker region representing the expanding area. Meanwhile, Fig.~\ref{fig1:takeuchi2012}b depicts the probability distribution of the rescaled local height for both interfaces, corresponding to the TW-GUE and TW-GOE distributions, respectively.

Moreover, regarding the height covariance $ C_1(r,t) $,
theoretical studies have demonstrated that, in the asymptotic limit, the covariance of the interface fluctuations in the one-dimensional KPZ universality class corresponds exactly to the time correlation of the stochastic Airy process. Specifically, it follows the Airy$_1$ for flat interfaces \cite{Borodin2008, Sasamoto2005} and the Airy$_2$ for curved interfaces \cite{Prahofer2002, Prolhac2011}. Bornemann \textit{et al.} \cite{Bornemann2008, Bornemann2010} have numerically estimated the correlation functions for the Airy processes. Figure~\ref{fig1:takeuchi2012}c presents the correlation functions of the Airy processes alongside experimental data from liquid-crystal turbulence for circular and flat interfaces, as reported in Ref.~\cite{Takeuchi2012}. Notably, this characteristic is also shared by the 1D EW universality class \cite{Carrasco2019}. Moreover, recent findings have shown that the covariance of the 1D EW and KPZ equations with columnar noise is identical, corresponding in these cases to that of the Larkin model for elastic interfaces in disordered media \cite{Gutierrez2023}. The universality of the two-dimensional KPZ universality class, along with its corresponding limit distributions (higher-dimensional counterparts to TW-GOE and TW-GUE), as well as the universal spatial correlations (analogous to the covariance of the Airy processes), have also been thoroughly characterized \cite{HalpinHealy2012, Oliveira2013, HalpinHealy2014}. For $d=3$, the KPZ radial class has been extensively examined in \cite{HalpinHealy2013}.

\subsection{Relation of the KPZ equation with other equations}

The significance of the KPZ equation is further underscored by its deep connections to other fundamental equations in physics. From Eq.~\eqref{eq1:KPZ}, and defining $\boldsymbol{v}=-\nabla h$ such that $\nabla\times \boldsymbol{v}=\boldsymbol{0}$, one obtains:
\begin{equation}
        \frac{\partial \boldsymbol{v}}{\partial t} + \lambda \boldsymbol{v} \cdot \nabla \boldsymbol{v}=\nu \nabla^2 \boldsymbol{v}-\nabla \eta(\boldsymbol{x},t).
        \label{eq1:burgers}
\end{equation}
This corresponds to the (stochastic) Burgers' equation, a fundamental partial differential equation commonly encountered in fluid dynamics \cite{Burgers1948}. If $\nu=0$ then this equation is known as the inviscid Burgers' (IB)\nomenclature{IB}{Inviscid Burgers} equation. Equation~\eqref{eq1:burgers} characterizes the dynamics of a viscous fluid and is frequently used to model complex phenomena such as shock waves, turbulence, and wave propagation. In this context, $v(\boldsymbol{x}, t)$ denotes the velocity field of the fluid as a function of position $\boldsymbol{x}$ and time $t$, while $\nu$ represents the fluid's kinematic viscosity. In particular, the inviscid limit is used to describe shock waves.

Moreover, from Eq.~\eqref{eq1:KPZ} and using the Cole-Hopf transformation $H(\boldsymbol{x},t)=\exp\left[ \frac{\lambda}{2\nu}h(\boldsymbol{x},t)\right]$, one gets a linear equation in $H$\cite{Kriecherbauer2010}:
    \begin{equation}
    \frac{\partial H}{\partial t}=\nu \nabla^2 H+ \left( \frac{\lambda}{2 \nu} \eta(\boldsymbol{x},t) \right)H,
    \label{eq1:cole}
    \end{equation}
which is a heat equation with a multiplicative stochastic force. In particular, if $\lambda = 0$, this equation reduces to a diffusion equation. Therefore, if $\lambda$ is not zero, the additional term can be understood as a term that creates or destroys particles, depending on the sign of the noise. These relationships highlight the pivotal role of the KPZ equation as a unifying framework in understanding complex dynamic systems across diverse fields. 

\subsection{Tensionless case of the Kardar-Parisi-Zhang equation}\label{sec1:TKPZ}

The Cole-Hopf transformation, which linearizes the KPZ equation, is not applicable in the tensionless case, i.e., when $\nu = 0$. Setting $\nu = 0$ in Eq.~\eqref{eq1:KPZ} leads to the so-called \textit{Tensionless Kardar-Parisi-Zhang} (TKPZ)\nomenclature{TKPZ}{Tensionless Kardar-Parisi-Zhang} \textit{equation}
\begin{equation}
    \frac{\partial h(\boldsymbol{x},t)}{\partial t}= \frac{\lambda}{2} (\nabla h)^2 +\eta(\boldsymbol{x},t),
    \label{eq1:KPZ_tensionless}
\end{equation}
where we have omitted the velocity term $v$ that previously appeared in Eq.~\eqref{eq1:KPZ}. Equation~\eqref{eq1:KPZ_tensionless} is marginally unstable to perturbations of a flat solution \cite{Cuerno1995}, making it particularly challenging to integrate numerically. It was only recently that numerical integration of this equation became feasible \cite{RodrguezFernndez2022} and has been found to define its own universality class, differing from the case with surface tension. Besides, it exhibits intrinsically anomalous scaling \cite{Lopez1997,Ramasco2000}. 

In $d=1$ the growth exponent has been found to be $\beta = 1$, except for the transient state in which $\beta = 1/2$, as in the RD model. Additionally, the values of the roughness and dynamic exponents are $\alpha = z = 1$, whereas the local roughness exponent has been found to be $\alpha_{\mathrm{loc}} = 1/2$. Thus, Eq.~\eqref{eq1:KPZaz} seems to hold also for the TKPZ equation.

On the other hand, the skewness and kurtosis of the TKPZ in $ d = 1 $ were found to grow over time from their Gaussian values ($ S_\mathrm{Gauss} = 0 $ and $ K_\mathrm{Gauss} = 3 $)\footnote{The definitions of skewness $S$ and kurtosis $K$ are provided in Sec.~\ref{sec3:observables}.} until reaching local maxima, at which point the PDF clearly deviates from both the Gaussian and TW-GOE distributions. After that, they reach stationary values, with the skewness returning to zero and the kurtosis settling below 3, mainly due to the distribution being flatter than the Gaussian in its central region. Remarkably, the universality class identified for Eq.~\eqref{eq1:KPZ_tensionless} has previously been reported for discrete growth models associated with isotropic percolation \cite{Asikainen2002}.

\subsection{KPZ Roughening transition}

\begin{figure}[hbt]
        \vspace{0.4cm}
        \centering
        \includegraphics[width=0.75\textwidth]{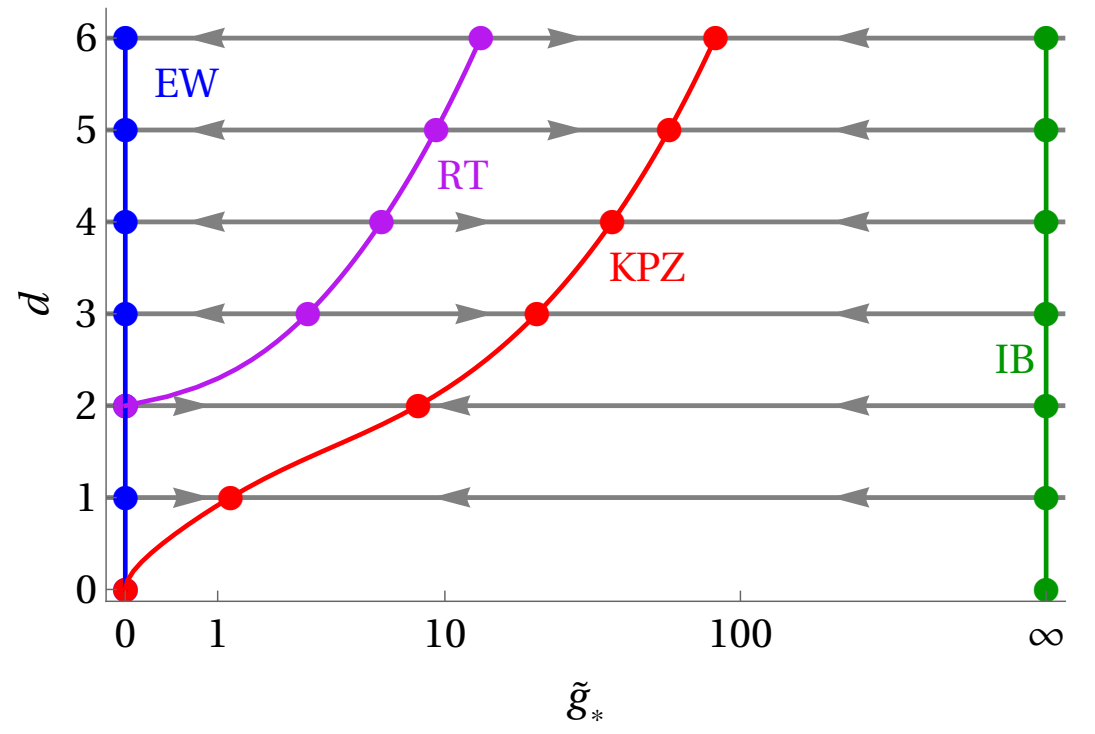}
        \caption[]{RG Flow diagram (i.e. fixed point values of $\tilde{g}_\ast$ as function of the dimension $d$) of the Burgers-KPZ equation. The dots represent the fixed points, and the arrows indicate the flows.  Reproduced from Ref.~\cite{Gosteva2024}.}
        \label{fig1:roughneningTransition}
        \vspace{0.3cm}
\end{figure}

One of the most important features of the KPZ equation is that it presents a transition between a smooth phase and a rough phase called non-equilibrium roughening transition \cite{Tang1990,Dashti-Naserabadi2017,Gosteva2024}, that have been studied through RG calculations. For $d \leq 2$, the interface always roughens and becomes scale-invariant at large distances, exhibiting a power-law behavior characterized by its critical exponents. In terms of the RG, this means that the rough phase is controlled by a non-Gaussian, fully attractive fixed point, namely the celebrated KPZ fixed point. In $d > 2$, the roughening transition (RT)\nomenclature{RT}{Roughening Transition} occurs, depending on the microscopic non-linearity $\lambda$. For $\lambda < \lambda_c$, the interface remains smooth and is described by the Gaussian EW fixed point \cite{Dashti-Naserabadi2017}. In other words, flat surfaces appear. This regime is known as the weak coupling regime. In contrast, for $\lambda > \lambda_c$, the interface becomes rough. The phase transition is continuous and controlled by a non-Gaussian fixed point with one relevant (unstable) direction. For $\lambda > \lambda_c$, the nonlinear term becomes relevant, and the system exhibits FV scaling, characterized by non-flat surfaces with KPZ exponents. The critical value $\lambda_c$ increases with the spatial dimension of the system, as shown in Fig.~\ref{fig1:roughneningTransition}.

In Fig.~\ref{fig1:roughneningTransition}, each point represents a fixed point of the RG that controls the behavior of the $d$-dimensional Burgers-KPZ equation at different scales and conditions, and $\tilde{g}_\ast$ is the rescaled (non-dimensional) coupling parameter defined as \cite{Gosteva2024}
\begin{equation}
    \tilde{g}_\ast\equiv\hat{g}_\kappa v_d,\hspace{1cm}
    v_d = \left(2^d \pi^{d/2} \Gamma(d/2)\right)^{-1},\hspace{1cm}
    \hat{g}_\kappa = \frac{\kappa^{d - 2} \lambda^2 D}{\nu^3},
\end{equation}
where $\kappa$ is the cut-off scale of the renormalization group. Arrows indicate the flow of the RG when moving from small to large scales. In this figure it can be seen that the KPZ fixed point is always attractive, and the RT and IB ($\tilde{g}_\ast\rightarrow\infty$) fixed points always unstable. The EW fixed point ($\tilde{g}_\ast\rightarrow 0$) changes stability in $d=2$, from unstable in $d \leq 2$ to stable in $d>2$~\cite{Gosteva2024}. The IB fixed point is widely acknowledged to be the same as that associated with the TKPZ equation. One of the main conclusions of Ref.~\cite{Gosteva2024} is that, in the case of the IB fixed point, the critical exponents are dimension-independent and take the values \(\alpha = 1 = z\). These exponents match those of the TKPZ equation in \(d = 1\), as discussed in the previous section. However, the analysis in Ref.~\cite{Gosteva2024} does not address the anomalous scaling observed in the TKPZ equation. Moreover, there exists a family of equations, to which the KPZ equation belongs, that, for a particular choice of parameters, also exhibit the same behavior, i.e., \(\alpha = 1 = z\) regardless of the dimension~\cite{Castro2012,Nicoli2009-2}, as well as non-anomalous (i.e., FV) scaling.

Two important conclusions can be drawn from this RG analysis. First, in high dimensions the nonlinearity of the KPZ equation manifests itself only for sufficiently large values of \(\lambda\). Second, since the IB fixed point is unstable, the only way to observe TKPZ behavior is by setting \(\nu = 0\). Any large, but finite, coupling \(\tilde{g}_\ast\) will instead lead to KPZ-like behavior.



\section{Surface growth in real-world physical systems} \label{sec1:experiments}

While this work primarily focuses on simulations of discrete models and the integration of continuum equations, it is worth emphasizing the significant connection between surface kinetic roughening theory and real growing interfaces. Numerous systems in nature demonstrate the complex dynamics of these interfaces. They are not exclusively formed through particle deposition or addition, but can also emerge via particle removal processes. As an illustration, the propagation of a burning front in paper \cite{Zhang1992} has been studied, revealing that its scaling properties in the long-time regime correspond to those of the 1D KPZ universality class \cite{Myllys2001}. Similarly, the surface growth of NiW alloy substrates obtained through electrochemical deposition \cite{Orrillo2017} was found to align with the 2D KPZ universality class. While numerous systems exhibit the behavior described in this chapter, we will focus here on a few detailed examples that are particularly relevant to the context of this thesis:


\begin{enumerate}[leftmargin=0.5cm]
    \item \textit{Spreading fronts of liquid droplets.}
    When nonvolatile liquid droplets spread over flat surfaces, they generate growth fronts that evolve over time. This complex dynamics, primarily governed by the interaction between the fluid and the substrate, leads to the emergence of a precursor film under complete wetting conditions. This film, only a few molecules thick, expands significantly faster than the macroscopic droplet and is believed to exhibit universal behavior.
    One of the main objectives of this thesis is to characterize the kinetic roughening properties of these films. A detailed discussion on the emergence of these precursor films and the various models that can be used to describe them will be presented in depth in the next chapter.

    \item \textit{Bacteria biofilms}
    Dervaux \textit{et al.} \cite{Dervaux2014} studied the formation and growth of a microbial community of the model organism \textit{Bacillus subtilis}. Figure~\ref{fig1:bacteria} illustrates the evolution of the bacterial community. The analysis of the roughness of these fronts reveals that they exhibit anomalous scaling, characterized by the critical exponents $\alpha_{\mathrm{loc}} = 0.6(1)$ and $\beta = 0.5(1)$. Although those exponents do not fall into any of the known university classes, similar exponents have been reported in the context of surface growth of metals \cite{Vazquez1996} and polymer films \cite{Zhao2000}.
    \begin{figure}[ht]
         \centering
         \includegraphics[width=0.3\textwidth]{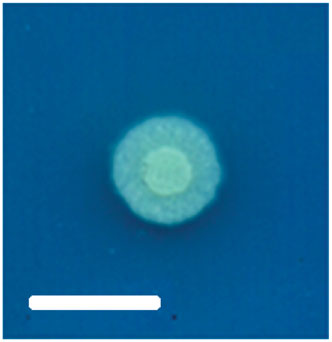}
         \includegraphics[width=0.3\textwidth]{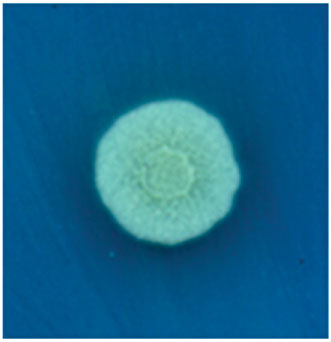}
         \includegraphics[width=0.305\textwidth]{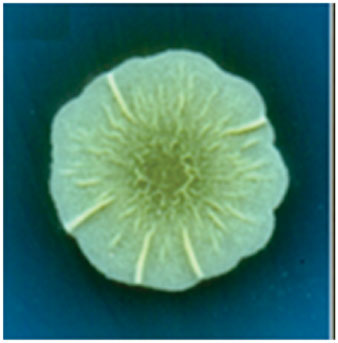}
         \caption{ Snapshots of the growth of a \textit{Bacillus subtilis} biofilm at different times. Reproduced from Ref.~\cite{Dervaux2014}.}
         \vspace{0.5cm}
            \label{fig1:bacteria}
    \end{figure}
    
    \item \textit{Silicon surfaces irradiated by ion-beam sputtering.}
    Vivo \textit{et al.} \cite{Vivo2012} studied the surfaces generated through erosion of silicon targets by ion-beam sputtering. The authors show that, by tuning the angle of incidence of the ion beam onto the surface or the average ion energy, it was possible to get surfaces with varying topographical properties, from disordered and rough to nanopatterned. Figure~\ref{fig1:silicon_surfaces} shows the morphologies of these surfaces when scanned by an electron microscope. Furthermore, the authors discovered that the surface kinetic roughening properties are spatially anisotropic, meaning the exponents associated with growth are not the same in every direction. This serves as an example of a two-dimensional ($d=2$) front. 
    \begin{figure}[h]
         \centering
         \includegraphics[width=0.49\textwidth]{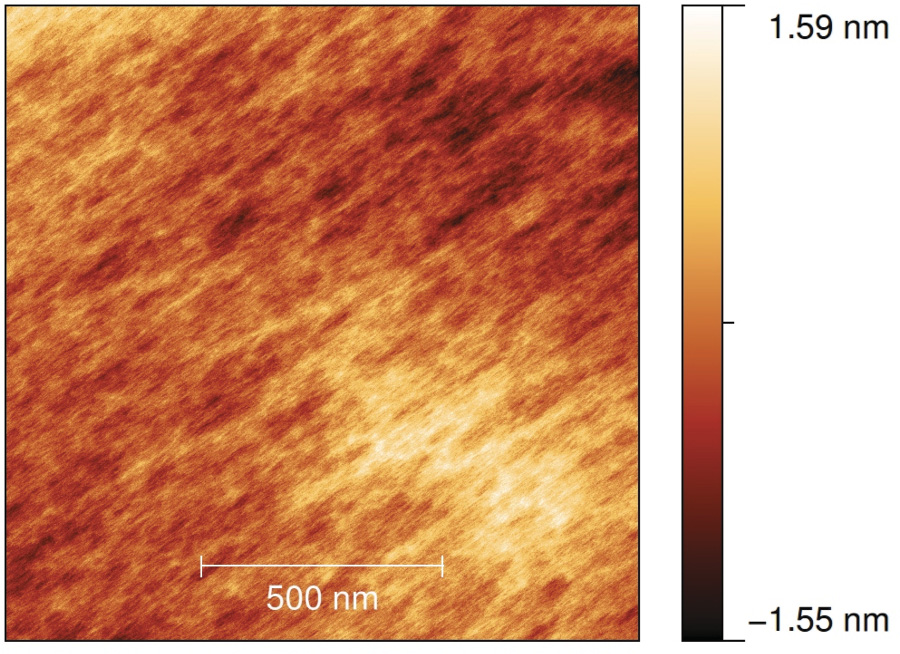}
         \includegraphics[width=0.4\textwidth]{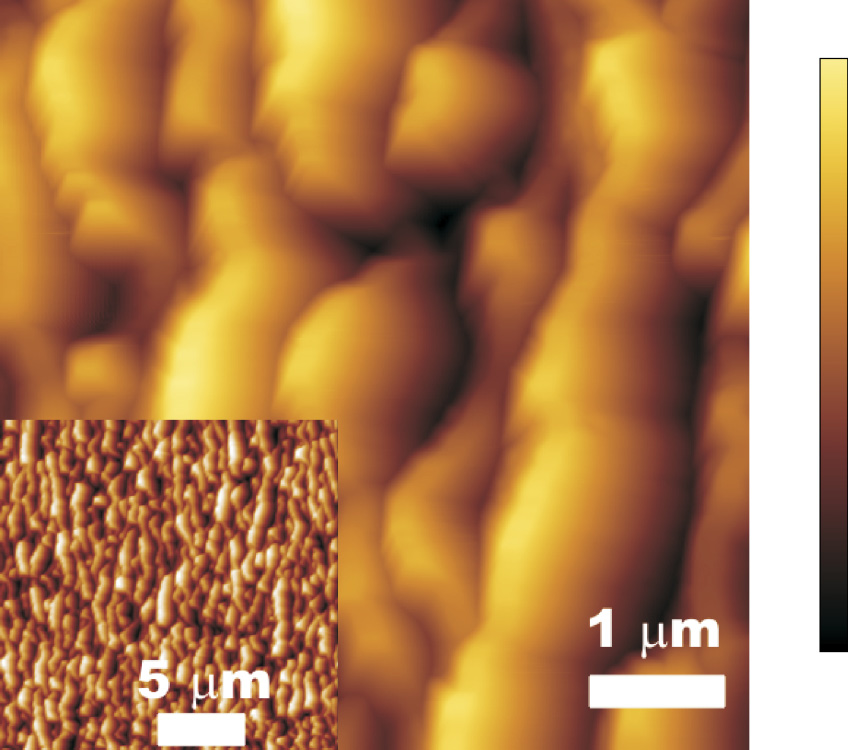}
         \caption{Morphologies of different silicon surfaces scanned by using scanning tunneling microscopy (left) and atomic force microscopy (right). Reproduced from Ref.~\cite{Vivo2012}.}
            \label{fig1:silicon_surfaces}
    \end{figure}
    \item \textit{Growth dynamics of cancer cell colonies.}
    Huergo \textit{et al.} \cite{Huergo2011,Huergo2012} studied the two-dimensional growth dynamics of HeLa cervix cancer cell colonies. The colonies spread linearly and radially in two dimensions, so in both cases the fronts are one-dimensional. Figure~\ref{fig1:cancer_cells} depicts the evolution of cell colonies spreading as a result of cell division. The front spreads along the normal directions, with fluctuations emerging due to the stochastic nature of cellular behavior. The snapshots also reveal the progressive increase in front roughness over time. The analysis of the roughness of these colony fronts yields the critical exponents \mbox{$\alpha=0.50(5)$}, $\beta=0.32(4)$, and $z=1.5(2)$ irrespective of the colony geometry. These exponents are consistent with those of the 1D KPZ universality class.
    \begin{figure}[t!]
        \centering
        \includegraphics[width=0.5\textwidth]{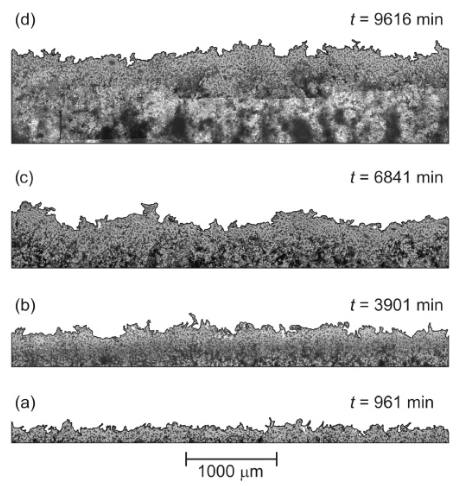}
        \includegraphics[width=0.5\textwidth]{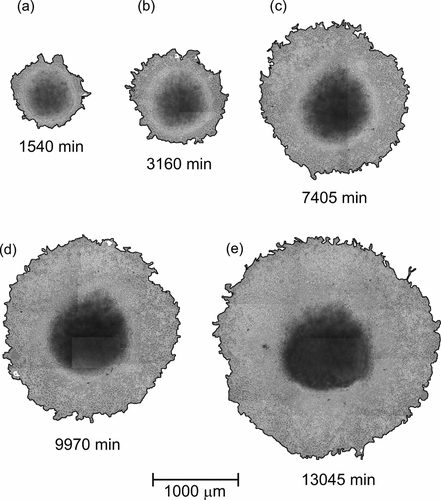}
        \caption{Snapshots of the growth of cancer cells colonies at different times for a linear and radial geometries. Reproduced from Ref.~\cite{Huergo2012}.}
        \label{fig1:cancer_cells}
    \end{figure}
    \newpage
\end{enumerate}

\graphicspath{{2_capitulo/fig2/}}

\chapter{Precursor films of wetting droplets}
\label{chap2:wetting}

In this chapter, we outline the fundamental physics of the model, whose front properties will be analyzed in Chapters \ref{chap4:band} and \ref{chap5:radial_spreading}.



Wetting and spreading play a crucial role in a wide range of applications. Wetting phenomena are ubiquitous, appearing in both natural systems and technological processes~\cite{Bonn2009}. On a large scale, the wetting or nonwetting behavior significantly impacts oil recovery \cite{Bertrand2002}, the effective deposition of pesticides on plant leaves \cite{Bergeron2000}, water drainage on highways \cite{Shahidzadeh2003}, and the cooling of industrial reactors. On a smaller scale, wetting-based solutions have been explored to address technological challenges in microfluidics, nanoprinting, and inkjet printing \cite{Tabeling2023}. Furthermore, wetting plays a vital role in the protective spin coating of various surfaces, including CDs, DVDs, glass lenses, car mirrors, and windows. It is also essential in the production of water-resistant fabrics, inkjet printing, and wall painting \cite{deGennes2004,Starov2019}. All these phenomena are primarily governed by surface and interfacial interactions, which typically act at small (a few nanometers for van der Waals or electrostatic forces) or even molecular-scale distances.

\section{Equilibrium properties}

The wetting process is governed by various surface forces, and the interactions among these forces define the possible wetting scenarios. Beyond surface chemistry, which plays a crucial role in wetting behavior, forces such as van der Waals and electrostatic interactions are essential in determining whether a fluid will wet a particular surface.

\begin{figure}[t!]
     \centering
     \begin{subfigure}[b]{0.4\textwidth}
         \centering
         \includegraphics[width=\textwidth]{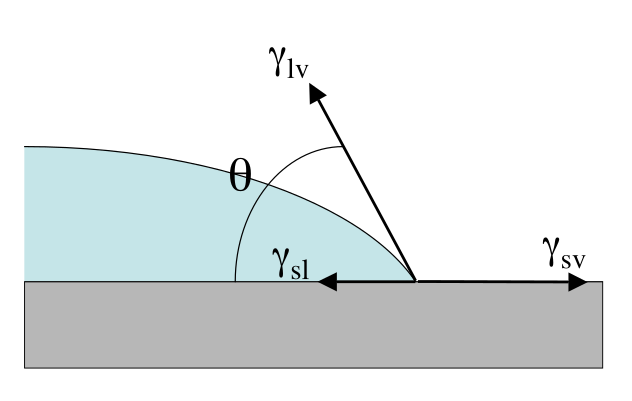}
         \caption{ Scheme illustrating the three surface tensions involved in Young's equation, along with the equilibrium contact angle.}
         \label{fig2:YoungEquation}
     \end{subfigure}
     \hspace{0.5cm}
     \begin{subfigure}[b]{0.5\textwidth}
         \centering
         \includegraphics[width=\textwidth]{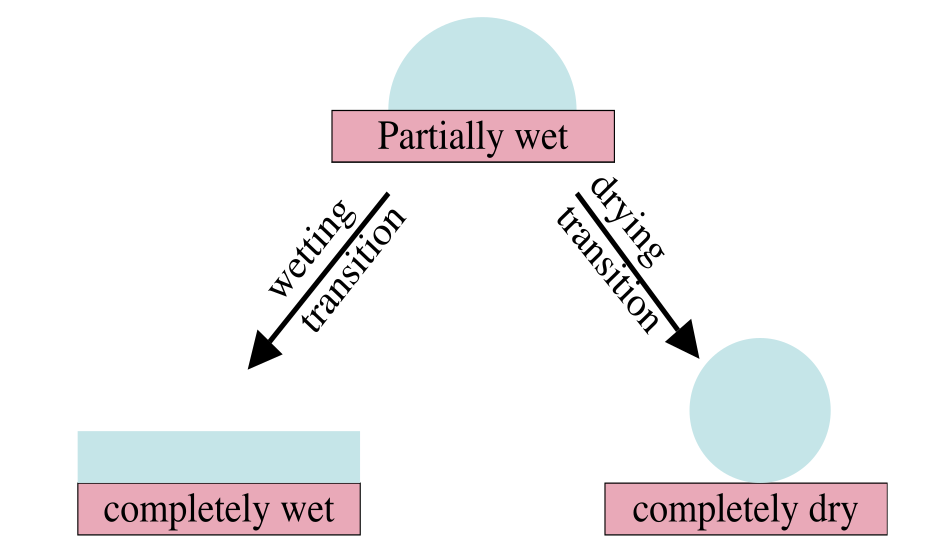}
         \caption{The three different possible wetting states according to Young’s equation.}
         \label{fig2:completeWetting}
     \end{subfigure}
        \caption{Diagram illustrating Young's equation (a) and the three possible wetting regimes (b). Reproduced from Ref.~\cite{Bonn2009}.}
        \label{fig2:statics}
\end{figure}

Beginning with the most fundamental description, which is that of the state of equilibrium, when a liquid droplet is placed on a solid substrate, three distinct phases coexist, as illustrated in Fig.~\ref{fig2:YoungEquation}. Consequently, three surface tensions must be taken into account: solid-liquid, liquid-gas, and solid-gas. The equilibrium contact angle, $ \theta_\mathrm{eq} $, formed by the droplet on the surface is governed by Young’s equation, which establishes the relationship between these surface tensions:
\begin{equation}
    \gamma_{sv} = \gamma_{sl} + \gamma \cos\theta_\mathrm{eq},
    \label{eq2:young}
\end{equation}
where $\gamma_{sv}$ and $\gamma_{sl}$ are the surface tension of the solid-vapor and solid-liquid interfaces respectively, and $\gamma_\equiv\gamma_{lv}$ is the surface tension of the liquid-vapor interface. Young’s equation can also be understood as a mechanical force balance at the three-phase contact line, where surface tension, expressed as energy per unit area, corresponds to a force per unit length acting along the contact line. In this context, the surface tensions are defined when the solid, liquid, and gas phases are in mechanical, chemical and thermal equilibrium, i.e. there is force balance, equal chemical potentials and the same temperatures for the three phases. In addition, $ \theta_\mathrm{eq} $ is understood to be measured on a macroscopic scale, beyond the influence of long-range intermolecular forces.

When the three surface tensions are known, the wetting state of the fluid can be determined directly. If $ \gamma_{sv} < \gamma_{sl} + \gamma $, the system minimizes its free energy by forming a droplet with a finite contact angle, a condition referred to as partial wetting. Conversely, if $ \gamma_{sv} = \gamma_{sl} + \gamma $, the contact angle becomes zero, leading to equilibrium when a uniform macroscopic liquid layer spreads across the entire solid surface, a state known as complete wetting. Furthermore, in a solid-liquid-vapor system, complete drying occurs when a macroscopic vapor layer intrudes between the solid and the liquid. From a thermodynamic perspective, wetting and drying are closely related, differing only in the exchange of liquid and vapor. However, in practice, drying is relatively uncommon because van der Waals forces generally work to form thin vapor layers. Figure~\ref{fig2:completeWetting} illustrates the three possible wetting regimes derived from Young's equation.

Partial wetting occurs when droplets form on the surface, surrounded by a microscopically thin adsorbed film, while complete wetting results in a macroscopically thick liquid layer. In the partial wetting state, the surface outside the droplet is never entirely dry. At thermodynamic equilibrium, at least some molecules will always be adsorbed onto the substrate.


However, when a droplet is placed on a dry substrate, it is rarely in equilibrium. Here, it is crucial to distinguish between volatile and nonvolatile liquids. For volatile liquids, thermodynamic equilibrium can be reached within a reasonable time, meaning that the substrate outside the droplet does not remain completely dry. Instead, it typically interacts with the saturated vapor phase through condensation onto the substrate, resulting in a partial wetting configuration. Even in the complete wetting regime, when volatile liquid layers evaporate under non-equilibrium conditions, both theory and experimental observations indicate that a two-phase state can emerge, where a molecularly thin film coexists with a macroscopically thick layer \cite{Bonn2009}.

This does not apply to nonvolatile liquids though, as they cannot reach thermodynamic equilibrium within the typical experimental time frame, which usually spans days or longer. Nonvolatile liquids become trapped in a metastable state with a contact angle $ \theta_i \neq \theta_{\mathrm{eq}} $ or undergo continuous spreading, flattening while maintaining their volume in the complete wetting regime. This constraint of volume conservation is crucial for understanding both the spreading dynamics and the final state of non-volatile liquid droplets. It explains why, in certain cases, a non-volatile liquid drop does not spread into an unbounded film of uniform thickness but instead halts its spreading, adopting a ‘pancake’-like shape. This occurs when short-range interactions promote dewetting, despite the overall system being in a complete wetting regime. Such structures were theoretically predicted and analyzed in \cite{deGennes1985} and were later confirmed experimentally \cite{Cazabat1994}. As a consequence of the preceding discussion, non-volatile liquids are constantly engaged in a purely non-equilibrium process.

As mentioned in the previous chapter, the liquids we are interested in analyzing statistical properties are those that are both nonvolatile and in the complete wetting regime.

\section{Spreading of nonvolatile droplets}

As discussed previously, when a droplet of nonvolatile liquid is placed on a solid surface, it is generally far from equilibrium. Consequently, a flow is initiated until the equilibrium contact angle, as given by Eq.~\eqref{eq2:young}, is achieved, provided the droplet is not trapped in a metastable state. The hydrodynamics of the macroscopic problem has been extensively studied both experimentally and theoretically (see, e.g., Refs.~\cite{Bonn2009, deGennes1985}). In the case of complete wetting, where $ \theta_{\mathrm{eq}} = 0 $, the droplet continues to spread indefinitely, eventually reaching a thickness determined by van der Waals forces.

Various spreading laws of the form $R(t) \sim t^n$, where $R(t)$ represents the radial extent of the droplet as a function of time, have been derived for different systems, depending on whether their behavior is dominated by gravity or surface tension and where dissipation occurs either at the contact line or within the bulk of the droplet \cite{Bonn2009}. The most well-known case is for $ n = 1/10 $, which describes the simplest scenario of a small, viscous droplet spreading on a completely wetting surface. In this case, the droplet is sufficiently small for gravity to be negligible, making surface tension the dominant driving force. This law is widely known as Tanner's Law \cite{Tanner1979}. For more information on the different macroscopic spreading mechanisms and the various scaling laws derived from them, see Tables I and II in Ref.~\cite{Bonn2009} and the references therein.

\begin{figure}[t]
        \vspace{0.4cm}
        \centering
        \includegraphics[width=0.75\textwidth]{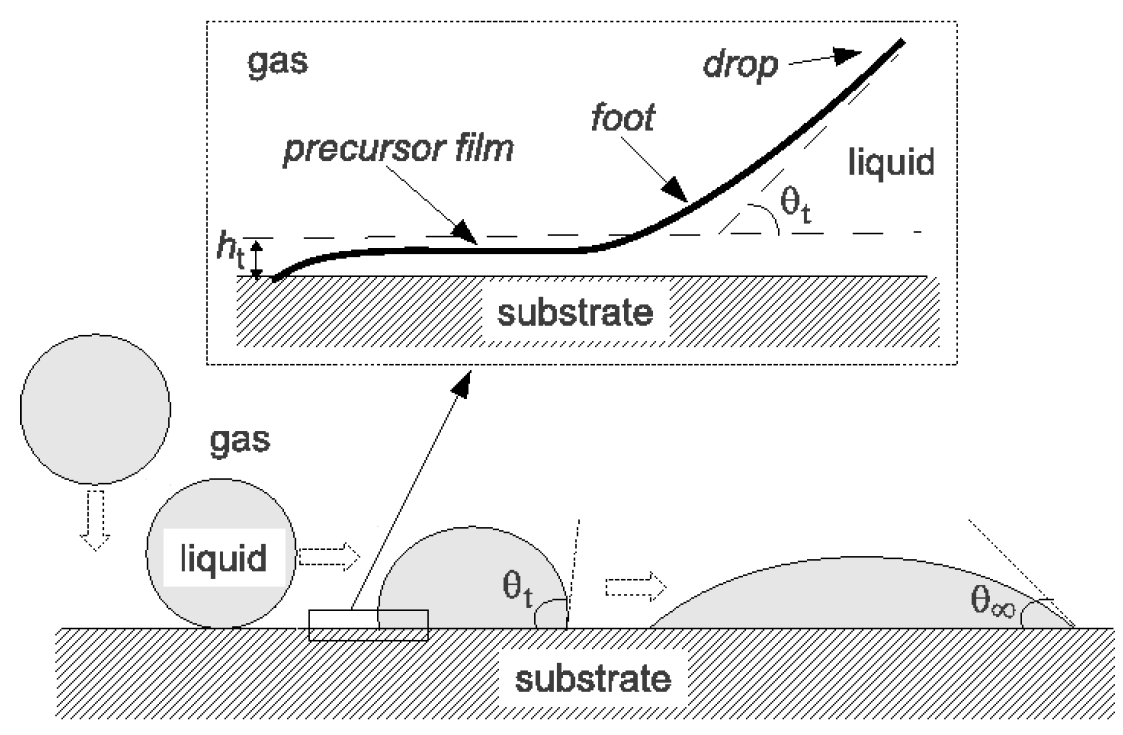}
        \caption[]{ Schematic representation of a spherically shaped droplet of a nonvolatile liquid spreading on an inert, flat, and unbounded substrate. The inset illustrates how the macroscopic spherical cap transitions into a film covering the substrate through a mesoscopic-sized ‘foot’ region, along with the emergence of a precursor film during the spreading process. Reproduced from Ref.~\cite{Popescu2012}.}
        \label{fig2:scheme}
        \vspace{0.3cm}
\end{figure}

However, beyond the macroscopic behavior of the droplet, extensive experimental evidence and theoretical insights suggest the formation of mesoscopic and microscopic films that spread ahead of the macroscopic droplet \cite{Bonn2009, Popescu2012}. Figure~\ref{fig2:scheme} shows a schematic representation of a typical configuration for a nonvolatile liquid droplet spreading on a solid substrate. The droplet can be categorized into two main regions: (i) the macroscopic bulk and (ii) the precursor film, whose thickness can range from a mesoscopic to a microscopic scale.

The earliest documented observation of an ‘invisible’ film spreading ahead of a macroscopic droplet dates back to the pioneering work of Hardy \cite{Hardy1919} more than a century ago. Studying droplets of water, acetic acid, and various polar organic liquids on clean glass and steel surfaces, Hardy discovered that a liquid film approximately one micron thick extends outward from the droplet. Notably, he observed that this process could occur independently of the droplet's own spreading.

Hardy acknowledged that he could not identify a mechanism responsible for the film being pushed out of the droplet and suggested that its spreading occurs through a steady condensation of vapor. Nearly 70 years later, compelling evidence for an evaporation–condensation mechanism was provided by Novotny \textit{et. al} in Ref.~\cite{Novotny1991}, where films with nanometer-scale thickness were observed on a plate separated by a narrow gap from the substrate, with the sessile drop in a partial wetting state. However, Hardy's evaporation–condensation mechanism is not the only possible process responsible for film formation. Experimental findings suggest that the primary film can also develop through the surface diffusion of molecules from the droplet’s edge \cite{Bangham1938,Bahadur2009}.

The thickness of the mesoscopic film has been found to vary significantly depending on the specific liquid–solid pair under study, typically reaching a few hundred angstroms. Notably, Ausserré \textit{et al.} \cite{Ausserr1986} were the first to directly visualize precursor films of this scale using polarized reflection microscopy while investigating the spreading of nonvolatile, high molecular weight polydimethylsiloxane (PDMS)\nomenclature{PDMS}{Polydimethylsiloxane} on smooth horizontal silicon wafers.

Theoretical approaches to study the mesoscopic film focus on scales ranging from approximately 30 \AA~to 1 $\mu m$ \cite{Popescu2012}. At these scales, a continuum description remains valid; however, long-range forces, primarily van der Waals interactions, become significant. As a result, interfacial tensions alone cannot fully account for the system's free energy, making it essential to consider interactions between the two interfaces, primarily solid–liquid in the case of liquid-on-solid spreading \cite{Popescu2012}. Specifically, for the case of complete wetting, a macroscopic non-volatile droplet spreads very slowly due to the balance between hydrodynamic viscous dissipation in the bulk and the driving force for spreading,  generated by surface tensions arising from the droplet's non-equilibrium shape \cite{Popescu2012}. As it spreads, the droplet gradually depletes into a mesoscopically thin film, which continues to flatten over time. Once the entire droplet volume transitions into the film, the spreading process ceases, resulting in the formation of an equilibrium ‘pancake’ \cite{deGennes1985}.

Thus, the phenomena at macroscopic and mesoscopic scales, where hydrodynamics is applicable, are well understood. However, experimental studies \cite{Beaglehole1989,Heslot1989} reveal that the spreading of non-volatile droplets of PDMS on silicon wafers is accompanied by the formation of a film with a microscopic rather than mesoscopic thickness. These phenomena will be explored in detail in the next section.

\section{Microscopic precursor films}

This section is structured into two subsections. The first presents experimental evidence on precursor films, exploring their properties and the interactions governing their behavior. The second discusses the most relevant models proposed to explain their emergence and dynamics.

\subsection{Experimental evidence}

The experimental study of films with thicknesses on the scale of just a few molecular diameters became feasible with the development of advanced techniques such as spatially resolved ellipsometry. This optical method enables precise measurement of the local thickness of ultra-thin films deposited on substrates with a refractive index $ n $ different from that of the film. When the contrast between refractive indices is significant—for instance, silicon oil ($ n = 1.4 $ for red light) on a silicon substrate ($ n = 3.8 $)—effective film thicknesses as small as 0.1 \AA~ can be detected \cite{Beaglehole1989}.

\begin{figure}[hbt]
        \vspace{0.4cm}
        \centering
        \includegraphics[width=0.75\textwidth]{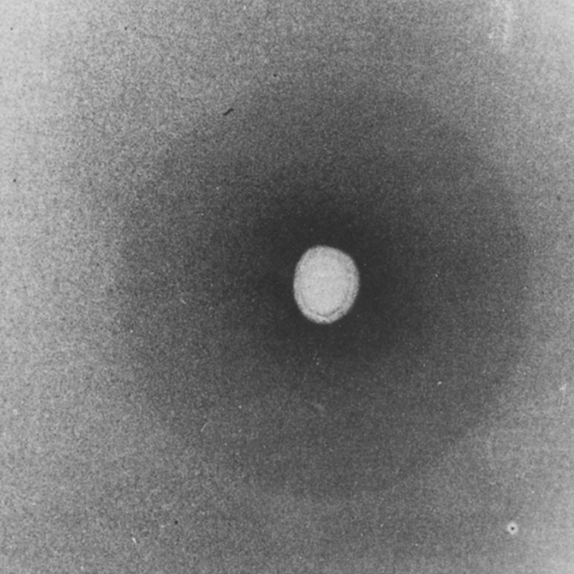}
        \caption[]{ Top view of a typical breath pattern (visible mark that appears when water vapor condenses on a substrate partially covered by a thin liquid film) observed on a spreading PDMS droplet a few days after deposition. The central spot represents the macroscopic part of the drop, while the outer ring corresponds to a thin oil film, approximately 6 to 7 \AA~ thick, at the edge. Reproduced from Ref.~\cite{Heslot1992}.}
        \label{fig2:TopView}
        \vspace{0.3cm}
\end{figure}

Using spatially resolved ellipsometry with modulated polarization, Heslot \textit{et al.} \cite{Heslot1989,Heslot1989-2,Heslot1989-3,Heslot1992} conducted a systematic analysis of the spreading dynamics of ultrathin precursor films. Their study focused on the temporal evolution of the shapes of small droplets (approximately $10^{-4} \mu l$ in volume) of non-volatile liquids, such as PDMS or squalane, spreading on silicon wafers under complete wetting conditions. For reference, Fig.~\ref{fig2:TopView} presents a top view of one of these experiments.

First, Heslot \textit{et al.} did not observe a ‘pancake’, i.e. a limited flat wetted spot with abrupt edges, as the final stage of spreading, as predicted by some theoretical works \cite{deGennes1985,Cazabat1994}; instead, they detected a gradual transition into a surface gas driven by molecular diffusion along the substrate. A surface gas is a state in which the molecules of a liquid are dispersed over a solid surface in a highly diluted form, resembling a two-dimensional gas. Second, their analysis revealed a precursor film with a nearly molecular thickness, whose radial extent increases over time as:
\begin{equation}
    R\sim\sqrt{t}.
    \label{eq2:sqrt}
\end{equation}
Moreover, Heslot \textit{et al.}\cite{Heslot1989-3} investigated the spreading speed and the number density profiles perpendicular to the substrate of a molecularly thin precursor film. This film originated from a macroscopic meniscus in a capillary rise setup, where a vertical silicon wafer covered with a natural oxide was immersed in a light silicon oil (PDMS).

\begin{figure}[hbt]
        \vspace{0.4cm}
        \centering
        \includegraphics[width=0.75\textwidth]{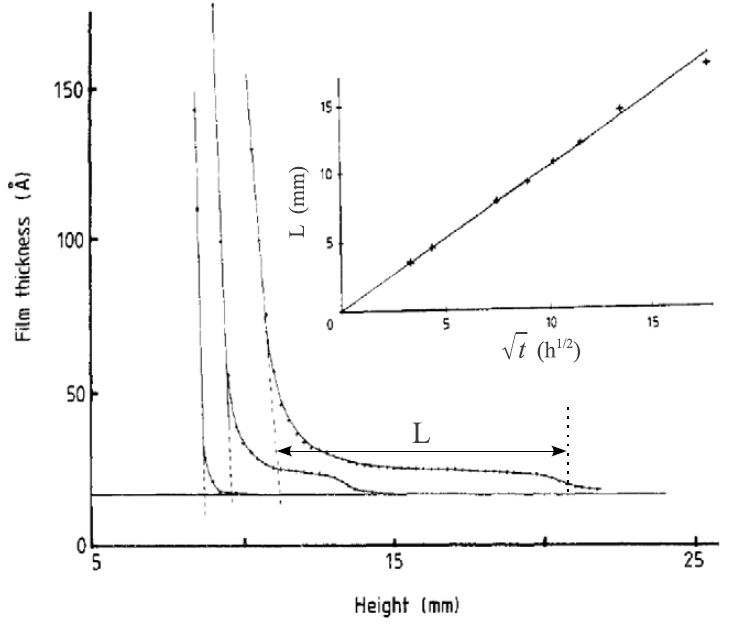}
        \caption[]{Ellipsometric profiles of films forming along a vertical wall. The thick region on the left marks the onset of the transition toward the macroscopic droplet. The $x$-axis represents the vertical distance (in mm), while the $y$-axis indicates the film thickness (in \AA). The curves, from left to right, correspond to ellipsometric thickness profiles recorded at 10 minutes, 10 hours, and 56 hours, respectively. Notably, while the macroscopic meniscus also moves upward along the vertical plate, its movement is significantly slower than that of the film. Inset: Film length $L$ (measured from the dashed vertical line) plotted as a function of the square root of time. Reproduced from Refs.~\cite{Popescu2012} and \cite{Heslot1989-3}.}
        \label{fig2:Profiles}
        \vspace{0.3cm}
\end{figure}

For most of the film in the lateral direction, the effective thickness remained nearly constant at approximately 6 \AA. Toward the tip, however, the effective thickness gradually decreased. Given that PDMS is a worm-like polymer with monomer sizes around 6 \AA, these observations suggested that the majority of the film consists of a compact monolayer of disentangled PDMS molecules lying flat on the solid surface. The lateral expansion of the film was measured over time and found to follow the scaling behavior described by Eq.~\eqref{eq2:sqrt}. Figure~\ref{fig2:Profiles} presents the ellipsometric profiles of these films, along with their growth rates observed in the experiment. Furthermore, the authors suggested that the region near the tip, where the measured effective thickness decreases to a submonolayer regime, can be interpreted as a surface gas composed of PDMS molecules. 

Furthermore, significant progress was made in \cite{Heslot1989-2}, which reported the remarkable phenomenon of 'terraced wetting'. Through spatially and time-resolved ellipsometry, it was demonstrated that liquid droplets of PDMS spreading on silicon wafers exhibit pronounced dynamic layering near the solid surface. In this process, a spreading droplet advances through a series of distinct molecular layers, each expanding proportionally to $ \sqrt{t} $ and characterized by its own diffusion coefficient. Similar phenomena, including terraced spreading or single monolayer precursor spreading with an expansion proportional to $ \sqrt{t} $, have also been observed in various complex liquids, such as liquid crystals and alkanes \cite{OuRamdane1998,Xu2000,Lazar2005}.

As previously discussed, most studies on precursor films have focused on the spreading of liquids on solid substrates. However, recent findings suggest that microscopically thin films extending over macroscopic distances also emerge in solid-on-solid wetting systems. A notable example, which has been the focus of several recent studies \cite{Moon2004,Monchoux2006}, involves metal films—such as Pb, Bi, or Pb–Bi alloys—spreading on metal substrates, including monocrystalline Cu(111), and Cu(100).

\subsection{Models}

Theoretical investigations into the physical mechanisms behind the seemingly universal $ \sqrt{t} $-law and the ‘terraced wetting’ phenomenon have taken various approaches. For clarity, we will present a selection of theoretical models, a few examples of Molecular Dynamics (MD)\nomenclature{MD}{Molecular Dynamics} simulations, and several cases of systems simulated using the Monte Carlo (MC)\nomenclature{MC}{Monte Carlo} algorithm. Among these, the discrete model of interest in this thesis is included.

De Gennes and Cazabat \cite{deGennes1991} proposed an analytical model to describe the ‘terraced wetting’ phenomenon, where a liquid drop on a solid surface is treated as a fully layered structure. In this model, the $ n $th layer is considered a quasi-two-dimensional, incompressible fluid with a molecular thickness $ a $ and a macroscopic radial extent $ R_n $. The interaction energy of a molecule within the $ n $th layer and the solid substrate is represented by a general negative function $ W_n $, which increases toward zero with the distance $ n \times a $ from the substrate.

They identified two types of flow: a horizontal, outwardly directed radial particle current and vertical permeation fluxes—one from the neighboring upper layer and another towards the adjacent lower layer. When the distinct layers expand laterally at comparable rates, their growth follows a $\sqrt{t}$ scaling. However, if the film closest to the solid substrate expands significantly faster than the layers above—effectively decoupling from the rest of the drop, which then acts as a reservoir—this model predicts that its growth follows $\sqrt{t/\mathrm{ln}(t)}$, which is slower than $\sqrt{t}$. In such a scenario, terraced wetting does not occur.

In this context, Abraham \textit{et al.} \cite{Abraham1990,Abraham1990-2} and De Coninck \textit{et al.} \cite{DeConinck1993} proposed alternative approaches to describe horizontal solid-on-solid (HSOS)\nomenclature{HSOS}{Horizontal Solid-on-Solid} layers models. However, these approaches also failed to capture the $\sqrt{t}$-law, as the growth instead follows a scaling law proportional to $\sqrt{t~\mathrm{ln}(t)}$.

\begin{figure}[t]
        \vspace{0.4cm}
        \centering
        \includegraphics[width=0.75\textwidth]{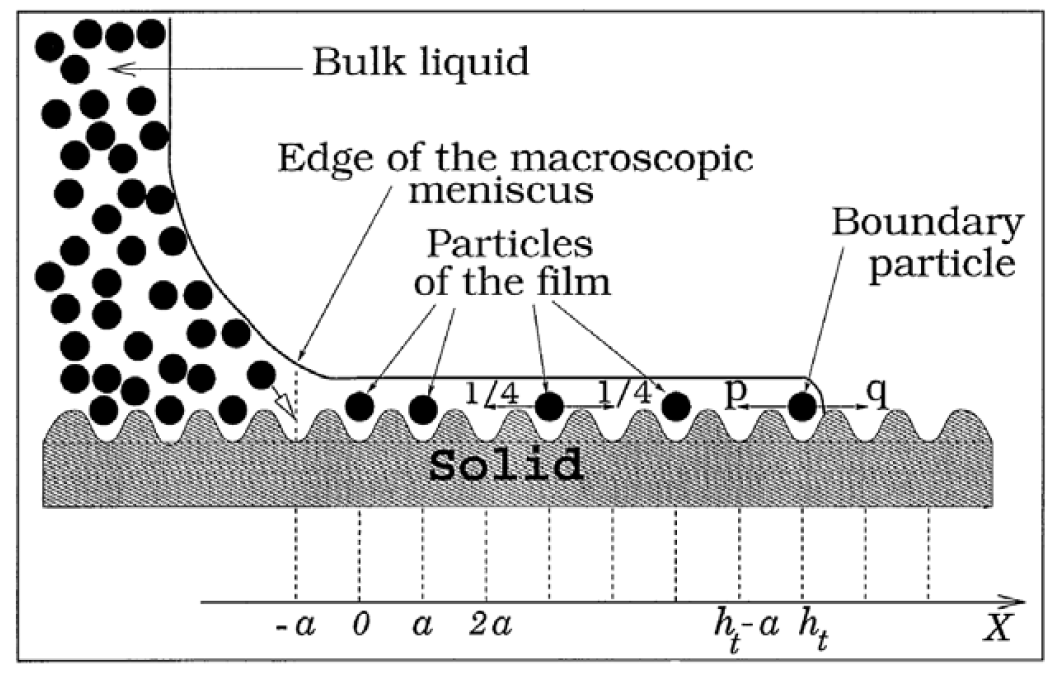}
        \caption[]{ Schematic representation of the molecularly thin precursor film spreading model proposed in Ref.~\cite{Burlatsky1996}. The setup corresponds to a capillary rise geometry, featuring a vertical two-dimensional wall immersed in a liquid bath. In this effectively one-dimensional setup, the X-coordinate represents the height above the edge of the macroscopic meniscus at the liquid–gas interface, while $ h_t $ denotes the position of the rightmost particle in the film, also referred to as the boundary particle. In this model, particles within the film are not subject to any mean force and have equal probabilities of hopping toward or away from the meniscus. For a square lattice, these probabilities were set to 1/4. Reproduced from Ref.~\cite{Burlatsky1996}.}
        \label{fig2:Burlatsky}
        \vspace{0.3cm}
\end{figure}

To address the issue arising from these models, Burlatsky \textit{et al.} \cite{Burlatsky1996,Burlatsky1996-2} introduced a microscopic stochastic model for the spreading of molecularly thin precursor films. In their approach, the film is treated as a two-dimensional hard-sphere fluid with particle–vacancy exchange dynamics. While attractive interactions between particles in the precursor film were not explicitly incorporated, they were accounted for in a mean-field-like manner. Figure~\ref{fig2:Burlatsky} presents a schematic representation of the model proposed by Burlatsky \textit{et al}. The film was assumed to be connected to an infinite reservoir, representing the bulk liquid or a macroscopic drop.

Unlike the model in Ref.~\cite{deGennes1991}, which considers hydrodynamic effects, the approach in Refs.~\cite{Burlatsky1996,Burlatsky1996-2} primarily focuses on molecular diffusion. This model assumes that the reservoir and the film are in mechanical equilibrium, eliminating any hydrodynamic pressure difference that could drive particle flow from the reservoir or push particles along the substrate away from the droplet. As a result, this approach successfully predicts the $\sqrt{t}$-law for the late-stage growth of the molecularly thin film. 

This work also clearly suggested that the physical mechanism underlying the $\sqrt{t}$-law is driven by the diffusive transport of vacancies from the edge of the advancing film to the contact line. There, these vacancies disrupt the equilibrium between the macroscopic drop and the film, leading to their filling with fluid particles from the drop.

Turning to MD simulations, this approach involves specifying particle–particle interactions, with the system's dynamics governed by the direct integration of Newton’s equations of motion. In the context of droplet spreading and precursor film formation, MD simulations have proven to be a valuable tool, providing direct insight into the molecular-scale mechanisms driving the spreading process.  

While the MD method eliminates the need for numerous assumptions, it is highly computationally demanding, requiring substantial memory and CPU time even for relatively small systems. Additionally, the exact interaction potentials are often not well known. Consequently, MD simulations of droplet propagation are generally regarded as capturing qualitative aspects rather than providing a precise quantitative description \cite{Popescu2012}.

Early MD studies successfully demonstrated the occurrence of terraced spreading but yielded conflicting results regarding the dynamics of precursor films \cite{Yang1991,Yang1992,Nieminen1992,Nieminen1994}. For instance, in Refs.~\cite{Yang1991,Yang1992}, fully atomistic MD simulations were conducted, where both the droplet and the substrate were modeled as atoms interacting via Lennard-Jones (LJ) potentials with a cut-off range comparable to the atomic diameter. 
By adjusting the strength of liquid–solid interactions while keeping fluid–fluid interactions constant, the simulation examined different wetting regimes.

These studies provided clear evidence of terraced spreading and layering within the drop's core while remaining in the liquid state. However, in all cases examined, the precursor film—corresponding to the liquid layer adjacent to the substrate—exhibited significantly slower spreading, following a $\sqrt{\ln(t)}$ scaling, rather than the behavior observed experimentally or predicted theoretically. This finding was particularly puzzling, as it did not appear to be a finite-size effect; simulations with twice as many fluid particles displayed the same behavior.

The MD studies in \cite{Nieminen1992, Nieminen1994} investigated both a pure atomic fluid and a binary mixture consisting of single-particle solvents and chain molecules. The chain molecules were composed of two, four, or eight single particles connected by a stiff, isotropic harmonic oscillator potential. All particles interacted via LJ potentials and were in contact with a homogeneous, impenetrable substrate. Additionally, the substrate exerted a van der Waals-type interaction on the particles at a distance $ z $, characterized by a potential of the form $ A/z^3 $, where $A$ is known as the Hamaker constant.

Simulations conducted at temperatures where evaporation was negligible revealed that, in most cases, a precursor film formed for atomic and diatomic molecules. However, for longer molecules with orientational degrees of freedom, layering and terraced spreading occurred only if the attractive component of the substrate potential exceeded a threshold value, which depended on the chain length. Whenever a precursor film was present, its dynamics exhibited $\sqrt{t}$ spreading behavior following a transient period associated with precursor formation.

\begin{figure}[t]
        \vspace{0.4cm}
        \centering
        \includegraphics[width=0.5\textwidth]{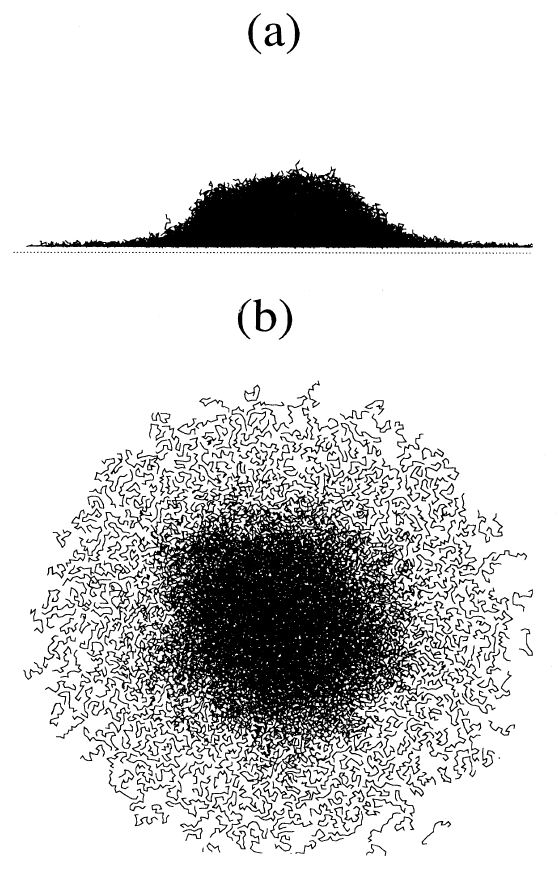}
        \caption[]{Side (a) and top (b) view of the MD simulations performed in Ref.~\cite{DOrtona1996}. The initial droplet consisted of 2000 16-atom molecules during spreading. Reproduced from Ref.~\cite{DOrtona1996}.}
        \label{fig2:MD}
        \vspace{0.3cm}
\end{figure}

The dynamics of precursor spreading has been further explored through MD simulations in Refs.~\cite{DeConinck1995, DOrtona1996}. These studies adopted the same atomistic representation of the substrate as in Refs.~\cite{Yang1991, Yang1992}, applying a cut-off at 2.5 times the fluid core size for all LJ pair potentials. Additionally, an $ A/z^3 $ term was included in the substrate potential. Figure~\ref{fig2:MD} presents top and lateral snapshots from the simulations conducted in Ref.~\cite{DOrtona1996}.
These studies, however, utilized chain molecules composed of eight or 16 atoms, bound together by a confining pair potential. This approach minimized evaporation and eliminated the size similarity between the solid and fluid species. The simulations demonstrated the formation of a well-defined first-layer precursor film, along with up to three additional layers that spread significantly more slowly. The dynamics of the first layer exhibited a clear $\sqrt{t}$ behavior, suggesting that the previously reported $\sqrt{\ln(t)}$ scaling must be linked to the size of the fluid particles used in those earlier studies.

As for MC simulations, numerous simulations have been performed to investigate droplet spreading, examining different wetting regimes, precursor film dynamics, and the effects of substrate interactions \cite{Abraham1991, Abraham1991-2, Heini1992, Cheng1993, DeConinck1993-2, Lukkarinen1995, Abraham2002}.

MC simulations of the HSOS model for a liquid wedge have been carried out in both two and three dimensions, exploring various values of surface tension and different cut-off ranges for a van der Waals-type substrate potential, which decays as $1/z^3$ with the distance $z$ above the planar continuum substrate \cite{Abraham1991, Abraham1991-2, Heini1992}. With the exception of specific parameter values where the results remain inconclusive due to extremely slow dynamics, the simulations provided strong evidence that the first layer spreads with a linear time dependence. These findings were seen as evidence of the limitations of the HSOS model, suggesting that it was overly simplistic. 

A few years later, Lukkarinen \textit{et al.} \cite{Lukkarinen1995} introduced a three-dimensional Ising model to describe droplet spreading upon contact with a planar substrate. This model is of special interest in our study, since it is the predecessor of the one we are going to study in this thesis. This model is a lattice gas representation that exhibits the spreading of an ultrathin precursor film. It can be viewed as a microscopic counterpart to the continuum model of permeation layers proposed by de Gennes and Cazabat \cite{deGennes1991}. The Ising lattice gas model incorporates nearest-neighbor interactions within an external field generated by the substrate potential and is defined on a cubic lattice of infinite extent in the $x$- and $y$-directions, with a finite extent along the positive z-direction.

In this model, the spins, characteristic of the Ising model, are replaced by occupancy numbers $ n(\boldsymbol{r}) $, which can take values of $ n(\boldsymbol{r}) = 0 $ if the lattice cell is empty or $ n(\boldsymbol{r}) = 1 $ if the cell is occupied. In fact it can be mapped back to the Ising model spins by performing the transformation $n_i=(s_i+1)/2$. Moreover, the sites with $ z < 0 $ correspond to a continuous substrate and cannot be occupied by fluid particles. 
The model is defined by the following Hamiltonian:
\begin{equation}
    \mathcal{H}= -J \sum_{\langle \boldsymbol{r}, \boldsymbol{s} \rangle} n(\boldsymbol{r},t)n(\boldsymbol{s},t) - A \sum_{\boldsymbol{r}}\frac{n(\boldsymbol{r},t)}{z^3}.
    \label{eq2:energy}
\end{equation}
The first term represents a strong nearest-neighbor attraction $\left( J/k_B T \gg 1 \right)$ to ensure low volatility, while the second term accounts for the van der Waals attraction exerted by the substrate. In the simulation, particle conservation is maintained through Kawasaki spin-exchange dynamics, where opposite spins on neighboring sites can swap positions, with a transition probability $P$ dependent on the energy change $\Delta \mathcal{H}$.

\begin{figure}[t]
        \vspace{0.4cm}
        \centering
        \includegraphics[width=0.75\textwidth]{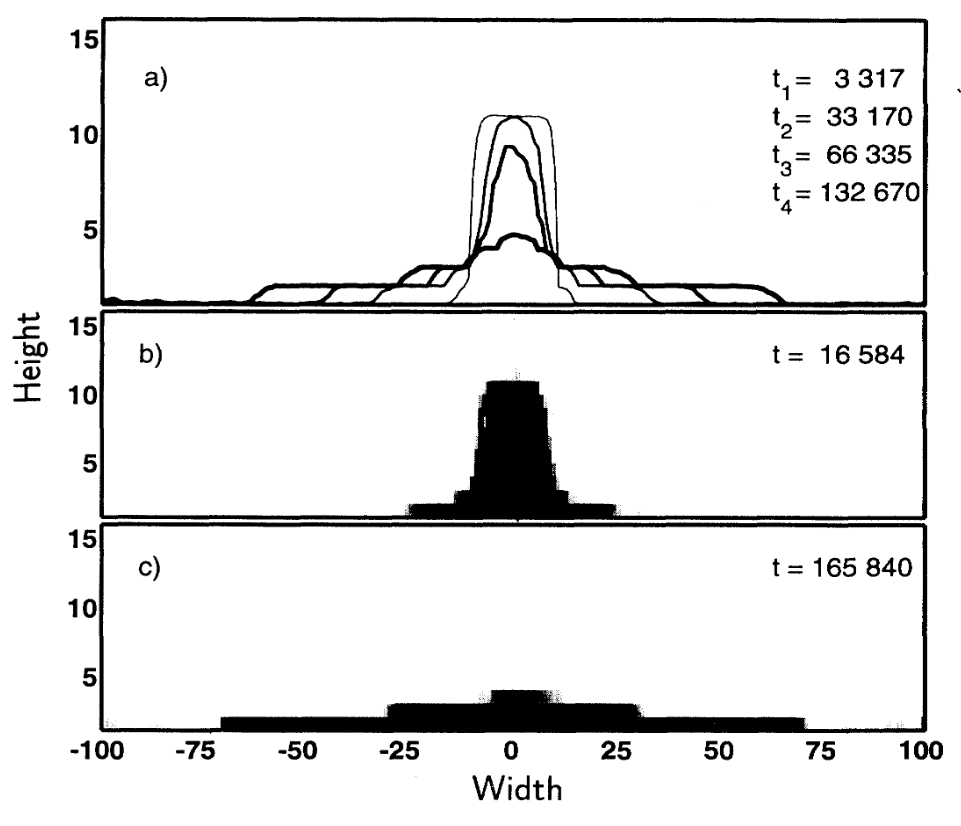}
        \caption[]{ Lateral views of the model proposed in Ref.~\cite{Lukkarinen1995}. Panel a) shows the temporal development of the droplet profile, while panels b) and c) shows the average densities profiles of the droplet for two different times. Reproduced from Ref.~\cite{Lukkarinen1995}.}
        \label{fig2:lukka}
        \vspace{0.3cm}
\end{figure}

The initial state consisted of a rectangular fluid ridge positioned at the center of the system, extending along the $y$-direction, with periodic boundary conditions (PBC)\nomenclature{PBC}{Periodic Boundary Conditions} imposed in this direction. The spreading occurs along the $x$-direction. Figure~\ref{fig2:lukka} shows various lateral views of the model.
The authors showed that, after a transient period dependent on the strength of the substrate potential, the spreading of each layer followed a $\sqrt{t}$ time dependence.

Additionally, the authors examined various mechanisms of particle transport. Near the surface of the droplet, particles tend to move downward toward the attractive substrate due to van der Waals interactions. Besides, particles initially bonded to the reservoir can detach either by evaporating into the vacuum, migrating across the substrate, or moving atop a molecular layer closer to the substrate.

The model also explored the mechanism by which the precursor film formed. Although it is possible that a particle may evaporate from the droplet and travel to the edge of the system and then fall (under the substrate attraction) into the growing edge, it is highly unlikely due to the strong nearest-neighbor interaction. 

The authors demonstrated that, at sufficiently long times, the growth of the precursor film is mainly driven by two key mechanisms:  
\begin{itemize}
    \item Holes in the precursor film that migrate backward toward the macroscopic droplet, where they were filled, driven by the van der Waals interaction.
    \item Particles in the second layer that diffuse until they reach either the edge of the precursor film or a hole within it, which they then fill.
\end{itemize}
Finally, the study revealed that, for long times, the third layer tends to shrink due to the finite number of particles in the system.

\begin{figure}[t]
        \vspace{0.4cm}
        \centering
        \includegraphics[width=0.75\textwidth]{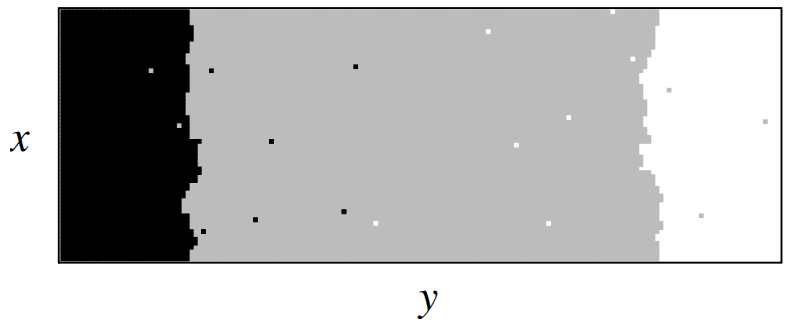}
        \caption[]{ Top view of a typical snapshot of the model proposed in Ref.~\cite{Abraham2002}. Occupied cells in $z=1$ (precursor layer) are in gray, while occupied cells in $z=2$ (supernatant layer) appear in black. Noncolored cells are empty. Parameters used were $A=10$, $J=1$, and $k_BT=1/3$. Reproduced from Ref.~\cite{Abraham2002}.}
        \label{fig2:PRL}
        \vspace{0.3cm}
\end{figure}

Building on this, Abraham \textit{et al.} \cite{Abraham2002} proposed a model that focused exclusively on the two dominant layers and incorporated a reservoir-like boundary condition (BC), akin to the previous model by Burlatsky \cite{Burlatsky1996}, to supply particles to the films. Moreover, the authors reported that the model follows the universal $\sqrt{t}$-law with even greater precision than observed in the 3D simulations of \cite{Lukkarinen1995}.
This model limits the system vertical layers to just two: $ z=1 $ (precursor layer) and $ z=2 $ (supernatant layer), making it a quasi-two-dimensional model. Figure~\ref{fig2:PRL} presents a top-view snapshot of the system showing these two layers. 

In this model, the reservoir that simulates the macroscopic droplet supplying particles to the films is represented by the first column at $ y = 1 $. Initially, only this column is occupied. If any cell in this column becomes empty due to an exchange in the Kawasaki algorithm, it is immediately refilled. Consequently, although the Kawasaki algorithm conserves particles, the BC introduces new particles into the system, driving the growth of the films.

To analyze the dynamics of the precursor edge, spin-percolative definitions were required. The precursor film was therefore defined as the set of neighboring particles at $z=1$ that are connected through nearest-neighbor bonds back to the reservoir at $y=1$. Thus, the precursor front was defined as the set of points $ y = h(x,t) $ for $ x = 1, \dots, L_x $, where $ h(x,t) $ corresponds to the maximum $ y $-value among the cells $(x, y)$ that are part of the precursor film.

The same reservoir definition will be used in Chapter \ref{chap4:band}, where we conduct a detailed study of the system in a band geometry. However, in Chapter \ref{chap5:radial_spreading}, where we examine the same system in a circular geometry, this BC will need to be reconsidered in a non-trivial manner. The fronts produced by this model are the ones that we are going to analyze in Chapters \ref{chap4:band} and \ref{chap5:radial_spreading} using the arguments and techniques presented in Chapter \ref{chap1:intro}, with a detailed examination provided in Chapter \ref{chap3:methods}. While previous studies have explored this or similar models using parameters that reflect realistic conditions, such as $ J/k_B T > 1 $ to ensure low volatility and $ A/k_B T > 1 $ to maintain the system in the complete wetting regime, we aim to simulate this system across a broader range of parameter values to uncover its universal properties. 

An important note regarding this type of fluid modeling is that it provides a statistical, rather than a fully atomistic, description of the fluid. In this context, a `particle' should be understood as a group of fluid molecules rather than a single one. The presence or absence of a particle at a lattice site corresponds to an increased or decreased probability of finding fluid molecules at that location. For a detailed discussion, see Ref.~\cite{Chalmers2017} and references therein. The main advantage of this modeling approach is that, despite its simplified nature, it allows for efficient investigation of the system’s interfacial scaling properties while still capturing the essential structural and thermodynamic characteristics of the fluid \cite{Areshi2019}.


\graphicspath{{3_capitulo/fig3/}}

\chapter{Methodology}\label{chap3:methods}


\section{Monte Carlo}\label{sec3:mc}

The Monte Carlo method \cite{Newman1999} is a fundamental tool for simulating complex systems in various fields of physics and other disciplines. Its ability to statistically sample system configurations allows for the precise study of thermodynamic and dynamic properties. However, it is important to distinguish between equilibrium and non-equilibrium simulations: while equilibrium simulations explore states consistently with the Boltzmann distribution, non-equilibrium systems require a specialized approach to capture temporal evolution and irreversible effects, posing additional challenges in interpreting the results.

In equilibrium, the usual goal of any MC simulation of a system is to compute the expected value $\langle Q \rangle$ of some quantity $Q$, such as the internal energy or the magnetization of a system. The ideal way to perform this calculation is by averaging the quantity of interest over all states $\mu$ of the system, weighting them according to their Boltzmann probability,

\begin{equation}
 \langle Q \rangle=\frac{\sum\limits_{\mu} Q_{\mu} e^{-\beta E_{\mu}}}{\sum\limits_{\mu} e^{-\beta E_{\mu}}},
 \label{eq3:qmedia}
\end{equation}
where $ Q_{\mu} $ and $ E_{\mu} $ are the values of the quantity $ Q $ and the energy in state $ \mu $, respectively, and $ \beta = \frac{1}{k_BT} $, where $ k_B $ is the Boltzmann constant and $ T $ is the temperature.

However, the expected value \eqref{eq3:qmedia} can only be calculated exactly for very small systems. If one considers a system that better represents real physical conditions, the number of states becomes so large that averaging over all states is not feasible. For example, in a simple two-dimensional spin system, where spins can only take two values, of size $10 \times 10$, the total number of states is on the order of $10^{30}$.

For these cases, the MC method approaches this problem through sampling. That is, the MC method attempts to evaluate the expression \eqref{eq3:qmedia} by averaging only over a subset of states $\left\{\mu_{1}, \ldots, \mu_{M}\right\}$. Obviously, this set of states cannot be chosen arbitrarily, as not all states of the system are equally probable. Any MC method must satisfy two fundamental conditions. The first is the ergodicity condition, meaning that all states of the system must be accessible by the algorithm. The second is that the sampling must generate configurations that follow the Boltzmann distribution, meaning that their probability is
\begin{equation}
p_{\mu}=\frac{1}{Z}e^{-\beta E_{\mu}},
\end{equation}  
where $ Z = \sum\limits_{\mu} e^{-\beta E_{\mu}} $ is the partition function. Once these $M$ states have been chosen, the estimation for $ \langle Q \rangle $ is simply
\begin{equation}  
Q_{M} = \frac{1}{M} \sum_{i=1}^{M} Q_{\mu_{i}} .
\end{equation}
The usual way to choose the states $\{\mu_{1}, \ldots, \mu_{M}\}$ is through Markov chains. A Markov chain is a sequence of random variables $ \{\mu_{t}\} $ such that the transitions $ \mu_{t} \rightarrow \mu_{t+1} $ and $ \mu_{t+1} \rightarrow \mu_{t+2} $ and so on are statistically independent. If the probability of transitioning from state $ \mu $ to state $ \nu $ is $ P(\mu \rightarrow \nu) \geq 0 $, with the normalization condition $ \sum\limits_{\nu} P(\mu \rightarrow \nu) = 1 $, then, it can be shown that under certain conditions, such as ergodicity, the chain will eventually reach a stationary distribution $ p_{\nu} $ if the following condition, known as detailed balance, is satisfied:
\begin{equation}
p_{\mu}P(\mu \rightarrow \nu)=p_{\nu}P(\nu \rightarrow \mu).
\label{eq3:balanceDetallado}
\end{equation}

If the probability distribution we want to sample is the Boltzmann distribution, i.e. $ p_{\mu} = e^{-\beta E_{\mu}} $, then the condition reduces to
\begin{equation}
\frac{P(\mu \rightarrow \nu)}{P(\nu \rightarrow \mu)}=\frac{p_{\nu}}{p_{\mu}}=e^{-\beta(E_{\nu}-E_{\mu})}.
\end{equation}
Thus, the transition probabilities $ P(\mu \rightarrow \nu) $ must be chosen to satisfy this condition. 
It is common to separate the probabilities $ P(\mu \rightarrow \nu) $ into the product of two others
\begin{equation}
P(\mu \rightarrow \nu)=g(\mu \rightarrow \nu) A(\mu \rightarrow \nu),
\label{eq3:tasasAcep}
\end{equation}
where $ g(\mu \rightarrow \nu) $ is the probability of proposing the transition $\mu \rightarrow \nu$, and $ A(\mu \rightarrow \nu) $ is the probability of accepting it.

Any choice that satisfies the detailed balance condition, Eq.~\eqref{eq3:balanceDetallado}, and is ergodic will correctly sample the Boltzmann distribution. However, the most standard way to choose them is known as the Metropolis algorithm. This approach can be applied whenever the transition selection probabilities are symmetric, meaning that $ g(\mu \rightarrow \nu) = g(\nu \rightarrow \mu) $. This condition is not very restrictive and is satisfied in most simple algorithms that simulate the behavior of physical systems, such as the spin-flip algorithm or the spin-exchange algorithm. In this case, the detailed balance condition simplifies to
\begin{equation}
\frac{P(\mu \rightarrow \nu)}{P(\nu \rightarrow \mu)}=\frac{A(\mu \rightarrow \nu)}{A(\nu \rightarrow \mu)}=e^{-\beta(E_{\nu}-E_{\mu})}.
\end{equation}
Metropolis' proposal for the acceptance rates is
\begin{equation}
A(\mu \rightarrow \nu)=\left\{\begin{array}{cc}{e^{-\beta \Delta E}} & {\Delta E>0}, \\ {1} & {\Delta E \leq 0},\end{array}\right.
\label{eq3:metropolis}
\end{equation}
that is, if the change reduces the system's energy, it is always accepted, and if it increases the energy, it is accepted with a certain probability. This probability decreases for large energy differences and for low temperatures.

\subsection{Kawasaki dynamics}

Kawasaki, or spin-exchange, dynamics arises from the study of the conserved order parameter (COP)\nomenclature{COP}{Conserved Order Parameter} Ising model, in which the magnetization remains constant. To achieve this, the system state is updated by selecting two cells (spins) and swapping their values. The proposal for the final state $ \nu $ satisfies
\begin{equation}
g(\mu \rightarrow \nu)=g(\nu \rightarrow \mu)=\frac{1}{N_{p}},
\end{equation}
where $ N_{p} $ represents the number of exchangeable pairs in the system. This number depends only on the geometry of the system and not on the values of the cells. For this reason, it is the same in both directions. 

There are two versions of this algorithm: the local and the non-local ones. In the non-local algorithm, any two cells in the system are randomly selected, whereas in the local algorithm, two neighboring cells are selected. Although both algorithms can be used to sample the Boltzmann distribution and thus perform equilibrium simulations, we will later see that for a realistic non-equilibrium simulation, only the local algorithm can be considered.

In the model we aim to study, where cells are either occupied by particles or empty, this dynamic preserves the total number of particles in the system rather than the magnetization.

\subsection{Continuous-time rejection-free algorithm}

In the context of MC simulations, it is common for the algorithm to become trapped in an energy minimum, specially for simulations performed at low temperatures, due to to very low acceptance rates, making the probability of escaping from a given state $\mu$ extremely small. In such cases, numerous change proposals are required before a transition is accepted and a new state is reached.

This type of scenario makes the algorithm extremely inefficient. In such situations, it is possible to introduce a continuous definition of time, which improves the simulation's performance. The main idea of the approach is to choose a possible future state $ \nu $, accessible from $ \mu $, and always accept it. Then, update the time continuously to account for the expected time the system would have remained trapped in state $ \mu $. The probability of staying in state $ \mu $ for $ t $ steps is
\begin{equation}  
\left[P(\mu \rightarrow \mu)\right]^t = e^{t \log P(\mu \rightarrow \mu)}  
\end{equation}  
and therefore, the timescale that determines how time should be updated is given by \cite{Newman1999}:
\begin{align}
\Delta t &= \frac{-1}{\log P(\mu \rightarrow \mu)} = \frac{-1}{\log \left[1-\sum_{\nu\ne\mu}P(\mu \rightarrow \nu)\right]} \approx \label{eq3:tiempocontinuo1} \\
&\approx \frac{1}{\sum_{\nu\ne\mu}P(\mu \rightarrow \nu)} = \frac{1}{\sum_{\nu\ne\mu}g(\mu \rightarrow \nu)A(\mu \rightarrow \nu)} = \frac{N_P}{\sum_{\nu\ne\mu}A(\mu \rightarrow \nu)}. \nonumber
\end{align}
where the logarithm has been approximated since, as $ P(\mu \rightarrow \mu) \approx 1 $, it follows that the net probability of escaping the $\mu$ state is $ \sum\limits_{\nu\ne\mu} P(\mu \rightarrow \nu) \ll 1 $. 

The interval $\Delta t$ gives an estimate of the typical number of MC steps the simulation will be trapped in that state. Although the MC steps are a discrete variable, it is reasonable to treat $ \Delta t $ as a continuous variable, given that this number is expected to be very large.

Note that the time advance has nothing to do with which transition is chosen. The time to escape the $ \mu $ state depends only on the transition probabilities. However, to satisfy the detailed balance condition \eqref{eq3:balanceDetallado}, the states $ \nu $ must be chosen in proportion to the probabilities $ P(\mu \rightarrow \nu) $. This implies that, in order to perform a continuous-time simulation, all possible transitions, and their probabilities, between the current state $ \mu $ and the potential future states $ \nu $ must be known at each step of the algorithm.

Although this may seem computationally expensive, it can often be significantly reduced by considering the characteristics and dynamics of the system being studied. For instance, in our case study, as we will detail below, the acceptance rates depend solely on the energy difference between states. Since we are going to consider only local transitions, most acceptance rates for potential exchanges remain unchanged after an exchange occurs. Therefore, it is possible to create a list where all possible transitions from state $ \mu $ to all potential future states $ \nu $ with their respective acceptance rates are stored. Then, when an exchange occurs, only a few transitions need to be added, removed, or updated in the list. This algorithm was originally proposed by Bortz, Kalos, and Lebowitz \cite{Bortz1975} to study the Ising model near the critical temperature, and it is therefore also known in the literature as the BKL algorithm.

The time update described by Eq.~\eqref{eq3:tiempocontinuo1} can be derived more formally by considering the independence of events in a Markov chain. A demonstration of this will be presented below, considering also the physical time.

\subsection{Kinetic Monte Carlo}

Everything we have explained so far is based on the fact that, to simulate a system in equilibrium, one knows the probability distribution that the states of the system will follow, namely the Boltzmann distribution. In equilibrium, it is sufficient to average over properly generated configurations of the system. However, outside equilibrium, it is necessary to examine carefully how the system evolves from one state to another and how the time is updated.

For out-of-equilibrium systems, there is no established physical theory to guide simulations. However, the approach known as kinetic Monte Carlo (kMC)\nomenclature{kMC}{kinetic Monte Carlo} has become a standard tool, provided that certain key characteristics are present in the algorithm for the simulation to be considered physically realistic. The first requirement is that the dynamics of the system must resemble those of the microscopic system. This, for example, rules out the use of non-local pair-exchange algorithms, as it is physically meaningless for two particles (spins) to be swapped at an arbitrary distance. Furthermore, arbitrary steps that lack a justification based on the system's physics should be avoided, such as those that leave the system unchanged. In others words, the algorithm should be rejection-free.

Additionally, the transition rates must be derived from the underlying physics. For instance, a common choice, based on Arrhenius theory, is to define the transition rates as
\begin{equation}  
    w_i= \nu_0 e^{-E_i / k_B T},
    \label{eq3:arrhenius}
\end{equation}  
where $\nu_0$ is the so-called attempt frequency, and $E_i$ represents an energy barrier for the transition $\mu\rightarrow\nu$. This energy barrier may or may not align with the energy difference between the two states of the transition and is heavily influenced by the microscopic dynamics of the system being studied. More generally, those transitions rates will be
\begin{equation}  
    w_i= \nu_0 A(\mu\rightarrow\nu),
    \label{eq3:arrhenius2}
\end{equation} 
where $A(\mu\rightarrow\nu)$ is a generic acceptance rate for the transition $\mu\rightarrow\nu$, that may or may not be the Metropolis acceptance rate.

Note that the transition rates defined in Eq.~\eqref{eq3:arrhenius} and \eqref{eq3:arrhenius2} are not probabilities. The probability of performing a transition is given by
\begin{equation}
    P(\mu\rightarrow\nu) = 
    \frac{A(\mu\rightarrow\nu)}{\sum\limits_{\nu\ne\mu} A(\mu\rightarrow\nu)},
\end{equation}
and is therefore independent of the attempt frequency. The attempt frequency only sets the global time scale of the algorithm. If $\tau$ is the typical physical time between transitions, then $\nu_0=1/\tau$.

Let us calculate the time with which one must update the algorithm. Since events in a Markov chain are independent, the probability of a transition occurring within a time interval $\Delta t$ follows a Poisson distribution. Therefore
\begin{equation}
    P(\Delta t)=\Omega e^{-\Omega\Delta t}
\end{equation}
is the probability of a transition occurring in the time interval $\Delta t$, where $\Omega$ is the total transition rate for the current state of the algorithm, i.e., $\Omega = \sum\limits_{i} \omega_i$.
Thus, to generate waiting times $\Delta t$, i.e. stochastic time between exchanges, that follow this distribution it is sufficient to compute
\begin{equation}
    r=\int_0^{\Delta t}\Omega e^{-\Omega t}dt,
\end{equation}
where $r$ is a uniform random number $r\in \left(0,1\right)$. Then, the time update is simply
\begin{equation}
    \Delta t=-\frac{\log(r)}{\Omega}
    =-\frac{\tau\log(r)}{\sum\limits_{\nu\ne\mu} A(\mu\rightarrow\nu)}.
     \label{eq3:tiempocontinuo2}
\end{equation}
Here, some remarks have to be made. The first point to note is that the choice of $ \tau $ only sets up the time scale of the algorithm, as mentioned earlier. Since this work focuses on studying how observables scale with time, one can safely ignore its value, or alternatively set it to 1 or to $N_P$, as all the results discussed on this work do not depend on its value. In general, the values of these frequencies or times can only be determined if the microscopic dynamics of the studied system are known, such as the typical vibration frequency of an atom in a crystal or the frequency at which a spin attempts to change its state in a real magnet.

The second point to consider is that, upon comparing Eq.~\eqref{eq3:tiempocontinuo1} and Eq.~\eqref{eq3:tiempocontinuo2}, one notices that they are not exactly the same, since Eq.~\eqref{eq3:tiempocontinuo2} contains an additional term, $ \log(r) $. This difference arises from the second derivation of the time update process, which, in many cases, is not significant when performing a MC simulation. In fact, it is easy to verify that the expected value of the time update is $\langle \Delta t \rangle = \frac{1}{\Omega}$, which is the time update formula shown in Eq.~\eqref{eq3:tiempocontinuo1}. What happens is that Eq.~\eqref{eq3:tiempocontinuo1} does not account for the randomness inherent in the transition process and instead updates the time using the expected value of the waiting time.

This difference can be significant in certain cases. However, in our study, where we focus on how different observables scale in the long-time regime—where both equations produce the same time values—either time equation can be used seamlessly. The main benefit of using Eq.~\eqref{eq3:tiempocontinuo1} is that it eliminates the need to generate an additional random number, thereby enhancing the overall speed of the algorithm. For all these reasons, this will be our choice for the time update in our algorithms.


\subsection{Simulations details}

As we are interested in simulating out of equilibrium a kMC method was used. Specifically, we used Kawasaki local dynamics with a continuous update of time.

The algorithm maintains a list of all possible pairs of neighbors nodes whose exchange alters the state of the system, which we refer to as a non-trivial exchanges. Trivial exchanges, on the other hand, involve either two filled cells or two empty cells—cases that would always be accepted in a scheme capable of rejecting exchanges but that, in practice, leave the system unchanged.

For each transition in this list, its transition probability is determined by the Metropolis acceptance criterion $A(\mu \rightarrow \nu)$ given by Eq.~\eqref{eq3:metropolis}. The starting point for the algorithm is to select one of these exchanges, proportionally to their acceptance ratio, and to carry out the exchange. The simulation time is then updated by adding the time interval given by Eq.~\eqref{eq3:tiempocontinuo1} \cite{Newman1999}.
Once an exchange is performed, the transition list is updated to be ready for the next step of the algorithm. Since the dynamics are local and the acceptance rates depend only on the energy variation, it is only necessary to remove, add, and update the transition rates involving the neighbors of the nodes involved in the exchange. This greatly reduces the computation time. 

In our simulations we do not fix the final time, but the total number of exchanges that will take place. As each run has a different seed, times between runs are not the same.

What we assume, as is standard in kMC algorithms, is that Eq.~\eqref{eq3:metropolis} remains valid even out of equilibrium. This allows us to use the described algorithm to simulate the evolution in time of the system.  


\section{Observables} \label{sec3:observables}

In this section, we introduce the main observables that will be analyzed throughout this work, along with the scaling laws they are expected to follow in the context of kinetic roughening. We will first introduce the observables in a general manner, and then, in various subsections, present some remarks that will be useful in the following chapters. In Chapters~\ref{chap4:band} and \ref{chap5:radial_spreading}, the front will consist of two films, one on the top layer and one on the bottom layer, as explained in the previous chapter. In this section and in the following chapters, we will not distinguish between the observables of the two layers in the notation; the distinction will be clear from the context.

The first step in surface growth studies is to establish a connection between a given particle model and the corresponding interface dynamics. In the cases considered herein, as will be discussed in Chapters~\ref{chap4:band} and \ref{chap5:radial_spreading}, this correspondence is straightforward, as particles visibly grow in a specific direction, naturally defining the interface as the boundary between the occupied and empty phases. However, in other scenarios, this relationship may be less direct, as the interface does not necessarily correspond to particle motion but instead represents other physical quantities. In Chapter~\ref{chap6:KPZ}, where we numerically integrate the KPZ equation, this step is not required.

In all cases, the interface, also called front position, is described by a set of local heights, $ h_i(t) $, where $ i $ represents the substrate positions at which the front is measured. This coordinate, $ i $, can correspond to a simple 1D regular front, where $ i \equiv x = 1, \dots, L $, or a more complex structure, as in Chapter~\ref{chap5:radial_spreading}. In a regular lattice $h_i(t)\equiv  h(\boldsymbol{x},t)$. Once the local heights are defined, all the others magnitudes of the front under study are derived from them.

Given the previously defined set of heights, the front position is naturally determined by the mean height, which is calculated as the average of the local heights $ h_i(t) $,
\begin{equation}
    \overline{h(t)}=\frac{1}{L}\sum_{i} h_i(t),
    \label{eq3:mean_height}
\end{equation}
where $L$ denotes the lateral length of the substrate. For a $ d $-dimensional interface, this formula can be generalized by dividing by $ L^d $, or, more generally, by dividing by the total number of positions $N$ of $ h_i(t) $ that define the front at a given time. In some cases, such as in Chapter~\ref{chap4:band}, the front length remains fixed and corresponds to the lateral size of the system, whereas in others, like in Chapter~\ref{chap5:radial_spreading}, this length evolves and grows over time as the system does.

The front width, or roughness, $ w(L,t) $, is defined as the standard deviation of the front heights, namely
\begin{equation}
    w^2(L,t)=\left\langle \overline{[h_i(t)-\overline{h(t)}]^2} \right\rangle .
    \label{eq3:w_definicion}
\end{equation}
Throughout this thesis, we will use distinct notations for different types of averaging. Spatial averages will be denoted by $ \overline{(\cdots)} $, while $ \langle \cdots \rangle $ will denote  an average over different realizations of the noise, i.e. simulations with different random number generator (RNG)\nomenclature{RNG}{Random Number Generator} seeds. We will just simply refer to this different simulations as runs.

The typical time evolution of $ w(L,t) $ has already been discussed in Sec.~\ref{sec1:scaling_properties}, but we restate it here for completeness. Under kinetic roughening conditions, the roughness $ w(L,t) $ follows the 
FV scaling law \cite{Barabasi1995, Krug1997}:
\begin{equation}
	\label{eq3:wFV}
	w(L,t)=t^{\beta}f\left( t/L^z \right),
\end{equation}
where $\beta$ and $z$ are the growth and dynamic exponents, respectively.

In Eq.~\eqref{eq3:wFV}, the scaling function exhibits two distinct asymptotic behaviors. For $ t \ll L^z $, the function follows $ f(y) \sim \text{const} $, leading to the relation $ w(L,t) \sim t^{\beta} $. Conversely, for $ t \gg L^z $, the function behaves as $ f(y) \sim y^{-\beta} $, resulting in a saturation of the roughness at a constant value, $ w = \text{const} \equiv w_{\mathrm{sat}} $. This saturation roughness, $ w_{\mathrm{sat}} $, scales with the lateral size of the system according to $ w_{\mathrm{sat}}(L) \sim L^{\alpha} $, where $ \alpha $ is the roughness exponent. We recall that only two exponents in the FV scaling framework are independent as $\alpha=\beta z$, see Eq.~\eqref{eq1:zalphabeta}. In the following chapters, and for simplicity, we will denote roughness simply as $w(t)$.

Additionally, the short-time and long-time regimes can be cast in terms of the lateral correlation length $\xi(t)$. As discussed in Sec.~\ref{sec1:scaling_properties} it is expected to scale as
\begin{equation}
    \label{eq3:correlation_length}
    \xi(t) \sim t^{1/z},
\end{equation}
in such way that $\xi(t) \ll L$ for short times and $\xi(t) \approx L$ for long times when the system has saturated. 

The rescaled front fluctuations, $\chi(\boldsymbol{x},t)$, calculated relative to the mean and normalized by the roughness, are given by
\begin{equation}
	\chi(\boldsymbol{x},t) = \frac {h(\boldsymbol{x},t) - \overline{h(t)}}{w(t)}\,.
    \label{eq3:flu2}
\end{equation}

If the system analyzed has not reached the saturation state, this formula can be simplified to
\begin{equation}
	\chi(\boldsymbol{x},t) = \frac {h(\boldsymbol{x},t) - \overline{h(t)}}{t^{\beta}}\,.
    \label{eq3:flu}
\end{equation}
Based on this, the skewness $ S $ and kurtosis $ K $ are defined as functions of the local height fluctuation $ \delta h = h(\boldsymbol{x},t) - \overline{h(t)} $. Specifically, $S=\langle \delta h^3 \rangle_c / \langle \delta h^2 \rangle_c^{3/2}$ and $K=\langle \delta h^4 \rangle_c / \langle \delta h^2 \rangle_c^{2}$, where $\langle \cdots \rangle_c$ denotes the cumulant average.

The analysis of these fluctuations is significant in this context, as the PDF of height fluctuations is recognized as another universal characteristic. For example, in the one-dimensional KPZ universality class, it follows either the TW-GOE or TW-GUE distributions, as discussed in Sec.~\ref{sec1:KPZ}.

Similar to equilibrium critical dynamics \cite{tauber2014}, in kinetic roughening systems, scaling behavior is reflected in the properties of correlation functions. To characterize the spatio-temporal evolution of the front, two additional spatial correlation functions are considered, specifically the height covariance $C_1(\boldsymbol{r}, t)$,
\begin{equation}
    \begin{gathered}
    	C_1(\boldsymbol{r}, t) = \frac 1{L^d} \sum_x \langle h(\boldsymbol{x}+\boldsymbol{r}, t) h(\boldsymbol{x}, t) \rangle-\langle \overline{h(t)}\rangle^2
    \end{gathered}
    \label{eq3:correlation_1}
\end{equation}
and the height-difference correlation function $C_2(r,t)$,
\begin{equation}	
	\label{eq3:correlation_2}
	\begin{gathered}
    	C_2(\boldsymbol{r}, t)=\frac 1{L^d} \sum_x \left\langle [h(\boldsymbol{x}+\boldsymbol{r}, t)-h(\boldsymbol{x}, t)]^2 \right\rangle  \\
	    = 2\langle \overline{h(t)^2}\rangle-\frac 2{L}\sum_x \langle h(\boldsymbol{r}+\boldsymbol{x}, t)h(\boldsymbol{x}, t) \rangle.
	\end{gathered}
\end{equation}


Analyzing the height covariance correlation function $ C_1(\boldsymbol{r},t) $ enables a deeper characterization of the two-point front statistics. As discussed in Sec.~\ref{sec1:KPZ}, within the KPZ universality class, the height covariance is anticipated to exhibit a universal behavior in front fluctuations. For one-dimensional KPZ interfaces, it is expected to converge to the Airy process covariance,
\begin{equation}
    C_1(r,t)=a_1 \, t^{2\beta} \mathrm{A}_i\left(a_2 r/t^{1/z} \right),
   \label{eq3:airy1}
\end{equation}
where $\mathrm{A}_i(u)$ denotes the covariance of the Airy$_i$ process, with $i=1$ for flat interfaces and $i=2$ for radial ones.


The parameters $ a_1 $ and $ a_2 $ in Eq.~\eqref{eq3:airy1} are numerical constants that must be determined to validate Eq.~\eqref{eq3:airy1} \cite{Alves2011,Oliveira2012,Nicoli2013,Barreales2020}. The value of $ a_1 $ can be computed as
\begin{equation}  
a_1 = \frac{C_1(0,t)}{t^{2\beta} \mathrm{A}_i(0)}.  
\label{eq3:a1}  
\end{equation}
The value of $ a_2 $ can be estimated by selecting a specific point on the graph of the $\mathrm{A}_i(u)$ function, $(\tilde{x}, \mathrm{A}_i(\tilde{x}))$. The relationship between $\tilde{x}$ and $ a_2 $ is given by $\tilde{x} \equiv a_2 r/t^{1/z}$. Substituting this into Eq.~\eqref{eq3:airy1}, we obtain  
\begin{equation}  
C_1\left(\tilde{x} t^{1/z} / a_2 \right)= a_1 t^{2\beta}\mathrm{A}_i\left( \tilde{x} \right).  
\label{eq3:airy1b}  
\end{equation}  
Given the value of $ C_1\left(\tilde{x} t^{1/z} / a_2 \right) $, a linear interpolation of the data allows us to determine its argument, thereby solving for $ a_2 $. Once the constants $a_1$ and $a_2$ have been determined, the collapsed function $R(\tilde{x}, t) \equiv \frac{C_1\left(\tilde{x} t^{1/z} / a_2\right)}{a_1 t^{2\beta}}$ should match the universal function for each geometry, $\mathrm{A}_i(\tilde{x})$, at all times for a given condition.

While $ C_1(\boldsymbol{r},t) $ serves as a tool for testing universal properties, $ C_2(\boldsymbol{r},t) $ enables the evaluation of quantities such as the correlation length $ \xi(t) $. Specifically, under kinetic roughening conditions, the FV dynamic scaling Ansatz suggests that $ C_2 $ follows the relation:
\begin{equation}
    C_2(\boldsymbol{r},t)=r^{2\alpha} g_{\mathrm{FV}}(r/\xi(t)) ,
    \label{eq3:c2_FV}
\end{equation}
where $g_{\mathrm{FV}}$ is a scaling function which behaves as $g_{\mathrm{FV}}(u) \sim u^{-2\alpha}$ for $u\gg 1$ and $g_{\mathrm{FV}}(u)\sim {\rm const}$ for $u\ll1$ \cite{Barabasi1995,Krug1997}. Thus, for $ r $ smaller than the correlation length, $ C_2(r,t) \sim r^{2\alpha} $. Conversely, when $ r $ exceeds the correlation length, $ C_2(r,t) $ reaches a plateau, $ C_{2,p}(t) $, becoming independent of $ r $, leading to
\begin{equation}
C_{2,p} \sim \xi^{2\alpha}   \ \ \ \ \ \mathrm{for} \ \  r \gg \xi(t)\,.
\label{eq3:c2p}
\end{equation}
Furthermore, the correlation length can be determined using  
\begin{equation}  
    C_2(\xi_a(t),t) = a C_{2,p}(t),  
    \label{eq3:xi_a}  
\end{equation}  
where $ a $ is a constant, typically chosen as $ a = 0.8 $ or $ 0.9 $. With this definition, the correlation length at a given time $ t $ corresponds to the distance along the front where the correlation function $ C_2 $ reaches 80\% or 90\% (respectively) of its plateau value $ C_{2,p}(t) $. It is important to note that the specific choice of $ a $ does not affect the scaling behavior of the correlation length.

We have observed that, under the FV scaling, for values of $r$ larger than the correlation length $\xi(t)$, the height-difference correlation function $C_2(r,t)$ reaches a plateau, which grows according to Eq.~\eqref{eq3:c2p}. Since the correlation length increases over time as a power law governed by the inverse of the dynamic exponent [see Eq.~\eqref{eq3:correlation_length}], the FV scaling behavior of $C_2(r,t)$ can alternatively be expressed as
\begin{equation}  
    \label{eq3:c2_FV_2}
    C_2(r,t) \sim \left\{ \begin{array}{l}
    r^{2\alpha}\; \mbox{if} \; r\ll\xi(t), \\
    t^{2\beta}\;  \mbox{if} \; r\gg\xi(t), \end{array}
    \right.
\end{equation}
where the scaling relation given by Eq.~\eqref{eq1:zalphabeta} has been employed.

In some kinetically rough systems, the height-difference correlation function exhibits anomalous behavior that deviates from the FV form described by Eq.~\eqref{eq3:c2_FV}. When this so-called anomalous scaling occurs, the FV scaling must be generalized as follows \cite{Lopez1997, Ramasco2000, Cuerno2004}:
\begin{equation}
    C_2(\boldsymbol{r},t)=r^{2\alpha} g(r/\xi(t))\,,
    \label{eq3:c2an}
\end{equation}
where $g(u) \sim u^{-2\alpha}$ for $u\gg 1$ and $g(u) \sim u^{-2(\alpha-\alpha_{\rm loc})}$ for $u\ll 1$. Now, $ \alpha_{\text{loc}} $ is the so-called \textit{local roughness exponent}, which characterizes the front fluctuations at distances smaller than the system size $ L $. Under FV scaling, the two roughness exponents are equal \cite{Barabasi1995, Krug1997}, $ \alpha = \alpha_{\text{loc}} $, so $ g(u) = g_{\mathrm{FV}}(u) $, and thus Eq.~\eqref{eq3:c2an} reduces to Eq.~\eqref{eq3:c2_FV}. However, in some cases, $ \alpha_{\text{loc}} \neq \alpha $, meaning that front fluctuations at small and large distances are governed by two distinct roughness exponents. In such cases, the curves of $ C_2(\boldsymbol{r},t) $ obtained at different times shift systematically over time and do not overlap at small $ r $, which is a characteristic feature of anomalous scaling. For convenience, we will denote $\alpha' = \alpha - \alpha_{\rm loc}$.

Anomalous scaling can arise from various mechanisms \cite{Lopez1997, Lopez1997-2}. One such mechanism is superroughening, which occurs when the global roughness exponent $ \alpha $ is greater than or equal to one. Another case arises when an independent local roughness exponent $ \alpha_{\text{loc}} $ governs small-scale fluctuations, leading to distinct scaling properties at different length scales. Systems exhibiting this behavior are said to display intrinsic anomalous kinetic roughening \cite{Lopez1997-2, Ramasco2000, Cuerno2004}.

Anomalous scaling can also be effectively characterized \cite{Lopez1997} using the front structure factor $ S(\boldsymbol{k},t) $, defined as
\begin{equation}
    S(\boldsymbol{k},t)=\langle |\mathcal{F}[h(\boldsymbol{x},t)]|^2 \rangle ,
    \label{eq3:sfactor}
\end{equation}
where $ \mathcal{F} $ represents the spatial Fourier transform, and $ \boldsymbol{k} $ is the $ d $-dimensional wave vector. In isotropic systems exhibiting intrinsic anomalous scaling, $ S(\boldsymbol{k},t) $ follows \cite{Lopez1997}
\begin{equation}
S(k,t) = k^{-(2 \alpha +d)} s(k t^{1/z}) ,
\label{eq3:Skanom}
\end{equation}
where $s(y) \propto y^{2(\alpha - \alpha_{\rm loc})}$ for $y\gg 1$, $s(y) \propto y^{2\alpha +d}$ for $y\ll 1$, and $k=~ |\boldsymbol{k}|$. Similarly to the height-difference correlation function, Eq.~\eqref{eq3:Skanom} extends the FV Ansatz for the structure factor \cite{Barabasi1995, Krug1997}, which is recovered when $ \alpha_{\text{loc}} = \alpha $. In the presence of intrinsic anomalous scaling (but not superroughening), Eq.~\eqref{eq3:Skanom} has two key implications that should be highlighted. First, the curves of $ S(k,t) $ as functions of $ k $ do not overlap for different times. Second, for large $ k \gg t^{-1/z} $, the scaling of the structure factor with $ k $ reveals the local roughness exponent, following  
\begin{equation}  
    S(k) \sim k^{-(2\alpha_{\text{loc}}+d)}.  
    \label{eq3:sfactor_scaling}  
\end{equation}  

\subsection{Growing fronts observables}\label{sec3:observables_growing}

As mentioned earlier, in Chapter~\ref{chap5:radial_spreading} the length of the front $L$ will be seen to evolve over time. This forces us to modify some of the definitions of the observables introduced above. Although the definitions of most observables, such as the mean front position or the roughness, can be naturally extended by taking spatial averages over an increasing number of front positions, the definitions of the correlation functions must be updated with care.

\begin{figure}[t]
\centering
\includegraphics[width=0.7\textwidth]{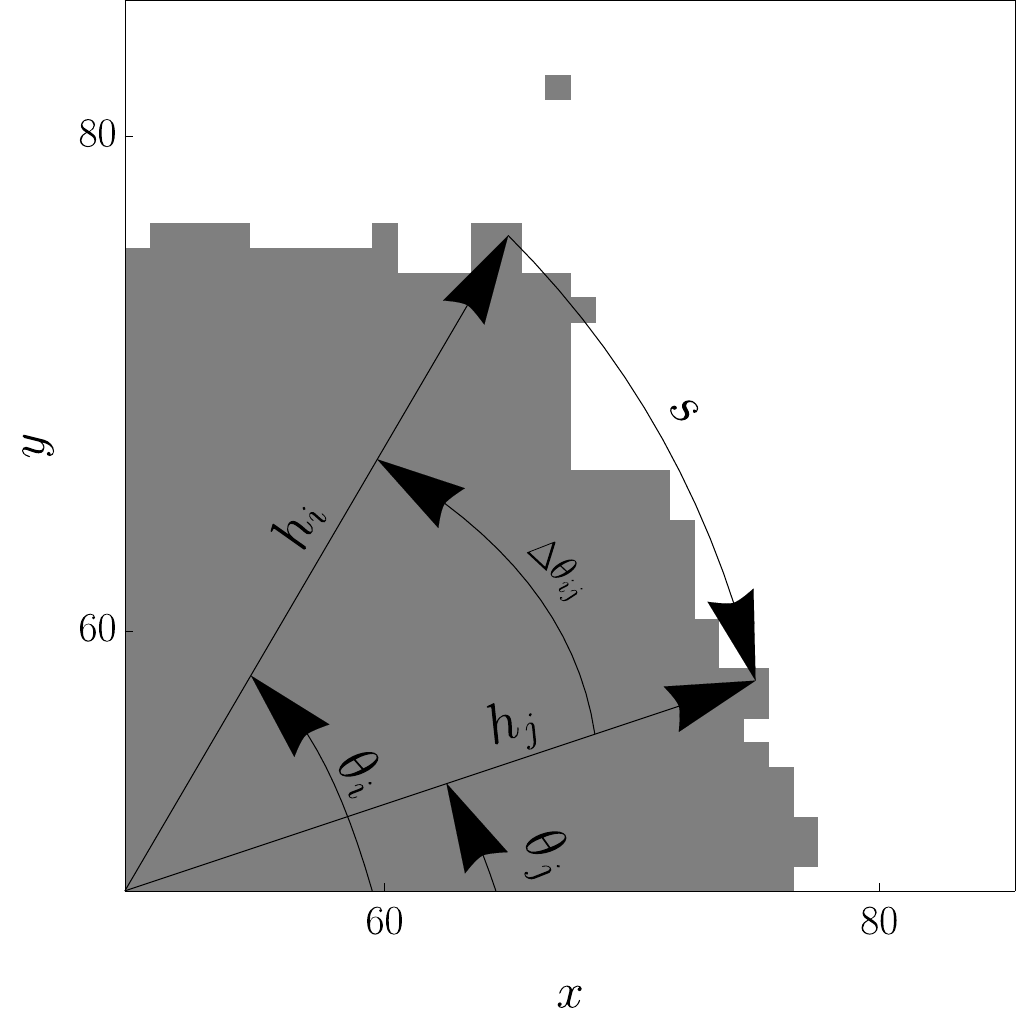}
\caption{Zoom of Fig.~\ref{fig5:Snapshot} showing the distances to the center of the reservoir $h_i$ and $h_j$, the angles $\theta_i$ and $\theta_j$, the angle difference $\Delta\theta_{ij}$ between two cells $i$ and $j$ belonging to the front of the precursor, and the arc length $s = \bar h \Delta\theta_{ij}$.}
\label{fig3:GrafIRS}
\end{figure}

The previous definitions of the correlation functions, Eqs.\ \eqref{eq3:correlation_1} and \eqref{eq3:correlation_2}, assume that the substrate of the front is a regular lattice. However, for a system that grows radially, such as the one studied in Chapter~\ref{chap5:radial_spreading}, the situation is different. For this case it is difficult to define a straightforward lateral distance, $\boldsymbol{r}$. Instead, the arc length $s$ must be used. In Chapter~\ref{chap5:radial_spreading}, the following definition of the height-difference correlation function will be used
\begin{equation}
	\label{eq3:correlation_2_circ}
    	C_2(s, t)=\frac 1{N} \sum_{\bar{h}\Delta\theta_{ij}\in s} \left\langle [h_{i}(t)-h_j(t)]^2 \right\rangle,
\end{equation}
where $\Delta\theta_{ij}=\left(\theta_i-\theta_j\right) \text{mod } 2\pi$ is the angular difference between the cells $i$ and $j$ and $s \equiv \bar{h}\Delta\theta_{ij}$ is therefore the arc length between these cells, see Fig.~\ref{fig3:GrafIRS}. The sum spans all the pairs of cells whose arc length is $s$; in Eq.~\eqref{eq3:correlation_2_circ}, $N$ is the number of those pairs. By definition, $s$ takes values between $0$ and $2 \pi \overline{h(t)}=L_f(t)$, where $L_f(t)$ is the average front length. As said earlier, for the spreading model in a circular geometry the length of the front $L\equiv L_f(t)$\footnote{We use this notation to emphasize that the front length evolves over time and to avoid confusion with other sizes.} grows with time, as $\overline{h(t)}$ also increases. 

As there are many possible arc differences between cells, we compute the value of the function $C_2(s,t)$ by discretizing the angle interval $[0,2\pi)$ (and thus the arc length interval) in boxes $(\theta-\delta\theta,\theta+\delta\theta)$ where $\delta\theta$ is a parameter that sets the width of the interval. In practice, we set $\delta\theta$ as $\delta\theta=2\pi/N_{\mathrm{A,B}}$, where $N_{\mathrm{A,B}}$ sets the number of angular boxes (bins) in which we discretize the interval $[0,2\pi)$. The particular choice for $N_{\mathrm{A,B}}$ does not change the results obtained. This analysis has already been used to study the radial growth of experimental cell colonies \cite{Galeano2003,Huergo2011,Huergo2012,Santalla2018} and tumors \cite{Bru1998,Bru2003,Block2007}, and for both continuous \cite{Santalla2014,Santalla2015,Santalla2018} and discrete \cite{Santalla2018b} models of surface kinetic roughening.

Once this definition has been established, the same analysis described above for the height-difference correlation function can be performed simply by replacing $r$ with $s$. A similar definition can be applied to the height covariance correlation function, $C_1$.

An important remark that needs to be made is that the definition of the arc-length used ($s \equiv \bar{h}\Delta\theta_{ij}$) assumes that all the cells of the front are, on average, at the same distance from the center of the system. This, as we will discuss in detail in Chapter~\ \ref{chap5:radial_spreading}, will not be always the case. The fact that not all cells are at the same distance from the center implies that the shape of the front is, on average, not circular. This effect will manifest itself in the form of the correlation function, which will not be the usual one in which a plateau is reached.

\subsection{Limit shape observables}\label{sec3:limit_shape}

\begin{figure}[t]
     \centering
         \includegraphics[width=0.4\textwidth]{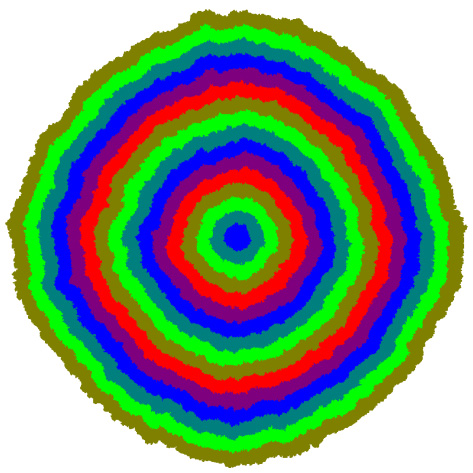}
         \hspace{5mm}
         \includegraphics[width=0.4\textwidth]{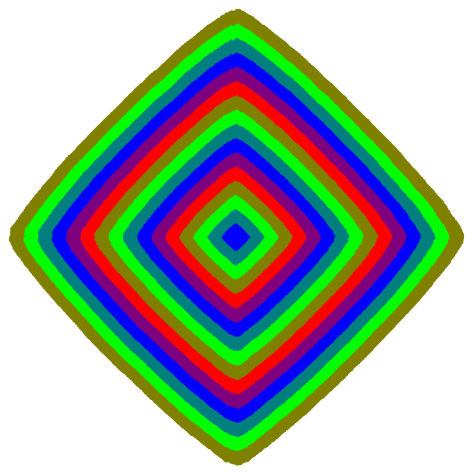}
        \caption{Isochrone curve morphology from simulations of a first-passage percolation model under varying noise levels: (a) High noise levels result in circular shapes. (b) Low noise levels produce a diamond-shaped pattern. Colors change at regular time intervals. Reproduced from Ref.~\cite{Domenech2024}.}
        \label{fig1:formaCaracteristica}
\end{figure}

As we will see in detail in the following chapters, certain discrete models develop a characteristic shape, also known as limit shape, during their growth which differs from the expected for such geometry, i.e. a straight line for a flat growing interface or a circle for a radially growing one. In these cases, fluctuations should be measured relative to these characteristic shapes rather than to the average front positions.

For instance, Domenech \textit{et al.} \cite{Domenech2024} recently studied isochrone curves in first-passage percolation on a 2D square lattice and observed that their instantaneous average shape transitions from a diamond to a circular form as noise levels increase. Figure~\ref{fig1:formaCaracteristica} illustrates the different morphologies of these isochrone curves. These can be interpreted as fluctuating interfaces with an inhomogeneous local width that reflects the underlying lattice structure. The authors demonstrate that, after accounting for these inhomogeneities, the fluctuations align remarkably well with the KPZ universality class, successfully reproducing the FV Ansatz with the expected exponents and the TW distribution for local radial fluctuations.


In particular, as previously discussed and as will be demonstrated in detail in Chapter~\ref{chap5:radial_spreading}, the spreading model in a circular geometry develops, for certain parameter conditions, a non-circular shape. For these conditions we define the roughness relative to a local front. Namely, the average front position in an angular box $\Omega$ is defined as:
\begin{equation}
    \overline{h_\Omega(t)}=\frac{1}{N(\Omega)}\sum_{i\in\Omega}h_{i}(t)\,,
    \label{eq3:distanciaPromANG}
\end{equation}
where the sum runs only over those front positions that lie within the $\Omega$ box, and $N(\Omega)$ is the number of points that belong to the corresponding front in the angular box $\Omega$. Then, the front width is defined as
\begin{equation}
	\label{eq3:width_shape}
	w^2_{\Omega}(L_f,t)=\left\langle \overline{\left[h_{i}(t)-\left\langle\overline{h_\Omega(t)}\right\rangle\right]^2} \right\rangle,
\end{equation}
where $\overline{h_\Omega(t)}$ is the average front position taken in the angular box into which the cell $i$ falls. This alternative definition of the front width will result into a different value for the growth exponent, that will denoted as $\beta_{\Omega}$. 

The front fluctuations must also be measured as deviations from the local average front in these cases. To do so, we define
\begin{equation}
	\label{eq3:Chi_local}
	{\chi_{\Omega}}_i(t)=\frac{h_i(t)-\overline{h_\Omega(t)}}{t^{\beta_{\Omega}}},
\end{equation}
where $\beta_\Omega$ quantifies the time increase of the local roughness $w_{\Omega}(t)$ defined in Eq.~\eqref{eq3:width_shape}.

\subsection{Specific observables for the Bethe lattice}\label{sec3:observables_bethe}

\begin{figure}[t]
\centering
\includegraphics[width=0.7\textwidth]{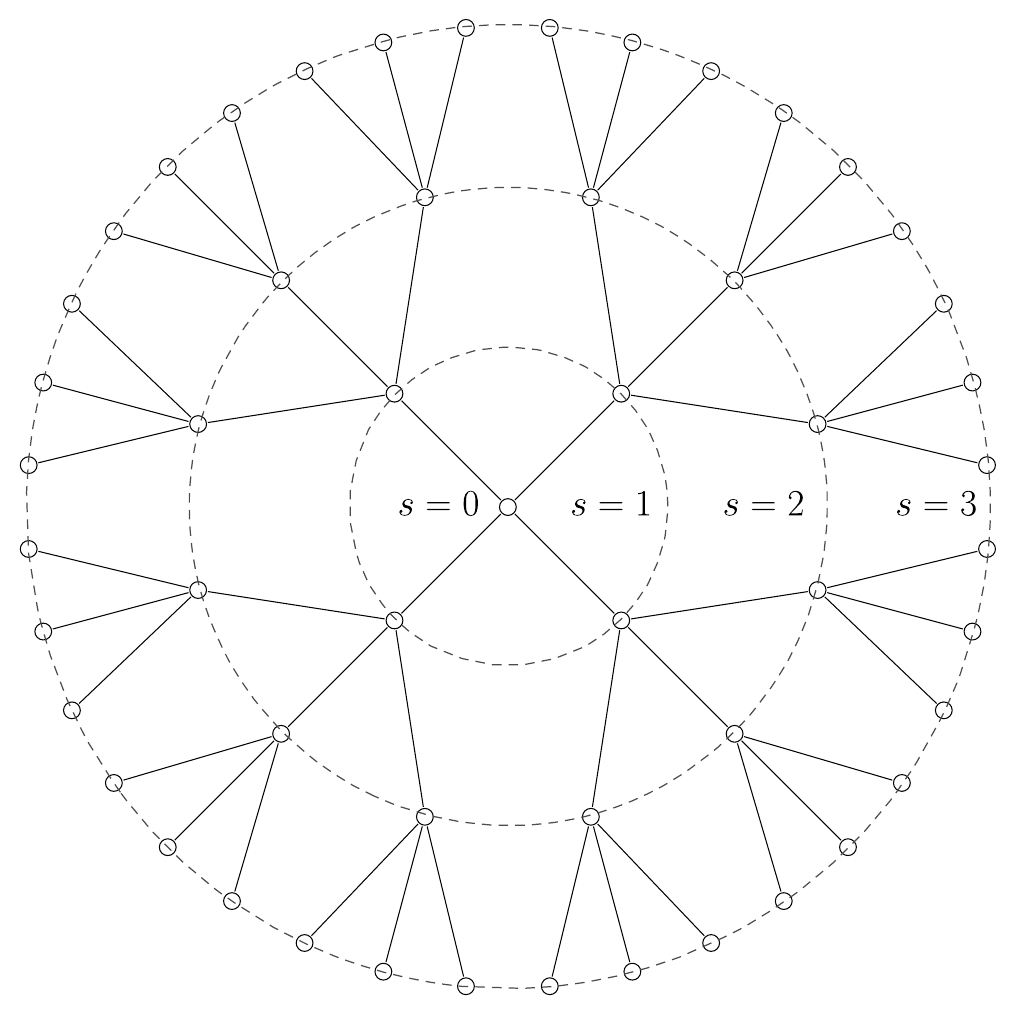}
\caption{Cayley tree with coordination number $q=4$ and three shells ($k=3$). Different sites belonging to the same shell are joined by dashed lines. Each shell is labeled with its respective $s$ value.}
\label{fig3:cayley}
\end{figure}

In Chapter~\ref{chap6:KPZ}, we will integrate the KPZ equation on networks, which are a clear case of non-regular lattices. In particular, we will analyze in depth the case of the Cayley tree\nomenclature{CT}{Cayley Tree}, whose topology is shown in Fig.~\ref{fig3:cayley}. The main observable will still be the global roughness of the front, $w(L,t)$, whose definition remains unchanged.

However, due to the characteristic topology of this lattice, we have also measured additional quantities. In particular, we have measured the local roughness, $w_0$, as defined by Oliveira in Ref. \cite{Oliveira2021}, as
\begin{equation}
	\label{eq3:local_roughness}
	w_0^2=\langle h_0^2\rangle-\langle h_0\rangle^2,
\end{equation}
where $h_0$ is the height of the central node of the Bethe lattice. Besides, we have computed the variance of the average height $\overline{h}(t)$, $w_{\overline{h}}$, defined as
\begin{equation}
    \label{eq3:mean_fluctuations}
    w_{\overline{h}}^2=\langle \overline{h}^2\rangle-\langle \overline{h}\rangle^2.
\end{equation}

To further analyze how the surface shape evolves in time, it is interesting to study how the layers grow relative to each other and to the global average of the front. In order to do it, we measured the difference between the mean heights at the center and the system border
\begin{equation}
	\label{eq3:delta_h}
	\Delta \langle h\rangle=|\langle \bar{h} \rangle_0-\langle \bar{h} \rangle_k|,
\end{equation}
where $\langle \bar{h} \rangle_k$ is the mean height value restricted to the outermost ($k$-th) shell or layer, averaged over different noise realizations. Moreover, we measured the average growth of the $s$-th layer relative to the global average of the front, i.e.,
\begin{equation}
	\label{eq3:capas}
	A(s,t) = \langle \overline{h_i - \bar{h}} \rangle_s ,
\end{equation}
where $s=0,1,\ldots, k$. Note that $\langle  \bar{h} \rangle_0\equiv \langle h_0 \rangle$.

Finally, we have also computed the height-difference correlation function $C_2(r,t)$ relative to the central node of the lattice, namely,
\begin{equation}
	\label{eq3:correlation}
	C_2(r,t)=\frac{1}{N_r}\sum_{i\in \text{shell(r)}}{\left\langle \left[h_i(t)-h_0(t)\right]^2\right\rangle},
\end{equation}
where $N_r=q(q - 1)^{r - 1}$ is the number of nodes belonging to the $r$-shell. As the system lacks PBC, this is a natural way of computing the correlations in the tree~\cite{Angelini2020}, as being in the $r$-shell is the same as being a distance $r$ away from the central node. 

\section{Computation of uncertainty} \label{sec3:JK}

The statistical errors for various observables have been calculated based on highly correlated raw numerical data. The approach for error estimation is detailed below, with additional information available in \cite{Yllanes2011,Michael1994,Lulli2016,Seibert1994,GarciaBarreales2024}. Following standard practice, we indicate the estimated uncertainty in the final digit(s) by enclosing them in parentheses. These error bars are included in the graphics, though they may be difficult to discern in some cases.

To conduct a statistical analysis of the system, multiple simulation runs are performed. Since the MC algorithm updates time continuously, the time intervals between runs are not uniform. To facilitate comparisons of a given quantity across different runs, we define temporal bins of width $\Delta t$, grouping data points from various simulations that fall within the interval $t \in (t, t+\Delta t)$. Typically, these temporal bins are chosen to be evenly spaced on a logarithmic time scale. We define the best estimate of a quantity $ x $ within the temporal box $ (t, t+\Delta t) $ for the $ i $-th run as the simple average of all data points within that interval, namely
\begin{equation}
\hat{x}_i=\frac{1}{n} \sum_{j=1}^n x_j \,,
\end{equation}
where $n$ is the number of points included in that particular box. The mean, $\bar{x}$, is defined then as:  
\begin{equation}  
  \bar{x} = \frac{1}{N} \sum_{i=1}^N \hat{x}_i \,,  
\end{equation}  
where $ N $ represents the total number of runs, corresponding to the number of simulations performed.

As a general practice, the errors for all results presented in the following sections have been computed using the jackknife (JK) procedure \cite{Young2015,Efron1982}. The $ i $-th jackknife estimate of a quantity $ x $ is obtained by averaging over all runs while excluding the data from the $ i $-th run:
\begin{equation}
x_i^{\mathrm{JK}}=\frac{1}{N-1}\sum_{k=1, k \neq i}^N \hat{x}_k\,.
\end{equation}
The variance of $\bar{x}$ is then defined as
\begin{equation}
    \sigma_\mathrm{JK}(\bar{x})=\frac{N-1}{N} \sum_{k=1}^N(\bar{x}-x_i^\mathrm{JK})^2\,. 
    \label{eq3:sigmaJK}
\end{equation}
Thus, for each temporal box, the estimated value is given by $ \bar{x} \pm \sqrt{\sigma_\mathrm{JK}} $ (within one standard deviation). It is important to note that, for a given set of $ {x_i} $, the standard error formula and Eq.~\eqref{eq3:sigmaJK} yield identical results. In this thesis, the jackknife method is employed due to nonlinear dependencies among the variables.

Typically, determining a critical exponent requires fitting data to a power-law. However, it is crucial to recognize that the data exhibit strong correlations (e.g., the $\xi \sim t^{1/z}$ data points are highly correlated). Therefore, to accurately compute an exponent using a least-squares fit, one should ideally employ the full covariance matrix for the global fit. The challenge arises because, in most cases, the full covariance matrix is singular or nearly singular (i.e., its determinant is close to zero) \cite{Yllanes2011,Michael1994,Lulli2016,Seibert1994}, making it impossible to compute its inverse, which is required for the fitting procedure.  

To address this issue, we also use the jackknife procedure as an alternative approach that accounts for the statistical correlations in the data. This method has proven highly effective in various contexts, such as the study of spin glasses and the computation of hadron masses in lattice QCD \cite{Yllanes2011,Michael1994,Lulli2016}. The details of this procedure are as follows: the mean value, $ \bar{z} $, of a given exponent is determined by using data from all runs. The statistical error for this exponent is estimated using Eq.~\eqref{eq3:sigmaJK}. In this approach, the $ i $-th run is omitted from the dataset, and the corresponding jackknife estimate for the exponent, $ z_i^\mathrm{JK} $, is computed. The error is then determined using the standard jackknife formula as:
\begin{equation}  
\sigma_\mathrm{JK}(\bar{z}) = \frac{N-1}{N} \sum_{k=1}^N (\bar{z} - z_i^\mathrm{JK})^2\,.  
\end{equation}
By employing the aforementioned procedure, we account for the strong correlations within the data, ensuring a more accurate estimation of the statistical error associated with the exponent.

Lastly, we have selected the fitting intervals to ensure that the reduced $\chi^2$ (calculated as $\chi^2$ divided by the number of degrees of freedom, where the degrees of freedom correspond to the number of data points minus the number of fitted parameters) is close to one. The $\chi^2$ values have been computed under the assumption of a diagonal covariance matrix \cite{Young2015}.

\graphicspath{{4_capitulo/fig4/}}

\chapter{Band Spreading}
\label{chap4:band}


In this chapter, we will conduct an in-depth analysis of the spreading model in a band geometry. We will begin by revisiting the key features of this model, previously introduced in Chapter.\ \ref{chap2:wetting}. Then we will provide all the necessary details to reproduce the simulations. This includes identifying the most relevant parameters from an experimental perspective. Finally, we will present the simulation results for this geometry along with some concluding remarks.

\section{Model and simulation details}

The microscopic driven Ising lattice gas model examined in this chapter consists of two overlapping 2D rectangular layers with dimensions $ L_x \times L_y $. Each node of the square lattice, denoted as $ \boldsymbol{r} = (x, y, Z) $,\footnote{To avoid confusion with the standard notation for the dynamic exponent $ z $, we will use an uppercase $ Z $ to represent the vertical coordinate in 3D space.} can be occupied by at most one particle at any given time. Consequently, the occupation number $ n(\boldsymbol{r},t) $ can take values of either 0 or 1. The lower layer $(Z = 1)$ and the upper layer $(Z = 2)$ are referred to as the precursor and supernatant, respectively, while the substrate on which the droplet expands is positioned at $Z = 0$. PBC are applied in the $y$-direction, following the approach in Refs.\ \cite{Abraham2002,Harel2018,Harel2021}. It is important to note that the choice of BC is not expected to affect universal properties, such as the values of exponents defining the kinetic roughening behavior that may arise in the system \cite{Barabasi1995,Krug1997}. The energy of the system, already presented in Eq.~\eqref{eq2:energy}, is given by: 
\begin{equation}
    \mathcal{H}= -J \sum_{\langle \boldsymbol{r}, \boldsymbol{s} \rangle} n(\boldsymbol{r},t)n(\boldsymbol{s},t) - A \sum_{\boldsymbol{r}}\frac{n(\boldsymbol{r},t)}{Z^3}.
    \label{eq4:energy}
\end{equation}
While the first term represents the interactions between liquid particles and their nearest neighbors, the second term accounts for the interaction with the substrate, which is characterized by a Hamaker constant $ A > 0 $.

The first column $(x=0)$ of both layers acts as the fluid reservoir, serving as a BC that supplies particles to the layers and represents the macroscopic droplet. Initially, only these cells are occupied. If, during the evolution of the system, any cell belonging to the reservoir becomes empty due to an exchange, it is immediately refilled. As previously mentioned in Chapter \ref{chap2:wetting}, while the Kawasaki algorithm conserves the number of particles, this BC is essential to the growth of the system as it is the only way new particles come into the system. Conversely, if a particle reaches the last column of the lattice at any point, it is assumed to escape from the system.

\begin{figure}[t]
\centering
\includegraphics[width=0.99\textwidth]{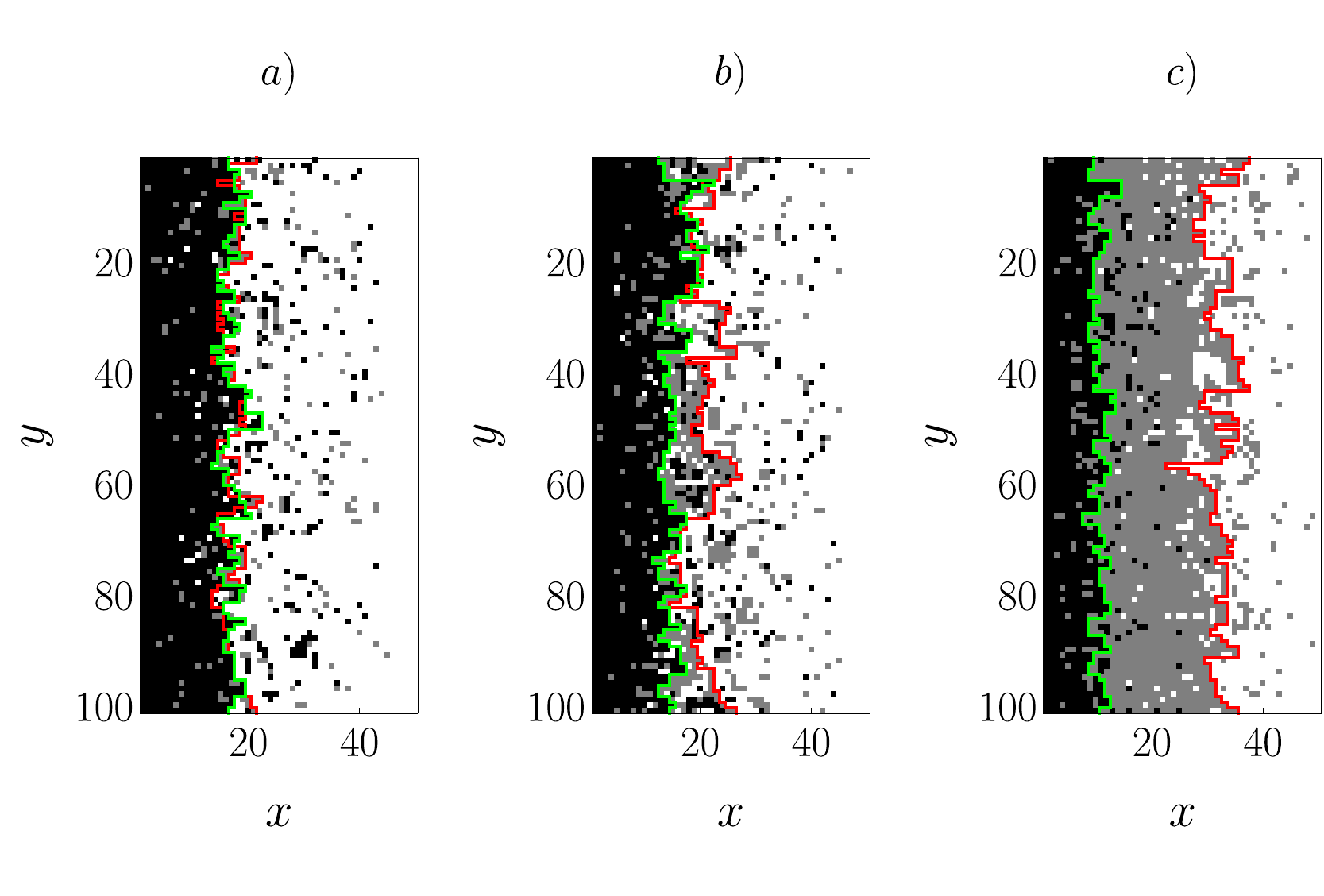}
\caption{Top views of three snapshots of the lattice gas model for increasing values of the Hamaker constant $A$, left to right. Occupied cells in the precursor and supernatant layers are in gray and black, respectively, with the red and green lines delimiting the corresponding fronts; empty cells are uncolored. Parameters used are $J = 1$, $T = 1$, $L_{y} = 100$, $L_{x} = 50$, and a) $A = 0.1$, b) $A = 1$, and c) $A = 10$. The three snapshots were taken at the same simulation time. All units are arbitrary.}
\label{fig4:grafCombT}
\end{figure}

\begin{figure}[t]
\centering
\includegraphics[width=0.99\textwidth]{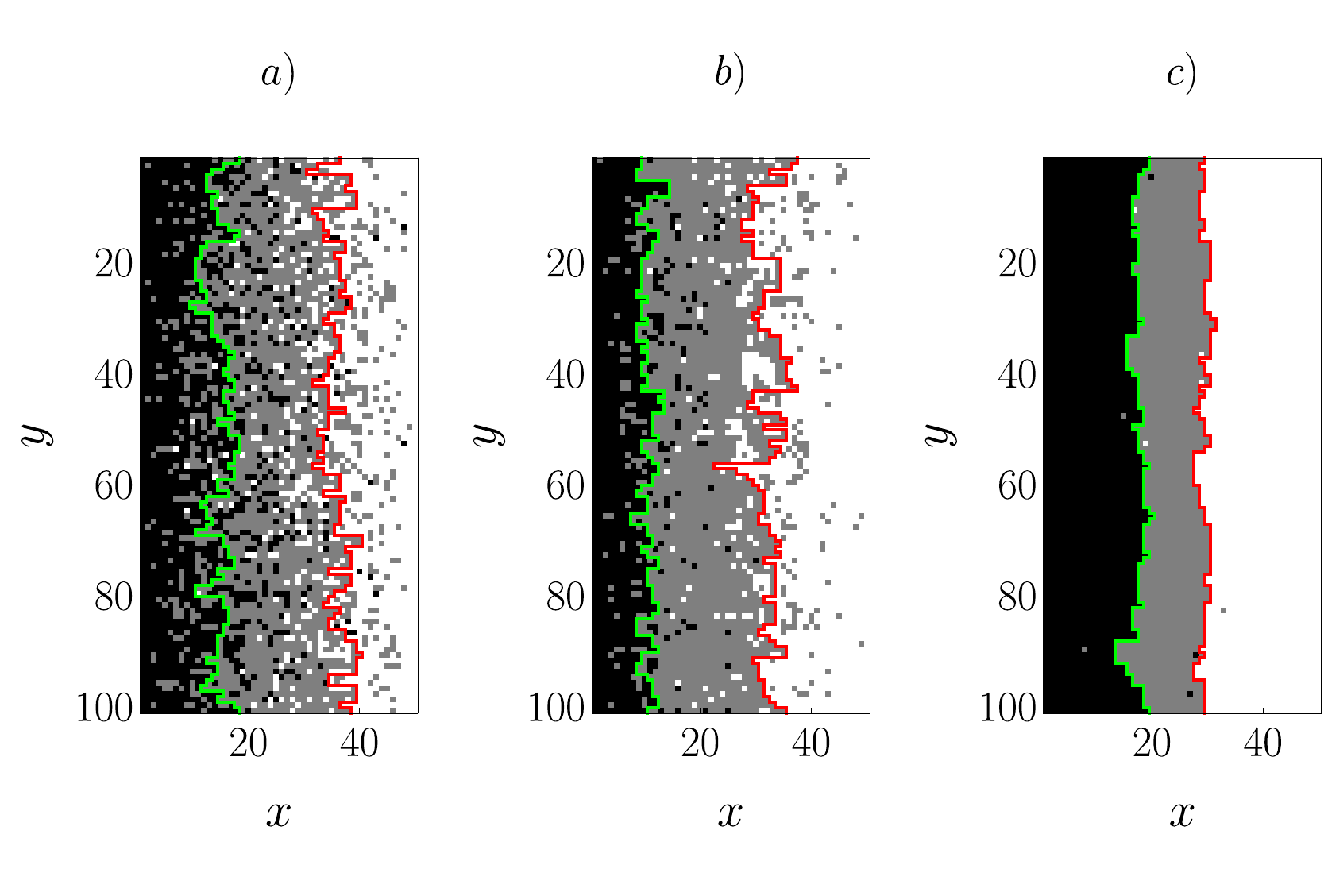}
\caption{The same as in Fig.~\ref{fig4:grafCombT} but for for decreasing values of the temperature $T$, left to right. Specifically, $J = 1$, $A=10$, $L_{y} = 100$, $L_{x} = 50$, and a) $T = 3$, b) $T = 1$, and c) $T = 1/3$. All units are arbitrary.}
\label{fig4:grafCombA}
\end{figure}

The evolution of the system has been simulated by continuous-time MC Kawasaki local dynamics, as described in the previous chapter. At any given time, a particle is considered part of the precursor (or supernatant) film if it is connected to the droplet reservoir through a continuous chain of nearest-neighbor occupied sites. For a fixed $y$, the front position, $ h(y,t,Z) $, is defined as the highest $x$-coordinate where a cell remains connected to the reservoir. Once this front definition is established, all the formulas from Section. \ref{sec3:observables} can be applied to analyze the kinetic roughening properties of the fronts generated by this model in this geometry. Examples illustrating the definition of the fronts are shown in Figs.\ \ref{fig4:grafCombT} and \ref{fig4:grafCombA}.

Since the Metropolis acceptance criterion $ A(\mu \rightarrow \nu) $ [see Eq.~\eqref{eq3:metropolis}] used in the MC algorithm depends on the ratio $ \Delta E/k_B T $, the exact values of the parameters are irrelevant; only the ratios $ J/k_{\mathrm{B}}T $ and $ A/k_{\mathrm{B}}T $ are relevant to the evolution of the system. In this and the following chapter, we adopt physical units such that $ k_B = 1 $, while other parameters remain arbitrary. Additionally, in all simulations, we fix $ J = 1 $, modifying only the Hamaker constant $ A $ and the temperature $ T $. The system size was set to $ L_x = 1000 $ in all runs, ensuring that the film does not reach the boundary of the system, whereas in most simulations we use $ L_y = 256 $. A summary of all the simulation conditions considered is provided in Table \ref{tab4:param}.

The total energy of the system, as defined by Eq.~\eqref{eq4:energy}, is expressed in terms of $ A $ and $ J $. From a physical perspective \cite{Bonn2009,Popescu2012}, the most relevant values for the pairs $(A,J)$ are those for which $ J/k_{\mathrm{B}}T $ is sufficiently large to ensure a high degree of involatility, and $ A/k_{\mathrm{B}}T $ is large enough to place the system in the complete wetting regime, as discussed in Ref.\ \cite{Abraham2002}. Among all the conditions reported in this chapter (see Table \ref{tab4:param}), the most physically realistic, and therefore closest to those observed in liquids exhibiting a precursor film, are those where $ A $ is large and $ T $ is low. However, we also present results for conditions that do not strictly meet these criteria, as our goal is to investigate the spreading model across a broad range of parameters. According to Eq.~\eqref{eq4:energy}, the lowest energy state of the system corresponds to the smallest value of $ Z $, indicating that occupying the precursor layer is energetically favorable. This preferential occupation becomes more pronounced when $ A \gg J $, in which case the bottom layer is expected to grow faster than the upper one. Conversely, when $ J $ dominates, both layers are likely to expand at the same rate. This effect is illustrated in Figure \ref{fig4:grafCombT}, which displays three top-view snapshots of the system obtained for a fixed $ T $ and three different values of the Hamaker constant. The effect of temperature on the system is illustrated in Figure \ref{fig4:grafCombA}. At higher temperatures, the generated fronts are noisier. This figure clearly shows how front roughness grows as the temperature of the system increases for a fixed value of $ A $. Moreover, the rightmost snapshot of this figure shows the most physically realistic condition simulated.

\begin{table}[p]
\centering
\renewcommand{\arraystretch}{0.9}
\begin{tabular}{@{}ccccccl@{}}
\toprule
$L_x$ & $L_y$ & $T$ & $A$ & $N_E$ & Runs \\
\midrule\midrule

\multirow{20}{*}{1000} & \multirow{20}{*}{256} & \multirow{5}{*}{10}
  & 10   & $1.5 \times 10^8$  & 100 \\
  &   &   & 5    & $1.5 \times 10^8$  & 100 \\
  &   &   & 1    & $1.0 \times 10^8$  & 100 \\
  &   &   & 0.1  & $1.0 \times 10^8$  & 100 \\
  &   &   & 0.01 & $1.0 \times 10^8$  & 100 \\
\cmidrule(l){3-6}
  &   & \multirow{5}{*}{3}
  & 10   & $1.0 \times 10^8$  & 100 \\
  &   &   & 5    & $1.0 \times 10^8$  & 100 \\
  &   &   & 1    & $1.0 \times 10^8$  & 100 \\
  &   &   & 0.1  & $1.0 \times 10^8$  & 100 \\
  &   &   & 0.01 & $1.0 \times 10^8$  & 100 \\
\cmidrule(l){3-6}
  &   & \multirow{5}{*}{1}
  & 10   & $2.0 \times 10^8$  & 100 \\
  &   &   & 5    & $2.0 \times 10^8$  & 100 \\
  &   &   & 1    & $2.0 \times 10^8$  & 1000 \\
  &   &   & 0.1  & $2.0 \times 10^8$  & 100 \\
  &   &   & 0.01 & $2.0 \times 10^8$  & 100 \\
\cmidrule(l){3-6}
  &   & \multirow{5}{*}{3/4}
  & 10   & $4.0 \times 10^8$  & 100 \\
  &   &   & 5    & $4.0 \times 10^8$  & 100 \\
  &   &   & 1    & $4.0 \times 10^8$  & 100 \\
  &   &   & 0.1  & $4.0 \times 10^8$  & 100 \\
  &   &   & 0.01 & $4.0 \times 10^8$  & 100 \\

\midrule

\multirow{5}{*}{1000} & \multirow{5}{*}{256} & \multirow{5}{*}{1/2}
  & 10   & $7.5 \times 10^8$  & 100 \\
  &   &   & 5    & $7.5 \times 10^8$  & 100 \\
  &   &   & 1    & $7.5 \times 10^8$  & 100 \\
  &   &   & 0.1  & $7.5 \times 10^8$  & 100 \\
  &   &   & 0.01 & $7.5 \times 10^8$  & 100 \\

\midrule

\multirow{3}{*}{1000} & \multirow{3}{*}{256} & \multirow{3}{*}{1/3}
  & 1    & $1.25 \times 10^{10}$ & 100 \\
  &   &   & 0.1  & $1.25 \times 10^{10}$ & 100 \\
  &   &   & 0.01 & $1.25 \times 10^{10}$ & 100 \\

\midrule

1000 & 64  & 1/3 & 10 & $5.0 \times 10^9$  & 100 \\
1000 & 64  & 1/3 & 5  & $5.0 \times 10^9$  & 100 \\
1000 & 128 & 1              & 1  & $1.0 \times 10^8$  & 250 \\
1000 & 512 & 1              & 1  & $4.0 \times 10^8$  & 250 \\

\bottomrule
\end{tabular}
\caption{Parameters used for the runs reported in this chapter. $N_E$ is the total number of exchanges performed, and the last column shows the number of runs simulated in each case.}
\label{tab4:param}
\end{table}

\section{Results}

All the figures in this section illustrate the dynamic evolution of the precursor layer. As we will show, both layers exhibit the same behavior. 

We have performed simulations for two different system sizes: $L_y = 64$ and $L_y = 256$, as seen in Table~\ref{tab4:param}. For the majority of conditions, we used $L_y = 256$ to ensure a sufficiently large front for robust statistical analysis. With that size, for most of the studied conditions, the front was able to grow long enough to explore the scaling behavior of the different observables. However, for very low temperature, namely $T = 1/3$, and especially with high Hamaker constants, the system exhibited a remarkably slow growth, even for a smaller system size ($L_y = 64$). Although, as can be seen in Table \ref{tab4:param}, some simulations were conducted with $T = 1/3$, we do not report exponent values for these conditions, as the scaling behavior was more difficult to be clearly observed.

However, for completeness, it is worth mentioning that for $T = 1/3$ and a small Hamaker constant, the results closely resemble those for $T = 1/2$ and the same Hamaker constant. For the cases with $T = 1/3$ and a higher Hamaker constant, the behavior differs from that at higher temperatures with the same Hamaker constant, as the roughness does not exhibit a clear growth phase. Although it does not reach saturation, despite its relatively small size, it appears to go through several transient stages. A similar phenomenon was reported by Abraham \textit{et al.}~in Ref.~\cite{Abraham2002}, where they studied only one condition, namely $T=1/3$ and $A=10$, and observed that the roughness exponent changed from $\beta = 1/6$ to $\beta = 1/8$. For this condition, as we will detail below, the results reported in Ref.~\cite{Abraham2002} can be recovered once the timescales in their work and ours are properly related.

\vspace{-5pt}
\subsection{Front position} 

\begin{figure}[h!]
\centering
\includegraphics[width=0.7\textwidth]{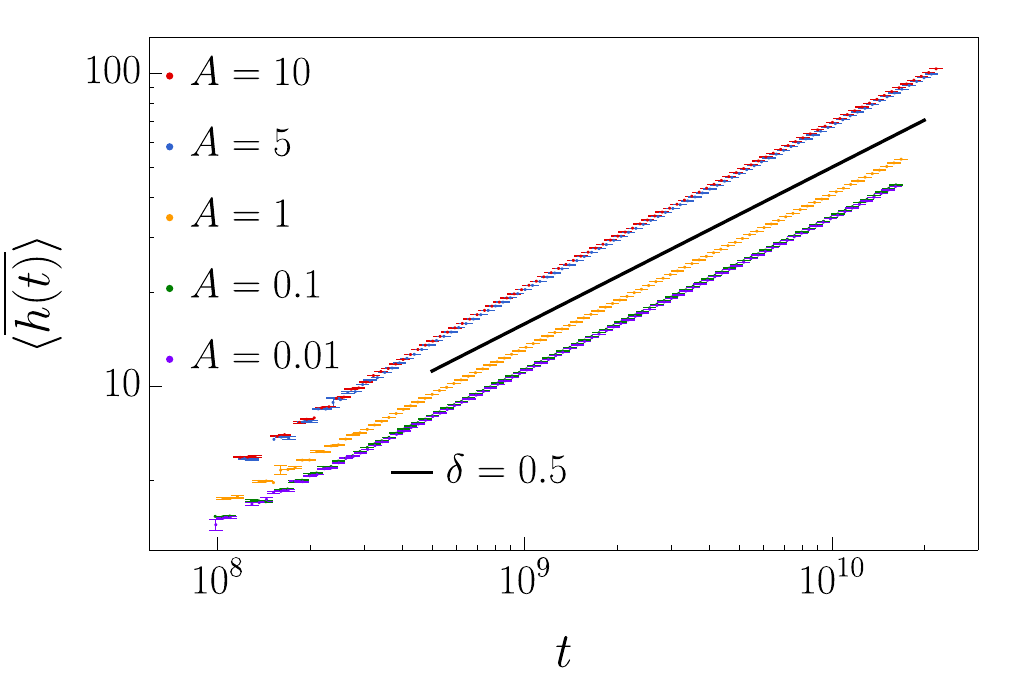}
\caption{Average front position $\langle \overline{h(t)}\rangle$ as a function of time for $T=1$ and several values of $A$. The solid black line corresponds to the reference scaling $\langle \overline{h(t)} \rangle \sim t^{1/2}$. All units are arbitrary in this and all figures in this chapter.}
\label{fig4:ht}
\end{figure}

Figure \ref{fig4:ht} illustrates the evolution of $\langle \overline{h(t)}\rangle$ for five distinct parameter sets. Regardless of the values of $A$ and $T$, the mean front position follows the expected growth law $\langle \overline{h(t)}\rangle \sim t^\delta$, with an exponent approximately $\delta \approx 1/2$. Table \ref{tab4:delta1} presents the values of the $\delta$ exponent for the precursor film under each parameter condition, while Table \ref{tab4:delta2} provides the corresponding values for the supernatant film.

\begin{table}[h!]
\centering
\small
\renewcommand{\arraystretch}{1.2}
\begin{tabular}{@{}lccccc@{}}
\toprule
\diagbox[width=4em,height=3em,dir=SE,trim=l]{$A$}{$T$} & 10 & 3 & 1 & 3/4 & 1/2 \\
\cmidrule(r){1-6}
10   & 0.4804(7) & 0.4911(4) & 0.5091(4) & 0.5165(4) & 0.5503(4) \\
5    & 0.4781(8) & 0.485(6)  & 0.5079(5) & 0.5169(3) & 0.5502(4) \\
1    & 0.4751(9) & 0.4799(5) & 0.489(1)  & 0.4891(8) & 0.4985(4) \\
0.1  & 0.474(1)  & 0.4798(6) & 0.4909(8) & 0.4897(6) & 0.5103(5) \\
0.01 & 0.4754(9) & 0.4766(8) & 0.4889(9) & 0.4931(6) & 0.5131(7) \\
\bottomrule
\end{tabular}
\caption{Values of the exponent $\delta$ for the precursor layer, for all the conditions studied.}
\label{tab4:delta1}
\end{table}
\begin{table}[h!]
\centering
\small
\renewcommand{\arraystretch}{1.2}
\begin{tabular}{@{}lccccc@{}}
\toprule
\diagbox[width=4em,height=3em,dir=SE,trim=l]{$A$}{$T$} & 10 & 3 & 1 & 3/4 & 1/2 \\
\cmidrule(r){1-6}
10   & 0.471(1)  & 0.4721(8) & 0.4887(9) & 0.4945(9) & 0.491(1) \\
5    & 0.472(1)  & 0.4719(9) & 0.4892(9) & 0.4952(9) & 0.490(1) \\
1    & 0.4742(9) & 0.4771(5) & 0.493(1)  & 0.5061(7) & 0.5181(4) \\
0.1  & 0.474(9)  & 0.4792(6) & 0.4929(8) & 0.4920(6) & 0.5113(5) \\
0.01 & 0.4753(9) & 0.4768(8) & 0.4890(9) & 0.4933(6) & 0.5132(7) \\
\bottomrule
\end{tabular}
\caption{Values of the exponent $\delta$ for the supernatant layer.}
\label{tab4:delta2}
\end{table}

\clearpage
\subsection{Roughness}

\begin{figure}[t]
\centering
\includegraphics[width=0.7\textwidth]{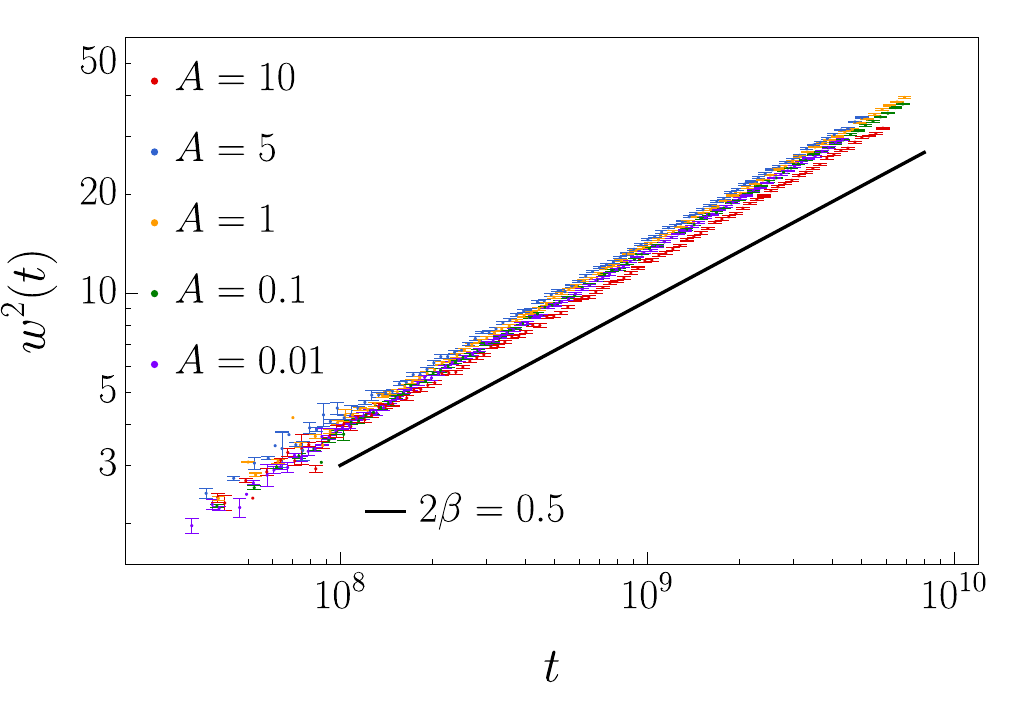}
\caption{Squared roughness $w^2(t)$ as a function of time for $T=3$ and several values of $A$. As a visual reference, the solid black line corresponds to $w^2(t) \sim t^{1/2}$.}
\label{fig4:w2t}
\end{figure}

Analogously, Figure \ref{fig4:w2t} shows the time evolution of the roughness $w^2(t)$ for five different parameter conditions. As expected, the roughness follows the FV growth law, $w^2(t) \sim t^{2\beta}$. However, due to the large lattice sizes used in our simulations, we did not observe any evidence of eventual saturation to a steady-state value \cite{Barabasi1995,Krug1997}. Furthermore, we remark that the exponent values reported show no significant time dependence at long times. Therefore, for those conditions in which we report exponents, we have avoided the very-long-time regime explored by Abraham \textit{et al.} in Ref. \cite{Abraham2002} in which the precursor film has grown so wide that diffusion is no longer able to communicate the front with the reservoir efficiently, causing the front to behave as if it were evolving without the external driving of the reservoir. The condition explored by Abraham \textit{et al.} in Ref. \cite{Abraham2002} will be discussed in more detail below.

Table \ref{tab4:beta1} presents the computed growth exponent for the precursor film across all studied conditions, while Table \ref{tab4:beta2} provides the corresponding values for the supernatant film. These tables indicate that the detailed value of $\beta$ depends on the physical parameters $A$ and specially, $T$. 

At high temperatures ($T \gtrsim 1$), the growth exponent remains approximately $\beta \approx 0.26$ for both the precursor and supernatant layers, showing no dependence on the Hamaker constant $A$.  At low temperatures ($T < 1$), the growth exponent differs slightly between the two layers and appears to be more sensitive to the value of $A$ for both layers. As a reference for the low-temperature regime, the kMC simulations by Abraham \textit{et al.} reported a growth exponent of $\beta \simeq 1/6$ for the precursor layer using $J=1$, $A=10$, and $T=1/3$, which aligns with our results.

As an overview, Tables \ref{tab4:delta1} to \ref{tab4:beta2} already suggest a non-trivial dependence of the scaling exponents on temperature, while their dependence on the Hamaker constant appears significantly weaker. This indicates the existence of two primary scaling regimes, a low-temperature and a high-temperature regime, with intermediate values of $T$ showing temperature-dependent exponents. As we will see below, additional exponent estimates further support this interpretation.

\begin{table}[h!]
\centering
\small
\renewcommand{\arraystretch}{1.2}
\begin{tabular}{@{}lccccc@{}}
\toprule
\diagbox[width=4em,height=3em,dir=SE,trim=l]{$A$}{$T$} & 10 & 3 & 1 & 3/4 & 1/2 \\
\cmidrule(r){1-6}
10   & 0.539(4) & 0.536(3) & 0.516(7) & 0.489(8) & 0.29(1) \\
5    & 0.537(4) & 0.533(3) & 0.517(7) & 0.483(6) & 0.28(2) \\
1    & 0.536(4) & 0.538(3) & 0.544(8) & 0.536(9) & 0.26(1) \\
0.1  & 0.543(4) & 0.538(3) & 0.475(9) & 0.343(9) & 0.30(2) \\
0.01 & 0.537(4) & 0.538(4) & 0.497(9) & 0.34(1)  & 0.33(3) \\
\bottomrule
\end{tabular}
\caption{Values of the exponent $2\beta$ for the precursor layer, for all the conditions studied.}
\label{tab4:beta1}
\end{table}
\begin{table}[h!]
\centering
\small
\renewcommand{\arraystretch}{1.2}
\begin{tabular}{@{}lccccc@{}}
\toprule
\diagbox[width=4em,height=3em,dir=SE,trim=l]{$A$}{$T$} & 10 & 3 & 1 & 3/4 & 1/2 \\
\cmidrule(r){1-6}
10   & 0.538(3) & 0.541(3) & 0.530(3) & 0.489(4) & 0.318(9) \\
5    & 0.537(3) & 0.539(3) & 0.526(4) & 0.479(4) & 0.314(8) \\
1    & 0.533(4) & 0.538(3) & 0.537(8) & 0.503(8) & 0.29(1)  \\
0.1  & 0.542(4) & 0.536(3) & 0.476(9) & 0.347(9) & 0.30(2)  \\
0.01 & 0.539(4) & 0.538(4) & 0.494(9) & 0.34(1)  & 0.34(3)  \\
\bottomrule
\end{tabular}
\caption{Values of the exponent $2\beta$ for the supernatant layer, for all the conditions studied.}
\label{tab4:beta2}
\end{table}
\clearpage


\begin{figure}[t!]
\centering
\includegraphics[width=0.7\textwidth]{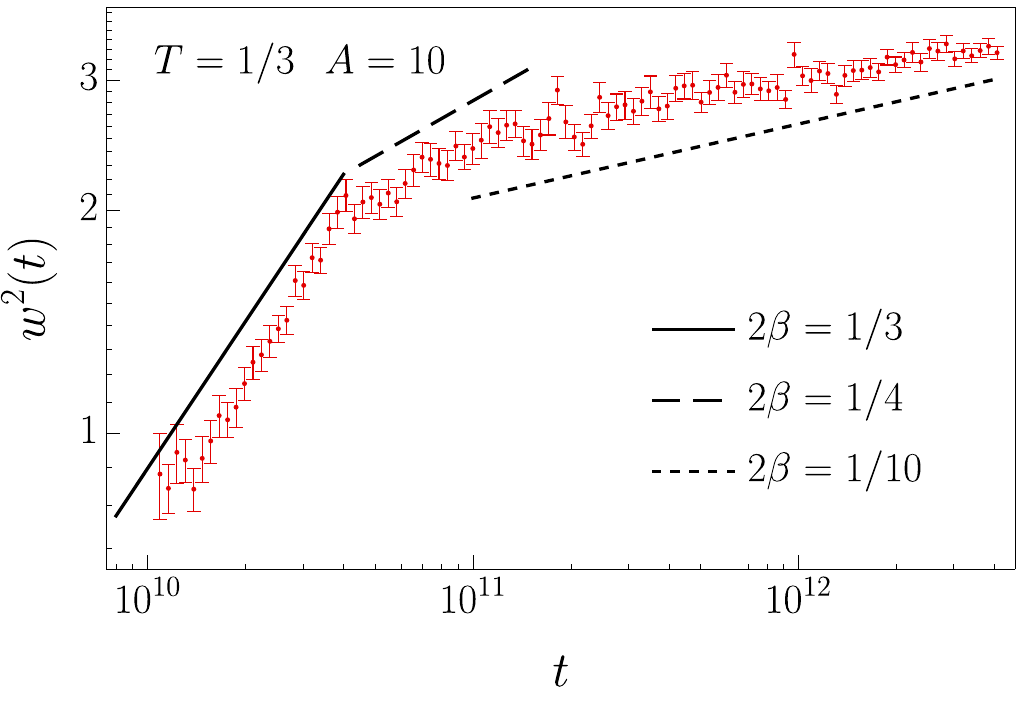}
\caption{Squared roughness $w^2(t)$ as a function of time for $T=1/3$ and $A=10$. As a visual reference, the solid line corresponds to $w^2(t) \sim t^{1/3}$, the dashed line corresponds to $w^2(t) \sim t^{1/4}$ and the dotted line corresponds to $w^2(t) \sim t^{1/10}$.}
\label{fig4:w2t_abraham}
\end{figure}

Figure \ref{fig4:w2t_abraham} shows the time evolution of the roughness $ w^2(t) $ for $ T=1/3 $ and $ A=10 $; this condition was the one simulated by Abraham \textit{et al.} in Ref.~\cite{Abraham2002}. As mentioned earlier, these authors found that the roughness grew as $ w \sim t^{1/6} $ for short times and then transitions to $ w \sim t^{1/8} $ at later times, although this later regime was somehow obscure. In this work, we also observe a behavior $ w^2 \sim t^{1/3}$ ($w \sim t^{1/6}$) for short times, as can be seen in Fig.~\ref{fig4:w2t_abraham}. However, the second, long-time regime is not as clear. Initially, it appears to follow $ w^2 \sim t^{1/4} $, but then it curves, suggesting that the exponent could be smaller. As a reference, we show the $ w^2(t) \sim t^{1/10} $ behavior with a dotted line in Fig.~\ref{fig4:w2t_abraham}. It is important to remark that the timescale of the data used in Ref.~\cite{Abraham2002} was up to $ 10^6 $, while in this work it extends beyond $ 10^{12} $. However, the way we update time and the way the authors of Ref.~\cite{Abraham2002} do it are different. While they update the time using the hopping rates $\omega_i=\nu e^{-\Delta H/(k_B T)}$, where $\nu$ was taken as the inverse of the number of destination sites, we set the time scale following Eq.~\eqref{eq3:tiempocontinuo1}. With that in mind, the ratio between the timescales can be estimated as  $N_p\nu\approx (5 L_x L_y)/5\sim 10^5$ so our $t\sim 10^{12}$ would be approximately equivalent to $t\sim 10^{7}$ in that reference. In summary, we access longer time scales than in Ref.~\cite{Abraham2002}, allowing for a deeper exploration of later times.



\subsection{Height-difference correlation function: computation of $\alpha$ and $z$ exponents}

\begin{figure}[b!]
\centering
\includegraphics[width=0.7\textwidth]{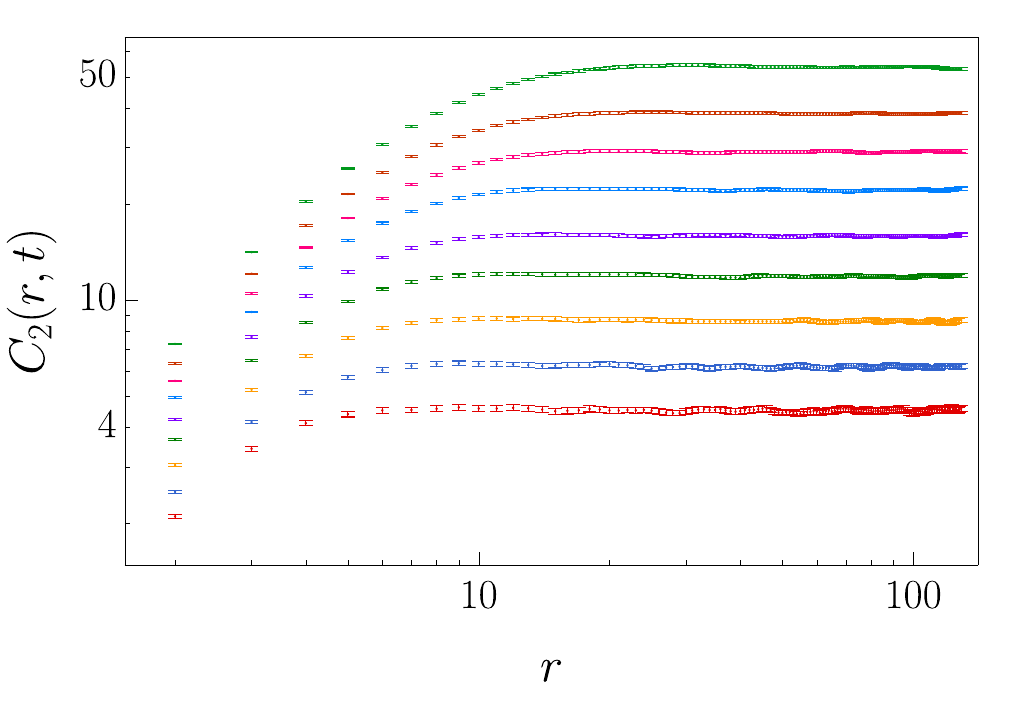}
\caption{Height-difference correlation function vs. $r$ for time boxes increasing from 20 to 100, bottom to top, at regular intervals for $T=1$ and $A=1$.}
\label{fig4:c2}
\end{figure}

Figure~\ref{fig4:c2} shows the height-difference correlation function as a function of $r$ for different times and a given parameter condition. In this figure, we plot the height-difference correlation function only up to $r=L_y/2=128$, as the function is symmetric by definition. From a physical perspective, since the front has PBC, the distances $r$ and $-r$ (or equivalently, $L_y-r$) exhibit the same correlation. 

\begin{figure}[t!]
\centering
\includegraphics[width=0.7\textwidth]{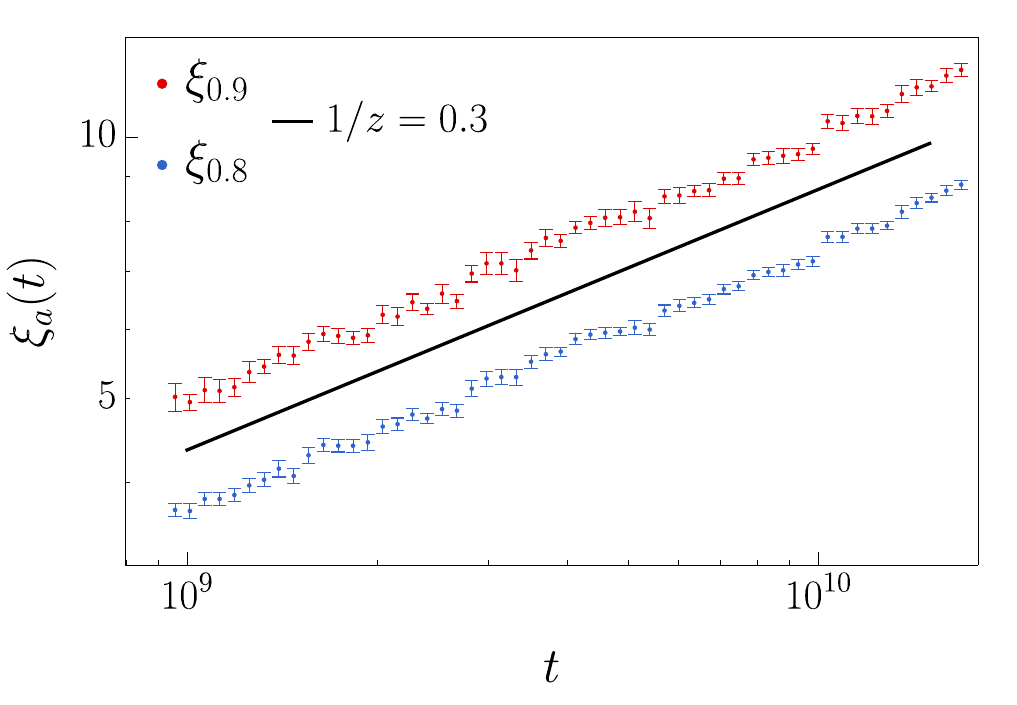}
\caption{Estimates $\xi_{0.8}(t)$ and $\xi_{0.9}(t)$ as functions of time, obtained for $T=1$ and $A=1$. As a visual reference, the solid black line corresponds to $\xi(t) \sim t^{0.3}$.}
\label{fig4:dc}
\end{figure}

As explained in Sec.\ \ref{sec3:observables}, the correlation length at a given time $t$, denoted as $\xi(t)$, can be estimated from the plateau of the $C_2(r,t)$ curves at sufficiently large $r$ for different values of $a$. According to Eq.~\eqref{eq3:correlation_length}, the double logarithmic plots of these correlation lengths as functions of time should yield straight lines, whose slopes correspond to the exponent $1/z$. Figure \ref{fig4:dc} presents log-log plots of $\xi_a(t)$ versus $t$ for the precursor layer, calculated for $a = 0.8$ and $a = 0.9$, with an estimated exponent of $1/z \sim 0.3$.

Since the distance values $ r $ at which the height-difference correlation function is evaluated are discrete, linear interpolation was applied in the estimation of $ \xi_a(t) $ to improve the accuracy of the correlation distance measurements, following Eq.~\eqref{eq3:xi_a}. Moreover, for simplicity, the correlation functions evaluated at $ r = L_y/2 $ were used as an approximation for the plateau value.

\begin{figure}[t!]
\centering
\includegraphics[width=0.7\textwidth]{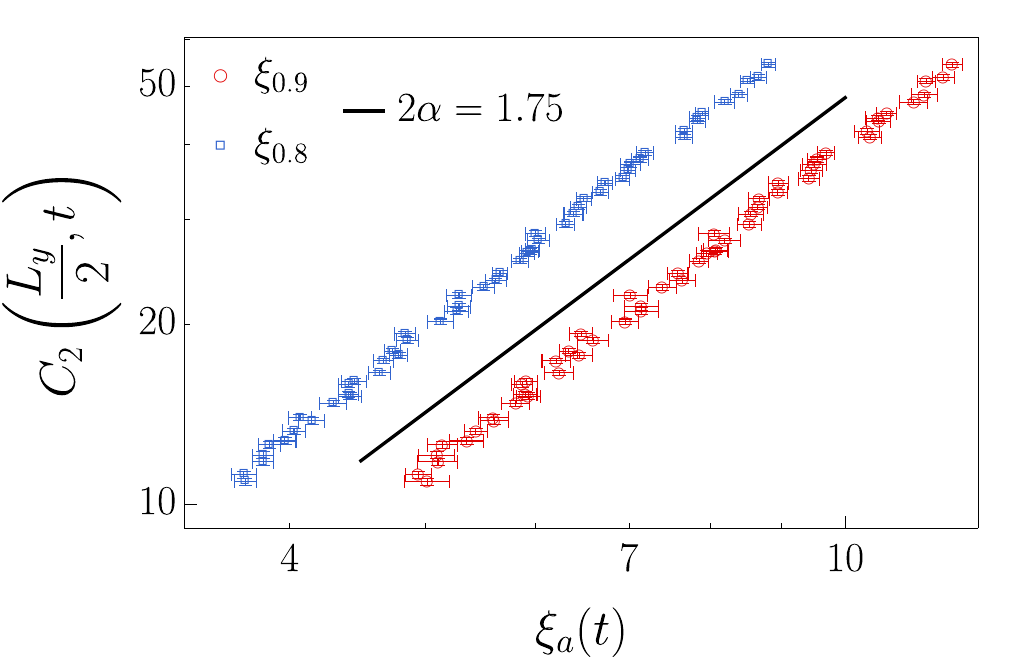}
\caption{Height-difference correlation function $C_2\left(L_y/2,t\right)$ versus $\xi_{0.8}(t)$ and $\xi_{0.9}(t)$ at different times. Conditions are $T=1$ and $A=1$ 
. As a visual reference, the solid black line corresponds to $C_2\left(L_y/2,t\right) \sim t^{1.75}$.}
\label{fig4:plvsdc}
\end{figure}

On the other hand, Eq.~\eqref{eq3:c2p} gives $ C_{2,p}(t) \sim \xi^{2\alpha}(t) $ for $ r \gg \xi(t) $. Consequently, the exponent $ \alpha $ can be determined from the slope of the best-fit lines in a log-log plot of $ C_2(r,t) $ versus $ \xi(t) $ at the plateau. In Fig.~\ref{fig4:plvsdc}, we plot $ C_2(L_y/2,t) \equiv C_{2,p}(t) $ against $ \xi_a(t) $ for the precursor layer, using the same values of $ a $, with an estimated exponent of $ 2\alpha \sim 1.75 $.
The full set of $ 1/z $ values, calculated for $ a = 0.8 $ and $ a = 0.9 $, is provided in Tables \ref{tab4:1overz} and \ref{tab4:1overz_a9}, respectively. Similarly, the set of $ 2\alpha $ exponents, also computed for $ a = 0.8 $ and $ a = 0.9 $, is presented in Tables \ref{tab4:2alpha} and \ref{tab4:2alpha_a9}, respectively. From these data, along with the previously presented values of the $ \beta $ exponent, one can easily verify that the expected scaling relation $ \alpha = \beta z $ holds for both the precursor and the supernatant layers. Moreover, one can easily verify that the exponents are independent of the specific value of the parameter $a$ used in the computation.

Similar to what was previously noted for $ \beta $, the dependence of $ \alpha $ and $ z $ on the Hamaker constant is relatively minor, whereas their dependence on temperature is much more pronounced. This trend also suggests a transition from a low-temperature to a high-temperature regime, with $ T $-dependent exponents for intermediate temperatures around $ T = 1 $. In general, both $ \alpha $ and $ z $ exhibit a sharp change with $ T $, shifting from their low-$ T $ values to approximately $ \alpha \approx 0.9 $ and $ z \approx 3.4 $ in the high-$ T $ regime.

\begin{table}[h]
\centering
\small
\renewcommand{\arraystretch}{1.2}
\begin{tabular}{@{}lccccc@{}}
\toprule
\diagbox[width=4em,height=3em,dir=SE,trim=l]{$A$}{$T$} & 10 & 3 & 1 & 3/4 & 1/2 \\
\cmidrule(r){1-6}
10   & 0.299(5) & 0.309(5) & 0.286(3) & 0.251(6) & 0.202(9) \\
5    & 0.299(5) & 0.304(6) & 0.289(3) & 0.224(6) & 0.203(8) \\
1    & 0.299(6) & 0.304(3) & 0.303(4) & 0.259(6) & 0.27(1)  \\
0.1  & 0.309(5) & 0.304(3) & 0.257(4) & 0.236(6) & 0.28(2)  \\
0.01 & 0.301(5) & 0.307(5) & 0.260(5) & 0.245(9) & 0.25(2)  \\
\bottomrule
\end{tabular}
\caption{Values of the exponent $1/z$ for the precursor layer, calculated with $a=0.8$, for all the conditions studied.}
\label{tab4:1overz}
\end{table}
\begin{table}[h]
\centering
\small
\renewcommand{\arraystretch}{1.2}
\begin{tabular}{@{}lccccc@{}}
\toprule
\diagbox[width=4em,height=3em,dir=SE,trim=l]{$A$}{$T$} & 10 & 3 & 1 & 3/4 & 1/2 \\
\cmidrule(r){1-6}
10   & 1.83(2) & 1.77(2) & 1.78(2) & 1.92(3) & 1.33(5) \\
5    & 1.83(2) & 1.79(3) & 1.76(2) & 1.95(4) & 1.35(4) \\
1    & 1.83(3) & 1.81(2) & 1.78(2) & 2.06(5) & 1.15(5) \\
0.1  & 1.80(2) & 1.81(1) & 1.89(3) & 1.43(3) & 1.07(5) \\
0.01 & 1.82(3) & 1.79(3) & 1.88(3) & 1.40(5) & 1.23(6) \\
\bottomrule
\end{tabular}
\caption{Values of the exponent $2\alpha$ for the precursor layer, calculated with $a=0.8$, for all the conditions studied.}
\label{tab4:2alpha}
\end{table}
\begin{table}[h]
\centering
\small
\renewcommand{\arraystretch}{1.2}
\begin{tabular}{@{}lccccc@{}}
\toprule
\diagbox[width=4em,height=3em,dir=SE,trim=l]{$A$}{$T$} & 10 & 3 & 1 & 3/4 & 1/2 \\
\cmidrule(r){1-6}
10   & 0.296(6) & 0.305(7) & 0.285(4) & 0.251(8) & 0.22(3) \\
5    & 0.294(7) & 0.301(6) & 0.288(5) & 0.247(7) & 0.21(1) \\
1    & 0.295(6) & 0.299(3) & 0.307(5) & 0.261(9) & 0.30(3) \\
0.1  & 0.306(8) & 0.301(4) & 0.258(5) & 0.245(9) & 0.3(1)  \\
0.01 & 0.296(7) & 0.304(7) & 0.263(7) & 0.25(2)  & 0.18(4) \\
\bottomrule
\end{tabular}
\caption{Values of the exponent $1/z$ for the precursor layer, calculated with $a=0.9$, for all the conditions studied.}
\label{tab4:1overz_a9}
\end{table}
\begin{table}[h!]
\centering
\small
\renewcommand{\arraystretch}{1.2}
\begin{tabular}{@{}lccccc@{}}
\toprule
\diagbox[width=4em,height=3em,dir=SE,trim=l]{$A$}{$T$} & 10 & 3 & 1 & 3/4 & 1/2 \\
\cmidrule(r){1-6}
10   & 1.85(4) & 1.79(3) & 1.79(3) & 1.91(6) & 1.2(1)  \\
5    & 1.85(4) & 1.80(3) & 1.76(3) & 1.93(5) & 1.30(7) \\
1    & 1.86(3) & 1.84(2) & 1.75(3) & 2.04(7) & 1.0(1)  \\
0.1  & 1.81(4) & 1.82(3) & 1.87(4) & 1.36(6) & 0.7(3)  \\
0.01 & 1.85(4) & 1.80(4) & 1.85(5) & 1.33(8) & 1.5(3)  \\
\bottomrule
\end{tabular}
\caption{Values of the exponent $2\alpha$ for the precursor layer, calculated with $a=0.9$, for all the conditions studied.}
\label{tab4:2alpha_a9}
\end{table}

\clearpage
\subsection{Anomalous scaling of the height-correlation function}

Global roughness exponent values of $ \alpha \lesssim 1 $, as obtained here for $ T \gtrsim 1/2 $, indicate large fluctuations in the front position. In this case, these fluctuations are associated to intrinsic anomalous scaling. This behavior is evident in Fig.~\ref{fig4:c2}, where the $ C_2(r,t) $ curves for different times systematically shift over time without overlapping. As discussed in Sec.\ \ref{sec3:observables}, this is a landmark behavior of anomalous scaling.

\begin{figure}[b!]
\centering
\includegraphics[width=0.7\textwidth]{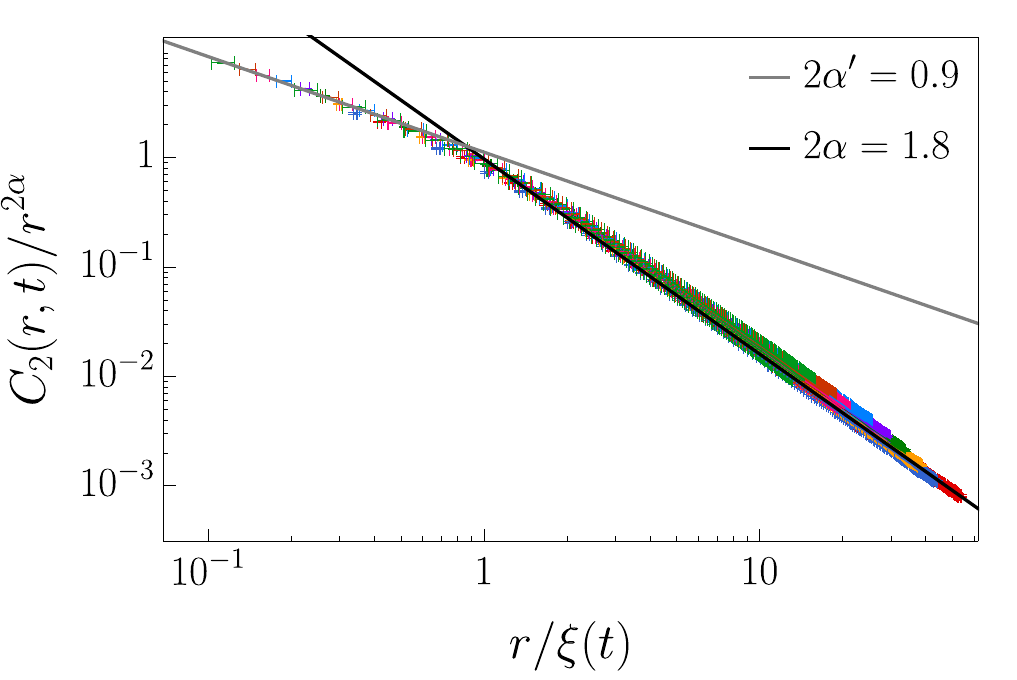}
\caption{Data collapse of the height-difference correlation function obtained for different values of time, for $T=1$ and $A=1$, using $\alpha=0.9$. The curve onto which collapse occurs is the function $g(r/\xi(t))$ of Eq.~\eqref{eq3:c2an}, the solid black line representing the theoretical behavior for large $u$, $g(u) \sim u^{-2\alpha}$,  and the solid gray line representing the behavior for small $u$, $g(u) \sim u^{-2\alpha'}$ (see Tables \ref{tab4:2alpha} and \ref{tab4:ap}).}
\label{fig4:c2_reescalado}
\end{figure}

Anomalous scaling can arise from various causes. One of which is the presence of large $ \alpha $ values, known as superroughening \cite{Lopez1997}. In the present case, it originates from the fact that $ \alpha_{\rm loc} \neq \alpha $, indicating the existence of two independent roughness exponents. This is clearly demonstrated in Fig.~\ref{fig4:c2_reescalado}, which shows a consistent data collapse of the height-difference correlation function following Eq.~\eqref{eq3:c2an} for a representative set of parameters. If the scaling behavior followed the standard FV type, the scaling function $ g(u) $ would be independent of $ u $ for small arguments (\mbox{$u \ll 1$}). However, our data instead align with a scaling law of the form $g(u)\sim u^{-2\alpha'} $, ($\alpha'=\alpha-\alpha_\mathrm{loc}$) with $ 2\alpha' \approx 0.9 $, leading to $ \alpha_\mathrm{loc}\approx0.45$ while $\alpha=0.89$. This confirms the presence of intrinsic anomalous scaling \cite{Lopez1997}. Similar behavior is observed for other parameter choices, with specific exponent values provided in Table \ref{tab4:ap}. As mentioned in Sec.\ \ref{sec3:observables}, the anomalous shift of the height-difference correlation function curves over time, as shown in Fig.~\ref{fig4:c2}, could, in principle, be attributed to a large roughness exponent. However, the data collapse in Fig.~\ref{fig4:c2_reescalado} with $ \alpha' \neq 0 $ clearly confirms that this behavior originates from intrinsic anomalous scaling.

\begin{table}[h]
\centering
\small
\renewcommand{\arraystretch}{1.2}
\begin{tabular}{@{}lccccc@{}}
\toprule
\diagbox[width=4em,height=3em,dir=SE,trim=l]{$A$}{$T$} & 10 & 3 & 1 & 3/4 & 1/2 \\
\cmidrule(r){1-6}
10   & 0.89(2) & 0.86(3) & 0.87(2) & 1.00(3) & 0.44(5) \\
5    & 0.90(2) & 0.86(2) & 0.85(2) & 1.03(4) & 0.47(3) \\
1    & 0.89(3) & 0.88(1) & 0.87(2) & 1.13(4) & 0.36(5) \\
0.1  & 0.87(2) & 0.88(1) & 1.01(3) & 0.61(3) & 0.37(5) \\
0.01 & 0.89(3) & 0.87(3) & 0.99(3) & 0.60(5) & 0.53(6) \\
\bottomrule
\end{tabular}
\caption{Values of the exponent $2\alpha'$ for the precursor layer, calculated with $a=0.8$, for all the conditions studied.}
\label{tab4:ap}
\end{table}

\vspace{-20pt}
\subsection{Structure factor}

\begin{figure}[!t]%
    \centering
    \includegraphics[width=0.7\textwidth]{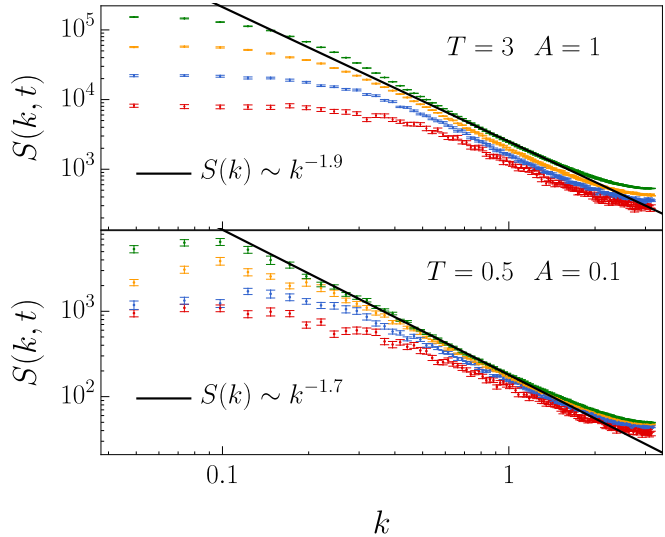}
    \label{subfig:structureFactor}
    \caption{Structure factor calculated for $T = 0.5$, $A=0.1$ (bottom panel) and $T = 3$, $A=1$ (top panel), for times increasing bottom to top in both panels. The scaling behavior at a fixed time is $S(k,t) \sim |k|^{-(2\alpha_{\rm loc}+1)}$, where $\alpha_{\rm loc}$ has been evaluated as $\alpha-\alpha'$, see Tables \ref{tab4:2alpha} and \ref{tab4:ap}. The power laws represented by the solid lines are indicated in the corresponding legends.}%
    \label{fig4:structureFactor}%
\end{figure}

To further understand the intrinsic anomalous scaling of the front, it is worth analyzing it by means of the structure factor. Figure \ref{fig4:structureFactor} presents the structure factor calculated at different times for two representative temperatures, $ T = 0.5 $ and $ T = 3 $. Notably, the $ S(k,t) $ curves systematically shift upward over time, consistent with Eq.~\eqref{eq3:Skanom}, which is another hallmark of intrinsic anomalous scaling \cite{Lopez1997}.

Indeed, in the presence of intrinsic anomalous scaling, the structure factor is expected to scale as $S(k,t)\sim |k|^{-(2\alpha_\mathrm{loc}+1)}$ for long enough times [see Eq.~\eqref{eq3:sfactor_scaling}], so that the roughness exponent derived from the power-law behavior of $ S(k,t) $ in Fig.~\ref{fig4:structureFactor} corresponds to $ \alpha_{\rm loc} $ rather than $ \alpha $.

This is particularly relevant to the original results reported by Abraham \textit{et al.} in Ref. \cite{Abraham2002}. While the systematic time shift of the structure factor is clearly visible in Figure 3(a) of that paper, the interpretation of the scaling exponents was overlooked there. Therefore, we infer that the low-temperature roughness exponent obtained in Ref. \cite{Abraham2002} corresponds to the local exponent rather than the global one.

\subsection{Front fluctuations}

As discussed in Chapter \ref{chap1:intro}, recent advances in surface kinetic roughening, particularly in the context of KPZ scaling, have shown that universality extends beyond just the values of critical exponents for many important universality classes. Specifically, by normalizing front fluctuations around their mean by their time-dependent amplitude [see Eq.~\eqref{eq3:flu}], the PDF of these $ \chi $ random variables becomes time-independent and is shared by all members of the same universality class \cite{Kriecherbauer2010,HalpinHealy2015,Carrasco2016,Takeuchi2018,Carrasco2019}.

\begin{figure}[!htbp]
\centering
\includegraphics[width=0.7\textwidth]{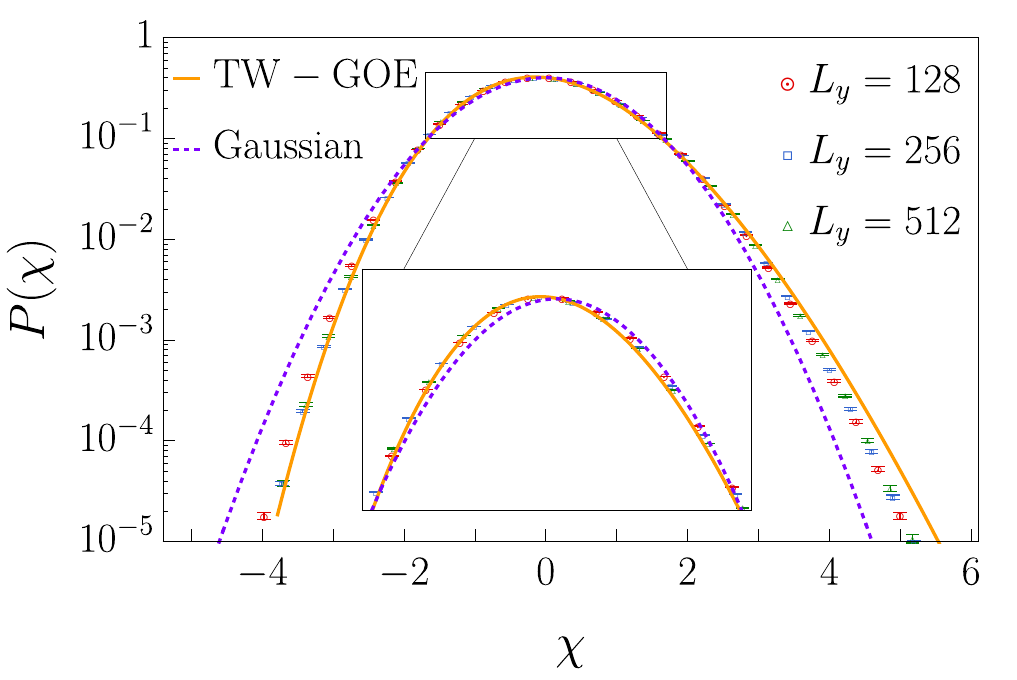}
\caption{Fluctuation histograms calculated according to Eq.~\eqref{eq3:flu} for $A = 1$, $T=1$ and several system sizes, as indicated in the legend. The solid orange line corresponds to the TW-GOE distribution while the dotted purple line corresponds to the Gaussian distribution.}
\label{fig4:histograma}
\end{figure}

Figure~\ref{fig4:histograma} shows the PDF corresponding to various system sizes for a given parameter condition, along with the Gaussian distribution and the TW-GOE distribution, which is expected for the KPZ class in one-dimensional flat fronts (as opposed to the TW-GUE distribution expected for a circular geometry). The agreement with the TW-GOE distribution is remarkable, especially considering that the exponents of the system do not match those of the KPZ universality class. Moreover, the agreement slightly improves for larger system sizes. However, simulations for system sizes larger than those shown in Fig.~\ref{fig4:histograma} are not computationally feasible. Given that the kinetic roughening of our kMC fronts exhibits intrinsic anomalous scaling (whereas the KPZ equation \cite{Kardar1986} follows the standard FV type) and features non-KPZ exponents, the agreement of our numerical PDF with the TW distribution is particularly striking.

In addition, we have computed the skewness and excess kurtosis for the PDF in Fig.~\ref{fig4:histograma}. Their values are: $ S = 0.221(3) $, $ K = 0.239(5) $ for $ L_y = 128 $; $ S = 0.236(2) $, $ K = 0.249(1) $ for $ L_y = 256 $; and $ S = 0.264(2) $, $ K = 0.239(4) $ for $ L_y = 512 $. These values suggest that, while $ K $ remains relatively stable with system size, $ S $ increases as $ L_y $ grows. \footnote{For reference, the exact skewness and excess kurtosis values of the TW-GOE distribution are $S=0.29346452408$ and $K=0.1652429384$, respectively \cite{Bornemann2010}.}

\subsection{Front covariance}

The front covariance $ C_1(r,t) $, defined in Eq.~\eqref{eq3:correlation_1}, also displays KPZ behavior. As discussed in Sec.\ \ref{sec3:observables}, this function is expected to behave as
\begin{equation}
    C_1(r,t)=a_1 \, t^{2\beta} f\left(a_2 r/t^{1/z} \right) ,
   \label{eq4:airy1}
\end{equation}
where $ f(u) $ is a universal function, and $ a_1 $ and $ a_2 $ are non-universal constants \cite{Alves2011,Oliveira2012,Nicoli2013} to be determined from simulations. As mentioned in Chapter \ref{chap1:intro} and detailed in Sec.\ \ref{sec3:observables}, for the one-dimensional KPZ equation with PBC, the function $ f(u) $ corresponds to $ \mathrm{Airy_1}(u) $, where $ \mathrm{Airy}_1(u) $ represents the covariance of the Airy$_1$ process \cite{Bornemann2009,HalpinHealy2015,Takeuchi2018}. Moreover, the procedure to compute $a_1$ and $a_2$ is detailed in Sec.\ \ref{sec3:observables} [see Eqs.\ \eqref{eq3:airy1}--\eqref{eq3:airy1b}].

\begin{figure}[h]
\centering
\includegraphics[width=0.7\textwidth]{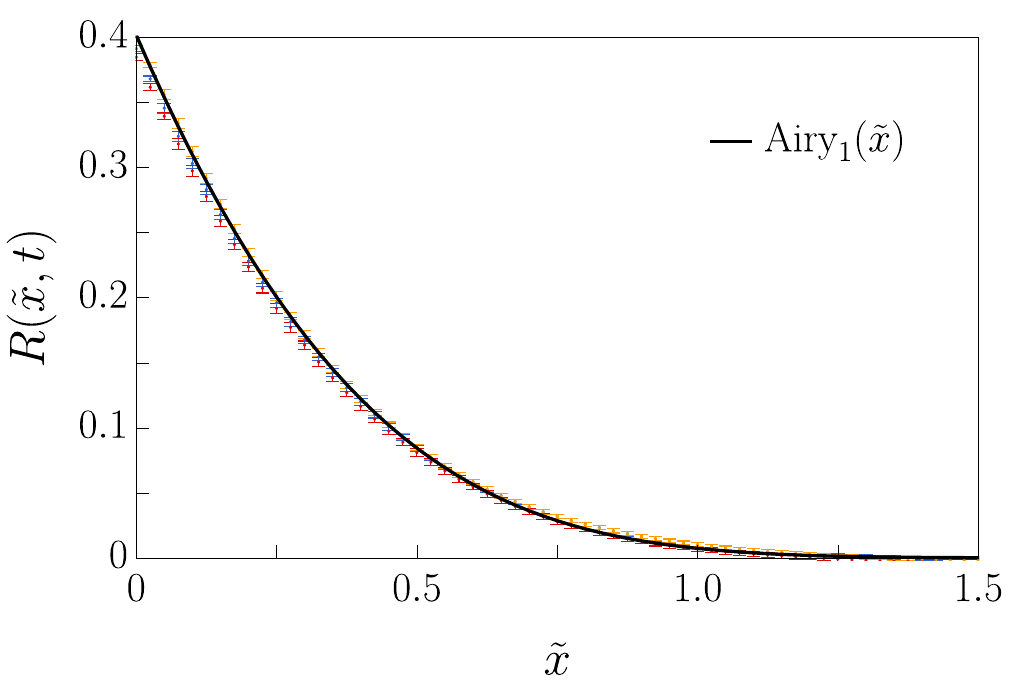}
\caption{$R(\tilde{x},t) \equiv \frac{C_1\left(\tilde{x} t^{1/z}/a_2\right)}{a_1t^{2\beta}}$ versus $\tilde{x}\equiv a_2 r/t^{1/z}$ for the time boxes $t_{\mathrm{BOX}}=60$, $80$, and $100$, calculated for the same conditions as in Fig.~\ref{fig4:histograma}, using $1/z=0.32 $, $2\beta=0.544 $, $a_1= 1.834\times10^{-4}$, and $a_2=8.985\times 10^{-3}$. The solid line corresponds to the exact $\mathrm{Airy}_1(\tilde{x})$ function.}
\label{fig4:c1}
\end{figure}

Figure \ref{fig4:c1} shows the collapsed height covariance function
\begin{equation}
    C_1(\tilde{x}t^{1/z}/a_2)/(a_1 t^{2\beta})\equiv~R(\tilde{x},t),
\end{equation}
plotted against $ \tilde{x} $ for various times. The figure confirms that the universal behavior predicted by Eq.~\eqref{eq4:airy1} holds with $ f(u) = \mathrm{Airy_1}(u) $, despite the exponent values differing from those of 1D KPZ and despite the fact that the front exhibits intrinsic anomalous scaling. Notably, this agreement deteriorates as the temperature decreases. For $T\lesssim 3/4$, the rescaled front covariance significantly deviates from the $\mathrm{Airy_1}$ form.

\section{Conclusions}

In summary, in this chapter we have investigated the spatiotemporal dynamics of liquid drop fronts spreading on planar substrates through numerical simulations of the Ising lattice gas model for the spreading described in detail in Chapter \ref{chap2:wetting}. We have analyzed its behavior under different parameter settings—such as the Hamaker constant (wettability) and temperature—using extensive kMC simulations. 

Across a broad range of model parameters, we have examined classical morphological observables, including the mean front position and roughness. Furthermore, we have systematically analyzed two-point correlation functions in both real and Fourier space, tracking their temporal evolution. Additionally, we have evaluated the statistical properties of front fluctuations through their PDF.

\begin{figure}[ht]
\centering
\includegraphics[width=0.7\textwidth]{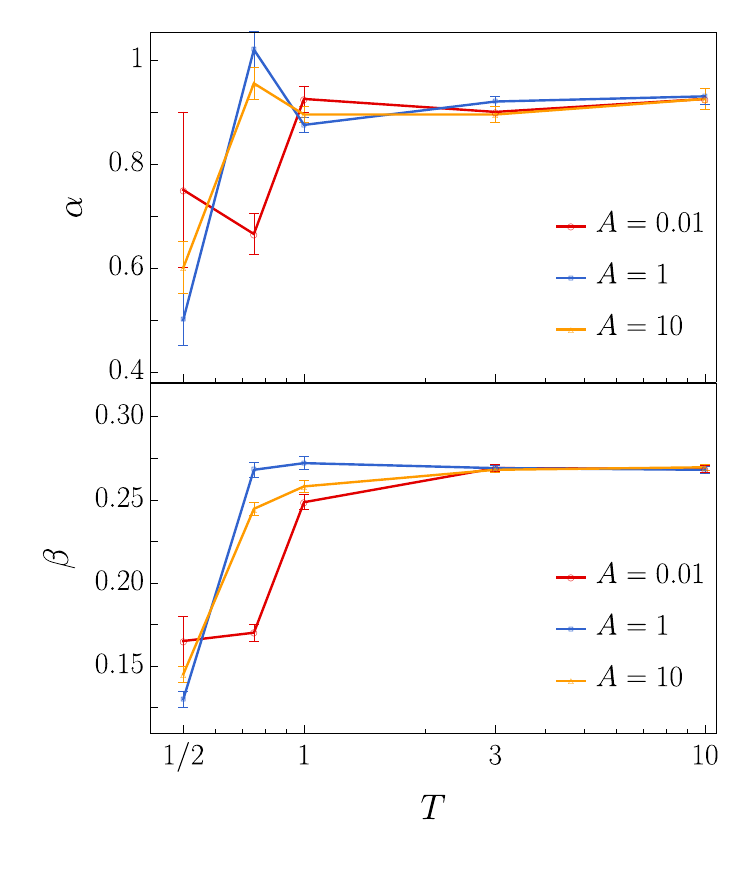}
\caption{Values of $\alpha$ (top) and $\beta$ (bottom) for the precursor film (taken from Tables \ref{tab4:delta1} and \ref{tab4:1overz}) vs $T$ for $A=0.01$ (red circles), $A=1$ (blue squares), and $A=10$ (orange triangles). Lines are guides to the eye.}
\label{fig4:TablaExponentes}
\end{figure}

We can summarize the main findings obtained for the discrete lattice gas model as follows. The scaling properties of the fronts in both the precursor and supernatant layers are identical. The exponent $\delta \approx 0.50$, which characterizes the mean position of the front, appears to be universal across all parameter values considered. Regardless of these parameter values, the front exhibits intrinsic anomalous scaling, meaning that the roughness exponents quantifying front fluctuations differ between large ($\alpha$) and small ($\alpha_{\rm loc}$) length scales. Moreover, the critical exponent values $\beta$, $\alpha$, and $z$ are more strongly influenced by temperature than by the Hamaker constant, showing a transition from a low-temperature to a high-temperature regime. This is clearly illustrated in Figure \ref{fig4:TablaExponentes}, which shows the temperature dependence of the $\alpha$ and $\beta$ exponents [note that $z$ is related to them through Eq.~\eqref{eq1:zalphabeta}] for several values of the Hamaker constant.

For the lowest temperatures studied, the exponent values align closely with those previously reported for the same model \cite{Abraham2002}, specifically \mbox{$\alpha\simeq 0.6$}, $\alpha_{\rm loc}\simeq 0.38$, $z\simeq3.3$, and $\beta\simeq0.18$. As temperature increases, the exponent values change rapidly and eventually become independent of $T$ for $T\gtrsim 1$, reaching approximately $\alpha\simeq 0.90$, $\alpha_\mathrm{loc} \simeq 0.45$, $z \simeq 3.3$, and $\beta\simeq 0.26$. Despite these variations in exponent values, the statistics of front fluctuations remain consistent with those characteristic of the one-dimensional KPZ universality class.


Overall, our simulations highlight the emergence of universal behavior in the spreading of thin fluid films, which becomes particularly evident at high temperatures. Interestingly, in this regime, the front fluctuations exhibit properties that classify them as another instance of 1D KPZ behavior, though not necessarily with KPZ exponents. This has been already observed in the dynamics of many low-dimensional, strongly correlated, nonequilibrium systems \cite{Takeuchi2018}. In particular, it has been found that the PDF of a continuous equation with the KPZ nonlinearity and a coefficient $\lambda \propto 1/t^{1/2}$ shows good agreement with the TW-GOE distribution \cite{Marcos2022}. Similarly, other systems exhibiting anomalous scaling and non-KPZ exponents have recently been found to possess a PDF that follows the TW distribution, such as the synchronization of oscillator lattices \cite{Gutierrez2023}. An especially interesting direction for future research would be to determine whether this conclusion, derived from the “microscopic” simulations reported here, can be corroborated by alternative simulation methods, such as MD or lattice-Boltzmann approaches, and/or by experimental studies on the spreading of precursor films.

At this stage, in relation to identifying universal scaling behavior, it is important to acknowledge that our kMC results are inherently constrained by the finite system sizes used and the statistical analysis performed. In our simulations, we have generally been unable to control the (subdominant) scaling corrections in the data, primarily due to the strong temporal correlation of the measured observables and the absence of precise theoretical predictions across the different parameter regions studied. Without such control, it is challenging to provide stronger numerical arguments in support of universal behavior. Nevertheless, the critical exponent values reported remain statistically consistent across a wide range of parameters, with differences generally within two standard deviations. Furthermore, the long simulation times give us confidence that the impact of subdominant terms can be reasonably neglected.


\graphicspath{{5_capitulo/fig5/}}

\chapter{Radial Spreading}
\label{chap5:radial_spreading}


In this chapter, we examine the model analyzed in the previous one, transitioning from a band geometry to a radial geometry. We will first discuss the necessary modifications for simulating the model, followed by the presentation of the results. Next, we will present some conclusions based on these results and compare them with those obtained for the band geometry. Although the system discussed in this chapter differs from the previous one only in terms of geometry, we will provide a comprehensive summary of all its characteristics for completeness.

\section{Model and simulation details}
\label{sec5:model}

The microscopic driven Ising lattice gas model discussed in this chapter also consists of two overlapping 2D rectangular layers, now defined with dimensions $ L_\mathrm{side} \times L_\mathrm{side} $. Again, each site on the square lattice, represented as $ \boldsymbol{r} = (x, y, Z) $, can be occupied by a maximum of one particle at any moment and the occupation number $ n(\boldsymbol{r},t) $ can only assume values of 0 or 1. The lower layer ($Z = 1$) is the precursor layer, while the upper layer ($Z = 2$) is referred to as the supernatant. The substrate, which serves as the base for the droplet's expansion, is located at $ Z = 0 $. In this geometry, PBC are not applied in any direction. As in the previous chapter, if a particle reaches any of the four borders of the lattice, it is assumed to escape from the system. The energy of the system remains the same as in the previous chapter, specifically:
\begin{equation}
    \mathcal{H}= -J \sum_{\langle \boldsymbol{r}, \boldsymbol{s} \rangle} n(\boldsymbol{r},t)n(\boldsymbol{s},t) - A \sum_{\boldsymbol{r}}\frac{n(\boldsymbol{r},t)}{Z^3}.
    \label{eq5:energy}
\end{equation}
As before, the first term represents the interactions between liquid particles and their nearest neighbors, while the second term accounts for the interaction with the substrate.

\begin{figure}[t]
\centering
\includegraphics[width=0.7\textwidth]{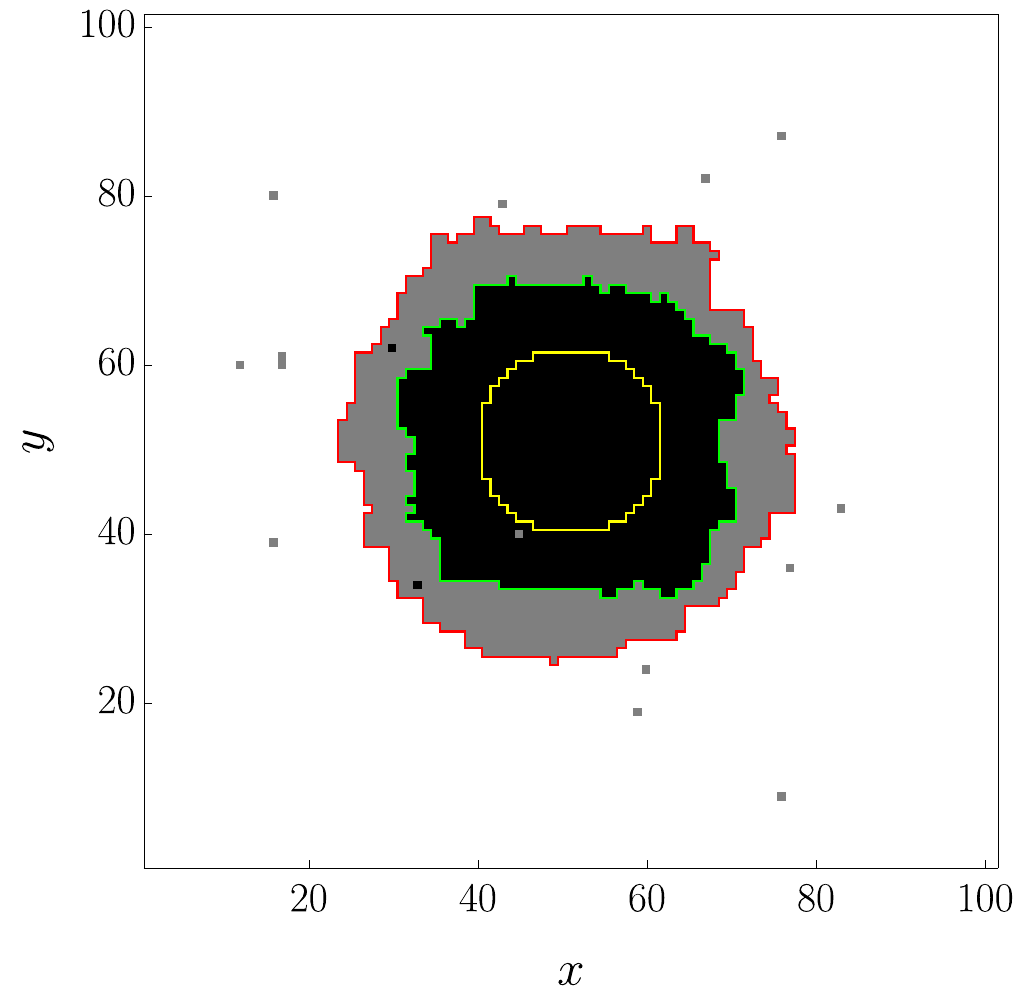}
\caption{Top view of a snapshot of the lattice gas model. The occupied cells in the precursor and supernatant layers are represented in gray and black, respectively, while empty cells are uncolored. The red and green lines delimit the corresponding fronts, while the yellow line delimits the reservoir. The conditions used were $T=1/3$, $A=10$, $R_R=11$, $J=1$, and $L_{\rm side}=101$.}
\label{fig5:Snapshot}
\end{figure}

While in the previous chapter the fluid reservoir was defined as the first column $(x=0)$ and the spreading occurs along the $x$-direction, in this chapter a more careful definition of the reservoir has to be made. Actually, in this context, the reservoir can be defined in multiple ways. The most straightforward approach is to define the central cell of the system as the reservoir. However, this choice results in extremely slow fluid film growth, since only a single cell per layer supplies material for the expansion, leading to prohibitively long simulation times. Specifically, after a few steps of the kMC algorithm a substantial number of particles become disconnected from the reservoir. As a result, most transitions fail to contribute to film growth, causing the algorithm to operate significantly slower. Moreover, a point-like reservoir may be an overly idealized representation of experimental conditions. To overcome this issue, an alternative approach was adopted, utilizing a larger reservoir that includes all cells within a specified radius $ R_{R} $ (the reservoir radius) from the center of the system. The shape of this reservoir is depicted in Figure \ref{fig5:Snapshot}. This selection significantly accelerates the system dynamics. Once again only these cells are initially occupied. If, during the evolution of the system, a cell in the reservoir becomes empty due to an exchange, it is instantly refilled.

This reservoir definition is not the only approach that can be used to address the slow growth problem. Besides the circular reservoir used in the simulations, several tests were conducted with alternative geometries, including a square and a hexagon. The results were comparable across all cases. The circular reservoir was selected for its simplicity and the ease with which relevant physical quantities can be computed. Furthermore, the reservoir size must be chosen carefully. If it is too small, the same issues observed in single-cell reservoirs may arise. Conversely, if the reservoir is excessively large, the front will require a considerable time to grow away from it, as it must traverse a greater area before reaching a given distance. The reservoir size adopted here strikes a balance between these two extremes.

As in the previous chapter, the evolution of the system has been simulated by continuous-time MC Kawasaki local dynamics. At each time, a particle is considered as belonging to the precursor (or the supernatant) film if there are nearest-neighbor connections filled with particles all the way back to the droplet reservoir. However, defining the front is not as straightforward as in the previous chapter. In the band geometry, for a fixed $y$, the front position $ h(y,t,Z) $ was determined as the highest $x$-coordinate where a cell remained connected to the reservoir. In the radial geometry, a particle is considered to be at the front if it belongs to the film, i.e,. if it is connected to the reservoir, and there exists an empty nearest-neighbor cell connected to the system boundary through empty nearest-neighbors.

This ``strict'' definition of the front is clearly computationally inefficient. To improve the efficiency of the algorithm, a simplified alternative definition has been adopted. Specifically, instead of verifying whether connections extend all the way to the system boundary, which is located at a distance much greater than the typical film sizes in our simulations, we only check for empty nearest-neighbor connections up to a distance equal to twice the last measured film size plus an offset of 10. It is assumed that if a path exists up to this distance, a connection to the system boundary is also present. In Figure \ref{fig5:Snapshot}, the resulting fronts of the precursor and supernatant films are depicted by red and green lines, respectively.


As in the previous chapter, we adopt physical units such that $ k_B = 1 $, while other parameters remain arbitrary. Additionally, in all simulations, we fix $ J = 1 $, modifying only the Hamaker constant $ A $ and the temperature $ T $. The system size was set to $ L_\mathrm{side}=1001 $ in all runs, ensuring that the film does not reach the system's boundary. A summary of all the simulation conditions considered is provided in Table \ref{tab5:parametros}.

Following the same reasoning as in the previous chapter, the exact parameter values are not relevant, only the ratios $ J/k_{\mathrm{B}}T $ and $ A/k_{\mathrm{B}}T $ determining the evolution of the system. Again, from a physical perspective, the most relevant conditions are those where $ J/k_{\mathrm{B}}T $ is sufficiently large to ensure a high degree of involatility and $ A/k_{\mathrm{B}}T $ is large enough to place the system in the complete wetting regime. However, as in the previous chapter, we also present results for conditions that do not strictly adhere to these criteria.


\begin{figure}[t!]
\centering
\includegraphics[width=1.0\textwidth]{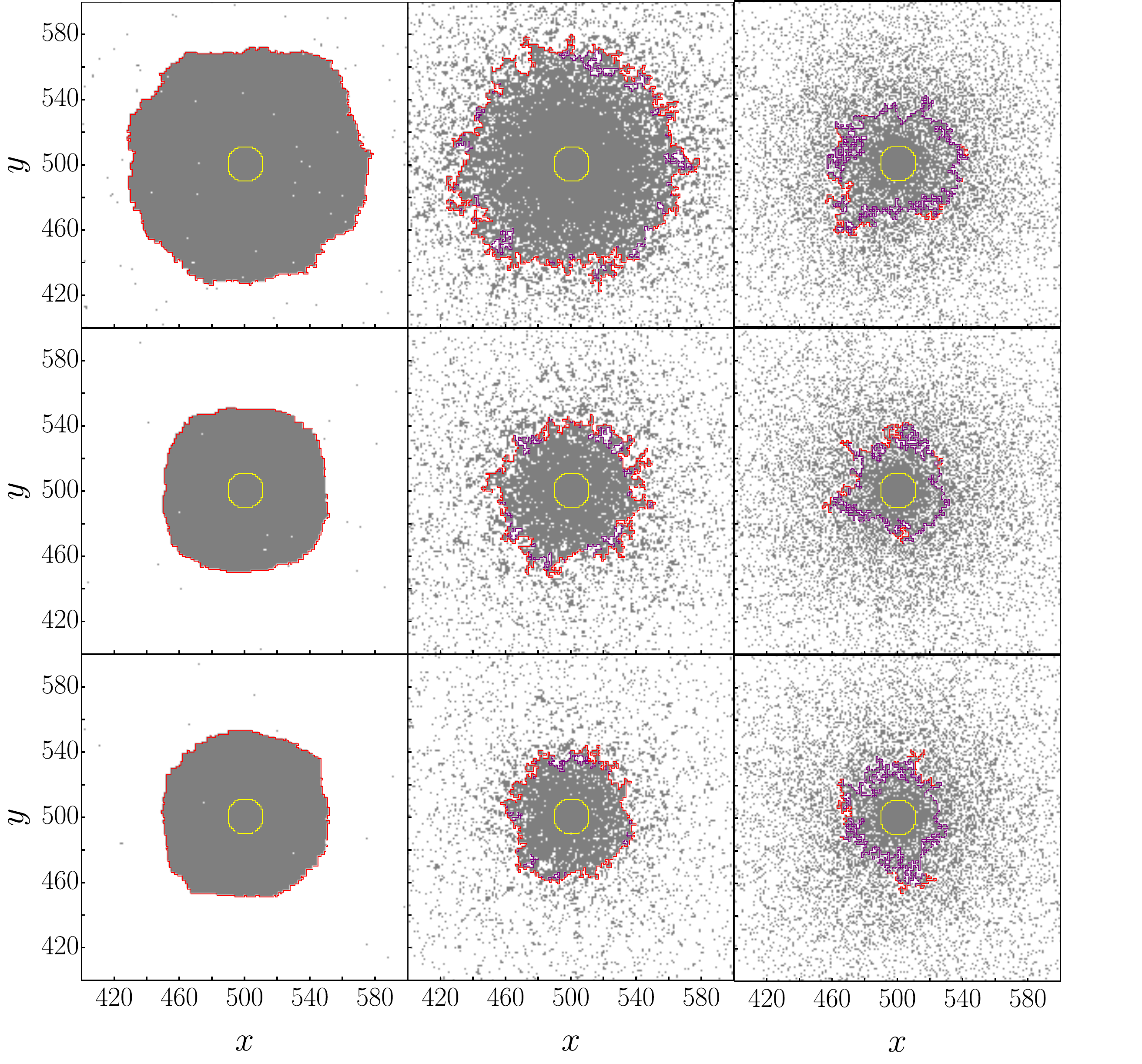}
\caption{Top views of snapshots of the bottom layer ($Z=1$) of the lattice gas model. Occupied cells are in gray, while empty cells are uncolored. The red line delimits the front of the precursor film computed in accordance with the strict definition, whereas the purple line delimits the front as computed through the eased alternative definition. The yellow line delimits the reservoir. The conditions used were $T=\{1/3,1,10\}$, $A=\{10,1,0.1\}$, $R_R=11$, $J=1$, and $L_{\rm side}=1001$. $T$ increases from left to right and $A$ increases from bottom to top.}
\label{fig5:morfologia}
 \end{figure}

Figure \ref{fig5:morfologia} illustrates the morphology of the expanding precursor film (i.e. the bottom layer) under several of the studied conditions. The front computed according to the ``strict'' definition detailed above is shown in red, whereas we show in purple the front computed by allowing diagonal neighbors in the process of searching for empty cells connected to the system boundary (i.e., ``eased'' definition of the front). This figure demonstrates that the behavior of the system undergoes a significant transformation as temperature increases. At high temperatures, thermal fluctuations become much greater than the cohesive energy of the liquid particles, causing the particles to diffuse rather than cluster together. Specifically, at very high temperatures, the system exhibits substantial noise, making it difficult to define the front unambiguously. In extreme cases, it may even be impossible to identify any point belonging to the front. For this reason, excessively high temperatures have been excluded from our study, and only results up to $ T = 3 $ will be presented and discussed. The front shape for the conditions closest to the experimental ones is shown in the upper left corner of Fig.~\ref{fig5:morfologia}. 

Figure \ref{fig5:morfologia} also demonstrates that at intermediate temperatures, where a front can still be defined but lacks connectivity and appears more dispersed, the front presents some gaps when computed from its strict definition. In contrast, the relaxed definition consistently generates a continuous front without gaps. Note, however, that achieving this requires including some extraneous points to the front.

Our simulation results and observables are based on the strict definition of the front. Since gaps in the front become more prevalent at higher temperatures, we explicitly verified that the exponents obtained using both definitions of the front remain the same for one such condition ($ T = A = 1 $). We are confident that the reported exponents hold consistently for other parameter values as well. It is important to emphasize that the gaps appearing in the front under the strict definition are relatively few and small, and their locations change over time. As a result, their effect is expected to diminish when averaging over multiple runs. Note that the conditions shown in the rightmost column of Fig.~\ref{fig5:morfologia} correspond to $ T = 10 $. Under these conditions, we have refrained from reporting results since the front could not be clearly defined.

Furthermore, Fig.~\ref{fig5:morfologia} also reveals that, at very low temperatures and a low Hamaker constant (bottom left corner of the figure), the shape of the film deviates from a circular form, adopting a more square-like configuration. Under these conditions, it will be necessary to measure the relevant variables, such as roughness and front fluctuations, locally, following the guidelines provided in Section \ref{sec3:observables}. Additionally, this distinctive shape will be reflected in the analysis of the height-difference correlation function, which will deviate from the typical behavior observed in the previous chapter, where it reaches a plateau. All of these aspects will be thoroughly discussed in the next section.

\begin{table}[p]
\centering
\renewcommand{\arraystretch}{1.0}
\begin{tabular}{@{}ccccc@{}}
\toprule
$L_{\rm side}$ & $T$ & $A$ & $N_E$ & Runs \\
\midrule\midrule

\multirow{5}{*}{1001} & \multirow{5}{*}{10}
  & 10   & $1.25 \times 10^8$ & 50 \\
  &   & 3     & $1.25 \times 10^8$ & 50 \\
  &   & 1     & $2.5 \times 10^8$ & 50 \\
  &   & 1/3 & $2.5 \times 10^8$ & 50 \\
  &   & 0.1   & $2.5 \times 10^8$ & 50 \\

\midrule

\multirow{5}{*}{1001} & \multirow{5}{*}{3}
  & 10   & $2.5 \times 10^8$ & 50 \\
  &   & 3     & $2.5 \times 10^8$ & 100 \\
  &   & 1     & $5 \times 10^8$   & 100 \\
  &   & 1/3 & $5 \times 10^8$   & 100 \\
  &   & 0.1   & $5 \times 10^8$   & 100 \\

\midrule

\multirow{5}{*}{1001} & \multirow{5}{*}{1}
  & 10   & $2.5 \times 10^8$ & 100 \\
  &   & 3     & $2.5 \times 10^8$ & 100 \\
  &   & 1     & $5 \times 10^8$   & 112 \\
  &   & 1/3 & $5 \times 10^8$   & 125 \\
  &   & 0.1   & $5 \times 10^8$   & 125 \\

\midrule

\multirow{5}{*}{1001} & \multirow{5}{*}{3/4}
  & 10   & $5 \times 10^8$   & 125 \\
  &   & 3     & $5 \times 10^8$   & 125 \\
  &   & 1     & $2.5 \times 10^9$ & 150 \\
  &   & 1/3 & $2.5 \times 10^9$ & 150 \\
  &   & 0.1   & $2.5 \times 10^9$ & 150 \\

\midrule

\multirow{5}{*}{1001} & \multirow{5}{*}{1/2}
  & 10   & $1.25 \times 10^9$ & 125 \\
  &   & 3     & $2.5 \times 10^9$  & 150 \\
  &   & 1     & $5 \times 10^9$    & 150 \\
  &   & 1/3 & $5 \times 10^9$    & 150 \\
  &   & 0.1   & $5 \times 10^9$    & 150 \\

\midrule

\multirow{5}{*}{1001} & \multirow{5}{*}{1/3}
  & 10   & $2.5 \times 10^9$  & 100 \\
  &   & 3     & $5 \times 10^9$    & 150 \\
  &   & 1     & $7.5 \times 10^9$  & 100 \\
  &   & 1/3 & $7.5 \times 10^9$  & 100 \\
  &   & 0.1   & $7.5 \times 10^9$  & 100 \\

\bottomrule
\end{tabular}
\caption{Parameters used for the runs reported in this chapter. Here, $N_E$ represents the total number of exchanges performed, and the last column shows the number of runs launched in each case.}
\label{tab5:parametros}
\end{table}

\newpage
\section{Results}


\subsection{Front position} 

The average distance of the precursor or supernatant front from the center of the system at a given time $t$ is defined, following Eq.~\eqref{eq3:mean_height}, as
\begin{equation} 
    \overline{h(t)}=\frac{1}{N}\sum_{i}h_{i}(t),
    \label{eq5:distanciaProm1}
\end{equation}
where $N$ is the number of cells that belongs to the precursor or supernatant front at a given time $t$, and $h_{i}(t)$ represents the Euclidean distance from each front cell to the center of the system, and the sum runs over all cells $i$ that belong to the corresponding front. However, the appropriate way to measure the front position in this configuration is by its distance to the fluid reservoir. This normalized average front position is given by \cite{Huergo2012}
\begin{equation}
    \overline{h_R(t)}=\frac{1}{N}\sum_{i}\left(h_{i}(t)-R_{R}\right)\,.
    \label{eq5:distanciaProm2}
\end{equation}
In Eq.~\eqref{eq5:distanciaProm2}, $\overline{h_R(t)}$ measures how much the front has grown from its starting position at $R_{R}$.

\begin{figure}[t!]
\centering
\includegraphics[width=0.7\textwidth]{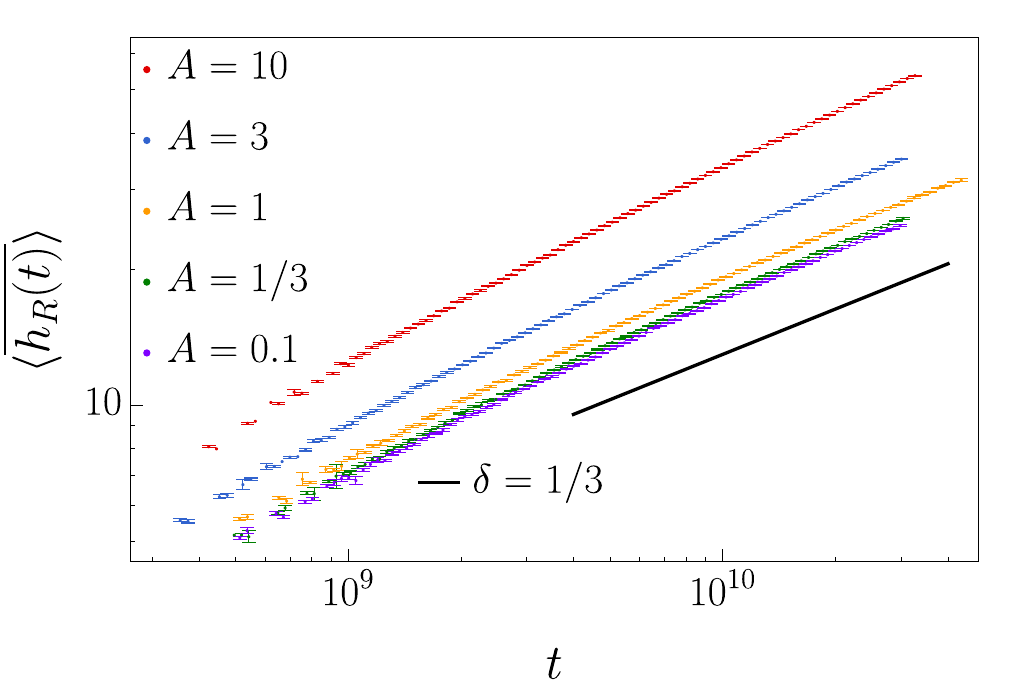}
\caption{Average front position $\langle \overline{h_R(t)}\rangle$ plotted as a function of time for $T=3$, $R_R=11$, $Z=1$, and several values of $A$. The solid black line corresponds to the reference scaling $\langle \overline{h_R(t)}\rangle \sim t^{1/3}$. All units are arbitrary in this and all figures below in this chapter.}
\label{fig5:ht}
\end{figure}

Figure \ref{fig5:ht} illustrates the evolution of $\langle \overline{h_R(t)}\rangle$ for five different values of the Hamaker constant. For nearly all combinations of $A$ and $T$, the mean front position exhibits a long-time power-law growth, $\langle \overline{h_R(t)} \rangle \sim t^\delta$, as expected. The exponent $\delta$ varies depending on the parameters, as detailed in Table \ref{tab5:delta_precursor} for the precursor layer and Table \ref{tab5:delta_supernatant} for the supernatant layer. We have also computed the average front distance as defined in Eq.~\eqref{eq5:distanciaProm1}. However, with this measure, the average front distance does not exhibit power-law behavior under any of the conditions studied.

\begin{table}[h!]
\centering
\small
\renewcommand{\arraystretch}{1.2}
\begin{tabular}{@{}lccccc@{}}
\toprule
\diagbox[width=4em,height=3em,dir=SE,trim=l]{$A$}{$T$} & 3 & 1 & 3/4 & 1/2 & 1/3 \\
\cmidrule(r){1-6}
10   & 0.394(2) & 0.426(1) & 0.4360(8) & 0.4672(8) & * \\
3    & 0.368(2) & 0.413(1) & 0.430(1)  & 0.446(1)  & * \\
1    & 0.336(3) & 0.374(2) & 0.367(4)  & 0.3488(9) & 0.3493(8) \\
1/3  & 0.345(2) & 0.369(2) & 0.378(3)  & 0.350(1)  & 0.3441(8) \\
0.1  & 0.341(3) & 0.372(2) & 0.382(2)  & 0.3534(8) & 0.3501(9) \\
\bottomrule
\end{tabular}
\caption{Values of the exponent $\delta$ for the precursor layer for all the conditions under study. The two conditions in which the average front position does not reach a regime governed by a power law are indicated with an asterisk.}
\label{tab5:delta_precursor}
\end{table}
\begin{table}[h!]
\centering
\small
\renewcommand{\arraystretch}{1.2}
\begin{tabular}{@{}lccccc@{}}
\toprule
\diagbox[width=4em,height=3em,dir=SE,trim=l]{$A$}{$T$} & 3 & 1 & 3/4 & 1/2 & 1/3 \\
\cmidrule(r){1-6}
10   & 0.285(4)  & 0.287(3)  & 0.278(3)  & 0.253(3)  & 0.17(3)  \\
3    & 0.315(3)  & 0.306(3)  & 0.291(3)  & 0.247(4)  & 0.168(5)  \\
1    & 0.312(4)  & 0.358(3)  & 0.382(4)  & 0.367(1)  & 0.3496(8) \\
1/3  & 0.340(3)  & 0.372(2)  & 0.387(2)  & 0.353(1)  & 0.3450(8) \\
0.1  & 0.341(3)  & 0.373(2)  & 0.385(2)  & 0.3542(8) & 0.350(1)  \\
\bottomrule
\end{tabular}
\caption{Values of the exponent $\delta$ for the supernatant layer for all the conditions studied.}
\label{tab5:delta_supernatant}
\end{table}

\begin{figure}[h!]
\centering
\includegraphics[width=0.7\textwidth]{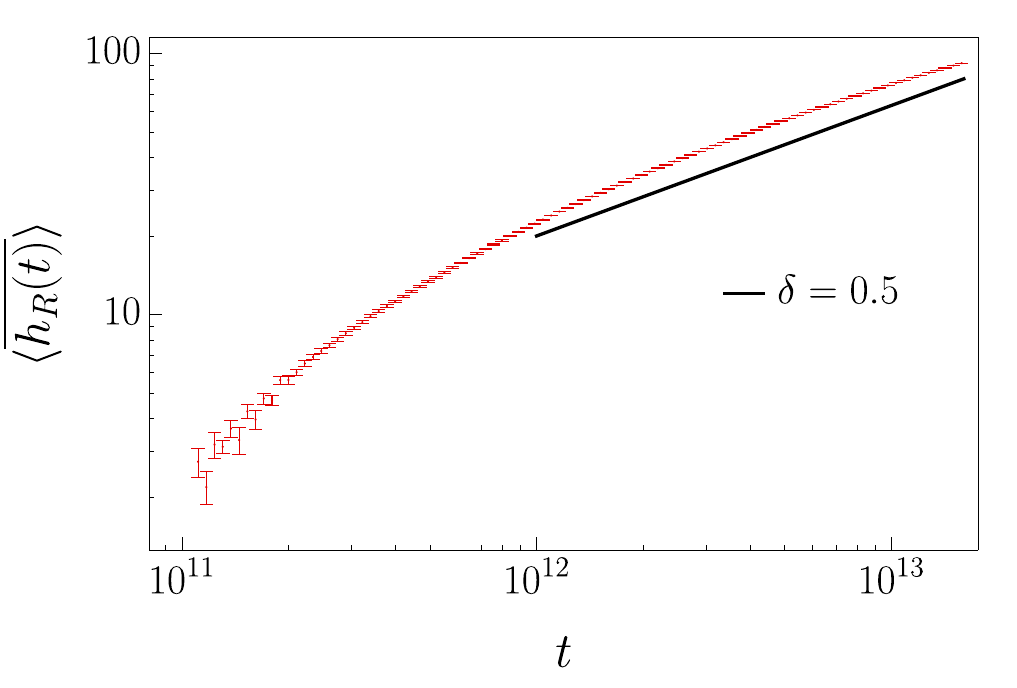}
\caption{Average front position $\langle \overline{h_R(t)}\rangle$ as a  function of time for $T=1/3$, $R_R=11$, $Z=1$ (precursor film), and $A=10$. The solid black line corresponds to the reference scaling $\langle \overline{h_R(t)} \rangle \sim t^{1/2}$.}
\label{fig5:ht_anomalo}
\end{figure}

In the previous chapter, where the same system was simulated using a band geometry, the scaling exponent $\delta$ was found to be approximately $1/2$ under all studied conditions and for both layers. In contrast, for the circular geometry considered in this chapter, $\delta$ takes values between $1/3$ and $1/2$ for most conditions. Subdiffusive, non-Tanner values of $\delta$ have also been observed in MD simulations of circular fluid droplets; see, for instance, Ref.\ \cite{Weng2017}. For the cases with $T = 1/3$ and $A = 10$ or $A = 3$, which correspond to relevant conditions for precursor spreading, as the ratios $J/k_{\mathrm{B}}T$ and $A/k_{\mathrm{B}}T$ reach their highest values, the average front position $\overline{h_R}(t)$ does not enter a clear power-law regime, as illustrated in Fig.~\ref{fig5:ht_anomalo} for $A = 10$. Consequently, no value for the $\delta$ exponent is reported for these two cases. Nonetheless, it is worth noting that the behavior of $\langle \overline{h_R}(t) \rangle$ does not deviate significantly from the expected $\delta \approx 1/2$ trend. When $A$ is small, both layers exhibit the same exponent, as expected, whereas for larger $A$, the precursor layer appears to grow with a higher exponent than the supernatant one.

\subsection{Roughness}

\begin{figure}[t]
\centering
\includegraphics[width=0.7\textwidth]{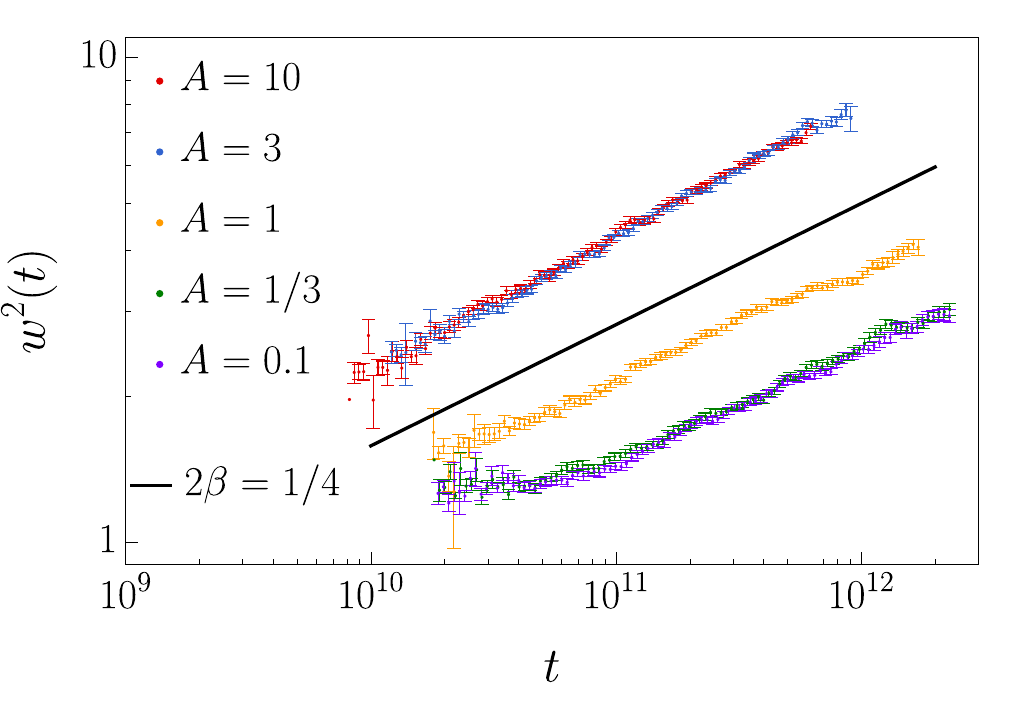}
\caption{Squared roughness $w^2(t)$ as a function of time for $T = 1/2$, $R_R = 11$, and several values of $A$. The solid black line corresponds to the reference scaling $w^2(t) \sim t^{1/4}$.}
\label{fig5:w2}
\end{figure}

With regard to the roughness, it scales as $w^2(t) \sim t^{2\beta}$, as expected. This behavior is shown in Fig.~\ref{fig5:w2} for a number of values of the Hamaker constant. No evidence of saturation to a steady-state value \cite{Barabasi1995,Krug1997} has been observed. In fact, steady-state saturation of the roughness is not expected in our system, as the length of the front increases more rapidly than the correlation length. As discussed in Sec.~\ref{sec3:observables}, roughness saturation occurs when the correlation length reaches the front size $L$. However, in this geometry, the front size is not fixed but grows with time, i.e., $L \equiv L_f(t)$. Consequently, since the correlation length never catches up to the expanding front, the system cannot reach saturation. As a result, the precursor film continues to grow indefinitely, and no finite equilibrium state is achieved \cite{Hocking1992}.

The value of $\beta$ depends on the physical parameters $A$ and $T$, as shown in Tables \ref{tab5:beta_precursor} and \ref{tab5:beta_supernatant} for the precursor and supernatant layers, respectively. At high temperatures (approximately $T > 1$), $\beta$ falls within the range of $1/4$ to $1/5$ and shows little sensitivity to the Hamaker constant $A$. At lower temperatures ($T < 1$), the growth exponent decreases, reaching values around $\beta \approx 1/10$. As before, both layers display similar exponents for small $A$, whereas the precursor layer shows a higher exponent as $A$ increases.

As explained above, for very low temperatures and low Hamaker constants the shape of the front is no longer circular. In these cases it is more appropriate to study its fluctuations locally, following the procedure described in Sec.~\ref{sec3:limit_shape}. Table \ref{tab5:beta_nueva_forma} shows the values of $\beta_\Omega$, i.e. the exponent computed from the scaling of the local roughness, $w^2_{\Omega}(L_f,t,Z)\sim t^{\beta_\Omega}$, where the local roughness $w^2_{\Omega}(L_f,t,Z)$ is computed according to Eq.~\eqref{eq3:width_shape}. As expected, both $\beta$ and $\beta_\Omega$ take similar values at high temperatures, where local and global fluctuations coincide. However, at low temperatures ($T = 1/3$), the exponent obtained through this method is significantly larger.

Overall, as in the scenario examined in the preceding chapter, the exponent values reveal a clear non-trivial dependence on temperature and a much weaker sensitivity to the Hamaker constant. Furthermore, two distinct scaling regimes emerge at low and high temperatures, with $T$-dependent exponents in the intermediate range. As will be discussed below, additional estimates of the exponents further support this interpretation.

\begin{table}[h!]
\centering
\small
\renewcommand{\arraystretch}{1.2}
\begin{tabular}{@{}lccccc@{}}
\toprule
\diagbox[width=4em,height=3em,dir=SE,trim=l]{$A$}{$T$} & 3 & 1 & 3/4 & 1/2 & 1/3 \\
\cmidrule(r){1-6}
10   & 0.463(9) & 0.512(8) & 0.50(1)  & 0.25(2)  & 0.20(6) \\
3    & 0.420(6) & 0.50(1)  & 0.51(1)  & 0.24(4)  & 0.17(6) \\
1    & 0.41(1)  & 0.48(1)  & 0.3(1)   & 0.24(3)  & 0.14(2) \\
1/3  & 0.400(7) & 0.44(1)  & 0.23(4)  & 0.20(3)  & 0.17(3) \\
0.1  & 0.398(6) & 0.44(1)  & 0.27(2)  & 0.24(4)  & 0.14(2) \\
\bottomrule
\end{tabular}
\caption{Values of the exponent $2\beta$ for the precursor layer for all the conditions under study.}
\label{tab5:beta_precursor}
\end{table}
\begin{table}[h!]
\centering
\small
\renewcommand{\arraystretch}{1.2}
\begin{tabular}{@{}lccccc@{}}
\toprule
\diagbox[width=4em,height=3em,dir=SE,trim=l]{$A$}{$T$} & 3 & 1 & 3/4 & 1/2 & 1/3 \\
\cmidrule(r){1-6}
10   & 0.335(6) & 0.315(6) & 0.273(7) & 0.142(9) & 0.13(2) \\
3    & 0.358(7) & 0.360(7) & 0.295(8) & 0.16(2)  & 0.14(2) \\
1    & 0.38(1)  & 0.459(9) & 0.41(9)  & 0.26(3)  & 0.14(2) \\
1/3  & 0.392(7) & 0.44(1)  & 0.24(4)  & 0.20(3)  & 0.17(3) \\
0.1  & 0.399(7) & 0.44(1)  & 0.27(2)  & 0.24(4)  & 0.14(2) \\
\bottomrule
\end{tabular}
\caption{Values of the exponent $2\beta$ for the supernatant layer for all the conditions studied.}
\label{tab5:beta_supernatant}
\end{table}
\begin{table}[t!]
\centering
\small
\renewcommand{\arraystretch}{1.2}
\begin{tabular}{@{}lccccc@{}}
\toprule
\diagbox[width=4em,height=3em,dir=SE,trim=l]{$A$}{$T$} & 3 & 1 & 3/4 & 1/2 & 1/3 \\
\cmidrule(r){1-6}
10   & 0.48(1)  & 0.512(8) & 0.50(1)  & 0.25(2)  & 0.20(3) \\
3    & 0.436(6) & 0.50(1)  & 0.50(1)  & 0.22(3)  & 0.21(3) \\
1    & 0.42(2)  & 0.49(1)  & 0.43(2)  & 0.22(3)  & 0.17(2) \\
1/3  & 0.415(8) & 0.44(1)  & 0.25(2)  & 0.21(3)  & 0.22(3) \\
0.1  & 0.416(7) & 0.44(1)  & 0.27(2)  & 0.23(3)  & 0.24(3) \\
\bottomrule
\end{tabular}
\caption{Values of the exponent $2\beta_{\Omega}$ for the precursor layer, computed using Eq.~\eqref{eq3:width_shape}, for all the conditions studied.}
\label{tab5:beta_nueva_forma}
\end{table}

\newpage
\subsection{Height-difference correlation function: computation of $\alpha$ and $z$ exponents}

\begin{figure}[h]
\centering
\includegraphics[width=0.7\textwidth]{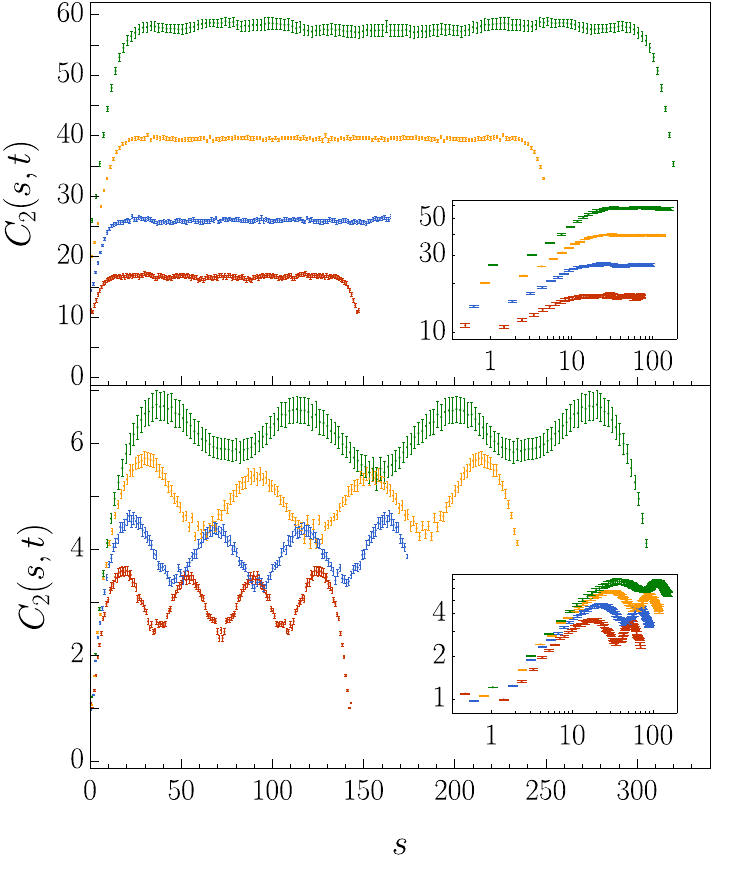}
\caption{Height-difference correlation function $C_2(s,t)$ as a function of $s$ for times increasing from 40 to 100 at regular intervals. Conditions are: $T=3$, $A=1$ (top), and $T=1/2$, $A=0.1$ (bottom). In both panels $Z=1$. Inset: Log-log plot of the same function for $s$ between $0$ and $L_f(t)/2$.}
\label{fig5:CorrelacionTotal}
\end{figure}

Figure \ref{fig5:CorrelacionTotal} shows the height-difference correlation function at various times for two representative sets of parameters, corresponding to high and low temperature conditions. As explained in Sec.~\ref{sec3:observables} and applied in the previous chapter, the correlation length $\xi(t)$ can be estimated from the plateau of the $C_2(s,t)$ curves at sufficiently large $s$, for a fixed value of $a$, when the height-difference correlation function reaches a plateau. Since the exponent values do not depend on the specific choice of the parameter $a$, as shown in the previous chapter and in Refs.\cite{Barreales2020,Marcos2022}, we only use $a = 0.9$ throughout this chapter.

\begin{figure}[t!]
\centering
\includegraphics[width=0.7\textwidth]{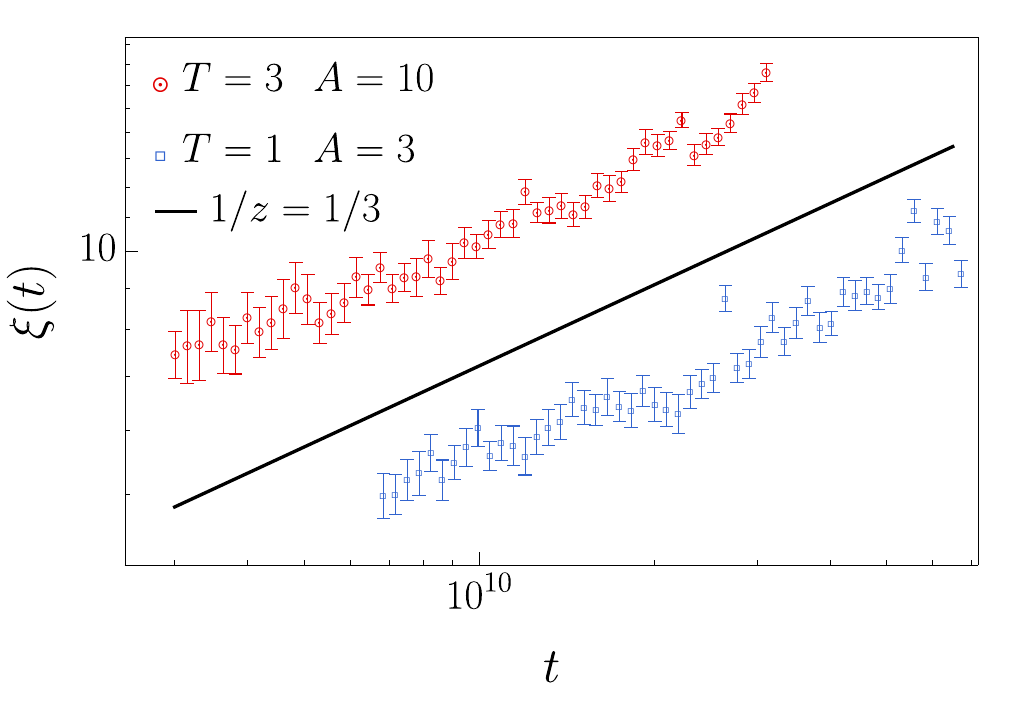}
\caption{Correlation length $\xi(t)$ for $T=3$, $A=10$ (red circles) and $T=1$, $A=3$ (blue squares) as functions of time. In both cases $Z=1$. As a visual reference, the solid black line corresponds to the reference scaling $\xi(t)\sim t^{1/z}$, with $1/z=1/3$.}
\label{fig5:xi}
\centering
\includegraphics[width=0.7\textwidth]{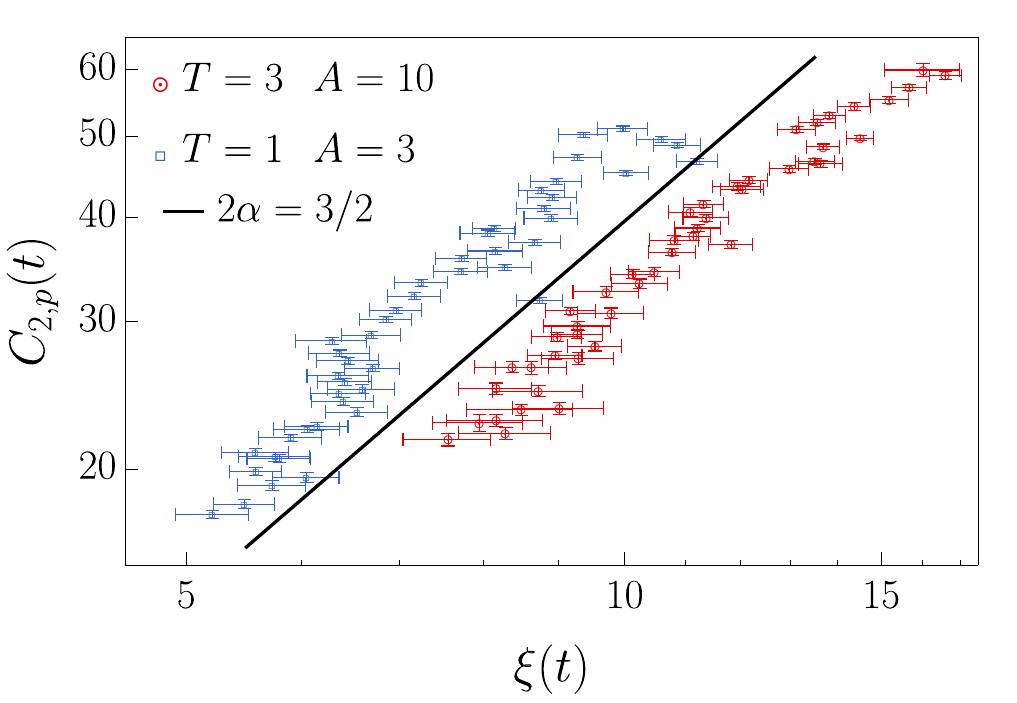}
\caption{Plateau of the height-difference correlation function $C_{2,p}(t)$ versus $\xi(t)$  for \mbox{$T=3$}, $A=10$ (red circles) and $T=1$, $A=3$ (blue squares) at different times. In both cases $Z=1$. As a visual reference, the solid black line corresponds to the reference scaling $C_{2,p}(t)\sim \xi(t)^{2\alpha}$, with $2\alpha=3/2$.}
\label{fig5:plateau}
\end{figure}

Using Eq.~\eqref{eq3:correlation_length} in the same way as in the previous chapter, the double logarithmic plots of these correlation lengths as functions of time should yield straight lines, whose slopes correspond to the exponent $1/z$. Figure \ref{fig5:xi} shows log-log plots of $\xi(t)$ vs. $t$ for two conditions of the precursor layer with $1/z\sim 1/3$. Furthermore, Eq.~\eqref{eq3:c2p} implies that $C_{2,p}(t) \sim \xi^{2\alpha}(t)$ for $s \gg \xi(t)$. Therefore, the exponent $\alpha$ can be determined from the slope of the best-fit lines in a log-log plot of $C_{2,p}(t)$ versus $\xi(t)$. In cases where the correlation function clearly reaches a plateau, like in the top panel of Fig.~\ref{fig5:CorrelacionTotal}, this value was evaluated at the center of the height-difference correlation function, following the approach used in the previous chapter. Figure~\ref{fig5:plateau} displays $C_{2,p}(t)$ as a function of $\xi(t)$ for the precursor layer under two different conditions; a reference line with slope $2\alpha = 3/2$ is included for comparison.

%

However, when the height-difference correlation function does not reach a plateau, like in the bottom panel of Fig.~\ref{fig5:CorrelacionTotal}, this approach cannot be used. Instead, the way the correlation length and plateau are calculated must be redefined. Since in all these cases the peaks of the height-difference correlation function exhibit an approximately parabolic shape, we proceed as follows: first, we fit the first peak of the correlation function $C_2(s,t)$ to a parabola of the form $f(s) = a + bs + cs^2$. The maximum value of this parabola is then taken as an estimate of the plateau value, i.e., we define $C_{2,p}(t) \equiv f_{\rm max} = a - \frac{b^2}{4c}$. The correlation length is subsequently calculated as the point to the left of the peak where the parabola reaches 90\% of its maximum value, that is, $\xi(t) \equiv x_{0.9}=[-b+\sqrt{0.1(b^2-4ac)}]/(2c)$.

\begin{figure}[t!]
\centering
\includegraphics[width=0.7\textwidth]{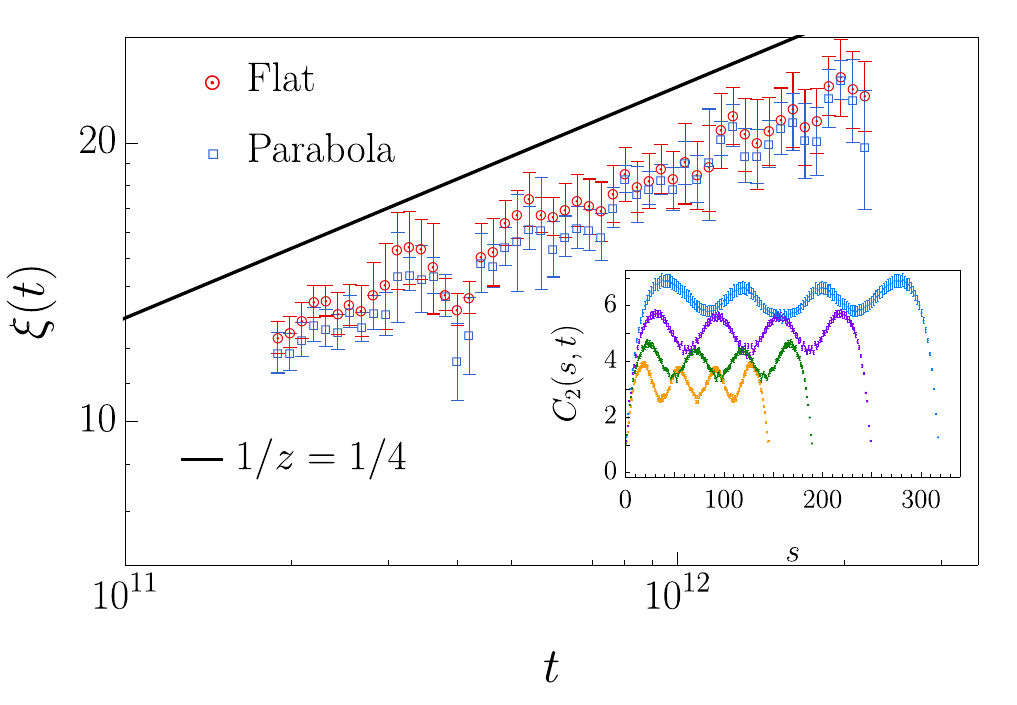}
\caption{Correlation length $\xi(t)$ as a function of time, calculated using the flat method (red circles) and the parabola method (blue squares) for \mbox{$T=1/2$} and $A=1/3$. In both cases $Z=1$. As a visual reference, the solid black line corresponds to $\xi(t)\sim t^{1/z}$, with $1/z=1/4$. The quoted values of the exponents $1/z$ are $1/z=0.25(3)$ (flat) and $1/z=0.26(2)$ (parabola). Inset: height-difference correlation function as a function of $s$ for times increasing from 60 to 90, bottom to top, at regular intervals.}
\label{fig5:equiv}
\end{figure}

We have verified that, in the parameter regimes where both methods are applicable (specifically, for $T = 1/2$ with low $A$, and $T = 1/3$), the resulting values for the correlation length and plateau coincide. Figure~\ref{fig5:equiv} shows the correlation length $\xi(t)$ computed using both approaches for a representative case ($T = 1/2$, $A = 1/3$). When both methods are valid, we have chosen the values corresponding to the smallest associated errors. 

Tables~\ref{tab5:1overz_precursor} and \ref{tab5:1overz_supernatant} list the complete set of $1/z$ exponents, while Tables~\ref{tab5:2alpha_precursor} and \ref{tab5:2alpha_supernatant} contain the corresponding $2\alpha$ values for the precursor and supernatant layers. Conditions where the parabolic approximation was used are highlighted in bold.

The data presented in these tables confirm that the expected scaling relation $\alpha = \beta z$ holds well across most of the conditions analyzed. As previously noted regarding the temperature dependence of $\beta$, the exponents $\alpha$ and $z$ also exhibit a clear dependence on temperature. This behavior further supports the existence of a transition from a low-temperature to a high-temperature regime, with temperature-dependent exponents for \mbox{$T < 1$}. Moreover, the influence of the Hamaker constant on $\alpha$ and $z$ appears to be more pronounced than in the band geometry, particularly at low temperatures.
\begin{table}[h!]
\centering
\small
\renewcommand{\arraystretch}{1.2}
\begin{tabular}{@{}lccccc@{}}
\toprule
\diagbox[width=4em,height=3em,dir=SE,trim=l]{$A$}{$T$} & 3 & 1 & 3/4 & 1/2 & 1/3 \\
\cmidrule(r){1-6}
10   & 0.33(4)  & 0.31(1)  & 0.21(2)  & 0.18(2)  & 0.26(5) \\
3    & 0.38(1)  & 0.32(1)  & 0.21(2)  & 0.24(2)  & 0.29(6) \\
1    & 0.37(1)  & 0.35(2)  & 0.25(2)  & 0.20(3)  & \textbf{0.28(1)} \\
1/3  & 0.38(1)  & 0.28(1)  & 0.18(2)  & \textbf{0.26(2)} & \textbf{0.28(1)} \\
0.1  & 0.40(1)  & 0.31(2)  & 0.19(2)  & \textbf{0.25(2)} & \textbf{0.28(1)} \\
\bottomrule
\end{tabular}
\caption{Values of the exponent $1/z$ for the precursor layer for all the conditions under study. In this and the next three tables, the values calculated approximating the peak as a parabola appear in bold.}
\label{tab5:1overz_precursor}
\end{table}
\begin{table}[h!]
\centering
\small
\renewcommand{\arraystretch}{1.2}
\begin{tabular}{@{}lccccc@{}}
\toprule
\diagbox[width=4em,height=3em,dir=SE,trim=l]{$A$}{$T$} & 3 & 1 & 3/4 & 1/2 & 1/3 \\
\cmidrule(r){1-6}
10   & 0.37(2)  & 0.29(2)  & 0.28(2)  & 0.17(5)  & 0.37(3) \\
3    & 0.47(2)  & 0.21(2)  & 0.13(2)  & 0.38(1)  & 0.38(2) \\
1    & 0.39(1)  & 0.35(2)  & 0.24(2)  & 0.20(2)  & \textbf{0.28(1)} \\
1/3  & 0.39(1)  & 0.28(1)  & 0.18(2)  & \textbf{0.26(2)} & \textbf{0.28(1)} \\
0.1  & 0.39(1)  & 0.31(2)  & 0.20(2)  & \textbf{0.25(2)} & \textbf{0.28(1)} \\
\bottomrule
\end{tabular}
\caption{Values of the exponent $1/z$ for the supernatant layer for all the conditions studied.}
\label{tab5:1overz_supernatant}
\end{table}
\clearpage
\begin{table}[h!]
\centering
\small
\renewcommand{\arraystretch}{1.2}
\begin{tabular}{@{}lccccc@{}}
\toprule
\diagbox[width=4em,height=3em,dir=SE,trim=l]{$A$}{$T$} & 3 & 1 & 3/4 & 1/2 & 1/3 \\
\cmidrule(r){1-6}
10   & 1.42(6) & 1.56(6) & 2.0(2)  & 1.4(2)   & 0.7(2) \\
3    & 1.13(3) & 1.48(7) & 2.0(2)  & 1.0(1)   & 0.7(2) \\
1    & 1.07(3) & 1.27(7) & 1.5(1)  & 0.8(1)   & \textbf{0.48(6)} \\
1/3  & 1.02(3) & 1.43(7) & 1.3(2)  & \textbf{0.80(6)} & \textbf{0.45(6)} \\
0.1  & 0.98(3) & 1.22(6) & 1.3(2)  & \textbf{0.83(5)} & \textbf{0.42(6)} \\
\bottomrule
\end{tabular}
\caption{Values of the exponent $2\alpha$ for the precursor layer for all the conditions under study.}
\label{tab5:2alpha_precursor}
\end{table}
\begin{table}[h!]
\centering
\small
\renewcommand{\arraystretch}{1.2}
\begin{tabular}{@{}lccccc@{}}
\toprule
\diagbox[width=4em,height=3em,dir=SE,trim=l]{$A$}{$T$} & 3 & 1 & 3/4 & 1/2 & 1/3 \\
\cmidrule(r){1-6}
10   & 0.80(7) & 0.77(7) & 0.56(6) & 0.4(1)   & 0.20(5) \\
3    & 0.76(3) & 1.0(1)  & 0.94(2) & 0.43(2)  & 0.21(4) \\
1    & 0.91(3) & 1.22(6) & 1.5(1)  & 1.0(2)   & \textbf{0.48(6)} \\
1/3  & 0.89(4) & 1.47(7) & 1.5(2)  & \textbf{0.81(6)} & \textbf{0.52(6)} \\
0.1  & 0.95(3) & 1.24(6) & 1.3(2)  & \textbf{0.83(6)} & \textbf{0.42(7)} \\
\bottomrule
\end{tabular}
\caption{Values of the exponents $2\alpha$ for the supernatant layer for all the conditions studied.}
\label{tab5:2alpha_supernatant}
\end{table}



\subsection{Shape of the height-difference correlation function at low temperature}

The bottom panel of Fig.~\ref{fig5:CorrelacionTotal} shows the height-difference correlation function at low temperature for several times; the presence of multiple maxima and minima is evident. Moreover, as illustrated in Fig.~\ref{fig5:morfologia}, the films, and in particular the precursor film, exhibit a square-like shape for low-temperature and low Hamaker constant conditions. These two observations are closely related. Unlike the expected plateau, the height-difference correlation function at low temperature displays four peaks and three local minima, indicating that points separated by arc-lengths of $\pi \bar{h}/2$, $\pi \bar{h}$, and $3\pi \bar{h}/2$ are less correlated than those at both shorter and longer distances. Notably, the correlation function of a perfect square (not shown here) features four perfectly symmetric peaks and three local minima that drop to zero at the same positions as the local minima observed in our case.

\begin{figure}[t]
\centering
\includegraphics[width=0.7\textwidth]{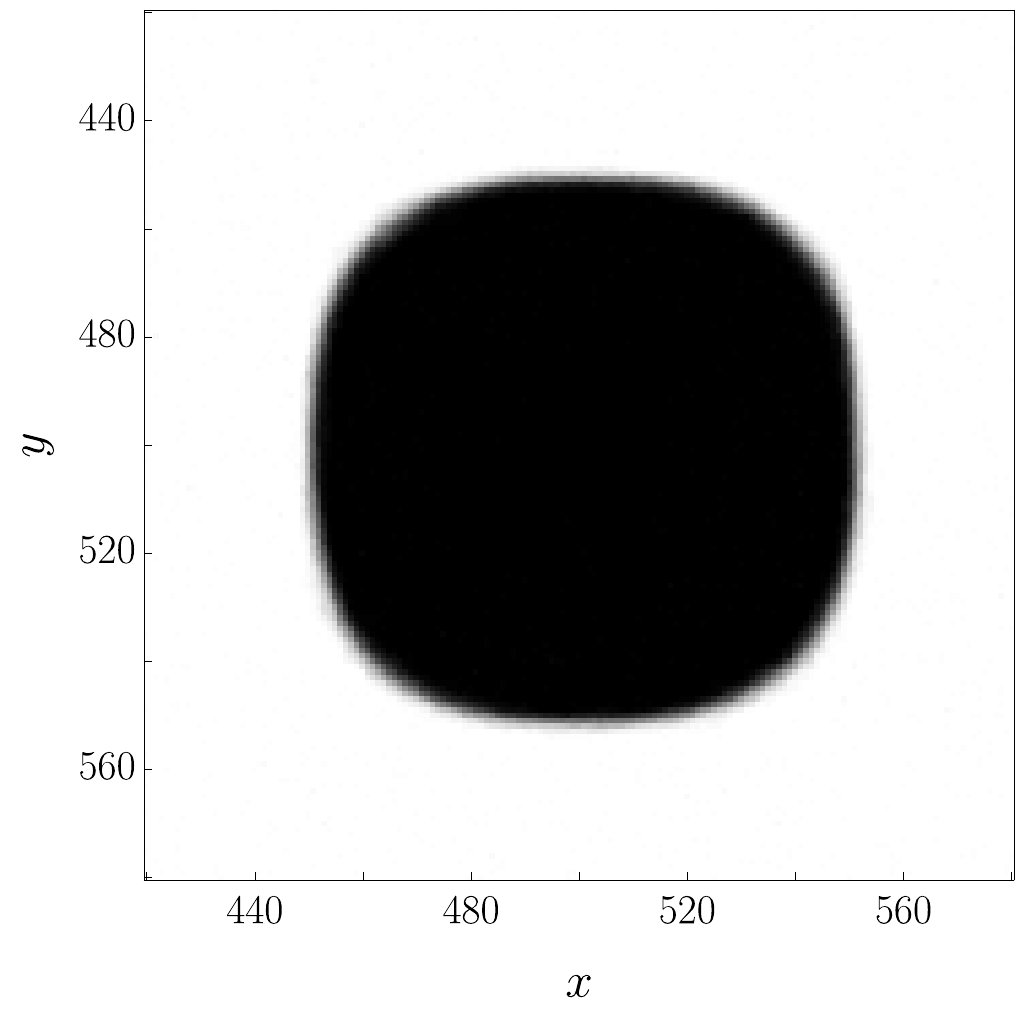}
\caption{Average of the last configurations measured for $T=1/3$, $A=1$, and $Z=1$ (i.e., the precursor film). The figure plots the gray level of the point density. In other words, a solid black cell (such as those belonging to the droplet reservoir at the center of the figure) indicates that the cell was occupied in all the runs. Conversely, a solid white cell is indicative that all the runs have this cell empty. Intermediate gray-level values represent varying degrees of density.}
\label{fig5:cuadrado}
\end{figure}

These results suggest that, at low temperatures, the film adopts a shape intermediate between a square and a circle. To test this hypothesis, we computed the average film shape under a single low-temperature condition across multiple simulations. The outcome, shown in Fig.~\ref{fig5:cuadrado}, corresponds to an average over runs with specific parameters ($T = 1/3$ and $A = 1$), which produce oscillations in the height-difference correlation function. The figure reveals that the film assumes a square-like shape with rounded corners. Notably, the transition between always-occupied and always-empty cells is sharp.

The emergence of this shape can be understood by examining the evolution in time of the system. In fact, our model employs the same Kawasaki dynamics as the COP Ising model, and thus exhibits similar behavior \cite{Newman1999}. In the COP Ising model, the domain shape at very low temperatures (around $T/T_c = 0.25$, with $T_c \approx 2.27$) resembles a square, while at higher temperatures it becomes more rounded (see, for example, Fig.~5.4 of Ref.\ \cite{Newman1999}). This behavior arises because the system minimizes its energy by reducing the perimeter of the domain. The dynamics of our model resemble those of the COP Ising model, as both are defined on a regular lattice and follow Kawasaki exchange rules. However, there are two important differences. The first is the second term in our Hamiltonian [Eq.~\eqref{eq5:energy}], which accounts for the interaction with the substrate. The second is the presence of a reservoir that continuously supplies particles to the system. Nevertheless, these differences are not significant in the context of the film morphology analysis. In particular, the primary effect of the substrate interaction is to promote the growth of the precursor film. However, it does not alter the energy of particles within the same layer and is therefore irrelevant when analyzing the shape adopted by the films. On the other hand, the continuous addition of particles from the reservoir becomes irrelevant at long times, as the rate at which particles reach the front decreases over time. During these late stages, the algorithm performs numerous steps without any change in the total number of particles in the system. Therefore, as in the COP Ising model, the system minimizes its energy by reducing the perimeter, leading to a square shape with rounded corners. This characteristic shape emerges from two main factors: the simplicity of the model and the choice of lattice. Other authors studying fluid droplets using kMC simulations of discrete models based on the Ising model have similarly observed that the rectangular shape disappears when interactions beyond nearest neighbors are included (see, e.g., Refs.~\cite{Areshi2019,Chalmers2017}). However, incorporating such interactions in our model would significantly hinder our computational ability to analyze scaling behavior, which requires sufficiently long simulation times and large system sizes. Moreover, it would prevent a meaningful comparison with the band geometry discussed in the previous chapter. In addition, other authors studying the Ising model with Kawasaki dynamics on a hexagonal lattice in the zero-temperature limit have found that the system's equilibrium state adopts a hexagonal shape \cite{Baldassarri2023}. This characteristic geometry should not be interpreted as a physically meaningful result, but rather as an inherent feature of the model.

\subsection{Anomalous scaling of the height-correlation function}

\begin{figure}[b!]
\centering
\includegraphics[width=0.7\textwidth]{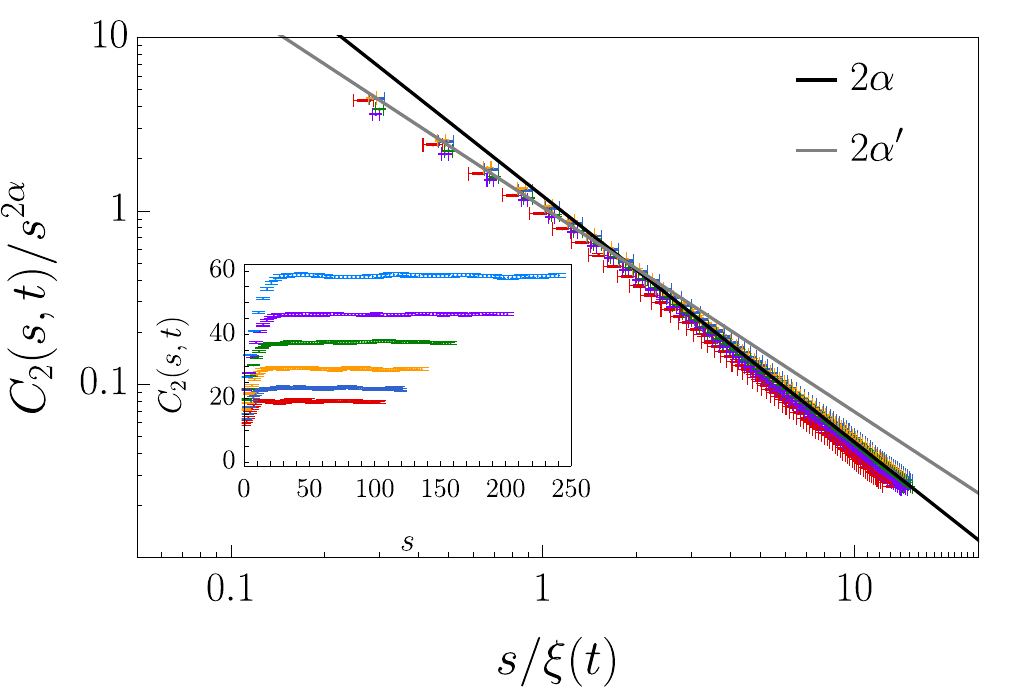}
\caption{Data collapse of the height-difference correlation function obtained for different values of time, $T = 3$, $A = 10$, and $Z=1$. The curve onto which collapse occurs is the function $g(s/\xi(t))$ of Eq.~\eqref{eq3:c2an}, with the solid black line representing the theoretical behavior for large argument, $g(u)\sim u^{-2\alpha}$ with $2\alpha=1.42$, and the solid gray line representing the behavior for small argument, $g(u)\sim u^{-2\alpha'}$  with $2\alpha'=1.18$. Inset: height-difference correlation function as a function of $s$ for times increasing from 50 to 100 bottom to top at regular intervals.}
\label{fig5:reescalada}
\end{figure}

As in the previous chapter, the systematic time-dependent shift of the $C_2(s,t)$ curves, without overlap for $s < \xi(t)$, as shown in Fig.~\ref{fig5:CorrelacionTotal}, is a clear indication of anomalous scaling behavior \cite{Lopez1997}. The presence of intrinsic anomalous scaling arises from the inequality $\alpha_{\rm loc} < \alpha$, indicating the existence of two independent roughness exponents. This temporal shift is clearly evidenced in the main panel of Fig.~\ref{fig5:reescalada}, which also demonstrates a consistent data collapse of the height-difference correlation function in accordance with Eq.~\eqref{eq3:c2an} for a representative set of parameters.

In particular, the fact that $g(u) \sim u^{-2\alpha'}$ for $u \ll 1$, rather than remaining constant at small arguments, is a clear indication of intrinsic anomalous scaling. We have computed the value of $2\alpha'$ by fitting the re-scaled height-difference correlation function $C_2(s,t)/s^{2\alpha}$ vs $s/\xi(t)$ for $s/\xi(t)<1$ and several times. Tables \ref{tab5:dap_precursor} and \ref{tab5:dap_supernatant} list the resulting $2\alpha'$ values for the precursor and supernatant layers, respectively, for all the conditions studied. According to these tables, the exponents depend heavily on the parameter conditions.

\begin{table}[h!]
\centering
\small
\renewcommand{\arraystretch}{1.2}
\begin{tabular}{@{}lccccc@{}}
\toprule
\diagbox[width=4em,height=3em,dir=SE,trim=l]{$A$}{$T$} & 3 & 1 & 3/4 & 1/2 & 1/3 \\
\cmidrule(r){1-6}
10   & 1.18(6) & 1.44(6) & 1.9(2)  & 0.9(2)   & 0.0(2)     \\
3    & 0.90(3) & 1.35(7) & 1.9(2)  & 0.7(1)   & 0.0(2)     \\
1    & 0.84(3) & 1.15(6) & 1.4(1)  & 0.4(1)   & $-0.66(6)$ \\
1/3  & 0.81(3) & 1.33(7) & 1.1(2)  & 0.30(6)  & $-0.62(6)$ \\
0.1  & 0.78(3) & 1.11(6) & 1.1(2)  & 0.31(5)  & $-0.73(6)$ \\
\bottomrule
\end{tabular}
\caption{Value of the exponent $2\alpha'$, for the precursor layer, for all the conditions studied.}
\label{tab5:dap_precursor}
\end{table}
\begin{table}[h!]
\centering
\small
\renewcommand{\arraystretch}{1.2}
\begin{tabular}{@{}lccccc@{}}
\toprule
\diagbox[width=4em,height=3em,dir=SE,trim=l]{$A$}{$T$} & 3 & 1 & 3/4 & 1/2 & 1/3 \\
\cmidrule(r){1-6}
10   & 0.70(7) & 0.73(7) & 0.52(6)  & *        & $-0.41(6)$ \\
3    & 0.58(3) & 1.0(1)  & 0.9(2)   & *        & $-0.41(4)$ \\
1    & 0.70(3) & 1.12(6) & 1.3(1)   & 0.6(2)   & $-0.66(6)$ \\
1/3  & 0.68(4) & 1.37(7) & 1.3(2)   & 0.30(7)  & $-0.60(6)$ \\
0.1  & 0.74(3) & 1.13(6) & 1.1(2)   & 0.31(6)  & $-0.72(6)$ \\
\bottomrule
\end{tabular}
\caption{Value of the exponent $2\alpha'$ for the supernatant layer, for all the conditions studied. We denote with an asterisk two conditions in which the collapse of the height-difference correlation function was so noisy that it was impossible to compute an exponent.}
\label{tab5:dap_supernatant}
\end{table}

\begin{figure}[h]
\centering
\includegraphics[width=0.7\textwidth]{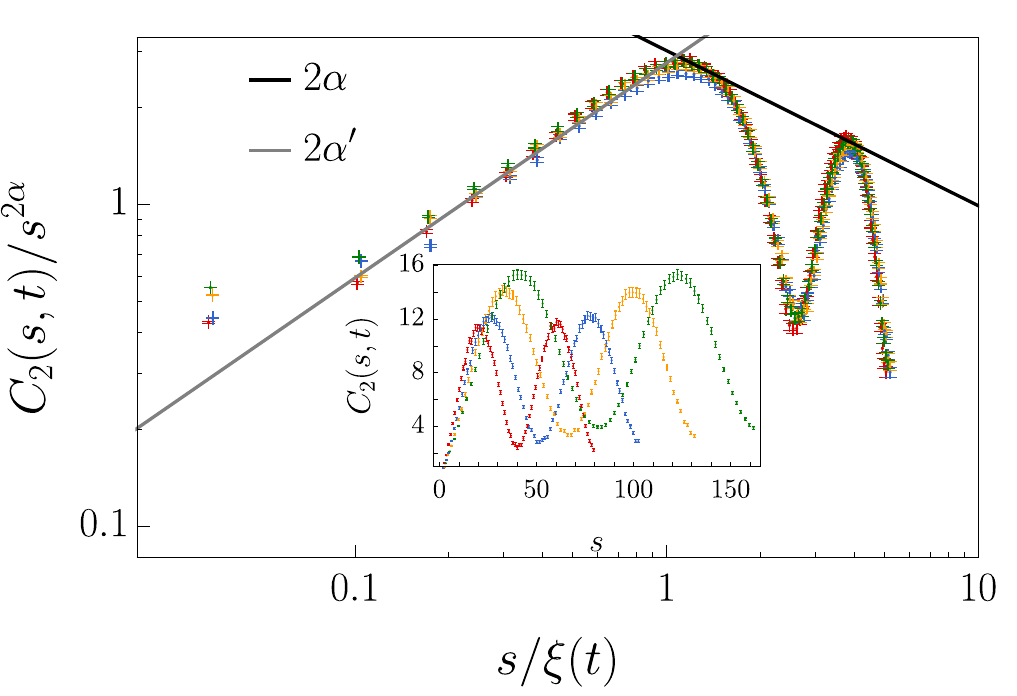}
\caption{Data collapse according to Eq.~\eqref{eq5:h} for the height-difference correlation function obtained for different values of time, for $T = 1/3$, $A = 1$, and $Z=1$. The solid black line corresponds to $g(u)\sim u^{-2\alpha}$ with $2\alpha=0.48$, and the solid gray line corresponds to $g(u)\sim u^{-2\alpha'}$  with $2\alpha'=-0.66$. Inset: height-difference correlation function as a function of $s$ for times increasing from 55 to 100 bottom to top at regular intervals.}
\label{fig5:reescaladaTbaja}
\end{figure}


As in the case of a band geometry, although the anomalous shift of the height-difference correlation function curves with increasing time, shown in the inset of Fig.~\ref{fig5:reescalada}, could be attributed to a large roughness exponent, the data collapse in Fig.~\ref{fig5:reescalada} with $\alpha' \neq 0$ clearly indicates that the origin of this behavior is intrinsic anomalous scaling. It is worth noting that there are a few cases ($T = 1/3$ with $A = 10$ and $A = 3$) where the condition $\alpha \neq \alpha'$ does not hold.

For those conditions in which the height-difference correlation function exhibits oscillations, it is still possible to achieve a data collapse analogous to that of Eq.~\eqref{eq3:c2an}. An illustrative example is shown in the main panel of Fig.~\ref{fig5:reescaladaTbaja}. In these low-temperature cases, however, the specific scaling function governing the collapse differs from that in Eq.~\eqref{eq3:c2an} and Fig.~\ref{fig5:reescalada}. Specifically, Eq.~\eqref{eq3:c2an} is modified to
\begin{equation} 
C_2(s,t) = s^{2\alpha} h(s/\xi(t)), 
\label{eq5:h} 
\end{equation}
where $h(u) \sim u^{-2\alpha'}$ for $u \ll 1$. For $u \gg 1$, the function $h(u)$ oscillates with an amplitude that decays as $1/u^{2\alpha}$ (see Fig.~\ref{fig5:reescaladaTbaja}). Note also that, in the figure, $2\alpha'$ takes a negative value.

\newpage
\subsection{Front fluctuations}

\begin{figure}[t]
\centering
\includegraphics[width=0.7\textwidth]{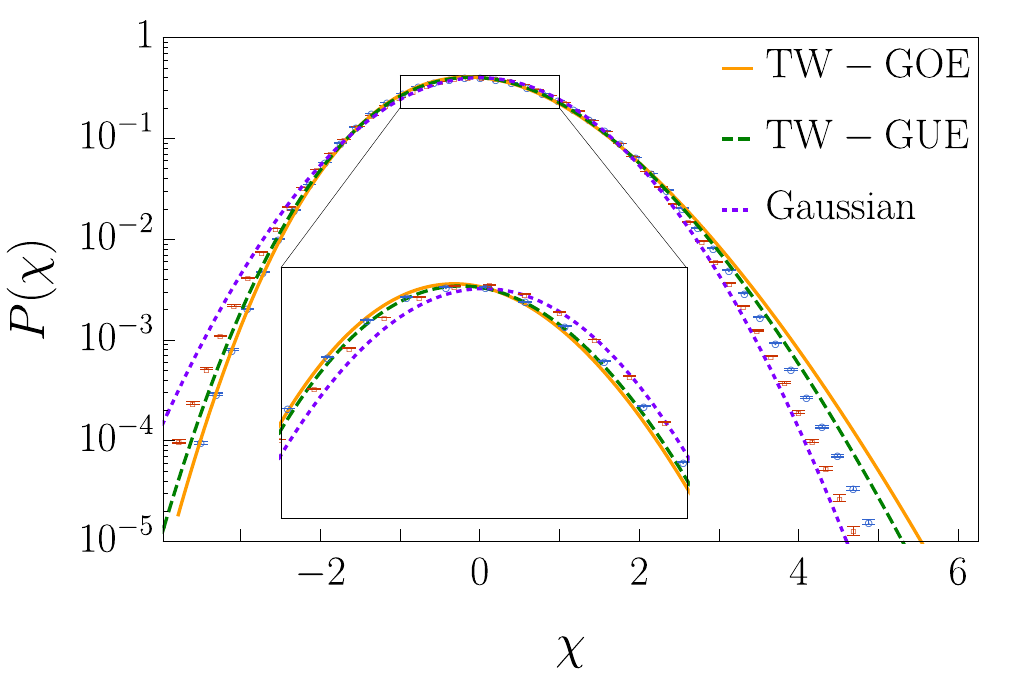}
\caption{Fluctuation histograms of the variable $\chi$ calculated according to Eq.~\eqref{eq3:flu} for $T = 1$, $A=1$ (blue circles) and $T = 3$, $A=10$ (red squares). In both cases $Z=1$. The solid orange line and the green dashed line correspond to the TW-GOE and TW-GUE distributions, respectively. The dotted purple line correspond to a Gaussian distribution. Inset: zoom for small $\chi$.}
\label{fig5:histograma}
\end{figure}

Figure \ref{fig5:histograma} displays the TW-GOE distribution, associated with the KPZ universality class in band geometry, alongside the TW-GUE distribution, which corresponds to circular geometry, as well as the Gaussian distribution and data from our numerical simulations under two relevant conditions.

The agreement with the TW-GUE distribution is remarkable, especially considering that the exponents of the system do not match those of the KPZ universality class. In the previous chapter, we showed that the correspondence between numerical data and the theoretical distribution improved with increasing system size, suggesting that the observed discrepancies were due to finite-size effects. In the present case, however, the film length $L_f(t)$ is not a parameter that can be controlled but a time-dependent quantity that grows as the system evolves.

We found the best agreement at high temperatures ($T \ge 3/4$), while noticeable discrepancies arise at lower temperatures. We also observe reduced agreement for smaller Hamaker constants, although this parameter appears to be less influential than temperature. At low temperatures, the tail of the distribution tends to approach the Gaussian more closely than the TW-GOE or TW-GUE distributions.

\begin{figure}[t]
\centering
\includegraphics[width=0.7\textwidth]{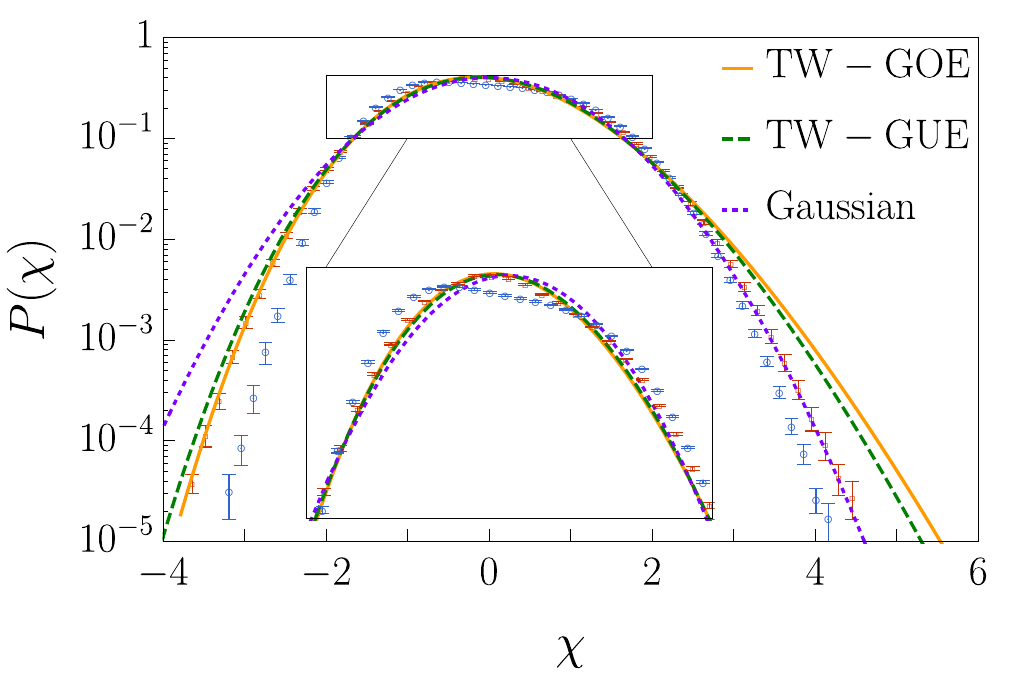}
\caption{Fluctuation histograms of the variable $\chi$ calculated for $T = 1/3$, $A=1$, and $Z=1$ according to Eq.~\eqref{eq3:flu}, i.e. $\chi$, (shown in blue circles) and according to Eq.~\eqref{eq3:Chi_local}, i.e.  $\chi_\Omega$, (shown in red squares). The solid orange and the green dashed lines correspond to the TW-GOE and TW-GUE distributions, respectively. The dotted purple line correspond to a Gaussian distribution. In each case, the growth exponent used was the one calculated with each method, i.e. the one appearing in Table \ref{tab5:beta_precursor}, $\beta$, for the first case and the one appearing in Table \ref{tab5:beta_nueva_forma}, $\beta_\Omega$, for the second case. Inset: zoom for small $\chi$ and $\chi_\Omega$.}
\label{fig5:histograma_Tbaja}
\end{figure}

Moreover, as outlined in Sec.\ \ref{sec3:limit_shape}, when a limit shape is present, front fluctuations must be measured relative to the local average front, i.e., using Eq.~\eqref{eq3:Chi_local} rather than Eq.~\eqref{eq3:flu}. This approach is expected to yield improved results in systems that develop characteristic shapes \cite{Domenech2024}, as occurs in our system at low temperatures. Figure \ref{fig5:histograma_Tbaja} presents the fluctuation PDFs obtained for a representative condition where the film shape plays a significant role ($T = 1/3$, $A = 1$), using both global [Eq.~\eqref{eq3:flu}] and local [Eq.~\eqref{eq3:Chi_local}] measurements of fluctuations. As shown in the figure, the local methodology is particularly well-suited for scenarios in which the film shape deviates from a circular configuration.

To complement the fluctuation PDF, we have also directly calculated its third and fourth-order cumulants, namely the skewness and excess kurtosis. In particular, for the cases displayed in Fig.~\ref{fig5:histograma}, we obtained $S = 0.207(2)$ and $K = 0.063(4)$ for $T = 1$ and $A = 1$, while for $T = 3$ and $A = 10$, the values were $S = 0.086(2)$ and $K = 0.061(3)$. \footnote{For reference, the exact skewness and excess kurtosis values are $S = 0.29346452408$ and $K = 0.1652429384$ for the TW-GOE, and $S = 0.224084203610$ and $K = 0.0934480876$ for the TW-GUE \cite{Bornemann2010}.} Overall, we found that the skewness and excess kurtosis generally lie within the ranges $0 < S < 1/3$ and $0 < K < 0.25$, except in cases where the cumulants were computed after subtracting the shape (as for $T = 1/3$ and $A \le 1$), in which case $K \approx -0.2$. This behavior is reflected in the tails of the distribution shown in Fig.~\ref{fig5:histograma_Tbaja}.

\subsection{Structure factor}

\begin{figure}[t]
\centering
\includegraphics[width=0.7\textwidth]{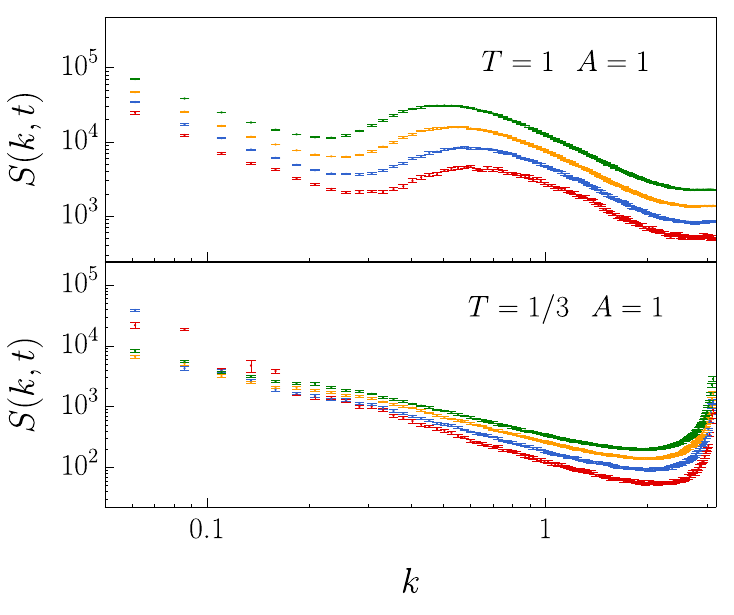}
\caption{Structure factor calculated for the precursor layer at $T = 1/3$, $A=1$ (bottom panel) and $T = 1$, $A=1$ (top panel), for times increasing bottom to top in both panels. In both panels $Z=1$.}
\label{fig5:factor_estructura}
\end{figure}

To gain deeper insight into the intrinsic anomalous scaling of the front, we also analyze its structure factor. Figure \ref{fig5:factor_estructura} shows the structure factor computed at various times for two representative parameter conditions: $T = 1/3$, $A = 1$, and $T = 1$, $A = 1$. Remarkably, the $S(k,t)$ curves exhibit a consistent upward shift over time, a distinctive signature of intrinsic anomalous scaling \cite{Lopez1997}.

When intrinsic anomalous scaling is present, the structure factor is expected to follow the scaling relation $S(k,t) \sim |k|^{-(2\alpha_\mathrm{loc} + 1)}$ at sufficiently long times and for large values of $k$ [see Eq.~\eqref{eq3:sfactor_scaling}]. However, the behavior of the structure factor is not so clear in this case, making it impossible to verify scaling laws. 

Nonetheless, we can still extract useful information from the behavior observed in Fig.~\ref{fig5:factor_estructura}. The presence of peaks in the structure factor reflects underlying features in the morphology of the system, indicating some degree of periodicity in the front. For instance, under conditions where the limit shape emerges, that is, under low-temperature conditions, the structure factor displays a pronounced peak at high $ k $ values, close to $ k \sim \pi $, as shown in the bottom panel of Fig.~\ref{fig5:factor_estructura}. This suggests that fluctuations are more prominent at small spatial scales. In other words, fluctuations at short arc lengths or small angular separations become dominant in these conditions. This is visually apparent in Fig.~\ref{fig5:morfologia}, where the morphology of the precursor film is depicted: in the corresponding case (bottom left of the figure), large portions of the front appear straight, and most deviations are caused by individual cells breaking away from this alignment.

For the remaining conditions, the behavior of the structure factor is less clear. Nevertheless, in most cases, a peak appears at intermediate $ k $ values. This may suggest a tendency for the front to develop bulges or ``fingers'' at certain preferred angles. However, this peak is considerably weaker than the one observed in the low-temperature case, indicating that conclusions drawn from this analysis should be approached with caution. Furthermore, the position of this intermediate peak appears to shift toward lower $ k $ values over time, suggesting that it may eventually disappear in the long-time regime. In addition, this peak becomes less pronounced at high temperatures.

\subsection{Front covariance}

We have also calculated the front covariance $ C_1(s,t) $, as defined in Eq.~\eqref{eq3:correlation_1}. In the previous chapter we showed that at high temperatures this function exhibits KPZ behavior. In particular, we demonstrated that through an appropriate rescaling, $ C_1(s,t) = a_1\, t^{2\beta} f\left(a_2 s / t^{1/z} \right) $, the covariance collapses onto a universal curve. In this expression, $ f(u) $ is a universal scaling function, while $ a_1 $ and $ a_2 $ are non-universal constants \cite{Alves2011,Oliveira2012,Nicoli2013}. As discussed in Sec.\ \ref{sec3:observables}, for the one-dimensional KPZ equation with PBC, the function $ f(u) $ corresponds to $ \mathrm{Airy}_1(u) $ when the front evolves in a band geometry. In contrast, for radial growth, $ f(u) $ corresponds to $ \mathrm{Airy}_2(u) $, which characterizes the covariance of the Airy$_2$ process \cite{Bornemann2009,HalpinHealy2015,Takeuchi2018}.

The procedure for determining $ a_1 $ and $ a_2 $, as described in Sec.\ \ref{sec3:observables}, cannot be applied in the present case, since we lack values at $ s = 0 $ required to use Eq.~\eqref{eq3:a1}, due to the necessity of defining angular boxes, as explained in Sec.~\ref{sec3:observables_growing}. Nevertheless, these constants can still be estimated by alternative methods, such as interpolating the function at two points for a fixed time.

\begin{figure}[t]
\centering
\includegraphics[width=0.7\textwidth]{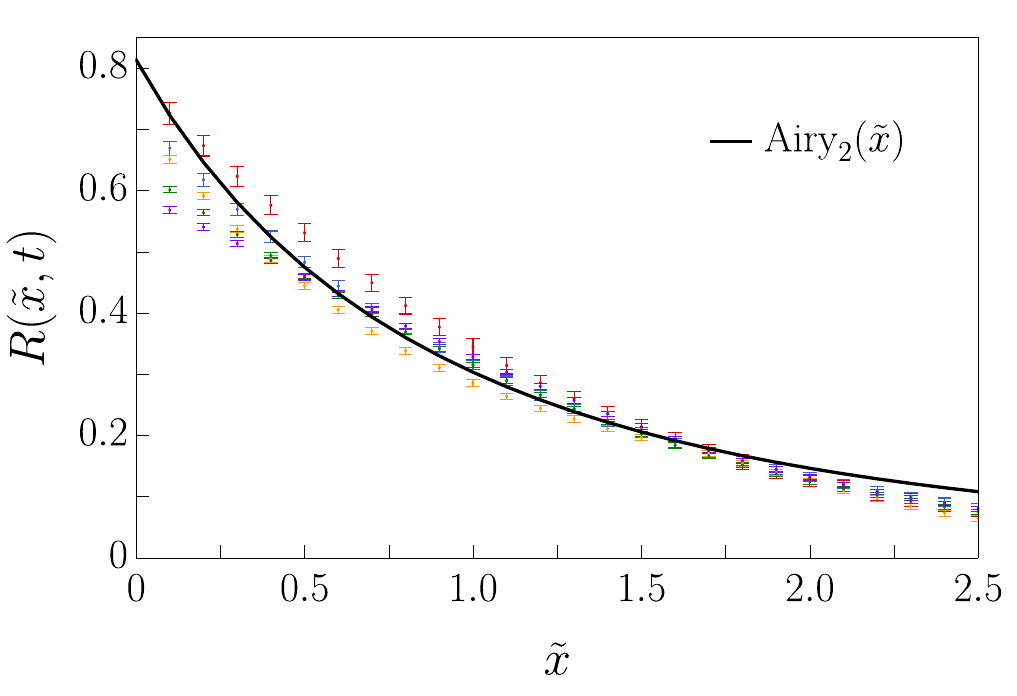}
\caption{$R(\tilde{x},t) \equiv \frac{C_1\left(\tilde{x} t^{1/z}/a_2\right)}{a_1t^{2\beta}}$ versus $\tilde{x}\equiv a_2 s/t^{1/z}$ for the time boxes $t_{\mathrm{BOX}}=\{50$, $60$, $70$, $80$, $90\}$, calculated for $T=3$, $A=1$, and $Z=1$ using $1/z=0.37 $, $2\beta=0.41 $, $a_1=0.008$, and $a_2=2400$. The solid line corresponds to the exact $\mathrm{Airy}_2(\tilde{x})$ function.}
\label{fig5:c1}
\end{figure}

Figure \ref{fig5:c1} displays the rescaled height covariance function
\begin{equation}
    C_1(\tilde{x}t^{1/z}/a_2)/(a_1 t^{2\beta}) \equiv R(\tilde{x},t)
\end{equation}
plotted as a function of $ \tilde{x} $ for different times. In contrast to the results presented in the previous chapter, differences with the universal behavior are significantly larger here. In fact, the collapse is quite poor for low $ \tilde{x} $ and the observed behavior deviates noticeably from the theoretical prediction.

Although this is not unexpected, particularly given that the exponent values differ from those of the KPZ class and the front exhibits intrinsic anomalous scaling, it remains noteworthy, especially when compared to the clearer collapse observed in the band geometry. At lower temperatures,  as observed in the previous chapter, the quality of the collapse further deteriorates, and the agreement with the $ \mathrm{Airy}_2 $ function becomes increasingly poor.

\section{Conclusions and comparison of both geometries}

In summary, we have investigated in this chapter the spatiotemporal dynamics of circular liquid droplet fronts spreading on flat substrates, using comprehensive kMC simulations of the Ising lattice gas model described in detail in Chapter \ref{chap2:wetting}. As in the previous chapter, we have analyzed the behavior of the system under varying parameter conditions, specifically the Hamaker constant (related to wettability) and temperature, through extensive kMC simulations.

We have explored a wide range of model parameters, focusing on classical morphological observables such as the mean front position and roughness. In addition, we conducted a systematic analysis of two-point correlation functions, both in real and Fourier space, following their evolution over time. To complement this, we also examined the statistical properties of front fluctuations by evaluating their PDF.

The exponent $ \delta \approx 1/2 $, which characterizes the mean position of the front of the precursor film, appears to be reached only under the most realistic conditions—specifically, low temperature and high Hamaker constant. In general, although the values obtained for $ \delta $ in circular geometry are somewhat smaller than those found in band geometry, they remain significantly larger than the classical Tanner law values associated with the spreading of macroscopic droplets.

\begin{figure}[t]
\centering
\includegraphics[width=0.7\textwidth]{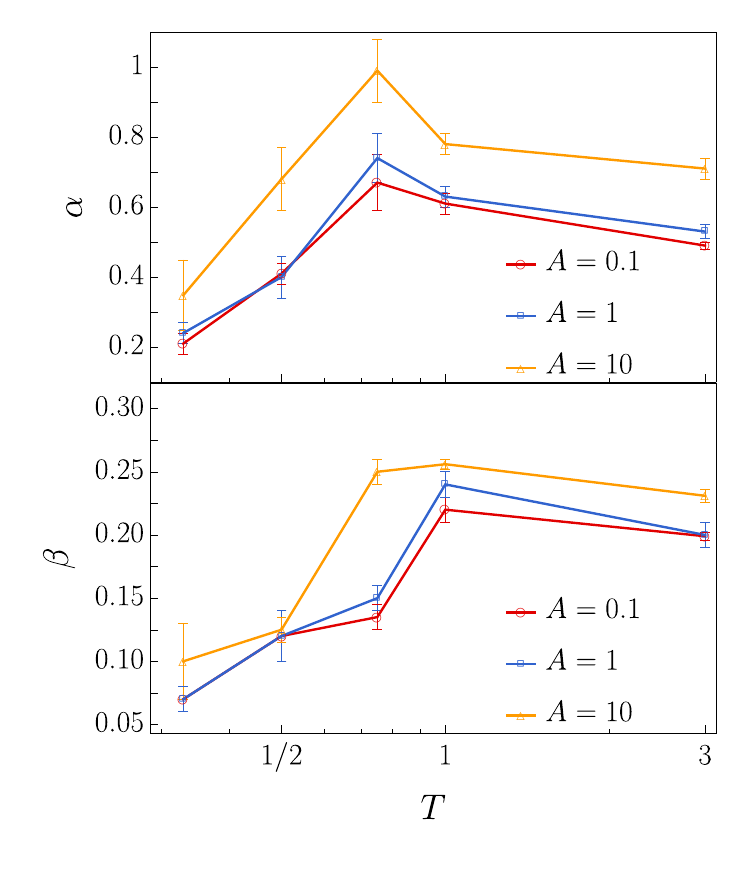}
\caption{Values of $\alpha$ (top) and $\beta$ (bottom) for the precursor film (taken from Tables \ref{tab5:beta_precursor} and \ref{tab5:1overz_precursor}) vs $T$ for $A=0.1$ (red circles), $A=1$ (blue squares), and $A=10$ (orange triangles). Lines are guides to the eye.}
\label{fig5:TablaExponentes}
\end{figure}

The critical exponents $ \alpha $, $ \beta $, and $ z $ exhibit a stronger dependence on temperature than on the Hamaker constant. They display a transition from a low-temperature to a high-temperature regime, beyond which they become largely temperature-independent. This trend is clearly illustrated in Fig.~\ref{fig5:TablaExponentes}, which shows the temperature dependence of the $ \alpha $ and $ \beta $ exponents [$ z $ being related to them through Eq.~\eqref{eq1:zalphabeta}] for several values of the Hamaker constant. Although minor quantitative differences are observed, the overall behavior closely resembles that reported for band geometry.

Regardless of the specific parameter values, the front consistently exhibits intrinsic anomalous scaling. This implies that the roughness exponents characterizing fluctuations at large ($ \alpha $) and small ($ \alpha_{\rm loc} $) length scales are different, a behavior that mirrors what is observed in band geometry. Moreover, although the Fourier analysis did not allow for a direct confirmation of the scaling laws, it clearly indicates the presence of intrinsic anomalous scaling, as the curves display a systematic upward shift over time.

At relatively low temperatures and Hamaker constants, the average shape of the film deviates noticeably from a circular to a square-like profile. This is evident both in the morphology of the films and in the corresponding two-point correlation functions under those conditions. To address such cases, in which the height-difference correlation function does not reach a plateau, we have developed a consistent method to compute the correlation length $ \xi(t) $.

Finally, despite the fact that the critical exponent values differ from those of the 1D KPZ universality class and that the dynamic scaling follows an intrinsically anomalous rather than TV form, the PDFs of the front fluctuations show a reasonable level of agreement with those of the 1D KPZ class in circular geometry, specifically the TW-GUE distribution. However, the behavior of the front covariance remains unclear, and no universal features can be reliably extracted from its analysis.

Admittedly, some quantitative, though not qualitative, differences remain between the results presented in this chapter and those obtained in the previous one, where the same model was analyzed in a band geometry. It should be noted, however, that the definition of the front differs between the two geometries. In the band geometry, a single-valued approximation was employed, whereas in the present chapter we have introduced a more complex and better-suited definition for the front of expanding circular droplets. Another key difference lies in the behavior of the front length $ L_f(t) $: in the band geometry, it remains fixed and equal to the reservoir size, while in the circular geometry, $ L_f(t) $ increases with time as the reservoir size stays constant. This leads to a slower effective growth rate, which poses additional challenges for studying the system in this geometry.

In any case, we believe that the combined results of both chapters support the existence of a well-defined universality class for this type of film spreading processes: one that features intrinsic anomalous scaling with temperature-dependent exponents, as well as a dependence on interface geometry, reflected in the subclass governing the statistics of front fluctuations, consistent with expectations for 1D KPZ-related interfaces.

\graphicspath{{6_capitulo/fig6/}}

\chapter{Numerical integration on networks}
\label{chap6:KPZ}


In this chapter, we integrate the KPZ equation [see Eq.~\eqref{eq1:KPZ}] and some related equations on the Bethe lattice. We begin with a brief overview of previous studies in which this equation has been integrated on regular lattices. We then demonstrate how the numerical integration method developed for regular networks can be extended to generic networks with arbitrary topology. Finally, we present the results obtained from this integration, along with some concluding remarks.

As discussed in Chapter~\ref{chap1:intro}, previous studies by Saberi~\cite{Saberi2013} and Oliveira~\cite{Oliveira2021} have investigated discrete models within the KPZ and EW universality classes on the Bethe lattice, aiming to shed light on the existence and nature of the upper critical dimension of the KPZ class. Building upon these foundational contributions, the primary objective here is to advance our understanding of how the associated continuum equations of these discrete models behave when defined on non-regular structures. In particular, we seek to determine whether the defining features of the KPZ and EW universality classes are preserved when moving from regular lattices to more complex, tree-like topologies such as the Bethe lattice. This inquiry constitutes the central motivation for the present work.

\section{Introduction}\label{sec3:intro}

As we have already mentioned, the numerical integration of the KPZ equation has been the subject of study for many years \cite{Amar1990, Moser1991, Forrest1993}. Using this approach, its critical exponents have been computed in one \cite{Moser1991, Forrest1993}, two \cite{Amar1990, Moser1991}, and three dimensions \cite{Moser1991}. Most previous studies addressing the numerical integration of the KPZ equation on finite-dimensional regular lattices have adopted the explicit Euler–Maruyama discretization scheme \cite{Amar1990, Moser1991}. Specifically, in one dimension, the following discretization (with $\Delta x$ lattice spacing) has been frequently used for the Laplacian and the square of the gradient:
\begin{equation}
	\label{eq6:discretizacion_1d}
        \begin{split}
        &\nabla^2 h\left(x_i,t\right)=\frac{1}{(\Delta x)^2}(h_{i+1}+h_{i-1}-2h_{i}), \\
        &\left(\nabla h\right)^2\left(x_i,t\right)=\frac{1}{(2\Delta x)^2} (h_{i+1}-h_{i-1})^2,
        \end{split}
\end{equation}
so that, using $n$ to denote time steps with time spacing $\Delta t$, the discretized KPZ equation reads \cite{Lam1998,Giada2002}
\begin{equation}
	\label{eq6:integracion_euler_1d}
    	h_{i}^{n+1}=h_{i}^n+\frac{\nu\Delta t}{(\Delta x)^2}(h_{i+1}^n+h_{i-1}^n-2h_{i}^n)+ \frac{\lambda\Delta t}{8(\Delta x)^2}(h_{i+1}^n-h_{i-1}^n)^2+\sqrt{2D\Delta t}\hspace{1mm}\eta_i^n\,.
\end{equation}
In Eq.~\eqref{eq6:integracion_euler_1d}, $\Delta x$ is typically set to 1 without loss of generality, and $\eta_i^n$ represents a Gaussian random variable with zero mean and unit variance. In higher dimensions, the discretized equation can be straightforwardly generalized by appropriately adding terms to both the Laplacian and the squared gradient to account for the additional dimensions.

However, the discrete nonlinear term in Eq.~\eqref{eq6:integracion_euler_1d} is highly unstable. For sufficiently large values of the nonlinear coefficient $\lambda$, the numerical integration diverges when noise-induced fluctuations grow more rapidly than they can be suppressed by the smoothing effect of the Laplacian \cite{Dasgupta1996, Dasgupta1997}. This instability is an inherent feature of the discretized KPZ equation, caused by the rapid temporal growth of isolated pillars. When the coupling constant exceeds a critical threshold, these pillars grow uncontrollably, leading to numerical blow-up \cite{Dasgupta1997}.

To address these issues, several improvements have been proposed, including refined real-space discretizations of the nonlinear term \cite{Lam1998}, as well as pseudospectral methods \cite{Giada2002, Gallego2011, Toral2014}. Although the latter require PBC and therefore cannot be directly adapted to a generic network, we will show that the discretization proposed by Lam and Shin in Ref.~\cite{Lam1998} can be reformulated for arbitrary network structures. The approach proposed by Lam and Shin 
involves improving the discretization of the squared gradient by introducing an additional cross term. Specifically, their proposal is
\begin{equation}
	\label{eq6:discretizacion_ls_1d}
    (\nabla h)^2\left(x_i,t\right)=\frac{1}{3(\Delta x)^2} \Big[(h_{i+1}-h_{i})^2+(h_{i}-h_{i-1})^2+(h_{i+1}-h_{i})(h_{i}-h_{i-1})\Big].
\end{equation}
In Ref.~\cite{Lam1998}, the authors demonstrate that establishing a correspondence between continuum growth equations and their discrete counterparts is a complex issue. They point out that, although many studies on the direct numerical integration of the KPZ equation have routinely employed finite difference discretizations, these methods are only accurate when the surface remains microscopically smooth, a condition not met in the case of the KPZ equation. 


Another effective approach to manage the intrinsic instability of the discretized equation, successfully applied to the KPZ equation and other kinetic roughening universality classes \cite{Dasgupta1996,Dasgupta1997,Miranda2008,Ales2019,Song2021}, is to replace the term $(\nabla h_i)^2$ with a function $f\left((\nabla h_i)^2\right)$ in the discretized equation. This discretization takes the form
\begin{eqnarray}\nonumber
	\label{eq6:discretizacion_control}
    	&&(\nabla h_i)^2\longrightarrow f\left((\nabla h_i)^2\right)\\
	    &&f(x) = (1 - e^{-cx}) / c,
\end{eqnarray}
where $c > 0$ is a tunable parameter and the squared gradient is typically discretized using the standard scheme, as given by Eq.~\eqref{eq6:discretizacion_1d}. This substitution effectively introduces an infinite series of higher-order terms in $(\nabla h_i)^2$, with coefficients that depend on the value of $c$. For values of $c$ above a certain critical threshold, the numerical instability is completely eliminated from the discretized equation, allowing for reliable estimation of scaling exponents \cite{Dasgupta1996,Dasgupta1997,Miranda2008,Ales2019,Song2021}. In contrast, when $c = 0$, and these additional terms are absent, the growth equation may exhibit numerical instabilities.

As explained in Chapter~\ref{chap1:intro}, the TKPZ equation was recently integrated numerically in Ref.\cite{RodrguezFernndez2022}. In that work, the authors employed a multistep predictor-corrector pseudospectral scheme, originally proposed in Ref.\cite{Gallego2011}, along with uniformly distributed noise of unit variance to carry out the integration. Since the method is pseudospectral and therefore requires PBC, it cannot be extended to generic networks. To integrate the TKPZ equation on such structures, alternative methods must be employed.

\section{Numerical integration schemes for PDEs on networks}\label{sec3:num_integra}



In the previous section, we discussed how numerically integrating the KPZ equation, and other stochastic differential equations, is far from straightforward, primarily due to the lack of a clear correspondence between the continuous formulation and its discrete counterpart. Building on this, and motivated by recent advances in the study of PDEs on discrete networks~\cite{Ortega2015,Chung2007}, we now explore how to adapt the previously introduced methods to the context of a generic network. Specifically, we present three approaches for extending the definitions of the Laplacian and the squared gradient from a regular lattice, as given by Eq.~\eqref{eq6:discretizacion_1d}, Eq.~\eqref{eq6:discretizacion_ls_1d}, and Eq.~\eqref{eq6:discretizacion_control}.

\subsection{Standard discretization}

A natural approach to extend the definitions of the Laplacian and the squared gradient on a network is to consider the contributions from all neighboring sites of a given node, i.e.,
\begin{equation}
	\label{eq6:discretizacion_grafos_st}
        \begin{split}
        &\nabla^2 h\left(\boldsymbol{x},t\right)=\sum_{j\sim i} h_{j}-\textrm{deg}(i)h_{i}, \\
        &\left(\nabla h\right)^2\left(\boldsymbol{x},t\right)=\sum_{j\sim i} (h_j-h_i)^2,
        \end{split}
\end{equation}
where $\textrm{deg}(i)$ denotes the degree of site $i$, i.e., the number of its neighbors, and the sum $\sum_{j \sim i}$ runs over all neighbors $j$ of the given site $i$. In the equation above, all neighbors in the network are assumed to be equidistant; however, these expressions can also be generalized to weighted networks~\cite{Ortega2015,Chung2007}. For simplicity we assume that all neighbors are at a distance of one unit. Applying Eq.~\eqref{eq6:discretizacion_grafos_st} to the KPZ equation leads to the following discretized equation:
\begin{equation}
	\label{eq6:integracion_euler_graf}
    	h_{i}^{n+1}=h_{i}^n+\nu\Delta t\left[\sum_{j\sim i} h_{j}^n-\textrm{deg}(i)h_{i}^n\right]+ \frac{\lambda\Delta t}{2}\sum_{j\sim i} (h_j^n-h_i^n)^2+\sqrt{2D\Delta t}\hspace{1mm}\eta_i^n\,,
\end{equation}
All the $h$-terms on the right-hand side of the equation correspond to the $n$-th time step; that is, the method is explicit. This basic extension of the Laplacian and the squared gradient will be referred to as the standard (ST)\nomenclature{ST}{Standard Integration} method throughout this chapter.

\subsection{LS discretization}

As mentioned in the previous section, Lam and Shin~\cite{Lam1998} proposed an alternative discretization for the squared gradient on regular lattices that enhances the stability of the numerical scheme. 
Their method can be applied to any finite-dimensional lattice by simply adding the additional terms for each additional spatial dimension. 
However, extending this discretization to a general network is not straightforward, as networks lack well-defined spatial directions, making it unclear how to select pairs of neighbors to represent the cross term in the LS discretization. The most straightforward approach would be to include all possible pairs of neighboring nodes in the discretization, namely,
\begin{equation}
(\nabla h)^2\left(\boldsymbol{x},t\right)=\sum_{j\sim i} (h_j-h_i)^2+\sum_{\langle j, k \rangle} (h_j-h_i)(h_i-h_k),
\label{eq6:lsn}
\end{equation}
where the sum $\sum_{\langle j, k \rangle}$ in the second term runs over all distinct pairs of neighbors of site $i$, without repetition. Despite its formal similarity to the regular-lattice case, Eq.~\eqref{eq6:lsn} fails to preserve the non-negativity of the squared gradient when the degree of node $i$ exceeds three. This issue renders the expression physically inconsistent and unsuitable for use in general network settings.

In the discretization above, the number of quadratic terms grows linearly with $\textrm{deg}(i)$, while the number of cross terms increases combinatorially as ${\textrm{deg}(i) \choose 2} = \textrm{deg}(i)\left(\textrm{deg}(i) - 1\right)/2$. In the LS discretization for a regular lattice, this ratio is fixed, as the number of quadratic terms is always twice the number of cross terms. To preserve this ratio and ensure that the discretized squared gradient remains positive-definite, we propose dividing the cross-term summation by $(\textrm{deg}(i) - 1)$. This adjustment maintains the desired 2:1 ratio between quadratic and cross terms for any node degree. The resulting discretized equation reads:
\begin{eqnarray}\nonumber
	\label{eq6:integracion_euler_ls}
    	&&h_{i}^{n+1}=h_{i}^n+\nu\Delta t\left[\sum_{j\sim i} h_{j}^n-\textrm{deg}(i)h_{i}^n\right]+\sqrt{2D\Delta t}\hspace{1mm}\eta_i^n+
        \\
	    &+&\frac{\lambda\Delta t}{2}\Bigg[\sum_{j\sim i} (h_j^n-h_i^n)^2+\frac{1}{\textrm{deg}(i)-1}\sum_{\langle j, k \rangle} (h_j^n-h_i^n)(h_i^n-h_k^n)\Bigg]\,.
\end{eqnarray}
In this chapter we refer to the above integration method as the Lam-Shin (LS)\nomenclature{LS}{Lam-Shin Integration} method.

\subsection{Controlled instability method using higher powers of the gradient}

As previously mentioned, the method proposed in Refs.~\cite{Dasgupta1996,Dasgupta1997} addresses the intrinsic instability of the discretized equation by replacing the term $(\nabla h_i)^2$ with a regularized function $f\left((\nabla h_i)^2\right)$, where $f(x) = (1 - e^{-c x})/c$, and $c$ is a tunable parameter. Applying this substitution to Eq.~\eqref{eq6:integracion_euler_graf}, the discretized KPZ equation becomes:
\begin{equation}
	\label{eq6:integracion_control}
    	h_{i}^{n+1}=h_{i}^n+\nu\Delta t\left[\sum_{j\sim i} h_{j}^n-\textrm{deg}(i)h_{i}^n\right] + \frac{\lambda\Delta t}{2}f\left(\sum_{j\sim i} (h_j^n-h_i^n)^2\right)+\sqrt{2D\Delta t}\hspace{1mm}\eta_i^n\,.
\end{equation}
Following Refs.~\cite{Dasgupta1996,Dasgupta1997}, we refer to the integration scheme defined by Eq.~\eqref{eq6:integracion_control} as the controlled instability (CI)\nomenclature{CI}{Controlled Instability Integration} method. The parameter $c$ should be chosen as small as possible to closely approximate the KPZ equation, while still being large enough to suppress numerical instabilities.

\section{Model and simulation details}

Although the schemes developed in the previous section can be applied to any network, following the works of Saberi~\cite{Saberi2013} and Oliveira~\cite{Oliveira2021}, we will focus in the rest of the chapter on the integration of the KPZ equation [Eq.~\eqref{eq1:KPZ}] and some related equations, namely the RD equation [Eq.~\eqref{eq1:eq_simple}], the EW equation [see Eq.~\eqref{eq1:EW}], and the TKPZ [Eq.~\eqref{eq1:KPZ_tensionless}], on the Bethe lattice. In the case of the RD equation, we set the average number of particles arriving at a given site per unit time, denoted by $F$, to zero, since this term can be absorbed via a Galilean transformation and does not influence the scaling behavior, as explained in Chapter~\ref{chap1:intro}.

We begin by recalling that a CT is a connected, loopless graph in which all interior vertices have the same degree (also referred to as the coordination number) $q$, while boundary vertices have only one neighbor. The simplest way to construct a CT is to start from a central site (assigned shell index $s = 0$), then add $q$ neighboring sites to form the first shell ($s = 1$). Subsequently, each site in the previous shell is connected to $q - 1$ new neighbors, continuing this process iteratively until the desired number of shells is reached. An example of a CT with coordination number $q = 4$ and three shells is shown in Fig.~\ref{fig3:cayley} in Chapter~\ref{chap3:methods}. The total number of sites in a CT can be calculated as
\begin{equation}
	\label{eq6:nt}
	N_T=1 + \frac{q[(q - 1)^k - 1]}{(q - 2)} \hspace{2mm} (q>2),
\end{equation}
where $k$ is the largest value of $s$ for the specific tree ($k=3$ for the tree shown in Fig.~\ref{fig3:cayley}). The number of sites belonging to the $s$-shell is
\begin{equation}
	\label{eq6:nk}
	N_s=q(q - 1)^{s - 1} \hspace{2mm} (s>0).
\end{equation}
As the number of shells increases, the ratio between the number of sites in the outermost shell (i.e., the boundary, whose sites have only one neighbor) and the total number of sites does not vanish, as it does in regular lattices, but instead approaches a finite value, $(q - 2)/(q - 1)$. This implies that, even in the thermodynamic limit $k \rightarrow \infty$, a macroscopic fraction of the sites belong to the boundary. As a result, models defined on CTs are strongly influenced by boundary effects. The core of an infinite CT, in which the central region lies at an infinite distance from the boundary and is therefore unaffected by it, is known as the Bethe lattice \cite{Bethe1935,Baxter1985}. As explained in Chapter~\ref{chap1:intro}, the Bethe lattice has been used, in certain contexts, as an approximation for an infinite-dimensional system. A huge variety of systems have been studied using the Bethe lattice as a substrate \cite{Dorogovtsev2008}, including percolation-related models \cite{Chae2012}, diffusion processes \cite{Sahimi1988,Hughes1982}, random aggregation \cite{Krug1988,Bradley1984}, and transport phenomena \cite{Sahimi1993}.

In the next section, we present integration results for the three schemes proposed in the previous section: the ST, LS, and CI methods. In all cases, the noise term $\eta_i^n$, which is a Gaussian random variable with zero mean and unit standard deviation, is generated using the standard Box–Muller method \cite{Box1958}.

Furthermore, the BC of the system must be specified. For both the ST and LS methods, we used Neumann BC; that is, at each time step, the height values of the sites in the last layer were set equal to those of their parent sites in the penultimate layer. For the CI method, we generally used Free BC, although some simulations were performed with Neumann BC. In the former case, the nodes in the last layer were allowed to evolve freely according to Eq.~\eqref{eq6:integracion_control} at each step. Free BC were not used with the ST and LS methods due to numerical instabilities.

Regarding the choice of the parameter $c$ in Eq.~\eqref{eq6:discretizacion_control}, we found that numerical instabilities emerge for values around $c \approx 0.001$ or lower. Although none of the simulations resulted in overflow, runs with such small $c$ values exhibited clear signs of numerical instability. For example, simulations performed with such small values of $c$ exhibit a global roughness that overshoots its saturation value and subsequently relaxes back in an irregular and unpredictable manner. Based on these observations, we fixed $c = 0.01$ for all simulations in which this method was employed.

Finally, a brief summary of the simulations performed in this chapter is presented. For both the ST and LS methods, we fixed the time step at \mbox{$\Delta t = 0.001$}, as it offered a reasonable balance between computational efficiency and numerical stability. However, due to instability issues, we were limited to simulating relatively small values of $\lambda$. Table~\ref{tab6:methods_LS_ST} summarizes the KPZ simulations carried out using the ST and LS methods. In addition, Table~\ref{tab6:EW_RD} outlines the conditions under which the EW equation (where \mbox{$\lambda = 0$}) and the RD equation (where both $\nu$ and $\lambda$ are zero) were simulated. In these cases, the choice of integration scheme for the nonlinear term becomes irrelevant.

On the other hand, the CI method significantly enhances the numerical stability of the discretized equation, allowing for the use of a larger time step, $\Delta t = 0.01$, which facilitates simulations over longer time scales. Table~\ref{tab6:methods_control} presents a summary of the KPZ and TKPZ simulations conducted using this method. Notably, only the CI method allowed successful integration of the TKPZ equation; attempts using the ST and LS methods invariably resulted in numerical overflow after just a few steps.

\begin{table}[t]
\centering
\renewcommand{\arraystretch}{1.2}
\begin{tabular}{@{}cccc@{}}
\toprule
\hspace{0.5em}$q$ & $k$ & $\lambda$ & Runs \\
\midrule\midrule
\hspace{0.5em}3 & 4, 6, 8, 10, 12, 14 and 16 & 0.5 & 50 \\
\midrule
\hspace{0.5em}4 & 4, 6, 8 and 10 & 0.5 & 50 \\
\bottomrule
\end{tabular}
\caption{Summary of simulations performed for the KPZ equation using the ST and LS methods. For all these simulations, $\Delta t = 0.001$, the number of time steps $N_{\textrm{steps}} = 10^7$, and $D = \nu = 1$.}
\label{tab6:methods_LS_ST}
\end{table}

\begin{table}[t]
\centering
\renewcommand{\arraystretch}{1.2}
\begin{tabular}{@{}cccc@{}}
\toprule
\hspace{0.5em}$q$ & $k$ & $\lambda$ & Runs \\
\midrule\midrule
\hspace{0.5em}3 & 4, 6, 8, 10, 12, 14 and 16 & 0 & 50 \\
\midrule
\hspace{0.5em}4 & 4, 6, 8 and 10 & 0 & 50 \\
\midrule
\hspace{0.5em}5 & 4, 6 and 8 & 0 & 50 \\
\midrule
\hspace{0.5em}6 & 4 and 6 & 0 & 50 \\
\bottomrule
\end{tabular}
\caption{Summary of simulations performed for the EW and RD equations. For all these simulations, $\Delta t = 0.001$, the number of time steps $N_{\textrm{steps}} = 10^7$, and $D=1$. For the simulations of the EW equation, $\nu=1$.}
\label{tab6:EW_RD}
\end{table}

\begin{table}[t]
\centering
\renewcommand{\arraystretch}{1.2}
\begin{tabular}{@{}cccc@{}}
\toprule
\hspace{0.5em}$q$ & $k$ & $\lambda$ & Runs \\
\midrule\midrule
\hspace{0.5em}3 & 4, 6, \textbf{8}, \textbf{10}, 12, 14 and 16 & 3.0 & 100 \\
\midrule
\hspace{0.5em}4 & 4, 6, \textbf{8} and 10 & 3.0 & 100 \\
\midrule
\hspace{0.5em}5 & 4, 5, 6, and 7 & 3.0 & 100 \\
\midrule
\hspace{0.5em}6 & 4, 5, \textbf{6}, and 7 & 3.0 & 100 \\
\midrule
\hspace{0.5em}7 & 4, 5 and 6 & 3.0 & 100 \\
\midrule
\hspace{0.5em}8 & \textbf{4}, 5 and 6 & 3.0 & 100 \\
\bottomrule
\end{tabular}
\caption{Summary of simulations performed for the KPZ equation using the CI method. For all these simulations, $\Delta t = 0.01$, the number of time steps $N_{\textrm{steps}} = 10^7$, and $D=\nu=1$. The five conditions that appear in bold type were also used to simulate the TKPZ equation in which $\nu = 0$.}
\label{tab6:methods_control}
\end{table}

For the reader’s convenience, we recall here the continuous equations to be integrated in this chapter, corresponding to the RD, EW, KPZ, and TKPZ universality classes, before discussing the results
\begin{equation}
\begin{aligned}
    \frac{\partial h(\boldsymbol{x},t)}{\partial t} &= \eta(\boldsymbol{x},t) && \text{(RD)} \\
    \frac{\partial h(\boldsymbol{x},t)}{\partial t} &= \nu \nabla^2 h + \eta(\boldsymbol{x},t) && \text{(EW)} \\
    \frac{\partial h(\boldsymbol{x},t)}{\partial t} &= \nu \nabla^2 h + \frac{\lambda}{2} (\nabla h)^2 + \eta(\boldsymbol{x},t) && \text{(KPZ)} \\
    \frac{\partial h(\boldsymbol{x},t)}{\partial t} &= \frac{\lambda}{2} (\nabla h)^2 + \eta(\boldsymbol{x},t) && \text{(TKPZ)}
\end{aligned}
\label{eq6:universality_classes}
\end{equation}

\section{Results}

In this section, we present the results of integrating the continuous equations described above on CTs. We begin with a comparison of the different integration methods, followed by an analysis of the results obtained under the two previously discussed BC. We then conduct a systematic study of all observables across the various equations, see Sec.~\ref{sec3:observables_bethe} for definitions. For the first set of observables, particularly the global and local roughnesses, we will focus on comparing our results with those reported by Saberi in Ref.~\cite{Saberi2013} and Oliveira in Ref.~\cite{Oliveira2021}. We will then present results for the height-difference correlation function and the statistics of the front, which, to the best of our knowledge, have not been previously studied for these equations on such lattices. While examining the outcomes of the height-difference correlation function, we briefly revisit the comparison between the two BC for this specific observable. Finally, at the end of this section, a detailed analysis of how each layer grows relative to the mean height will be provided.

\subsection{Comparison of integration methods}

\begin{figure}[t]
\centering
\includegraphics[width=0.7\textwidth]{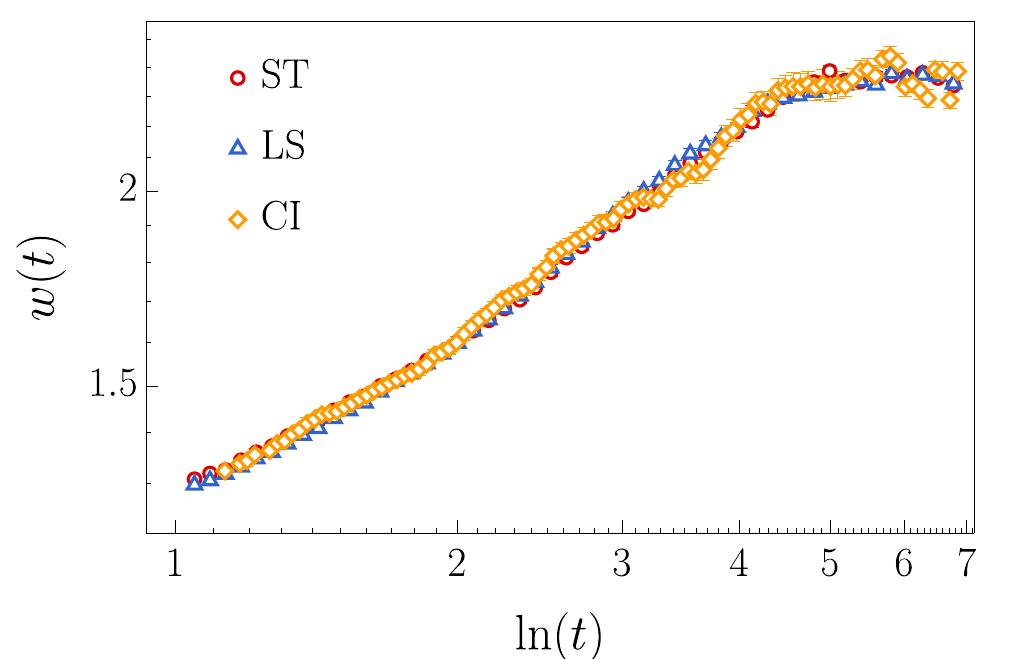}
\caption{Global roughness as a function of time for the KPZ equation computed using the three integration methods. The ST method is shown in red circles, the LS method is shown in blue triangles and the CI method is shown in yellow triangles. In this figure $q=3$, $k=8$, $\nu=D=1$, and $\lambda=0.5$. In all cases Neumann BC were applied.}
\label{fig6:compara_metodos}
\end{figure}

Figure~\ref{fig6:compara_metodos} illustrates the evolution of $w(t)$ for the KPZ equation using the three integration methods 
In all cases, Neumann BC are applied. It is evident that all three approaches produce very similar results within the parameter range where they are simultaneously applicable. Other observables, such as the local roughness $w_0(t)$, exhibit very similar behavior across the different integration methods.

While the LS method offers a slight improvement in integration stability compared to the ST one, this advantage is less pronounced in the present case than it is for regular lattices. The main distinction between the ST and LS methods, beyond stability, lies in the average front value, $\overline{h}$, which is lower in the LS method than in the ST one. Nevertheless, the two remain proportional. This behavior is expected, as the squared gradient term is generally smaller in the LS method, resulting in a reduced average front value. For the CI method, $\overline{h}$ takes an intermediate value between those of the ST and LS methods, while still maintaining proportionality with both. Additionally, it is observed that the numerical stability of the ST and LS methods decreases as the coordination number $q$ increases.

For the ST and LS methods, we have only been able to simulate the KPZ equation for small values of $\lambda$. Assuming the CT provides an appropriate framework for exploring the upper critical dimension of the KPZ class, this probably places our ST and LS simulation results in the weak-coupling regime of the KPZ equation, where only the smooth phase is accessible. Moreover, neither the ST nor the LS methods support simulations at higher $\lambda$ values, and both fail in the TKPZ limit. This limitation is the main reason for using the CI method throughout this chapter. 
\vspace{-5pt}
\subsection{Comparison of boundary conditions}\label{sec6:comparaBC1}

Figure~\ref{fig6:compara_BC} shows the evolution of $w(t)$ for the two BC choices
, namely Free BC and Neumann BC, both implemented using the CI integration method. We recall here that the CI method was the only one of the three introduced in this chapter for which Free BC were used, as they further worsened the stability issues of both the ST and LS methods. 

Although modifying the BC introduces some quantitative differences, the overall qualitative behavior of the system remains unchanged. This suggests that the observed phenomena are relatively robust to the choice of BC and are not merely artifacts of a specific setup. This conclusion also applies to other observables, such as the height-difference correlation function, whose specific dependence on the BC will be analyzed in detail below.

\vspace{-5pt}
\subsection{Average front position}

\begin{figure}[t!]
\centering
\includegraphics[width=0.7\textwidth]{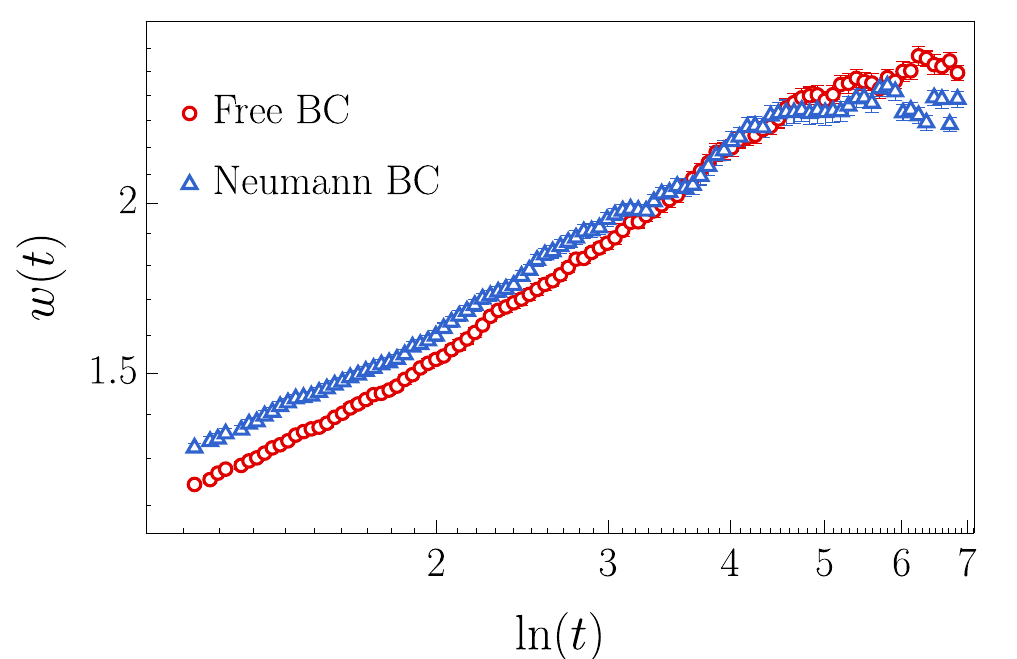}
\caption{Global roughness as a function of time for the KPZ equation computed for the two choices of BC. Free BC results are shown in red circles while Neumann BC are shown in blue triangles. In this figure $q=3$, $k=8$, $\nu=D=1$, and $\lambda=0.5$. The integration method used was CI.}
\label{fig6:compara_BC}
\vspace{1cm}
\centering
\includegraphics[width=0.7\textwidth]{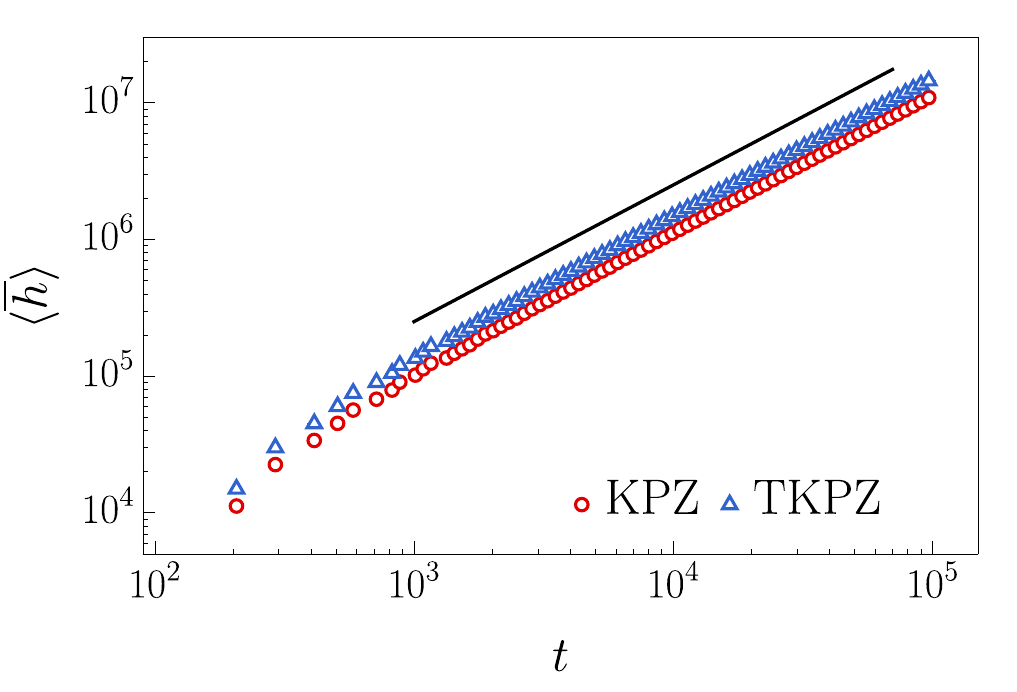}
\caption{Time evolution of the average front position $\langle\overline{h}\rangle$ for the KPZ and TKPZ equations. In this figure $q=3$ and $k=10$. As visual reference, the solid black line correspond to $\langle\overline{h}\rangle\sim t$. The integration method used was CI.}
\label{fig6:h_KPZ_TKPZ}
\end{figure}

Figure~\ref{fig6:h_KPZ_TKPZ} shows the average front position, $\langle\overline{h}\rangle$, for the KPZ and TKPZ equations, both evaluated using the same set of parameters. In both cases, the average front position increases linearly with time, as expected from analogous behavior on regular lattices. The front grows faster for the TKPZ equation, which is consistent with the absence of a relaxation term. Furthermore, we have verified (not shown here) that the mean height restricted to each shell (or layer), $\langle\overline{h}\rangle_s$, also grows linearly in time across all shells. Some differences in the growth of individual layers arise depending on the specific equation considered; these will be examined in detail below. 
For both the RD and EW equations, all these averages are zero, as expected given the nature of those equations.

\subsection{Global and local roughness. Variance of the mean height}

\begin{figure}[t!]
\centering
\includegraphics[width=0.7\textwidth]{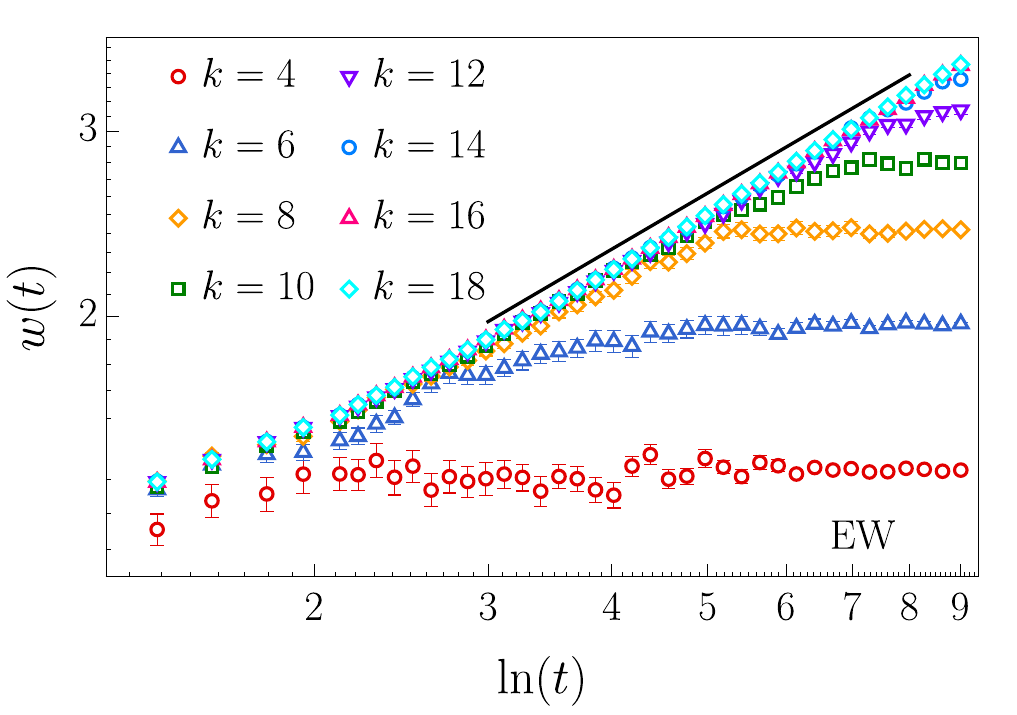}
\caption{Log-log plot of the global roughness $w(t)$ as a function of $\ln t$ for the EW equation on CTs with coordination number $q=3$. Data are provided for different system sizes, $k$, of the trees, see legends. As visual reference, the solid black line correspond to $w\sim (\ln t)^{0.55}$.}
\label{fig6:w_integra_EW}
\vspace{1.0cm}
\includegraphics[width=0.7\textwidth]{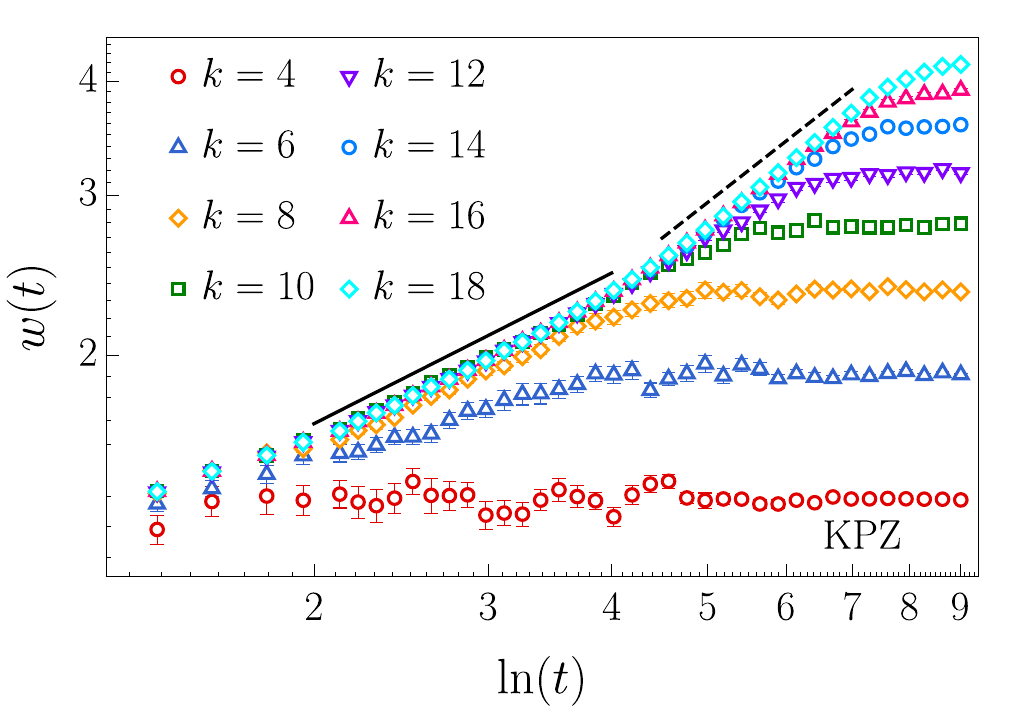}
\caption{Log-log plots of the global roughness $w(t)$ as a function of $\ln t$ for the KPZ equation on CTs with coordination number $q=3$. Data are provided for different system sizes, $k$, of the trees, see legends. As visual references, the solid black line corresponds to $w\sim (\ln t)^{0.55}$ and the dashed black line corresponds to $w\sim (\ln t)^{0.85}$. The integration method used was LS.}
\label{fig6:w_integra_KPZ}
\end{figure}

Figure~\ref{fig6:w_integra_EW} displays the evolution of the global roughness $w(t)$  as a function of the logarithm of time for trees of increasing size $k$ for the EW equation, whereas Fig.~\ref{fig6:w_integra_KPZ} presents the corresponding results for the KPZ equation. In Ref.~\cite{Saberi2013}, Saberi reported a similar logarithmic scaling behavior across several discrete models. Specifically, within the KPZ universality class, he found $ w \sim (\ln t)^{0.75} $ for the BD model and $ w \sim (\ln t)^{0.57} $ for the RSOS model. Additionally, for the RDSR model, belonging to the EW class, he found $ w \sim (\ln t)^{0.51} $.

\begin{figure}[t!]
\centering
\includegraphics[width=0.7\textwidth]{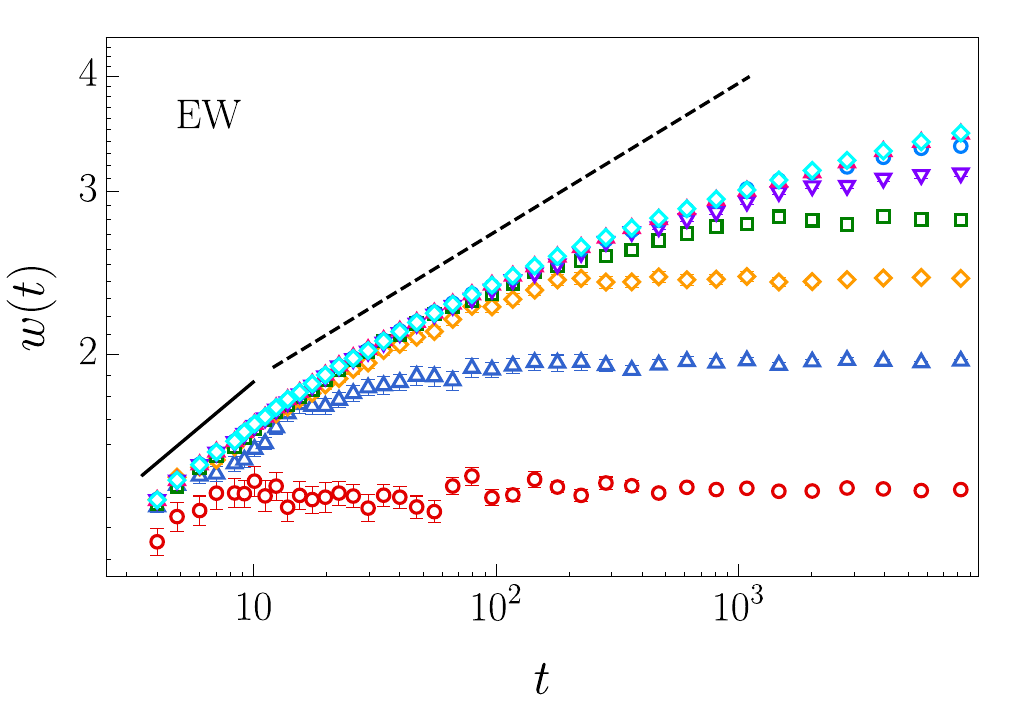}
\caption{Same data as in Fig.~\ref{fig6:w_integra_EW}, but with $t$, rather than $\ln t$, on the horizontal axis. As visual reference, the solid black line now corresponds to $w\sim t^{0.22}$ and the dashed black line corresponds to $w\sim t^{0.16}$.}
\label{fig6:w_noLog_EW}
\vspace{1.0cm}
\includegraphics[width=0.7\textwidth]{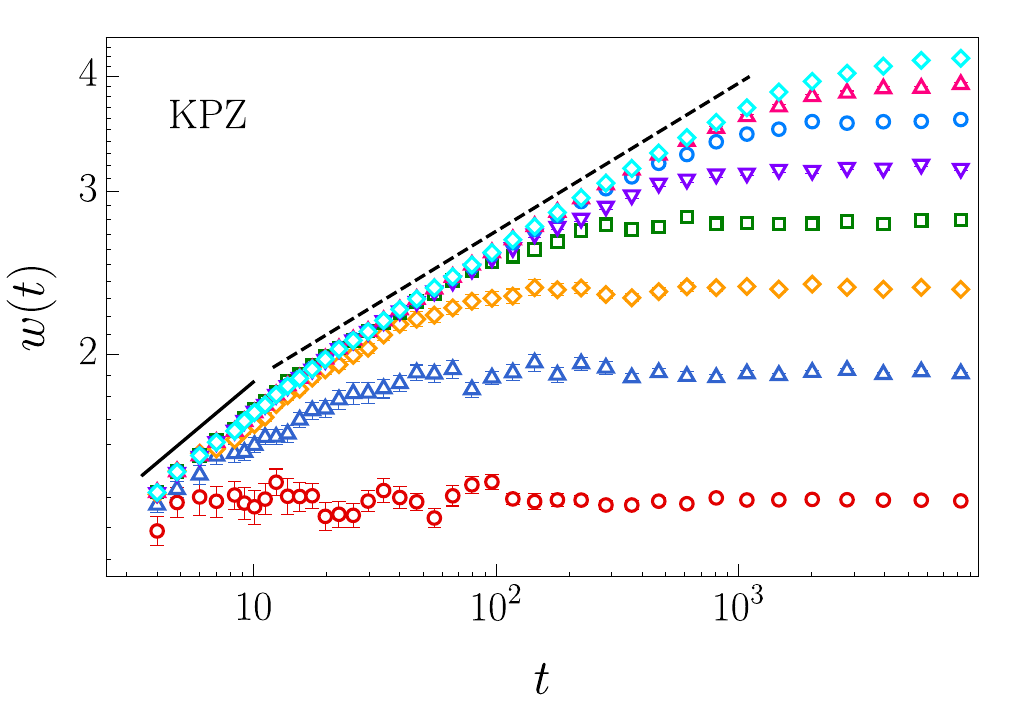}
\caption{Same data as in Fig.~\ref{fig6:w_integra_KPZ}, but with $t$, rather than $\ln t$, on the horizontal axis. As visual reference, the solid black line now corresponds to $w\sim t^{0.22}$ and the dashed black line corresponds to $w\sim t^{0.16}$.}
\label{fig6:w_noLog_KPZ}
\end{figure}

In our case, the numerical integration of the EW equation clearly exhibits logarithmic scaling, with an exponent close to that reported by Saberi. In contrast, the integration of the KPZ equation does not clearly exhibit this behavior, as the roughness deviates from the logarithmic scaling at sufficiently long times, as shown in Fig.~\ref{fig6:w_integra_KPZ}. More precisely, the roughness in the KPZ equation initially grows in a manner similar to that of the EW case during a transient regime, before deviating from it. This type of time crossover behavior is familiar for the KPZ equation. For example, in $ d = 1 $, the roughness exhibits different growth exponents at successive time scales: $\beta = 1/2$ (as in RD) at very short times, followed by $\beta = 1/4$ (as in EW) at intermediate times, and eventually $\beta = 1/3$, which characterizes the KPZ growth regime, before saturation to a steady state \cite{Forrest1993}.
In fact, for the KPZ case, we find that the growth of the global roughness $ w(t) $ is more accurately described by a power-law in $ t $, rather than in $ \ln (t) $. Figure~\ref{fig6:w_noLog_EW} shows the evolution of the global roughness $w(t)$ as a function of time instead of its logarithm for the EW equation, whereas Fig.~\ref{fig6:w_noLog_KPZ} presents the corresponding results for the KPZ equation. The behavior of the EW equation is clearly better described by a logarithmic scaling, as shown in Fig.~\ref{fig6:w_integra_EW}. In contrast, for the KPZ case, although the roughness follows a similar logarithmic trend for small values of $ k $, it is better described by a power-law $ w \sim t^{\beta} $ with $ \beta \approx 0.16 $ for larger sizes $ k > 14 $, albeit within a limited time window before saturation. This crossover in time appears reminiscent of the EW-to-KPZ transition observed in low dimensions, as previously discussed \cite{Forrest1993}. To further clarify this behavior, it would be desirable to perform simulations with larger system sizes. However, our study is constrained by the relatively small sizes that we are able to simulate. It is important to recall that the number of nodes in a CT grows exponentially with the number of layers $ k $ [see Eq.~\eqref{eq6:nt}], which severely limits the feasibility of simulating significantly larger systems.

\begin{figure}[t!]
\centering
\includegraphics[width=0.7\textwidth]{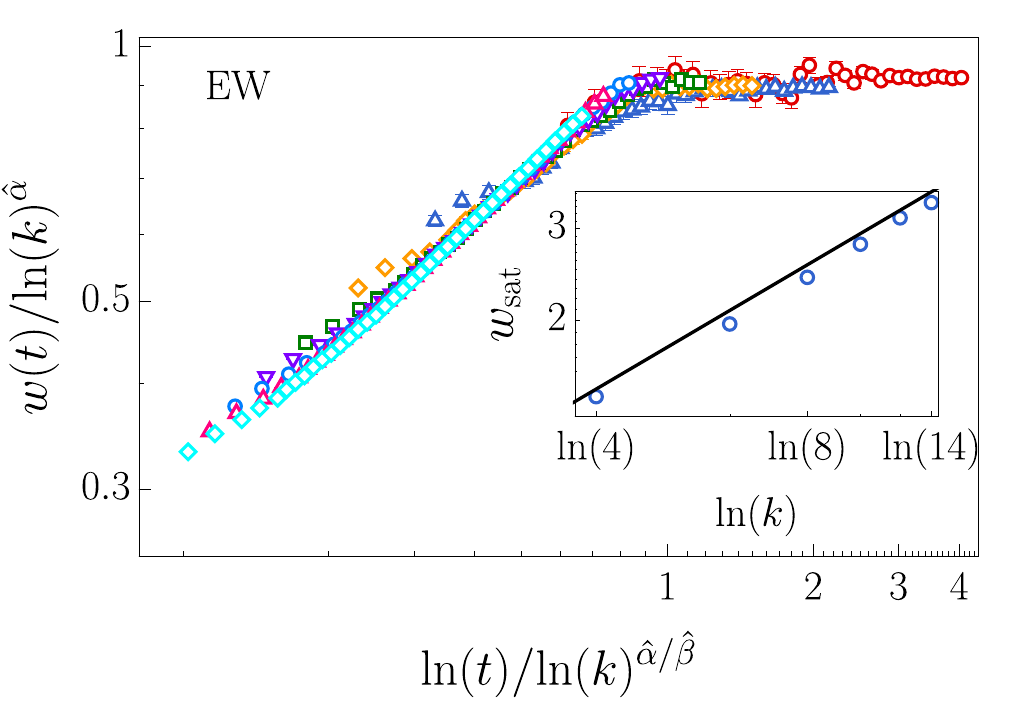}
\caption{Data collapse $w(t)/\mathrm{ln}(k)^{\hat{\alpha}}$ vs $\mathrm{ln}(t)/\mathrm{ln}(k)^{\hat{\alpha}/\hat{\beta}}$ for the EW simulations addressed in Figs.\ \ref{fig6:w_integra_EW} and \ref{fig6:w_noLog_EW}, using $\hat{\alpha} = 1.4$ and $\hat{\beta}=0.85$. Inset: Saturation value of the global roughness $w_{\rm sat}$ as a function of the logarithm of the number of layers $\mathrm{ln}(k)$. As a visual reference, the solid black line corresponds to $w_{\rm sat}\sim \mathrm{ln}(k)^{1.4}$.}
\label{fig6:w_colapso_EW}
\includegraphics[width=0.7\textwidth]{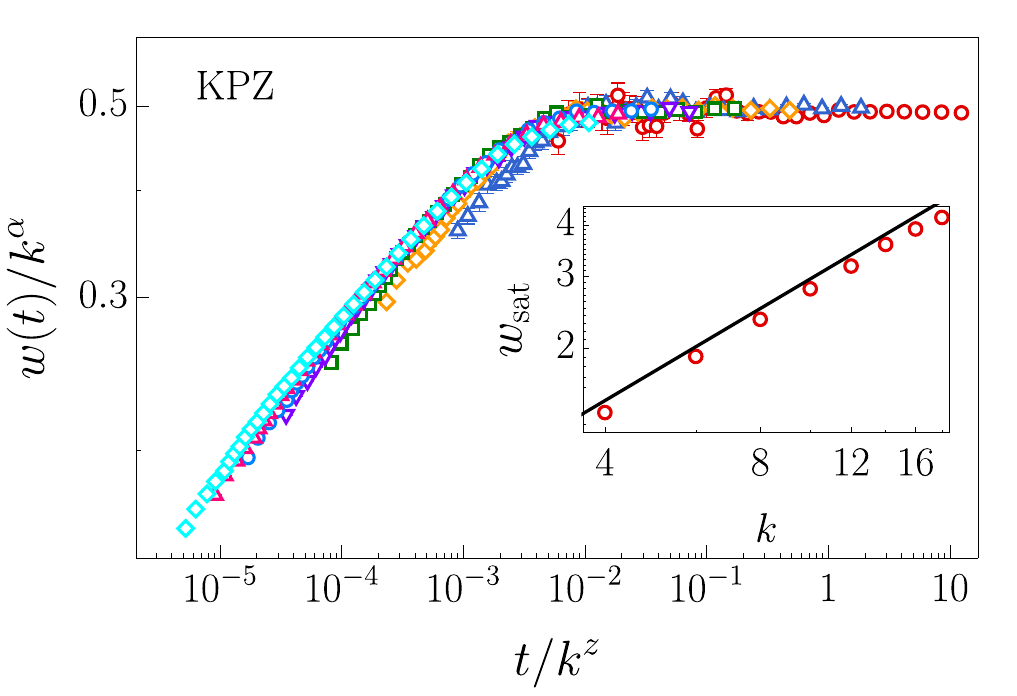}
\caption{Data collapse $w(t)/k^\alpha$ vs $t/k^z$ for the KPZ simulations addressed in Figs.\ \ref{fig6:w_integra_KPZ} and \ref{fig6:w_noLog_KPZ}, using $\alpha = 0.75$ and $z=0.75/0.16=4.69$. Inset: Saturation value of the global roughness $w_{\rm sat}$ as a function of the number of layers $k$. As a visual reference, the solid black line corresponds to $w_{\rm sat}\sim k^{0.75}$.}
\label{fig6:w_colapso_KPZ}
\end{figure}

Regarding the system-size dependence of the global roughness, Saberi also reported a similar logarithmic scaling for the saturation value, specifically \mbox{$w_{\rm sat}\sim(\ln k)^{\hat{\alpha}}$}, in the BD and RSOS models, while a more conventional power-law scaling, $ w_{\rm sat} \sim k^{\alpha} $, was observed for the RDSR model. In our simulations of the EW equation, we found that the saturation value follows a power-law scaling, $ w_{\rm sat} \sim (\ln k)^{\hat{\alpha}} $, with $ \hat{\alpha} \approx 1.4 $; see the inset of Fig.~\ref{fig6:w_colapso_EW}. Furthermore, by using the scaling $ w(t) \sim (\ln t)^{\hat{\beta}} $ observed in Fig.~\ref{fig6:w_integra_EW}, with $ \hat{\beta} = 0.55 $, the EW data for $ w(t) $ approximately collapse onto a single master curve, in analogy with the FV data collapse for global roughness \cite{Barabasi1995}.

A similar data collapse was achieved by Saberi for the BD model. In our case, we were unable to collapse the data using the standard power-law forms $ k^\alpha $ and $ k^z $, instead relying on the logarithmic scalings $ (\ln k)^{\hat{\alpha}} $ and $ (\ln k)^{\hat{z}} $ employed in Fig.~\ref{fig6:w_colapso_EW}. 
Furthermore, the values obtained for $ \hat{\alpha} $ and $ \hat{\beta} $ appear to be parameter-dependent. Notably, $ \hat{\alpha} $ decreases as the coordination number $ q $ increases, which is consistent with Saberi’s findings and supports the expectation that the condition $ d > d_u $ is better approximated for larger $ q $.

For the KPZ equation, the saturation value of the global roughness as a function of system size $ k $ (see the inset of Fig.~\ref{fig6:w_colapso_KPZ}) is consistent with a power-law scaling, $ w_{\rm sat} \sim k^{\alpha} $, with $ \alpha \approx 0.75 $. Combining this with the time-dependent behavior $ w(t) \sim t^{\beta} $, obtained from Fig.~\ref{fig6:w_noLog_KPZ} with $ \beta \approx 0.16 $, yields a well-defined FV data collapse, as shown in Fig.~\ref{fig6:w_colapso_KPZ}, using the dynamic exponent $ z = \alpha/\beta \approx 4.69 $. In this figure, deviations from the data collapse are only observed at short times, corresponding to the initial EW-like transient.

These results are in good agreement with those reported by Saberi in Ref.~\cite{Saberi2013}. However, some notable differences exist. Most prominently, one of the key findings in that work is the logarithmic scaling of the global roughness for models within the KPZ class. In contrast, our results do not unambiguously support this behavior. Instead, the data collapse of the global roughness for the KPZ equation is more consistent with standard FV scaling. Furthermore, there are inconsistencies in how the saturation value of the roughness scales with system size. While our results indicate that, for the KPZ equation, the best scaling is with the number of shells $ k $, and for the EW equation, with its logarithm $ \ln(k) $, Saberi’s work reports the opposite trend. However, in our case, and likely in Saberi’s as well, both scaling forms may be compatible due to the limited range of $ k $ values analyzed. Additionally, the behavior we observe as the coordination number $ q $ increases differs from the findings reported by Saberi. While Saberi, based on simulations of the BD model, observed that the saturation value increases with $ q $, our numerical integration of the KPZ equation reveals a decreasing trend in the saturation value as $ q $ increases. Interestingly, we also identified cases in which the saturation value does not exhibit a monotonic dependence on $ q $. For example, at $ k = 6 $, the value for $ q = 7 $ is higher than that for $ q = 6 $, and comparable to the value for $ q = 5 $.

\begin{figure}[t!]
\centering
\includegraphics[width=0.7\textwidth]{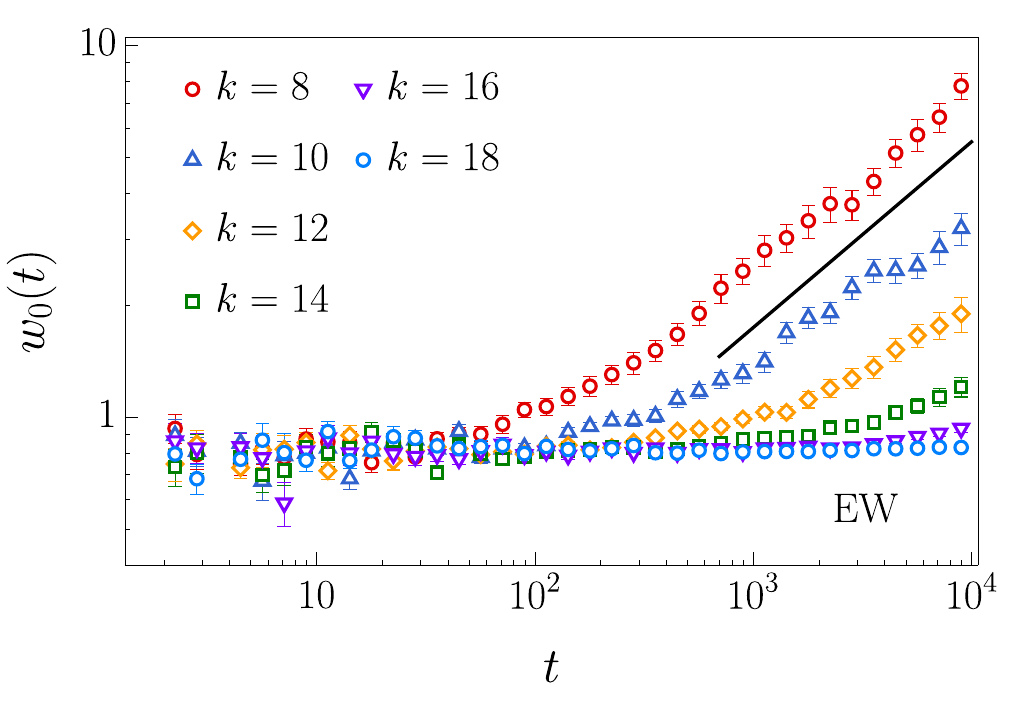}
\caption{Local roughness $w_0(t)$ as a function of time for $q=3$ and several values of $k$ (see legends) for the EW equation. As a visual reference, the solid black line corresponds to $w_0\sim t^{0.5}$. The integration method used was LS.}
\label{fig6:w0_integra_EW}
\includegraphics[width=0.7\textwidth]{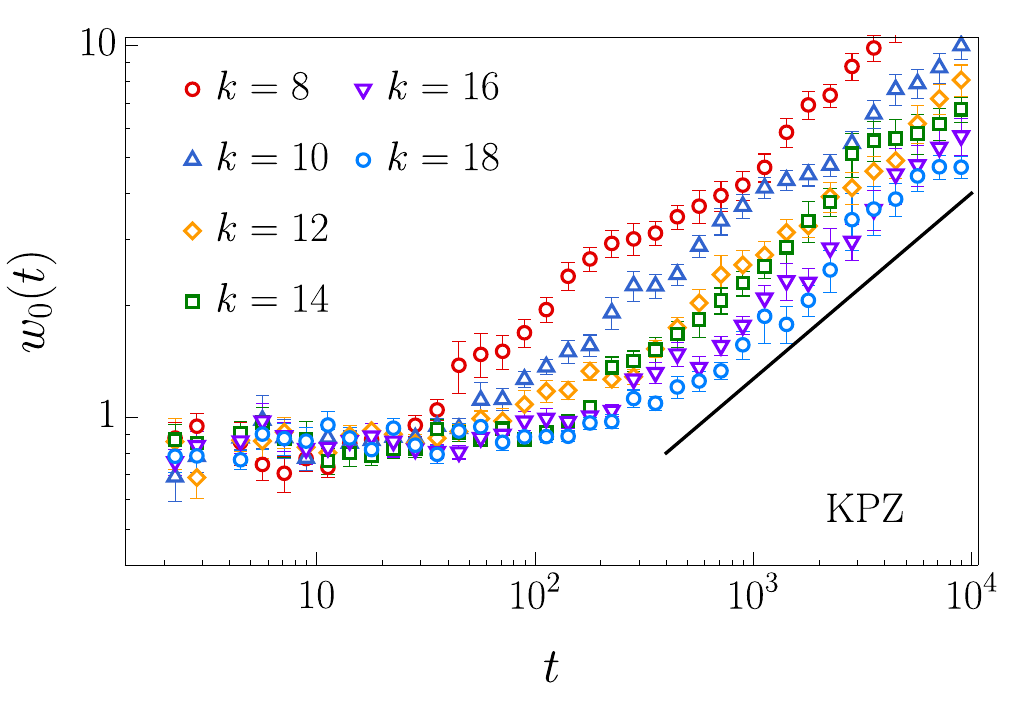}
\caption{Local roughness $w_0(t)$ as a function of time for $q=3$ and several values of $k$ (see legends) for the KPZ equation. As a visual reference, the solid black line corresponds to $w_0\sim t^{0.5}$. The integration method used was LS.}
\label{fig6:w0_integra_KPZ}
\end{figure}

We further investigate the local height fluctuations by analyzing the behavior of the local roughness $ w_0(t) $ [see Eq.~\eqref{eq3:local_roughness}]. Figure~\ref{fig6:w0_integra_EW} shows the evolution in time of this quantity for various networks with increasing size for the EW equation, whereas Fig.~\ref{fig6:w0_integra_KPZ} presents the corresponding results for the KPZ equation. In both cases, the local roughness grows as $ w_0 \sim t^{1/2} $ following an initial transient during which it remains approximately constant. The duration of this transient increases with system size, particularly in the EW case.

In Ref.~\cite{Oliveira2021}, Oliveira performed simulations of the RDSR model (belonging to the EW class), two versions of the RSOS model (KPZ class) that differ in their time-update schemes, and the BD model (also in the KPZ class). He found flat surfaces, $ w_0 \sim \mathrm{const.} $, for the RDSR model and for one of the RSOS variants (after a short transient), while the BD model and the other RSOS variant, referred to by Oliveira as the “commonest” version (RSOSc), exhibited scaling behavior $ w_0 \sim t^{1/2} $. Our results are consistent with those reported by Oliveira in Ref.~\cite{Oliveira2021}, where he explained that saturation was not observed in the BD and RSOSc models because the variance of the average height is zero, as $ \bar{h} $ is deterministic in those models. The underlying argument is that for flat substrates, such as a $ d $-dimensional regular lattices, the presence of spatial translation invariance allows the one-point height fluctuations to be expressed as  
\begin{equation}
    w_0^2 = w^2 + w_{\overline{h}}^2.
    \label{eq6:Oliveira}
\end{equation}
This relation does not hold exactly for non-flat substrates like the CT, but a similar behavior is expected. As pointed out by Oliveira, it is well known that $ w_{\overline{h}} \sim t^{1/2} $ in the stationary regime of one-dimensional KPZ and EW systems \cite{Oliveira2021}. In our case, due to the presence of white noise, $\overline{h}$ is never deterministic in the RD, EW, KPZ, or TKPZ equations. 
Consequently, this stochastic contribution is always present in the measurement of $w_0$. We have verified that, for all the equations and conditions studied, the local roughness follows the scaling $w_0 \sim t^{1/2}$, and that the variance of the average height also grows as $w_{\overline{h}} \sim t^{1/2}$.

\begin{figure}[t!]
\centering
\includegraphics[width=0.7\textwidth]{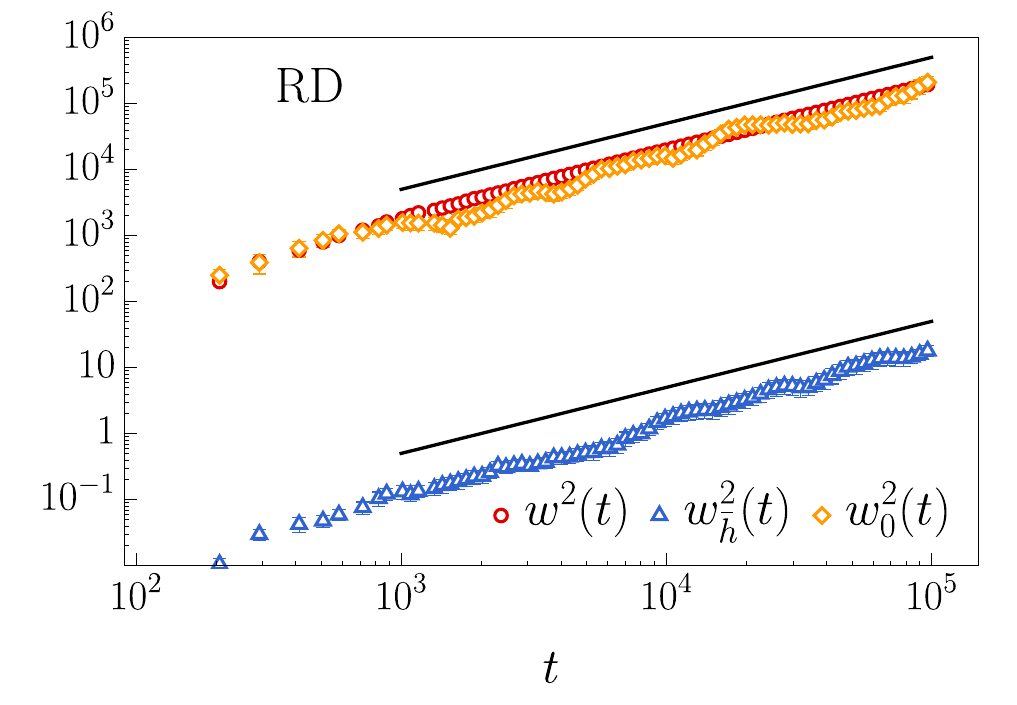}
\caption{Global roughness $w^2(t)$ (red circles), local roughness $w_0(t)$ (yellow diamonds), and the variance of the average height $w_{\bar{h}}(t)$ (blue triangles), for one condition of RD equation. As a visual reference, the solid blacks lines correspond to linear scaling with $t$. In this figure $q=3$ and $k=12$.}
\label{fig6:compara_3w_RD}
\vspace{1cm}
\includegraphics[width=0.7\textwidth]{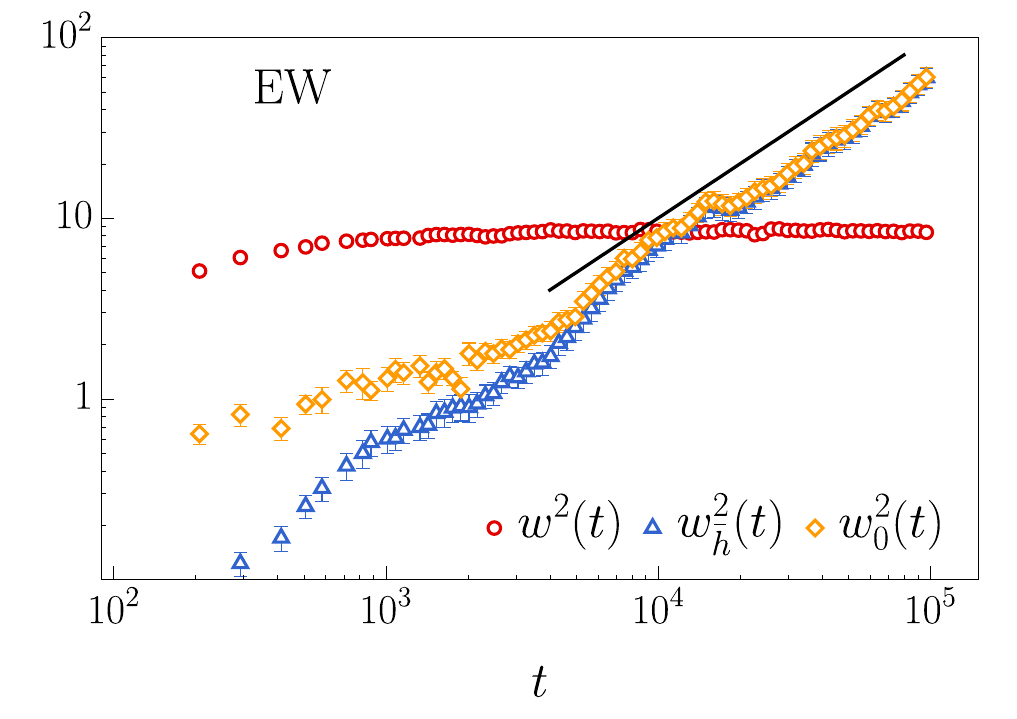}
\caption{Global roughness $w^2(t)$ (red circles), local roughness $w_0(t)$ (yellow diamonds), and the variance of the average height $w_{\bar{h}}(t)$ (blue triangles), for one condition of the EW equation. As a visual reference, the solid black line corresponds to linear scaling with $t$. In this figure $q=3$ and $k=10$.}
\label{fig6:compara_3w_EW}
\end{figure}

\begin{figure}[t!]
\centering
\includegraphics[width=0.7\textwidth]{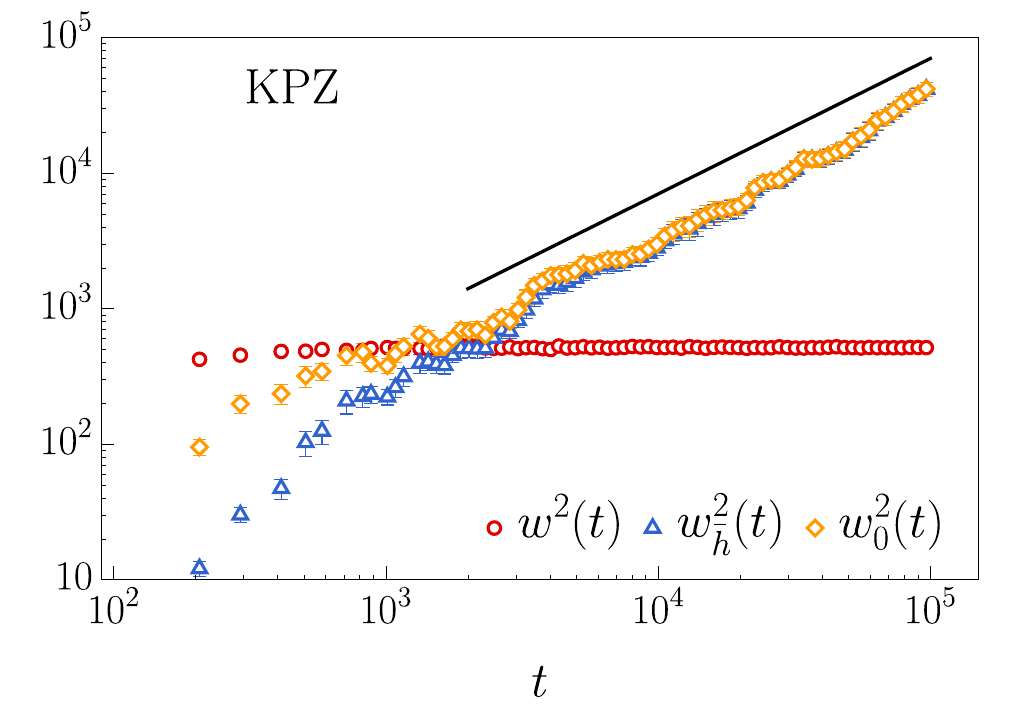}
\caption{Global roughness $w^2(t)$ (red circles), local roughness $w_0(t)$ (yellow diamonds), and the variance of the average height $w_{\bar{h}}(t)$ (blue triangles), for one condition of the KPZ equation. As a visual reference, the solid black line corresponds to linear scaling with $t$. In this figure $q=3$ and $k=10$. The integration method used was CI.}
\label{fig6:compara_3w_KPZ}
\vspace{1cm}
\includegraphics[width=0.7\textwidth]{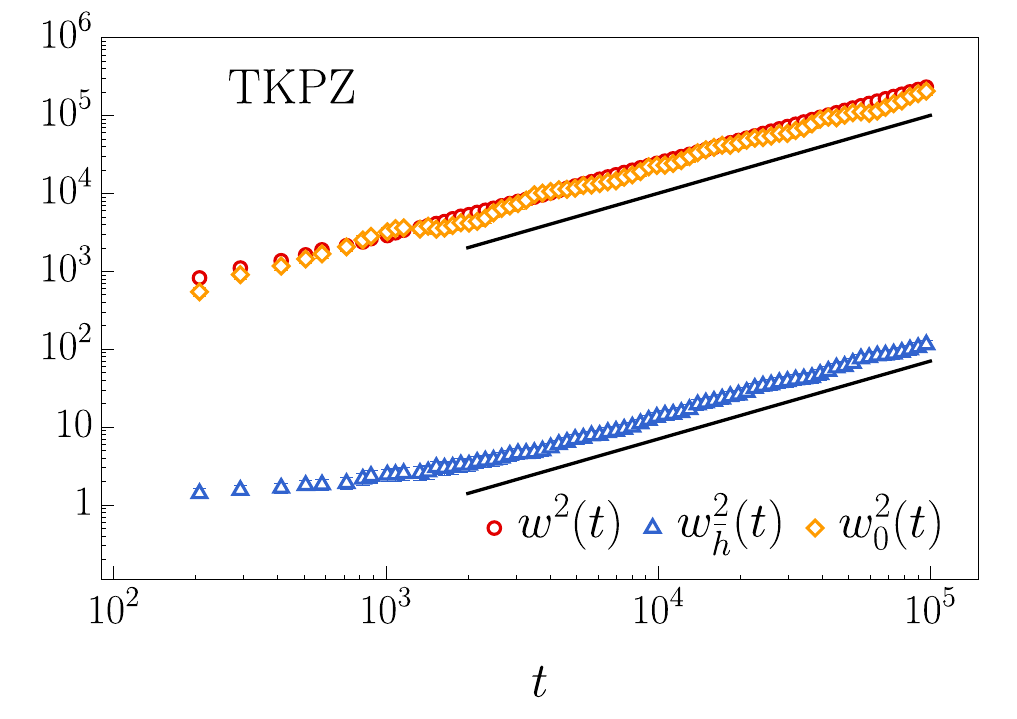}
\caption{Global roughness $w^2(t)$ (red circles), local roughness $w_0(t)$ (yellow diamonds), and the variance of the average height $w_{\bar{h}}(t)$ (blue triangles), for one condition of TKPZ equation. As a visual reference, the solid blacks lines correspond to linear scaling with $t$. In this figure $q=3$ and $k=10$. The integration method used was CI.}
\label{fig6:compara_3w_TKPZ}
\end{figure}

Figures~\ref{fig6:compara_3w_RD} through~\ref{fig6:compara_3w_TKPZ} show the time evolution of $ w^2(t) $, $ w_0^2(t) $, and $ w_{\bar{h}}^2(t) $ from simulations of the RD, EW, KPZ, and TKPZ equations, respectively. As shown in those figures, both $ w_0(t) $ and $ w_{\bar{h}}(t) $ exhibit a consistent $ t^{1/2} $ growth across all simulations, regardless of the integration method employed. However, for both the KPZ and EW equations, $ w \sim \text{const.} $ because the system conditions shown in the figure correspond to the saturated regime of the global roughness. Nonetheless, $ w_0 \approx w_{\bar{h}} \sim t^{1/2} $ continues to hold. In contrast, for the RD and TKPZ equations, $ w \approx w_0 \sim t^{1/2} \sim w_{\bar{h}} $, with the variance of the average height being noticeably smaller than the other two quantities. This indicates that, in these cases, the $ t^{1/2} $ growth of the global roughness is intrinsic, and not merely a consequence of the growth of $ w_{\bar{h}}(t) $.

The results for the KPZ and EW equations are consistent with those reported by Oliveira in Ref.~\cite{Oliveira2021}. In the case of the RD equation, the observed behavior is the one expected~\cite{Barabasi1995} and reflects the typical growth dynamics of this model, which is characterized by the absence of spatial correlations. The behavior of the TKPZ equation is particularly interesting, as it mirrors that of the RD equation. In particular, the $ w^2 \sim t $ growth of the global roughness does not appear to be followed by saturation to a steady state. Remarkably, in one dimension, the correlation structure of the TKPZ equation is analogous, though not identical, to that of the RD model \cite{RodrguezFernndez2022}; see also below.

We do not think that the results observed in the TKPZ case arise from numerical instabilities related to the integration scheme used. However, they may be influenced by boundary effects, as is also the case for the KPZ and EW equations. As previously noted by Oliveira, and in our case, from the perspective of the corresponding stochastic equations, the unusual behavior observed for the EW equation, whose upper critical dimension is known to be 2, suggests that the Bethe lattice, or more precisely its approximation by CTs, may not provide a suitable substrate for studying the mean-field limit of these universality classes.

\clearpage
\subsection{Height-difference correlation function}

We now proceed to examine the behavior of real-space correlation functions, a topic that has received little attention in previous studies of surface growth models on CTs. By analyzing these correlations, we hope to gain additional insight into spatial dependencies and fluctuations that may not be fully captured by global or averaged quantities.

\begin{figure}[t!]
\centering
\includegraphics[width=0.7\textwidth]{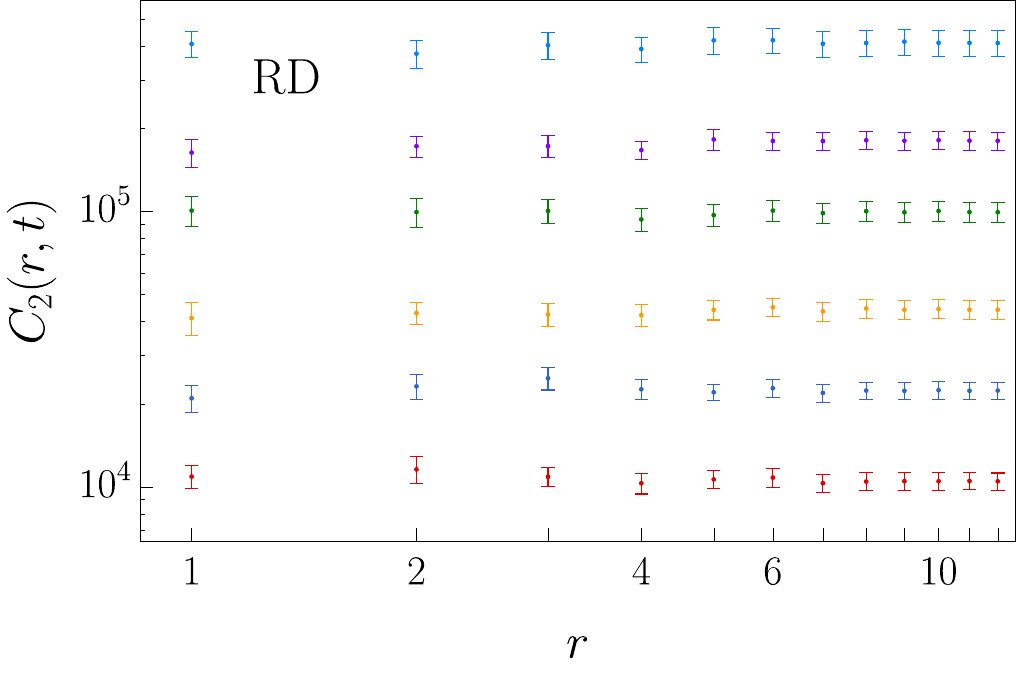}
\caption{Height-difference correlation function $C_2(r,t)$ as a function of $r$ for the following time-boxes: $\{25,40,55,70,85,100\}$, for the RD equation using $q=3$ and $k=12$.}
\label{fig6:C2_vs_r_RD}
\centering
\vspace{1cm}
\includegraphics[width=0.7\textwidth]{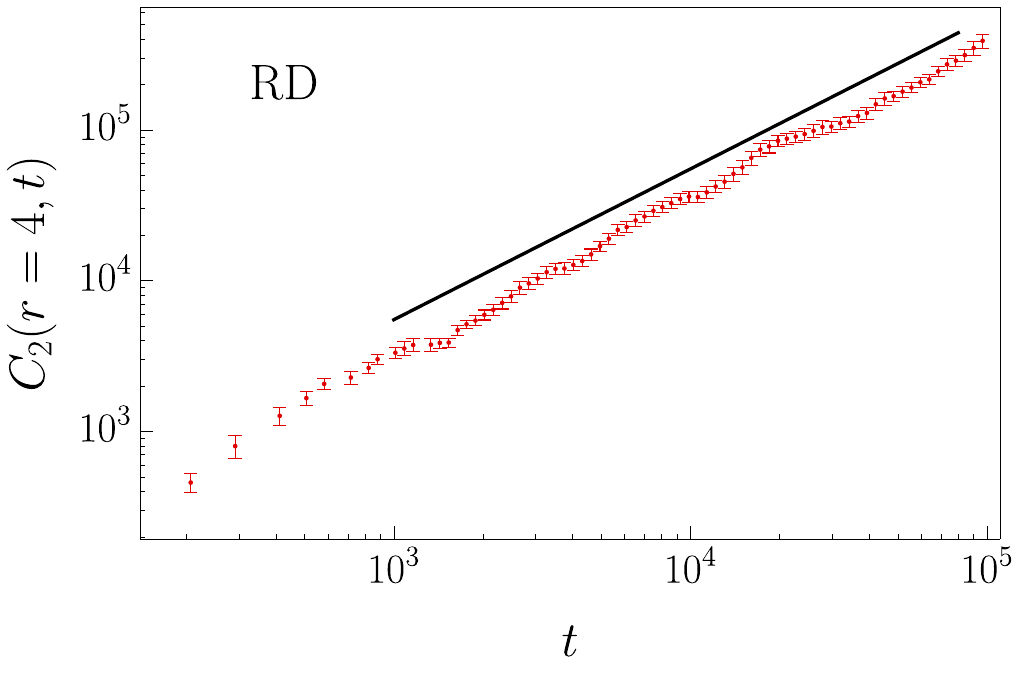}
\caption{Time evolution of the height-difference correlation function $C_2(r,t)$ for fixed $r=4$, for the RD equation using $q=3$, and $k=12$. As a visual reference, the solid black line corresponds to $C_2(4,t) \sim t$.}
\label{fig6:C2_rfijo_RD}
\end{figure}

Figure~\ref{fig6:C2_vs_r_RD} displays the height-difference correlation function, $C_2(r,t)$, plotted against the distance from the center of the tree, $r$, for the RD equation and for several times. In this plot, higher values of $C_2(r,t)$ correspond to later times. Conversely, Fig.~\ref{fig6:C2_rfijo_RD} presents the time evolution of the correlation function at a fixed distance from the center, specifically at \mbox{$r = 4$}. As expected for the RD equation, the correlation function remains constant with respect to $r$ at fixed time $t$, confirming the absence of spatial correlations. Additionally, $C_2(r,t)$ exhibits a linear increase with time, which is a characteristic feature of the RD universality class.

\begin{figure}[t!]
\centering
\includegraphics[width=0.7\textwidth]{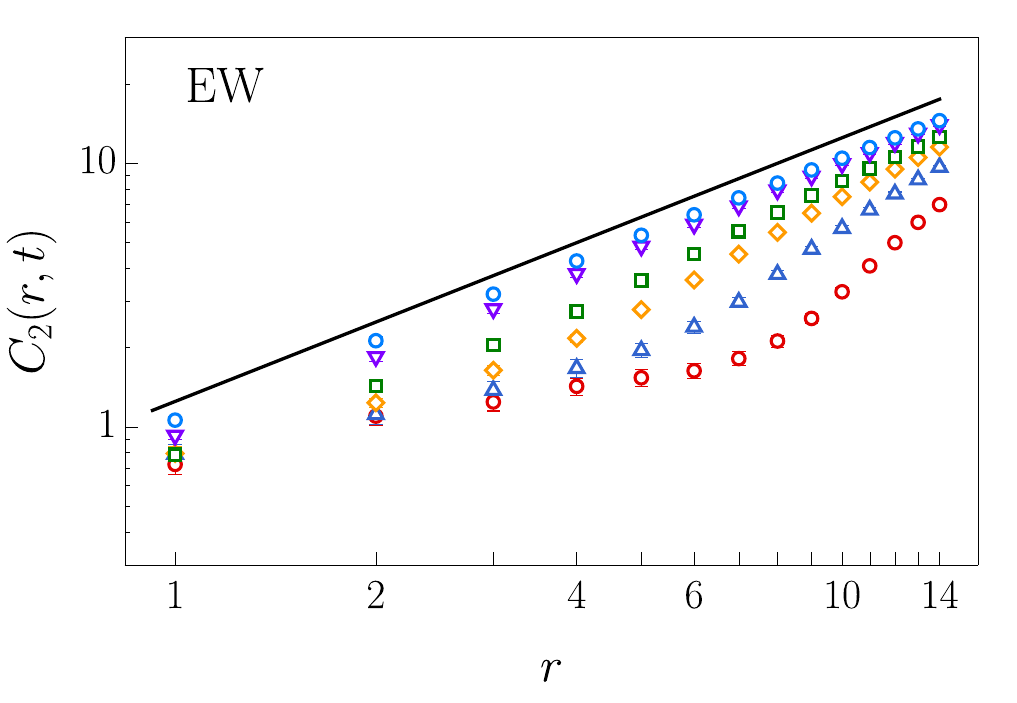}
\caption{Height-difference correlation function $C_2(r,t)$ as a function of $r$ for the following time-boxes: $\{10,30,50,60,80,100\}$, for the EW equation using $q=3$ and $k=14$. As a visual reference, the solid black line corresponds to $C_2(r,t)\sim r$.}
\label{fig6:C2_vs_r_EW}
\centering
\vspace{1cm}
\includegraphics[width=0.7\textwidth]{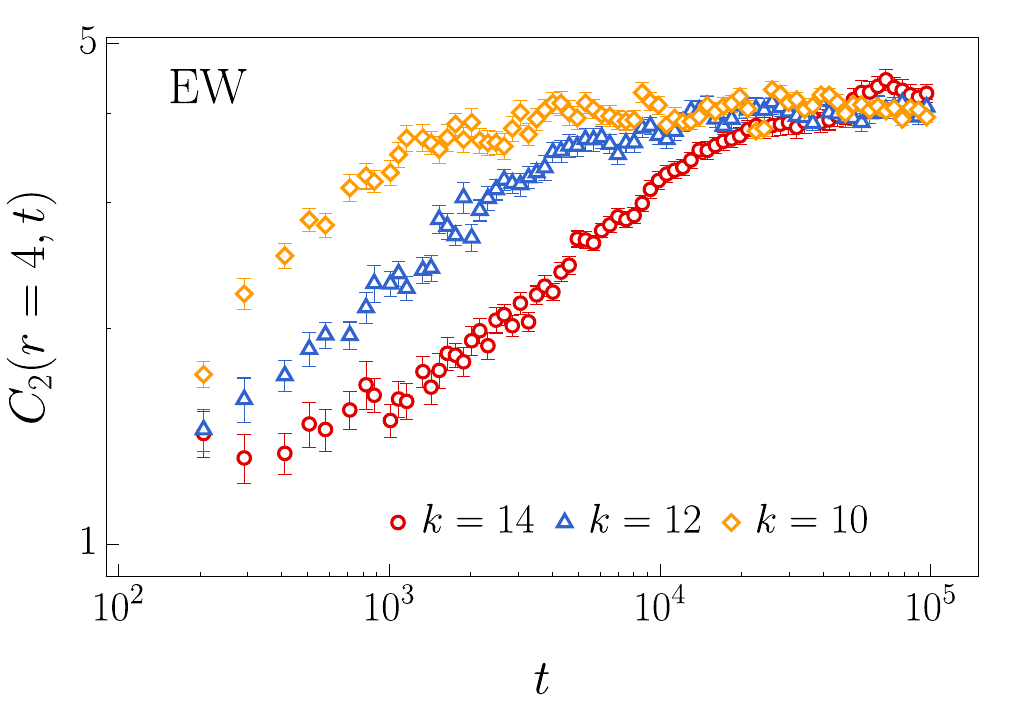}
\caption{Time evolution of the height-difference correlation function $C_2(r,t)$ for fixed $r=4$, for the EW equation using $q=3$, and three system sizes, namely $k=10$, $k=12$ and $k=14$.}
\label{fig6:C2_rfijo_EW}
\end{figure}

Figure~\ref{fig6:C2_vs_r_EW} shows the height-difference correlation function, $C_2(r,t)$, as a function of the distance from the center of the tree, $r$, for the EW equation at various times. On the other hand, Fig.~\ref{fig6:C2_rfijo_EW} depicts how the correlation function evolves over time at a fixed distance from the center, namely at \mbox{$r = 4$}, for various system sizes. In Fig.~\ref{fig6:C2_vs_r_EW}, higher values of $C_2(r,t)$ correspond to later times. In all cases studied for the EW equation, the correlation function saturates at long times to the form $C_2(r,t) \sim r$. Unlike the RD case, the behavior is clearly nontrivial, indicating the presence of spatial correlations.

We recall here Eq.~\eqref{eq3:c2_FV_2}, which indicates that, below the upper critical dimension, the height-difference correlation function $C_2(r,t)$ is expected to scale as 
\begin{equation}  
    \label{eq6:FV_c2}
    C_2(r,t) \sim \left\{ \begin{array}{l}
    r^{2\alpha}\; \mbox{if} \; r\ll\xi(t), \\
    t^{2\beta}\;  \mbox{if} \; r\gg\xi(t). \end{array}
    \right.
\end{equation}
Assuming that the system has reached saturation, i.e., that $r \ll \xi(t)$ at the longest times shown in Fig.~\ref{fig6:C2_vs_r_EW}, leads to an estimated roughness exponent $\alpha = 1/2$, which corresponds to the known value for the EW universality class in $d = 1$. This exponent is consistently observed across all our simulations of the EW equation. It is worth noting, however, that the detailed time evolution of the correlation data does not fully align with the expected FV behavior given by Eq.~\eqref{eq6:FV_c2} at short times. In particular, Fig.~\ref{fig6:C2_rfijo_EW}, which shows the time evolution of $C_2(r,t)$ at a fixed distance, indicates that the function reaches a saturation value that appears largely independent of the system size $k$. Moreover, the time at which the correlation function saturates coincides with the saturation time of the global roughness $w(t)$, which is expected to scale as $k^z$. In contrast, according to Eq.~\eqref{eq6:FV_c2}, $C_2(r,t)$ should saturate at a scale-dependent time proportional to $r^z$. From this perspective, the time-dependent behavior observed in the numerical data for $C_2(r,t)$ resembles the features of anomalous surface kinetic roughening \cite{Krug1988,Lopez1997}.

In our system, the behavior observed in Fig.~\ref{fig6:C2_vs_r_EW} and Fig.~\ref{fig6:C2_rfijo_EW} can be understood as follows. After the front has saturated, and considering that correlations are measured from the center, each branch of the network, from the central node to the outermost layer, effectively behaves like a one-dimensional chain with free BC.

Given that CTs contain no loops, we argue that the branching structure does not significantly affect the behavior of the correlation function. This interpretation will be further supported in the next section, where the fluctuation distribution exhibits the Gaussian profile characteristic of the EW universality class. Similar behavior has been observed in other systems; for example, correlations in the Ising model on a CT are known to follow those of a one-dimensional system \cite{Dorogovtsev2008}, and likewise for bond percolation \cite{Christensen2005}.

\begin{figure}[t!]
\centering
\includegraphics[width=0.7\textwidth]{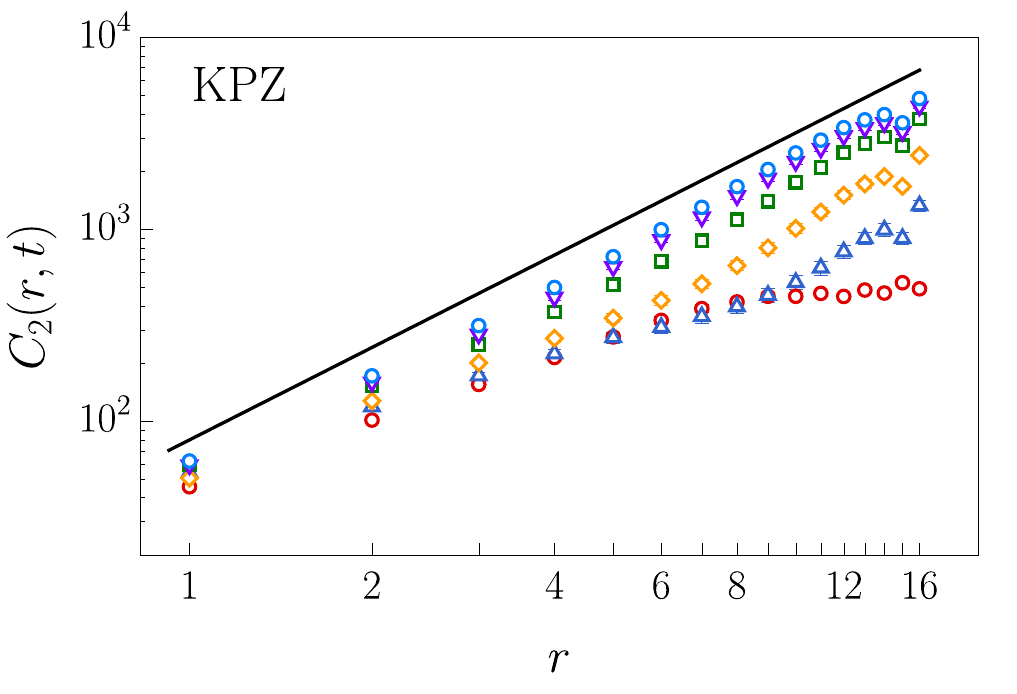}
\caption{Height-difference correlation function $C_2(r,t)$ as a function of $r$ for the following time-boxes: $\{10,20,30,40,50,100\}$, for the KPZ equation using $q=3$ and $k=16$. As a visual reference, the solid black line corresponds to $C_2(r,t)\sim r^{1.6}$. The integration method used was CI.}
\label{fig6:C2_vs_r_KPZ}
\vspace{1cm}
\includegraphics[width=0.7\textwidth]{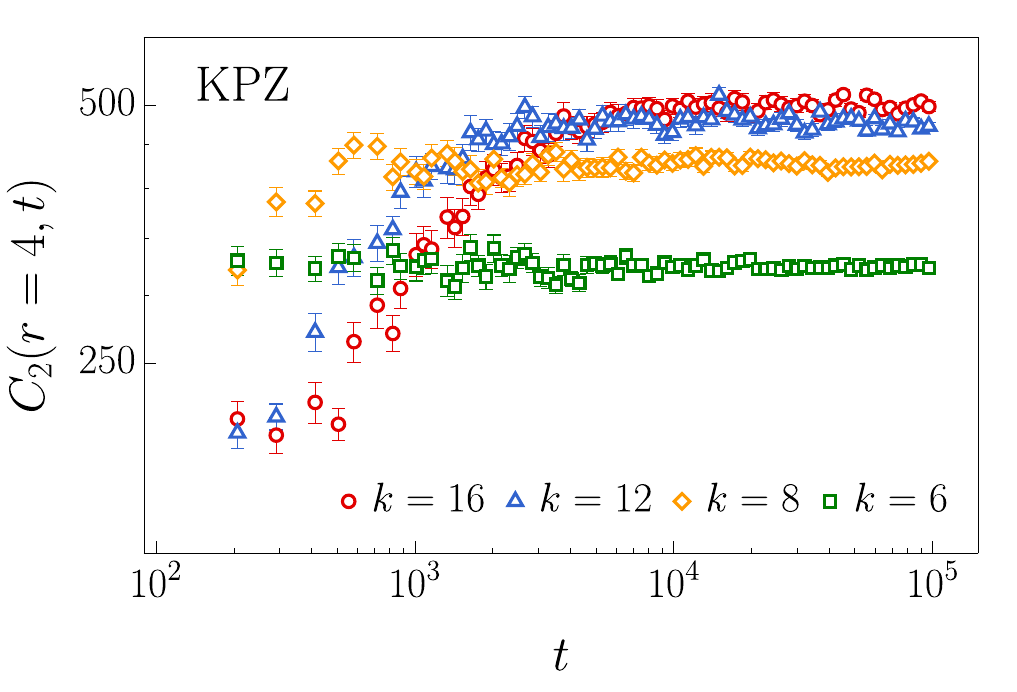}
\caption{Time evolution of the height-difference correlation function $C_2(r,t)$ for fixed $r=4$, for the KPZ equation using $q=3$, and various system sizes, namely $k=6$, $k=8$, $k=12$, and $k=16$. The integration method used was CI.}
\label{fig6:C2_rfijo_KPZ}
\end{figure}

We now turn to the behavior of the height-difference correlation function for the KPZ equation, as shown in Figures~\ref{fig6:C2_vs_r_KPZ} and~\ref{fig6:C2_rfijo_KPZ}. In particular, Fig.~\ref{fig6:C2_vs_r_KPZ} shows $C_2(r,t)$ as a function of the distance from the center, $r$, at different times. As in previous cases, larger values of $C_2(r,t)$ correspond to later times. The behavior $C_2(r,t) \sim r^{1.6}$ at the longest times, shown in Fig.~\ref{fig6:C2_vs_r_KPZ}, suggests a roughness exponent of approximately $\alpha \approx 0.8$, which is reasonably close to the value $\alpha \approx 0.75$ obtained from the data collapse of the global roughness shown in Fig.~\ref{fig6:w_colapso_KPZ} for the KPZ equation. For reference, we recall that the roughness exponent for the 1D KPZ universality class is $\alpha = 1/2$.

Figure~\ref{fig6:C2_rfijo_KPZ} shows the time evolution of the height-difference correlation function at a fixed distance from the center, $r = 4$. The correlation function is seen to rapidly saturate after a brief transient. Additionally, the saturation value at this fixed distance increases with the system size $k$. As in the case of the EW equation, the saturation time of the correlation function matches that of the global roughness $w(t)$.

Although the shape of the correlation function in Fig.~\ref{fig6:C2_vs_r_KPZ} is qualitatively similar to that observed for the EW equation, several notable differences arise. First, the effective roughness exponent varies across the different conditions studied and is consistently greater than one. Consequently, under no parameter set does $C_2(r,t)$ reproduce the behavior expected for one-dimensional KPZ scaling, i.e. $\alpha=1/2$. Furthermore, this effective exponent increases with both the coordination number $q$ and the number of layers $k$. This trend can be attributed to the influence of the nonlinear term in the KPZ equation, which becomes increasingly relevant as $q$ and $k$ grow. In such cases, the contribution of external branches to the local dynamics becomes more pronounced, driving the system further away from the EW-like behavior. A second key difference is the appearance of a discontinuity in the penultimate layer, where the value of the correlation function is systematically lower than expected. This behavior is examined in greater detail in Sec.~\ref{sec6:layers}, where we analyze the growth dynamics of individual layers across the different stochastic equations.

\begin{figure}[t!]
\centering
\includegraphics[width=0.7\textwidth]{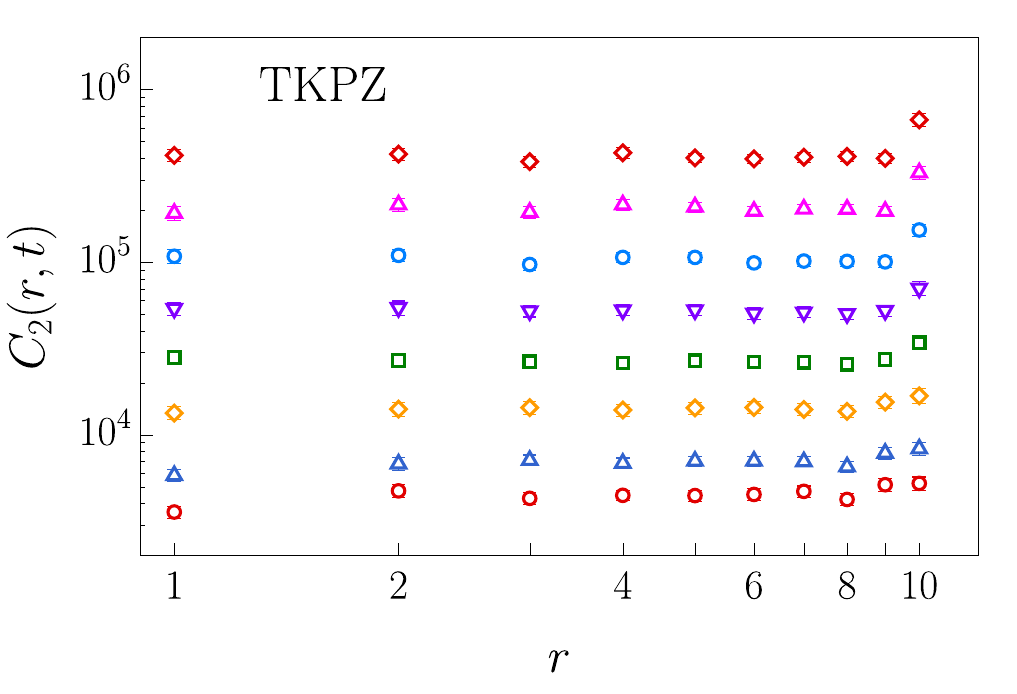}
\caption{Height-difference correlation function $C_2(r,t)$ as a function of $r$ for the following time-boxes: $\{30,40,50,60,70,80,90,100\}$, for the TKPZ equation using $q=3$ and $k=10$. The integration method used was CI.}
\label{fig6:C2_vs_r_TKPZ}
\vspace{1cm}
\includegraphics[width=0.7\textwidth]{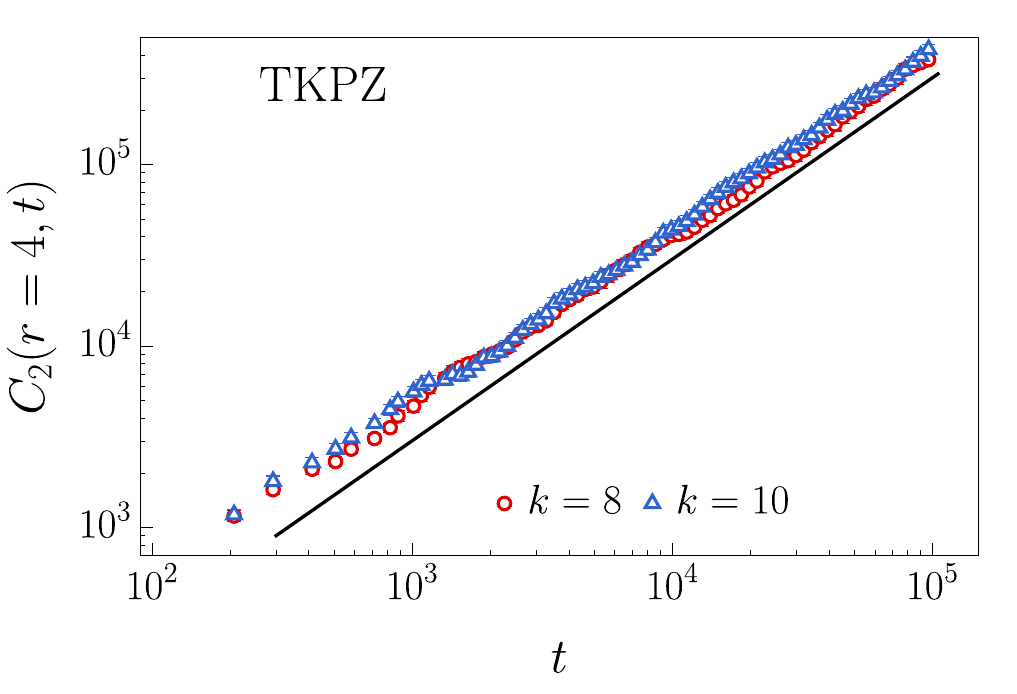}
\caption{Time evolution of the height-difference correlation function $C_2(r,t)$ for fixed $r=4$, for the TKPZ equation using $q=3$, and two system sizes, namely $k=8$ and $k=10$. As a visual reference, the solid black line corresponds to $C_2(4,t) \sim t$. The integration method used was CI.}
\label{fig6:C2_rfijo_TKPZ}
\end{figure}

Finally, Figures~\ref{fig6:C2_vs_r_TKPZ} and~\ref{fig6:C2_rfijo_TKPZ} present the results for the height-difference correlation function in the case of the TKPZ equation. Figure~\ref{fig6:C2_vs_r_TKPZ} shows $C_2(r,t)$ as a function of the distance from the center, $r$, at different times, with larger values of $C_2(r,t)$ corresponding, once again, to later times. On the other hand, Figure~\ref{fig6:C2_rfijo_TKPZ} illustrates the time evolution of the height-difference correlation function at a fixed distance, $r = 4$, from the center.

The behavior observed in these figures closely resembles that of the RD equation. For instance, in Fig.~\ref{fig6:C2_rfijo_TKPZ}, the correlation function increases continuously over time without reaching saturation. In fact, it grows at the same rate as the squared global roughness for the TKPZ equation, shown in Fig.~\ref{fig6:compara_3w_TKPZ}, following the relation $C_2(r,t) \sim t \sim w^2(t)$, i.e. the same scaling observed in the case of RD (see Fig.~\ref{fig6:C2_rfijo_RD}). Moreover, the $r$-independent (uncorrelated) profile of $C_2(r,t)$ observed in Fig.~\ref{fig6:C2_vs_r_TKPZ} further supports this interpretation in terms of the RD model, which likewise exhibits no spatial correlations and does not reach saturation.

As previously discussed in Sec.~\ref{sec1:TKPZ}, the height-difference correlation function for the TKPZ equation in one dimension exhibits nontrivial scaling with $r$, characterized by a local roughness exponent $\alpha_{\rm loc} = 1/2$ \cite{RodrguezFernndez2022}. Interestingly, for the 
derivative of the 1D TKPZ equation, namely the stochastic IB equation, $C_2(r,t)$ displays behavior similar to that shown in Fig.~\ref{fig6:C2_vs_r_TKPZ}, with the notable distinction that, in this case, the system does reach saturation in the steady state \cite{RodrguezFernndez2022}.

However, there is a important deviation compared to the results for the RD case, namely the appearance of a discontinuity in the last layer of the $C_2(r,t)$ function, as shown in Fig.~\ref{fig6:C2_vs_r_TKPZ}. The origin of this discontinuity will be examined in more detail in Sec.\ref{sec6:layers}.

\clearpage
\subsubsection{Effect of boundary conditions}

\begin{figure}[t!]
\centering
\includegraphics[width=0.7\textwidth]{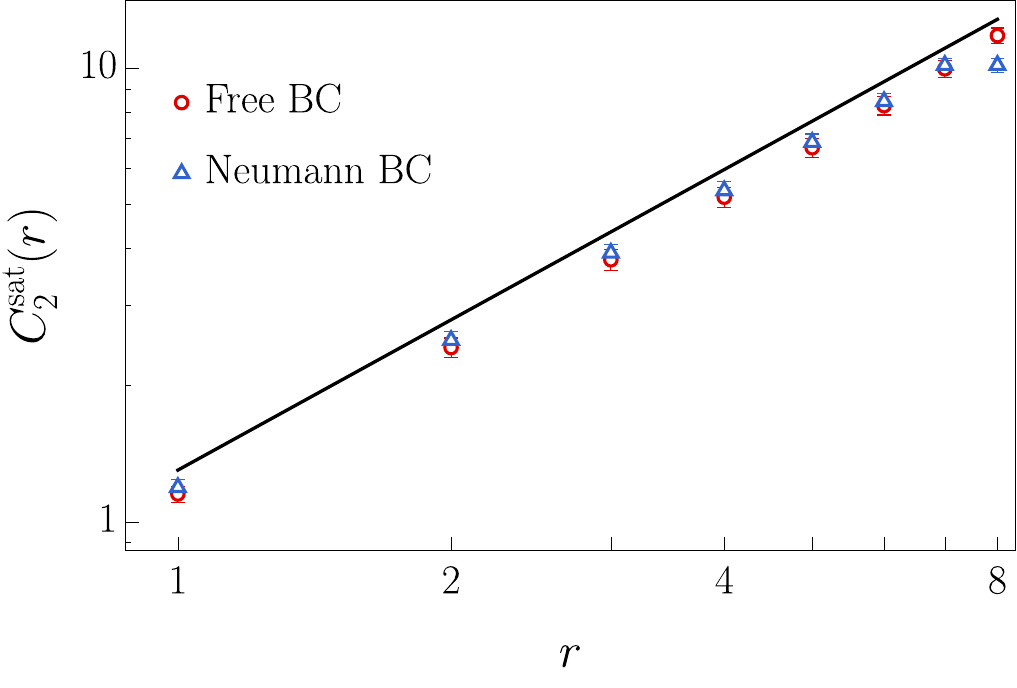}
\caption{
Comparison of saturation value of the height difference correlation function, $C^{\rm sat}_2(r)$, as a function of $r$ for two different BC, as in the legend. As a visual reference, the solid black line corresponds to $C_2^{\rm sat}(r)\sim r^{1.1}$. In this figure $q=3$, $k=8$, $\nu=1$, $\lambda=0.5$, and $D=1$. The integration method used was CI.}
\label{fig6:C2_vs_r_ComparaBC}
\vspace{1cm}
\includegraphics[width=0.7\textwidth]{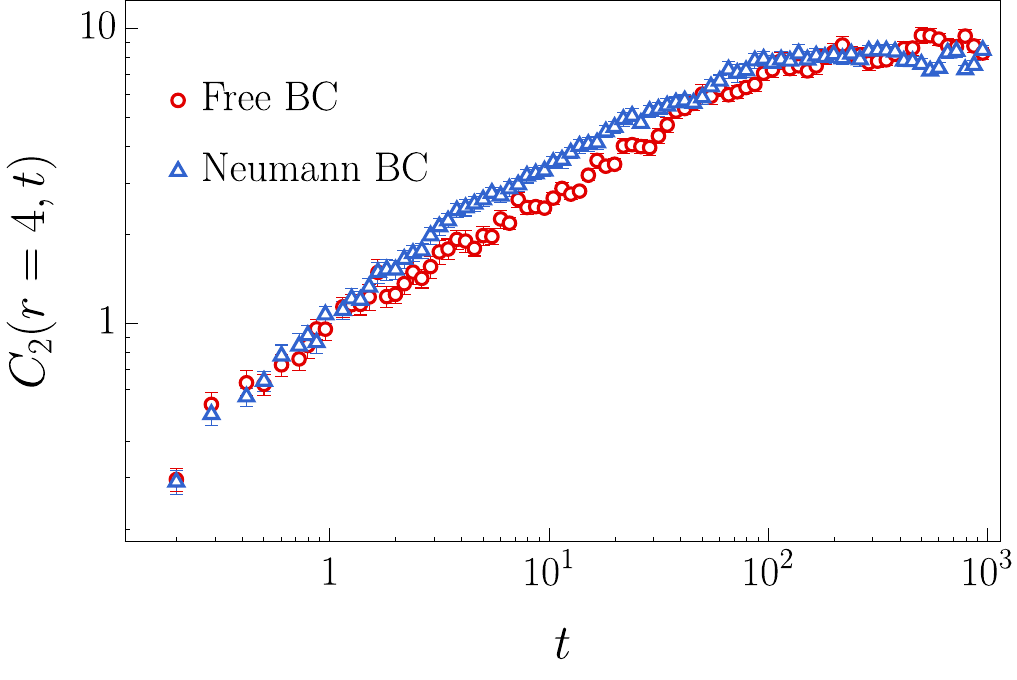}
\caption{Correlation function vs BC for the KPZ equation. Time evolution of the height difference correlation function $C_2(r,t)$ for fixed $r=4$ and the two different BC, as in the legend. In this figure $q=3$, $k=8$, $\nu=1$, $\lambda=0.5$, and $D=1$. The integration method used was CI.}
\label{fig6:C2_rfijo_ComparaBC}
\end{figure}

Having discussed the general behavior of the height-difference correlation function, we now examine how it is influenced by changes in the BC used during the integration of the equations on the tree. This effect is illustrated in Figures~\ref{fig6:C2_vs_r_ComparaBC} and~\ref{fig6:C2_rfijo_ComparaBC}. The study will focus on the KPZ equation.

Figure~\ref{fig6:C2_vs_r_ComparaBC} shows the saturation value of the height-difference correlation function, $C^{\rm sat}(r)$, as a function of the distance from the center $r$. In contrast, Fig.~\ref{fig6:C2_rfijo_ComparaBC} illustrates the time evolution of $C_2(r,t)$ at a fixed distance, $r = 4$, from the center.

In the latter figure, the correlation function is seen to grow in a nearly identical manner for both BC, ultimately reaching the same saturation value. Furthermore, as shown in Fig.~\ref{fig6:C2_vs_r_ComparaBC}, the steady-state profile of 
$C_2$ as a function of the distance from the center is identical for both cases, except at the outermost layer. While the Neumann BC constrains the height values at this layer to match those of the penultimate one, the Free BC allows them to evolve independently, leading to differences in $C^{\rm sat}(r)$ only at $r = k$.

Based on the results presented in this section, together with those discussed in Sec.~\ref{sec6:comparaBC1}, we believe that the influence of the BC on the system is well understood. Our analysis indicates that the main features and conclusions of the study remain largely unaffected by the specific choice of BC.

\clearpage
\subsection{Statistics of height fluctuations}

The results presented in the previous section for the height-difference correlation function are complemented and more clearly interpreted when examined in conjunction with the analysis of height fluctuations. This analysis will provide a more complete understanding of the kinetic roughening universality classes associated of the RD, EW, KPZ, and TKPZ equations on CTs.

\begin{figure}[t!]
\centering
\includegraphics[width=0.7\textwidth]{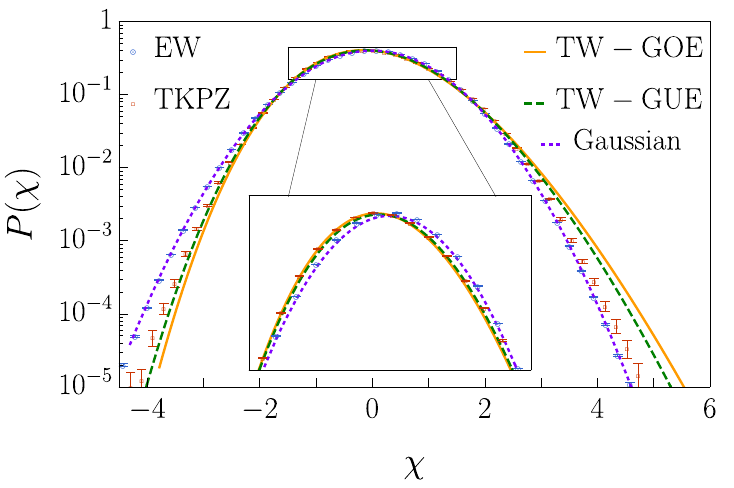}
\caption{Fluctuation histograms of the rescaled height fluctuations, $\chi$ [see Eq.~\eqref{eq3:flu2}] for two conditions of the EW and the TKPZ equations, for $q=3$ and $k=10$. The inset shows a zoom of the boxed area for the central part of the distributions in the $-1.5 < \chi < 1.5$ interval. The solid orange and green dashed lines correspond to the TW-GOE and TW-GUE distributions, respectively, while the dotted purple line corresponds to a Gaussian distribution. The integration method used for the TKPZ equation was CI.}
\label{fig6:chi_EW_TKPZ}
\includegraphics[width=0.7\textwidth]{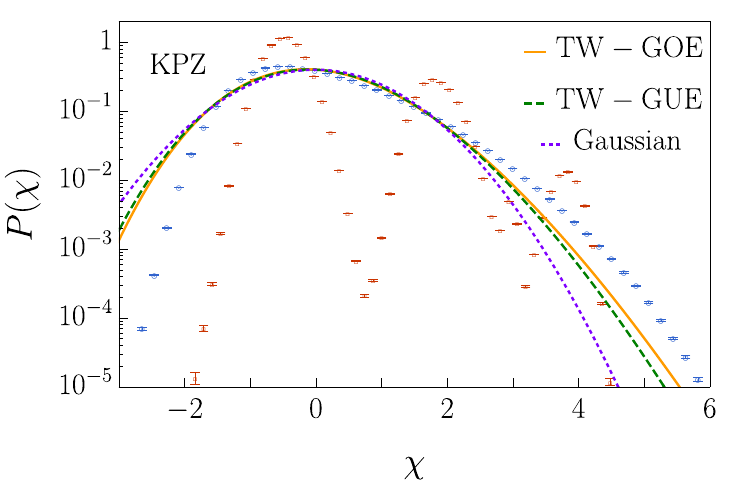}
\caption{Fluctuation histograms of the rescaled height fluctuations, $\chi$ [see Eq.~\eqref{eq3:flu2}] for two conditions of the KPZ equation, namely $k=10$ for $q=3$ (blue circles) and $k=4$ for $q=6$ (red squares). The solid orange and green dashed lines correspond tto the TW-GOE and TW-GUE distributions, respectively, while the dotted purple line corresponds to a Gaussian distribution. The integration method used was CI.}
\label{fig6:chi_KPZ}
\end{figure}

\begin{figure}[t!]
\centering
\includegraphics[width=0.7\textwidth]{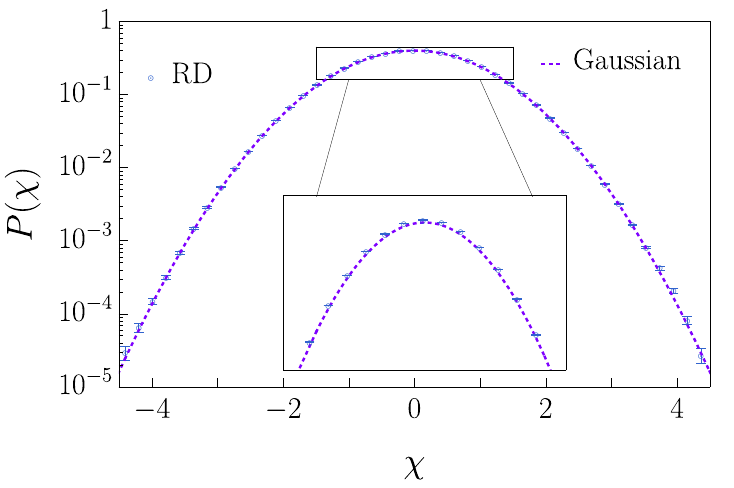}
\caption{Fluctuation histograms of the rescaled height fluctuations, $\chi$ [see Eq.~\eqref{eq3:flu2}] for one condition of the RD equation, namely $k=12$ and $q=3$. The dotted purple line corresponds to a Gaussian distribution.}
\label{fig6:chi_RD}
\end{figure}

Figures~\ref{fig6:chi_EW_TKPZ} to~\ref{fig6:chi_RD} present histograms of the rescaled height fluctuations $\chi$, defined in Eq.~\eqref{eq3:flu2}, for the four equations studied in this chapter. Specifically, Fig.~\ref{fig6:chi_EW_TKPZ} presents the results for two cases of the EW and TKPZ equations. Figure~\ref{fig6:chi_KPZ} displays the results for two cases of the KPZ equation, corresponding to $q = 3$, $k = 10$ and $q = 6$, $k = 4$. Finally, Fig.~\ref{fig6:chi_RD} shows the results for a single case of the RD equation. All figures include the Gaussian distribution for comparison. Additionally, Fig.~\ref{fig6:chi_EW_TKPZ} and~\ref{fig6:chi_KPZ} also displays the TW-GOE and TW-GUE distributions.

We recall that the fluctuation PDF for the EW class is known to be Gaussian for any dimension $d$ below the upper critical dimension $d_u$ \cite{Barabasi1995,Krug1988,Prolhac2011}. The same holds for the RD class, for which the PDF is Gaussian in all dimensions \cite{Barabasi1995}. In contrast, for the one-dimensional KPZ equation, the fluctuation PDF follows the TW-GOE or TW-GUE distributions, depending on the geometry of the interface, as discussed in Sec.~\ref{sec1:KPZ}. For the 1D TKPZ equation with PBC, the fluctuation PDF is non-Gaussian, but it is also known not to follow the TW-GOE distribution, see Sec.~\ref{sec1:TKPZ} and Ref.~\cite{RodrguezFernndez2022} for more details.

Figures~\ref{fig6:chi_EW_TKPZ} and~\ref{fig6:chi_RD} clearly confirm that the fluctuation PDFs are Gaussian for our simulations of the EW and RD equations on the CT, respectively. This behavior remains consistent across all tree configurations studied. On the contrary, the PDFs of height fluctuations for the TKPZ equation, shown in Fig.~\ref{fig6:chi_EW_TKPZ}, are clearly non-Gaussian. Although no specific TW behavior is expected for the TKPZ equation \cite{RodrguezFernndez2022}, the tails of the distribution are, to some extent, not far from TW, particularly the TW-GUE distribution. In any case, the present comparison highlights the non-Gaussian nature of the PDFs obtained on CTs. Interestingly, this resemblance to TW behavior diminishes as the coordination number $q$ increases, although the asymmetry in the distribution persists.

Hence, the TKPZ equation differs from the RD equation in this regard, despite the fact that the analysis of the roughness and the height-difference correlation function yielded similar results for both equations. It is worth noting that, in the context of the KPZ equation, the non-zero skewness (asymmetry) of the fluctuation PDF is a hallmark of the nonlinear term \cite{Takeuchi2018}. In this light, the PDF behavior observed for the TKPZ equation may serve as strong evidence that the trends identified in the global roughness $w(t)$ and the height-difference correlation function are genuinely nontrivial, even though they are significantly influenced by boundary effects. This interpretation will be further supported by the analysis of the growth of layers presented in the next section.

Figure~\ref{fig6:chi_KPZ} presents results for two different parameter sets of the KPZ equation. In both cases, the fluctuation PDFs do not match either the Gaussian or the TW distributions. Moreover, the behavior varies significantly between conditions. For instance, in one of the cases shown in Fig.~\ref{fig6:chi_KPZ}, the distribution exhibits noticeable oscillations. As will be shown in the next section, these oscillations can be understood by examining the growth dynamics of each layer for those conditions.

To complement the analysis, we have also computed the skewness and excess kurtosis of the fluctuation PDFs shown in these figures. These statistical moments are known analytically for the Gaussian, TW-GOE, and TW-GUE distributions\footnote{For reference, the precise values of skewness and excess kurtosis are \mbox{$S = 0.29346452408$} and $K = 0.1652429384$ for TW-GOE, and $S = 0.224084203610$ and $K = 0.0934480876$ for TW-GUE \cite{Bornemann2010}.}. Specifically, for the cases presented in Fig.~\ref{fig6:chi_EW_TKPZ}, we find $S = 0.00(6)$ and $K = 0.08(1)$ for the EW equation, consistent with Gaussian statistics. For the TKPZ case, we obtain $S = 0.172(4)$ and $K = 0.055(7)$. Finally, for the KPZ equation results shown in Fig.~\ref{fig6:chi_KPZ}, the values are $S = 0.74(1)$ and $K = 0.52(2)$ for the case with $q = 3$, $k = 10$, and $S = 1.471(1)$ and $K = 1.022(1)$ for $q = 6$, $k = 4$, respectively, both showing clear non-Gaussian features and increasing asymmetry. 

\subsection{Analysis of the growth of layers}\label{sec6:layers}

In this final section, we examine the growth dynamics of each layer within the tree. Specifically, we present results for two observables: the difference between the average height at the center and at the system boundary, $\Delta \langle h \rangle$ [see Eq.~\eqref{eq3:delta_h}], and the relative growth of each layer compared to the global average front height, $A(s,t)$ [see Eq.~\eqref{eq3:capas}]. As previously mentioned, analyzing these quantities will help us understand the origin of the jumps observed in the correlation function, as well as the oscillations that occasionally appear in the fluctuation distribution.

Let us begin with the second observable. Before proceeding, it is important to note that, in the following figures showing the evolution of $A(s,t)$, the error bars have been estimated differently from those in the previous plots. Instead of applying the jackknife method discussed in Sec.~\ref{sec3:JK}, we chose to define the error as the average (over time and runs) of the standard deviations within each time-box. This approach proves useful for explaining specific features of our system, particularly the oscillations observed in the PDF of the KPZ case, as whether or not the layers overlap plays a key role in understanding this behavior.

\begin{figure}[t!]
\centering
\includegraphics[width=0.7\textwidth]{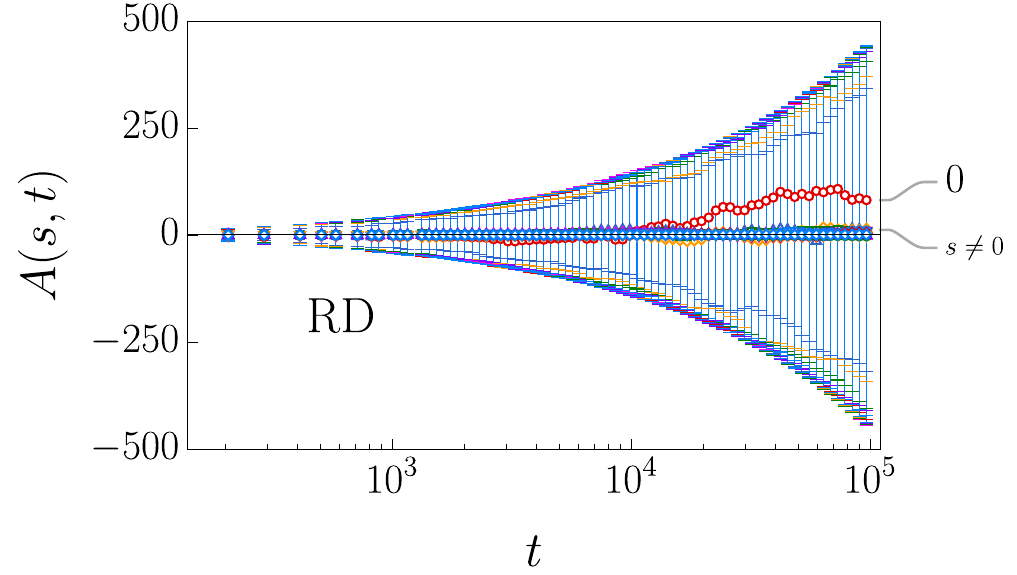}
\caption{Evolution in time of the average of each layer relative to the average front position of the system, $A(s,t)$, for one condition of the RD equation. The labels on the right margin of the figure identify the central node ($s=0$) and all other layers ($s\ne 0$). In this figure, $q=3$ and $k=12$.}
\label{fig6:capas_RD}
\end{figure}

\begin{figure}[t!]
\centering
\includegraphics[width=0.7\textwidth]{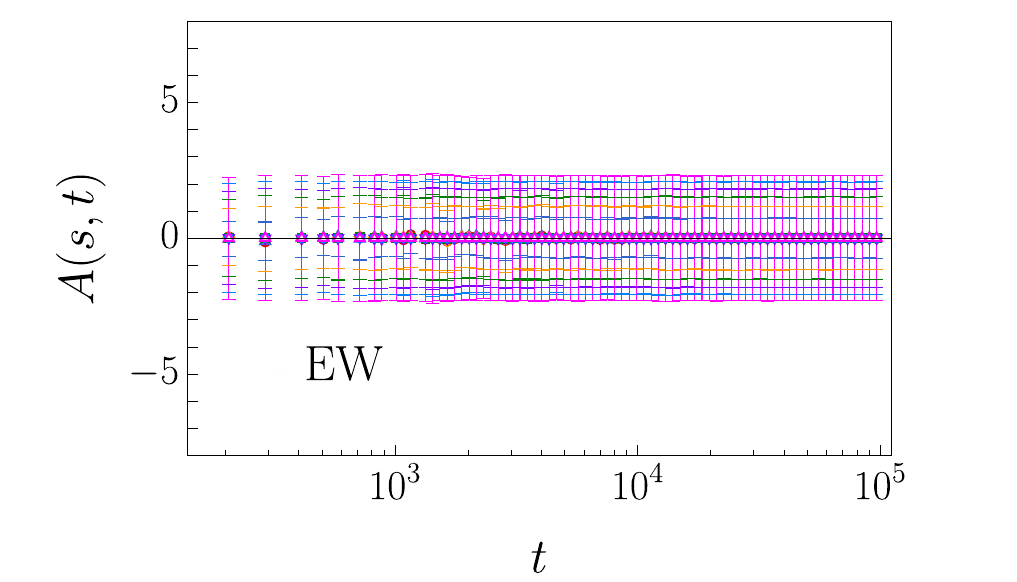}
\caption{Evolution in time of the average of each layer relative to the average front position of the system, $A(s,t)$, for one condition of the EW equation. In this figure, $q=3$ and $k=8$.}
\label{fig6:capas_EW}
\vspace{0.5cm}
\includegraphics[width=0.7\textwidth]{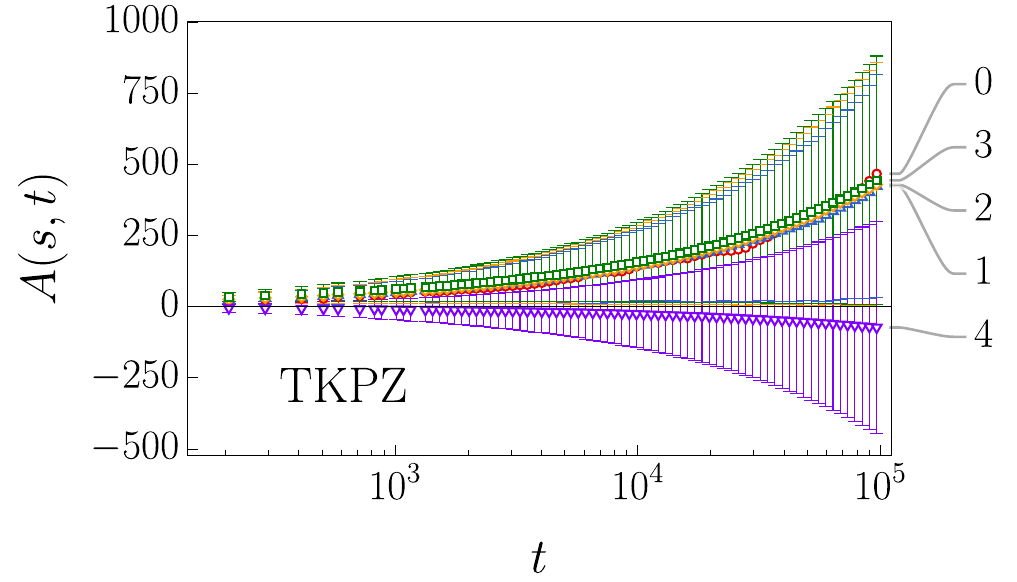}
\caption{Evolution in time of the average of each layer relative to the average front position of the system, $A(s,t)$, for one condition of the TKPZ equation. In this figure $q=6$ and $k=4$. The labels on the right margin of the figure identify each layer $s$ in each case. The integration method used was CI.}
\label{fig6:capas_TKPZ}
\end{figure}

Figure~\ref{fig6:capas_RD} shows the time evolution of $A(s,t)$ for a condition of the RD equation, while Fig.~\ref{fig6:capas_EW} presents the corresponding results for the EW equation. In both cases, the layers remain clustered around zero, with the notable exception of the shell $s = 0$, i.e., the central node, in the RD case. Since in this model each node evolves independently as a Brownian motion, it is expected that individual cells, such as the central one, may occasionally deviate significantly from the mean. A key distinction is that, due to the lack of relaxation in the RD equation, the error bars increase continuously over time. Recall that these error bars represent the standard deviations of the layer values at each time. This effect is what causes the roughness to grow in time as $w(t) \sim t^{1/2}$ in the RD model, while in the EW case it quickly saturates. This behavior observed in the RD and EW equations remains consistent across all the trees analyzed. 

\begin{figure}[t!]
\centering
\includegraphics[width=0.7\textwidth]{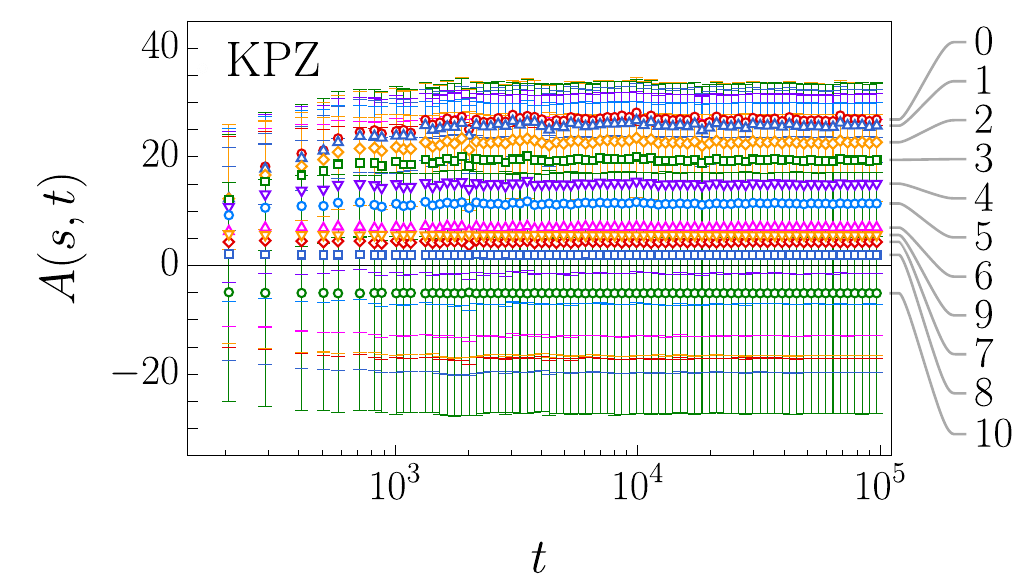}
\caption{Evolution in time of the average of each layer relative to the average front position of the system, $A(s,t)$, for one condition of the KPZ equation. In this figure $q=3$ and $k=10$. The labels on the right margin of the figure identify each layer $s$ in each case. The integration method used was CI.}
\label{fig6:capas_KPZ1}
\includegraphics[width=0.7\textwidth]{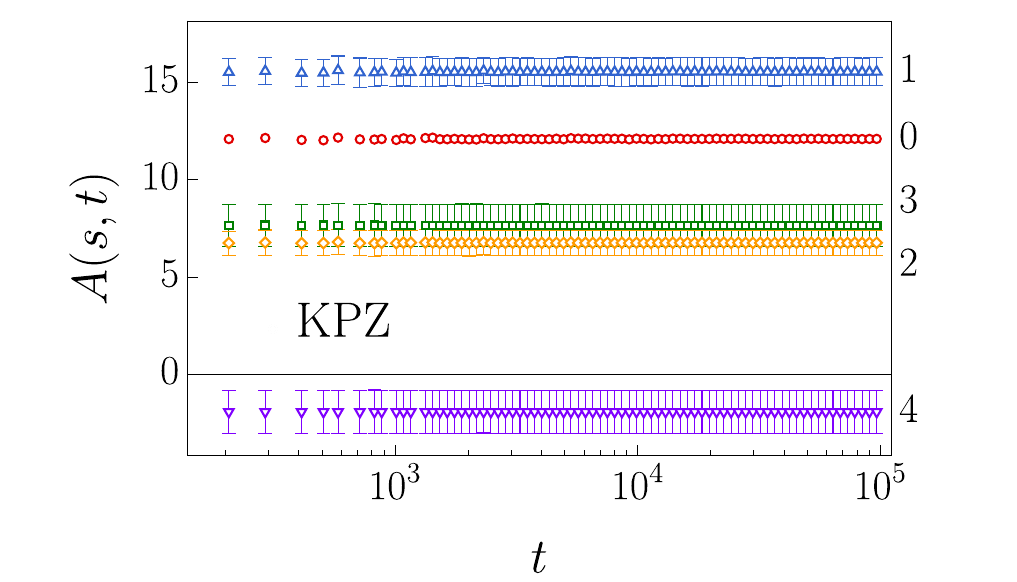}
\caption{Evolution in time of the average of each layer relative to the average front position of the system, $A(s,t)$, for one condition of the KPZ equation. In this figure $q=6$ and $k=4$. The labels on the right margin of the figure identify each layer $s$ in each case. The integration method used was CI.}
\label{fig6:capas_KPZ2}
\end{figure}

Figure~\ref{fig6:capas_TKPZ} displays the time evolution of $A(s,t)$ for a condition of the TKPZ equation. Figures~\ref{fig6:capas_KPZ1} and~\ref{fig6:capas_KPZ2} show the corresponding results for two different conditions of the KPZ equation. Unlike the behavior exhibited by the EW and RD equations, both the KPZ and TKPZ equations display a markedly different evolution for each layer. In both cases, the outermost layer of the system lies below the average front height, i.e. $A(s,t)<0$. This trend is consistently observed across all studied conditions as well. The underlying reason is that nodes in the outermost layer, having only one neighbor, experience significantly less growth, driven by the squared gradient term, than the nodes in the inner layers.

This effect propagates inward, layer by layer, ultimately reaching the central node of the network, which typically exhibits the fastest growth relative to the average front height. Consequently, the closer a layer is to the center, the faster it tends to grow. However, this trend does not generally apply to the penultimate layer, which is often taller than the one immediately adjacent to the center. We attribute this to the influence of the outermost layer that, being significantly lower than the rest, effectively enhances the growth of the penultimate layer. As a result, when correlations are measured from the center, the penultimate layer in the KPZ case appears more correlated than the layers located closer to the center.

A similar effect is observed in simulations of the TKPZ equation. However, in this case, the differences between layers do not saturate over time, due to the absence of smoothening mechanisms stemming from the lack of surface tension. As a consequence, the inner layers tend to grow closely together, while the outermost layer increasingly deviates from the rest. This behavior explains the jump observed in the last layer when computing the correlation function for the TKPZ equation. In addition, we attribute the observed growth of the roughness in the TKPZ case, $w \sim t^{1/2}$, to the fact that the standard deviation of the layer values (i.e., the displayed error bars) does not saturate over time, in a manner analogous to what is observed in the RD case. In this sense, we consider that the roughness growth in the TKPZ case is dominated by noise, and that the nonlinear term does not play a significant role in its behavior. However, a comparison between Figs.~\ref{fig6:capas_RD} and~\ref{fig6:capas_TKPZ} clearly shows that the nonlinear term has a non-negligible effect on the system.


Moreover, by comparing Figures~\ref{fig6:capas_KPZ1} and~\ref{fig6:capas_KPZ2}, one can understand why oscillations appear in the PDFs under certain KPZ conditions, but not under others. In some cases, particularly when the system has few layers, as in Fig.~\ref{fig6:capas_KPZ2}, there are noticeable gaps (absence of layer overlap) in $A(s,t)$. These gaps correspond to regions where fluctuations are highly unlikely, leading to low-density areas in the distribution. This is clearly reflected in the case shown with red squares in Fig.~\ref{fig6:chi_KPZ}, which corresponds to the same conditions as the aforementioned figure. However, when the system contains many layers, these tend to overlap more significantly. As a result, the distribution of fluctuations exhibits a more continuous decay, without clear gaps, as seen for the conditions marked with blue circles in Fig.~\ref{fig6:chi_KPZ}. In such cases, the distribution deviates from known universal forms and displays non-standard behavior.

\begin{figure}[t!]
\centering
\includegraphics[width=0.7\textwidth]{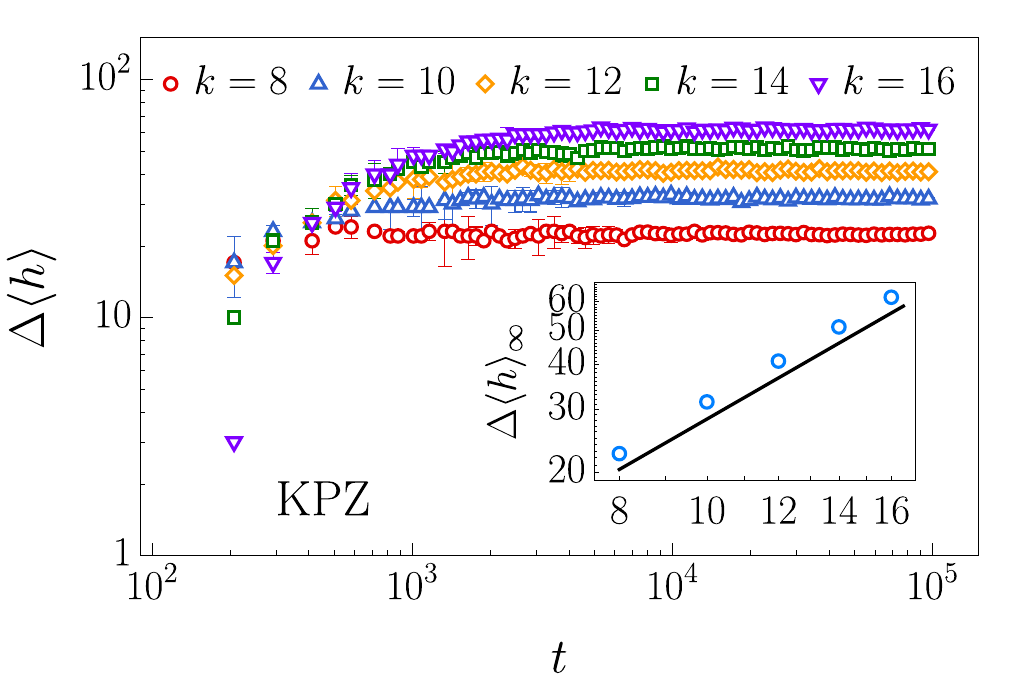}
\caption{Difference between the value of the central node and the average of the last shell $k$, $\Delta\langle h \rangle$, for various system sizes $k$ for the KPZ equation, see legend. In this figure $q=3$. Inset: Saturation value $\Delta\langle h \rangle_\infty$ versus the system size $k$. The solid black line corresponds to the slope $\Delta\langle h \rangle_\infty\sim k^{1.45}$. The integration method used was CI.}
\label{fig6:deltaH_KPZ}
\includegraphics[width=0.7\textwidth]{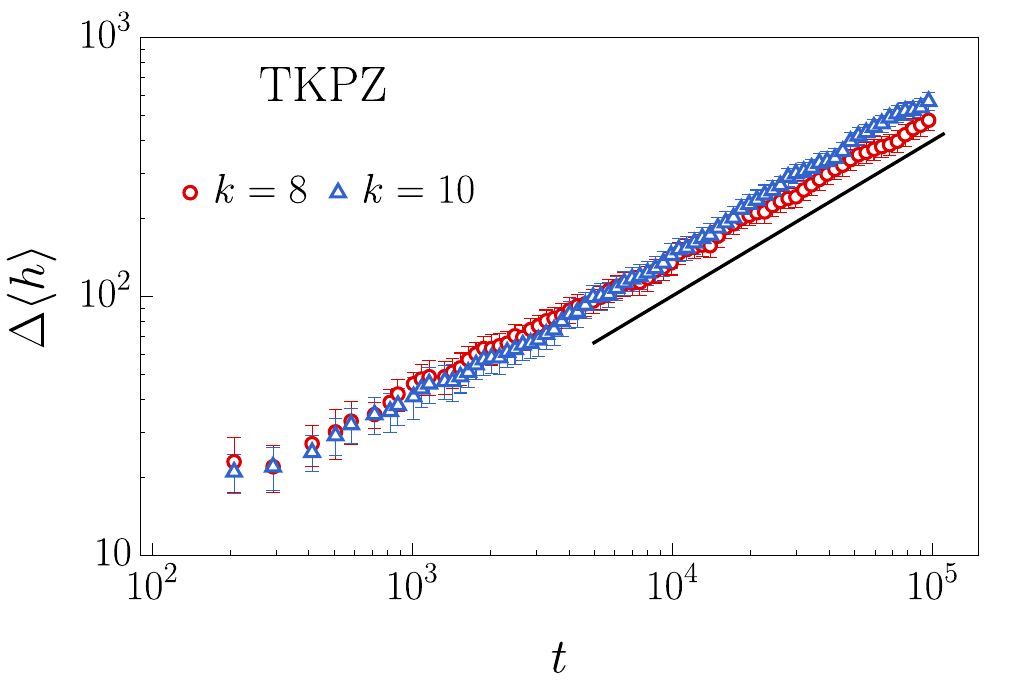}
\caption{Difference between the value of the central node and the average of the last shell $k$, $\Delta\langle h \rangle$, for various system sizes $k$ for the TKPZ equation, see legend. In this figure $q=3$. As a visual reference, the solid black line corresponds to the slope $\Delta\langle h \rangle\sim t^{0.6}$. The integration method used was CI.}
\label{fig6:deltaH_TKPZ}
\end{figure}

We now turn to the analysis of the temporal evolution of $\Delta\langle h\rangle$, as defined in Eq.~\eqref{eq3:delta_h}, following the approach introduced by Oliveira~\cite{Oliveira2021}. This observable provides complementary insight into the growth dynamics across the system. It is important to note that, in the following figures, we return to estimating the error bars using the jackknife procedure.

Figure~\ref{fig6:deltaH_KPZ} shows the time evolution of $\Delta\langle h\rangle$ for one condition of the KPZ equation across different system sizes, while Fig.~\ref{fig6:deltaH_TKPZ} presents the corresponding results for the TKPZ equation. In the case of the KPZ equation, the difference between the central and outermost layers saturates over time. In contrast, for the TKPZ equation, this difference continues to grow indefinitely, following a power-law behavior $\Delta\langle h \rangle \sim t^{0.6}$.

Moreover, for the KPZ equation, the saturation value of $\Delta\langle h \rangle$ increases with the system size, following a power law $\Delta\langle h \rangle_\infty \sim k^{1.45}$. As previously noted by Oliveira in Ref.~\cite{Oliveira2021}, this provides clear evidence that, in the thermodynamic limit ($k \rightarrow \infty$), the surfaces become macroscopically curved. As a result, boundary effects inevitably hinder accurate measurements of the global roughness. The same behavior is observed in the case of the TKPZ equation. However, in this case, it is not even necessary to reach the thermodynamic limit for such effects to become apparent.

The results for $\Delta\langle h\rangle$ are not shown for the EW and RD equations, as this quantity remains essentially zero at all times in both cases.

\section{Conclusions}

In this chapter, we have numerically integrated the RD, EW, KPZ, and TKPZ equations on CTs. Tables~\ref{table6:RD_EW} and~\ref{table6:KPZ_TKPZ} summarize the results obtained in our present study.

\begin{table}[t!]
\centering
\renewcommand{\arraystretch}{1.3}
\begin{tabular}{@{}ccc@{}}
\toprule
\textbf{Observable} & \textbf{RD} & \textbf{EW} \\
\midrule\midrule

$\langle\overline{h} \rangle_s$ & $\langle\overline{h} \rangle_s = 0$ & $\langle\overline{h} \rangle_s = 0$ \\
\midrule

$\langle\bar{h}\rangle$ & $\langle\bar{h}\rangle = 0$ & $\langle\bar{h}\rangle = 0$ \\
\midrule

$w(t)$ & $w \sim t^{1/2}$ & 
\begin{tabular}[c]{@{}c@{}} $w \sim (\ln t)^{\hat{\beta}}$ \\ $\hat{\beta}$ depends on $q$ and $k$ \end{tabular} \\
\midrule

$w_{\rm sat}$ & No saturation & 
\begin{tabular}[c]{@{}c@{}} $w_{\rm sat} \sim (\ln k)^{\hat{\alpha}}$ \\ $\hat{\alpha}$ depends on $q$ \end{tabular} \\
\midrule

$w_0(t)$ & $w_0 \sim t^{1/2}$ & $w_0 \sim t^{1/2}$ \\
\midrule

$w_{\bar{h}}(t)$ & $w_{\bar{h}} \sim t^{1/2}$ & $w_{\bar{h}} \sim t^{1/2}$ \\
\midrule

$C_2(r,t)$ & $C_2(r,t) \sim \text{const.}$ & $C_2(r,t) \to r$ \\
\midrule

$P(\chi)$ & Gaussian & Gaussian \\
\midrule

$\Delta \langle h\rangle$ & 
\begin{tabular}[c]{@{}c@{}} $\Delta \langle h\rangle \approx 0$ \\ for all layers \end{tabular} &
\begin{tabular}[c]{@{}c@{}} $\Delta \langle h\rangle \approx 0$ \\ for all layers \end{tabular} \\
\midrule

$A(s,t)$ & 
\begin{tabular}[c]{@{}c@{}} $A(s,t) \approx 0$ \\ for all layers \end{tabular} &
\begin{tabular}[c]{@{}c@{}} $A(s,t) \approx 0$ \\ for all layers \end{tabular} \\

\bottomrule
\end{tabular}
\caption{Summary of results for the RD and EW equations on CTs.}
\label{table6:RD_EW}
\end{table}

Furthermore, we compared several discretization methods, evaluating both their numerical stability and their effectiveness in capturing the growth dynamics. Our analysis showed that the ST and LS methods successfully reproduce the behavior expected from discrete models in the KPZ class but suffer from numerical instabilities at high values of the nonlinear parameter $\lambda$. In contrast, the CI method was crucial for stabilizing the numerical integration under these conditions, enabling us to explore a wider parameter space, including the TKPZ equation. This method also supports longer simulation times, which is essential for investigating the behavior of the system at large time scales. Moreover, the results obtained from the three methods were largely indistinguishable. We also examined how the main observables depend on the BC. While some differences were found between Free and Neumann BC, the key results and conclusions of the study are mostly insensitive to the choice of BC.

Our results closely reproduce earlier simulations of discrete KPZ and EW models on Bethe lattices. For the EW equation, the global roughness shows logarithmic scaling, in agreement with the findings of Saberi \cite{Saberi2013}. On the other hand, the KPZ equation displays a more complex behavior: the global roughness initially grows similarly to the EW case during a brief transient regime, but subsequently transitions to a different growth regime characterized by a power-law scaling. Analyzing this later regime in detail is difficult, as the system quickly reaches saturation. Notably, the roughness in the TKPZ equation increases indefinitely, displaying a growth pattern reminiscent of the RD equation.

These findings, particularly those for the EW equation, suggest that the Bethe lattice, or more precisely finite CTs, cannot be regarded as a straightforward infinite-dimensional limit of hypercubic lattices for these stochastic PDEs. Instead, they exhibit strong finite-size and boundary effects. If the Bethe lattice were a good approximation of the infinite-dimensional limit, the surface should remain smooth, since the upper critical dimension for the EW equation is $d_u^{\text{EW}} = 2$.

A central aspect of our analysis is the role of boundary effects in the growth process. The distinctive structure of CTs causes the outermost layer to grow more slowly than the average interface when the non-linear term is present, as each node in this layer has only a single neighbor. Conversely, the central node exhibits the fastest growth. This asymmetry propagates across the layers, giving rise to non-trivial correlations and deviations from standard scaling behavior. 


With regard to fluctuation distributions, our results confirm that height fluctuations in the RD and EW equations follow a Gaussian distribution, as expected. In the case of the KPZ equation, the PDF depends on system conditions; in certain scenarios, oscillations appear due to the relative growth dynamics between layers. For the TKPZ equation, the fluctuations resemble those of the TW distribution, although noticeable deviations arise at higher coordination numbers. These findings suggest that fluctuation behavior in KPZ growth processes on network-like structures differ substantially from that observed on regular lattices.

The analysis of the saturation of height differences between the center and the boundary reveals that KPZ surfaces on CTs remain macroscopically curved in the thermodynamic limit ($k \rightarrow \infty$). This finding supports Oliveira’s conclusion that boundary effects hinder reliable measurements of global roughness \cite{Oliveira2021}. Moreover, for the TKPZ equation, the height differences between successive layers fail to saturate entirely.

\begin{table}[t!]
\centering
\renewcommand{\arraystretch}{1.3}
\begin{tabular}{@{}ccc@{}}
\toprule
\textbf{Observable} & \textbf{KPZ} & \textbf{TKPZ} \\
\midrule\midrule

$\langle\overline{h} \rangle_s$ & $\langle\overline{h} \rangle_s \sim t$ & $\langle\overline{h} \rangle_s \sim t$ \\
\midrule

$\langle\bar{h}\rangle$ & $\langle\bar{h}\rangle \sim t$ & $\langle\bar{h}\rangle \sim t$ \\
\midrule

$w(t)$ & 
\begin{tabular}[c]{@{}c@{}} $w \sim (\ln t)^{\hat{\beta}}$ (EW transient) \\ $w \sim t^{\beta}$ (prior to saturation) \\ $\hat{\beta}$ and $\beta$ depend on $q$ and $k$ \end{tabular} &
\begin{tabular}[c]{@{}c@{}} $w \sim t^{\beta}$ \\ $\beta = 1/2$ \end{tabular} \\
\midrule

$w_{\rm sat}$ &
\begin{tabular}[c]{@{}c@{}} $w_{\rm sat} \sim k^{\alpha}$ \\ $\alpha$ depends on $q$ \end{tabular} &
No saturation \\
\midrule

$w_0(t)$ & $w_0 \sim t^{1/2}$ & $w_0 \sim t^{1/2}$ \\
\midrule

$w_{\bar{h}}(t)$ & $w_{\bar{h}} \sim t^{1/2}$ & $w_{\bar{h}} \sim t^{1/2}$ \\
\midrule

$C_2(r,t)$ &
\begin{tabular}[c]{@{}c@{}} $C_2(r,t) \to r^{2\alpha}$ \\ Jump in penultimate layer \end{tabular} &
\begin{tabular}[c]{@{}c@{}} $C_2(r,t) \sim \text{const.}$ \\ Jump in last layer \end{tabular} \\
\midrule

$P(\chi)$ &
\begin{tabular}[c]{@{}c@{}} No clear shape \\ Oscillations for some conditions \\ Positive skewness \end{tabular} &
\begin{tabular}[c]{@{}c@{}} Non-Gaussian \\ Positive skewness \end{tabular} \\
\midrule

$\Delta \langle h\rangle$ &
\begin{tabular}[c]{@{}c@{}} $\Delta \langle h\rangle$ saturates \\ $\Delta \langle h\rangle_\infty \sim k^{\lambda}$ \end{tabular} &
$\Delta \langle h\rangle \sim t^{\delta}$ \\
\midrule

$A(s,t)$ &
\begin{tabular}[c]{@{}c@{}} Last layer slower than $\bar{h}$ \\ Other layers faster than $\bar{h}$ \\ Stationary value reached \\ Stratified \end{tabular} &
\begin{tabular}[c]{@{}c@{}} Last layer slower than $\bar{h}$ \\ Other layers faster than $\bar{h}$ \\ No stationary value \end{tabular} \\

\bottomrule
\end{tabular}
\caption{Summary of results for the KPZ and TKPZ equations on CTs.}
\label{table6:KPZ_TKPZ}
\end{table}

Our study underscores the challenges of employing CTs as a substrate for probing the infinite-dimensional limit of KPZ growth. Although our numerical integration methods offer a robust framework for studying growth dynamics on networked structures, the pronounced influence of boundary effects requires careful consideration when interpreting the results. Future investigations might explore alternative network topologies that more faithfully capture high-dimensional behavior while reducing artifacts introduced by BC.






\graphicspath{{7_capitulo/fig7/}}

\chapter[MC Modeling of Oil Extraction via SAWs]{Monte Carlo Modeling of Oil Extraction via Surface Acoustic Waves}\label{chap7:lou}

This chapter is dedicated to modeling, via the Monte Carlo method, a complex physical scenario of significant practical interest: the extraction of oil from an oil-in-water emulsions using Surface Acoustic Waves (SAWs)\nomenclature{SAW}{Surface Acoustic Wave}. Although the analysis of this system is somewhat detached from the main focus of the thesis, since it does not involve a study of the kinetic roughening properties of the system, it remains conceptually related to the dynamics explored in Chapters \ref{chap4:band} and \ref{chap5:radial_spreading}. An important note to clarify is that, due to the complexity of the system, the objective of this study is primarily qualitative rather than quantitative. The focus lies in modeling the SAW itself, aiming to develop the simplest model that still captures its fundamental properties.

The structure of this chapter is as follows. We first provide an overview of the experimental insights into this phenomenon, which has been investigated only in recent years \cite{paper_experimento}. We then present the discrete model developed to study this process, with a focus on the modeling of the SAW. Finally, we present the results and conclusions derived from the analysis of this model.

\section{Introduction}

Conventional commercial techniques for oil-water separation, such as high-power distillation~\cite{Distillation} and the coagulation of oil droplets~\cite{Zouboulis2000,Ahmad2005}, both of which have been employed for nearly two centuries, are energy-intensive and often rely on the use of additional chemical agents. More recently, it has been demonstrated that interfacial (surface) effects can play a significant role in oil-water separation, offering promising implications for reducing energy consumption.

Oil is generally characterized by low surface tension. For example, commercial silicone oil at ambient conditions exhibits a surface tension of around 20 mN/m at the air interface. In contrast, water is known for its comparatively high surface tension, with pure water under the same conditions exhibiting a value of approximately 70 mN/m. Moreover, the addition of surfactants to water reduces the surface tension of the mixture. This provides a means to tune the surface tension, a key parameter for performing certain experiments. Therefore, oil typically exhibits a small three-phase (vapor/liquid/solid) contact angle on most solid surfaces. Silicone oil, in particular, often displays a near-zero equilibrium contact angle, enabling it to spontaneously spread over surfaces, a property that underpins many of its practical applications. In opposition, water and water/surfactant solutions generally sustain finite contact angles on most substrates, leading to the formation of discrete droplets~\cite{mittal_guide_2009}.

Recent experimental studies have expanded on the concept of employing surface effects to improve oil-water separation by introducing acoustic stress into the mixture. This approach creates a capillary–acoustic stress balance that promotes the displacement of oil from the emulsion~\cite{paper_experimento}. A mechanism that has recently gained significant attention in the scientific literature is acoustic streaming. An acoustic wave propagating through a fluid, or through a solid in contact with a fluid, induces stress and fluid motion. This results in the formation of a boundary layer flow near the solid–fluid interfaces~\cite{LordRayleigh1884, Schlichting:1932p447}, as well as a bulk flow within the fluid~\cite{Eckart1948,Lighthill1978,Nyborg1953}. The bulk flow, whose steady-state component at long times is referred to as Eckart streaming~\cite{Eckart1948}, arises from spatial variations in the acoustic wave intensity within the fluid.

These flows can exhibit complex behavior; for example, attenuation due to viscous and thermal dissipation may also lead to diffraction, resulting in the formation of leaky waves. In such waves, part of the acoustic energy radiates into the adjacent fluid at an angle known as the Rayleigh angle (see, e.g. Ref. ~\cite{Yabe2014}). This energy leakage plays a crucial role in coupling the acoustic field to the fluid, enabling phenomena such as bulk streaming and enhanced transport. Moreover, the interaction of acoustic waves with an interface, specifically, in this case, the vapor/liquid interface of droplets and thin films, gives rise to a net force known as acoustic radiation pressure~\cite{hamilton1998nonlinear}. This phenomenon is well known to exert stress on the surfaces of particles~\cite{King1934} and other solid objects~\cite{Chu1982,Borgnis1953,Hasegawa2000,Karlsen2016,Rajendran2023}, and has also been shown to deform and displace soft interfaces~\cite{Biwersi2000,Alzuaga2005,Issenmann2006,Rajendran2022}.

It has been demonstrated that MHz-frequency SAWs, propagating along a solid substrate, can drive the dynamic wetting of both oil~\cite{Rezk2012,Rezk2014,Manor2015} and water~\cite{Altshuler2015,Altshuler2016} films in both directions, along and opposite to the wave propagation. The interaction of acoustic stress with the liquid film depends strongly on the surface tension of the fluid, resulting in distinct behaviors for oil and water. When the acoustic stress exceeds the opposing capillary stress within the film, the liquid can dynamically wet the substrate in either direction relative to the SAW. This condition is easily met in the case of silicone oil due to its inherently low surface tension. In contrast, for water or water–surfactant mixtures, a higher SAW intensity is required to overcome capillary forces.

Horesh \textit{et al.}~\cite{Horesh2019} investigated this difference by incorporating gravitational effects into the balance between acoustic and capillary stresses. Their results showed that oil films were able to continuously climb a vertical SAW actuator against gravity, while water and water–surfactant films only rose a few millimeters before reaching an equilibrium height, determined by the interplay among gravitational, capillary, and acoustic forces.

In a recent study~\cite{paper_experimento}, this previous work was extended by investigating the extraction of oil films from oil-in-water emulsions under laboratory conditions. It was observed that the oil phase migrated in the direction opposite to the SAW propagation. This behavior was attributed to the acoustic stress exceeding the capillary stress associated with the low surface tension of oil. In contrast, the water phase remained stationary, as its higher surface tension resulted in a capillary stress that dominated over the applied acoustic stress. As a result, the oil separated from the emulsion and moved away, while the water phase was left behind.

The experimental setup used in the study was designed to extract oil from oil-in-water emulsions using SAWs. At the core of the system there is a SAW actuator composed of a transparent lithium niobate (LiNbO$_3$) substrate. This material exhibits piezoelectric properties, allowing it to convert electrical signals into mechanical surface waves. On top of this substrate, a series of interdigitated metal electrodes, known as an Interdigital Transducer or IDT, are fabricated. These electrodes receive a high-frequency (20 MHz) electrical signal from a signal generator and amplifier, producing a SAW that propagates along the surface of the solid substrate.

\begin{figure}[t!]%
\centering
\includegraphics[width=0.75\textwidth]{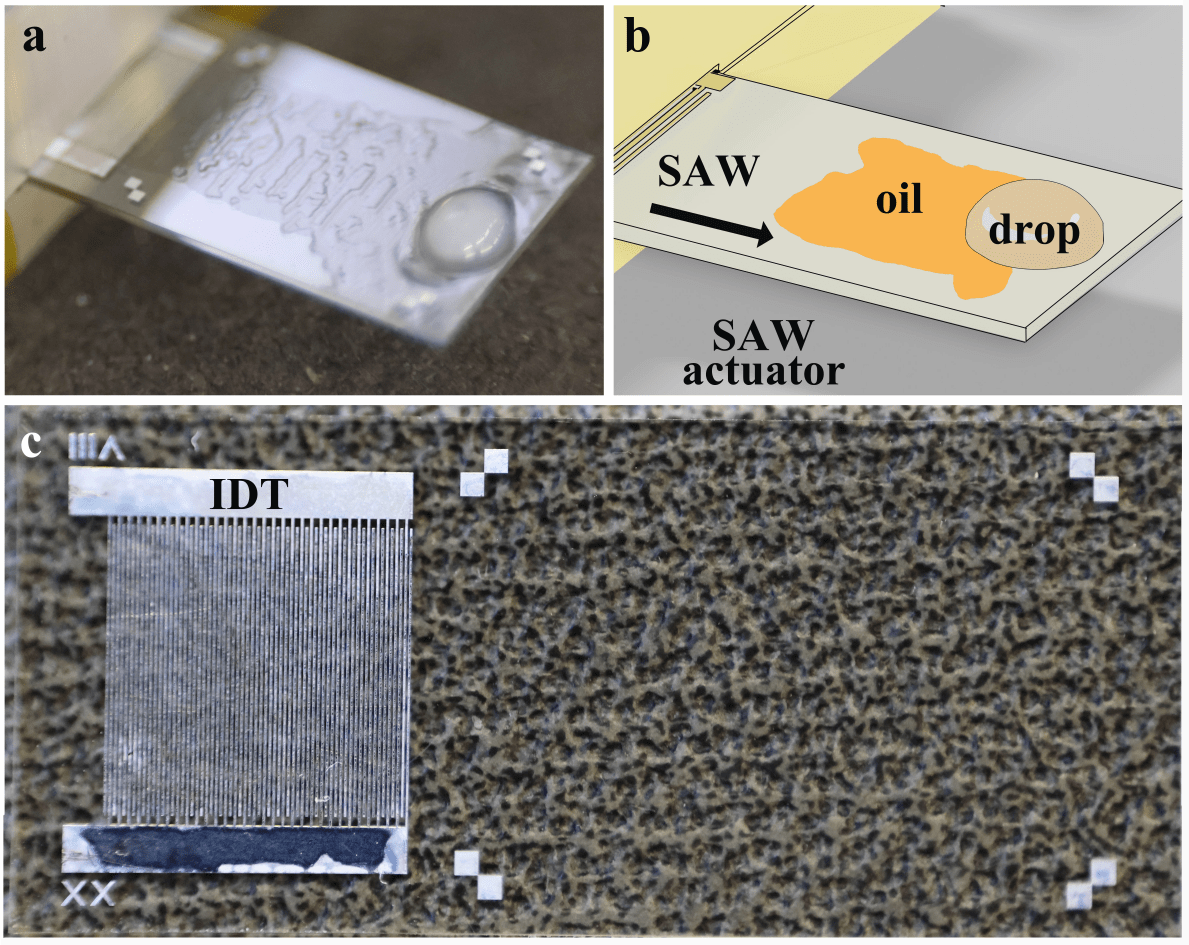}
\caption{(a) Top view of the experimental setup showing the SAW actuator, which supports the emulsion droplet. The actuator is placed in a 3D-printed plastic case that connects it to a power source. (b) A schematic sketch (view from above) of the same system, further illustrating the oil film emerging from the emulsion sessile drop under SAW excitation. (c) The SAW actuator (placed on a checkered surface) is comprised of inter-digital metal electrodes (referred to as IDT) fabricated on the top of a transparent piezoelectric lithium-niobate (LN) substrate. The sides of the metal squares fabricated atop the LN substrate, away from the IDT, are 0.5 mm long. Reproduced from Ref.~\cite{paper_experimento}.}
\label{fig7:setup}
\end{figure}

Once the SAW is generated, a small drop (10 $\mu$L) of an 40\% oil-in-water emulsion, stabilized with surfactants, is deposited on the piezoelectric surface, away from the IDT. The SAW interacts with the sessile emulsion drop, exerting an acoustic stress at the solid–liquid interface. This interaction leads to a phenomenon known as \textit{acoustowetting}, in which the lower-surface-tension oil phase responds to the acoustic excitation by forming thin oil films that spread across the solid substrate. In contrast, the water phase, with higher surface tension and a finite contact angle on lithium niobate (30–60°), remains pinned and largely unaffected under the same acoustic conditions. Figure~\ref{fig7:setup} shows the experimental setup used in the study, along with a schematic sketch of the oil film extraction and a detailed image of the SAW actuator.

\begin{figure}[t!]
\centering
\begin{tabular}{cccc}
{\includegraphics[width=0.22\textwidth]{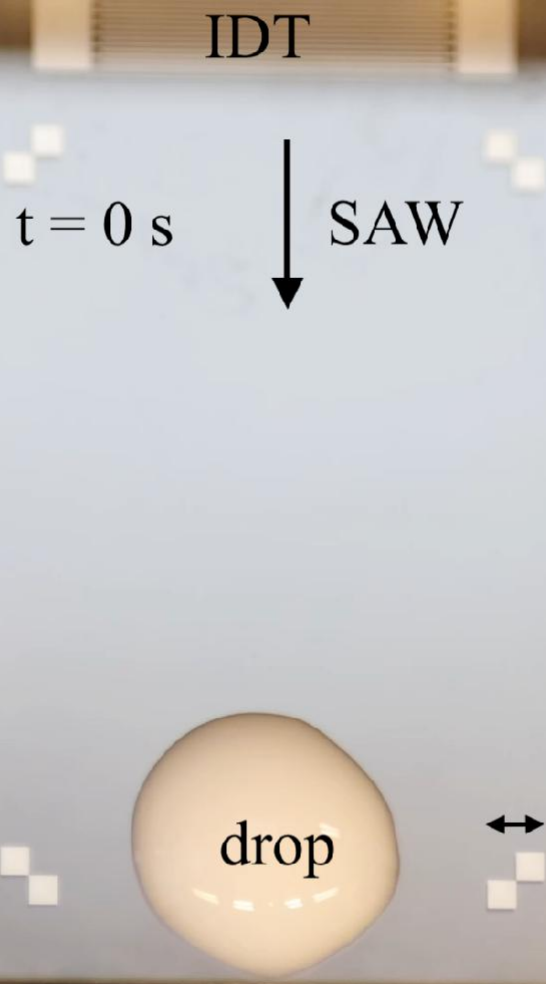}}
&
{\includegraphics[width=0.22\textwidth]{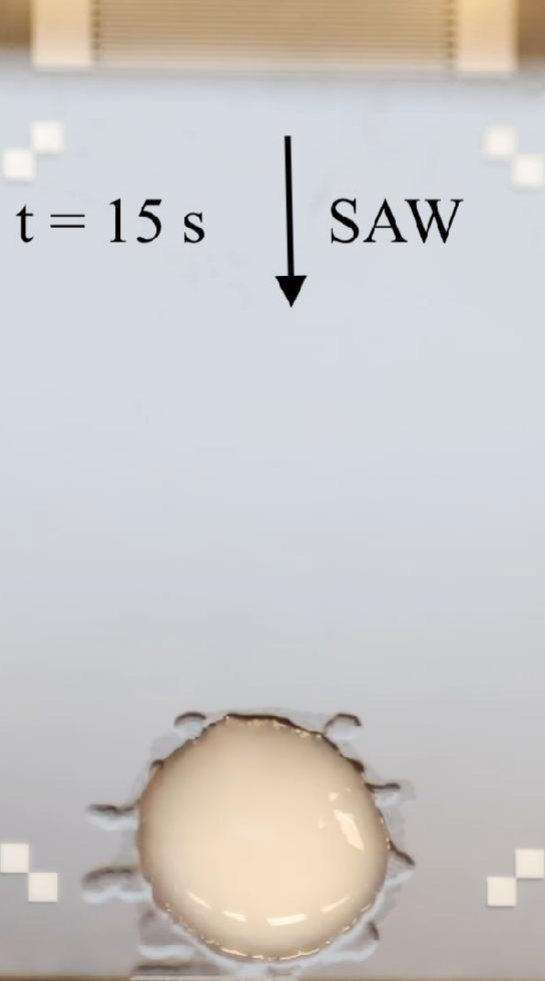}}
&
{\includegraphics[width=0.22\textwidth]{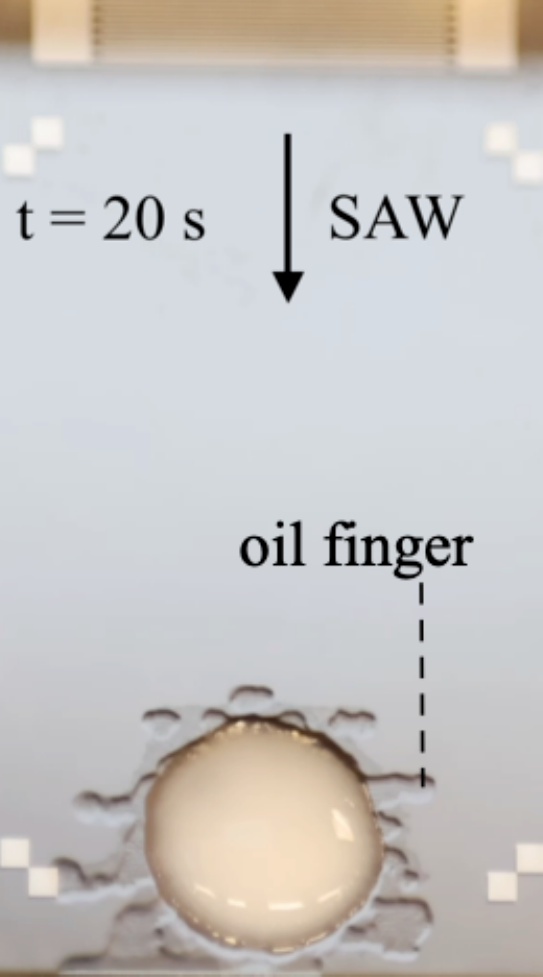}}
&
{\includegraphics[width=0.22\textwidth]{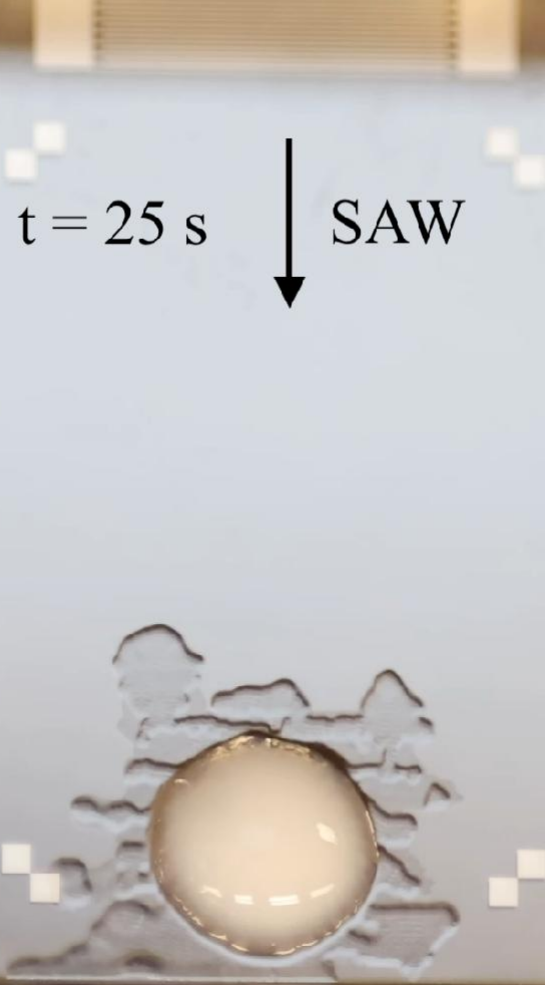}}
\\
{\includegraphics[width=0.22\textwidth]{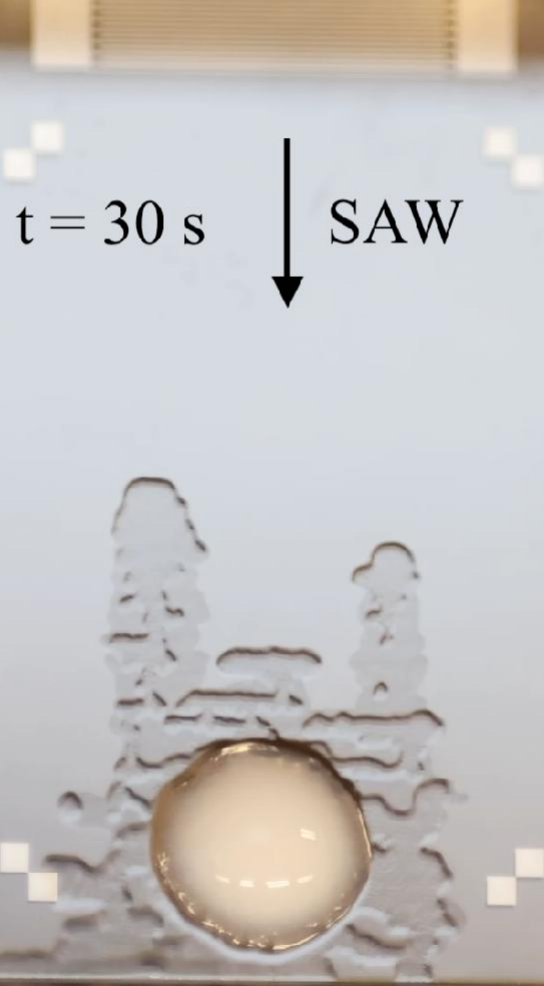}}
&
{\includegraphics[width=0.22\textwidth]{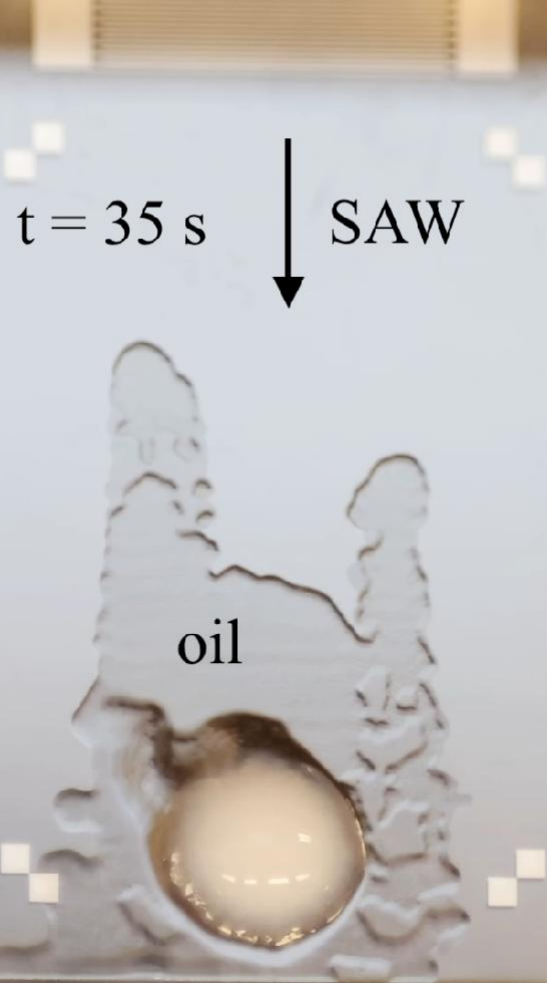}}
&
{\includegraphics[width=0.22\textwidth]{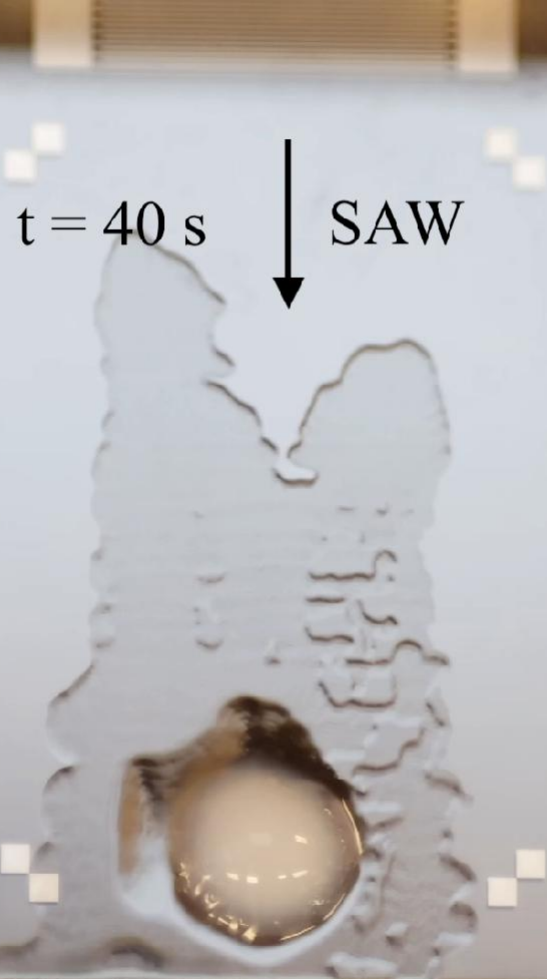}}
&
{\includegraphics[width=0.22\textwidth]{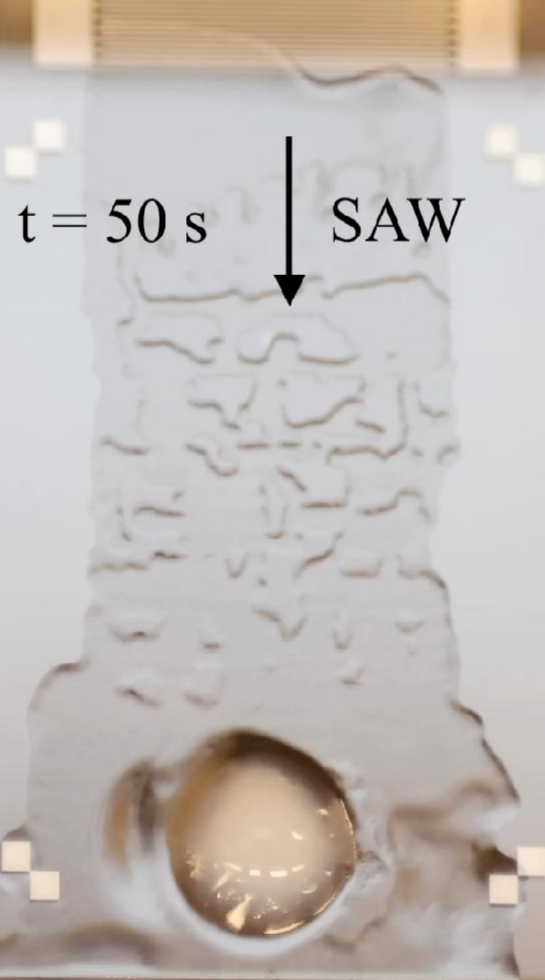}}
\end{tabular}
\caption{Top view of a typical experiment. The emulsion has a volume of 10 $\mu$L (40\% oil-in-water emulsion, 230 nm oil droplets diameter) at lab ambient conditions (50\% humidity, 20$^\circ$ C). The  SAW excitation amplitude is $A=1.8$ nm. Time $t=0$ corresponds to the moment oil is observed at the drop circumference, which here occurs after a wait-time period of $t_w = 190$~s after the onset of SAW excitation. Initially ($t=0-20$~s), fingers of oil leave the drop transverse to the path of the SAW. After $t = 20$~s  the oil fingers that have emerged from the drop change direction and spread in the direction opposite the SAW propagation. The double arrow in the $t = 0$ image is 1 mm long.  Reproduced from Ref.~\cite{paper_experimento}.}
\label{fig7:experiment_top_view}
\end{figure}

Furthermore, Fig.~\ref{fig7:experiment_top_view} presents a top-view time-lapse sequence of a typical oil extraction experiment. The figure illustrates the evolution of the oil phase in response to the SAW excitation. Time zero, defined as the moment when oil first appears at the edge of the drop, follows a waiting period of approximately 190 seconds from the onset of SAW excitation. Initially (from $t = 0$ to $20$ s), transparent oil ``fingers'' begin to emerge from the edge of the drop, moving in a direction transverse to the SAW propagation path. As time progresses beyond $20$ seconds, the direction of the oil spreading shifts, and the film begins to expand in the direction opposite to that of the SAW, a hallmark of acoustowetting. In the later images ($t = 25–50$ s), the oil film shows wavy surface patterns, with thickness variations around \mbox{$0.5$ mm} wide. These patterns differ from the $200$ $\mu$m wavelength of the SAW, which suggests that they are caused by another effect—likely a mix of acoustic and capillary forces. The figure captures the key steps of oil extraction: the first appearance of fingers, a change in spreading direction, and the start of surface instabilities.

Simulating the dynamics of oil–water separation under SAW excitation can offer valuable insights into the underlying mechanisms of the process. Although previous studies have attempted to model acoustowetting experiments using continuum theory~\cite{Altshuler2015,Altshuler2016}, they offer limited insight into the mechanisms by which SAWs enhance phase separation in oil–in–water emulsions. Moreover, these works do not address the dynamics of either the phase separation process or the extraction of the oil film. Therefore, to gain new insight into these issues, we introduce in this chapter a simple Ising-lattice gas model aimed at clarifying the key factors underlying the experimentally observed behavior. MC–based methods have already been applied to simulate droplet dynamics, as discussed in Chapter~\ref{chap2:wetting} in the context of precursor film spreading. Here, we briefly review studies that are particularly relevant to the present problem \cite{Nussbaumer2008, Chalmers2017, Areshi2019} and differ from those covered in Chapter~\ref{chap2:wetting}. 

Notably, Refs.\cite{Cheng1993} and \cite{Lukkarinen1995}, which provide significant contributions to the modeling of spreading phenomena, also serve as a useful reference point for our approach. Building upon these foundational studies, other researchers have employed similar lattice-based models to investigate a broader range of droplet behaviors beyond spreading. For instance, Nussbaumer \textit{et al.}~\cite{Nussbaumer2008} examined the droplet formation–dissolution transition using a simple two-dimensional Ising-lattice gas model that includes only nearest-neighbor interactions. Expanding on this framework, Chalmers \textit{et al.}~\cite{Chalmers2017} studied the evaporation of nanoparticle suspensions using a discrete model with extended pair interactions, beyond nearest neighbors, for all key interaction types: liquid–liquid, nanoparticle–nanoparticle, and liquid–nanoparticle. This model also incorporates a chemical potential to govern the vapor–liquid phase transition and introduces distinct substrate interactions for liquid and nanoparticle species. More recently, Areshi \textit{et al.}~\cite{Areshi2019} applied a similar extended model to explore several aspects of droplet dynamics on solid surfaces, including the analysis of density profiles, how two droplets come together and merge, and the behavior of droplets under a constant lateral driving force parallel to the substrate.

Our goal is to develop a discrete model that includes the minimal set of interactions necessary to retain the essential features of the system. This will be the focus of the next section. As in the spreading models discussed in detail in Chapters~\ref{chap4:band} and \ref{chap5:radial_spreading}, we will employ Kawasaki dynamics to capture the particle exchange mechanisms.

\section{Model}

The model used in this thesis represents a simplified version of an oil-in-water emulsion, consisting of oil and water regions represented as discrete cells on a two-dimensional regular lattice. This system mimics a sessile drop of an oil-in-water emulsion placed on a solid horizontal substrate and subjected to a SAW, where the SAW-induced stress in the liquid is modeled as an external force. Each cell in the lattice may be occupied or unoccupied, as described in detail below. The system evolves according to a MC scheme with Kawasaki local dynamics \cite{Newman1999}. It is worth clarifying that, in this chapter, time is not updated continuously, as our primary interest lies in capturing the qualitative behavior of the system rather than obtaining precise quantitative results. To perform the simulations, we randomly select a pair of neighboring cells and attempt to exchange them according to the Metropolis acceptance criterion [see Eq.~\eqref{eq3:metropolis}].

The model is defined on a rectangular grid consisting of $L_x$ cells along the $x$-axis (parallel to the solid substrate) and $L_y$ cells along the $y$-axis (perpendicular to the substrate). While this framework can be readily extended to three dimensions, we focus on a two-dimensional system for simplicity. All cells are uniform in size and may be occupied by water, oil, or remain empty, representing air. As in the spreading model discussed in Chapter~\ref{chap2:wetting}, it is important to emphasize that this fluid representation is statistical in nature, rather than atomistic. For further details, refer to the final paragraph of Chapter~\ref{chap2:wetting}.

We define $ o_{\bm{i}} $ and $ w_{\bm{i}} $ as the occupation numbers for oil and water, respectively. For instance, $ o_{\bm{i}} = 1 $ indicates that cell $ \bm{i} $ is occupied by an oil particle, while $ o_{\bm{i}} = 0 $ means the cell contains water or is empty. The 2D position vector is denoted by $\bm{i}=(x_{\bm{i}},y_{\bm{i}})$. We do not assign an occupation number to air, since it neither interacts with the other particles nor responds to the governing forces. A cell is considered to represent air when both $ o_{\bm{i}} = 0 $ and $ w_{\bm{i}} = 0 $.


Initially, we consider a sessile drop composed of a random mixture of water and oil particles. The initial oil volume fraction, $c = \frac{\sum_{\bm{i}} o_{\bm{i}}}{\sum_{\bm{i}} \left( o_{\bm{i}} + w_{\bm{i}} \right)}$, is set to $c = 0.4$, in accordance with the experimental setup. The total energy of the system is obtained by summing contributions from all cells, incorporating close-neighbor interactions, external forcing due to gravity, and acoustic stress induced by the SAW propagating through the solid substrate. Each cell is assumed to interact with four nearest neighbors in the $x$ and $y$ directions, as well as four next-nearest neighbors along the diagonals. The model includes interactions between water-water, oil-oil, and oil-water particle pairs. Although it is possible to restrict interactions only to nearest neighbors, we include diagonal neighbors to prevent the formation of  unrealistic, rectangular-shaped droplets, as observed in Chapter~\ref{chap5:radial_spreading} and discussed in Refs.~\cite{Areshi2019,Chalmers2017}. The total energy of the system is given by the following Hamiltonian,
\begin{equation}
    \begin{aligned}
        \mathcal{H} = & -\sum_{\langle \bm{i}, \bm{j} \rangle}c_{\bm{i}\bm{j}}\left( J_{oo} o_{\bm{i}}o_{\bm{j}} + J_{ww} w_{\bm{i}}w_{\bm{j}} + J_{ow} o_{\bm{i}}w_{\bm{j}}\right) \\
        & + \sum_{\bm{i}} \rho_{\bm{i}} g y_{\bm{i}} +\sum_{\bm{i}}\rho_{\bm{i}} p_{RS} \mathcal{U}(x_{\bm{i}},x_{B},\alpha).
    \end{aligned}
    \label{eq7:energy}
\end{equation}
The first term in Eq.~\eqref{eq7:energy} represents the close-neighbor interaction energy between cells, the second term accounts for gravitational effects, and the third term corresponds to the acoustic stress. As in Chapters~\ref{chap4:band} and \ref{chap5:radial_spreading}, we adopt physical units such that $ k_B = 1 $, while other parameters remain dimensionless or arbitrary. Furthermore, we fix the temperature at $ T = 1 $ in our simulations. Consequently, each term in the Hamiltonian is expressed in units of $ k_B T $.


In Eq.~\eqref{eq7:energy}, the first term accounts for the attractive interactions between water and oil particles through a near-neighbor scheme, analogous to van der Waals forces acting between water and oil, as well as between water particles and oil particles themselves. Interactions with air are neglected, as we assume that the low (taken to be zero in this model) density of air does not contribute to the total interaction energy. Specifically, the interaction energies between water-water, oil-oil, and water-oil particle pairs are denoted by $ J_{ww} $, $ J_{oo} $, and $ J_{wo} \equiv J_{ow} $, respectively. These parameters are positive, and larger values correspond to stronger attractive forces, resulting in greater cohesion between the interacting particles. The interaction strength between two particles located at lattice sites $ \bm{i} $ and $ \bm{j} $ depends on their relative distance and is represented by the coefficient $ c_{\bm{i}\bm{j}} $, defined as in Ref.~\cite{Areshi2019}.
\begin{equation}
    c_{\bm{i}\bm{j}}=\left\{
\begin{array}{ll}
      1 & \text{if $\bm{j}\in$ NN$\bm{i}$} \\
      1/2 & \text{if $\bm{j}\in$ NNN$\bm{i}$} \\
      0 & \text{otherwise} \\
\end{array} \right.
    \label{eq7:cij}
\end{equation}
Here, NN$\bm{i}$ and NNN$\bm{i}$ denote the sets of nearest and next-nearest neighbors of the lattice site $\bm{i}$, respectively. Due to the negative sign preceding this term in the Hamiltonian, the system tends to evolve toward configurations that maximize the number of favorable interactions (or bonds), particularly those associated with the largest interaction strengths $ J_{kl} $ ($k,l = o, w$). Since cells at the droplet surface of the droplet have fewer neighboring cells to bond with, the species exhibiting stronger interactions will preferentially occupy the interior of the droplet, whereas species with weaker interactions will tend to migrate toward the surface. Consequently, by adjusting the parameters governing the near-neighbor interactions (as discussed below), we effectively modify the interfacial energy, i.e. the surface tension, between oil and water.

The second summation in Eq.~\eqref{eq7:energy} accounts for the gravitational contribution to the potential energy of each cell $\bm{i}$, where $\rho_{\bm{i}}$ denotes the density, $g$ is the acceleration of gravity, and $1 \le y_{\bm{i}} \le L_y$ is the vertical coordinate of the cell. The density of air is taken to be zero, while the densities of oil and water are assumed to be equal. This approximation allows us to neglect buoyancy effects arising from the small density difference between oil and water, as our primary focus is on acoustic forcing. For simplicity, we set $\rho_{\bm{i}} = 1$ for all non-air cells.

The third summation represents the novel component of our model, capturing the contribution of acoustic stress, also known as the Reynolds stress, within the liquid phase. This stress is assumed to be proportional to the fluid density, and is therefore negligible in air, as well as to the power of the SAW propagating through the solid substrate. The term includes a factor $ p_{RS} $, which denotes the acoustic stress experienced by a cell containing either water or oil (assuming both exhibit similar acoustic impedance) in the presence of an unattenuated SAW. Additionally, it includes a spatially dependent factor $ \mathcal{U} $, which accounts for the attenuation of the SAW beneath the droplet.

\begin{figure}[t!]
\centering
\includegraphics[width=1.0\textwidth]{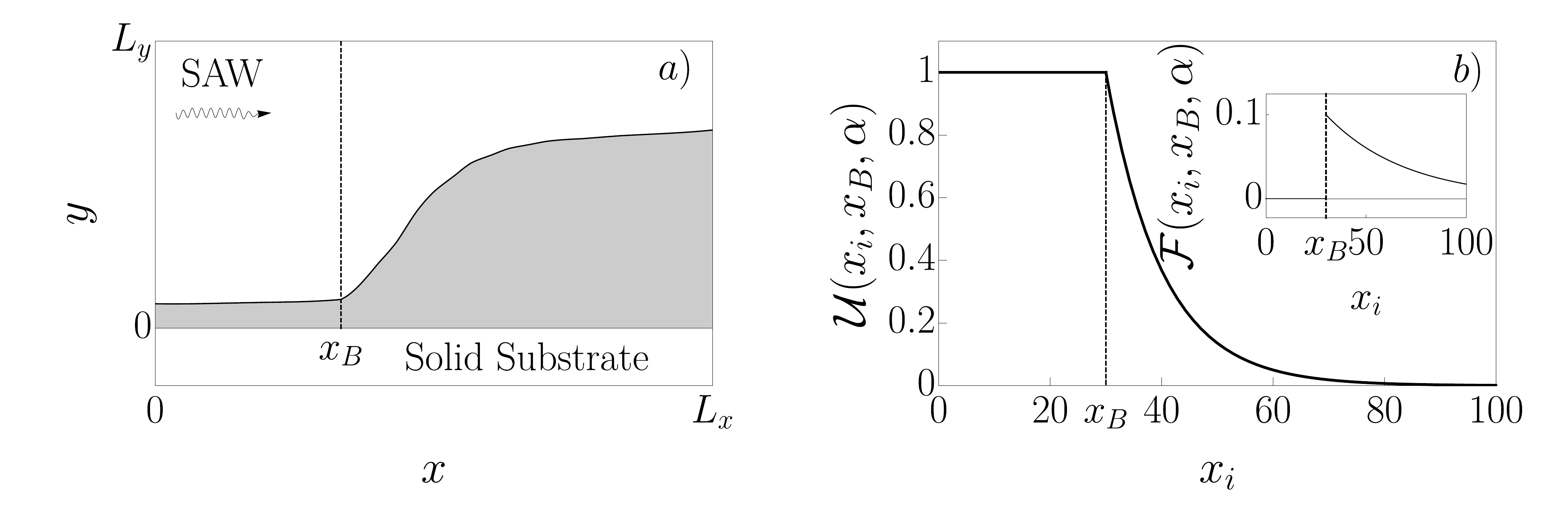}
\caption{(a) A sketch of the liquid (oil and water) geometry studied, where $x_B$ indicates the transition between an oil film ($x<x_B$) and the emulsion drop ($x>x_B$). (b) Spatial variation of the acoustic potential in the liquid, given by $\mathcal{U}(x_{\bm{i}},x_B,\alpha)$ in Eq.~\eqref{eq7:potential_energy_1}, where the dashed line indicates the position of $x_B$. Inset: Spatial variation of the force  $\mathcal{F}=-\frac{\partial\mathcal{U}}{\partial x}$ generated in the liquid by the acoustic stress.}
\label{fig7:extForce}
\end{figure}

As shown in experiments (see, e.g. Ref.~\cite{paper_experimento}), a SAW decays exponentially beneath a sufficiently thick fluid layer. To capture this behavior, we model the attenuation factor $ \mathcal{U} $ as a function of $ x_{\bm{i}} $, the discrete horizontal coordinate of cell $ \bm{i} $ along the solid substrate. In our simplified analysis, we do not account for the Rayleigh angle at which ultrasonic waves leak from the SAW into the liquid. Instead, we adopt the approximation that the acoustic stress within the liquid is directly proportional to the local SAW intensity in the solid substrate directly beneath each fluid cell. This assumption is generally valid when the liquid film is sufficiently thin compared to the wavelength of the acoustic waves leaking from the SAW, as supported by previous studies~\cite{Rezk2012,Rezk2014,Manor2015}. However, for larger-scale systems, such as the macroscopic droplet studied in Ref.~\cite{paper_experimento}, this approximation can become less accurate. In such cases, the acoustic energy does not propagate vertically but instead leaks into the fluid at the Rayleigh angle, generating a more complex stress distribution throughout the liquid. Nevertheless, we adopt the simplified approach in our model, as it is expected to be sufficient for capturing the qualitative behavior of interest and achieving the main objectives of our study. Therefore, we expect the potential associated with the acoustic stress in the liquid phase to take the following form
\begin{equation}
    \mathcal{U}(x_{\bm{i}},x_B,\alpha) =\left\{\begin{matrix}
1 & x_{\bm{i}}<x_B ,\\ 
e^{-\alpha(x_{\bm{i}}-x_B)} & x_{\bm{i}}>x_B,
\end{matrix}\right.
    \label{eq7:potential_energy_1}
\end{equation}
where $ x_B $ denotes the position of the droplet edge (see Fig.~\ref{fig7:extForce}), and $ \alpha $ is the attenuation coefficient associated with the SAW. Recalling that the SAW attenuation only becomes significant when the wave propagates beneath a macroscopic droplet, but remains negligible elsewhere~\cite{Rezk2012,Rezk2014}, we define the extent of the macroscopic drop in our simulations as the region where the fluid layer maintains a thickness greater than two cells. Accordingly, the droplet edge $ x_B $ is identified as the position $ x_{\bm{i}} $ where the film thickness transitions from two to three cells and remains at least three cells thick as one moves further into the droplet. To determine $ x_B $, the algorithm scans the system from left to right at each time step, locating the first position where this thickness criterion is consistently satisfied. The specific threshold used to define the drop edge is not critical, provided it remains small relative to the maximum thickness of the droplet. Alternative, similarly small choices yield consistent results. Based on this definition, we neglect SAW attenuation in the region $ x_{\bm{i}} < x_B $ by setting $ \mathcal{U} = 1 $, and assume that attenuation only occurs for $ x_{\bm{i}} > x_B $, where the SAW propagates beneath the thicker portion of the droplet. In this region, we model the attenuation of the acoustic stress by setting $ \mathcal{U} = e^{-\alpha(x_{\bm{i}} - x_B)} $, where $ 1/\alpha $ represents the characteristic attenuation length of the SAW.

Figure~\ref{fig7:extForce} illustrates the attenuation factor $ \mathcal{U} $ and the corresponding force exerted by the SAW on the liquid, given by $ \mathcal{F} = -\partial \mathcal{U} / \partial x $, as defined in Eq.~\eqref{eq7:potential_energy_1}. It is important to note that the position $ x_B $, which marks the onset of attenuation beneath the droplet, evolves over time as the droplet deforms, and must therefore be dynamically tracked throughout the simulation. Regarding the attenuation coefficient $ \alpha $, we adopt a value of $ \alpha = 0.01 $, which ensures that the attenuation is appreciable across the computational domain, whose size is on the order of $ 1/\alpha $. We have verified that variations in $ \alpha $ within a similar range produce only minor changes in the simulation outcomes, indicating that the results are not strongly sensitive to this parameter.

The definition of $\mathcal{U}$ in Eq.~\eqref{eq7:potential_energy_1} implies that the SAW acts uniformly on all cells, regardless of their contents that is, it exerts the same effect on oil, water, and air. This assumption is not physically realistic, as it fails to distinguish between the different acoustic responses of each phase and, more importantly, does not account for the contribution of acoustic radiation pressure. In experimental observations, the interaction of the SAW with a droplet generates an excess pressure on the free surface, resulting in a normal stress on the liquid~\cite{Chu1982}. To incorporate this effect into our model, we refine the definition of $ \mathcal{U} $. A simple choice is
\begin{equation}
    \mathcal{U}(x_{\bm{i}},x_B,\alpha) =\left\{\begin{matrix}
        0 & \text{liquid cells detached from solid},\\ \\ 
        1 &  \begin{tabular}{@{}c@{}} 
            $x_{\bm{i}}<x_B$  \\  
            \textrm{liquid cells connected to solid,}
        \end{tabular}\\ \\ 
        e^{-\alpha(x_{\bm{i}}-x_B)} & \begin{tabular}{@{}c@{}} 
            $x_{\bm{i}}>x_B$  \\ 
            \textrm{liquid cells connected to solid.}
        \end{tabular}\\
\end{matrix}\right.
    \label{eq7:potential_energy_2}
\end{equation}
Here, ``detached from the solid'' refers to a situation in which a given cell has no continuous connection to the solid substrate through adjacent liquid-filled cells (either oil or water). Equation~\eqref{eq7:potential_energy_2} is based on the physical observation that the SAW does not act on air, resulting in a discontinuity in $ \mathcal{U} $ at the free surface of the droplet. This discontinuity serves as a simple mechanism to model the effect of acoustic radiation pressure. To illustrate this, consider a liquid cell (oil or water) located at the droplet surface. This cell is subject to acoustic stress due to the SAW. However, if it detaches from the droplet and becomes airborne, it is no longer influenced by the SAW, leading to a reduction in its potential energy. This energy drop translates into an effective outward force acting on surface cells, thereby mimicking the acoustic radiation pressure observed in physical experiments.

While Eq.~\eqref{eq7:potential_energy_2} successfully incorporates the desired effect of acoustic radiation pressure, it is convenient to introduce a parameter that controls the strength of this contribution relative to other physical effects. This consideration leads us to the final form of our definition,
\begin{equation}
    \mathcal{U}(x_{\bm{i}},x_B,\alpha)
    =\left\{
    \begin{array}{ll}
        p & \begin{tabular}{@{}c@{}} 
            $x_{\bm{i}}<x_B$  \\  
            \textrm{liquid cells} \\
            \textrm{detached from solid,}
        \end{tabular}\\ \\
        p\, e^{-\alpha(x_{\bm{i}}-x_B)}  & \begin{tabular}{@{}c@{}} 
            $x_{\bm{i}}>x_B$  \\  
            \textrm{liquid cells} \\
            \textrm{detached from solid,}
        \end{tabular}\\ \\
        1 & \begin{tabular}{@{}c@{}} 
            $x_{\bm{i}}<x_B$  \\  
            \textrm{liquid cells} \\
            \textrm{connected to solid,}
        \end{tabular}\\ \\
        e^{-\alpha(x_{\bm{i}}-x_B)} & \begin{tabular}{@{}c@{}} 
            $x_{\bm{i}}<x_B$  \\  
            \textrm{liquid cells} \\
            \textrm{connected to solid,}
        \end{tabular}\\
    \end{array}
    \right.
    \label{eq7:potential_energy_3}
\end{equation}
i.e., we assume that cells detached from the solid substrate experience an acoustic stress equal to a fraction $ p $ of the stress they would receive if they were connected to the solid. When $ p = 0 $, we recover the simpler approximation given by Eq.~\eqref{eq7:potential_energy_2}, which fully incorporates the effect of acoustic radiation pressure. In contrast, setting $ p = 1 $ corresponds to the original formulation in Eq.~\eqref{eq7:potential_energy_1}, where no distinction is made between attached and detached cells, and acoustic radiation pressure is neglected. The parameter $ p $ thus provides a means to tune the strength of the acoustic radiation pressure in our simulations. In the Results section, we present simulations based on Eq.~\eqref{eq7:potential_energy_3}, which incorporates radiation pressure effects, as well as results obtained using Eq.~\eqref{eq7:potential_energy_1}, where such effects are excluded.

In summary, acoustic stress in the liquid phase (oil or water) gives rise to two distinct yet related effects: acoustic streaming within the liquid~\cite{Shiokawa1989,Shiokawa1994}, and acoustic radiation pressure at the free surface~\cite{Chu1982}. Spatial variations in the acoustic stress within the bulk of the liquid, caused by the attenuation of the SAW along the substrate, generate a net body force that drives internal flow. Moreover, the reduced acoustic stress experienced by cells that are disconnected from the solid substrate results in a net outward force at the free surface, an effect characteristic of acoustic radiation pressure. These two mechanisms represent complementary manifestations of the acoustic stress induced by the SAW.

Previous studies (see, e.g., Ref.~\cite{Areshi2019}) have investigated the interaction between liquid particles and the substrate, highlighting its influence on the equilibrium configuration of the droplet, particularly the contact angle. In the present work, we omit such interactions from our model, as they are not essential for capturing the extraction mechanism under consideration. This mechanism is primarily governed by the effects of the SAW and the surface tension differences between the two liquid phases. Similarly, we do not consider gravity to play a significant role in the extraction dynamics. Its main purpose in our model is to ensure that the droplet remains adhered to the substrate. While substrate adhesion could alternatively be modeled through explicit liquid–solid interactions, we opt to include gravity as a simpler and more convenient representation.

To carry out the MC simulations, it is necessary to specify the values of the model parameters. These include $ J_{oo} $, $ J_{ww} $, $ J_{ow} $, $ g $, $ p_{RS} $, $ p $, and $ \alpha $. The parameters $ p_{RS} $, $ p $, and $ \alpha $ are associated with the SAW and, therefore, may vary depending on the experimental conditions. In contrast, the interaction parameters $ J_{kl} $ characterize the intermolecular forces between the different fluid phases. 
Moreover, the three coupling constants $ J_{ww} $, $ J_{oo} $, and $ J_{ow} $ are not independent. Since they represent short-range interactions between neighboring cells, they can be related to the surface tensions of the corresponding fluid phases. Consequently, knowing the surface tension values for water, oil, and their interface allows us to establish relationships among the $ J_{kl} $ parameters. A simple calculation, provided in Appendix~\ref{app:coupling}, yields
\begin{equation}
    J_{oo}\approx 0.28 J_{ww} , \hspace{8mm} J_{ow}\approx 0.4 J_{ww} 
\end{equation}
These ratios will be used to determine the specific values of the coupling constants $ J_{kl} $, as discussed in the following section. The gravitational parameter $ g $ is set to 20, primarily for convenience, as further explained below.

\section{Results}

This section is divided into three subsections. In the first one, we present simulations of the system in the absence of the SAW, which will allow us to set realistic interaction coupling parameters $ J_{kl} $. In the second one, we analyze qualitatively the behavior of the system when the SAW is present, focusing on the differences that arise depending on whether the acoustic radiation pressure is included or not. Finally, in the third subsection, we present some quantitative results that support the insights introduced in the second subsection.

\subsection{Simulations in the absence of SAW: setting the interaction energies}

Although the ratios of the coupling constants $ J_{kl} $ are known, their individual values still need to be determined. To achieve this, we performed trial simulations without the presence of the SAW, considering both pure liquids and the emulsion. Experimental observations~\cite{paper_experimento} show that, in the absence of the SAW, a pure water droplet retains its shape, whereas a pure oil droplet spreads completely over the substrate. Additionally, an emulsion droplet also retains its shape, with oil migrating to the free surface of the droplet~\cite{paper_experimento}.

In the absence of the SAW, the energy of the system [see Eq.~\eqref{eq7:energy}] has only two contributions: gravity and close-neighbor interactions. When gravity dominates, the droplet tends to spread out; conversely, when close-neighbor interactions dominate, the droplet maintains its initial form. Therefore, for a given value of $ g $, the coupling constants $ J_{kl} $ must be chosen such that the simulations reproduce the following behaviors: the oil droplet spreads over the entire substrate, the water droplet preserves its shape, and the emulsion droplet both retains its shape and exhibits oil accumulation at the surface.

In all simulations presented, the computational domain has dimensions $ L_x = 300 $ and $ L_y = 50 $ along the $ x $- and $ y $-directions, respectively. We set $ g = 20 $ and $ \rho_w = \rho_o = 1 $, neglecting buoyancy effects in order to isolate the influence of acoustic forcing and intermolecular interactions, as previously discussed. The value of $ g $ is chosen for convenience, since its absolute value is relevant only in relation to the coupling constants $ J_{kl} $. The simulation results are not sensitive to the system size; comparable outcomes are expected for both larger and smaller domains.

\begin{figure}[t!]
\centering
\includegraphics[width=1.0\textwidth]{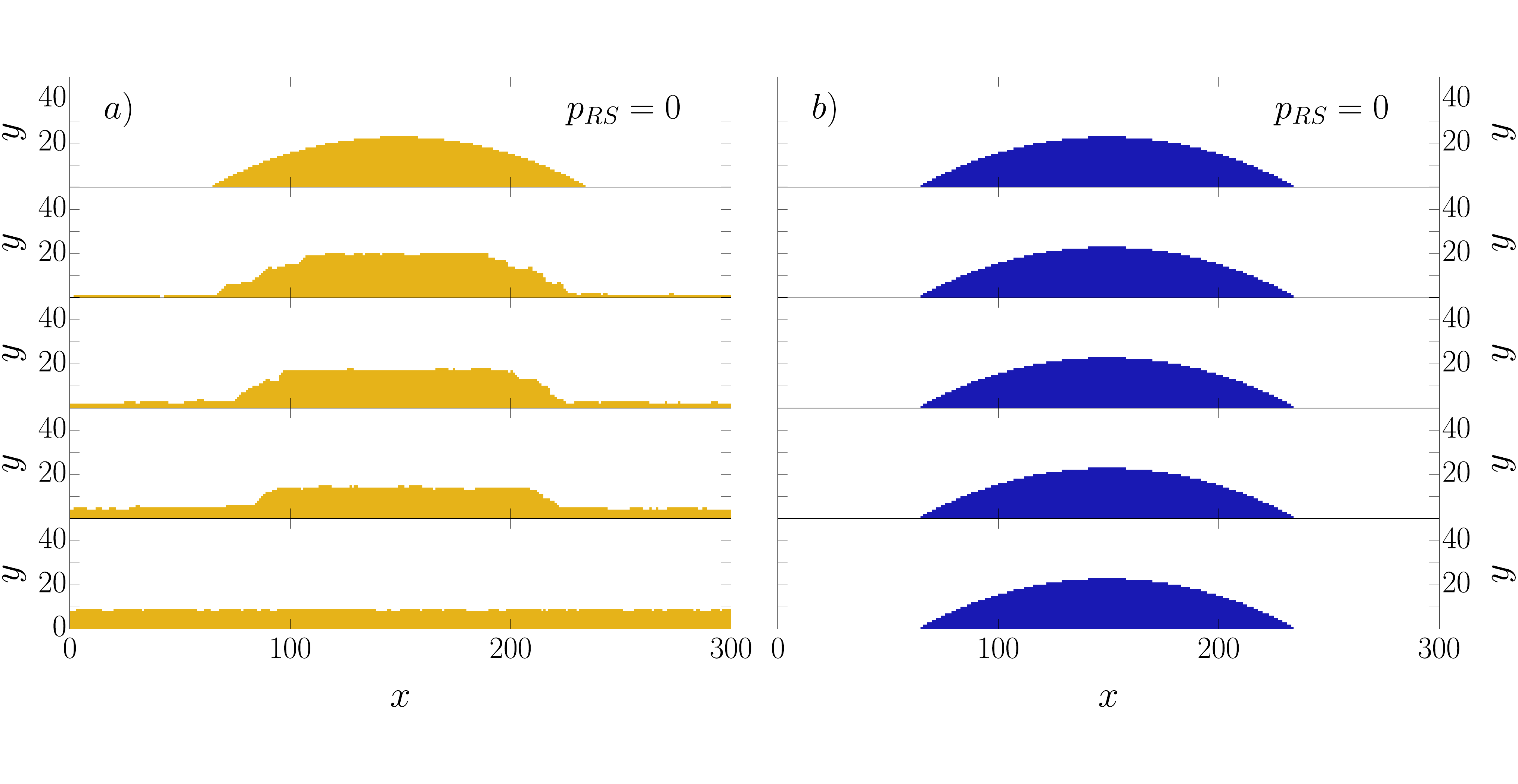}
\caption{Snapshots showing the evolution of a pure system in the absence of SAW ($ p_{RS} = 0 $): (a) oil and (b) water. The interaction parameters are $ J_{ww} = 12.6 $ and $ J_{oo} = 3.5 $, with a total of $ N_{\text{steps}} = 10^{11} $ MC steps. In this and all subsequent figures, $ g = 20 $. Time progresses from top to bottom in regular intervals, with the top row representing the initial condition. Water and oil cells are shown in blue and yellow, respectively, while air cells are omitted for clarity.}
\label{fig7:Graf_pure_liquids_NO_SAW}
\centering
\includegraphics[width=1.0\textwidth]{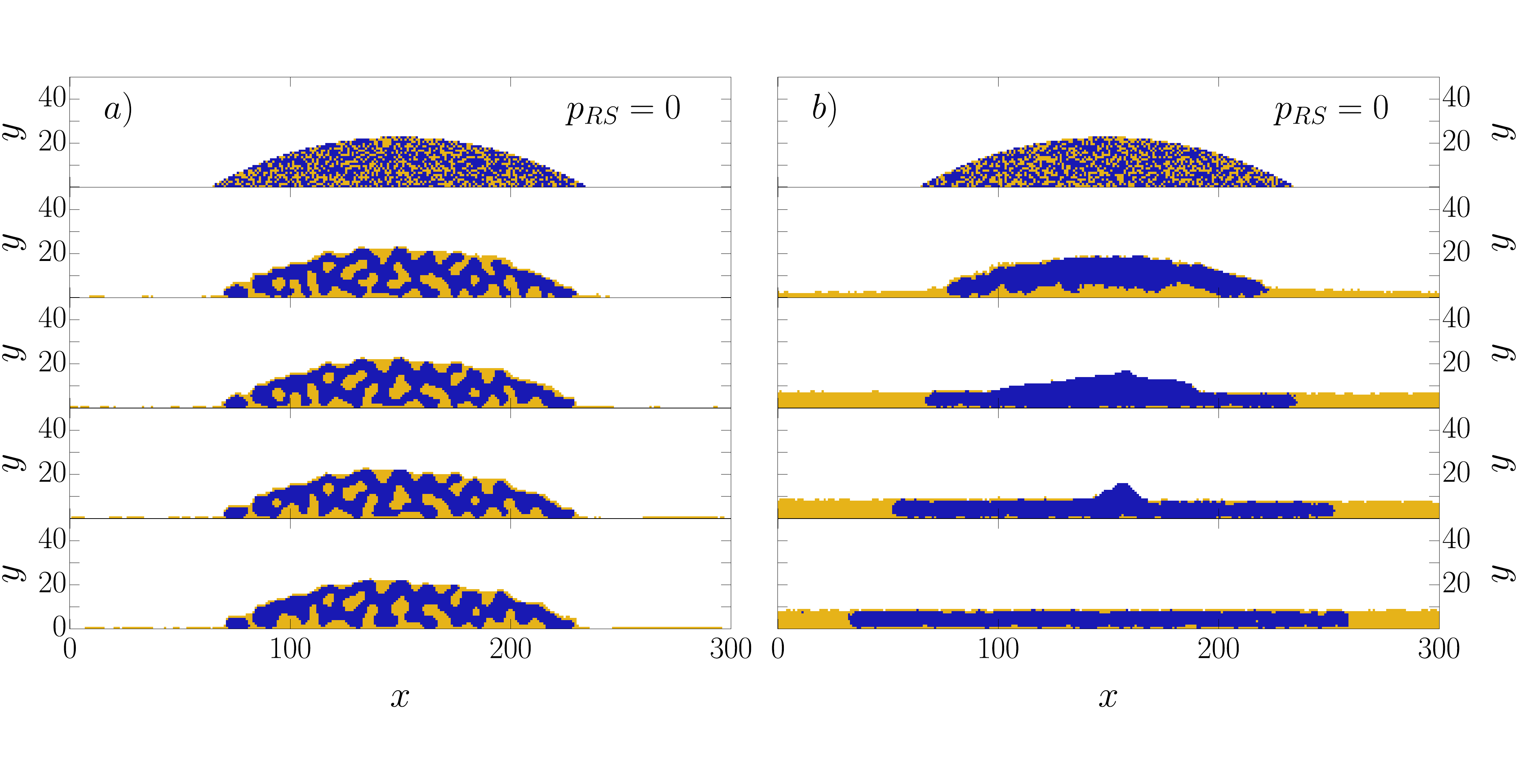}
\caption{Snapshots showing the evolution of the emulsion in the absence of SAW ($ p_{RS} = 0 $) for two different sets of parameters $J_{kl}$: (a)  $J_{ww}=12.6 $, $J_{oo}=3.5$ and $J_{ow}=5.1$,  and (b) $J_{ww}=5.4 $, $J_{oo}=1.5$ and $J_{ow}=2.2$. The number of MC steps was, $N_{\text{steps}}=10^{11}$ (a) and $N_{\text{steps}}=5\cdot10^{10}$ (b).}
\label{fig7:Graf_no_SAW}
\end{figure}

As mentioned in the previous section, all interaction constants $ J_{kl} $ are related, so fixing a single value is sufficient to determine the entire set. We begin by examining single-phase simulations, i.e., pure systems. Figure~\ref{fig7:Graf_pure_liquids_NO_SAW} displays the results for the pure components, oil (a) and water (b), in the absence of the SAW. With the selected values of $ J_{oo} $ and $ J_{ww} $, the oil droplet spreads completely across the substrate, while the water droplet retains its shape throughout the simulation, as expected.
At this stage, we emphasize that these results are intended solely to establish reasonable values for $ J_{ww} $ and $ J_{oo} $, rather than to fully capture the physics of wetting. Accurately modeling wetting phenomena would require the inclusion of substrate interactions, which play a critical role in determining droplet equilibrium properties~\cite{Areshi2019} and spreading dynamics~\cite{Abraham2002,Lukkarinen1995,Cheng1993}.

Returning to the choice of appropriate values for the $ J_{kl} $, we recall that an emulsion droplet should retain its shape in the absence of SAW. Figure~\ref{fig7:Graf_no_SAW} presents simulation results for an emulsion droplet without SAW. In Fig.~\ref{fig7:Graf_no_SAW}a, the selected $ J_{kl} $ values are sufficiently strong to preserve the shape of the droplet (these are the same parameters used in Fig.~\ref{fig7:Graf_pure_liquids_NO_SAW}). In contrast, Fig.~\ref{fig7:Graf_no_SAW}b shows a case where gravity dominates, leading the droplet to spread across the entire domain.

To ensure that our simulations reflect realistic behavior, we select the coupling constants $ J_{kl} $ to be sufficiently strong to keep the droplet cohesive. In both panels of Fig.~\ref{fig7:Graf_no_SAW}, the reported values respect the ratios established in Appendix~\ref{app:coupling}. We find that the set $ J_{ww} = 12.6 $, $ J_{oo} = 3.5 $, and $ J_{ow} = 5.1 $, the same used in Figs.~\ref{fig7:Graf_pure_liquids_NO_SAW} and~\ref{fig7:Graf_no_SAW}(a), successfully reproduces the expected behavior. Accordingly, we adopt these values for the remainder of the study.

Before proceeding, we make two brief remarks. First, we observe that extending interactions beyond the nearest neighbors aids in preserving the shape of the emulsion droplet. In simulations limited to interactions with only the four nearest neighbors (not shown here), oil particles are more likely to escape from the emulsion. Second, we note in passing the coarsening process that occurs over time, along with the migration of oil particles toward the interfaces, both liquid–air and liquid–solid.

\subsection{Simulations with SAW: importance of Acoustic Radiation Pressure}

Once the parameters $J_{kl}$ are set, we proceed to run simulations that also incorporate the SAW. As a reminder, the SAW propagates from left to right, corresponding to the positive direction of the $x$-axis. Since one of our objectives is to identify the effect of acoustic radiation pressure on the overall dynamics of the system, and specifically on oil extraction, we conducted a series of simulations varying the parameter $p$. In particular, we present results for $p = 1$, which corresponds to the absence of acoustic radiation pressure, and for $p < 1$, where the acoustic radiation pressure is present.

\begin{figure}[t!]
\centering
\includegraphics[width=1.0\textwidth]{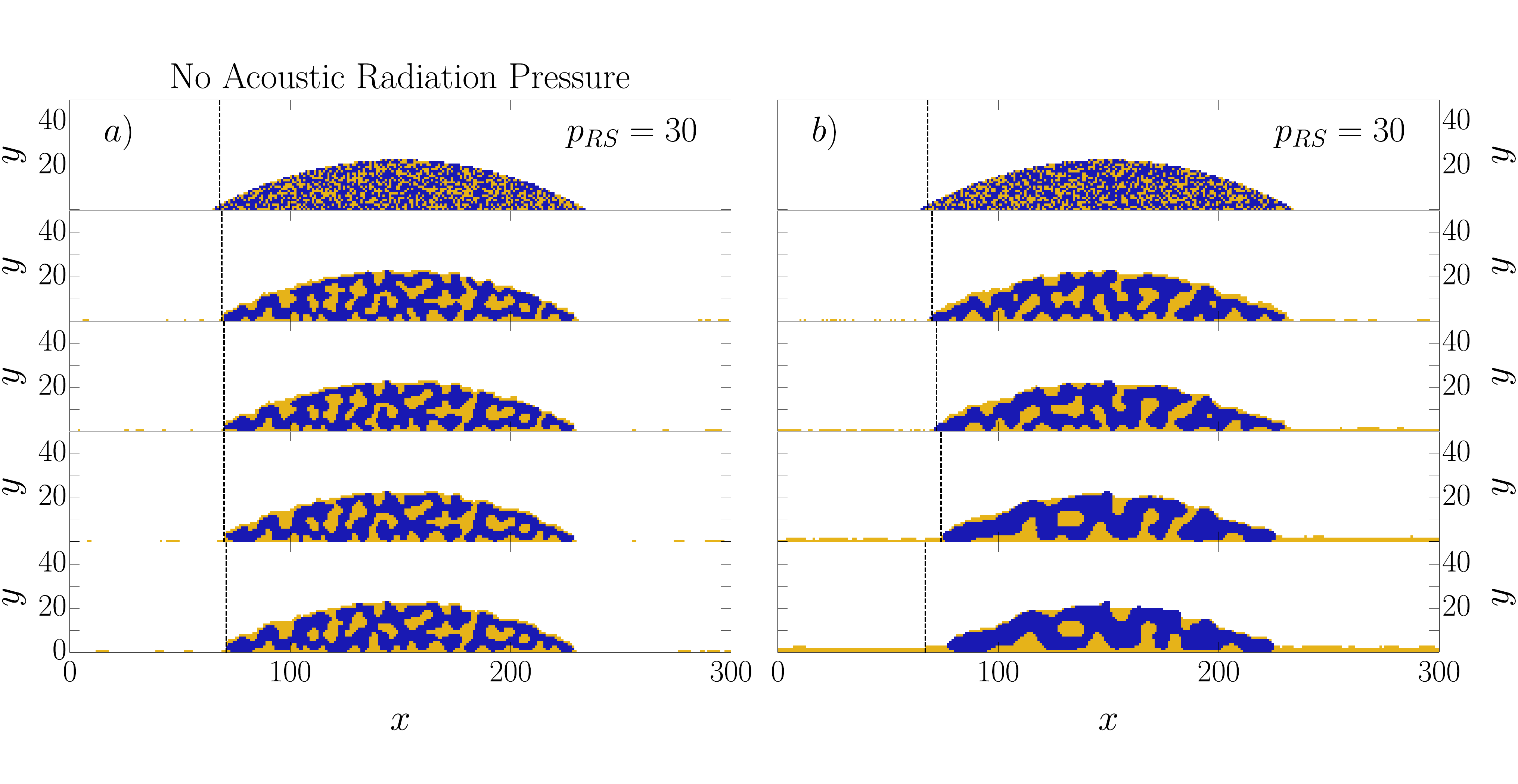}
\caption{Snapshots showing the evolution of the emulsion in the presence of SAW. In panel (a), acoustic radiation pressure is not considered, only the acoustic stress within the bulk of the liquid is included ($p = 1.0$). In panel (b), both the bulk acoustic stress and the acoustic radiation pressure at the free surface of the drop are taken into account ($p = 0.9$). The dashed black line indicates the position of the effective contact line, $x_B$. Note the formation of a thin oil film at later times in (b). The SAW intensity is set to $p_{RS} = 30$, and the total number of steps is $N_{\text{steps}} = 2 \cdot 10^{11}$}
\label{fig7:Graf_small_SAW}
\vspace{0.5cm}
\centering
\includegraphics[width=1.0\textwidth]{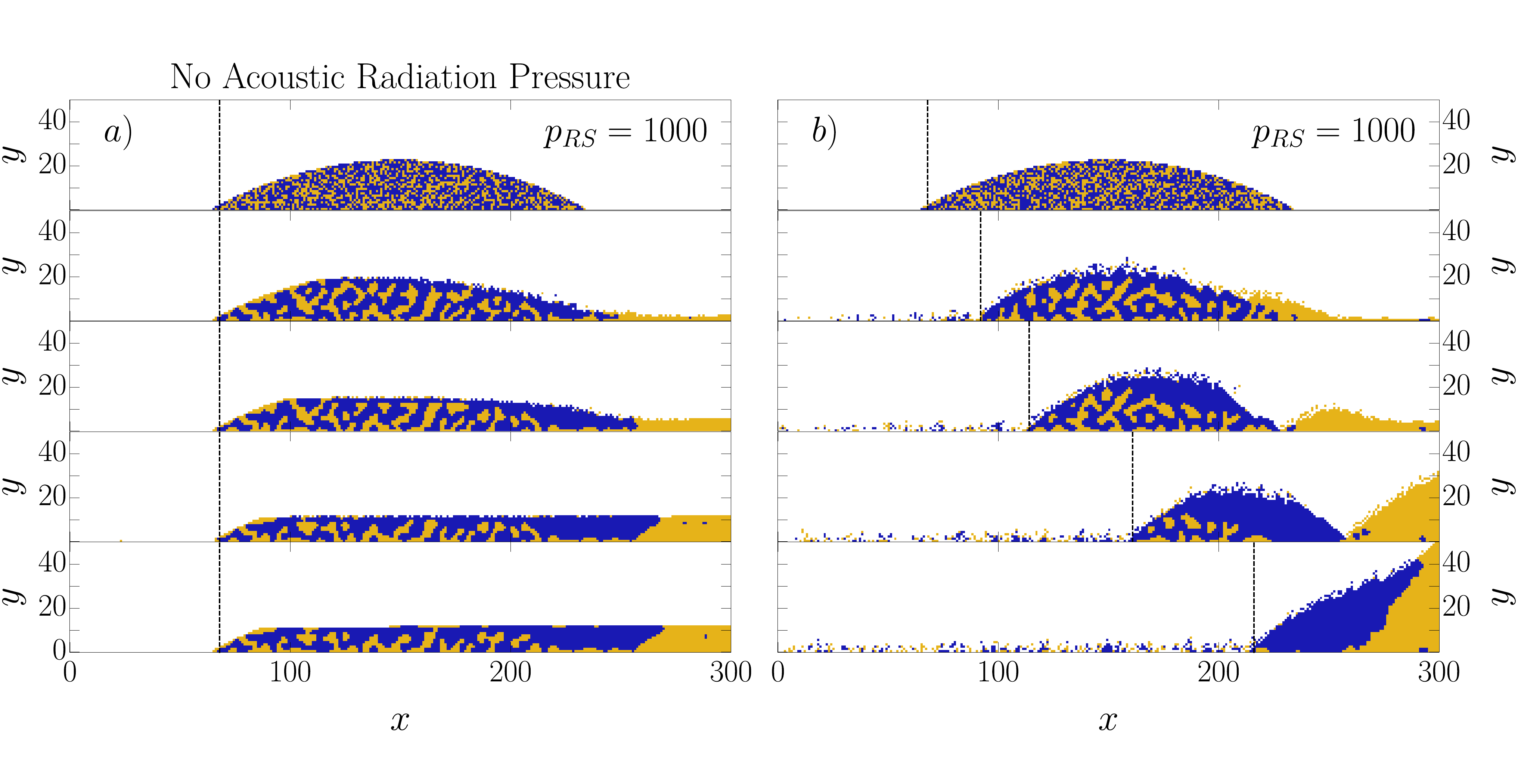}
\caption{Same as in Fig.~\ref{fig7:Graf_small_SAW}, but with $p_{RS} = 1000$ and $N_{\text{steps}} = 5 \cdot 10^{8}$.}
\label{fig7:Graf_high_SAW}
\end{figure}

Figures~\ref{fig7:Graf_small_SAW} and~\ref{fig7:Graf_high_SAW} show the results for weak and strong SAW intensities, characterized by the values of $p_{RS}$, and illustrate that, in both cases, there is a clear difference in the behavior of the system depending on the presence or absence of acoustic radiation pressure. In particular, Fig.~\ref{fig7:Graf_small_SAW} (corresponding to weak SAW intensity) highlights the crucial role of acoustic radiation pressure in enabling oil film formation. In the absence of this pressure, as shown in Fig.~\ref{fig7:Graf_small_SAW}a, virtually no oil particles escape from the droplet, and the droplet itself remains stationary, indicating minimal influence from the SAW. In contrast, Fig.~\ref{fig7:Graf_small_SAW}b, which includes acoustic radiation pressure, shows the emergence of a thin oil film on both sides of the droplet. This behavior is similar to that observed in physical experiments~\cite{paper_experimento}, although in those cases the oil film typically does not form in the direction of the SAW source (i.e., along the positive $x$-axis).

Figure~\ref{fig7:Graf_high_SAW}, which considers a substantially higher SAW intensity, also shows that acoustic radiation pressure plays an important role in the streaming of the simulated droplet. In Fig.~\ref{fig7:Graf_high_SAW}a, where there is no acoustic radiation pressure, although the particles are displaced in the direction of SAW propagation, the vertical line at $x = x_B$, which marks the transition between the film and the macroscopic droplet, and where the SAW begins to attenuate, remains stationary. In contrast, when acoustic radiation pressure is included, as in Fig.~\ref{fig7:Graf_high_SAW}b, the entire droplet is pushed forward in the direction of SAW propagation.

A close examination of both figures reveals that oil tends to migrate toward the surface of the macroscopic droplet. From there, the oil is extracted or pushed by the SAW. Additionally, Fig.~\ref{fig7:Graf_high_SAW} shows that oil is displaced more easily than water. Both effects are driven by its lower interaction energy $J_{oo}$, which corresponds to a lower surface tension.

As mentioned in the previous section, the parameter $p$ modulates the intensity of the acoustic radiation pressure; this contribution is maximal at $p = 0$ and vanishes when $p = 1$. For any value of $p \in [0,1)$, acoustic radiation pressure is present in the system, and our simulations display qualitatively similar behavior across this range. In the simulations presented here, we choose $p = 0.9$ for two main reasons. First, when $p$ is close to zero, the system behavior becomes noisy, making it more difficult to clearly identify each of the effects of the SAW. Second, simulations with $p$ values very close to unity, while qualitatively similar to those with $p = 0.9$, are more computationally expensive. Therefore, we adopt $p = 0.9$ as a compromise.

\begin{figure}[t!]
\centering
\includegraphics[width=0.5\textwidth]{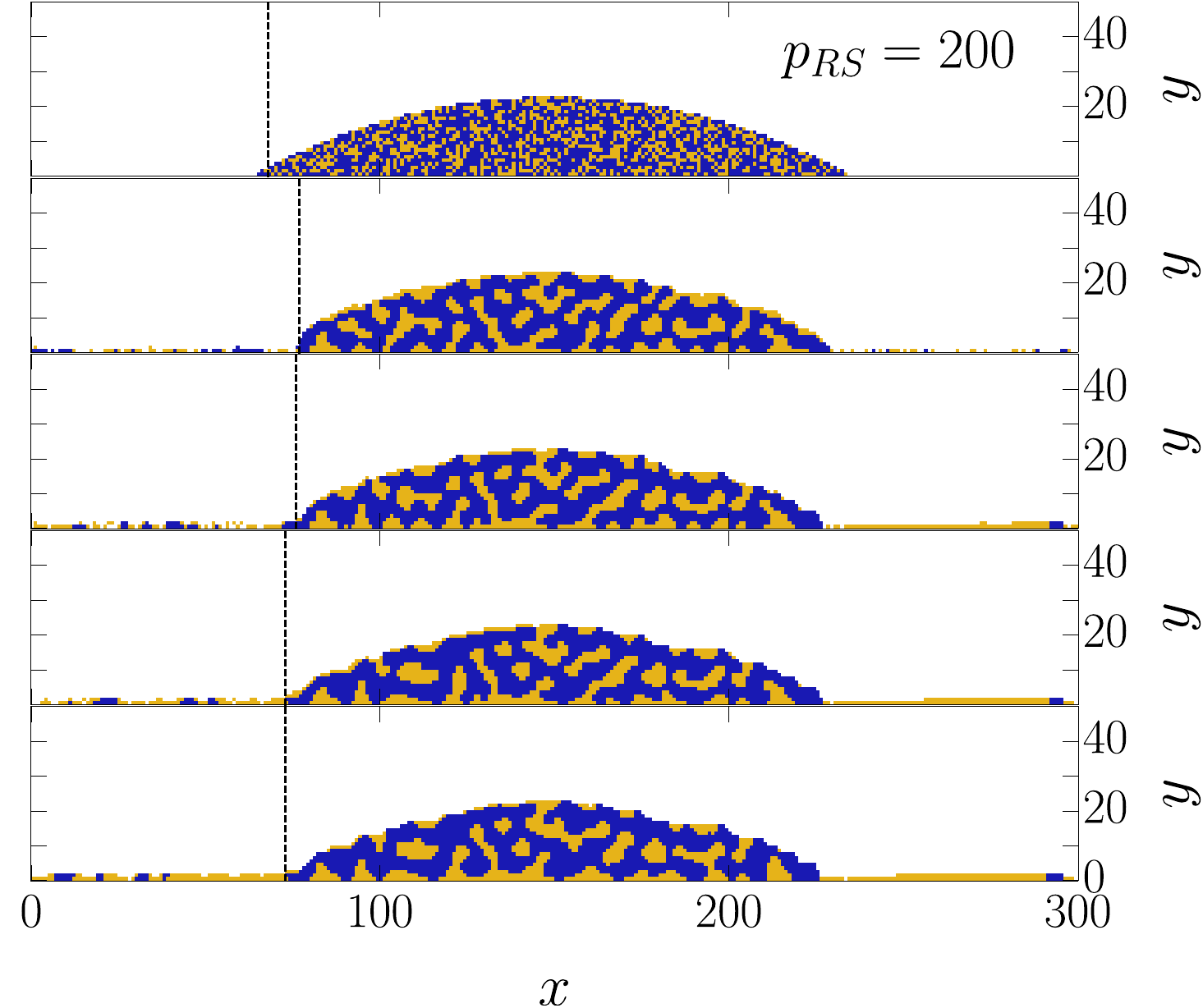}
\caption{Snapshots showing the evolution of the emulsion in the presence of SAW for $p_{RS}=200$, $p=0.9$, i.e. acoustic radiation pressure is present, and $N_{\text{steps}}=10^{9}$.}
\label{fig7:Graf_prs200}
\centering
\includegraphics[width=1.0\textwidth]{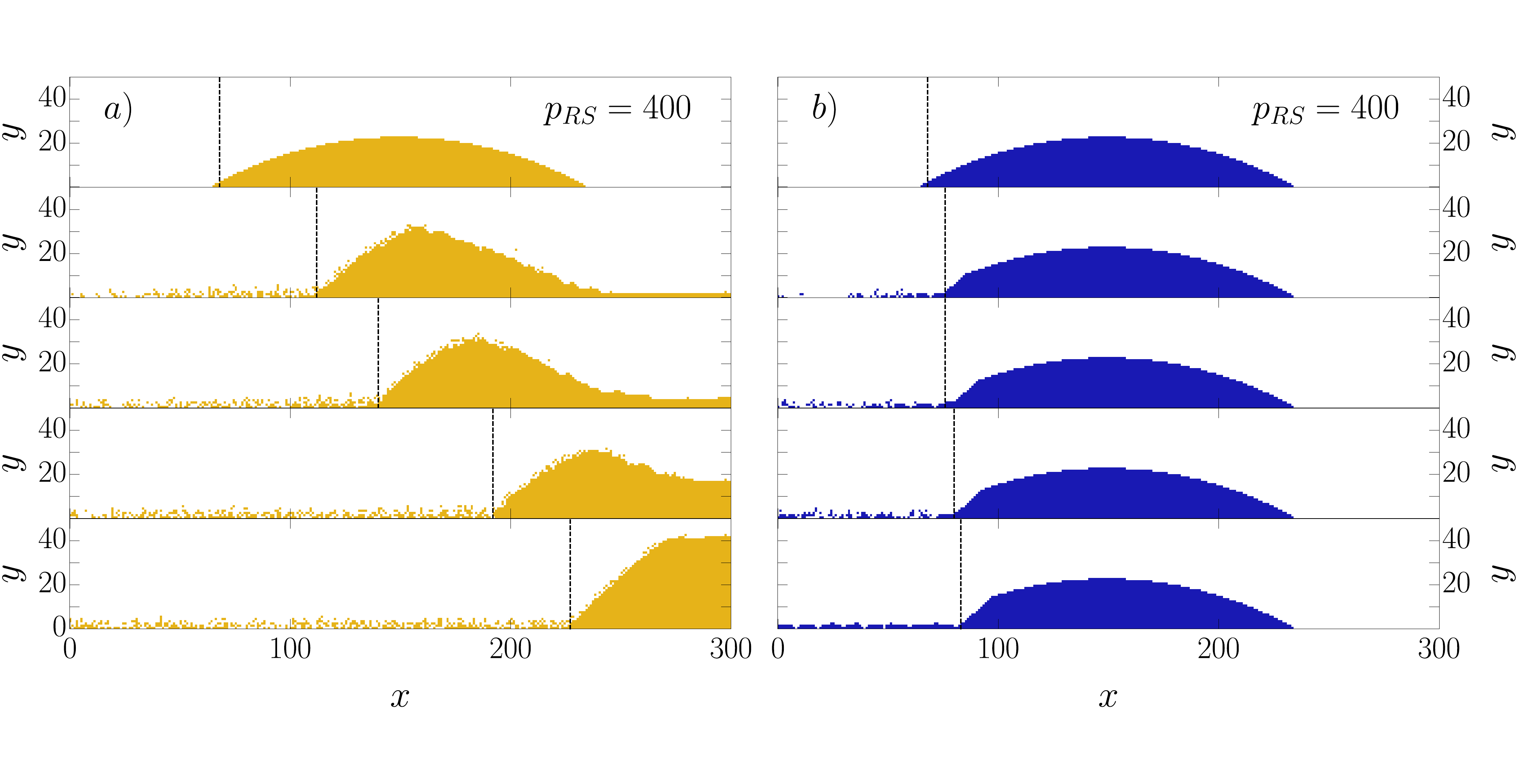}
\caption{Snapshots of the evolution of a pure system in the presence of SAW, for (a) pure oil, and (b) pure water. Here, $p_{RS}=400$, $p=0.9$ and $N_{\text{steps}}=10^{9}$.}
\label{fig7:Graf_pure_liquids_SAW}
\end{figure}

The SAW induces two distinct effects on the emulsion droplet, each occurring within different intensity regimes: the extraction of an oil film and the streaming of the macroscopic droplet away from the SAW source. At low SAW intensities (\mbox{$p_{RS} \sim 30$}, Fig.~\ref{fig7:Graf_small_SAW}), oil is extracted from the droplet, forming thin surface films. However, it is only at higher intensities (\mbox{$p_{RS} \sim 1000$}, Fig.~\ref{fig7:Graf_high_SAW}) that the streaming motion of the entire macroscopic emulsion droplet becomes apparent. Moreover, at intermediate intensities (\mbox{$p_{RS} \sim 100$}), it becomes possible to extract both oil and water, leading to films containing both components. This is illustrated in Fig.~\ref{fig7:Graf_prs200}, which presents a simulation with $p_{RS} = 200$.

For completeness, we briefly discuss the results obtained for pure oil and pure water droplets exposed to SAW. Figure~\ref{fig7:Graf_pure_liquids_SAW} illustrates a representative case, showing that, for the same value of $p_{RS}$, the SAW induces motion in the oil droplet, while the water droplet remains stationary, experiencing only a slight deformation. This contrasting behavior can be attributed to the different interaction energies, $J_{oo}$ and $J_{ww}$, which correspond to differences in surface tension. In both cases, a thin film of particles is observed moving in the direction opposite to the SAW propagation, an effect consistent with experimental observations for pure substances subjected to SAW~\cite{Rezk2012}.

\subsection{Time-dependent global results highlighting SAW influence}

Finally, we examine some quantitative, time-dependent results that reveal the global behavior of evolving emulsion drops. For simplicity, we focus on sufficiently small values of $ p_{RS} \leq 200 $ (with $ p = 0.9 $) to ensure that the drop remains stationary.

\begin{figure}[t]
\centering
\includegraphics[width=0.7\textwidth]{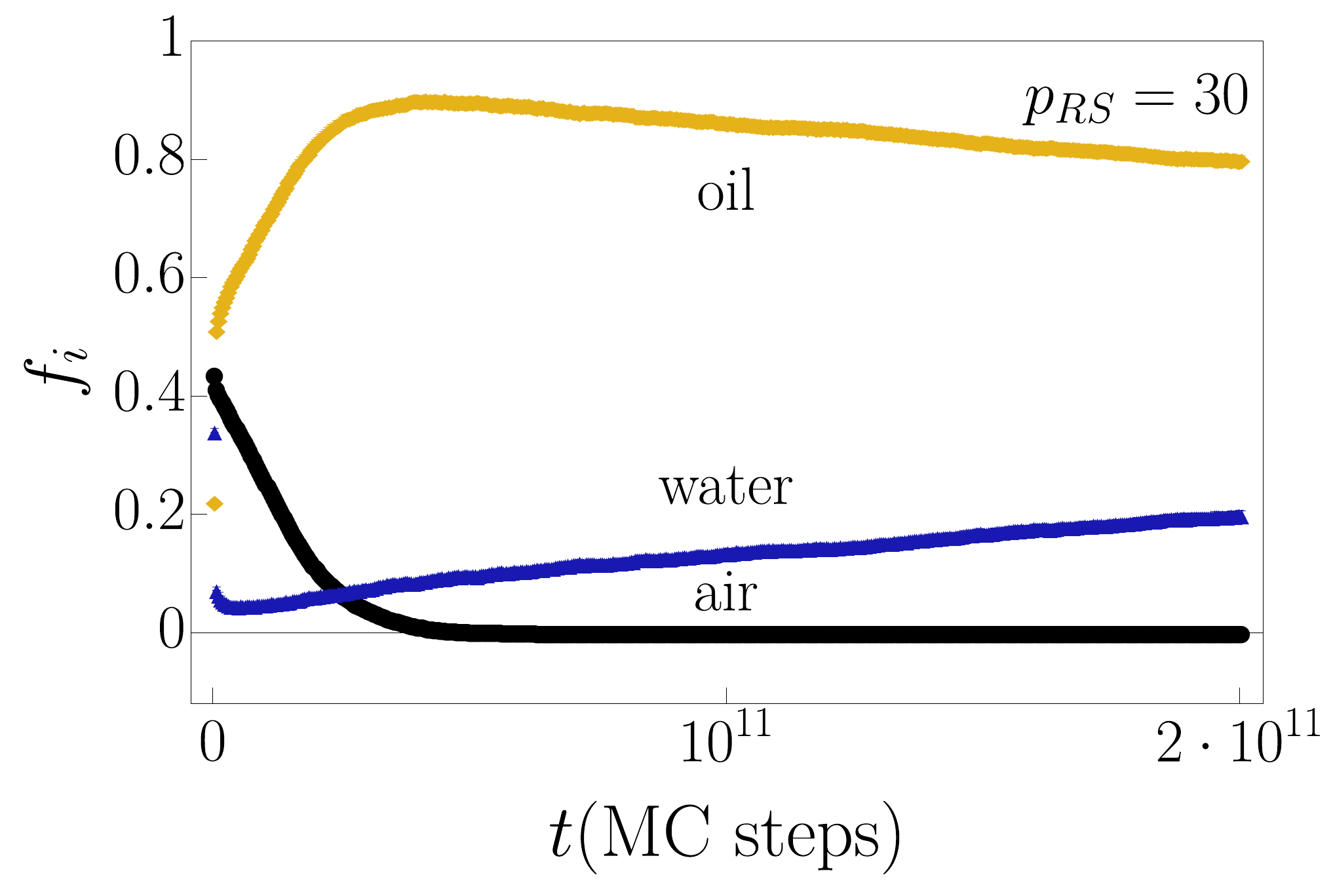}
\caption{Time evolution of the fractions $f_i$ (see text for the definition) of oil, water, and air, averaged over 50 realizations. Water is represented by blue triangles, oil by yellow diamonds, and air by black circles. The error bars are typically smaller than the symbol size. The parameters used are those of Fig.~\ref{fig7:Graf_small_SAW}b.}
\label{fig7:surface}
\end{figure}

To begin, Fig.~\ref{fig7:surface} illustrates how the composition at the droplet surface/interface evolves over time for $ p_{RS} = 30 $. The figure displays the surface fractions $ f_o $, $ f_w $, and $ f_a $, corresponding to oil, water, and air, respectively. To obtain the results shown in Fig.~\ref{fig7:surface}, we proceed as follows: at each time step, we scan each column of the cell array from top to bottom until encountering a liquid cell, either water or oil. The first cell encountered is designated as the surface cell, regardless of whether it belongs to the film or the macroscopic droplet. If no liquid cell is found before reaching the substrate, the column is classified as an air column. Although these air columns do not correspond to cells that are part of the droplet surface, we choose to include them in our analysis, as they offer additional insight into the dynamics of the system.

As a result of the initial condition (see top snapshots of Fig.~\ref{fig7:Graf_no_SAW}--\ref{fig7:Graf_prs200}), at $ t = 0 $ roughly half of the columns are classified as air. Additionally, the water fraction at the free surface of the droplet is approximately 1.5 times greater than that of oil. This occurs because the oil concentration in the droplet is $ c = 0.4 $, and the liquid cells are initially distributed at random.

At very early times, oil cells, with a lower interaction energy, quickly migrate to the surface, displacing water cells. This process takes place during the initial steps of the MC simulation and is illustrated in Fig.~\ref{fig7:surface}. As a result, the surface fraction of water drops sharply, appearing nearly discontinuous on the time scale shown in the figure. At intermediate times, the oil fraction at the interface continues to rise, while the air fraction begins to decline as oil is extracted from the droplet and spreads over the previously dry regions of the surface. Finally, once nearly all air columns have vanished from the system, its behavior shifts: the oil fraction begins to gradually decrease, while the water fraction increases. This trend is also visible in the final snapshots of Fig.~\ref{fig7:Graf_small_SAW}b, which use the same parameters as those of Fig.~\ref{fig7:surface}. This effect arises because the SAW extracts oil from the surface of the droplet, a process in which acoustic radiation pressure plays a leading role. At longer times, as this extraction continues, the oil content at the droplet surface declines and is progressively replaced by water, since there are not enough remaining oil cells to replenish those being removed.

\begin{figure}[t]
\centering
\includegraphics[width=0.7\textwidth]{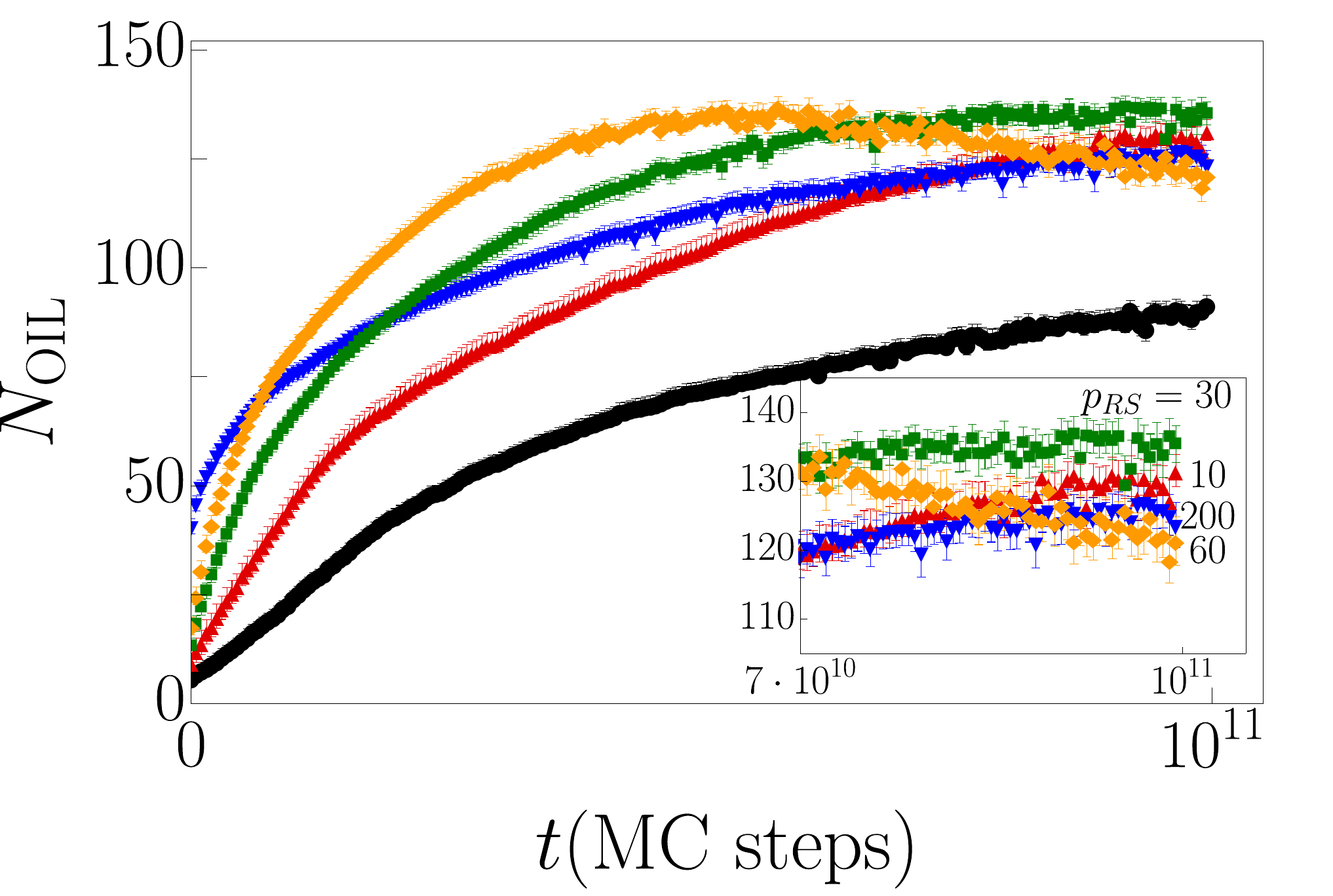}
\caption{Number of oil particles, $N_{\mathrm{OIL}}$, located to the left of the initial $x_B$ line (i.e., $x_B|_{t=0}$; see, e.g., the top snapshots of Fig.~\ref{fig7:Graf_small_SAW}), averaged over 50 simulation realizations for different values of $p_{RS}$, with $p = 0.9$. Inset: zoomed view at later times.}
\label{fig7:noil}
\end{figure}

Figure~\ref{fig7:noil} displays the number of oil particles located to the left of the initial position of the $ x_B $ line ($ x_B|_{t=0} $). This initial position is shown in the top snapshots of Figs.~\ref{fig7:Graf_small_SAW}--\ref{fig7:Graf_prs200}. In Fig.~\ref{fig7:noil}, we observe that, for $ p_{RS} = 0 $ (black line), a significant number of oil particles escape from the droplet, particularly at longer times. At early times, the amount of oil leaving the droplet consistently increases with $ p_{RS} $ (with $ p = 0.9 $ in all cases). This trend is expected, as once oil has accumulated at the droplet surface it becomes more easily extractable with increasing SAW intensity. However, at longer times, this trend breaks down. For higher values of $ p_{RS} $, such as 200, the system also begins to extract water, which hinders the extraction of oil from the droplet. This effect is not observed at moderate values of $ p_{RS} $, such as 30, where only oil is removed from the droplet.

\begin{figure}[t]
\centering
\includegraphics[width=0.7\textwidth]{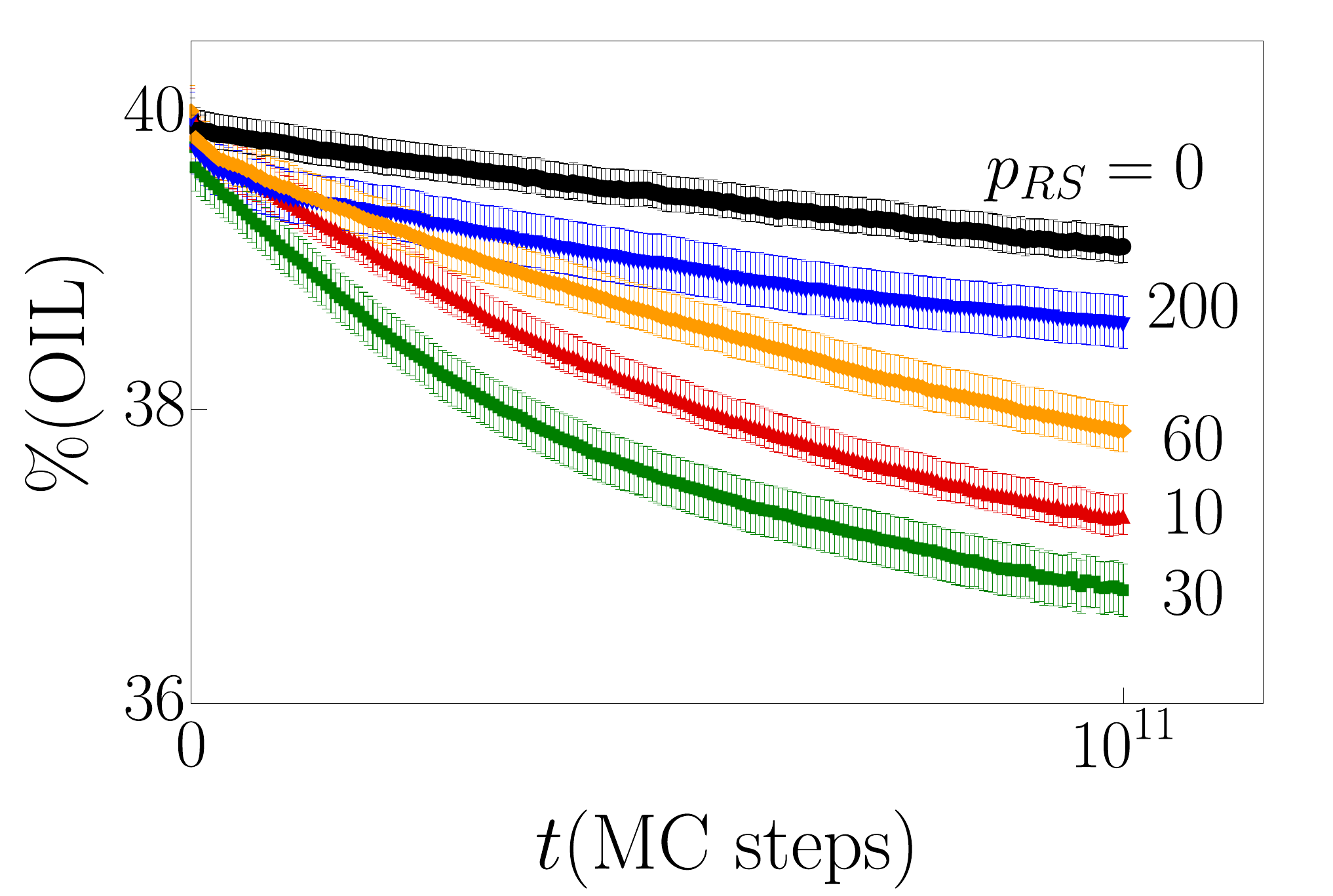}
\caption{Evolution of the oil content in the droplet, calculated as the percentage of oil cells located to the right of the (dynamical) boundary $ x_B (t)$, relative to the total number of liquid cells. Results are averaged over 50 simulation runs for various values of $ p_{RS} $ with $ p = 0.9 $. The color scheme follows that of Fig.~\ref{fig7:noil}.}
\label{fig7:poroil}
\end{figure}

Figure~\ref{fig7:poroil} presents the time evolution of the oil concentration within the macroscopic droplet, defined as the collection of all liquid cells located to the right of the dynamic (time-dependent) $ x_B\equiv x_B(t)$ line, i.e. the region where the SAW attenuates, for various values of $ p_{RS} $ (with $ p = 0.9 $ in all cases). At $ t = 0 $, all cases begin with an oil concentration of $ c = 0.4 $. As time advances and oil is extracted, this concentration decreases. Even in the absence of SAW excitation ($ p_{RS} = 0 $), the concentration drops slightly, since some oil particles can still escape from the droplet, as shown in Fig.~\ref{fig7:Graf_no_SAW}a. For small values of $ p_{RS} $, increasing its value enhances the efficiency of oil extraction. However, beyond a certain SAW intensity, around $ p_{RS} \sim 50 $, the oil concentration to the right of $ x_B(t) $ decreases more slowly compared to lower intensities. This is because, at these higher values, the SAW also begins to extract water from the macroscopic droplet. As a result, the ordering of the curves in Fig.~\ref{fig7:poroil} is no longer monotonic with $ p_{RS} $.

When interpreting both of these figures, it is important to note that, although we measure either the number of oil particles to the left of the initial $ x_B|_{t=0} $ line or the percentage of oil to the right of the time-dependent $ x_B(t) $ line, both metrics exclusively capture oil extraction in the direction opposite to the SAW, i.e. along the negative $ x $-axis. However, as illustrated in Fig.~\ref{fig7:Graf_small_SAW}b, a similar oil film also develops on the right side of the droplet, in the direction of the SAW source, along the positive $ x $-axis.

The time dependence of the $ x_B(t) $ line in Fig.~\ref{fig7:poroil} is a key factor to consider when comparing these results with those in Fig.~\ref{fig7:noil}, where the number of oil particles is measured relative to the initial position $ x_B|_{t=0} $. For instance, the case with $ p_{RS} = 200 $ shows a slightly higher percentage of extracted oil than the $ p_{RS} = 0 $ case when evaluated using the time-dependent $ x_B(t) $ reference. In contrast, Fig.~\ref{fig7:noil} shows a much larger number of oil particles for $ p_{RS} = 200 $ when measured with respect to the original $ x_B $ line. This discrepancy arises because the $ x_B $ boundary tends to shift leftward in the long time regime, as shown in Fig.~\ref{fig7:Graf_small_SAW}b, implying that some oil particles may be located to the left of the initial $ x_B $ position but still fall to the right of the dynamically updated one.

\clearpage
\section{Conclusions}
\vspace{-10pt}

In this chapter, we investigated oil extraction from an oil-in-water emulsion using a MC-based discrete model that accounts for interactions among oil, water, and air, as well as external forces such as gravity and SAW forcing. The results shed light on the underlying mechanisms driving oil separation under acoustic excitation and identify key factors influencing the observed dynamics.

Our simulations confirm that, irrespective of external forces, oil naturally moves to the droplet surface due to its lower surface energy. Upon introducing SAW forcing, both acoustic streaming within the liquid and acoustic radiation pressure at the free surface of the droplet are generated, promoting the formation of an oil film on and ahead of the droplet. This behavior is primarily governed by the contrast in surface tensions between oil and water, captured through appropriately defined interaction energies.

A key result is the pivotal role of acoustic radiation pressure in facilitating oil extraction. In simulations where this effect is omitted, oil remains confined within the droplet, and no film formation is observed. In contrast, when acoustic radiation pressure is included, an oil film detaches and spreads along the solid substrate, which is consistent with experimental observations~\cite{paper_experimento}. Furthermore, as the SAW intensity increases, oil extraction becomes progressively more efficient up to a critical threshold, beyond which water also begins to be extracted. At sufficiently high intensities, not only are oil and water extracted, but the entire droplet is set into motion, marking a transition from selective oil removal to bulk fluid transport.

The primary mechanism enabling oil extraction is the accumulation of oil at the free surface of the droplet, which serves as a reservoir from which oil is drawn into the film under SAW forcing. While acoustic stress within the droplet bulk drives flow along the SAW propagation direction, the detachment of the oil film and the advancement of the oil meniscus along the solid substrate result from an acoustic-capillary balance. This balance involves the acoustic radiation pressure exerted at the free surface opposing the capillary stress. For the oil phase, the acoustic stress dominates, allowing it to detach and spread. In contrast, for water, capillary forces prevail, keeping the phase stationary. Notably, explicit modeling of liquid–solid interaction forces is not required to capture the oil–water separation mechanism.

It is clear that a more rigorous treatment of the relative magnitudes of the various physical effects involved will be necessary for accurately modeling the quantitative details of any specific experimental setup. Capturing the complexities of real systems requires careful consideration of factors such as interfacial tensions, fluid viscosities, and the precise nature of acoustic interactions, which may vary significantly depending on the experimental conditions. Nevertheless, we are confident that the qualitative behavior observed in our simulations will remain robust under these more refined conditions.

In conclusion, our simulations demonstrate the effectiveness of discrete modeling in capturing the fundamental physics of oil extraction driven by SAWs, and emphasize the central role of acoustic radiation pressure in this process. Future work may aim to refine the model by incorporating additional experimental parameters, such as substrate interactions and more intricate fluid dynamics, with the objective of further narrowing the gap between simulation and real-world applications.
\graphicspath{{8_capitulo/fig8/}}

\chapter[Thesis summary and future work]{Thesis summary and \\ future work}


The thesis has pursued two main objectives: first, the analysis of growing fronts from the perspective of kinetic roughening, particularly in non-equilibrium systems where a surface can be defined and its dynamics analyzed through scaling hypotheses; and secondly, the Monte Carlo modeling of fluids, aimed at studying phenomena such as thin film spreading and the separation of target compounds from emulsions.

The relevance of this research lies in its focus on uncovering the fundamental mechanisms through which universal behaviors emerge in systems governed by disorder and fluctuations. This thesis quantifies critical exponents, examines universal scaling functions, and investigates a variety of systems where randomness plays a pivotal role, whether stemming from the intrinsic noise of stochastic partial differential equations or arising naturally from the probabilistic nature of Monte Carlo simulations.

A central goal throughout the work has been to deepen our understanding of how universality classes arise: how vastly different microscopic systems can exhibit the same macroscopic behavior, sharing scaling laws and statistical properties despite their structural and dynamical differences. By systematically analyzing these properties across distinct geometries and modeling approaches, the thesis contributes to a broader theoretical framework that helps explain why such universality holds, even in the presence of competing sources of randomness and complex BC.


This thesis is supported by four research works. In the first two, discussed in Chapters~\ref{chap4:band} and~\ref{chap5:radial_spreading}, we study the kinetic roughening properties of the precursor thin films of wetting band and circular droplets, and compare the results of both geometries. In these chapters, we have systematically determined the critical exponents—$\beta$, $\alpha$, and $z$—and consistently computed the universal functions that characterize the interface dynamics, including the probability distribution function of height fluctuations and the height-height covariance. Specifically, we proposed and implemented a novel method to estimate the correlation length directly from the height-difference correlation function in real space
, for cases where the function does not reach a clear saturation and exhibits oscillatory behavior. We hope this approach can be adopted in future studies that encounter similar features.

While some quantitative differences persist between the results obtained for the two geometries, the qualitative behavior remains consistent. Overall, we argue that the findings from both chapters strongly support the existence of a well-defined universality class governing these film spreading processes: one characterized by intrinsic anomalous scaling with temperature-dependent critical exponents, and a sensitivity to interface geometry. This geometrical dependence manifests in the subclass that governs the statistics of front fluctuations, aligning with theoretical expectations for one-dimensional KPZ-like interfaces. Interestingly, while the fluctuation statistics exhibit features reminiscent of 1D KPZ behavior, the associated critical exponents do not correspond to those of the standard KPZ class.

The third research work is presented in Chapter~\ref{chap6:KPZ}, where we investigate the integration of various stochastic partial differential equations, such as the KPZ equation, on the Bethe lattice. This chapter addresses a more analytical challenge, as it first requires a careful discussion of how such equations can be meaningfully integrated on network structures. To this end, we propose and compare novel numerical schemes for performing these integrations, focusing particularly on their numerical stability and ability to capture the correct growth dynamics. We tested several discretization methods and found that, despite their structural differences, all three produced largely indistinguishable results. In addition, we explored how key observables depend on the choice of BC. Although some differences emerged between Free and Neumann BC, the overall conclusions remained robust, with most observables showing small sensitivity to these variations.

Notably, our results, particularly those for the EW equation, show that the Bethe lattice (or more precisely, finite CT) cannot be considered a straightforward infinite-dimensional limit of hypercubic lattices in the context of stochastic growth models. The strong finite-size and boundary effects observed significantly influence the dynamics, posing challenges for using such structures to explore the infinite-dimensional limit of KPZ-type equations. While the numerical methods developed in this chapter offer a solid framework for studying growth on networked substrates, our findings underscore the need for caution when interpreting results in the presence of boundary-induced artifacts.

The final study presented in this thesis is detailed in Chapter~\ref{chap7:lou}, where we investigate oil extraction from an oil-in-water emulsion using a discrete Monte Carlo model that incorporates interactions among oil, water, and air, as well as external forces such as surface acoustic wave forcing. Our simulations reveal that, even in the absence of external forces, oil naturally migrates toward the droplet surface due to its lower surface energy. When SAW forcing is introduced, it generates both acoustic streaming within the droplet and acoustic radiation pressure at the free surface. These combined effects promote the formation of a thin oil film on and ahead of the droplet. This behavior is primarily driven by the difference in surface tensions between oil and water, which are effectively modeled through appropriately defined interaction energies. The most significant finding of this study is the central role played by acoustic radiation pressure in enabling oil extraction. In simulations where this effect is excluded, the oil remains trapped within the droplet, and no film formation occurs, underscoring its critical importance in the extraction mechanism.


Regarding open questions, several promising directions for future research remain within the systems discussed. In the context of the spreading model, a particularly compelling avenue would be to assess whether the conclusions drawn from the ``microscopic'' simulations presented here can be validated using alternative computational approaches, such as molecular dynamics or lattice-Boltzmann methods, or through experimental investigations of precursor film spreading. Additionally, it would be highly valuable to repeat the simulations in the band geometry using a front definition analogous to that employed in the radial case. While this would require a redefinition of certain observables, such as the height-difference correlation function, it could potentially resolve the quantitative discrepancies observed between the two geometries.


As for the integration of the KPZ and related equations on networks, it would be highly interesting to explore their behavior on other types of complex networks. For instance, integrating these equations on a Watts-Strogatz network, which enables tuning between regular and random topologies through a single parameter, or a complete graph, would provide an opportunity to investigate their properties in ``small-world'' regimes. Such a study could offer new insights into the behavior of these equations in the infinite-dimensional limit.

These integrations would introduce several challenges, particularly due to the coexistence of two sources of randomness: one arising from the intrinsic noise of the stochastic partial differential equation, and the other from the disorder inherent in the network topology itself. Nevertheless, we believe that techniques developed in the context of spin glasses, where similar dual sources of randomness are present, could be adapted and effectively applied to address these complexities.


With regard to the discrete model developed to study oil extraction from emulsions via surface acoustic waves, numerous avenues for further investigation remain open. One natural extension would be to perform analogous simulations within a fully three-dimensional framework, which could reveal additional features of the extraction process not captured in two dimensions.

To enhance the realism of the model, it would also be worthwhile to remove the effect of gravity and introduce interactions with the substrate. This would require a careful discussion on how best to represent such interactions, either through a simplified approach, such as assigning a constant negative energy to the first layer (as in Ref.~\cite{Areshi2019}), or through a more sophisticated model like the one presented in Chapter~\ref{chap2:wetting}, where the interaction energy depends on the distance from the substrate. Additionally, it would also be necessary to discuss how the substrate interaction constants of the water and the oil scale with each other.

Another improvement would be to refine the spatial profile of the SAW itself, i.e. changing the definition of the function $\mathcal{U}$. For instance, it is reasonable to assume that attenuation should not occur on the back side of the droplet if the film thickness is comparable to that of the front. This issue becomes even more complex in three dimensions, where the SAW influence would need to be defined along the entire perimeter of the macroscopic droplet. Taken together, these extensions could provide deeper insight into the mechanisms governing the extraction process. However, they also raise concerns regarding computational cost and, consequently, the practical feasibility of their implementation.

In conclusion, we hope that this thesis serves as a valuable reference for future researchers interested in the study of surface kinetic roughening and the modeling of fluid systems through Monte Carlo methods. The work provides a solid theoretical foundation, a clear and reproducible methodology, and a series of model systems that exemplify the complex and often subtle behaviors characteristic of non-equilibrium statistical physics. The models proposed here, due to their simplicity and adaptability, offer a flexible platform for extensions into more realistic and physically relevant scenarios. Alongside these models, we present a set of practical tools, such as the jackknife method for estimating statistical errors in highly correlated data, the novel approach developed here to estimate correlation lengths in oscillatory regimes, and the numerical integration methods designed for studying stochastic growth equations on networked structures, which together offer a versatile toolkit applicable across a wide range of problems in statistical physics and computational modeling.

\backmatter

\appendix
\chapter{Appendices}
\markboth{Appendices}{}

\renewcommand{\theequation}{\thesection.\arabic{equation}}
\renewcommand{\thesection}{\Alph{section}}

\section{Solution of the RD continuum equation} \label{app:RD}

We start from Eq.~\eqref{eq1:eq_simple}. Integrating over time we have
\begin{equation}
    h(\boldsymbol{x},t)=Ft+\int_0^t\eta(\boldsymbol{x},t')dt',
    \label{eqA:RD1}
\end{equation}
and thus
\begin{equation}
    \langle h(\boldsymbol{x},t)\rangle=Ft.
    \label{eqA:RD2}
\end{equation}
The mean of the square of Eq.~\eqref{eqA:RD1} can be computed as
\begin{equation}
    \langle h^2(\boldsymbol{x},t)\rangle=F^2t^2+2Dt,
    \label{eqA:RD3}
\end{equation}
so
\begin{equation}
    w^2(t)=\langle h^2(\boldsymbol{x},t)\rangle-\langle h(\boldsymbol{x},t)\rangle^2=2Dt,
    \label{eqA:RD4}
\end{equation}
therefore $w(t)\sim t^{1/2}$ indicating that the roughness exponent is
\begin{equation}
    \beta=\dfrac{1}{2}.
    \label{eqA:RD5}
\end{equation}

\section{Exact critical exponents in the EW equation} \label{app:Ew_exp}

The critical exponents of the EW equation can be determined using symmetry arguments. Given that the interface is self-affine, it remains invariant under the transformations in Eq.~\eqref{eq1:self_affine}. Consequently, the EW equation [Eq.~\eqref{eq1:EW}] must also remain invariant when time is rescaled as $t \rightarrow b^z t $ as well. Based on these transformations, we can write the EW equation as
\begin{equation}
\begin{aligned}
    \frac{\partial (b^\alpha h(\boldsymbol{x},t))}{\partial( b^z t)}=\nu \nabla^2 (b^\alpha h)+\eta(b\boldsymbol{x},b^zt),\\
    b^{\alpha-z}\frac{\partial h(\boldsymbol{x},t)}{\partial t}=\nu b^{\alpha-2}\nabla^2  h+\eta(b\boldsymbol{x},b^zt),
    \end{aligned}
    \label{eqA:eqEW1}
\end{equation}
and to analyze the second moment of the noise we use the relation 
\begin{equation}
    \delta^d(b\boldsymbol{x})=\frac{1}{b^d}\delta^d(\boldsymbol{x}),
    \label{eqA:eqEW2}
\end{equation}
and therefore
\begin{equation}
    \begin{aligned}
    \langle\eta(b\boldsymbol{x},b^zt)\eta(b\boldsymbol{x'},b^zt')\rangle=2D\delta^d(b(\boldsymbol{x}-\boldsymbol{x'}))\delta(b^z(t-t'))=\\
    2D b^{-(d+z)}\delta^d(\boldsymbol{x}-\boldsymbol{x'})\delta(t-t').
    \end{aligned}
    \label{eqA:eqEW3}
\end{equation}
Consequently, the scaled equation becomes
\begin{equation}
\begin{aligned}
    b^{\alpha-z}\frac{\partial h(\boldsymbol{x},t)}{\partial t}=\nu b^{\alpha-2}\nabla^2  h+b^{-(d+z)/2}\eta(\boldsymbol{x},t),\\
    \frac{\partial h(\boldsymbol{x},t)}{\partial t}=\nu b^{z-2}\nabla^2  h+b^{(z-d)/2-\alpha}\eta(\boldsymbol{x},t).\\
    \end{aligned}
    \label{eqA:eqEW4}
\end{equation}
For the equation to remain invariant, the following condition must hold:
\begin{equation}
\begin{aligned}
    z=2,\hspace{1cm}\alpha=\frac{2-d}{2},\hspace{1cm}\beta=\frac{\alpha}{z}=\frac{2-d}{4}.
    \end{aligned}
    \label{eqA:eqEW5}
\end{equation}

\section{Relation between \texorpdfstring{$\alpha$}{TEXT} and \texorpdfstring{$z$}{TEXT} in the KPZ equation} \label{app:KPZalphaz}

We start from the KPZ equation [Eq.~\eqref{eq1:KPZ}]. The effect of a small fluctuation $\delta \eta$ results in the formation of a bump or hole with length $\xi$ and height $\delta h$. Considering these as perturbations, we can rewrite the equation as 
\begin{equation}
    \frac{\delta h}{t} \approx \nu \frac{\delta h}{\xi^2} + \frac{\lambda}{2} \frac{(\delta h)^2}{\xi^2}.
    \label{eqA:eq1}
\end{equation}
Assuming the scaling of the width, Eq.~\eqref{eq1:wcomportamiento}, $ w \sim \langle \delta h \rangle  \sim L^{\alpha} $, and the scaling of the correlation length, Eq.~\eqref{eq1:xi}, $\xi \sim t^{1/z}$, we obtain that 

\begin{equation}
    t^{\frac{\alpha}{z}-1} \sim \nu t^{\frac{\alpha}{z}-\frac{2}{z}} + \frac{\lambda}{2} t^{\frac{2\alpha}{z}-\frac{2}{z}}.
    \label{eq:lambda}
\end{equation}
Since $\alpha/z>0$, the term proportional to $\lambda$ dominates: $\frac{2\alpha}{z}-\frac{2}{z}>\frac{\alpha}{z}-\frac{2}{z}$, then, $t^{\frac{2\alpha}{z}-\frac{2}{z}} \gg t^{\frac{\alpha}{z}-\frac{2}{z}}$. Note that, when $\lambda$ is absent, the universality class is EW. We can equate the exponent on the left hand side of Eq.~\eqref{eq:lambda} with the one in the term carrying $\lambda$ resulting into $\alpha + z =2$, as in Eq.~\eqref{eq1:KPZaz}.

\section[Relation between the coupling constants \texorpdfstring{$J_{kl}$}{TEXT} and the surface tensions of the liquids]{Relation between the coupling constants \texorpdfstring{$J_{kl}$}{TEXT} and the surface tensions of the liquids}\label{app:coupling}
In this appendix, we provide a straightforward calculation that connects the coupling constants of the particles in the model we formulate and study in Chapter~\ref{chap7:lou} to the known macroscopic surface tensions: water ($\gamma_w$), oil ($\gamma_o$), and the oil-water interfacial tension ($\gamma_{ow}$).

If $J_{ww}$ represents the typical binding energy between two water particles, then the binding energy per particle of water in the bulk is approximately
\begin{equation}
    E_{w,b}=-\frac{1}{2}J_{ww}Z_b,
\end{equation}
where $Z_b$ is the average number of neighbors and the factor 1/2 appears to avoid double-counting of interactions. Likewise, the binding energy per water particle at the surface is given by
\begin{equation}
    E_{w,s}=-\frac{1}{2}J_{ww}Z_s,
\end{equation}
where $Z_s$ is the average number of neighbors for a particle of water in the surface. As the surface tension is the energy required to create an interface per unit area, then
\begin{equation}
    \gamma_w=\frac{1}{2a}J_{ww}\left(Z_b-Z_s\right),
    \label{eq:surface_tension_water}
\end{equation}
where $a$ is the typical area occupied by a particle in the surface. Likewise, for oil we have
\begin{equation}
    \gamma_o=\frac{1}{2a}J_{oo}\left(Z_b-Z_s\right),
    \label{eq:surface_tension_oil}
\end{equation}
where the parameters $a$, $Z_s$, and $Z_b$ are, generally, not the same for oil and water. However, if we assume that they are similar, we find
\begin{equation}
    \frac{\gamma_w}{\gamma_o}\approx \frac{J_{ww}}{J_{oo}}.
    \label{eq:surface_tension_1}
\end{equation}
The interfacial tension $\gamma_{ow}$ between oil and water can be computed as
\begin{equation}
    \gamma_{ow}+\Delta W_{ow}=\gamma_w+\gamma_o,
\end{equation}
where $\Delta W_{ow}$ denotes the work per unit area required to split an oil-water interface into two separate interfaces: one between water and air, and the other between oil and air. This work can be estimated as
\begin{equation}
    \Delta W_{ow}=\frac{Z_b-Z_s}{a}J_{ow},
\end{equation}
where we assume that the number of pairs per unit area of the oil-water interface is the same as at the free surface of water and oil, i.e. $Z_s$ and $Z_b$ are the same as in Eq.~\eqref{eq:surface_tension_water} and \eqref{eq:surface_tension_oil}. This leads to
\begin{equation}
    \gamma_{ow}=\frac{Z_b-Z_s}{a}\left[\frac{1}{2}\left(J_{ww}+J_{oo}\right)-J_{ow}\right].
\end{equation}
From here it can be easily seen that
\begin{equation}
    \frac{\gamma_{ow}}{\gamma_{w}}=1+\frac{J_{oo}}{J_{ww}}-2\frac{J_{ow}}{J_{ww}},
\end{equation}
and
\begin{equation}
    J_{ow}=\frac{J_{ww}}{2}\left[1+\frac{J_{oo}}{J_{ww}}-\frac{\gamma_{ow}}{\gamma_{w}}\right]\approx\frac{J_{ww}}{2}\left[1+\frac{\gamma_{o}}{\gamma_{w}}-\frac{\gamma_{ow}}{\gamma_{w}}\right].
    \label{eq:surface_tension_2}
\end{equation}
If we take into account the experimental values of  $\gamma_w$, $\gamma_o$, and $\gamma_{ow}$ \cite{Peters2013}
$\gamma_o/\gamma_w\approx 0.28$ and $\gamma_{ow}/\gamma_w\approx0.5$, then the values of the $J_{oo}$ and $J_{ow}$ in terms of $J_{ww}$ that follow from Eq.~\eqref{eq:surface_tension_1} and \eqref{eq:surface_tension_2} are
\begin{equation}
    J_{oo}\approx 0.28 J_{ww} ,\hspace{8mm} J_{ow}\approx 0.4 J_{ww}.
    \label{eq:estimate}
\end{equation}
Assuming that $a$, $Z_s$, and $Z_b$ are identical across the three interfaces—water-air, oil-air, and water-oil—may be a rather rough approximation. However, since the water-water interaction ($J_{ww}$), dominated by hydrogen bonding, is significantly stronger than the oil-oil interaction ($J_{oo}$), governed by van der Waals forces, the estimates provided by Eq.~\eqref{eq:estimate} remain reasonable for our purposes.

\addcontentsline{toc}{chapter}{Bibliography}
\bibliographystyle{modified-apsrev4-2.bst}
\bibliography{ThinFilm_mod}

\begin{thebibliography}{213}%
\makeatletter
\providecommand \@ifxundefined [1]{%
 \@ifx{#1\undefined}
}%
\providecommand \@ifnum [1]{%
 \ifnum #1\expandafter \@firstoftwo
 \else \expandafter \@secondoftwo
 \fi
}%
\providecommand \@ifx [1]{%
 \ifx #1\expandafter \@firstoftwo
 \else \expandafter \@secondoftwo
 \fi
}%
\providecommand \natexlab [1]{#1}%
\providecommand \enquote  [1]{``#1''}%
\providecommand \bibnamefont  [1]{#1}%
\providecommand \bibfnamefont [1]{#1}%
\providecommand \citenamefont [1]{#1}%
\providecommand \href@noop [0]{\@secondoftwo}%
\providecommand \href [0]{\begingroup \@sanitize@url \@href}%
\providecommand \@href[1]{\@@startlink{#1}\@@href}%
\providecommand \@@href[1]{\endgroup#1\@@endlink}%
\providecommand \@sanitize@url [0]{\catcode `\\12\catcode `\$12\catcode
  `\&12\catcode `\#12\catcode `\^12\catcode `\_12\catcode `\%12\relax}%
\providecommand \@@startlink[1]{}%
\providecommand \@@endlink[0]{}%
\providecommand \url  [0]{\begingroup\@sanitize@url \@url }%
\providecommand \@url [1]{\endgroup\@href {#1}{\urlprefix }}%
\providecommand \urlprefix  [0]{URL }%
\providecommand \Eprint [0]{\href }%
\providecommand \doibase [0]{https://doi.org/}%
\providecommand \selectlanguage [0]{\@gobble}%
\providecommand \bibinfo  [0]{\@secondoftwo}%
\providecommand \bibfield  [0]{\@secondoftwo}%
\providecommand \translation [1]{[#1]}%
\providecommand \BibitemOpen [0]{}%
\providecommand \bibitemStop [0]{}%
\providecommand \bibitemNoStop [0]{.\EOS\space}%
\providecommand \EOS [0]{\spacefactor3000\relax}%
\providecommand \BibitemShut  [1]{\csname bibitem#1\endcsname}%
\let\auto@bib@innerbib\@empty
\bibitem [{\citenamefont {Barab{\'{a}}si}\ and\ \citenamefont
  {Stanley}(1995)}]{Barabasi1995}%
  \BibitemOpen
  \bibfield  {author} {\bibinfo {author} {\bibfnamefont {A.-L.}\ \bibnamefont
  {Barab{\'{a}}si}}\ and\ \bibinfo {author} {\bibfnamefont {H.~E.}\
  \bibnamefont {Stanley}},\ }\href {https://doi.org/10.1017/cbo9780511599798}
  {{\textit{\href {https://doi.org/10.1017/cbo9780511599798} {}}}\bibinfo
  {title} {\textit{Fractal Concepts in Surface Growth}}}\ (\bibinfo
  {publisher} {Cambridge University Press, Cambridge},\ \bibinfo {address}
  {Cambridge},\ \bibinfo {year} {1995})\BibitemShut {NoStop}%
\bibitem [{\citenamefont {Evans}\ \emph {et~al.}(2006)\citenamefont {Evans},
  \citenamefont {Thiel},\ and\ \citenamefont {Bartelt}}]{Evans2006}%
  \BibitemOpen
  \bibfield  {author} {\bibinfo {author} {\bibfnamefont {J.}~\bibnamefont
  {Evans}}, \bibinfo {author} {\bibfnamefont {P.}~\bibnamefont {Thiel}},\ and\
  \bibinfo {author} {\bibfnamefont {M.}~\bibnamefont {Bartelt}},\ }\bibfield
  {title} {\bibinfo {title} {{\textit{Morphological evolution during epitaxial
  thin film growth: Formation of 2D islands and 3D mounds}}},\ }\href
  {https://doi.org/10.1016/j.surfrep.2005.08.004} {\bibfield  {journal}
  {\bibinfo  {journal} {Surf. Sci. Rep.}\ }\textbf {\bibinfo {volume} {61}},\
  \bibinfo {pages} {1–128} (\bibinfo {year} {2006})}\BibitemShut {NoStop}%
\bibitem [{\citenamefont {Krug}(1997)}]{Krug1997}%
  \BibitemOpen
  \bibfield  {author} {\bibinfo {author} {\bibfnamefont {J.}~\bibnamefont
  {Krug}},\ }\bibfield  {title} {\bibinfo {title} {{\textit{Origins of scale
  invariance in growth processes}}},\ }\href
  {https://doi.org/10.1080/00018739700101498} {\bibfield  {journal} {\bibinfo
  {journal} {Adv. Phys.}\ }\textbf {\bibinfo {volume} {46}},\ \bibinfo {pages}
  {139} (\bibinfo {year} {1997})}\BibitemShut {NoStop}%
\bibitem [{\citenamefont {Family}(1986)}]{Family1986}%
  \BibitemOpen
  \bibfield  {author} {\bibinfo {author} {\bibfnamefont {F.}~\bibnamefont
  {Family}},\ }\bibfield  {title} {\bibinfo {title} {{\textit{Scaling of rough
  surfaces: effects of surface diffusion}}},\ }\href
  {https://doi.org/10.1088/0305-4470/19/8/006} {\bibfield  {journal} {\bibinfo
  {journal} {J. Phys. A: Math. Gen.}\ }\textbf {\bibinfo {volume} {19}},\
  \bibinfo {pages} {L441} (\bibinfo {year} {1986})}\BibitemShut {NoStop}%
\bibitem [{\citenamefont {Family}\ and\ \citenamefont
  {Vicsek}(1991)}]{Family1991}%
  \BibitemOpen
  \bibfield  {author} {\bibinfo {author} {\bibfnamefont {F.}~\bibnamefont
  {Family}}\ and\ \bibinfo {author} {\bibfnamefont {T.}~\bibnamefont
  {Vicsek}},\ }\href {https://doi.org/10.1142/1452} {{\textit{\href
  {https://doi.org/10.1142/1452} {}}}\bibinfo {title} {\textit{Dynamics of
  Fractal Surfaces}}}\ (\bibinfo  {publisher} {World Scientific, Singapore},\
  \bibinfo {year} {1991})\BibitemShut {NoStop}%
\bibitem [{\citenamefont {Meakin}(1998)}]{Meakin1998}%
  \BibitemOpen
  \bibfield  {author} {\bibinfo {author} {\bibfnamefont {P.}~\bibnamefont
  {Meakin}},\ }\href
  {https://www.cambridge.org/es/universitypress/subjects/physics/nonlinear-science-and-fluid-dynamics/fractals-scaling-and-growth-far-equilibrium?format=PB&isbn=9780521189811}
  {{\textit{\href
  {https://www.cambridge.org/es/universitypress/subjects/physics/nonlinear-science-and-fluid-dynamics/fractals-scaling-and-growth-far-equilibrium?format=PB&isbn=9780521189811}
  {}}}\bibinfo {title} {\textit{Fractals, Scaling and Growth Far from
  Equilibrium}}}\ (\bibinfo  {publisher} {Cambridge University Press,
  Cambridge},\ \bibinfo {address} {Cambridge},\ \bibinfo {year}
  {1998})\BibitemShut {NoStop}%
\bibitem [{\citenamefont {Kriecherbauer}\ and\ \citenamefont
  {Krug}(2010)}]{Kriecherbauer2010}%
  \BibitemOpen
  \bibfield  {author} {\bibinfo {author} {\bibfnamefont {T.}~\bibnamefont
  {Kriecherbauer}}\ and\ \bibinfo {author} {\bibfnamefont {J.}~\bibnamefont
  {Krug}},\ }\bibfield  {title} {\bibinfo {title} {{\textit{A
  pedestrian{\textquotesingle}s view on interacting particle systems, {KPZ}
  universality and random matrices}}},\ }\href
  {https://doi.org/10.1088/1751-8113/43/40/403001} {\bibfield  {journal}
  {\bibinfo  {journal} {J. Phys. A: Math. Theor.}\ }\textbf {\bibinfo {volume}
  {43}},\ \bibinfo {pages} {403001} (\bibinfo {year} {2010})}\BibitemShut
  {NoStop}%
\bibitem [{\citenamefont {Takeuchi}(2018)}]{Takeuchi2018}%
  \BibitemOpen
  \bibfield  {author} {\bibinfo {author} {\bibfnamefont {K.~A.}\ \bibnamefont
  {Takeuchi}},\ }\bibfield  {title} {\bibinfo {title} {{\textit{An appetizer to
  modern developments on the Kardar{\textendash}Parisi{\textendash}Zhang
  universality class}}},\ }\href {https://doi.org/10.1016/j.physa.2018.03.009}
  {\bibfield  {journal} {\bibinfo  {journal} {Physica A: Stat. Mech. Appl.}\
  }\textbf {\bibinfo {volume} {504}},\ \bibinfo {pages} {77} (\bibinfo {year}
  {2018})}\BibitemShut {NoStop}%
\bibitem [{\citenamefont {T\"{a}uber}(2014)}]{tauber2014}%
  \BibitemOpen
  \bibfield  {author} {\bibinfo {author} {\bibfnamefont {U.~C.}\ \bibnamefont
  {T\"{a}uber}},\ }\href {https://doi.org/10.1017/cbo9781139046213}
  {{\textit{\href {https://doi.org/10.1017/cbo9781139046213} {}}}\bibinfo
  {title} {\textit{Critical Dynamics: A Field Theory Approach to Equilibrium
  and Non-Equilibrium Scaling Behavior}}}\ (\bibinfo  {publisher} {Cambridge
  University Press, Cambridge},\ \bibinfo {year} {2014})\BibitemShut {NoStop}%
\bibitem [{\citenamefont {Wilson}\ and\ \citenamefont
  {Kogut}(1974)}]{Wilson1974}%
  \BibitemOpen
  \bibfield  {author} {\bibinfo {author} {\bibfnamefont {K.~G.}\ \bibnamefont
  {Wilson}}\ and\ \bibinfo {author} {\bibfnamefont {J.}~\bibnamefont {Kogut}},\
  }\bibfield  {title} {\bibinfo {title} {{\textit{The renormalization group and
  the $\epsilon$ expansion}}},\ }\href
  {https://doi.org/https://doi.org/10.1016/0370-1573(74)90023-4} {\bibfield
  {journal} {\bibinfo  {journal} {Phys. Rep.}\ }\textbf {\bibinfo {volume}
  {12}},\ \bibinfo {pages} {75} (\bibinfo {year} {1974})}\BibitemShut {NoStop}%
\bibitem [{\citenamefont {Sethna}(2006)}]{Sethna2006}%
  \BibitemOpen
  \bibfield  {author} {\bibinfo {author} {\bibfnamefont {J.}~\bibnamefont
  {Sethna}},\ }\href
  {https://doi.org/https://doi.org/10.1093/oso/9780198865247.001.0001}
  {{\textit{\href
  {https://doi.org/https://doi.org/10.1093/oso/9780198865247.001.0001}
  {}}}\bibinfo {title} {\textit{Statistical Mechanics: Entropy, Order
  Parameters and Complexity}}}\ (\bibinfo  {publisher} {Oxford University
  Press, Oxford},\ \bibinfo {year} {2006})\BibitemShut {NoStop}%
\bibitem [{\citenamefont {\rm Barreales}(2024)}]{GarciaBarreales2024}%
  \BibitemOpen
  \bibfield  {author} {\href {https://dehesa.unex.es/handle/10662/18744}
  {{\textit{\bibinfo {author} {\bibfnamefont {B.~G.}\ \bibnamefont {\rm
  Barreales}}}}},\ }\bibinfo {title} {\textit{Universality and kinetic
  roughening properties in non-equilibrium fronts}},\ \href
  {https://dehesa.unex.es/handle/10662/18744} {Ph.D. thesis},\ \bibinfo
  {school} {Universidad de Extremadura} (\bibinfo {year} {2024})\BibitemShut
  {NoStop}%
\bibitem [{\citenamefont {L\'opez}\ \emph {et~al.}(1997)\citenamefont
  {L\'opez}, \citenamefont {Rodr\'{\i}guez},\ and\ \citenamefont
  {Cuerno}}]{Lopez1997}%
  \BibitemOpen
  \bibfield  {author} {\bibinfo {author} {\bibfnamefont {J.~M.}\ \bibnamefont
  {L\'opez}}, \bibinfo {author} {\bibfnamefont {M.~A.}\ \bibnamefont
  {Rodr\'{\i}guez}},\ and\ \bibinfo {author} {\bibfnamefont {R.}~\bibnamefont
  {Cuerno}},\ }\bibfield  {title} {\bibinfo {title} {{\textit{Superroughening
  versus intrinsic anomalous scaling of surfaces}}},\ }\href
  {https://doi.org/10.1103/PhysRevE.56.3993} {\bibfield  {journal} {\bibinfo
  {journal} {Phys. Rev. E}\ }\textbf {\bibinfo {volume} {56}},\ \bibinfo
  {pages} {3993} (\bibinfo {year} {1997})}\BibitemShut {NoStop}%
\bibitem [{\citenamefont {Baiod}\ \emph {et~al.}(1988)\citenamefont {Baiod},
  \citenamefont {Kessler}, \citenamefont {Ramanlal}, \citenamefont {Sander},\
  and\ \citenamefont {Savit}}]{Baiod1988}%
  \BibitemOpen
  \bibfield  {author} {\bibinfo {author} {\bibfnamefont {R.}~\bibnamefont
  {Baiod}}, \bibinfo {author} {\bibfnamefont {D.}~\bibnamefont {Kessler}},
  \bibinfo {author} {\bibfnamefont {P.}~\bibnamefont {Ramanlal}}, \bibinfo
  {author} {\bibfnamefont {L.}~\bibnamefont {Sander}},\ and\ \bibinfo {author}
  {\bibfnamefont {R.}~\bibnamefont {Savit}},\ }\bibfield  {title} {\bibinfo
  {title} {{\textit{Dynamical scaling of the surface of finite-density
  ballistic aggregation}}},\ }\href {https://doi.org/10.1103/PhysRevA.38.3672}
  {\bibfield  {journal} {\bibinfo  {journal} {Phys. Rev. A}\ }\textbf {\bibinfo
  {volume} {38}},\ \bibinfo {pages} {3672} (\bibinfo {year}
  {1988})}\BibitemShut {NoStop}%
\bibitem [{\citenamefont {Meakin}\ \emph {et~al.}(1986)\citenamefont {Meakin},
  \citenamefont {Ramanlal}, \citenamefont {Sander},\ and\ \citenamefont
  {Ball}}]{Meakin1986}%
  \BibitemOpen
  \bibfield  {author} {\bibinfo {author} {\bibfnamefont {P.}~\bibnamefont
  {Meakin}}, \bibinfo {author} {\bibfnamefont {P.}~\bibnamefont {Ramanlal}},
  \bibinfo {author} {\bibfnamefont {L.~M.}\ \bibnamefont {Sander}},\ and\
  \bibinfo {author} {\bibfnamefont {R.~C.}\ \bibnamefont {Ball}},\ }\bibfield
  {title} {\bibinfo {title} {{\textit{Ballistic deposition on surfaces}}},\
  }\href {https://doi.org/10.1103/PhysRevA.34.5091} {\bibfield  {journal}
  {\bibinfo  {journal} {Phys. Rev. A}\ }\textbf {\bibinfo {volume} {34}},\
  \bibinfo {pages} {5091} (\bibinfo {year} {1986})}\BibitemShut {NoStop}%
\bibitem [{\citenamefont {Kim}\ and\ \citenamefont
  {Kosterlitz}(1989)}]{Kim1989}%
  \BibitemOpen
  \bibfield  {author} {\bibinfo {author} {\bibfnamefont {J.~M.}\ \bibnamefont
  {Kim}}\ and\ \bibinfo {author} {\bibfnamefont {J.~M.}\ \bibnamefont
  {Kosterlitz}},\ }\bibfield  {title} {\bibinfo {title} {{\textit{Growth in a
  restricted solid-on-solid model}}},\ }\href
  {https://doi.org/10.1103/physrevlett.62.2289} {\bibfield  {journal} {\bibinfo
   {journal} {Phys. Rev. Lett.}\ }\textbf {\bibinfo {volume} {62}},\ \bibinfo
  {pages} {2289–2292} (\bibinfo {year} {1989})}\BibitemShut {NoStop}%
\bibitem [{\citenamefont {Kelling}\ \emph {et~al.}(2016)\citenamefont
  {Kelling}, \citenamefont {Ódor},\ and\ \citenamefont
  {Gemming}}]{Kelling2016}%
  \BibitemOpen
  \bibfield  {author} {\bibinfo {author} {\bibfnamefont {J.}~\bibnamefont
  {Kelling}}, \bibinfo {author} {\bibfnamefont {G.}~\bibnamefont {Ódor}},\
  and\ \bibinfo {author} {\bibfnamefont {S.}~\bibnamefont {Gemming}},\
  }\bibfield  {title} {\bibinfo {title} {{\textit{Universality of
  (2+1)-dimensional restricted solid-on-solid models}}},\ }\href
  {http://dx.doi.org/10.1103/PhysRevE.94.022107} {\bibfield  {journal}
  {\bibinfo  {journal} {Phys. Rev. E}\ }\textbf {\bibinfo {volume} {94}}
  (\bibinfo {year} {2016})}\BibitemShut {NoStop}%
\bibitem [{\citenamefont {Sudijono}(2023)}]{Sudijono2023}%
  \BibitemOpen
  \bibfield  {author} {\bibinfo {author} {\bibfnamefont {T.}~\bibnamefont
  {Sudijono}},\ }\href {https://arxiv.org/abs/2304.07160} {\bibinfo {title}
  {{\textit{Fluctuation Bounds for the Restricted Solid-on-Solid Model of
  Surface Growth}}}} (\bibinfo {year} {2023}),\ \Eprint
  {https://arxiv.org/abs/2304.07160} {arXiv:2304.07160 [math.PR]} \BibitemShut
  {NoStop}%
\bibitem [{\citenamefont {Eden}(1961)}]{Eden1961}%
  \BibitemOpen
  \bibfield  {author} {\bibinfo {author} {\bibfnamefont {M.}~\bibnamefont
  {Eden}},\ }\bibfield  {title} {\bibinfo {title} {{\textit{A Two-dimensional
  Growth Process}}},\ }in\ \href
  {https://projecteuclid.org/euclid.bsmsp/1200512888} {\bibinfo {booktitle}
  {Proceedings of the Fourth Berkeley Symposium on Mathematical Statistics and
  Probability}},\ Vol.~\bibinfo {volume} {4}\ (\bibinfo  {publisher}
  {University of California Press},\ \bibinfo {year} {1961})\ pp.\ \bibinfo
  {pages} {223--239}\BibitemShut {NoStop}%
\bibitem [{\citenamefont {Mansfield}\ and\ \citenamefont
  {Klushin}(1994)}]{Mansfield1994}%
  \BibitemOpen
  \bibfield  {author} {\bibinfo {author} {\bibfnamefont {M.~L.}\ \bibnamefont
  {Mansfield}}\ and\ \bibinfo {author} {\bibfnamefont {L.~I.}\ \bibnamefont
  {Klushin}},\ }\bibfield  {title} {\bibinfo {title} {{\textit{A generalized
  Eden model of polymer crystallization}}},\ }\href
  {https://doi.org/10.1016/0032-3861(94)90403-0} {\bibfield  {journal}
  {\bibinfo  {journal} {Polymer}\ }\textbf {\bibinfo {volume} {35}},\ \bibinfo
  {pages} {2937–2943} (\bibinfo {year} {1994})}\BibitemShut {NoStop}%
\bibitem [{\citenamefont {Tian}\ and\ \citenamefont {Xia}(2024)}]{Tian2024}%
  \BibitemOpen
  \bibfield  {author} {\bibinfo {author} {\bibfnamefont {X.}~\bibnamefont
  {Tian}}\ and\ \bibinfo {author} {\bibfnamefont {H.}~\bibnamefont {Xia}},\
  }\bibfield  {title} {\bibinfo {title} {{\textit{Crossover effects and dynamic
  scaling properties from Eden growth to diffusion-limited aggregation}}},\
  }\href {https://doi.org/10.1016/j.physleta.2024.129494} {\bibfield  {journal}
  {\bibinfo  {journal} {Phys. Lett. A}\ }\textbf {\bibinfo {volume} {508}},\
  \bibinfo {pages} {129494} (\bibinfo {year} {2024})}\BibitemShut {NoStop}%
\bibitem [{\citenamefont {Witten}\ and\ \citenamefont
  {Sander}(1981)}]{Witten1981}%
  \BibitemOpen
  \bibfield  {author} {\bibinfo {author} {\bibfnamefont {T.~A.}\ \bibnamefont
  {Witten}}\ and\ \bibinfo {author} {\bibfnamefont {L.~M.}\ \bibnamefont
  {Sander}},\ }\bibfield  {title} {\bibinfo {title} {{\textit{Diffusion-Limited
  Aggregation, a Kinetic Critical Phenomenon}}},\ }\href
  {https://doi.org/10.1103/physrevlett.47.1400} {\bibfield  {journal} {\bibinfo
   {journal} {Phys. Rev. Lett.}\ }\textbf {\bibinfo {volume} {47}},\ \bibinfo
  {pages} {1400–1403} (\bibinfo {year} {1981})}\BibitemShut {NoStop}%
\bibitem [{\citenamefont {Castro}\ \emph {et~al.}(2000)\citenamefont {Castro},
  \citenamefont {Cuerno}, \citenamefont {S{\'{a}}nchez},\ and\ \citenamefont
  {Dom{\'{i}}nguez-Adame}}]{Castro2000}%
  \BibitemOpen
  \bibfield  {author} {\bibinfo {author} {\bibfnamefont {M.}~\bibnamefont
  {Castro}}, \bibinfo {author} {\bibfnamefont {R.}~\bibnamefont {Cuerno}},
  \bibinfo {author} {\bibfnamefont {A.}~\bibnamefont {S{\'{a}}nchez}},\ and\
  \bibinfo {author} {\bibfnamefont {F.}~\bibnamefont {Dom{\'{i}}nguez-Adame}},\
  }\bibfield  {title} {\bibinfo {title} {{\textit{{Multiparticle biased
  diffusion-limited aggregation with surface diffusion: A comprehensive model
  of electrodeposition}}}},\ }\href {https://doi.org/10.1103/PhysRevE.62.161}
  {\bibfield  {journal} {\bibinfo  {journal} {Phys. Rev. E}\ }\textbf {\bibinfo
  {volume} {62}},\ \bibinfo {pages} {161} (\bibinfo {year} {2000})}\BibitemShut
  {NoStop}%
\bibitem [{\citenamefont {Kardar}\ \emph {et~al.}(1986)\citenamefont {Kardar},
  \citenamefont {Parisi},\ and\ \citenamefont {Zhang}}]{Kardar1986}%
  \BibitemOpen
  \bibfield  {author} {\bibinfo {author} {\bibfnamefont {M.}~\bibnamefont
  {Kardar}}, \bibinfo {author} {\bibfnamefont {G.}~\bibnamefont {Parisi}},\
  and\ \bibinfo {author} {\bibfnamefont {Y.-C.}\ \bibnamefont {Zhang}},\
  }\bibfield  {title} {\bibinfo {title} {{\textit{Dynamic Scaling of Growing
  Interfaces}}},\ }\href {https://doi.org/10.1103/physrevlett.56.889}
  {\bibfield  {journal} {\bibinfo  {journal} {Phys. Rev. Lett.}\ }\textbf
  {\bibinfo {volume} {56}},\ \bibinfo {pages} {889–892} (\bibinfo {year}
  {1986})}\BibitemShut {NoStop}%
\bibitem [{\citenamefont {Halpin-Healy}\ and\ \citenamefont
  {Takeuchi}(2015)}]{HalpinHealy2015}%
  \BibitemOpen
  \bibfield  {author} {\bibinfo {author} {\bibfnamefont {T.}~\bibnamefont
  {Halpin-Healy}}\ and\ \bibinfo {author} {\bibfnamefont {K.~A.}\ \bibnamefont
  {Takeuchi}},\ }\bibfield  {title} {\bibinfo {title} {{\textit{{A KPZ
  cocktail-shaken, not stirred... Toasting 30 years of kinetically roughened
  surfaces}}}},\ }\href {https://doi.org/10.1007/s10955-015-1282-1} {\bibfield
  {journal} {\bibinfo  {journal} {J. Stat. Phys.}\ }\textbf {\bibinfo {volume}
  {160}},\ \bibinfo {pages} {794} (\bibinfo {year} {2015})}\BibitemShut
  {NoStop}%
\bibitem [{\citenamefont {Oliveira}(2022)}]{Oliveira2022}%
  \BibitemOpen
  \bibfield  {author} {\bibinfo {author} {\bibfnamefont {T.~J.}\ \bibnamefont
  {Oliveira}},\ }\bibfield  {title} {\bibinfo {title}
  {{\textit{Kardar-Parisi-Zhang universality class in $(d+1)$-dimensions}}},\
  }\href {https://doi.org/10.1103/PhysRevE.106.L062103} {\bibfield  {journal}
  {\bibinfo  {journal} {Phys. Rev. E}\ }\textbf {\bibinfo {volume} {106}},\
  \bibinfo {pages} {L062103} (\bibinfo {year} {2022})}\BibitemShut {NoStop}%
\bibitem [{\citenamefont {Colaiori}\ and\ \citenamefont
  {Moore}(2001)}]{Colaiori2001}%
  \BibitemOpen
  \bibfield  {author} {\bibinfo {author} {\bibfnamefont {F.}~\bibnamefont
  {Colaiori}}\ and\ \bibinfo {author} {\bibfnamefont {M.~A.}\ \bibnamefont
  {Moore}},\ }\bibfield  {title} {\bibinfo {title} {{\textit{Upper Critical
  Dimension, Dynamic Exponent, and Scaling Functions in the Mode-Coupling
  Theory for the Kardar-Parisi-Zhang Equation}}},\ }\href
  {https://doi.org/10.1103/PhysRevLett.86.3946} {\bibfield  {journal} {\bibinfo
   {journal} {Phys. Rev. Lett.}\ }\textbf {\bibinfo {volume} {86}},\ \bibinfo
  {pages} {3946} (\bibinfo {year} {2001})}\BibitemShut {NoStop}%
\bibitem [{\citenamefont {Bouchaud}\ and\ \citenamefont
  {Cates}(1993)}]{Bouchaud1993}%
  \BibitemOpen
  \bibfield  {author} {\bibinfo {author} {\bibfnamefont {J.~P.}\ \bibnamefont
  {Bouchaud}}\ and\ \bibinfo {author} {\bibfnamefont {M.~E.}\ \bibnamefont
  {Cates}},\ }\bibfield  {title} {\bibinfo {title} {{\textit{Self-consistent
  approach to the Kardar-Parisi-Zhang equation}}},\ }\href
  {https://doi.org/10.1103/PhysRevE.47.R1455} {\bibfield  {journal} {\bibinfo
  {journal} {Phys. Rev. E}\ }\textbf {\bibinfo {volume} {47}},\ \bibinfo
  {pages} {R1455} (\bibinfo {year} {1993})}\BibitemShut {NoStop}%
\bibitem [{\citenamefont {Doherty}\ \emph {et~al.}(1994)\citenamefont
  {Doherty}, \citenamefont {Moore}, \citenamefont {Kim},\ and\ \citenamefont
  {Bray}}]{Doherty1994}%
  \BibitemOpen
  \bibfield  {author} {\bibinfo {author} {\bibfnamefont {J.~P.}\ \bibnamefont
  {Doherty}}, \bibinfo {author} {\bibfnamefont {M.~A.}\ \bibnamefont {Moore}},
  \bibinfo {author} {\bibfnamefont {J.~M.}\ \bibnamefont {Kim}},\ and\ \bibinfo
  {author} {\bibfnamefont {A.~J.}\ \bibnamefont {Bray}},\ }\bibfield  {title}
  {\bibinfo {title} {{\textit{Generalizations of the Kardar-Parisi-Zhang
  equation}}},\ }\href {https://doi.org/10.1103/PhysRevLett.72.2041} {\bibfield
   {journal} {\bibinfo  {journal} {Phys. Rev. Lett.}\ }\textbf {\bibinfo
  {volume} {72}},\ \bibinfo {pages} {2041} (\bibinfo {year}
  {1994})}\BibitemShut {NoStop}%
\bibitem [{\citenamefont {Halpin-Healy}(1990)}]{HalpinHealy1990}%
  \BibitemOpen
  \bibfield  {author} {\bibinfo {author} {\bibfnamefont {T.}~\bibnamefont
  {Halpin-Healy}},\ }\bibfield  {title} {\bibinfo {title}
  {{\textit{Disorder-induced roughening of diverse manifolds}}},\ }\href
  {https://doi.org/10.1103/PhysRevA.42.711} {\bibfield  {journal} {\bibinfo
  {journal} {Phys. Rev. A}\ }\textbf {\bibinfo {volume} {42}},\ \bibinfo
  {pages} {711} (\bibinfo {year} {1990})}\BibitemShut {NoStop}%
\bibitem [{\citenamefont {Kloss}\ \emph
  {et~al.}(2014{\natexlab{a}})\citenamefont {Kloss}, \citenamefont {Canet},
  \citenamefont {Delamotte},\ and\ \citenamefont {Wschebor}}]{Kloss2014}%
  \BibitemOpen
  \bibfield  {author} {\bibinfo {author} {\bibfnamefont {T.}~\bibnamefont
  {Kloss}}, \bibinfo {author} {\bibfnamefont {L.}~\bibnamefont {Canet}},
  \bibinfo {author} {\bibfnamefont {B.}~\bibnamefont {Delamotte}},\ and\
  \bibinfo {author} {\bibfnamefont {N.}~\bibnamefont {Wschebor}},\ }\bibfield
  {title} {\bibinfo {title} {{\textit{Kardar-Parisi-Zhang equation with
  spatially correlated noise: A unified picture from nonperturbative
  renormalization group}}},\ }\href
  {https://doi.org/10.1103/PhysRevE.89.022108} {\bibfield  {journal} {\bibinfo
  {journal} {Phys. Rev. E}\ }\textbf {\bibinfo {volume} {89}},\ \bibinfo
  {pages} {022108} (\bibinfo {year} {2014}{\natexlab{a}})}\BibitemShut
  {NoStop}%
\bibitem [{\citenamefont {Kloss}\ \emph
  {et~al.}(2014{\natexlab{b}})\citenamefont {Kloss}, \citenamefont {Canet},\
  and\ \citenamefont {Wschebor}}]{Kloss2014-2}%
  \BibitemOpen
  \bibfield  {author} {\bibinfo {author} {\bibfnamefont {T.}~\bibnamefont
  {Kloss}}, \bibinfo {author} {\bibfnamefont {L.}~\bibnamefont {Canet}},\ and\
  \bibinfo {author} {\bibfnamefont {N.}~\bibnamefont {Wschebor}},\ }\bibfield
  {title} {\bibinfo {title} {{\textit{Strong-coupling phases of the anisotropic
  Kardar-Parisi-Zhang equation}}},\ }\href
  {https://doi.org/10.1103/PhysRevE.90.062133} {\bibfield  {journal} {\bibinfo
  {journal} {Phys. Rev. E}\ }\textbf {\bibinfo {volume} {90}},\ \bibinfo
  {pages} {062133} (\bibinfo {year} {2014}{\natexlab{b}})}\BibitemShut
  {NoStop}%
\bibitem [{\citenamefont {Castellano}\ \emph
  {et~al.}(1998{\natexlab{a}})\citenamefont {Castellano}, \citenamefont
  {Marsili},\ and\ \citenamefont {Pietronero}}]{Castellano1998}%
  \BibitemOpen
  \bibfield  {author} {\bibinfo {author} {\bibfnamefont {C.}~\bibnamefont
  {Castellano}}, \bibinfo {author} {\bibfnamefont {M.}~\bibnamefont
  {Marsili}},\ and\ \bibinfo {author} {\bibfnamefont {L.}~\bibnamefont
  {Pietronero}},\ }\bibfield  {title} {\bibinfo {title}
  {{\textit{Nonperturbative Renormalization of the Kardar-Parisi-Zhang Growth
  Dynamics}}},\ }\href {https://doi.org/10.1103/PhysRevLett.80.3527} {\bibfield
   {journal} {\bibinfo  {journal} {Phys. Rev. Lett.}\ }\textbf {\bibinfo
  {volume} {80}},\ \bibinfo {pages} {3527} (\bibinfo {year}
  {1998}{\natexlab{a}})}\BibitemShut {NoStop}%
\bibitem [{\citenamefont {Castellano}\ \emph
  {et~al.}(1998{\natexlab{b}})\citenamefont {Castellano}, \citenamefont
  {Gabrielli}, \citenamefont {Marsili}, \citenamefont {Mu\~noz},\ and\
  \citenamefont {Pietronero}}]{Castellano1998-2}%
  \BibitemOpen
  \bibfield  {author} {\bibinfo {author} {\bibfnamefont {C.}~\bibnamefont
  {Castellano}}, \bibinfo {author} {\bibfnamefont {A.}~\bibnamefont
  {Gabrielli}}, \bibinfo {author} {\bibfnamefont {M.}~\bibnamefont {Marsili}},
  \bibinfo {author} {\bibfnamefont {M.~A.}\ \bibnamefont {Mu\~noz}},\ and\
  \bibinfo {author} {\bibfnamefont {L.}~\bibnamefont {Pietronero}},\ }\bibfield
   {title} {\bibinfo {title} {{\textit{High dimensional behavior of the
  Kardar-Parisi-Zhang growth dynamics}}},\ }\href
  {https://doi.org/10.1103/PhysRevE.58.R5209} {\bibfield  {journal} {\bibinfo
  {journal} {Phys. Rev. E}\ }\textbf {\bibinfo {volume} {58}},\ \bibinfo
  {pages} {R5209} (\bibinfo {year} {1998}{\natexlab{b}})}\BibitemShut {NoStop}%
\bibitem [{\citenamefont {Alves}\ and\ \citenamefont
  {Ferreira}(2016)}]{Alves2016}%
  \BibitemOpen
  \bibfield  {author} {\bibinfo {author} {\bibfnamefont {S.~G.}\ \bibnamefont
  {Alves}}\ and\ \bibinfo {author} {\bibfnamefont {S.~C.}\ \bibnamefont
  {Ferreira}},\ }\bibfield  {title} {\bibinfo {title} {{\textit{Scaling,
  cumulant ratios, and height distribution of ballistic deposition in $3+1$ and
  $4+1$ dimensions}}},\ }\href {https://doi.org/10.1103/PhysRevE.93.052131}
  {\bibfield  {journal} {\bibinfo  {journal} {Phys. Rev. E}\ }\textbf {\bibinfo
  {volume} {93}},\ \bibinfo {pages} {052131} (\bibinfo {year}
  {2016})}\BibitemShut {NoStop}%
\bibitem [{\citenamefont {Kim}\ and\ \citenamefont {Kim}(2014)}]{Kim2014}%
  \BibitemOpen
  \bibfield  {author} {\bibinfo {author} {\bibfnamefont {S.-W.}\ \bibnamefont
  {Kim}}\ and\ \bibinfo {author} {\bibfnamefont {J.~M.}\ \bibnamefont {Kim}},\
  }\bibfield  {title} {\bibinfo {title} {{\textit{A restricted solid-on-solid
  model in higher dimensions}}},\ }\href
  {https://doi.org/10.1088/1742-5468/2014/07/P07005} {\bibfield  {journal}
  {\bibinfo  {journal} {J. Stat. Mech.: Theor. Exp.}\ }\textbf {\bibinfo
  {volume} {2014}},\ \bibinfo {pages} {P07005} (\bibinfo {year}
  {2014})}\BibitemShut {NoStop}%
\bibitem [{\citenamefont {Alves}\ \emph {et~al.}(2014)\citenamefont {Alves},
  \citenamefont {Oliveira},\ and\ \citenamefont {Ferreira}}]{Alves2014}%
  \BibitemOpen
  \bibfield  {author} {\bibinfo {author} {\bibfnamefont {S.~G.}\ \bibnamefont
  {Alves}}, \bibinfo {author} {\bibfnamefont {T.~J.}\ \bibnamefont
  {Oliveira}},\ and\ \bibinfo {author} {\bibfnamefont {S.~C.}\ \bibnamefont
  {Ferreira}},\ }\bibfield  {title} {\bibinfo {title} {{\textit{Universality of
  fluctuations in the Kardar-Parisi-Zhang class in high dimensions and its
  upper critical dimension}}},\ }\href
  {http://dx.doi.org/10.1103/PhysRevE.90.020103} {\bibfield  {journal}
  {\bibinfo  {journal} {Phys. Rev. E}\ }\textbf {\bibinfo {volume} {90}}
  (\bibinfo {year} {2014})}\BibitemShut {NoStop}%
\bibitem [{\citenamefont {Saberi}(2013)}]{Saberi2013}%
  \BibitemOpen
  \bibfield  {author} {\bibinfo {author} {\bibfnamefont {A.~A.}\ \bibnamefont
  {Saberi}},\ }\bibfield  {title} {\bibinfo {title} {{\textit{Growth models on
  the Bethe lattice}}},\ }\href {https://doi.org/10.1209/0295-5075/103/10005}
  {\bibfield  {journal} {\bibinfo  {journal} {Europhys. Lett.}\ }\textbf
  {\bibinfo {volume} {103}},\ \bibinfo {pages} {10005} (\bibinfo {year}
  {2013})}\BibitemShut {NoStop}%
\bibitem [{\citenamefont {Baxter}(1985)}]{Baxter1985}%
  \BibitemOpen
  \bibfield  {author} {\bibinfo {author} {\bibfnamefont {R.~J.}\ \bibnamefont
  {Baxter}},\ }\bibinfo {title} {{\textit{Exactly Solved Models in Statistical
  Mechanics}}},\ in\ \href {https://doi.org/10.1142/9789814415255_0002}
  {\bibinfo {booktitle} {Integrable Systems in Statistical Mechanics}}\
  (\bibinfo  {publisher} {World Scientific, Singapore},\ \bibinfo {year}
  {1985})\ p.\ \bibinfo {pages} {5–63}\BibitemShut {NoStop}%
\bibitem [{\citenamefont {Kurata}\ \emph {et~al.}(2004)\citenamefont {Kurata},
  \citenamefont {Kikuchi},\ and\ \citenamefont {Watari}}]{Kurata1953}%
  \BibitemOpen
  \bibfield  {author} {\bibinfo {author} {\bibfnamefont {M.}~\bibnamefont
  {Kurata}}, \bibinfo {author} {\bibfnamefont {R.}~\bibnamefont {Kikuchi}},\
  and\ \bibinfo {author} {\bibfnamefont {T.}~\bibnamefont {Watari}},\
  }\bibfield  {title} {\bibinfo {title} {{\textit{{A Theory of Cooperative
  Phenomena. III. Detailed Discussions of the Cluster Variation Method}}}},\
  }\href {https://doi.org/10.1063/1.1698926} {\bibfield  {journal} {\bibinfo
  {journal} {J. Chem. Phys.}\ }\textbf {\bibinfo {volume} {21}},\ \bibinfo
  {pages} {434} (\bibinfo {year} {2004})}\BibitemShut {NoStop}%
\bibitem [{\citenamefont {Oliveira}(2021)}]{Oliveira2021}%
  \BibitemOpen
  \bibfield  {author} {\bibinfo {author} {\bibfnamefont {T.~J.}\ \bibnamefont
  {Oliveira}},\ }\bibfield  {title} {\bibinfo {title} {{\textit{Surface growth
  on tree-like lattices and the upper critical dimension of the KPZ class}}},\
  }\href {https://doi.org/10.1209/0295-5075/133/28001} {\bibfield  {journal}
  {\bibinfo  {journal} {Europhys. Lett.}\ }\textbf {\bibinfo {volume} {133}},\
  \bibinfo {pages} {28001} (\bibinfo {year} {2021})}\BibitemShut {NoStop}%
\bibitem [{\citenamefont {Alves}\ \emph {et~al.}(2013)\citenamefont {Alves},
  \citenamefont {Oliveira},\ and\ \citenamefont {Ferreira}}]{Alves2013}%
  \BibitemOpen
  \bibfield  {author} {\bibinfo {author} {\bibfnamefont {S.~G.}\ \bibnamefont
  {Alves}}, \bibinfo {author} {\bibfnamefont {T.~J.}\ \bibnamefont
  {Oliveira}},\ and\ \bibinfo {author} {\bibfnamefont {S.~C.}\ \bibnamefont
  {Ferreira}},\ }\bibfield  {title} {\bibinfo {title} {{\textit{Non-universal
  parameters, corrections and universality in Kardar–Parisi–Zhang
  growth}}},\ }\href {https://doi.org/10.1088/1742-5468/2013/05/P05007}
  {\bibfield  {journal} {\bibinfo  {journal} {Journal of Statistical Mechanics:
  Theory and Experiment}\ }\textbf {\bibinfo {volume} {2013}},\ \bibinfo
  {pages} {P05007} (\bibinfo {year} {2013})}\BibitemShut {NoStop}%
\bibitem [{\citenamefont {Saberi}\ \emph {et~al.}(2019)\citenamefont {Saberi},
  \citenamefont {Dashti-Naserabadi},\ and\ \citenamefont {Krug}}]{Saberi2019}%
  \BibitemOpen
  \bibfield  {author} {\bibinfo {author} {\bibfnamefont {A.~A.}\ \bibnamefont
  {Saberi}}, \bibinfo {author} {\bibfnamefont {H.}~\bibnamefont
  {Dashti-Naserabadi}},\ and\ \bibinfo {author} {\bibfnamefont
  {J.}~\bibnamefont {Krug}},\ }\bibfield  {title} {\bibinfo {title}
  {{\textit{Competing Universalities in Kardar-Parisi-Zhang Growth Models}}},\
  }\href {https://doi.org/10.1103/PhysRevLett.122.040605} {\bibfield  {journal}
  {\bibinfo  {journal} {Phys. Rev. Lett.}\ }\textbf {\bibinfo {volume} {122}},\
  \bibinfo {pages} {040605} (\bibinfo {year} {2019})}\BibitemShut {NoStop}%
\bibitem [{\citenamefont {Oliveira}\ \emph {et~al.}(2013)\citenamefont
  {Oliveira}, \citenamefont {Alves},\ and\ \citenamefont
  {Ferreira}}]{Oliveira2013}%
  \BibitemOpen
  \bibfield  {author} {\bibinfo {author} {\bibfnamefont {T.~J.}\ \bibnamefont
  {Oliveira}}, \bibinfo {author} {\bibfnamefont {S.~G.}\ \bibnamefont
  {Alves}},\ and\ \bibinfo {author} {\bibfnamefont {S.~C.}\ \bibnamefont
  {Ferreira}},\ }\bibfield  {title} {\bibinfo {title}
  {{\textit{Kardar-Parisi-Zhang universality class in ($2+1$) dimensions:
  Universal geometry-dependent distributions and finite-time corrections}}},\
  }\href {https://doi.org/10.1103/PhysRevE.87.040102} {\bibfield  {journal}
  {\bibinfo  {journal} {Phys. Rev. E}\ }\textbf {\bibinfo {volume} {87}},\
  \bibinfo {pages} {040102} (\bibinfo {year} {2013})}\BibitemShut {NoStop}%
\bibitem [{\citenamefont {Anderson}\ \emph {et~al.}(2009)\citenamefont
  {Anderson}, \citenamefont {Guionnet},\ and\ \citenamefont
  {Zeitouni}}]{Anderson2009}%
  \BibitemOpen
  \bibfield  {author} {\bibinfo {author} {\bibfnamefont {G.~W.}\ \bibnamefont
  {Anderson}}, \bibinfo {author} {\bibfnamefont {A.}~\bibnamefont {Guionnet}},\
  and\ \bibinfo {author} {\bibfnamefont {O.}~\bibnamefont {Zeitouni}},\ }\href
  {https://doi.org/10.1017/CBO9780511801334} {{\textit{\href
  {https://doi.org/10.1017/CBO9780511801334} {}}}\bibinfo {title} {\textit{An
  Introduction to Random Matrices}}}\ (\bibinfo  {publisher} {Cambridge
  University Press, Cambridge},\ \bibinfo {year} {2009})\BibitemShut {NoStop}%
\bibitem [{\citenamefont {Tracy}\ and\ \citenamefont
  {Widom}(1994)}]{Tracy1994}%
  \BibitemOpen
  \bibfield  {author} {\bibinfo {author} {\bibfnamefont {C.~A.}\ \bibnamefont
  {Tracy}}\ and\ \bibinfo {author} {\bibfnamefont {H.}~\bibnamefont {Widom}},\
  }\bibfield  {title} {\bibinfo {title} {{\textit{Level-spacing distributions
  and the Airy kernel}}},\ }\href {https://doi.org/10.1007/BF02100489}
  {\bibfield  {journal} {\bibinfo  {journal} {Commun. Math. Phys.}\ }\textbf
  {\bibinfo {volume} {159}},\ \bibinfo {pages} {151} (\bibinfo {year}
  {1994})}\BibitemShut {NoStop}%
\bibitem [{\citenamefont {Tracy}\ and\ \citenamefont
  {Widom}(1996)}]{Tracy1996}%
  \BibitemOpen
  \bibfield  {author} {\bibinfo {author} {\bibfnamefont {C.~A.}\ \bibnamefont
  {Tracy}}\ and\ \bibinfo {author} {\bibfnamefont {H.}~\bibnamefont {Widom}},\
  }\bibfield  {title} {\bibinfo {title} {{\textit{On orthogonal and symplectic
  matrix ensembles}}},\ }\href {https://doi.org/10.1007/BF02099545} {\bibfield
  {journal} {\bibinfo  {journal} {Commun. Math. Phys.}\ }\textbf {\bibinfo
  {volume} {177}},\ \bibinfo {pages} {727} (\bibinfo {year}
  {1996})}\BibitemShut {NoStop}%
\bibitem [{\citenamefont {Takeuchi}\ and\ \citenamefont
  {Sano}(2012)}]{Takeuchi2012}%
  \BibitemOpen
  \bibfield  {author} {\bibinfo {author} {\bibfnamefont {K.~A.}\ \bibnamefont
  {Takeuchi}}\ and\ \bibinfo {author} {\bibfnamefont {M.}~\bibnamefont
  {Sano}},\ }\bibfield  {title} {\bibinfo {title} {{\textit{{Evidence for
  Geometry-Dependent Universal Fluctuations of the Kardar-Parisi-Zhang
  Interfaces in Liquid-Crystal Turbulence}}}},\ }\href
  {https://doi.org/10.1007/s10955-012-0503-0} {\bibfield  {journal} {\bibinfo
  {journal} {J. Stat. Phys.}\ }\textbf {\bibinfo {volume} {147}},\ \bibinfo
  {pages} {853} (\bibinfo {year} {2012})}\BibitemShut {NoStop}%
\bibitem [{\citenamefont {Borodin}\ \emph {et~al.}(2008)\citenamefont
  {Borodin}, \citenamefont {Ferrari},\ and\ \citenamefont
  {Sasamoto}}]{Borodin2008}%
  \BibitemOpen
  \bibfield  {author} {\bibinfo {author} {\bibfnamefont {A.}~\bibnamefont
  {Borodin}}, \bibinfo {author} {\bibfnamefont {P.~L.}\ \bibnamefont
  {Ferrari}},\ and\ \bibinfo {author} {\bibfnamefont {T.}~\bibnamefont
  {Sasamoto}},\ }\bibfield  {title} {\bibinfo {title} {{\textit{Large time
  asymptotics of growth models on space-like paths II: PNG and parallel
  TASEP}}},\ }\href {https://doi.org/10.1007/s00220-008-0515-4} {\bibfield
  {journal} {\bibinfo  {journal} {Commun. Math. Phys.}\ }\textbf {\bibinfo
  {volume} {283}},\ \bibinfo {pages} {417} (\bibinfo {year}
  {2008})}\BibitemShut {NoStop}%
\bibitem [{\citenamefont {Sasamoto}(2005)}]{Sasamoto2005}%
  \BibitemOpen
  \bibfield  {author} {\bibinfo {author} {\bibfnamefont {T.}~\bibnamefont
  {Sasamoto}},\ }\bibfield  {title} {\bibinfo {title} {{\textit{Spatial
  correlations of the 1D KPZ surface on a flat substrate}}},\ }\href
  {https://doi.org/10.1088/0305-4470/38/33/L01} {\bibfield  {journal} {\bibinfo
   {journal} {J. Phys. A: Math. Gen.}\ }\textbf {\bibinfo {volume} {38}},\
  \bibinfo {pages} {L549} (\bibinfo {year} {2005})}\BibitemShut {NoStop}%
\bibitem [{\citenamefont {Pr{\"{a}}hofer}\ and\ \citenamefont
  {Spohn}(2002)}]{Prahofer2002}%
  \BibitemOpen
  \bibfield  {author} {\bibinfo {author} {\bibfnamefont {M.}~\bibnamefont
  {Pr{\"{a}}hofer}}\ and\ \bibinfo {author} {\bibfnamefont {H.}~\bibnamefont
  {Spohn}},\ }\bibfield  {title} {\bibinfo {title} {{\textit{{Scale invariance
  of the PNG droplet and the airy process}}}},\ }\href
  {https://doi.org/https://doi.org/10.1023/A:1019791415147} {\bibfield
  {journal} {\bibinfo  {journal} {J. Stat. Phys.}\ }\textbf {\bibinfo {volume}
  {108}},\ \bibinfo {pages} {1071} (\bibinfo {year} {2002})}\BibitemShut
  {NoStop}%
\bibitem [{\citenamefont {Prolhac}\ and\ \citenamefont
  {Spohn}(2011)}]{Prolhac2011}%
  \BibitemOpen
  \bibfield  {author} {\bibinfo {author} {\bibfnamefont {S.}~\bibnamefont
  {Prolhac}}\ and\ \bibinfo {author} {\bibfnamefont {H.}~\bibnamefont
  {Spohn}},\ }\bibfield  {title} {\bibinfo {title} {{\textit{Height
  distribution of the Kardar-Parisi-Zhang equation with sharp-wedge initial
  condition: Numerical evaluations}}},\ }\href
  {https://doi.org/10.1103/PhysRevE.84.011119} {\bibfield  {journal} {\bibinfo
  {journal} {Phys. Rev. E}\ }\textbf {\bibinfo {volume} {84}},\ \bibinfo
  {pages} {011119} (\bibinfo {year} {2011})}\BibitemShut {NoStop}%
\bibitem [{\citenamefont {Bornemann}\ \emph {et~al.}(2008)\citenamefont
  {Bornemann}, \citenamefont {Ferrari},\ and\ \citenamefont
  {Pr{\"{a}}hofer}}]{Bornemann2008}%
  \BibitemOpen
  \bibfield  {author} {\bibinfo {author} {\bibfnamefont {F.}~\bibnamefont
  {Bornemann}}, \bibinfo {author} {\bibfnamefont {P.~L.}\ \bibnamefont
  {Ferrari}},\ and\ \bibinfo {author} {\bibfnamefont {M.}~\bibnamefont
  {Pr{\"{a}}hofer}},\ }\bibfield  {title} {\bibinfo {title} {{\textit{{The
  Airy$_1$ process is not the limit of the largest eigenvalue in GOE matrix
  diffusion}}}},\ }\href
  {https://doi.org/https://doi.org/10.1007/s10955-008-9621-0} {\bibfield
  {journal} {\bibinfo  {journal} {J. Stat. Phys.}\ }\textbf {\bibinfo {volume}
  {133}},\ \bibinfo {pages} {405} (\bibinfo {year} {2008})}\BibitemShut
  {NoStop}%
\bibitem [{\citenamefont {Bornemann}(2010)}]{Bornemann2010}%
  \BibitemOpen
  \bibfield  {author} {\bibinfo {author} {\bibfnamefont {F.}~\bibnamefont
  {Bornemann}},\ }\bibfield  {title} {\bibinfo {title} {{\textit{On the
  Numerical Evaluation of Distributions in Random Matrix Theory: A Review}}},\
  }\href {https://math-mprf.org/journal/articles/id1229/} {\bibfield  {journal}
  {\bibinfo  {journal} {Markov. Process. Relat. Fields}\ }\textbf {\bibinfo
  {volume} {16}},\ \bibinfo {pages} {803} (\bibinfo {year} {2010})}\BibitemShut
  {NoStop}%
\bibitem [{\citenamefont {Carrasco}\ and\ \citenamefont
  {Oliveira}(2019)}]{Carrasco2019}%
  \BibitemOpen
  \bibfield  {author} {\bibinfo {author} {\bibfnamefont {I.~S.~S.}\
  \bibnamefont {Carrasco}}\ and\ \bibinfo {author} {\bibfnamefont {T.~J.}\
  \bibnamefont {Oliveira}},\ }\bibfield  {title} {\bibinfo {title}
  {{\textit{{Geometry dependence in linear interface growth}}}},\ }\href
  {https://doi.org/10.1103/PhysRevE.100.042107} {\bibfield  {journal} {\bibinfo
   {journal} {Phys. Rev. E}\ }\textbf {\bibinfo {volume} {100}},\ \bibinfo
  {pages} {042107} (\bibinfo {year} {2019})}\BibitemShut {NoStop}%
\bibitem [{\citenamefont {Guti\'errez}\ and\ \citenamefont
  {Cuerno}(2023)}]{Gutierrez2023}%
  \BibitemOpen
  \bibfield  {author} {\bibinfo {author} {\bibfnamefont {R.}~\bibnamefont
  {Guti\'errez}}\ and\ \bibinfo {author} {\bibfnamefont {R.}~\bibnamefont
  {Cuerno}},\ }\bibfield  {title} {\bibinfo {title} {{\textit{Nonequilibrium
  criticality driven by Kardar-Parisi-Zhang fluctuations in the synchronization
  of oscillator lattices}}},\ }\href
  {https://doi.org/10.1103/PhysRevResearch.5.023047} {\bibfield  {journal}
  {\bibinfo  {journal} {Phys. Rev. Res.}\ }\textbf {\bibinfo {volume} {5}},\
  \bibinfo {pages} {023047} (\bibinfo {year} {2023})}\BibitemShut {NoStop}%
\bibitem [{\citenamefont {Halpin-Healy}(2012)}]{HalpinHealy2012}%
  \BibitemOpen
  \bibfield  {author} {\bibinfo {author} {\bibfnamefont {T.}~\bibnamefont
  {Halpin-Healy}},\ }\bibfield  {title} {\bibinfo {title}
  {{\textit{($2\mathbf{+}1$)-dimensional directed polymer in a random medium:
  scaling phenomena and universal distributions}}},\ }\href
  {https://doi.org/10.1103/PhysRevLett.109.170602} {\bibfield  {journal}
  {\bibinfo  {journal} {Phys. Rev. Lett.}\ }\textbf {\bibinfo {volume} {109}},\
  \bibinfo {pages} {170602} (\bibinfo {year} {2012})}\BibitemShut {NoStop}%
\bibitem [{\citenamefont {Halpin-Healy}\ and\ \citenamefont
  {Palasantzas}(2014)}]{HalpinHealy2014}%
  \BibitemOpen
  \bibfield  {author} {\bibinfo {author} {\bibfnamefont {T.}~\bibnamefont
  {Halpin-Healy}}\ and\ \bibinfo {author} {\bibfnamefont {G.}~\bibnamefont
  {Palasantzas}},\ }\bibfield  {title} {\bibinfo {title} {{\textit{Universal
  correlators and distributions as experimental signatures of (2 +
  1)-dimensional Kardar-Parisi-Zhang growth}}},\ }\href
  {https://doi.org/10.1209/0295-5075/105/50001} {\bibfield  {journal} {\bibinfo
   {journal} {Europhys. Lett.}\ }\textbf {\bibinfo {volume} {105}},\ \bibinfo
  {pages} {50001} (\bibinfo {year} {2014})}\BibitemShut {NoStop}%
\bibitem [{\citenamefont {Halpin-Healy}(2013)}]{HalpinHealy2013}%
  \BibitemOpen
  \bibfield  {author} {\bibinfo {author} {\bibfnamefont {T.}~\bibnamefont
  {Halpin-Healy}},\ }\bibfield  {title} {\bibinfo {title} {{\textit{Extremal
  paths, the stochastic heat equation, and the three-dimensional
  Kardar-Parisi-Zhang universality class}}},\ }\href
  {https://doi.org/10.1103/PhysRevE.88.042118} {\bibfield  {journal} {\bibinfo
  {journal} {Phys. Rev. E}\ }\textbf {\bibinfo {volume} {88}},\ \bibinfo
  {pages} {042118} (\bibinfo {year} {2013})}\BibitemShut {NoStop}%
\bibitem [{\citenamefont {Burgers}(1948)}]{Burgers1948}%
  \BibitemOpen
  \bibfield  {author} {\bibinfo {author} {\bibfnamefont {J.}~\bibnamefont
  {Burgers}},\ }\bibinfo {title} {{\textit{A Mathematical Model Illustrating
  the Theory of Turbulence}}},\ in\ \href
  {https://doi.org/10.1016/s0065-2156(08)70100-5} {\bibinfo {booktitle} {Adv.
  Appl. Mech. 1}}\ (\bibinfo  {publisher} {Elsevier},\ \bibinfo {year} {1948})\
  p.\ \bibinfo {pages} {171–199}\BibitemShut {NoStop}%
\bibitem [{\citenamefont {Cuerno}\ and\ \citenamefont
  {Lauritsen}(1995)}]{Cuerno1995}%
  \BibitemOpen
  \bibfield  {author} {\bibinfo {author} {\bibfnamefont {R.}~\bibnamefont
  {Cuerno}}\ and\ \bibinfo {author} {\bibfnamefont {K.~B.}\ \bibnamefont
  {Lauritsen}},\ }\bibfield  {title} {\bibinfo {title}
  {{\textit{Renormalization-group analysis of a noisy Kuramoto-Sivashinsky
  equation}}},\ }\href {https://doi.org/10.1103/physreve.52.4853} {\bibfield
  {journal} {\bibinfo  {journal} {Phys. Rev. E}\ }\textbf {\bibinfo {volume}
  {52}},\ \bibinfo {pages} {4853–4859} (\bibinfo {year} {1995})}\BibitemShut
  {NoStop}%
\bibitem [{\citenamefont {Rodríguez-Fernández}\ \emph
  {et~al.}(2022)\citenamefont {Rodríguez-Fernández}, \citenamefont
  {Santalla}, \citenamefont {Castro},\ and\ \citenamefont
  {Cuerno}}]{RodrguezFernndez2022}%
  \BibitemOpen
  \bibfield  {author} {\bibinfo {author} {\bibfnamefont {E.}~\bibnamefont
  {Rodríguez-Fernández}}, \bibinfo {author} {\bibfnamefont {S.~N.}\
  \bibnamefont {Santalla}}, \bibinfo {author} {\bibfnamefont {M.}~\bibnamefont
  {Castro}},\ and\ \bibinfo {author} {\bibfnamefont {R.}~\bibnamefont
  {Cuerno}},\ }\bibfield  {title} {\bibinfo {title} {{\textit{Anomalous
  ballistic scaling in the tensionless or inviscid Kardar-Parisi-Zhang
  equation}}},\ }\href {http://dx.doi.org/10.1103/PhysRevE.106.024802}
  {\bibfield  {journal} {\bibinfo  {journal} {Phys. Rev. E}\ }\textbf {\bibinfo
  {volume} {106}} (\bibinfo {year} {2022})}\BibitemShut {NoStop}%
\bibitem [{\citenamefont {Ramasco}\ \emph {et~al.}(2000)\citenamefont
  {Ramasco}, \citenamefont {López},\ and\ \citenamefont
  {Rodríguez}}]{Ramasco2000}%
  \BibitemOpen
  \bibfield  {author} {\bibinfo {author} {\bibfnamefont {J.~J.}\ \bibnamefont
  {Ramasco}}, \bibinfo {author} {\bibfnamefont {J.~M.}\ \bibnamefont
  {López}},\ and\ \bibinfo {author} {\bibfnamefont {M.~A.}\ \bibnamefont
  {Rodríguez}},\ }\bibfield  {title} {\bibinfo {title} {{\textit{Generic
  Dynamic Scaling in Kinetic Roughening}}},\ }\href
  {https://doi.org/10.1103/physrevlett.84.2199} {\bibfield  {journal} {\bibinfo
   {journal} {Phys. Rev. Lett.}\ }\textbf {\bibinfo {volume} {84}},\ \bibinfo
  {pages} {2199–2202} (\bibinfo {year} {2000})}\BibitemShut {NoStop}%
\bibitem [{\citenamefont {Asikainen}\ \emph {et~al.}(2002)\citenamefont
  {Asikainen}, \citenamefont {Majaniemi}, \citenamefont {Dubé},\ and\
  \citenamefont {Ala-Nissila}}]{Asikainen2002}%
  \BibitemOpen
  \bibfield  {author} {\bibinfo {author} {\bibfnamefont {J.}~\bibnamefont
  {Asikainen}}, \bibinfo {author} {\bibfnamefont {S.}~\bibnamefont
  {Majaniemi}}, \bibinfo {author} {\bibfnamefont {M.}~\bibnamefont {Dubé}},\
  and\ \bibinfo {author} {\bibfnamefont {T.}~\bibnamefont {Ala-Nissila}},\
  }\bibfield  {title} {\bibinfo {title} {{\textit{Interface dynamics and
  kinetic roughening in fractals}}},\ }\href
  {http://dx.doi.org/10.1103/PhysRevE.65.052104} {\bibfield  {journal}
  {\bibinfo  {journal} {Phys. Rev. E}\ }\textbf {\bibinfo {volume} {65}}
  (\bibinfo {year} {2002})}\BibitemShut {NoStop}%
\bibitem [{\citenamefont {Tang}\ \emph {et~al.}(1990)\citenamefont {Tang},
  \citenamefont {Nattermann},\ and\ \citenamefont {Forrest}}]{Tang1990}%
  \BibitemOpen
  \bibfield  {author} {\bibinfo {author} {\bibfnamefont {L.-H.}\ \bibnamefont
  {Tang}}, \bibinfo {author} {\bibfnamefont {T.}~\bibnamefont {Nattermann}},\
  and\ \bibinfo {author} {\bibfnamefont {B.~M.}\ \bibnamefont {Forrest}},\
  }\bibfield  {title} {\bibinfo {title} {{\textit{Multicritical and crossover
  phenomena in surface growth}}},\ }\href
  {https://doi.org/10.1103/PhysRevLett.65.2422} {\bibfield  {journal} {\bibinfo
   {journal} {Phys. Rev. Lett.}\ }\textbf {\bibinfo {volume} {65}},\ \bibinfo
  {pages} {2422} (\bibinfo {year} {1990})}\BibitemShut {NoStop}%
\bibitem [{\citenamefont {Dashti-Naserabadi}\ \emph {et~al.}(2017)\citenamefont
  {Dashti-Naserabadi}, \citenamefont {Saberi},\ and\ \citenamefont
  {Rouhani}}]{Dashti-Naserabadi2017}%
  \BibitemOpen
  \bibfield  {author} {\bibinfo {author} {\bibfnamefont {H.}~\bibnamefont
  {Dashti-Naserabadi}}, \bibinfo {author} {\bibfnamefont {A.~A.}\ \bibnamefont
  {Saberi}},\ and\ \bibinfo {author} {\bibfnamefont {S.}~\bibnamefont
  {Rouhani}},\ }\bibfield  {title} {\bibinfo {title} {{\textit{Roughening
  transition and universality of single step growth models in
  (2+1)-dimensions}}},\ }\href {https://doi.org/10.1088/1367-2630/aa7474}
  {\bibfield  {journal} {\bibinfo  {journal} {New J. Phys.}\ }\textbf {\bibinfo
  {volume} {19}},\ \bibinfo {pages} {063035} (\bibinfo {year}
  {2017})}\BibitemShut {NoStop}%
\bibitem [{\citenamefont {Gosteva}\ \emph {et~al.}(2024)\citenamefont
  {Gosteva}, \citenamefont {Tarpin}, \citenamefont {Wschebor},\ and\
  \citenamefont {Canet}}]{Gosteva2024}%
  \BibitemOpen
  \bibfield  {author} {\bibinfo {author} {\bibfnamefont {L.}~\bibnamefont
  {Gosteva}}, \bibinfo {author} {\bibfnamefont {M.}~\bibnamefont {Tarpin}},
  \bibinfo {author} {\bibfnamefont {N.}~\bibnamefont {Wschebor}},\ and\
  \bibinfo {author} {\bibfnamefont {L.}~\bibnamefont {Canet}},\ }\bibfield
  {title} {\bibinfo {title} {{\textit{Inviscid fixed point of the
  multidimensional Burgers–Kardar-Parisi-Zhang equation}}},\ }\href
  {http://dx.doi.org/10.1103/PhysRevE.110.054118} {\bibfield  {journal}
  {\bibinfo  {journal} {Phys. Rev. E}\ }\textbf {\bibinfo {volume} {110}}
  (\bibinfo {year} {2024})}\BibitemShut {NoStop}%
\bibitem [{\citenamefont {Castro}\ \emph {et~al.}(2012)\citenamefont {Castro},
  \citenamefont {Cuerno}, \citenamefont {Nicoli}, \citenamefont
  {V{\'{a}}zquez},\ and\ \citenamefont {Buijnsters}}]{Castro2012}%
  \BibitemOpen
  \bibfield  {author} {\bibinfo {author} {\bibfnamefont {M.}~\bibnamefont
  {Castro}}, \bibinfo {author} {\bibfnamefont {R.}~\bibnamefont {Cuerno}},
  \bibinfo {author} {\bibfnamefont {M.}~\bibnamefont {Nicoli}}, \bibinfo
  {author} {\bibfnamefont {L.}~\bibnamefont {V{\'{a}}zquez}},\ and\ \bibinfo
  {author} {\bibfnamefont {J.~G.}\ \bibnamefont {Buijnsters}},\ }\bibfield
  {title} {\bibinfo {title} {{\textit{{Universality of cauliflower-like fronts:
  From nanoscale thin films to macroscopic plants}}}},\ }\href
  {https://doi.org/10.1088/1367-2630/14/10/103039} {\bibfield  {journal}
  {\bibinfo  {journal} {New J. Phys.}\ }\textbf {\bibinfo {volume} {14}},\
  \bibinfo {pages} {103039} (\bibinfo {year} {2012})}\BibitemShut {NoStop}%
\bibitem [{\citenamefont {Nicoli}\ \emph {et~al.}(2009)\citenamefont {Nicoli},
  \citenamefont {Cuerno},\ and\ \citenamefont {Castro}}]{Nicoli2009-2}%
  \BibitemOpen
  \bibfield  {author} {\bibinfo {author} {\bibfnamefont {M.}~\bibnamefont
  {Nicoli}}, \bibinfo {author} {\bibfnamefont {R.}~\bibnamefont {Cuerno}},\
  and\ \bibinfo {author} {\bibfnamefont {M.}~\bibnamefont {Castro}},\
  }\bibfield  {title} {\bibinfo {title} {{\textit{Unstable Nonlocal Interface
  Dynamics}}},\ }\href {http://dx.doi.org/10.1103/PhysRevLett.102.256102}
  {\bibfield  {journal} {\bibinfo  {journal} {Phys. Rev. Lett.}\ }\textbf
  {\bibinfo {volume} {102}} (\bibinfo {year} {2009})}\BibitemShut {NoStop}%
\bibitem [{\citenamefont {Zhang}\ \emph {et~al.}(1992)\citenamefont {Zhang},
  \citenamefont {Zhang}, \citenamefont {Alstrøm},\ and\ \citenamefont
  {Levinsen}}]{Zhang1992}%
  \BibitemOpen
  \bibfield  {author} {\bibinfo {author} {\bibfnamefont {J.}~\bibnamefont
  {Zhang}}, \bibinfo {author} {\bibfnamefont {Y.~C.}\ \bibnamefont {Zhang}},
  \bibinfo {author} {\bibfnamefont {P.}~\bibnamefont {Alstrøm}},\ and\
  \bibinfo {author} {\bibfnamefont {M.~T.}\ \bibnamefont {Levinsen}},\
  }\bibfield  {title} {\bibinfo {title} {{\textit{Modeling forest fire by a
  paper-burning experiment, a realization of the interface growth
  mechanism}}},\ }\href {https://doi.org/10.1016/0378-4371(92)90050-Z}
  {\bibfield  {journal} {\bibinfo  {journal} {Physica A: Stat. Mech. Appl.}\
  }\textbf {\bibinfo {volume} {189}},\ \bibinfo {pages} {383} (\bibinfo {year}
  {1992})}\BibitemShut {NoStop}%
\bibitem [{\citenamefont {Myllys}\ \emph {et~al.}(2001)\citenamefont {Myllys},
  \citenamefont {Maunuksela}, \citenamefont {Alava}, \citenamefont
  {Ala-Nissila}, \citenamefont {Merikoski},\ and\ \citenamefont
  {Timonen}}]{Myllys2001}%
  \BibitemOpen
  \bibfield  {author} {\bibinfo {author} {\bibfnamefont {M.}~\bibnamefont
  {Myllys}}, \bibinfo {author} {\bibfnamefont {J.}~\bibnamefont {Maunuksela}},
  \bibinfo {author} {\bibfnamefont {M.}~\bibnamefont {Alava}}, \bibinfo
  {author} {\bibfnamefont {T.}~\bibnamefont {Ala-Nissila}}, \bibinfo {author}
  {\bibfnamefont {J.}~\bibnamefont {Merikoski}},\ and\ \bibinfo {author}
  {\bibfnamefont {J.}~\bibnamefont {Timonen}},\ }\bibfield  {title} {\bibinfo
  {title} {{\textit{Kinetic roughening in slow combustion of paper}}},\ }\href
  {https://doi.org/10.1103/PhysRevE.64.036101} {\bibfield  {journal} {\bibinfo
  {journal} {Phys. Rev. E}\ }\textbf {\bibinfo {volume} {64}},\ \bibinfo
  {pages} {12} (\bibinfo {year} {2001})}\BibitemShut {NoStop}%
\bibitem [{\citenamefont {Orrillo}\ \emph {et~al.}(2017)\citenamefont
  {Orrillo}, \citenamefont {Santalla}, \citenamefont {Cuerno}, \citenamefont
  {V{\'{a}}zquez}, \citenamefont {Ribotta}, \citenamefont {Gassa},
  \citenamefont {Mompean}, \citenamefont {Salvarezza},\ and\ \citenamefont
  {Vela}}]{Orrillo2017}%
  \BibitemOpen
  \bibfield  {author} {\bibinfo {author} {\bibfnamefont {P.~A.}\ \bibnamefont
  {Orrillo}}, \bibinfo {author} {\bibfnamefont {S.~N.}\ \bibnamefont
  {Santalla}}, \bibinfo {author} {\bibfnamefont {R.}~\bibnamefont {Cuerno}},
  \bibinfo {author} {\bibfnamefont {L.}~\bibnamefont {V{\'{a}}zquez}}, \bibinfo
  {author} {\bibfnamefont {S.~B.}\ \bibnamefont {Ribotta}}, \bibinfo {author}
  {\bibfnamefont {L.~M.}\ \bibnamefont {Gassa}}, \bibinfo {author}
  {\bibfnamefont {F.~J.}\ \bibnamefont {Mompean}}, \bibinfo {author}
  {\bibfnamefont {R.~C.}\ \bibnamefont {Salvarezza}},\ and\ \bibinfo {author}
  {\bibfnamefont {M.~E.}\ \bibnamefont {Vela}},\ }\bibfield  {title} {\bibinfo
  {title} {{\textit{{Morphological stabilization and KPZ scaling by
  electrochemically induced co-deposition of nanostructured NiW alloy
  films}}}},\ }\href {https://doi.org/10.1038/s41598-017-18155-7} {\bibfield
  {journal} {\bibinfo  {journal} {Sci. Rep.}\ }\textbf {\bibinfo {volume}
  {7}},\ \bibinfo {pages} {17997} (\bibinfo {year} {2017})}\BibitemShut
  {NoStop}%
\bibitem [{\citenamefont {Dervaux}\ \emph {et~al.}(2014)\citenamefont
  {Dervaux}, \citenamefont {Magniez},\ and\ \citenamefont
  {Libchaber}}]{Dervaux2014}%
  \BibitemOpen
  \bibfield  {author} {\bibinfo {author} {\bibfnamefont {J.}~\bibnamefont
  {Dervaux}}, \bibinfo {author} {\bibfnamefont {J.~C.}\ \bibnamefont
  {Magniez}},\ and\ \bibinfo {author} {\bibfnamefont {A.}~\bibnamefont
  {Libchaber}},\ }\bibfield  {title} {\bibinfo {title} {{\textit{On growth and
  form of Bacillus subtilis biofilms}}},\ }\href
  {https://doi.org/10.1098/rsfs.2013.0051} {\bibfield  {journal} {\bibinfo
  {journal} {Interface Focus}\ }\textbf {\bibinfo {volume} {4}},\ \bibinfo
  {pages} {20130051} (\bibinfo {year} {2014})}\BibitemShut {NoStop}%
\bibitem [{\citenamefont {Vázquez}\ \emph {et~al.}(1996)\citenamefont
  {Vázquez}, \citenamefont {Albella}, \citenamefont {Salvarezza},
  \citenamefont {Arvia}, \citenamefont {Levy},\ and\ \citenamefont
  {Perese}}]{Vazquez1996}%
  \BibitemOpen
  \bibfield  {author} {\bibinfo {author} {\bibfnamefont {L.}~\bibnamefont
  {Vázquez}}, \bibinfo {author} {\bibfnamefont {J.~M.}\ \bibnamefont
  {Albella}}, \bibinfo {author} {\bibfnamefont {R.~C.}\ \bibnamefont
  {Salvarezza}}, \bibinfo {author} {\bibfnamefont {A.~J.}\ \bibnamefont
  {Arvia}}, \bibinfo {author} {\bibfnamefont {R.~A.}\ \bibnamefont {Levy}},\
  and\ \bibinfo {author} {\bibfnamefont {D.}~\bibnamefont {Perese}},\
  }\bibfield  {title} {\bibinfo {title} {{\textit{Roughening kinetics of
  chemical vapor deposited copper films on Si(100)}}},\ }\href
  {https://doi.org/10.1063/1.115954} {\bibfield  {journal} {\bibinfo  {journal}
  {Appl. Phys. Lett.}\ }\textbf {\bibinfo {volume} {68}},\ \bibinfo {pages}
  {1285–1287} (\bibinfo {year} {1996})}\BibitemShut {NoStop}%
\bibitem [{\citenamefont {Zhao}\ \emph {et~al.}(2000)\citenamefont {Zhao},
  \citenamefont {Fortin}, \citenamefont {Bonvallet}, \citenamefont {Wang},\
  and\ \citenamefont {Lu}}]{Zhao2000}%
  \BibitemOpen
  \bibfield  {author} {\bibinfo {author} {\bibfnamefont {Y.-P.}\ \bibnamefont
  {Zhao}}, \bibinfo {author} {\bibfnamefont {J.~B.}\ \bibnamefont {Fortin}},
  \bibinfo {author} {\bibfnamefont {G.}~\bibnamefont {Bonvallet}}, \bibinfo
  {author} {\bibfnamefont {G.-C.}\ \bibnamefont {Wang}},\ and\ \bibinfo
  {author} {\bibfnamefont {T.-M.}\ \bibnamefont {Lu}},\ }\bibfield  {title}
  {\bibinfo {title} {{\textit{Kinetic Roughening in Polymer Film Growth by
  Vapor Deposition}}},\ }\href {https://doi.org/10.1103/physrevlett.85.3229}
  {\bibfield  {journal} {\bibinfo  {journal} {Phys. Rev. Lett.}\ }\textbf
  {\bibinfo {volume} {85}},\ \bibinfo {pages} {3229–3232} (\bibinfo {year}
  {2000})}\BibitemShut {NoStop}%
\bibitem [{\citenamefont {Vivo}\ \emph {et~al.}(2012)\citenamefont {Vivo},
  \citenamefont {Nicoli}, \citenamefont {Engler}, \citenamefont {Michely},
  \citenamefont {Vázquez},\ and\ \citenamefont {Cuerno}}]{Vivo2012}%
  \BibitemOpen
  \bibfield  {author} {\bibinfo {author} {\bibfnamefont {E.}~\bibnamefont
  {Vivo}}, \bibinfo {author} {\bibfnamefont {M.}~\bibnamefont {Nicoli}},
  \bibinfo {author} {\bibfnamefont {M.}~\bibnamefont {Engler}}, \bibinfo
  {author} {\bibfnamefont {T.}~\bibnamefont {Michely}}, \bibinfo {author}
  {\bibfnamefont {L.}~\bibnamefont {Vázquez}},\ and\ \bibinfo {author}
  {\bibfnamefont {R.}~\bibnamefont {Cuerno}},\ }\bibfield  {title} {\bibinfo
  {title} {{\textit{Strong anisotropy in surface kinetic roughening: Analysis
  and experiments}}},\ }\href {http://dx.doi.org/10.1103/PhysRevB.86.245427}
  {\bibfield  {journal} {\bibinfo  {journal} {Phys. Rev. B}\ }\textbf {\bibinfo
  {volume} {86}} (\bibinfo {year} {2012})}\BibitemShut {NoStop}%
\bibitem [{\citenamefont {Huergo}\ \emph {et~al.}(2011)\citenamefont {Huergo},
  \citenamefont {Pasquale}, \citenamefont {Gonz\'alez}, \citenamefont
  {Bolz\'an},\ and\ \citenamefont {Arvia}}]{Huergo2011}%
  \BibitemOpen
  \bibfield  {author} {\bibinfo {author} {\bibfnamefont {M.~A.~C.}\
  \bibnamefont {Huergo}}, \bibinfo {author} {\bibfnamefont {M.~A.}\
  \bibnamefont {Pasquale}}, \bibinfo {author} {\bibfnamefont {P.~H.}\
  \bibnamefont {Gonz\'alez}}, \bibinfo {author} {\bibfnamefont {A.~E.}\
  \bibnamefont {Bolz\'an}},\ and\ \bibinfo {author} {\bibfnamefont {A.~J.}\
  \bibnamefont {Arvia}},\ }\bibfield  {title} {\bibinfo {title}
  {{\textit{Dynamics and morphology characteristics of cell colonies with
  radially spreading growth fronts}}},\ }\href
  {https://doi.org/10.1103/PhysRevE.84.021917} {\bibfield  {journal} {\bibinfo
  {journal} {Phys. Rev. E}\ }\textbf {\bibinfo {volume} {84}},\ \bibinfo
  {pages} {021917} (\bibinfo {year} {2011})}\BibitemShut {NoStop}%
\bibitem [{\citenamefont {Huergo}\ \emph {et~al.}(2012)\citenamefont {Huergo},
  \citenamefont {Pasquale}, \citenamefont {Gonz\'alez}, \citenamefont
  {Bolz\'an},\ and\ \citenamefont {Arvia}}]{Huergo2012}%
  \BibitemOpen
  \bibfield  {author} {\bibinfo {author} {\bibfnamefont {M.~A.~C.}\
  \bibnamefont {Huergo}}, \bibinfo {author} {\bibfnamefont {M.~A.}\
  \bibnamefont {Pasquale}}, \bibinfo {author} {\bibfnamefont {P.~H.}\
  \bibnamefont {Gonz\'alez}}, \bibinfo {author} {\bibfnamefont {A.~E.}\
  \bibnamefont {Bolz\'an}},\ and\ \bibinfo {author} {\bibfnamefont {A.~J.}\
  \bibnamefont {Arvia}},\ }\bibfield  {title} {\bibinfo {title}
  {{\textit{Growth dynamics of cancer cell colonies and their comparison with
  noncancerous cells}}},\ }\href {https://doi.org/10.1103/PhysRevE.85.011918}
  {\bibfield  {journal} {\bibinfo  {journal} {Phys. Rev. E}\ }\textbf {\bibinfo
  {volume} {85}},\ \bibinfo {pages} {011918} (\bibinfo {year}
  {2012})}\BibitemShut {NoStop}%
\bibitem [{\citenamefont {Bonn}\ \emph {et~al.}(2009)\citenamefont {Bonn},
  \citenamefont {Eggers}, \citenamefont {Indekeu}, \citenamefont {Meunier},\
  and\ \citenamefont {Rolley}}]{Bonn2009}%
  \BibitemOpen
  \bibfield  {author} {\bibinfo {author} {\bibfnamefont {D.}~\bibnamefont
  {Bonn}}, \bibinfo {author} {\bibfnamefont {J.}~\bibnamefont {Eggers}},
  \bibinfo {author} {\bibfnamefont {J.}~\bibnamefont {Indekeu}}, \bibinfo
  {author} {\bibfnamefont {J.}~\bibnamefont {Meunier}},\ and\ \bibinfo {author}
  {\bibfnamefont {E.}~\bibnamefont {Rolley}},\ }\bibfield  {title} {\bibinfo
  {title} {{\textit{Wetting and spreading}}},\ }\href
  {https://doi.org/10.1103/revmodphys.81.739} {\bibfield  {journal} {\bibinfo
  {journal} {Rev. Mod. Phys.}\ }\textbf {\bibinfo {volume} {81}},\ \bibinfo
  {pages} {739} (\bibinfo {year} {2009})}\BibitemShut {NoStop}%
\bibitem [{\citenamefont {Bertrand}\ \emph {et~al.}(2002)\citenamefont
  {Bertrand}, \citenamefont {Bonn}, \citenamefont {Broseta}, \citenamefont
  {Dobbs}, \citenamefont {Indekeu}, \citenamefont {Meunier}, \citenamefont
  {Ragil},\ and\ \citenamefont {Shahidzadeh}}]{Bertrand2002}%
  \BibitemOpen
  \bibfield  {author} {\bibinfo {author} {\bibfnamefont {E.}~\bibnamefont
  {Bertrand}}, \bibinfo {author} {\bibfnamefont {D.}~\bibnamefont {Bonn}},
  \bibinfo {author} {\bibfnamefont {D.}~\bibnamefont {Broseta}}, \bibinfo
  {author} {\bibfnamefont {H.}~\bibnamefont {Dobbs}}, \bibinfo {author}
  {\bibfnamefont {J.}~\bibnamefont {Indekeu}}, \bibinfo {author} {\bibfnamefont
  {J.}~\bibnamefont {Meunier}}, \bibinfo {author} {\bibfnamefont
  {K.}~\bibnamefont {Ragil}},\ and\ \bibinfo {author} {\bibfnamefont
  {N.}~\bibnamefont {Shahidzadeh}},\ }\bibfield  {title} {\bibinfo {title}
  {{\textit{Wetting of alkanes on water}}},\ }\href
  {https://doi.org/10.1016/s0920-4105(01)00191-7} {\bibfield  {journal}
  {\bibinfo  {journal} {J. Pet. Sci. Eng.}\ }\textbf {\bibinfo {volume} {33}},\
  \bibinfo {pages} {217–222} (\bibinfo {year} {2002})}\BibitemShut {NoStop}%
\bibitem [{\citenamefont {Bergeron}\ \emph {et~al.}(2000)\citenamefont
  {Bergeron}, \citenamefont {Bonn}, \citenamefont {Martin},\ and\ \citenamefont
  {Vovelle}}]{Bergeron2000}%
  \BibitemOpen
  \bibfield  {author} {\bibinfo {author} {\bibfnamefont {V.}~\bibnamefont
  {Bergeron}}, \bibinfo {author} {\bibfnamefont {D.}~\bibnamefont {Bonn}},
  \bibinfo {author} {\bibfnamefont {J.~Y.}\ \bibnamefont {Martin}},\ and\
  \bibinfo {author} {\bibfnamefont {L.}~\bibnamefont {Vovelle}},\ }\bibfield
  {title} {\bibinfo {title} {{\textit{Controlling droplet deposition with
  polymer additives}}},\ }\href {https://doi.org/10.1038/35015525} {\bibfield
  {journal} {\bibinfo  {journal} {Nature}\ }\textbf {\bibinfo {volume} {405}},\
  \bibinfo {pages} {772–775} (\bibinfo {year} {2000})}\BibitemShut {NoStop}%
\bibitem [{\citenamefont {Shahidzadeh}(2003)}]{Shahidzadeh2003}%
  \BibitemOpen
  \bibfield  {author} {\bibinfo {author} {\bibfnamefont {N.}~\bibnamefont
  {Shahidzadeh}},\ }\bibfield  {title} {\bibinfo {title} {{\textit{Effect of
  Wetting on Gravity Drainage in Porous Media}}},\ }\href
  {https://doi.org/10.1023/a:1023597130973} {\bibfield  {journal} {\bibinfo
  {journal} {Transp. Porous Media}\ }\textbf {\bibinfo {volume} {52}},\
  \bibinfo {pages} {213–227} (\bibinfo {year} {2003})}\BibitemShut {NoStop}%
\bibitem [{\citenamefont {Tabeling}(2023)}]{Tabeling2023}%
  \BibitemOpen
  \bibfield  {author} {\bibinfo {author} {\bibfnamefont {P.}~\bibnamefont
  {Tabeling}},\ }\href {https://doi.org/10.1093/oso/9780192845306.001.0001}
  {{\textit{\href {https://doi.org/10.1093/oso/9780192845306.001.0001}
  {}}}\bibinfo {title} {\textit{Introduction to Microfluidics}}}\ (\bibinfo
  {publisher} {Oxford University Press, Oxford},\ \bibinfo {year}
  {2023})\BibitemShut {NoStop}%
\bibitem [{\citenamefont {de~Gennes}\ \emph {et~al.}(2004)\citenamefont
  {de~Gennes}, \citenamefont {Brochard-Wyart},\ and\ \citenamefont
  {Quéré}}]{deGennes2004}%
  \BibitemOpen
  \bibfield  {author} {\bibinfo {author} {\bibfnamefont {P.-G.}\ \bibnamefont
  {de~Gennes}}, \bibinfo {author} {\bibfnamefont {F.}~\bibnamefont
  {Brochard-Wyart}},\ and\ \bibinfo {author} {\bibfnamefont {D.}~\bibnamefont
  {Quéré}},\ }\href {https://doi.org/10.1007/978-0-387-21656-0}
  {{\textit{\href {https://doi.org/10.1007/978-0-387-21656-0} {}}}\bibinfo
  {title} {\textit{Capillarity and Wetting Phenomena}}}\ (\bibinfo  {publisher}
  {Springer New York},\ \bibinfo {year} {2004})\BibitemShut {NoStop}%
\bibitem [{\citenamefont {Starov}\ and\ \citenamefont
  {Velarde}(2019)}]{Starov2019}%
  \BibitemOpen
  \bibfield  {author} {\bibinfo {author} {\bibfnamefont {V.~M.}\ \bibnamefont
  {Starov}}\ and\ \bibinfo {author} {\bibfnamefont {M.~G.}\ \bibnamefont
  {Velarde}},\ }\href {https://doi.org/10.1201/9780429506246} {{\textit{\href
  {https://doi.org/10.1201/9780429506246} {}}}\bibinfo {title} {\textit{Wetting
  and Spreading Dynamics}}}\ (\bibinfo  {publisher} {CRC Press, Boca Raton},\
  \bibinfo {year} {2019})\BibitemShut {NoStop}%
\bibitem [{\citenamefont {de~Gennes}(1985)}]{deGennes1985}%
  \BibitemOpen
  \bibfield  {author} {\bibinfo {author} {\bibfnamefont {P.~G.}\ \bibnamefont
  {de~Gennes}},\ }\bibfield  {title} {\bibinfo {title} {{\textit{Wetting:
  statics and dynamics}}},\ }\href {https://doi.org/10.1103/revmodphys.57.827}
  {\bibfield  {journal} {\bibinfo  {journal} {Rev. Mod. Phys.}\ }\textbf
  {\bibinfo {volume} {57}},\ \bibinfo {pages} {827–863} (\bibinfo {year}
  {1985})}\BibitemShut {NoStop}%
\bibitem [{\citenamefont {Cazabat}\ \emph {et~al.}(1994)\citenamefont
  {Cazabat}, \citenamefont {Fraysse}, \citenamefont {Heslot}, \citenamefont
  {Levinson}, \citenamefont {Marsh}, \citenamefont {Tiberg},\ and\
  \citenamefont {Valignat}}]{Cazabat1994}%
  \BibitemOpen
  \bibfield  {author} {\bibinfo {author} {\bibfnamefont {A.}~\bibnamefont
  {Cazabat}}, \bibinfo {author} {\bibfnamefont {N.}~\bibnamefont {Fraysse}},
  \bibinfo {author} {\bibfnamefont {F.}~\bibnamefont {Heslot}}, \bibinfo
  {author} {\bibfnamefont {P.}~\bibnamefont {Levinson}}, \bibinfo {author}
  {\bibfnamefont {J.}~\bibnamefont {Marsh}}, \bibinfo {author} {\bibfnamefont
  {F.}~\bibnamefont {Tiberg}},\ and\ \bibinfo {author} {\bibfnamefont
  {M.}~\bibnamefont {Valignat}},\ }\bibfield  {title} {\bibinfo {title}
  {{\textit{Pancakes}}},\ }\href {https://doi.org/10.1016/0001-8686(94)80003-0}
  {\bibfield  {journal} {\bibinfo  {journal} {Adv. Colloid Interface Sci.}\
  }\textbf {\bibinfo {volume} {48}},\ \bibinfo {pages} {1–17} (\bibinfo
  {year} {1994})}\BibitemShut {NoStop}%
\bibitem [{\citenamefont {Tanner}(1979)}]{Tanner1979}%
  \BibitemOpen
  \bibfield  {author} {\bibinfo {author} {\bibfnamefont {L.~H.}\ \bibnamefont
  {Tanner}},\ }\bibfield  {title} {\bibinfo {title} {{\textit{The spreading of
  silicone oil drops on horizontal surfaces}}},\ }\href
  {https://doi.org/10.1088/0022-3727/12/9/009} {\bibfield  {journal} {\bibinfo
  {journal} {J. Phys. D. Appl. Phys.}\ }\textbf {\bibinfo {volume} {12}},\
  \bibinfo {pages} {1473} (\bibinfo {year} {1979})}\BibitemShut {NoStop}%
\bibitem [{\citenamefont {Popescu}\ \emph {et~al.}(2012)\citenamefont
  {Popescu}, \citenamefont {Oshanin}, \citenamefont {Dietrich},\ and\
  \citenamefont {Cazabat}}]{Popescu2012}%
  \BibitemOpen
  \bibfield  {author} {\bibinfo {author} {\bibfnamefont {M.~N.}\ \bibnamefont
  {Popescu}}, \bibinfo {author} {\bibfnamefont {G.}~\bibnamefont {Oshanin}},
  \bibinfo {author} {\bibfnamefont {S.}~\bibnamefont {Dietrich}},\ and\
  \bibinfo {author} {\bibfnamefont {A.-M.}\ \bibnamefont {Cazabat}},\
  }\bibfield  {title} {\bibinfo {title} {{\textit{Precursor films in wetting
  phenomena}}},\ }\href {https://doi.org/10.1088/0953-8984/24/24/243102}
  {\bibfield  {journal} {\bibinfo  {journal} {J. Phys. Condens. Matter}\
  }\textbf {\bibinfo {volume} {24}},\ \bibinfo {pages} {243102} (\bibinfo
  {year} {2012})}\BibitemShut {NoStop}%
\bibitem [{\citenamefont {Hardy}(1919)}]{Hardy1919}%
  \BibitemOpen
  \bibfield  {author} {\bibinfo {author} {\bibfnamefont {W.}~\bibnamefont
  {Hardy}},\ }\bibfield  {title} {\bibinfo {title} {{\textit{III. The spreading
  of fluids on glass}}},\ }\href {https://doi.org/10.1080/14786440708635928}
  {\bibfield  {journal} {\bibinfo  {journal} {The London, Edinburgh, and Dublin
  Philosophical Magazine and Journal of Science}\ }\textbf {\bibinfo {volume}
  {38}},\ \bibinfo {pages} {49–55} (\bibinfo {year} {1919})}\BibitemShut
  {NoStop}%
\bibitem [{\citenamefont {Novotny}\ and\ \citenamefont
  {Marmur}(1991)}]{Novotny1991}%
  \BibitemOpen
  \bibfield  {author} {\bibinfo {author} {\bibfnamefont {V.}~\bibnamefont
  {Novotny}}\ and\ \bibinfo {author} {\bibfnamefont {A.}~\bibnamefont
  {Marmur}},\ }\bibfield  {title} {\bibinfo {title} {{\textit{Wetting
  autophobicity}}},\ }\href {https://doi.org/10.1016/0021-9797(91)90367-h}
  {\bibfield  {journal} {\bibinfo  {journal} {J. Colloid Interface Sci.}\
  }\textbf {\bibinfo {volume} {145}},\ \bibinfo {pages} {355–361} (\bibinfo
  {year} {1991})}\BibitemShut {NoStop}%
\bibitem [{\citenamefont {Bangham}\ and\ \citenamefont
  {Saweris}(1938)}]{Bangham1938}%
  \BibitemOpen
  \bibfield  {author} {\bibinfo {author} {\bibfnamefont {D.~H.}\ \bibnamefont
  {Bangham}}\ and\ \bibinfo {author} {\bibfnamefont {Z.}~\bibnamefont
  {Saweris}},\ }\bibfield  {title} {\bibinfo {title} {{\textit{The behaviour of
  liquid drops and adsorbed films at cleavage surfaces of mica}}},\ }\href
  {https://doi.org/10.1039/tf9383400554} {\bibfield  {journal} {\bibinfo
  {journal} {Trans. Faraday Soc.}\ }\textbf {\bibinfo {volume} {34}},\ \bibinfo
  {pages} {554} (\bibinfo {year} {1938})}\BibitemShut {NoStop}%
\bibitem [{\citenamefont {Bahadur}\ \emph {et~al.}(2009)\citenamefont
  {Bahadur}, \citenamefont {Yadav}, \citenamefont {Chaurasia}, \citenamefont
  {Leh},\ and\ \citenamefont {Tadmor}}]{Bahadur2009}%
  \BibitemOpen
  \bibfield  {author} {\bibinfo {author} {\bibfnamefont {P.}~\bibnamefont
  {Bahadur}}, \bibinfo {author} {\bibfnamefont {P.~S.}\ \bibnamefont {Yadav}},
  \bibinfo {author} {\bibfnamefont {K.}~\bibnamefont {Chaurasia}}, \bibinfo
  {author} {\bibfnamefont {A.}~\bibnamefont {Leh}},\ and\ \bibinfo {author}
  {\bibfnamefont {R.}~\bibnamefont {Tadmor}},\ }\bibfield  {title} {\bibinfo
  {title} {{\textit{Chasing drops: Following escaper and pursuer drop couple
  system}}},\ }\href {https://doi.org/10.1016/j.jcis.2008.12.050} {\bibfield
  {journal} {\bibinfo  {journal} {J. Colloid Interface Sci.}\ }\textbf
  {\bibinfo {volume} {332}},\ \bibinfo {pages} {455–460} (\bibinfo {year}
  {2009})}\BibitemShut {NoStop}%
\bibitem [{\citenamefont {Ausserré}\ \emph {et~al.}(1986)\citenamefont
  {Ausserré}, \citenamefont {Picard},\ and\ \citenamefont
  {Léger}}]{Ausserr1986}%
  \BibitemOpen
  \bibfield  {author} {\bibinfo {author} {\bibfnamefont {D.}~\bibnamefont
  {Ausserré}}, \bibinfo {author} {\bibfnamefont {A.~M.}\ \bibnamefont
  {Picard}},\ and\ \bibinfo {author} {\bibfnamefont {L.}~\bibnamefont
  {Léger}},\ }\bibfield  {title} {\bibinfo {title} {{\textit{Existence and
  Role of the Precursor Film in the Spreading of Polymer Liquids}}},\ }\href
  {https://doi.org/10.1103/physrevlett.57.2671} {\bibfield  {journal} {\bibinfo
   {journal} {Phys. Rev. Lett.}\ }\textbf {\bibinfo {volume} {57}},\ \bibinfo
  {pages} {2671–2674} (\bibinfo {year} {1986})}\BibitemShut {NoStop}%
\bibitem [{\citenamefont {Beaglehole}(1989)}]{Beaglehole1989}%
  \BibitemOpen
  \bibfield  {author} {\bibinfo {author} {\bibfnamefont {D.}~\bibnamefont
  {Beaglehole}},\ }\bibfield  {title} {\bibinfo {title} {{\textit{Profiles of
  the precursor of spreading drops of siloxane oil on glass, fused silica, and
  mica}}},\ }\href {https://doi.org/10.1021/j100339a067} {\bibfield  {journal}
  {\bibinfo  {journal} {J. Phys. Chem.}\ }\textbf {\bibinfo {volume} {93}},\
  \bibinfo {pages} {893–899} (\bibinfo {year} {1989})}\BibitemShut {NoStop}%
\bibitem [{\citenamefont {Heslot}\ \emph
  {et~al.}(1989{\natexlab{a}})\citenamefont {Heslot}, \citenamefont {Cazabat},\
  and\ \citenamefont {Levinson}}]{Heslot1989}%
  \BibitemOpen
  \bibfield  {author} {\bibinfo {author} {\bibfnamefont {F.}~\bibnamefont
  {Heslot}}, \bibinfo {author} {\bibfnamefont {A.~M.}\ \bibnamefont
  {Cazabat}},\ and\ \bibinfo {author} {\bibfnamefont {P.}~\bibnamefont
  {Levinson}},\ }\bibfield  {title} {\bibinfo {title} {{\textit{Dynamics of
  wetting of tiny drops: Ellipsometric study of the late stages of
  spreading}}},\ }\href {https://doi.org/10.1103/PhysRevLett.62.1286}
  {\bibfield  {journal} {\bibinfo  {journal} {Phys. Rev. Lett.}\ }\textbf
  {\bibinfo {volume} {62}},\ \bibinfo {pages} {1286} (\bibinfo {year}
  {1989}{\natexlab{a}})}\BibitemShut {NoStop}%
\bibitem [{\citenamefont {Heslot}\ \emph
  {et~al.}(1989{\natexlab{b}})\citenamefont {Heslot}, \citenamefont {Fraysse},\
  and\ \citenamefont {Cazabat}}]{Heslot1989-2}%
  \BibitemOpen
  \bibfield  {author} {\bibinfo {author} {\bibfnamefont {F.}~\bibnamefont
  {Heslot}}, \bibinfo {author} {\bibfnamefont {N.}~\bibnamefont {Fraysse}},\
  and\ \bibinfo {author} {\bibfnamefont {A.~M.}\ \bibnamefont {Cazabat}},\
  }\bibfield  {title} {\bibinfo {title} {{\textit{Molecular layering in the
  spreading of wetting liquid drops}}},\ }\href
  {https://doi.org/10.1038/338640a0} {\bibfield  {journal} {\bibinfo  {journal}
  {Nature}\ }\textbf {\bibinfo {volume} {338}},\ \bibinfo {pages} {640–642}
  (\bibinfo {year} {1989}{\natexlab{b}})}\BibitemShut {NoStop}%
\bibitem [{\citenamefont {Heslot}\ \emph
  {et~al.}(1989{\natexlab{c}})\citenamefont {Heslot}, \citenamefont {Cazabat},\
  and\ \citenamefont {Fraysse}}]{Heslot1989-3}%
  \BibitemOpen
  \bibfield  {author} {\bibinfo {author} {\bibfnamefont {F.}~\bibnamefont
  {Heslot}}, \bibinfo {author} {\bibfnamefont {A.~M.}\ \bibnamefont
  {Cazabat}},\ and\ \bibinfo {author} {\bibfnamefont {N.}~\bibnamefont
  {Fraysse}},\ }\bibfield  {title} {\bibinfo {title}
  {{\textit{Diffusion-controlled wetting films}}},\ }\href
  {https://doi.org/10.1088/0953-8984/1/33/024} {\bibfield  {journal} {\bibinfo
  {journal} {J. Phys. Condens. Matter}\ }\textbf {\bibinfo {volume} {1}},\
  \bibinfo {pages} {5793–5798} (\bibinfo {year}
  {1989}{\natexlab{c}})}\BibitemShut {NoStop}%
\bibitem [{\citenamefont {Heslot}\ \emph {et~al.}(1992)\citenamefont {Heslot},
  \citenamefont {Cazabat}, \citenamefont {Fraysse},\ and\ \citenamefont
  {Levinson}}]{Heslot1992}%
  \BibitemOpen
  \bibfield  {author} {\bibinfo {author} {\bibfnamefont {F.}~\bibnamefont
  {Heslot}}, \bibinfo {author} {\bibfnamefont {A.~M.}\ \bibnamefont {Cazabat}},
  \bibinfo {author} {\bibfnamefont {N.}~\bibnamefont {Fraysse}},\ and\ \bibinfo
  {author} {\bibfnamefont {P.}~\bibnamefont {Levinson}},\ }\bibfield  {title}
  {\bibinfo {title} {{\textit{Experiments on spreading droplets and thin
  films}}},\ }\href {https://doi.org/10.1016/0001-8686(92)80058-6} {\bibfield
  {journal} {\bibinfo  {journal} {Adv. Colloid Interface Sci.}\ }\textbf
  {\bibinfo {volume} {39}},\ \bibinfo {pages} {129–145} (\bibinfo {year}
  {1992})}\BibitemShut {NoStop}%
\bibitem [{\citenamefont {Ou~Ramdane}\ \emph {et~al.}(1998)\citenamefont
  {Ou~Ramdane}, \citenamefont {Auroy},\ and\ \citenamefont
  {Silberzan}}]{OuRamdane1998}%
  \BibitemOpen
  \bibfield  {author} {\bibinfo {author} {\bibfnamefont {O.}~\bibnamefont
  {Ou~Ramdane}}, \bibinfo {author} {\bibfnamefont {P.}~\bibnamefont {Auroy}},\
  and\ \bibinfo {author} {\bibfnamefont {P.}~\bibnamefont {Silberzan}},\
  }\bibfield  {title} {\bibinfo {title} {{\textit{Wetting of Polymer Brushes by
  a Nematogenic Compound}}},\ }\href
  {https://doi.org/10.1103/physrevlett.80.5141} {\bibfield  {journal} {\bibinfo
   {journal} {Phys. Rev. Lett.}\ }\textbf {\bibinfo {volume} {80}},\ \bibinfo
  {pages} {5141–5144} (\bibinfo {year} {1998})}\BibitemShut {NoStop}%
\bibitem [{\citenamefont {Xu}\ \emph {et~al.}(2000)\citenamefont {Xu},
  \citenamefont {Salmeron},\ and\ \citenamefont {Bardon}}]{Xu2000}%
  \BibitemOpen
  \bibfield  {author} {\bibinfo {author} {\bibfnamefont {L.}~\bibnamefont
  {Xu}}, \bibinfo {author} {\bibfnamefont {M.}~\bibnamefont {Salmeron}},\ and\
  \bibinfo {author} {\bibfnamefont {S.}~\bibnamefont {Bardon}},\ }\bibfield
  {title} {\bibinfo {title} {{\textit{Wetting and Molecular Orientation of 8CB
  on Silicon Substrates}}},\ }\href
  {https://doi.org/10.1103/physrevlett.84.1519} {\bibfield  {journal} {\bibinfo
   {journal} {Phys. Rev. Lett.}\ }\textbf {\bibinfo {volume} {84}},\ \bibinfo
  {pages} {1519–1522} (\bibinfo {year} {2000})}\BibitemShut {NoStop}%
\bibitem [{\citenamefont {Lazar}\ \emph {et~al.}(2005)\citenamefont {Lazar},
  \citenamefont {Schollmeyer},\ and\ \citenamefont {Riegler}}]{Lazar2005}%
  \BibitemOpen
  \bibfield  {author} {\bibinfo {author} {\bibfnamefont {P.}~\bibnamefont
  {Lazar}}, \bibinfo {author} {\bibfnamefont {H.}~\bibnamefont {Schollmeyer}},\
  and\ \bibinfo {author} {\bibfnamefont {H.}~\bibnamefont {Riegler}},\
  }\bibfield  {title} {\bibinfo {title} {{\textit{Spreading and Two-Dimensional
  Mobility of Long-Chain Alkanes at Solid/Gas Interfaces}}},\ }\href
  {http://dx.doi.org/10.1103/PhysRevLett.94.116101} {\bibfield  {journal}
  {\bibinfo  {journal} {Phys. Rev. Lett.}\ }\textbf {\bibinfo {volume} {94}}
  (\bibinfo {year} {2005})}\BibitemShut {NoStop}%
\bibitem [{\citenamefont {Moon}\ \emph {et~al.}(2004)\citenamefont {Moon},
  \citenamefont {Wynblatt}, \citenamefont {Garoff},\ and\ \citenamefont
  {Suter}}]{Moon2004}%
  \BibitemOpen
  \bibfield  {author} {\bibinfo {author} {\bibfnamefont {J.}~\bibnamefont
  {Moon}}, \bibinfo {author} {\bibfnamefont {P.}~\bibnamefont {Wynblatt}},
  \bibinfo {author} {\bibfnamefont {S.}~\bibnamefont {Garoff}},\ and\ \bibinfo
  {author} {\bibfnamefont {R.}~\bibnamefont {Suter}},\ }\bibfield  {title}
  {\bibinfo {title} {{\textit{Diffusion kinetics of Bi and Pb–Bi monolayer
  precursing films on Cu(1 1 1)}}},\ }\href
  {https://doi.org/10.1016/j.susc.2004.04.018} {\bibfield  {journal} {\bibinfo
  {journal} {Surf. Sci.}\ }\textbf {\bibinfo {volume} {559}},\ \bibinfo {pages}
  {149–157} (\bibinfo {year} {2004})}\BibitemShut {NoStop}%
\bibitem [{\citenamefont {Monchoux}\ \emph {et~al.}(2006)\citenamefont
  {Monchoux}, \citenamefont {Chatain},\ and\ \citenamefont
  {Wynblatt}}]{Monchoux2006}%
  \BibitemOpen
  \bibfield  {author} {\bibinfo {author} {\bibfnamefont {J.}~\bibnamefont
  {Monchoux}}, \bibinfo {author} {\bibfnamefont {D.}~\bibnamefont {Chatain}},\
  and\ \bibinfo {author} {\bibfnamefont {P.}~\bibnamefont {Wynblatt}},\
  }\bibfield  {title} {\bibinfo {title} {{\textit{Impact of surface phase
  transitions and structure on surface diffusion profiles of Pb and Bi over
  Cu(100)}}},\ }\href {https://doi.org/10.1016/j.susc.2006.01.012} {\bibfield
  {journal} {\bibinfo  {journal} {Surf. Sci.}\ }\textbf {\bibinfo {volume}
  {600}},\ \bibinfo {pages} {1265–1276} (\bibinfo {year} {2006})}\BibitemShut
  {NoStop}%
\bibitem [{\citenamefont {de~Gennes}\ and\ \citenamefont
  {Cazabat}(1991)}]{deGennes1991}%
  \BibitemOpen
  \bibfield  {author} {\bibinfo {author} {\bibfnamefont {P.~G.}\ \bibnamefont
  {de~Gennes}}\ and\ \bibinfo {author} {\bibfnamefont {A.~M.}\ \bibnamefont
  {Cazabat}},\ }\bibinfo {title} {{\textit{Spreading of a stratified,
  incompressible, droplet}}},\ in\ \href
  {https://doi.org/10.1007/3-540-54367-8_37} {\bibinfo {booktitle} {Capillarity
  Today}}\ (\bibinfo  {publisher} {Springer Berlin Heidelberg},\ \bibinfo
  {year} {1991})\ p.\ \bibinfo {pages} {33–39}\BibitemShut {NoStop}%
\bibitem [{\citenamefont {Abraham}\ \emph
  {et~al.}(1990{\natexlab{a}})\citenamefont {Abraham}, \citenamefont {Collet},
  \citenamefont {De~Coninck},\ and\ \citenamefont {Dunlop}}]{Abraham1990}%
  \BibitemOpen
  \bibfield  {author} {\bibinfo {author} {\bibfnamefont {D.~B.}\ \bibnamefont
  {Abraham}}, \bibinfo {author} {\bibfnamefont {P.}~\bibnamefont {Collet}},
  \bibinfo {author} {\bibfnamefont {J.}~\bibnamefont {De~Coninck}},\ and\
  \bibinfo {author} {\bibfnamefont {F.}~\bibnamefont {Dunlop}},\ }\bibfield
  {title} {\bibinfo {title} {{\textit{Langevin dynamics of spreading and
  wetting}}},\ }\href {https://doi.org/10.1103/physrevlett.65.195} {\bibfield
  {journal} {\bibinfo  {journal} {Phys. Rev. Lett.}\ }\textbf {\bibinfo
  {volume} {65}},\ \bibinfo {pages} {195–198} (\bibinfo {year}
  {1990}{\natexlab{a}})}\BibitemShut {NoStop}%
\bibitem [{\citenamefont {Abraham}\ \emph
  {et~al.}(1990{\natexlab{b}})\citenamefont {Abraham}, \citenamefont {Collet},
  \citenamefont {De~Coninck},\ and\ \citenamefont {Dunlop}}]{Abraham1990-2}%
  \BibitemOpen
  \bibfield  {author} {\bibinfo {author} {\bibfnamefont {D.}~\bibnamefont
  {Abraham}}, \bibinfo {author} {\bibfnamefont {P.}~\bibnamefont {Collet}},
  \bibinfo {author} {\bibfnamefont {J.}~\bibnamefont {De~Coninck}},\ and\
  \bibinfo {author} {\bibfnamefont {F.}~\bibnamefont {Dunlop}},\ }\bibfield
  {title} {\bibinfo {title} {{\textit{Langevin dynamics of an interface near a
  wall}}},\ }\href {https://doi.org/10.1007/bf01027290} {\bibfield  {journal}
  {\bibinfo  {journal} {J. Stat. Phys.}\ }\textbf {\bibinfo {volume} {61}},\
  \bibinfo {pages} {509–532} (\bibinfo {year}
  {1990}{\natexlab{b}})}\BibitemShut {NoStop}%
\bibitem [{\citenamefont {De~Coninck}\ \emph
  {et~al.}(1993{\natexlab{a}})\citenamefont {De~Coninck}, \citenamefont
  {Dunlop},\ and\ \citenamefont {Menu}}]{DeConinck1993}%
  \BibitemOpen
  \bibfield  {author} {\bibinfo {author} {\bibfnamefont {J.}~\bibnamefont
  {De~Coninck}}, \bibinfo {author} {\bibfnamefont {F.}~\bibnamefont {Dunlop}},\
  and\ \bibinfo {author} {\bibfnamefont {F.}~\bibnamefont {Menu}},\ }\bibfield
  {title} {\bibinfo {title} {{\textit{Spreading of a solid-on-solid drop}}},\
  }\href {https://doi.org/10.1103/physreve.47.1820} {\bibfield  {journal}
  {\bibinfo  {journal} {Phys. Rev. E}\ }\textbf {\bibinfo {volume} {47}},\
  \bibinfo {pages} {1820–1823} (\bibinfo {year}
  {1993}{\natexlab{a}})}\BibitemShut {NoStop}%
\bibitem [{\citenamefont {Burlatsky}\ \emph
  {et~al.}(1996{\natexlab{a}})\citenamefont {Burlatsky}, \citenamefont
  {Oshanin}, \citenamefont {Cazabat},\ and\ \citenamefont
  {Moreau}}]{Burlatsky1996}%
  \BibitemOpen
  \bibfield  {author} {\bibinfo {author} {\bibfnamefont {S.~F.}\ \bibnamefont
  {Burlatsky}}, \bibinfo {author} {\bibfnamefont {G.}~\bibnamefont {Oshanin}},
  \bibinfo {author} {\bibfnamefont {A.~M.}\ \bibnamefont {Cazabat}},\ and\
  \bibinfo {author} {\bibfnamefont {M.}~\bibnamefont {Moreau}},\ }\bibfield
  {title} {\bibinfo {title} {{\textit{Microscopic Model of Upward Creep of an
  Ultrathin Wetting Film}}},\ }\href
  {https://doi.org/10.1103/physrevlett.76.86} {\bibfield  {journal} {\bibinfo
  {journal} {Phys. Rev. Lett.}\ }\textbf {\bibinfo {volume} {76}},\ \bibinfo
  {pages} {86–89} (\bibinfo {year} {1996}{\natexlab{a}})}\BibitemShut
  {NoStop}%
\bibitem [{\citenamefont {Burlatsky}\ \emph
  {et~al.}(1996{\natexlab{b}})\citenamefont {Burlatsky}, \citenamefont
  {Oshanin}, \citenamefont {Cazabat}, \citenamefont {Moreau},\ and\
  \citenamefont {Reinhardt}}]{Burlatsky1996-2}%
  \BibitemOpen
  \bibfield  {author} {\bibinfo {author} {\bibfnamefont {S.~F.}\ \bibnamefont
  {Burlatsky}}, \bibinfo {author} {\bibfnamefont {G.}~\bibnamefont {Oshanin}},
  \bibinfo {author} {\bibfnamefont {A.~M.}\ \bibnamefont {Cazabat}}, \bibinfo
  {author} {\bibfnamefont {M.}~\bibnamefont {Moreau}},\ and\ \bibinfo {author}
  {\bibfnamefont {W.~P.}\ \bibnamefont {Reinhardt}},\ }\bibfield  {title}
  {\bibinfo {title} {{\textit{Spreading of a thin wetting film: Microscopic
  approach}}},\ }\href {https://doi.org/10.1103/physreve.54.3832} {\bibfield
  {journal} {\bibinfo  {journal} {Phys. Rev. E}\ }\textbf {\bibinfo {volume}
  {54}},\ \bibinfo {pages} {3832–3845} (\bibinfo {year}
  {1996}{\natexlab{b}})}\BibitemShut {NoStop}%
\bibitem [{\citenamefont {Yang}\ \emph {et~al.}(1991)\citenamefont {Yang},
  \citenamefont {Koplik},\ and\ \citenamefont {Banavar}}]{Yang1991}%
  \BibitemOpen
  \bibfield  {author} {\bibinfo {author} {\bibfnamefont {J.-x.}\ \bibnamefont
  {Yang}}, \bibinfo {author} {\bibfnamefont {J.}~\bibnamefont {Koplik}},\ and\
  \bibinfo {author} {\bibfnamefont {J.~R.}\ \bibnamefont {Banavar}},\
  }\bibfield  {title} {\bibinfo {title} {{\textit{Molecular dynamics of drop
  spreading on a solid surface}}},\ }\href
  {https://doi.org/10.1103/physrevlett.67.3539} {\bibfield  {journal} {\bibinfo
   {journal} {Phys. Rev. Lett.}\ }\textbf {\bibinfo {volume} {67}},\ \bibinfo
  {pages} {3539–3542} (\bibinfo {year} {1991})}\BibitemShut {NoStop}%
\bibitem [{\citenamefont {Yang}\ \emph {et~al.}(1992)\citenamefont {Yang},
  \citenamefont {Koplik},\ and\ \citenamefont {Banavar}}]{Yang1992}%
  \BibitemOpen
  \bibfield  {author} {\bibinfo {author} {\bibfnamefont {J.-x.}\ \bibnamefont
  {Yang}}, \bibinfo {author} {\bibfnamefont {J.}~\bibnamefont {Koplik}},\ and\
  \bibinfo {author} {\bibfnamefont {J.~R.}\ \bibnamefont {Banavar}},\
  }\bibfield  {title} {\bibinfo {title} {{\textit{Terraced spreading of simple
  liquids on solid surfaces}}},\ }\href
  {https://doi.org/10.1103/physreva.46.7738} {\bibfield  {journal} {\bibinfo
  {journal} {Phys. Rev. A}\ }\textbf {\bibinfo {volume} {46}},\ \bibinfo
  {pages} {7738–7749} (\bibinfo {year} {1992})}\BibitemShut {NoStop}%
\bibitem [{\citenamefont {Nieminen}\ \emph {et~al.}(1992)\citenamefont
  {Nieminen}, \citenamefont {Abraham}, \citenamefont {Karttunen},\ and\
  \citenamefont {Kaski}}]{Nieminen1992}%
  \BibitemOpen
  \bibfield  {author} {\bibinfo {author} {\bibfnamefont {J.~A.}\ \bibnamefont
  {Nieminen}}, \bibinfo {author} {\bibfnamefont {D.~B.}\ \bibnamefont
  {Abraham}}, \bibinfo {author} {\bibfnamefont {M.}~\bibnamefont {Karttunen}},\
  and\ \bibinfo {author} {\bibfnamefont {K.}~\bibnamefont {Kaski}},\ }\bibfield
   {title} {\bibinfo {title} {{\textit{Molecular dynamics of a microscopic
  droplet on solid surface}}},\ }\href
  {https://doi.org/10.1103/physrevlett.69.124} {\bibfield  {journal} {\bibinfo
  {journal} {Phys. Rev. Lett.}\ }\textbf {\bibinfo {volume} {69}},\ \bibinfo
  {pages} {124–127} (\bibinfo {year} {1992})}\BibitemShut {NoStop}%
\bibitem [{\citenamefont {Nieminen}\ and\ \citenamefont
  {Ala-Nissila}(1994)}]{Nieminen1994}%
  \BibitemOpen
  \bibfield  {author} {\bibinfo {author} {\bibfnamefont {J.~A.}\ \bibnamefont
  {Nieminen}}\ and\ \bibinfo {author} {\bibfnamefont {T.}~\bibnamefont
  {Ala-Nissila}},\ }\bibfield  {title} {\bibinfo {title} {{\textit{Spreading
  dynamics of polymer microdroplets: A molecular-dynamics study}}},\ }\href
  {https://doi.org/10.1103/physreve.49.4228} {\bibfield  {journal} {\bibinfo
  {journal} {Phys. Rev. E}\ }\textbf {\bibinfo {volume} {49}},\ \bibinfo
  {pages} {4228–4236} (\bibinfo {year} {1994})}\BibitemShut {NoStop}%
\bibitem [{\citenamefont {De~Coninck}\ \emph {et~al.}(1995)\citenamefont
  {De~Coninck}, \citenamefont {D’Ortona}, \citenamefont {Koplik},\ and\
  \citenamefont {Banavar}}]{DeConinck1995}%
  \BibitemOpen
  \bibfield  {author} {\bibinfo {author} {\bibfnamefont {J.}~\bibnamefont
  {De~Coninck}}, \bibinfo {author} {\bibfnamefont {U.}~\bibnamefont
  {D’Ortona}}, \bibinfo {author} {\bibfnamefont {J.}~\bibnamefont {Koplik}},\
  and\ \bibinfo {author} {\bibfnamefont {J.~R.}\ \bibnamefont {Banavar}},\
  }\bibfield  {title} {\bibinfo {title} {{\textit{Terraced Spreading of Chain
  Molecules via Molecular Dynamics}}},\ }\href
  {https://doi.org/10.1103/physrevlett.74.928} {\bibfield  {journal} {\bibinfo
  {journal} {Phys. Rev. Lett.}\ }\textbf {\bibinfo {volume} {74}},\ \bibinfo
  {pages} {928–931} (\bibinfo {year} {1995})}\BibitemShut {NoStop}%
\bibitem [{\citenamefont {D’Ortona}\ \emph {et~al.}(1996)\citenamefont
  {D’Ortona}, \citenamefont {De~Coninck}, \citenamefont {Koplik},\ and\
  \citenamefont {Banavar}}]{DOrtona1996}%
  \BibitemOpen
  \bibfield  {author} {\bibinfo {author} {\bibfnamefont {U.}~\bibnamefont
  {D’Ortona}}, \bibinfo {author} {\bibfnamefont {J.}~\bibnamefont
  {De~Coninck}}, \bibinfo {author} {\bibfnamefont {J.}~\bibnamefont {Koplik}},\
  and\ \bibinfo {author} {\bibfnamefont {J.~R.}\ \bibnamefont {Banavar}},\
  }\bibfield  {title} {\bibinfo {title} {{\textit{Terraced spreading mechanisms
  for chain molecules}}},\ }\href {https://doi.org/10.1103/physreve.53.562}
  {\bibfield  {journal} {\bibinfo  {journal} {Phys. Rev. E}\ }\textbf {\bibinfo
  {volume} {53}},\ \bibinfo {pages} {562–569} (\bibinfo {year}
  {1996})}\BibitemShut {NoStop}%
\bibitem [{\citenamefont {Abraham}\ \emph
  {et~al.}(1991{\natexlab{a}})\citenamefont {Abraham}, \citenamefont {Heinio},\
  and\ \citenamefont {Kaski}}]{Abraham1991}%
  \BibitemOpen
  \bibfield  {author} {\bibinfo {author} {\bibfnamefont {D.~B.}\ \bibnamefont
  {Abraham}}, \bibinfo {author} {\bibfnamefont {J.}~\bibnamefont {Heinio}},\
  and\ \bibinfo {author} {\bibfnamefont {K.}~\bibnamefont {Kaski}},\ }\bibfield
   {title} {\bibinfo {title} {{\textit{Computer simulation studies of fluid
  spreading}}},\ }\href {https://doi.org/10.1088/0305-4470/24/6/010} {\bibfield
   {journal} {\bibinfo  {journal} {J. Phys. A: Math. Gen.}\ }\textbf {\bibinfo
  {volume} {24}},\ \bibinfo {pages} {L309–L315} (\bibinfo {year}
  {1991}{\natexlab{a}})}\BibitemShut {NoStop}%
\bibitem [{\citenamefont {Abraham}\ \emph
  {et~al.}(1991{\natexlab{b}})\citenamefont {Abraham}, \citenamefont {Collet},
  \citenamefont {De~Coninck}, \citenamefont {Dunlop}, \citenamefont
  {Heini\"{o}}, \citenamefont {Kaski},\ and\ \citenamefont
  {Ko}}]{Abraham1991-2}%
  \BibitemOpen
  \bibfield  {author} {\bibinfo {author} {\bibfnamefont {D.}~\bibnamefont
  {Abraham}}, \bibinfo {author} {\bibfnamefont {P.}~\bibnamefont {Collet}},
  \bibinfo {author} {\bibfnamefont {J.}~\bibnamefont {De~Coninck}}, \bibinfo
  {author} {\bibfnamefont {F.}~\bibnamefont {Dunlop}}, \bibinfo {author}
  {\bibfnamefont {J.}~\bibnamefont {Heini\"{o}}}, \bibinfo {author}
  {\bibfnamefont {K.}~\bibnamefont {Kaski}},\ and\ \bibinfo {author}
  {\bibfnamefont {L.-F.}\ \bibnamefont {Ko}},\ }\bibfield  {title} {\bibinfo
  {title} {{\textit{Theory of wetting and spreading}}},\ }\href
  {https://doi.org/10.1016/0378-4371(91)90315-4} {\bibfield  {journal}
  {\bibinfo  {journal} {Physica A: Stat. Mech. Appl.}\ }\textbf {\bibinfo
  {volume} {172}},\ \bibinfo {pages} {125–136} (\bibinfo {year}
  {1991}{\natexlab{b}})}\BibitemShut {NoStop}%
\bibitem [{\citenamefont {Heini\"{o}}\ \emph {et~al.}(1992)\citenamefont
  {Heini\"{o}}, \citenamefont {Kaski},\ and\ \citenamefont
  {Abraham}}]{Heini1992}%
  \BibitemOpen
  \bibfield  {author} {\bibinfo {author} {\bibfnamefont {J.}~\bibnamefont
  {Heini\"{o}}}, \bibinfo {author} {\bibfnamefont {K.}~\bibnamefont {Kaski}},\
  and\ \bibinfo {author} {\bibfnamefont {D.~B.}\ \bibnamefont {Abraham}},\
  }\bibfield  {title} {\bibinfo {title} {{\textit{Dynamics of a microscopic
  droplet on a solid surface: Theory and simulation}}},\ }\href
  {https://doi.org/10.1103/physrevb.45.4409} {\bibfield  {journal} {\bibinfo
  {journal} {Phys. Rev. B}\ }\textbf {\bibinfo {volume} {45}},\ \bibinfo
  {pages} {4409–4416} (\bibinfo {year} {1992})}\BibitemShut {NoStop}%
\bibitem [{\citenamefont {Cheng}\ and\ \citenamefont
  {Ebner}(1993)}]{Cheng1993}%
  \BibitemOpen
  \bibfield  {author} {\bibinfo {author} {\bibfnamefont {E.}~\bibnamefont
  {Cheng}}\ and\ \bibinfo {author} {\bibfnamefont {C.}~\bibnamefont {Ebner}},\
  }\bibfield  {title} {\bibinfo {title} {{\textit{Dynamics of liquid-droplet
  spreading: A Monte Carlo study}}},\ }\href
  {https://doi.org/10.1103/PhysRevB.47.13808} {\bibfield  {journal} {\bibinfo
  {journal} {Phys. Rev. B}\ }\textbf {\bibinfo {volume} {47}},\ \bibinfo
  {pages} {13808} (\bibinfo {year} {1993})}\BibitemShut {NoStop}%
\bibitem [{\citenamefont {De~Coninck}\ \emph
  {et~al.}(1993{\natexlab{b}})\citenamefont {De~Coninck}, \citenamefont
  {Hoorelbeke}, \citenamefont {Valignat},\ and\ \citenamefont
  {Cazabat}}]{DeConinck1993-2}%
  \BibitemOpen
  \bibfield  {author} {\bibinfo {author} {\bibfnamefont {J.}~\bibnamefont
  {De~Coninck}}, \bibinfo {author} {\bibfnamefont {S.}~\bibnamefont
  {Hoorelbeke}}, \bibinfo {author} {\bibfnamefont {M.~P.}\ \bibnamefont
  {Valignat}},\ and\ \bibinfo {author} {\bibfnamefont {A.~M.}\ \bibnamefont
  {Cazabat}},\ }\bibfield  {title} {\bibinfo {title} {{\textit{Effective
  microscopic model for the dynamics of spreading}}},\ }\href
  {https://doi.org/10.1103/physreve.48.4549} {\bibfield  {journal} {\bibinfo
  {journal} {Phys. Rev. E}\ }\textbf {\bibinfo {volume} {48}},\ \bibinfo
  {pages} {4549–4555} (\bibinfo {year} {1993}{\natexlab{b}})}\BibitemShut
  {NoStop}%
\bibitem [{\citenamefont {Lukkarinen}\ \emph {et~al.}(1995)\citenamefont
  {Lukkarinen}, \citenamefont {Kaski},\ and\ \citenamefont
  {Abraham}}]{Lukkarinen1995}%
  \BibitemOpen
  \bibfield  {author} {\bibinfo {author} {\bibfnamefont {A.}~\bibnamefont
  {Lukkarinen}}, \bibinfo {author} {\bibfnamefont {K.}~\bibnamefont {Kaski}},\
  and\ \bibinfo {author} {\bibfnamefont {D.~B.}\ \bibnamefont {Abraham}},\
  }\bibfield  {title} {\bibinfo {title} {{\textit{Mechanisms of fluid
  spreading: Ising model simulations}}},\ }\href
  {https://doi.org/10.1103/PhysRevE.51.2199} {\bibfield  {journal} {\bibinfo
  {journal} {Phys. Rev. E}\ }\textbf {\bibinfo {volume} {51}},\ \bibinfo
  {pages} {2199} (\bibinfo {year} {1995})}\BibitemShut {NoStop}%
\bibitem [{\citenamefont {Abraham}\ \emph {et~al.}(2002)\citenamefont
  {Abraham}, \citenamefont {Cuerno},\ and\ \citenamefont {Moro}}]{Abraham2002}%
  \BibitemOpen
  \bibfield  {author} {\bibinfo {author} {\bibfnamefont {D.~B.}\ \bibnamefont
  {Abraham}}, \bibinfo {author} {\bibfnamefont {R.}~\bibnamefont {Cuerno}},\
  and\ \bibinfo {author} {\bibfnamefont {E.}~\bibnamefont {Moro}},\ }\bibfield
  {title} {\bibinfo {title} {{\textit{Microscopic Model for Thin Film
  Spreading}}},\ }\href {https://doi.org/10.1103/physrevlett.88.206101}
  {\bibfield  {journal} {\bibinfo  {journal} {Phys. Rev. Lett.}\ }\textbf
  {\bibinfo {volume} {88}},\ \bibinfo {pages} {206101} (\bibinfo {year}
  {2002})}\BibitemShut {NoStop}%
\bibitem [{\citenamefont {Chalmers}\ \emph {et~al.}(2017)\citenamefont
  {Chalmers}, \citenamefont {Smith},\ and\ \citenamefont
  {Archer}}]{Chalmers2017}%
  \BibitemOpen
  \bibfield  {author} {\bibinfo {author} {\bibfnamefont {C.}~\bibnamefont
  {Chalmers}}, \bibinfo {author} {\bibfnamefont {R.}~\bibnamefont {Smith}},\
  and\ \bibinfo {author} {\bibfnamefont {A.~J.}\ \bibnamefont {Archer}},\
  }\bibfield  {title} {\bibinfo {title} {{\textit{Modelling the evaporation of
  nanoparticle suspensions from heterogeneous surfaces}}},\ }\href
  {https://doi.org/10.1088/1361-648X/aa76fd} {\bibfield  {journal} {\bibinfo
  {journal} {J. Phys. Condens. Matter}\ }\textbf {\bibinfo {volume} {29}},\
  \bibinfo {pages} {295102} (\bibinfo {year} {2017})}\BibitemShut {NoStop}%
\bibitem [{\citenamefont {Areshi}\ \emph {et~al.}(2019)\citenamefont {Areshi},
  \citenamefont {Tseluiko},\ and\ \citenamefont {Archer}}]{Areshi2019}%
  \BibitemOpen
  \bibfield  {author} {\bibinfo {author} {\bibfnamefont {M.}~\bibnamefont
  {Areshi}}, \bibinfo {author} {\bibfnamefont {D.}~\bibnamefont {Tseluiko}},\
  and\ \bibinfo {author} {\bibfnamefont {A.~J.}\ \bibnamefont {Archer}},\
  }\bibfield  {title} {\bibinfo {title} {{\textit{Kinetic Monte Carlo and
  hydrodynamic modeling of droplet dynamics on surfaces, including evaporation
  and condensation}}},\ }\href {https://doi.org/10.1103/PhysRevFluids.4.104006}
  {\bibfield  {journal} {\bibinfo  {journal} {Phys. Rev. Fluids}\ }\textbf
  {\bibinfo {volume} {4}},\ \bibinfo {pages} {104006} (\bibinfo {year}
  {2019})}\BibitemShut {NoStop}%
\bibitem [{\citenamefont {Newman}\ and\ \citenamefont
  {Barkema}(1999)}]{Newman1999}%
  \BibitemOpen
  \bibfield  {author} {\bibinfo {author} {\bibfnamefont {M.~E.~J.}\
  \bibnamefont {Newman}}\ and\ \bibinfo {author} {\bibfnamefont {G.~T.}\
  \bibnamefont {Barkema}},\ }\href
  {https://global.oup.com/academic/product/monte-carlo-methods-in-statistical-physics-9780198517979?cc=es&lang=en&}
  {{\textit{\href
  {https://global.oup.com/academic/product/monte-carlo-methods-in-statistical-physics-9780198517979?cc=es&lang=en&}
  {}}}\bibinfo {title} {\textit{Monte Carlo Methods in Statistical Physics}}}\
  (\bibinfo  {publisher} {Oxford University Press},\ \bibinfo {address}
  {Oxford},\ \bibinfo {year} {1999})\BibitemShut {NoStop}%
\bibitem [{\citenamefont {Bortz}\ \emph {et~al.}(1975)\citenamefont {Bortz},
  \citenamefont {Kalos},\ and\ \citenamefont {Lebowitz}}]{Bortz1975}%
  \BibitemOpen
  \bibfield  {author} {\bibinfo {author} {\bibfnamefont {A.}~\bibnamefont
  {Bortz}}, \bibinfo {author} {\bibfnamefont {M.}~\bibnamefont {Kalos}},\ and\
  \bibinfo {author} {\bibfnamefont {J.}~\bibnamefont {Lebowitz}},\ }\bibfield
  {title} {\bibinfo {title} {{\textit{A new algorithm for Monte Carlo
  simulation of Ising spin systems}}},\ }\href
  {https://doi.org/10.1016/0021-9991(75)90060-1} {\bibfield  {journal}
  {\bibinfo  {journal} {J. Comput. Phys.}\ }\textbf {\bibinfo {volume} {17}},\
  \bibinfo {pages} {10–18} (\bibinfo {year} {1975})}\BibitemShut {NoStop}%
\bibitem [{\citenamefont {Alves}\ \emph {et~al.}(2011)\citenamefont {Alves},
  \citenamefont {Oliveira},\ and\ \citenamefont {Ferreira}}]{Alves2011}%
  \BibitemOpen
  \bibfield  {author} {\bibinfo {author} {\bibfnamefont {S.~G.}\ \bibnamefont
  {Alves}}, \bibinfo {author} {\bibfnamefont {T.~J.}\ \bibnamefont
  {Oliveira}},\ and\ \bibinfo {author} {\bibfnamefont {S.~C.}\ \bibnamefont
  {Ferreira}},\ }\bibfield  {title} {\bibinfo {title} {{\textit{Universal
  fluctuations in radial growth models belonging to the {KPZ} universality
  class}}},\ }\href {https://doi.org/10.1209/0295-5075/96/48003} {\bibfield
  {journal} {\bibinfo  {journal} {Europhys. Lett.}\ }\textbf {\bibinfo {volume}
  {96}},\ \bibinfo {pages} {48003} (\bibinfo {year} {2011})}\BibitemShut
  {NoStop}%
\bibitem [{\citenamefont {Oliveira}\ \emph {et~al.}(2012)\citenamefont
  {Oliveira}, \citenamefont {Ferreira},\ and\ \citenamefont
  {Alves}}]{Oliveira2012}%
  \BibitemOpen
  \bibfield  {author} {\bibinfo {author} {\bibfnamefont {T.~J.}\ \bibnamefont
  {Oliveira}}, \bibinfo {author} {\bibfnamefont {S.~C.}\ \bibnamefont
  {Ferreira}},\ and\ \bibinfo {author} {\bibfnamefont {S.~G.}\ \bibnamefont
  {Alves}},\ }\bibfield  {title} {\bibinfo {title} {{\textit{Universal
  fluctuations in Kardar-Parisi-Zhang growth on one-dimensional flat
  substrates}}},\ }\href {https://doi.org/10.1103/physreve.85.010601}
  {\bibfield  {journal} {\bibinfo  {journal} {Phys. Rev. E}\ }\textbf {\bibinfo
  {volume} {85}},\ \bibinfo {pages} {010601(R)} (\bibinfo {year}
  {2012})}\BibitemShut {NoStop}%
\bibitem [{\citenamefont {Nicoli}\ \emph {et~al.}(2013)\citenamefont {Nicoli},
  \citenamefont {Cuerno},\ and\ \citenamefont {Castro}}]{Nicoli2013}%
  \BibitemOpen
  \bibfield  {author} {\bibinfo {author} {\bibfnamefont {M.}~\bibnamefont
  {Nicoli}}, \bibinfo {author} {\bibfnamefont {R.}~\bibnamefont {Cuerno}},\
  and\ \bibinfo {author} {\bibfnamefont {M.}~\bibnamefont {Castro}},\
  }\bibfield  {title} {\bibinfo {title} {{\textit{Dimensional fragility of the
  Kardar{\textendash}Parisi{\textendash}Zhang universality class}}},\ }\href
  {https://doi.org/10.1088/1742-5468/2013/11/p11001} {\bibfield  {journal}
  {\bibinfo  {journal} {J. Stat. Mech.: Theor. Exp.}\ }\textbf {\bibinfo
  {volume} {2013}},\ \bibinfo {pages} {P11001} (\bibinfo {year}
  {2013})}\BibitemShut {NoStop}%
\bibitem [{\citenamefont {Barreales}\ \emph {et~al.}(2020)\citenamefont
  {Barreales}, \citenamefont {Mel{\'{e}}ndez}, \citenamefont {Cuerno},\ and\
  \citenamefont {Ruiz-Lorenzo}}]{Barreales2020}%
  \BibitemOpen
  \bibfield  {author} {\bibinfo {author} {\bibfnamefont {B.~G.}\ \bibnamefont
  {Barreales}}, \bibinfo {author} {\bibfnamefont {J.~J.}\ \bibnamefont
  {Mel{\'{e}}ndez}}, \bibinfo {author} {\bibfnamefont {R.}~\bibnamefont
  {Cuerno}},\ and\ \bibinfo {author} {\bibfnamefont {J.~J.}\ \bibnamefont
  {Ruiz-Lorenzo}},\ }\bibfield  {title} {\bibinfo {title}
  {{\textit{Kardar{\textendash}Parisi{\textendash}Zhang universality class for
  the critical dynamics of reaction{\textendash}diffusion fronts}}},\ }\href
  {https://doi.org/10.1088/1742-5468/ab6a03} {\bibfield  {journal} {\bibinfo
  {journal} {J. Stat. Mech.: Theor. Exp.}\ }\textbf {\bibinfo {volume}
  {2020}},\ \bibinfo {pages} {023203} (\bibinfo {year} {2020})}\BibitemShut
  {NoStop}%
\bibitem [{\citenamefont {Cuerno}\ and\ \citenamefont
  {V\'{a}zquez}(2004)}]{Cuerno2004}%
  \BibitemOpen
  \bibfield  {author} {\bibinfo {author} {\bibfnamefont {R.}~\bibnamefont
  {Cuerno}}\ and\ \bibinfo {author} {\bibfnamefont {L.}~\bibnamefont
  {V\'{a}zquez}},\ }\bibfield  {title} {\bibinfo {title} {{\textit{Universality
  issues in surface kinetic roughening of thin solid films}}},\ }in\ \href
  {https://books.google.es/books?id=lIoZeb_domwC&lpg=PP1&hl=es&pg=PP1#v=onepage&q&f=false}
  {\bibinfo {booktitle} {\textit{Advances in Condensed Matter and Statistical
  Physics}}},\ \bibinfo {editor} {edited by\ \bibinfo {editor} {\bibfnamefont
  {E.}~\bibnamefont {Korutcheva}}\ and\ \bibinfo {editor} {\bibfnamefont
  {R.}~\bibnamefont {Cuerno}}}\ (\bibinfo  {publisher} {Nova Science
  Publishers},\ \bibinfo {address} {New York},\ \bibinfo {year}
  {2004})\BibitemShut {NoStop}%
\bibitem [{\citenamefont {L{\'{o}}pez}\ \emph {et~al.}(1997)\citenamefont
  {L{\'{o}}pez}, \citenamefont {Rodr{\'{i}}guez},\ and\ \citenamefont
  {Cuerno}}]{Lopez1997-2}%
  \BibitemOpen
  \bibfield  {author} {\bibinfo {author} {\bibfnamefont {J.~M.}\ \bibnamefont
  {L{\'{o}}pez}}, \bibinfo {author} {\bibfnamefont {M.~A.}\ \bibnamefont
  {Rodr{\'{i}}guez}},\ and\ \bibinfo {author} {\bibfnamefont {R.}~\bibnamefont
  {Cuerno}},\ }\bibfield  {title} {\bibinfo {title} {{\textit{{Power spectrum
  scaling in anomalous kinetic roughening of surfaces}}}},\ }\href
  {https://doi.org/10.1016/S0378-4371(97)00375-0} {\bibfield  {journal}
  {\bibinfo  {journal} {Physica A: Stat. Mech. Appl.}\ }\textbf {\bibinfo
  {volume} {246}},\ \bibinfo {pages} {329} (\bibinfo {year}
  {1997})}\BibitemShut {NoStop}%
\bibitem [{\citenamefont {Galeano}\ \emph {et~al.}(2003)\citenamefont
  {Galeano}, \citenamefont {Buceta}, \citenamefont {Juarez}, \citenamefont
  {Pumari{\~{n}}o}, \citenamefont {de~la Torre},\ and\ \citenamefont
  {Iriondo}}]{Galeano2003}%
  \BibitemOpen
  \bibfield  {author} {\bibinfo {author} {\bibfnamefont {J.}~\bibnamefont
  {Galeano}}, \bibinfo {author} {\bibfnamefont {J.}~\bibnamefont {Buceta}},
  \bibinfo {author} {\bibfnamefont {K.}~\bibnamefont {Juarez}}, \bibinfo
  {author} {\bibfnamefont {B.}~\bibnamefont {Pumari{\~{n}}o}}, \bibinfo
  {author} {\bibfnamefont {J.}~\bibnamefont {de~la Torre}},\ and\ \bibinfo
  {author} {\bibfnamefont {F.}~\bibnamefont {Iriondo}},\ }\bibfield  {title}
  {\bibinfo {title} {{\textit{{Dynamical scaling analysis of plant callus
  growth}}}},\ }\href {https://doi.org/10.1209/epl/i2005-10093-3} {\bibfield
  {journal} {\bibinfo  {journal} {Europhys. Lett.}\ }\textbf {\bibinfo {volume}
  {63}},\ \bibinfo {pages} {83} (\bibinfo {year} {2003})}\BibitemShut {NoStop}%
\bibitem [{\citenamefont {Santalla}\ \emph {et~al.}(2018)\citenamefont
  {Santalla}, \citenamefont {Rodr{\'{i}}guez-Laguna}, \citenamefont {Abad},
  \citenamefont {Mar{\'{i}}n}, \citenamefont {Espinosa}, \citenamefont
  {Mu{\~{n}}oz-Garc{\'{i}}a}, \citenamefont {V{\'{a}}zquez},\ and\
  \citenamefont {Cuerno}}]{Santalla2018}%
  \BibitemOpen
  \bibfield  {author} {\bibinfo {author} {\bibfnamefont {S.~N.}\ \bibnamefont
  {Santalla}}, \bibinfo {author} {\bibfnamefont {J.}~\bibnamefont
  {Rodr{\'{i}}guez-Laguna}}, \bibinfo {author} {\bibfnamefont {J.~P.}\
  \bibnamefont {Abad}}, \bibinfo {author} {\bibfnamefont {I.}~\bibnamefont
  {Mar{\'{i}}n}}, \bibinfo {author} {\bibfnamefont {M.~M.}\ \bibnamefont
  {Espinosa}}, \bibinfo {author} {\bibfnamefont {J.}~\bibnamefont
  {Mu{\~{n}}oz-Garc{\'{i}}a}}, \bibinfo {author} {\bibfnamefont
  {L.}~\bibnamefont {V{\'{a}}zquez}},\ and\ \bibinfo {author} {\bibfnamefont
  {R.}~\bibnamefont {Cuerno}},\ }\bibfield  {title} {\bibinfo {title}
  {{\textit{{Non-universality of front fluctuations for compact colonies of
  non-motile bacteria}}}},\ }\href {https://doi.org/10.1103/PhysRevE.98.012407}
  {\bibfield  {journal} {\bibinfo  {journal} {Phys. Rev. E}\ }\textbf {\bibinfo
  {volume} {98}},\ \bibinfo {pages} {012407} (\bibinfo {year}
  {2018})}\BibitemShut {NoStop}%
\bibitem [{\citenamefont {Br\'u}\ \emph {et~al.}(1998)\citenamefont {Br\'u},
  \citenamefont {Pastor}, \citenamefont {Fernaud}, \citenamefont {Br\'u},
  \citenamefont {Melle},\ and\ \citenamefont {Berenguer}}]{Bru1998}%
  \BibitemOpen
  \bibfield  {author} {\bibinfo {author} {\bibfnamefont {A.}~\bibnamefont
  {Br\'u}}, \bibinfo {author} {\bibfnamefont {J.~M.}\ \bibnamefont {Pastor}},
  \bibinfo {author} {\bibfnamefont {I.}~\bibnamefont {Fernaud}}, \bibinfo
  {author} {\bibfnamefont {I.}~\bibnamefont {Br\'u}}, \bibinfo {author}
  {\bibfnamefont {S.}~\bibnamefont {Melle}},\ and\ \bibinfo {author}
  {\bibfnamefont {C.}~\bibnamefont {Berenguer}},\ }\bibfield  {title} {\bibinfo
  {title} {{\textit{Super-Rough Dynamics on Tumor Growth}}},\ }\href
  {https://doi.org/10.1103/PhysRevLett.81.4008} {\bibfield  {journal} {\bibinfo
   {journal} {Phys. Rev. Lett.}\ }\textbf {\bibinfo {volume} {81}},\ \bibinfo
  {pages} {4008} (\bibinfo {year} {1998})}\BibitemShut {NoStop}%
\bibitem [{\citenamefont {Br{\'{u}}}\ \emph {et~al.}(2003)\citenamefont
  {Br{\'{u}}}, \citenamefont {Albertos}, \citenamefont {Subiza}, \citenamefont
  {Garc{\'{\i}}a-Asenjo},\ and\ \citenamefont {Br{\'{u}}}}]{Bru2003}%
  \BibitemOpen
  \bibfield  {author} {\bibinfo {author} {\bibfnamefont {A.}~\bibnamefont
  {Br{\'{u}}}}, \bibinfo {author} {\bibfnamefont {S.}~\bibnamefont {Albertos}},
  \bibinfo {author} {\bibfnamefont {J.~L.}\ \bibnamefont {Subiza}}, \bibinfo
  {author} {\bibfnamefont {J.~L.}\ \bibnamefont {Garc{\'{\i}}a-Asenjo}},\ and\
  \bibinfo {author} {\bibfnamefont {I.}~\bibnamefont {Br{\'{u}}}},\ }\bibfield
  {title} {\bibinfo {title} {{\textit{The Universal Dynamics of Tumor
  Growth}}},\ }\href {https://doi.org/10.1016/s0006-3495(03)74715-8} {\bibfield
   {journal} {\bibinfo  {journal} {Biophys. J.}\ }\textbf {\bibinfo {volume}
  {85}},\ \bibinfo {pages} {2948} (\bibinfo {year} {2003})}\BibitemShut
  {NoStop}%
\bibitem [{\citenamefont {Block}\ \emph {et~al.}(2007)\citenamefont {Block},
  \citenamefont {Sch{\"{o}}ll},\ and\ \citenamefont {Drasdo}}]{Block2007}%
  \BibitemOpen
  \bibfield  {author} {\bibinfo {author} {\bibfnamefont {M.}~\bibnamefont
  {Block}}, \bibinfo {author} {\bibfnamefont {E.}~\bibnamefont
  {Sch{\"{o}}ll}},\ and\ \bibinfo {author} {\bibfnamefont {D.}~\bibnamefont
  {Drasdo}},\ }\bibfield  {title} {\bibinfo {title} {{\textit{{Classifying the
  expansion kinetics and critical surface dynamics of growing cell
  populations}}}},\ }\href {https://doi.org/10.1103/PhysRevLett.99.248101}
  {\bibfield  {journal} {\bibinfo  {journal} {Phys. Rev. Lett.}\ }\textbf
  {\bibinfo {volume} {99}},\ \bibinfo {pages} {248101} (\bibinfo {year}
  {2007})}\BibitemShut {NoStop}%
\bibitem [{\citenamefont {Santalla}\ \emph {et~al.}(2014)\citenamefont
  {Santalla}, \citenamefont {Rodr{\'{i}}guez-Laguna},\ and\ \citenamefont
  {Cuerno}}]{Santalla2014}%
  \BibitemOpen
  \bibfield  {author} {\bibinfo {author} {\bibfnamefont {S.}~\bibnamefont
  {Santalla}}, \bibinfo {author} {\bibfnamefont {J.}~\bibnamefont
  {Rodr{\'{i}}guez-Laguna}},\ and\ \bibinfo {author} {\bibfnamefont
  {R.}~\bibnamefont {Cuerno}},\ }\bibfield  {title} {\bibinfo {title}
  {{\textit{{Circular Kardar-Parisi-Zhang equation as an inflating,
  self-avoiding ring polymer}}}},\ }\href
  {https://doi.org/10.1103/PhysRevE.89.010401} {\bibfield  {journal} {\bibinfo
  {journal} {Phys. Rev. E}\ }\textbf {\bibinfo {volume} {89}},\ \bibinfo
  {pages} {010401(R)} (\bibinfo {year} {2014})}\BibitemShut {NoStop}%
\bibitem [{\citenamefont {Santalla}\ \emph {et~al.}(2015)\citenamefont
  {Santalla}, \citenamefont {Rodr{\'{i}}guez-Laguna}, \citenamefont {Lagatta},\
  and\ \citenamefont {Cuerno}}]{Santalla2015}%
  \BibitemOpen
  \bibfield  {author} {\bibinfo {author} {\bibfnamefont {S.}~\bibnamefont
  {Santalla}}, \bibinfo {author} {\bibfnamefont {J.}~\bibnamefont
  {Rodr{\'{i}}guez-Laguna}}, \bibinfo {author} {\bibfnamefont {T.}~\bibnamefont
  {Lagatta}},\ and\ \bibinfo {author} {\bibfnamefont {R.}~\bibnamefont
  {Cuerno}},\ }\bibfield  {title} {\bibinfo {title} {{\textit{{Random geometry
  and the Kardar-Parisi-Zhang universality class}}}},\ }\href
  {https://doi.org/10.1088/1367-2630/17/3/033018} {\bibfield  {journal}
  {\bibinfo  {journal} {New J. Phys.}\ }\textbf {\bibinfo {volume} {17}},\
  \bibinfo {pages} {033018} (\bibinfo {year} {2015})}\BibitemShut {NoStop}%
\bibitem [{\citenamefont {Santalla}\ and\ \citenamefont
  {Ferreira}(2018)}]{Santalla2018b}%
  \BibitemOpen
  \bibfield  {author} {\bibinfo {author} {\bibfnamefont {S.~N.}\ \bibnamefont
  {Santalla}}\ and\ \bibinfo {author} {\bibfnamefont {S.~C.}\ \bibnamefont
  {Ferreira}},\ }\bibfield  {title} {\bibinfo {title} {{\textit{{Eden model
  with nonlocal growth rules and kinetic roughening in biological systems}}}},\
  }\href {https://doi.org/10.1103/PhysRevE.98.022405} {\bibfield  {journal}
  {\bibinfo  {journal} {Phys. Rev. E}\ }\textbf {\bibinfo {volume} {98}},\
  \bibinfo {pages} {022405} (\bibinfo {year} {2018})}\BibitemShut {NoStop}%
\bibitem [{\citenamefont {\'Alvarez~Domenech}\ \emph
  {et~al.}(2024)\citenamefont {\'Alvarez~Domenech}, \citenamefont
  {Rodr\'{\i}guez-Laguna}, \citenamefont {Cuerno}, \citenamefont
  {C\'ordoba-Torres},\ and\ \citenamefont {Santalla}}]{Domenech2024}%
  \BibitemOpen
  \bibfield  {author} {\bibinfo {author} {\bibfnamefont {I.}~\bibnamefont
  {\'Alvarez~Domenech}}, \bibinfo {author} {\bibfnamefont {J.}~\bibnamefont
  {Rodr\'{\i}guez-Laguna}}, \bibinfo {author} {\bibfnamefont {R.}~\bibnamefont
  {Cuerno}}, \bibinfo {author} {\bibfnamefont {P.}~\bibnamefont
  {C\'ordoba-Torres}},\ and\ \bibinfo {author} {\bibfnamefont {S.~N.}\
  \bibnamefont {Santalla}},\ }\bibfield  {title} {\bibinfo {title}
  {{\textit{Shape effects in the fluctuations of random isochrones on a square
  lattice}}},\ }\href {https://doi.org/10.1103/PhysRevE.109.034104} {\bibfield
  {journal} {\bibinfo  {journal} {Phys. Rev. E}\ }\textbf {\bibinfo {volume}
  {109}},\ \bibinfo {pages} {034104} (\bibinfo {year} {2024})}\BibitemShut
  {NoStop}%
\bibitem [{\citenamefont {Angelini}\ \emph {et~al.}(2020)\citenamefont
  {Angelini}, \citenamefont {Lucibello}, \citenamefont {Parisi}, \citenamefont
  {Ricci-Tersenghi},\ and\ \citenamefont {Rizzo}}]{Angelini2020}%
  \BibitemOpen
  \bibfield  {author} {\bibinfo {author} {\bibfnamefont {M.~C.}\ \bibnamefont
  {Angelini}}, \bibinfo {author} {\bibfnamefont {C.}~\bibnamefont {Lucibello}},
  \bibinfo {author} {\bibfnamefont {G.}~\bibnamefont {Parisi}}, \bibinfo
  {author} {\bibfnamefont {F.}~\bibnamefont {Ricci-Tersenghi}},\ and\ \bibinfo
  {author} {\bibfnamefont {T.}~\bibnamefont {Rizzo}},\ }\bibfield  {title}
  {\bibinfo {title} {{\textit{Loop expansion around the Bethe solution for the
  random magnetic field Ising ferromagnets at zero temperature}}},\ }\href
  {https://doi.org/10.1073/pnas.1909872117} {\bibfield  {journal} {\bibinfo
  {journal} {Proc. Natl. Acad. Sci. U.S.A.}\ }\textbf {\bibinfo {volume}
  {117}},\ \bibinfo {pages} {2268–2274} (\bibinfo {year} {2020})}\BibitemShut
  {NoStop}%
\bibitem [{\citenamefont {\rm Yllanes}(2011)}]{Yllanes2011}%
  \BibitemOpen
  \bibfield  {author} {\href
  {https://doi.org/https://doi.org/10.48550/arXiv.1111.0266} {{\textit{\bibinfo
  {author} {\bibfnamefont {D.}~\bibnamefont {\rm Yllanes}}}}},\ }\bibinfo
  {title} {\textit{Rugged Free-Energy Landscapes in Disordered Spin Systems}},\
  \href {https://doi.org/https://doi.org/10.48550/arXiv.1111.0266} {Ph.D.
  thesis},\ \bibinfo  {school} {Universidad Complutense de Madrid} (\bibinfo
  {year} {2011})\BibitemShut {NoStop}%
\bibitem [{\citenamefont {Michael}(1994)}]{Michael1994}%
  \BibitemOpen
  \bibfield  {author} {\bibinfo {author} {\bibfnamefont {C.}~\bibnamefont
  {Michael}},\ }\bibfield  {title} {\bibinfo {title} {{\textit{Fitting
  correlated data}}},\ }\href {https://doi.org/10.1103/physrevd.49.2616}
  {\bibfield  {journal} {\bibinfo  {journal} {Phys. Rev. D}\ }\textbf {\bibinfo
  {volume} {49}},\ \bibinfo {pages} {2616} (\bibinfo {year}
  {1994})}\BibitemShut {NoStop}%
\bibitem [{\citenamefont {Lulli}\ \emph {et~al.}(2016)\citenamefont {Lulli},
  \citenamefont {Parisi},\ and\ \citenamefont {Pelissetto}}]{Lulli2016}%
  \BibitemOpen
  \bibfield  {author} {\bibinfo {author} {\bibfnamefont {M.}~\bibnamefont
  {Lulli}}, \bibinfo {author} {\bibfnamefont {G.}~\bibnamefont {Parisi}},\ and\
  \bibinfo {author} {\bibfnamefont {A.}~\bibnamefont {Pelissetto}},\ }\bibfield
   {title} {\bibinfo {title} {{\textit{Out-of-equilibrium finite-size method
  for critical behavior analyses}}},\ }\href
  {https://doi.org/10.1103/PhysRevE.93.032126} {\bibfield  {journal} {\bibinfo
  {journal} {Phys. Rev. E}\ }\textbf {\bibinfo {volume} {93}},\ \bibinfo
  {pages} {032126} (\bibinfo {year} {2016})}\BibitemShut {NoStop}%
\bibitem [{\citenamefont {Seibert}(1994)}]{Seibert1994}%
  \BibitemOpen
  \bibfield  {author} {\bibinfo {author} {\bibfnamefont {D.}~\bibnamefont
  {Seibert}},\ }\bibfield  {title} {\bibinfo {title} {{\textit{Undesirable
  effects of covariance matrix techniques for error analysis}}},\ }\href
  {https://doi.org/10.1103/physrevd.49.6240} {\bibfield  {journal} {\bibinfo
  {journal} {Phys. Rev. D}\ }\textbf {\bibinfo {volume} {49}},\ \bibinfo
  {pages} {6240} (\bibinfo {year} {1994})}\BibitemShut {NoStop}%
\bibitem [{\citenamefont {Young}(2015)}]{Young2015}%
  \BibitemOpen
  \bibfield  {author} {\bibinfo {author} {\bibfnamefont {P.}~\bibnamefont
  {Young}},\ }\href {https://doi.org/10.1007/978-3-319-19051-8} {{\textit{\href
  {https://doi.org/10.1007/978-3-319-19051-8} {}}}\bibinfo {title}
  {\textit{Everything You Wanted to Know About Data Analysis and Fitting but
  Were Afraid to Ask}}}\ (\bibinfo  {publisher} {Springer International
  Publishing, Berlin},\ \bibinfo {year} {2015})\BibitemShut {NoStop}%
\bibitem [{\citenamefont {Efron}(1982)}]{Efron1982}%
  \BibitemOpen
  \bibfield  {author} {\bibinfo {author} {\bibfnamefont {B.}~\bibnamefont
  {Efron}},\ }\href {https://doi.org/10.1137/1.9781611970319} {{\textit{\href
  {https://doi.org/10.1137/1.9781611970319} {}}}\bibinfo {title} {\textit{The
  Jackknife, the Bootstrap and Other Resampling Plans}}}\ (\bibinfo
  {publisher} {Society for Industrial and Applied Mathematics, Philadelphia},\
  \bibinfo {year} {1982})\BibitemShut {NoStop}%
\bibitem [{\citenamefont {Harel}\ and\ \citenamefont
  {Taitelbaum}(2018)}]{Harel2018}%
  \BibitemOpen
  \bibfield  {author} {\bibinfo {author} {\bibfnamefont {M.}~\bibnamefont
  {Harel}}\ and\ \bibinfo {author} {\bibfnamefont {H.}~\bibnamefont
  {Taitelbaum}},\ }\bibfield  {title} {\bibinfo {title} {{\textit{Non-universal
  dynamic exponents for thin-film spreading}}},\ }\href
  {https://doi.org/10.1209/0295-5075/122/26002} {\bibfield  {journal} {\bibinfo
   {journal} {Europhys. Lett.}\ }\textbf {\bibinfo {volume} {122}},\ \bibinfo
  {pages} {26002} (\bibinfo {year} {2018})}\BibitemShut {NoStop}%
\bibitem [{\citenamefont {Harel}\ and\ \citenamefont
  {Taitelbaum}(2021)}]{Harel2021}%
  \BibitemOpen
  \bibfield  {author} {\bibinfo {author} {\bibfnamefont {M.}~\bibnamefont
  {Harel}}\ and\ \bibinfo {author} {\bibfnamefont {H.}~\bibnamefont
  {Taitelbaum}},\ }\bibfield  {title} {\bibinfo {title}
  {{\textit{{Non-monotonic dynamics of thin film spreading}}}},\ }\href
  {https://doi.org/10.1140/epje/s10189-021-00017-w} {\bibfield  {journal}
  {\bibinfo  {journal} {Eur. Phys. J. E}\ }\textbf {\bibinfo {volume} {44}},\
  \bibinfo {pages} {69} (\bibinfo {year} {2021})}\BibitemShut {NoStop}%
\bibitem [{\citenamefont {Carrasco}\ and\ \citenamefont
  {Oliveira}(2016)}]{Carrasco2016}%
  \BibitemOpen
  \bibfield  {author} {\bibinfo {author} {\bibfnamefont {I.~S.~S.}\
  \bibnamefont {Carrasco}}\ and\ \bibinfo {author} {\bibfnamefont {T.~J.}\
  \bibnamefont {Oliveira}},\ }\bibfield  {title} {\bibinfo {title}
  {{\textit{{Universality and geometry dependence in the class of the nonlinear
  molecular beam epitaxy equation}}}},\ }\href
  {https://doi.org/10.1103/PhysRevE.94.050801} {\bibfield  {journal} {\bibinfo
  {journal} {Phys. Rev. E}\ }\textbf {\bibinfo {volume} {94}},\ \bibinfo
  {pages} {050801(R)} (\bibinfo {year} {2016})}\BibitemShut {NoStop}%
\bibitem [{\citenamefont {Bornemann}(2009)}]{Bornemann2009}%
  \BibitemOpen
  \bibfield  {author} {\bibinfo {author} {\bibfnamefont {F.}~\bibnamefont
  {Bornemann}},\ }\bibfield  {title} {\bibinfo {title} {{\textit{On the
  numerical evaluation of Fredholm determinants}}},\ }\href
  {https://doi.org/10.1090/s0025-5718-09-02280-7} {\bibfield  {journal}
  {\bibinfo  {journal} {Math. Comput.}\ }\textbf {\bibinfo {volume} {79}},\
  \bibinfo {pages} {871} (\bibinfo {year} {2009})}\BibitemShut {NoStop}%
\bibitem [{\citenamefont {Marcos}\ \emph {et~al.}(2022)\citenamefont {Marcos},
  \citenamefont {Rodr{\'{i}}guez-L{\'{o}}pez}, \citenamefont {Mel{\'{e}}ndez},
  \citenamefont {Cuerno},\ and\ \citenamefont {Ruiz-Lorenzo}}]{Marcos2022}%
  \BibitemOpen
  \bibfield  {author} {\bibinfo {author} {\bibfnamefont {J.~M.}\ \bibnamefont
  {Marcos}}, \bibinfo {author} {\bibfnamefont {P.}~\bibnamefont
  {Rodr{\'{i}}guez-L{\'{o}}pez}}, \bibinfo {author} {\bibfnamefont {J.~J.}\
  \bibnamefont {Mel{\'{e}}ndez}}, \bibinfo {author} {\bibfnamefont
  {R.}~\bibnamefont {Cuerno}},\ and\ \bibinfo {author} {\bibfnamefont {J.~J.}\
  \bibnamefont {Ruiz-Lorenzo}},\ }\bibfield  {title} {\bibinfo {title}
  {{\textit{{Spreading fronts of wetting liquid droplets: Microscopic
  simulations and universal fluctuations}}}},\ }\href
  {https://doi.org/10.1103/PhysRevE.105.054801} {\bibfield  {journal} {\bibinfo
   {journal} {Phys. Rev. E}\ }\textbf {\bibinfo {volume} {105}},\ \bibinfo
  {pages} {054801} (\bibinfo {year} {2022})}\BibitemShut {NoStop}%
\bibitem [{\citenamefont {Weng}\ \emph {et~al.}(2017)\citenamefont {Weng},
  \citenamefont {Wu}, \citenamefont {Tsao},\ and\ \citenamefont
  {Sheng}}]{Weng2017}%
  \BibitemOpen
  \bibfield  {author} {\bibinfo {author} {\bibfnamefont {Y.~H.}\ \bibnamefont
  {Weng}}, \bibinfo {author} {\bibfnamefont {C.~J.}\ \bibnamefont {Wu}},
  \bibinfo {author} {\bibfnamefont {H.~K.}\ \bibnamefont {Tsao}},\ and\
  \bibinfo {author} {\bibfnamefont {Y.~J.}\ \bibnamefont {Sheng}},\ }\bibfield
  {title} {\bibinfo {title} {{\textit{{Spreading dynamics of a precursor film
  of nanodrops on total wetting surfaces}}}},\ }\href
  {https://doi.org/10.1039/c7cp04979j} {\bibfield  {journal} {\bibinfo
  {journal} {Phys. Chem. Chem. Phys.}\ }\textbf {\bibinfo {volume} {19}},\
  \bibinfo {pages} {27786} (\bibinfo {year} {2017})}\BibitemShut {NoStop}%
\bibitem [{\citenamefont {Hocking}(1992)}]{Hocking1992}%
  \BibitemOpen
  \bibfield  {author} {\bibinfo {author} {\bibfnamefont {L.~M.}\ \bibnamefont
  {Hocking}},\ }\bibfield  {title} {\bibinfo {title} {{\textit{Rival
  contact-angle models and the spreading of drops}}},\ }\href
  {https://doi.org/10.1017/S0022112092004579} {\bibfield  {journal} {\bibinfo
  {journal} {J. Fluid Mech.}\ }\textbf {\bibinfo {volume} {239}},\ \bibinfo
  {pages} {671–681} (\bibinfo {year} {1992})}\BibitemShut {NoStop}%
\bibitem [{\citenamefont {Baldassarri}\ and\ \citenamefont
  {Jacquier}(2023)}]{Baldassarri2023}%
  \BibitemOpen
  \bibfield  {author} {\bibinfo {author} {\bibfnamefont {S.}~\bibnamefont
  {Baldassarri}}\ and\ \bibinfo {author} {\bibfnamefont {V.}~\bibnamefont
  {Jacquier}},\ }\bibfield  {title} {\bibinfo {title} {{\textit{Metastability
  for Kawasaki Dynamics on the Hexagonal Lattice}}},\ }\href
  {http://dx.doi.org/10.1007/s10955-022-03061-8} {\bibfield  {journal}
  {\bibinfo  {journal} {J. Stat. Phys.}\ }\textbf {\bibinfo {volume} {190}}
  (\bibinfo {year} {2023})}\BibitemShut {NoStop}%
\bibitem [{\citenamefont {Amar}\ and\ \citenamefont {Family}(1990)}]{Amar1990}%
  \BibitemOpen
  \bibfield  {author} {\bibinfo {author} {\bibfnamefont {J.~G.}\ \bibnamefont
  {Amar}}\ and\ \bibinfo {author} {\bibfnamefont {F.}~\bibnamefont {Family}},\
  }\bibfield  {title} {\bibinfo {title} {{\textit{Numerical solution of a
  continuum equation for interface growth in 2+1 dimensions}}},\ }\href
  {https://doi.org/10.1103/PhysRevA.41.3399} {\bibfield  {journal} {\bibinfo
  {journal} {Phys. Rev. A}\ }\textbf {\bibinfo {volume} {41}},\ \bibinfo
  {pages} {3399} (\bibinfo {year} {1990})}\BibitemShut {NoStop}%
\bibitem [{\citenamefont {Moser}\ \emph {et~al.}(1991)\citenamefont {Moser},
  \citenamefont {Kertész},\ and\ \citenamefont {Wolf}}]{Moser1991}%
  \BibitemOpen
  \bibfield  {author} {\bibinfo {author} {\bibfnamefont {K.}~\bibnamefont
  {Moser}}, \bibinfo {author} {\bibfnamefont {J.}~\bibnamefont {Kertész}},\
  and\ \bibinfo {author} {\bibfnamefont {D.~E.}\ \bibnamefont {Wolf}},\
  }\bibfield  {title} {\bibinfo {title} {{\textit{Numerical solution of the
  Kardar-Parisi-Zhang equation in one, two and three dimensions}}},\ }\href
  {https://doi.org/https://doi.org/10.1016/0378-4371(91)90017-7} {\bibfield
  {journal} {\bibinfo  {journal} {Physica A: Stat. Mech. Appl.}\ }\textbf
  {\bibinfo {volume} {178}},\ \bibinfo {pages} {215} (\bibinfo {year}
  {1991})}\BibitemShut {NoStop}%
\bibitem [{\citenamefont {Forrest}\ and\ \citenamefont
  {Toral}(1993)}]{Forrest1993}%
  \BibitemOpen
  \bibfield  {author} {\bibinfo {author} {\bibfnamefont {B.~M.}\ \bibnamefont
  {Forrest}}\ and\ \bibinfo {author} {\bibfnamefont {R.}~\bibnamefont
  {Toral}},\ }\bibfield  {title} {\bibinfo {title} {{\textit{Crossover and
  finite-size effects in the (1+1)-dimensional Kardar-Parisi-Zhang
  equation}}},\ }\href {https://doi.org/10.1007/bf01053591} {\bibfield
  {journal} {\bibinfo  {journal} {J. Stat. Phys.}\ }\textbf {\bibinfo {volume}
  {70}},\ \bibinfo {pages} {703} (\bibinfo {year} {1993})}\BibitemShut
  {NoStop}%
\bibitem [{\citenamefont {Lam}\ and\ \citenamefont {Shin}(1998)}]{Lam1998}%
  \BibitemOpen
  \bibfield  {author} {\bibinfo {author} {\bibfnamefont {C.-H.}\ \bibnamefont
  {Lam}}\ and\ \bibinfo {author} {\bibfnamefont {F.~G.}\ \bibnamefont {Shin}},\
  }\bibfield  {title} {\bibinfo {title} {{\textit{Improved discretization of
  the Kardar-Parisi-Zhang equation}}},\ }\href
  {https://doi.org/10.1103/PhysRevE.58.5592} {\bibfield  {journal} {\bibinfo
  {journal} {Phys. Rev. E}\ }\textbf {\bibinfo {volume} {58}},\ \bibinfo
  {pages} {5592} (\bibinfo {year} {1998})}\BibitemShut {NoStop}%
\bibitem [{\citenamefont {Giada}\ \emph {et~al.}(2002)\citenamefont {Giada},
  \citenamefont {Giacometti},\ and\ \citenamefont {Rossi}}]{Giada2002}%
  \BibitemOpen
  \bibfield  {author} {\bibinfo {author} {\bibfnamefont {L.}~\bibnamefont
  {Giada}}, \bibinfo {author} {\bibfnamefont {A.}~\bibnamefont {Giacometti}},\
  and\ \bibinfo {author} {\bibfnamefont {M.}~\bibnamefont {Rossi}},\ }\bibfield
   {title} {\bibinfo {title} {{\textit{Pseudospectral method for the
  Kardar-Parisi-Zhang equation}}},\ }\href
  {https://doi.org/10.1103/PhysRevE.65.036134} {\bibfield  {journal} {\bibinfo
  {journal} {Phys. Rev. E}\ }\textbf {\bibinfo {volume} {65}},\ \bibinfo
  {pages} {036134} (\bibinfo {year} {2002})}\BibitemShut {NoStop}%
\bibitem [{\citenamefont {Dasgupta}\ \emph {et~al.}(1996)\citenamefont
  {Dasgupta}, \citenamefont {Das~Sarma},\ and\ \citenamefont
  {Kim}}]{Dasgupta1996}%
  \BibitemOpen
  \bibfield  {author} {\bibinfo {author} {\bibfnamefont {C.}~\bibnamefont
  {Dasgupta}}, \bibinfo {author} {\bibfnamefont {S.}~\bibnamefont
  {Das~Sarma}},\ and\ \bibinfo {author} {\bibfnamefont {J.~M.}\ \bibnamefont
  {Kim}},\ }\bibfield  {title} {\bibinfo {title} {{\textit{Controlled
  instability and multiscaling in models of epitaxial growth}}},\ }\href
  {https://doi.org/10.1103/PhysRevE.54.R4552} {\bibfield  {journal} {\bibinfo
  {journal} {Phys. Rev. E}\ }\textbf {\bibinfo {volume} {54}},\ \bibinfo
  {pages} {R4552} (\bibinfo {year} {1996})}\BibitemShut {NoStop}%
\bibitem [{\citenamefont {Dasgupta}\ \emph {et~al.}(1997)\citenamefont
  {Dasgupta}, \citenamefont {Kim}, \citenamefont {Dutta},\ and\ \citenamefont
  {Das~Sarma}}]{Dasgupta1997}%
  \BibitemOpen
  \bibfield  {author} {\bibinfo {author} {\bibfnamefont {C.}~\bibnamefont
  {Dasgupta}}, \bibinfo {author} {\bibfnamefont {J.~M.}\ \bibnamefont {Kim}},
  \bibinfo {author} {\bibfnamefont {M.}~\bibnamefont {Dutta}},\ and\ \bibinfo
  {author} {\bibfnamefont {S.}~\bibnamefont {Das~Sarma}},\ }\bibfield  {title}
  {\bibinfo {title} {{\textit{Instability, intermittency, and multiscaling in
  discrete growth models of kinetic roughening}}},\ }\href
  {https://doi.org/10.1103/PhysRevE.55.2235} {\bibfield  {journal} {\bibinfo
  {journal} {Phys. Rev. E}\ }\textbf {\bibinfo {volume} {55}},\ \bibinfo
  {pages} {2235} (\bibinfo {year} {1997})}\BibitemShut {NoStop}%
\bibitem [{\citenamefont {Gallego}(2011)}]{Gallego2011}%
  \BibitemOpen
  \bibfield  {author} {\bibinfo {author} {\bibfnamefont {R.}~\bibnamefont
  {Gallego}},\ }\bibfield  {title} {\bibinfo {title}
  {{\textit{{Predictor–corrector pseudospectral methods for stochastic
  partial differential equations with additive white noise}}}},\ }\href
  {https://doi.org/10.1016/j.amc.2011.09.038} {\bibfield  {journal} {\bibinfo
  {journal} {Appl. Math. Comp.}\ }\textbf {\bibinfo {volume} {218}},\ \bibinfo
  {pages} {3905} (\bibinfo {year} {2011})}\BibitemShut {NoStop}%
\bibitem [{\citenamefont {Toral}\ and\ \citenamefont
  {Colet}(2014)}]{Toral2014}%
  \BibitemOpen
  \bibfield  {author} {\bibinfo {author} {\bibfnamefont {R.}~\bibnamefont
  {Toral}}\ and\ \bibinfo {author} {\bibfnamefont {P.}~\bibnamefont {Colet}},\
  }\href {https://doi.org/10.1002/9783527683147} {{\textit{\href
  {https://doi.org/10.1002/9783527683147} {}}}\bibinfo {title} {Stochastic
  Numerical Methods: An Introduction for Students and Scientists}}\ (\bibinfo
  {publisher} {Wiley},\ \bibinfo {year} {2014})\BibitemShut {NoStop}%
\bibitem [{\citenamefont {Miranda}\ and\ \citenamefont {Aar\~ao
  Reis}(2008)}]{Miranda2008}%
  \BibitemOpen
  \bibfield  {author} {\bibinfo {author} {\bibfnamefont {V.~G.}\ \bibnamefont
  {Miranda}}\ and\ \bibinfo {author} {\bibfnamefont {F.~D.~A.}\ \bibnamefont
  {Aar\~ao Reis}},\ }\bibfield  {title} {\bibinfo {title} {{\textit{Numerical
  study of the Kardar-Parisi-Zhang equation}}},\ }\href
  {https://doi.org/10.1103/PhysRevE.77.031134} {\bibfield  {journal} {\bibinfo
  {journal} {Phys. Rev. E}\ }\textbf {\bibinfo {volume} {77}},\ \bibinfo
  {pages} {031134} (\bibinfo {year} {2008})}\BibitemShut {NoStop}%
\bibitem [{\citenamefont {Al{\'{e}}s}\ and\ \citenamefont
  {L{\'{o}}pez}(2019)}]{Ales2019}%
  \BibitemOpen
  \bibfield  {author} {\bibinfo {author} {\bibfnamefont {A.}~\bibnamefont
  {Al{\'{e}}s}}\ and\ \bibinfo {author} {\bibfnamefont {J.~M.}\ \bibnamefont
  {L{\'{o}}pez}},\ }\bibfield  {title} {\bibinfo {title} {{\textit{{Faceted
  patterns and anomalous surface roughening driven by long-range temporally
  correlated noise}}}},\ }\href {https://doi.org/10.1103/PhysRevE.99.062139}
  {\bibfield  {journal} {\bibinfo  {journal} {Phys. Rev. E}\ }\textbf {\bibinfo
  {volume} {99}},\ \bibinfo {pages} {062139} (\bibinfo {year}
  {2019})}\BibitemShut {NoStop}%
\bibitem [{\citenamefont {Song}\ and\ \citenamefont {Xia}(2021)}]{Song2021}%
  \BibitemOpen
  \bibfield  {author} {\bibinfo {author} {\bibfnamefont {T.}~\bibnamefont
  {Song}}\ and\ \bibinfo {author} {\bibfnamefont {H.}~\bibnamefont {Xia}},\
  }\bibfield  {title} {\bibinfo {title} {{\textit{{Kinetic roughening and
  nontrivial scaling in the Kardar-Parisi-Zhang growth with long-range temporal
  correlations}}}},\ }\href {https://doi.org/10.1088/1742-5468/ac06c3}
  {\bibfield  {journal} {\bibinfo  {journal} {J. Stat. Mech.: Theor. Exp.}\
  }\textbf {\bibinfo {volume} {2021}},\ \bibinfo {pages} {073203} (\bibinfo
  {year} {2021})}\BibitemShut {NoStop}%
\bibitem [{\citenamefont {Ortega}(2015)}]{Ortega2015}%
  \BibitemOpen
  \bibfield  {author} {\href@noop {} {{\textit{\bibinfo {author} {\bibfnamefont
  {A.}~\bibnamefont {Ortega}}}}},\ }\bibinfo {title} {Ecuaciones en derivadas
  parciales sobre grafos},\ \href@noop {} {Master's thesis},\ \bibinfo
  {school} {Universidad Carlos III de Madrid} (\bibinfo {year}
  {2015})\BibitemShut {NoStop}%
\bibitem [{\citenamefont {Chung}\ \emph {et~al.}(2007)\citenamefont {Chung},
  \citenamefont {Chung},\ and\ \citenamefont {Kim}}]{Chung2007}%
  \BibitemOpen
  \bibfield  {author} {\bibinfo {author} {\bibfnamefont {S.-Y.}\ \bibnamefont
  {Chung}}, \bibinfo {author} {\bibfnamefont {Y.-S.}\ \bibnamefont {Chung}},\
  and\ \bibinfo {author} {\bibfnamefont {J.-H.}\ \bibnamefont {Kim}},\
  }\bibfield  {title} {\bibinfo {title} {{\textit{Diffusion and Elastic
  Equations on Networks}}},\ }\href {https://doi.org/10.2977/prims/1201012039}
  {\bibfield  {journal} {\bibinfo  {journal} {Publ. Res. Ins. Math. Sci.}\
  }\textbf {\bibinfo {volume} {43}},\ \bibinfo {pages} {699} (\bibinfo {year}
  {2007})}\BibitemShut {NoStop}%
\bibitem [{\citenamefont {Bethe}(1935)}]{Bethe1935}%
  \BibitemOpen
  \bibfield  {author} {\bibinfo {author} {\bibfnamefont {H.~A.}\ \bibnamefont
  {Bethe}},\ }\bibfield  {title} {\bibinfo {title} {{\textit{Statistical theory
  of superlattices}}},\ }\href {https://doi.org/10.1098/rspa.1935.0122}
  {\bibfield  {journal} {\bibinfo  {journal} {Proc. R. Soc. Lond. A}\ }\textbf
  {\bibinfo {volume} {150}},\ \bibinfo {pages} {552} (\bibinfo {year}
  {1935})}\BibitemShut {NoStop}%
\bibitem [{\citenamefont {Dorogovtsev}\ \emph {et~al.}(2008)\citenamefont
  {Dorogovtsev}, \citenamefont {Goltsev},\ and\ \citenamefont
  {Mendes}}]{Dorogovtsev2008}%
  \BibitemOpen
  \bibfield  {author} {\bibinfo {author} {\bibfnamefont {S.~N.}\ \bibnamefont
  {Dorogovtsev}}, \bibinfo {author} {\bibfnamefont {A.~V.}\ \bibnamefont
  {Goltsev}},\ and\ \bibinfo {author} {\bibfnamefont {J.~F.~F.}\ \bibnamefont
  {Mendes}},\ }\bibfield  {title} {\bibinfo {title} {{\textit{Critical
  phenomena in complex networks}}},\ }\href
  {https://doi.org/10.1103/revmodphys.80.1275} {\bibfield  {journal} {\bibinfo
  {journal} {Rev. Mod. Phys.}\ }\textbf {\bibinfo {volume} {80}},\ \bibinfo
  {pages} {1275–1335} (\bibinfo {year} {2008})}\BibitemShut {NoStop}%
\bibitem [{\citenamefont {Chae}\ \emph {et~al.}(2012)\citenamefont {Chae},
  \citenamefont {Yook},\ and\ \citenamefont {Kim}}]{Chae2012}%
  \BibitemOpen
  \bibfield  {author} {\bibinfo {author} {\bibfnamefont {H.}~\bibnamefont
  {Chae}}, \bibinfo {author} {\bibfnamefont {S.-H.}\ \bibnamefont {Yook}},\
  and\ \bibinfo {author} {\bibfnamefont {Y.}~\bibnamefont {Kim}},\ }\bibfield
  {title} {\bibinfo {title} {{\textit{Explosive percolation on the Bethe
  lattice}}},\ }\href {http://dx.doi.org/10.1103/PhysRevE.85.051118} {\bibfield
   {journal} {\bibinfo  {journal} {Phys. Rev. E}\ }\textbf {\bibinfo {volume}
  {85}} (\bibinfo {year} {2012})}\BibitemShut {NoStop}%
\bibitem [{\citenamefont {Sahimi}(1988)}]{Sahimi1988}%
  \BibitemOpen
  \bibfield  {author} {\bibinfo {author} {\bibfnamefont {M.}~\bibnamefont
  {Sahimi}},\ }\bibfield  {title} {\bibinfo {title}
  {{\textit{Diffusion-controlled reactions in disordered porous media—I.
  Uniform distribution of reactants}}},\ }\href
  {https://doi.org/https://doi.org/10.1016/0009-2509(88)80051-4} {\bibfield
  {journal} {\bibinfo  {journal} {Chem. Eng. Sci.}\ }\textbf {\bibinfo {volume}
  {43}},\ \bibinfo {pages} {2981} (\bibinfo {year} {1988})}\BibitemShut
  {NoStop}%
\bibitem [{\citenamefont {Hughes}\ and\ \citenamefont
  {Sahimi}(1982)}]{Hughes1982}%
  \BibitemOpen
  \bibfield  {author} {\bibinfo {author} {\bibfnamefont {B.~D.}\ \bibnamefont
  {Hughes}}\ and\ \bibinfo {author} {\bibfnamefont {M.}~\bibnamefont
  {Sahimi}},\ }\bibfield  {title} {\bibinfo {title} {{\textit{Random walks on
  the Bethe lattice}}},\ }\href {https://doi.org/10.1007/bf01011791} {\bibfield
   {journal} {\bibinfo  {journal} {J. Stat. Phys.}\ }\textbf {\bibinfo {volume}
  {29}},\ \bibinfo {pages} {781–794} (\bibinfo {year} {1982})}\BibitemShut
  {NoStop}%
\bibitem [{\citenamefont {Krug}(1988)}]{Krug1988}%
  \BibitemOpen
  \bibfield  {author} {\bibinfo {author} {\bibfnamefont {J.}~\bibnamefont
  {Krug}},\ }\bibfield  {title} {\bibinfo {title} {{\textit{Surface structure
  of random aggregates on the Cayley tree}}},\ }\href
  {https://doi.org/10.1088/0305-4470/21/24/017} {\bibfield  {journal} {\bibinfo
   {journal} {J. Phys. A: Math. Gen.}\ }\textbf {\bibinfo {volume} {21}},\
  \bibinfo {pages} {4637–4647} (\bibinfo {year} {1988})}\BibitemShut
  {NoStop}%
\bibitem [{\citenamefont {Bradley}\ and\ \citenamefont
  {Strenski}(1984)}]{Bradley1984}%
  \BibitemOpen
  \bibfield  {author} {\bibinfo {author} {\bibfnamefont {R.~M.}\ \bibnamefont
  {Bradley}}\ and\ \bibinfo {author} {\bibfnamefont {P.~N.}\ \bibnamefont
  {Strenski}},\ }\bibfield  {title} {\bibinfo {title} {{\textit{Directed
  diffusion-limited aggregation on the Bethe lattice: Exact results}}},\ }\href
  {https://doi.org/10.1103/physrevb.30.6788} {\bibfield  {journal} {\bibinfo
  {journal} {Phys. Rev. B}\ }\textbf {\bibinfo {volume} {30}},\ \bibinfo
  {pages} {6788–6790} (\bibinfo {year} {1984})}\BibitemShut {NoStop}%
\bibitem [{\citenamefont {Sahimi}(1993)}]{Sahimi1993}%
  \BibitemOpen
  \bibfield  {author} {\bibinfo {author} {\bibfnamefont {M.}~\bibnamefont
  {Sahimi}},\ }\bibfield  {title} {\bibinfo {title} {{\textit{Nonlinear
  transport processes in disordered media}}},\ }\href
  {https://doi.org/10.1002/aic.690390302} {\bibfield  {journal} {\bibinfo
  {journal} {AIChE J.}\ }\textbf {\bibinfo {volume} {39}},\ \bibinfo {pages}
  {369–386} (\bibinfo {year} {1993})}\BibitemShut {NoStop}%
\bibitem [{\citenamefont {Box}\ and\ \citenamefont {Muller}(1958)}]{Box1958}%
  \BibitemOpen
  \bibfield  {author} {\bibinfo {author} {\bibfnamefont {G.~E.~P.}\
  \bibnamefont {Box}}\ and\ \bibinfo {author} {\bibfnamefont {M.~E.}\
  \bibnamefont {Muller}},\ }\bibfield  {title} {\bibinfo {title} {{\textit{A
  Note on the Generation of Random Normal Deviates}}},\ }\href
  {https://doi.org/10.1214/aoms/1177706645} {\bibfield  {journal} {\bibinfo
  {journal} {Ann. Math. Stat.}\ }\textbf {\bibinfo {volume} {29}},\ \bibinfo
  {pages} {610–611} (\bibinfo {year} {1958})}\BibitemShut {NoStop}%
\bibitem [{\citenamefont {Christensen}\ and\ \citenamefont
  {Moloney}(2005)}]{Christensen2005}%
  \BibitemOpen
  \bibfield  {author} {\bibinfo {author} {\bibfnamefont {K.}~\bibnamefont
  {Christensen}}\ and\ \bibinfo {author} {\bibfnamefont {N.~R.}\ \bibnamefont
  {Moloney}},\ }\href {https://doi.org/10.1142/p365} {{\textit{\href
  {https://doi.org/10.1142/p365} {}}}\bibinfo {title} {Complexity and
  Criticality}}\ (\bibinfo  {publisher} {Imperial College Press},\ \bibinfo
  {year} {2005})\BibitemShut {NoStop}%
\bibitem [{\citenamefont {Li}\ \emph {et~al.}(2025)\citenamefont {Li},
  \citenamefont {Marcos}, \citenamefont {Fasano}, \citenamefont {Diez},
  \citenamefont {Cummings}, \citenamefont {Kondic},\ and\ \citenamefont
  {Manor}}]{paper_experimento}%
  \BibitemOpen
  \bibfield  {author} {\bibinfo {author} {\bibfnamefont {Y.}~\bibnamefont
  {Li}}, \bibinfo {author} {\bibfnamefont {J.~M.}\ \bibnamefont {Marcos}},
  \bibinfo {author} {\bibfnamefont {M.}~\bibnamefont {Fasano}}, \bibinfo
  {author} {\bibfnamefont {J.}~\bibnamefont {Diez}}, \bibinfo {author}
  {\bibfnamefont {L.~J.}\ \bibnamefont {Cummings}}, \bibinfo {author}
  {\bibfnamefont {L.}~\bibnamefont {Kondic}},\ and\ \bibinfo {author}
  {\bibfnamefont {O.}~\bibnamefont {Manor}},\ }\href
  {https://arxiv.org/abs/2502.08800} {\bibinfo {title} {{\textit{Using wetting
  and ultrasonic waves to extract oil from oil/water mixtures}}}} (\bibinfo
  {year} {2025}),\ \Eprint {https://arxiv.org/abs/2502.08800} {arXiv:2502.08800
  [physics.flu-dyn]} \BibitemShut {NoStop}%
\bibitem [{\citenamefont {Gorak}\ and\ \citenamefont
  {Sorensen}(2014)}]{Distillation}%
  \BibitemOpen
  \bibfield  {author} {\bibinfo {author} {\bibfnamefont {A.}~\bibnamefont
  {Gorak}}\ and\ \bibinfo {author} {\bibfnamefont {E.}~\bibnamefont
  {Sorensen}},\ }\href {http://dx.doi.org/10.1016/C2010-0-66923-9}
  {{\textit{\href {http://dx.doi.org/10.1016/C2010-0-66923-9} {}}}\bibinfo
  {title} {Distillation: Fundamentals and Principles}}\ (\bibinfo  {publisher}
  {Academic Press, New York},\ \bibinfo {year} {2014})\BibitemShut {NoStop}%
\bibitem [{\citenamefont {Zouboulis}\ and\ \citenamefont
  {Avranas}(2000)}]{Zouboulis2000}%
  \BibitemOpen
  \bibfield  {author} {\bibinfo {author} {\bibfnamefont {A.}~\bibnamefont
  {Zouboulis}}\ and\ \bibinfo {author} {\bibfnamefont {A.}~\bibnamefont
  {Avranas}},\ }\bibfield  {title} {\bibinfo {title} {{\textit{Treatment of
  oil-in-water emulsions by coagulation and dissolved-air flotation}}},\ }\href
  {https://doi.org/10.1016/s0927-7757(00)00561-6} {\bibfield  {journal}
  {\bibinfo  {journal} {Colloids Surf. A}\ }\textbf {\bibinfo {volume} {172}},\
  \bibinfo {pages} {153–161} (\bibinfo {year} {2000})}\BibitemShut {NoStop}%
\bibitem [{\citenamefont {Ahmad}\ \emph {et~al.}(2005)\citenamefont {Ahmad},
  \citenamefont {Ismail},\ and\ \citenamefont {Bhatia}}]{Ahmad2005}%
  \BibitemOpen
  \bibfield  {author} {\bibinfo {author} {\bibfnamefont {A.~L.}\ \bibnamefont
  {Ahmad}}, \bibinfo {author} {\bibfnamefont {S.}~\bibnamefont {Ismail}},\ and\
  \bibinfo {author} {\bibfnamefont {S.}~\bibnamefont {Bhatia}},\ }\bibfield
  {title} {\bibinfo {title} {{\textit{Optimization of Coagulation-Flocculation
  Process for Palm Oil Mill Effluent Using Response Surface Methodology}}},\
  }\href {https://doi.org/10.1021/es0498080} {\bibfield  {journal} {\bibinfo
  {journal} {Environ. Sci. Technol.}\ }\textbf {\bibinfo {volume} {39}},\
  \bibinfo {pages} {2828–2834} (\bibinfo {year} {2005})}\BibitemShut
  {NoStop}%
\bibitem [{\citenamefont {Marmur}(2009)}]{mittal_guide_2009}%
  \BibitemOpen
  \bibfield  {author} {\bibinfo {author} {\bibfnamefont {A.}~\bibnamefont
  {Marmur}},\ }\bibfield  {title} {\bibinfo {title} {{\textit{A {Guide} {To}
  {The} {Equilibrium} {Contact} {Angles} {Maze}}}},\ }in\ \href
  {https://doi.org/10.1163/ej.9789004169326.i-400.5} {\bibinfo {booktitle}
  {Contact {Angle}, {Wettability} and {Adhesion}, {Volume} 6}},\ \bibinfo
  {editor} {edited by\ \bibinfo {editor} {\bibfnamefont {K.~L.}\ \bibnamefont
  {Mittal}}}\ (\bibinfo  {publisher} {Brill Academic Publishers, Leiden},\
  \bibinfo {year} {2009})\ pp.\ \bibinfo {pages} {1--18}\BibitemShut {NoStop}%
\bibitem [{\citenamefont {Rayleigh}(1884)}]{LordRayleigh1884}%
  \BibitemOpen
  \bibfield  {author} {\bibinfo {author} {\bibfnamefont {L.}~\bibnamefont
  {Rayleigh}},\ }\bibfield  {title} {\bibinfo {title} {{\textit{On the
  Circulation of Air Observed in Kundt’s Tubes, and on Some Allied Acoustical
  Problems}}},\ }\href {https://doi.org/10.1098/rstl.1884.0002} {\bibfield
  {journal} {\bibinfo  {journal} {Phil. Trans. Royal Soc. London}\ }\textbf
  {\bibinfo {volume} {175}},\ \bibinfo {pages} {1} (\bibinfo {year}
  {1884})}\BibitemShut {NoStop}%
\bibitem [{\citenamefont {Schlichting}(1932)}]{Schlichting:1932p447}%
  \BibitemOpen
  \bibfield  {author} {\bibinfo {author} {\bibfnamefont {H.}~\bibnamefont
  {Schlichting}},\ }\bibfield  {title} {\bibinfo {title} {{\textit{Calculation
  of even periodic barrier currents}}},\ }\href@noop {} {\bibfield  {journal}
  {\bibinfo  {journal} {Phys. Z.}\ }\textbf {\bibinfo {volume} {33}},\ \bibinfo
  {pages} {327} (\bibinfo {year} {1932})}\BibitemShut {NoStop}%
\bibitem [{\citenamefont {Eckart}(1948)}]{Eckart1948}%
  \BibitemOpen
  \bibfield  {author} {\bibinfo {author} {\bibfnamefont {C.}~\bibnamefont
  {Eckart}},\ }\bibfield  {title} {\bibinfo {title} {{\textit{Vortices and
  Streams Caused by Sound Waves}}},\ }\href
  {https://doi.org/10.1103/physrev.73.68} {\bibfield  {journal} {\bibinfo
  {journal} {Phys. Rev.}\ }\textbf {\bibinfo {volume} {73}},\ \bibinfo {pages}
  {68–76} (\bibinfo {year} {1948})}\BibitemShut {NoStop}%
\bibitem [{\citenamefont {Lighthill}(1978)}]{Lighthill1978}%
  \BibitemOpen
  \bibfield  {author} {\bibinfo {author} {\bibfnamefont {S.~J.}\ \bibnamefont
  {Lighthill}},\ }\bibfield  {title} {\bibinfo {title} {{\textit{Acoustic
  streaming}}},\ }\href {https://doi.org/10.1016/0022-460x(78)90388-7}
  {\bibfield  {journal} {\bibinfo  {journal} {J. Sound Vib.}\ }\textbf
  {\bibinfo {volume} {61}},\ \bibinfo {pages} {391–418} (\bibinfo {year}
  {1978})}\BibitemShut {NoStop}%
\bibitem [{\citenamefont {Nyborg}(1953)}]{Nyborg1953}%
  \BibitemOpen
  \bibfield  {author} {\bibinfo {author} {\bibfnamefont {W.~L.}\ \bibnamefont
  {Nyborg}},\ }\bibfield  {title} {\bibinfo {title} {{\textit{Acoustic
  Streaming due to Attenuated Plane Waves}}},\ }\href
  {https://doi.org/10.1121/1.1907010} {\bibfield  {journal} {\bibinfo
  {journal} {J. Acoust. Soc. Am.}\ }\textbf {\bibinfo {volume} {25}},\ \bibinfo
  {pages} {68–75} (\bibinfo {year} {1953})}\BibitemShut {NoStop}%
\bibitem [{\citenamefont {Yabe}\ \emph {et~al.}(2014)\citenamefont {Yabe},
  \citenamefont {Hamate}, \citenamefont {Hara}, \citenamefont {Oguchi},
  \citenamefont {Nagasawa},\ and\ \citenamefont {Kuwano}}]{Yabe2014}%
  \BibitemOpen
  \bibfield  {author} {\bibinfo {author} {\bibfnamefont {A.}~\bibnamefont
  {Yabe}}, \bibinfo {author} {\bibfnamefont {Y.}~\bibnamefont {Hamate}},
  \bibinfo {author} {\bibfnamefont {M.}~\bibnamefont {Hara}}, \bibinfo {author}
  {\bibfnamefont {H.}~\bibnamefont {Oguchi}}, \bibinfo {author} {\bibfnamefont
  {S.}~\bibnamefont {Nagasawa}},\ and\ \bibinfo {author} {\bibfnamefont
  {H.}~\bibnamefont {Kuwano}},\ }\bibfield  {title} {\bibinfo {title}
  {{\textit{A self-converging atomized mist spray device using surface acoustic
  wave}}},\ }\href {https://doi.org/10.1007/s10404-014-1358-2} {\bibfield
  {journal} {\bibinfo  {journal} {Microfluid. Nanofluidics.}\ }\textbf
  {\bibinfo {volume} {17}},\ \bibinfo {pages} {701–710} (\bibinfo {year}
  {2014})}\BibitemShut {NoStop}%
\bibitem [{\citenamefont {Hamilton}\ and\ \citenamefont
  {Blackstock}(1998)}]{hamilton1998nonlinear}%
  \BibitemOpen
  \bibfield  {author} {\bibinfo {author} {\bibfnamefont {M.}~\bibnamefont
  {Hamilton}}\ and\ \bibinfo {author} {\bibfnamefont {D.}~\bibnamefont
  {Blackstock}},\ }\href {http://dx.doi.org/10.1007/978-3-031-58963-8}
  {{\textit{\href {http://dx.doi.org/10.1007/978-3-031-58963-8} {}}}\bibinfo
  {title} {Nonlinear Acoustics}}\ (\bibinfo  {publisher} {Academic Press, New
  York},\ \bibinfo {year} {1998})\BibitemShut {NoStop}%
\bibitem [{\citenamefont {King}(1934)}]{King1934}%
  \BibitemOpen
  \bibfield  {author} {\bibinfo {author} {\bibfnamefont {L.}~\bibnamefont
  {King}},\ }\bibfield  {title} {\bibinfo {title} {{\textit{On the acoustic
  radiation pressure on spheres}}},\ }\href
  {https://doi.org/10.1098/rspa.1934.0215} {\bibfield  {journal} {\bibinfo
  {journal} {Proc. R. Soc. Lond. A}\ }\textbf {\bibinfo {volume} {147}},\
  \bibinfo {pages} {212} (\bibinfo {year} {1934})}\BibitemShut {NoStop}%
\bibitem [{\citenamefont {Chu}\ and\ \citenamefont {Apfel}(1982)}]{Chu1982}%
  \BibitemOpen
  \bibfield  {author} {\bibinfo {author} {\bibfnamefont {B.-T.}\ \bibnamefont
  {Chu}}\ and\ \bibinfo {author} {\bibfnamefont {R.~E.}\ \bibnamefont
  {Apfel}},\ }\bibfield  {title} {\bibinfo {title} {{\textit{Acoustic radiation
  pressure produced by a beam of sound}}},\ }\href
  {https://doi.org/10.1121/1.388660} {\bibfield  {journal} {\bibinfo  {journal}
  {J. Acoust. Soc. Am.}\ }\textbf {\bibinfo {volume} {72}},\ \bibinfo {pages}
  {1673–1687} (\bibinfo {year} {1982})}\BibitemShut {NoStop}%
\bibitem [{\citenamefont {Borgnis}(1953)}]{Borgnis1953}%
  \BibitemOpen
  \bibfield  {author} {\bibinfo {author} {\bibfnamefont {F.~E.}\ \bibnamefont
  {Borgnis}},\ }\bibfield  {title} {\bibinfo {title} {{\textit{Acoustic
  Radiation Pressure of Plane Compressional Waves}}},\ }\href
  {https://doi.org/10.1103/revmodphys.25.653} {\bibfield  {journal} {\bibinfo
  {journal} {Rev. Mod. Phys.}\ }\textbf {\bibinfo {volume} {25}},\ \bibinfo
  {pages} {653–664} (\bibinfo {year} {1953})}\BibitemShut {NoStop}%
\bibitem [{\citenamefont {Hasegawa}\ \emph {et~al.}(2000)\citenamefont
  {Hasegawa}, \citenamefont {Kido}, \citenamefont {Iizuka},\ and\ \citenamefont
  {Matsuoka}}]{Hasegawa2000}%
  \BibitemOpen
  \bibfield  {author} {\bibinfo {author} {\bibfnamefont {T.}~\bibnamefont
  {Hasegawa}}, \bibinfo {author} {\bibfnamefont {T.}~\bibnamefont {Kido}},
  \bibinfo {author} {\bibfnamefont {T.}~\bibnamefont {Iizuka}},\ and\ \bibinfo
  {author} {\bibfnamefont {C.}~\bibnamefont {Matsuoka}},\ }\bibfield  {title}
  {\bibinfo {title} {{\textit{A general theory of Rayleigh and Langevin
  radiation pressures}}},\ }\href {https://doi.org/10.1250/ast.21.145}
  {\bibfield  {journal} {\bibinfo  {journal} {J. Acoust. Soc. Jpn. (E)}\
  }\textbf {\bibinfo {volume} {21}},\ \bibinfo {pages} {145–152} (\bibinfo
  {year} {2000})}\BibitemShut {NoStop}%
\bibitem [{\citenamefont {Karlsen}\ \emph {et~al.}(2016)\citenamefont
  {Karlsen}, \citenamefont {Augustsson},\ and\ \citenamefont
  {Bruus}}]{Karlsen2016}%
  \BibitemOpen
  \bibfield  {author} {\bibinfo {author} {\bibfnamefont {J.~T.}\ \bibnamefont
  {Karlsen}}, \bibinfo {author} {\bibfnamefont {P.}~\bibnamefont
  {Augustsson}},\ and\ \bibinfo {author} {\bibfnamefont {H.}~\bibnamefont
  {Bruus}},\ }\bibfield  {title} {\bibinfo {title} {{\textit{Acoustic Force
  Density Acting on Inhomogeneous Fluids in Acoustic Fields}}},\ }\href
  {http://dx.doi.org/10.1103/PhysRevLett.117.114504} {\bibfield  {journal}
  {\bibinfo  {journal} {Phys. Rev. Lett.}\ }\textbf {\bibinfo {volume} {117}}
  (\bibinfo {year} {2016})}\BibitemShut {NoStop}%
\bibitem [{\citenamefont {Rajendran}\ \emph {et~al.}(2023)\citenamefont
  {Rajendran}, \citenamefont {Aravind~Ram},\ and\ \citenamefont
  {Subramani}}]{Rajendran2023}%
  \BibitemOpen
  \bibfield  {author} {\bibinfo {author} {\bibfnamefont {V.~K.}\ \bibnamefont
  {Rajendran}}, \bibinfo {author} {\bibfnamefont {S.}~\bibnamefont
  {Aravind~Ram}},\ and\ \bibinfo {author} {\bibfnamefont {K.}~\bibnamefont
  {Subramani}},\ }\bibfield  {title} {\bibinfo {title} {{\textit{On the
  stability of inhomogeneous fluids under acoustic fields}}},\ }\href
  {http://dx.doi.org/10.1017/jfm.2023.371} {\bibfield  {journal} {\bibinfo
  {journal} {J. Fluid Mech.}\ }\textbf {\bibinfo {volume} {964}} (\bibinfo
  {year} {2023})}\BibitemShut {NoStop}%
\bibitem [{\citenamefont {Biwersi}\ \emph {et~al.}(2000)\citenamefont
  {Biwersi}, \citenamefont {Manceau},\ and\ \citenamefont
  {Bastien}}]{Biwersi2000}%
  \BibitemOpen
  \bibfield  {author} {\bibinfo {author} {\bibfnamefont {S.}~\bibnamefont
  {Biwersi}}, \bibinfo {author} {\bibfnamefont {J.~F.}\ \bibnamefont
  {Manceau}},\ and\ \bibinfo {author} {\bibfnamefont {F.}~\bibnamefont
  {Bastien}},\ }\bibfield  {title} {\bibinfo {title} {{\textit{Displacement of
  droplets and deformation of thin liquid layers using flexural vibrations of
  structures. Influence of acoustic radiation pressure}}},\ }\href
  {https://doi.org/10.1121/1.428566} {\bibfield  {journal} {\bibinfo  {journal}
  {J. Acoust. Soc. Am.}\ }\textbf {\bibinfo {volume} {107}},\ \bibinfo {pages}
  {661–664} (\bibinfo {year} {2000})}\BibitemShut {NoStop}%
\bibitem [{\citenamefont {Alzuaga}\ \emph {et~al.}(2005)\citenamefont
  {Alzuaga}, \citenamefont {Manceau},\ and\ \citenamefont
  {Bastien}}]{Alzuaga2005}%
  \BibitemOpen
  \bibfield  {author} {\bibinfo {author} {\bibfnamefont {S.}~\bibnamefont
  {Alzuaga}}, \bibinfo {author} {\bibfnamefont {J.-F.}\ \bibnamefont
  {Manceau}},\ and\ \bibinfo {author} {\bibfnamefont {F.}~\bibnamefont
  {Bastien}},\ }\bibfield  {title} {\bibinfo {title} {{\textit{Motion of
  droplets on solid surface using acoustic radiation pressure}}},\ }\href
  {https://doi.org/10.1016/j.jsv.2004.02.020} {\bibfield  {journal} {\bibinfo
  {journal} {J. Sound Vib.}\ }\textbf {\bibinfo {volume} {282}},\ \bibinfo
  {pages} {151–162} (\bibinfo {year} {2005})}\BibitemShut {NoStop}%
\bibitem [{\citenamefont {Issenmann}\ \emph {et~al.}(2006)\citenamefont
  {Issenmann}, \citenamefont {Wunenburger}, \citenamefont {Manneville},\ and\
  \citenamefont {Delville}}]{Issenmann2006}%
  \BibitemOpen
  \bibfield  {author} {\bibinfo {author} {\bibfnamefont {B.}~\bibnamefont
  {Issenmann}}, \bibinfo {author} {\bibfnamefont {R.}~\bibnamefont
  {Wunenburger}}, \bibinfo {author} {\bibfnamefont {S.}~\bibnamefont
  {Manneville}},\ and\ \bibinfo {author} {\bibfnamefont {J.-P.}\ \bibnamefont
  {Delville}},\ }\bibfield  {title} {\bibinfo {title} {{\textit{Bistability of
  a Compliant Cavity Induced by Acoustic Radiation Pressure}}},\ }\href
  {http://dx.doi.org/10.1103/PhysRevLett.97.074502} {\bibfield  {journal}
  {\bibinfo  {journal} {Phys. Rev. Lett.}\ }\textbf {\bibinfo {volume} {97}}
  (\bibinfo {year} {2006})}\BibitemShut {NoStop}%
\bibitem [{\citenamefont {Rajendran}\ \emph {et~al.}(2022)\citenamefont
  {Rajendran}, \citenamefont {Jayakumar}, \citenamefont {Azharudeen},\ and\
  \citenamefont {Subramani}}]{Rajendran2022}%
  \BibitemOpen
  \bibfield  {author} {\bibinfo {author} {\bibfnamefont {V.~K.}\ \bibnamefont
  {Rajendran}}, \bibinfo {author} {\bibfnamefont {S.}~\bibnamefont
  {Jayakumar}}, \bibinfo {author} {\bibfnamefont {M.}~\bibnamefont
  {Azharudeen}},\ and\ \bibinfo {author} {\bibfnamefont {K.}~\bibnamefont
  {Subramani}},\ }\bibfield  {title} {\bibinfo {title} {{\textit{Theory of
  nonlinear acoustic forces acting on inhomogeneous fluids}}},\ }\href
  {http://dx.doi.org/10.1017/jfm.2022.257} {\bibfield  {journal} {\bibinfo
  {journal} {J. Fluid Mech.}\ }\textbf {\bibinfo {volume} {940}} (\bibinfo
  {year} {2022})}\BibitemShut {NoStop}%
\bibitem [{\citenamefont {Rezk}\ \emph {et~al.}(2012)\citenamefont {Rezk},
  \citenamefont {Manor}, \citenamefont {Friend},\ and\ \citenamefont
  {Yeo}}]{Rezk2012}%
  \BibitemOpen
  \bibfield  {author} {\bibinfo {author} {\bibfnamefont {A.~R.}\ \bibnamefont
  {Rezk}}, \bibinfo {author} {\bibfnamefont {O.}~\bibnamefont {Manor}},
  \bibinfo {author} {\bibfnamefont {J.~R.}\ \bibnamefont {Friend}},\ and\
  \bibinfo {author} {\bibfnamefont {L.~Y.}\ \bibnamefont {Yeo}},\ }\bibfield
  {title} {\bibinfo {title} {{\textit{Unique fingering instabilities and
  soliton-like wave propagation in thin acoustowetting films}}},\ }\href
  {http://dx.doi.org/10.1038/ncomms2168} {\bibfield  {journal} {\bibinfo
  {journal} {Nat. Commun.}\ }\textbf {\bibinfo {volume} {3}} (\bibinfo {year}
  {2012})}\BibitemShut {NoStop}%
\bibitem [{\citenamefont {Rezk}\ \emph {et~al.}(2014)\citenamefont {Rezk},
  \citenamefont {Manor}, \citenamefont {Yeo},\ and\ \citenamefont
  {Friend}}]{Rezk2014}%
  \BibitemOpen
  \bibfield  {author} {\bibinfo {author} {\bibfnamefont {A.~R.}\ \bibnamefont
  {Rezk}}, \bibinfo {author} {\bibfnamefont {O.}~\bibnamefont {Manor}},
  \bibinfo {author} {\bibfnamefont {L.~Y.}\ \bibnamefont {Yeo}},\ and\ \bibinfo
  {author} {\bibfnamefont {J.~R.}\ \bibnamefont {Friend}},\ }\bibfield  {title}
  {\bibinfo {title} {{\textit{Double flow reversal in thin liquid films driven
  by megahertz-order surface vibration}}},\ }\href
  {https://doi.org/10.1098/rspa.2013.0765} {\bibfield  {journal} {\bibinfo
  {journal} {Proc. R. Soc. A}\ }\textbf {\bibinfo {volume} {470}},\ \bibinfo
  {pages} {20130765} (\bibinfo {year} {2014})}\BibitemShut {NoStop}%
\bibitem [{\citenamefont {Manor}\ \emph {et~al.}(2015)\citenamefont {Manor},
  \citenamefont {Rezk}, \citenamefont {Friend},\ and\ \citenamefont
  {Yeo}}]{Manor2015}%
  \BibitemOpen
  \bibfield  {author} {\bibinfo {author} {\bibfnamefont {O.}~\bibnamefont
  {Manor}}, \bibinfo {author} {\bibfnamefont {A.~R.}\ \bibnamefont {Rezk}},
  \bibinfo {author} {\bibfnamefont {J.~R.}\ \bibnamefont {Friend}},\ and\
  \bibinfo {author} {\bibfnamefont {L.~Y.}\ \bibnamefont {Yeo}},\ }\bibfield
  {title} {\bibinfo {title} {{\textit{Dynamics of liquid films exposed to
  high-frequency surface vibration}}},\ }\href
  {http://dx.doi.org/10.1103/PhysRevE.91.053015} {\bibfield  {journal}
  {\bibinfo  {journal} {Phys. Rev. E}\ }\textbf {\bibinfo {volume} {91}}
  (\bibinfo {year} {2015})}\BibitemShut {NoStop}%
\bibitem [{\citenamefont {Altshuler}\ and\ \citenamefont
  {Manor}(2015)}]{Altshuler2015}%
  \BibitemOpen
  \bibfield  {author} {\bibinfo {author} {\bibfnamefont {G.}~\bibnamefont
  {Altshuler}}\ and\ \bibinfo {author} {\bibfnamefont {O.}~\bibnamefont
  {Manor}},\ }\bibfield  {title} {\bibinfo {title} {{\textit{Spreading dynamics
  of a partially wetting water film atop a MHz substrate vibration}}},\ }\href
  {http://dx.doi.org/10.1063/1.4932086} {\bibfield  {journal} {\bibinfo
  {journal} {Phys. Fluids}\ }\textbf {\bibinfo {volume} {27}} (\bibinfo {year}
  {2015})}\BibitemShut {NoStop}%
\bibitem [{\citenamefont {Altshuler}\ and\ \citenamefont
  {Manor}(2016)}]{Altshuler2016}%
  \BibitemOpen
  \bibfield  {author} {\bibinfo {author} {\bibfnamefont {G.}~\bibnamefont
  {Altshuler}}\ and\ \bibinfo {author} {\bibfnamefont {O.}~\bibnamefont
  {Manor}},\ }\bibfield  {title} {\bibinfo {title} {{\textit{Free films of a
  partially wetting liquid under the influence of a propagating MHz surface
  acoustic wave}}},\ }\href {http://dx.doi.org/10.1063/1.4955414} {\bibfield
  {journal} {\bibinfo  {journal} {Phys. Fluids}\ }\textbf {\bibinfo {volume}
  {28}} (\bibinfo {year} {2016})}\BibitemShut {NoStop}%
\bibitem [{\citenamefont {Horesh}\ \emph {et~al.}(2019)\citenamefont {Horesh},
  \citenamefont {Khaikin}, \citenamefont {Karnilaw}, \citenamefont {Zigelman},\
  and\ \citenamefont {Manor}}]{Horesh2019}%
  \BibitemOpen
  \bibfield  {author} {\bibinfo {author} {\bibfnamefont {A.}~\bibnamefont
  {Horesh}}, \bibinfo {author} {\bibfnamefont {D.}~\bibnamefont {Khaikin}},
  \bibinfo {author} {\bibfnamefont {M.}~\bibnamefont {Karnilaw}}, \bibinfo
  {author} {\bibfnamefont {A.}~\bibnamefont {Zigelman}},\ and\ \bibinfo
  {author} {\bibfnamefont {O.}~\bibnamefont {Manor}},\ }\bibfield  {title}
  {\bibinfo {title} {{\textit{Acoustogravitational balance in climbing
  films}}},\ }\href {http://dx.doi.org/10.1103/PhysRevFluids.4.022001}
  {\bibfield  {journal} {\bibinfo  {journal} {Phys. Rev. Fluids}\ }\textbf
  {\bibinfo {volume} {4}} (\bibinfo {year} {2019})}\BibitemShut {NoStop}%
\bibitem [{\citenamefont {Nu\ss{}baumer}\ \emph {et~al.}(2008)\citenamefont
  {Nu\ss{}baumer}, \citenamefont {Bittner},\ and\ \citenamefont
  {Janke}}]{Nussbaumer2008}%
  \BibitemOpen
  \bibfield  {author} {\bibinfo {author} {\bibfnamefont {A.}~\bibnamefont
  {Nu\ss{}baumer}}, \bibinfo {author} {\bibfnamefont {E.}~\bibnamefont
  {Bittner}},\ and\ \bibinfo {author} {\bibfnamefont {W.}~\bibnamefont
  {Janke}},\ }\bibfield  {title} {\bibinfo {title} {{\textit{Monte Carlo study
  of the droplet formation-dissolution transition on different two-dimensional
  lattices}}},\ }\href {https://link.aps.org/doi/10.1103/PhysRevE.77.041109}
  {\bibfield  {journal} {\bibinfo  {journal} {Phys. Rev. E}\ }\textbf {\bibinfo
  {volume} {77}},\ \bibinfo {pages} {041109} (\bibinfo {year}
  {2008})}\BibitemShut {NoStop}%
\bibitem [{\citenamefont {Shiokawa}\ \emph {et~al.}(1989)\citenamefont
  {Shiokawa}, \citenamefont {Matsui},\ and\ \citenamefont
  {Ueda}}]{Shiokawa1989}%
  \BibitemOpen
  \bibfield  {author} {\bibinfo {author} {\bibfnamefont {S.}~\bibnamefont
  {Shiokawa}}, \bibinfo {author} {\bibfnamefont {Y.}~\bibnamefont {Matsui}},\
  and\ \bibinfo {author} {\bibfnamefont {T.}~\bibnamefont {Ueda}},\ }\bibfield
  {title} {\bibinfo {title} {{\textit{Liquid streaming and droplet formation
  caused by leaky Rayleigh waves}}},\ }in\ \href
  {https://doi.org/10.1109/ULTSYM.1989.67063} {\bibinfo {booktitle}
  {Proceedings., IEEE Ultrasonics Symposium,}}\ (\bibinfo {year} {1989})\ pp.\
  \bibinfo {pages} {643--646 vol.1}\BibitemShut {NoStop}%
\bibitem [{\citenamefont {Shiokawa}\ and\ \citenamefont
  {Matsui}(1994)}]{Shiokawa1994}%
  \BibitemOpen
  \bibfield  {author} {\bibinfo {author} {\bibfnamefont {S.}~\bibnamefont
  {Shiokawa}}\ and\ \bibinfo {author} {\bibfnamefont {Y.}~\bibnamefont
  {Matsui}},\ }\bibfield  {title} {\bibinfo {title} {{\textit{The Dynamics of
  SAW Streaming and its Application to Fluid Devices}}},\ }\href
  {http://dx.doi.org/10.1557/PROC-360-53} {\bibfield  {journal} {\bibinfo
  {journal} {MRS Proceedings}\ }\textbf {\bibinfo {volume} {360}} (\bibinfo
  {year} {1994})}\BibitemShut {NoStop}%
\bibitem [{\citenamefont {Peters}\ and\ \citenamefont
  {Arabali}(2013)}]{Peters2013}%
  \BibitemOpen
  \bibfield  {author} {\bibinfo {author} {\bibfnamefont {F.}~\bibnamefont
  {Peters}}\ and\ \bibinfo {author} {\bibfnamefont {D.}~\bibnamefont
  {Arabali}},\ }\bibfield  {title} {\bibinfo {title} {{\textit{Interfacial
  tension between oil and water measured with a modified contour method}}},\
  }\href {https://doi.org/https://doi.org/10.1016/j.colsurfa.2013.03.010}
  {\bibfield  {journal} {\bibinfo  {journal} {Colloids Surf. A}\ }\textbf
  {\bibinfo {volume} {426}},\ \bibinfo {pages} {1} (\bibinfo {year}
  {2013})}\BibitemShut {NoStop}%
\end{thebibliography}%

\end{document}